\documentclass{article}
\usepackage{arxiv}

\usepackage[dvipsnames,table]{xcolor}
\usepackage[utf8]{inputenc} 
\usepackage[T1]{fontenc}    
\usepackage{hyperref}       
\usepackage{url}            
\usepackage{booktabs}       
\usepackage{amsfonts}       
\usepackage{nicefrac}       
\usepackage{microtype}      
\usepackage{lipsum}		
\usepackage{graphicx}
\usepackage{natbib}
\usepackage{doi}

\usepackage{adjustbox}
\usepackage{makecell}
\usepackage{verbatim}
\usepackage{tabularx}
\usepackage{amsmath}
\usepackage{listings}
\usepackage{booktabs}
\usepackage{hhline}
\usepackage{multirow}
\usepackage{subcaption}

\usepackage{xparse} 

\usepackage{comment}

\usepackage{graphicx}%
\usepackage{tikz}
\usepackage{svg}

\usepackage{hyperref}%
\hypersetup{
    colorlinks=true,
    citecolor=blue,
    linkcolor=blue,
    filecolor=blue,      
    urlcolor=blue,
    breaklinks=true,
    pdfpagemode=FullScreen,
    }
\usepackage{makecell}
\usepackage{longtable, array}
\usepackage{array}
\usepackage{multirow}
\usepackage{enumitem}

\usepackage{rotating}
\usepackage{xstring}

\usepackage[T1]{fontenc} 
\usepackage{lmodern}     
\usepackage{kpfonts}
\usepackage{scalefnt}

\usepackage{geometry}

\newcommand{\authorORCID}[1]{%
    \href{#1}{\includegraphics[width=10pt]{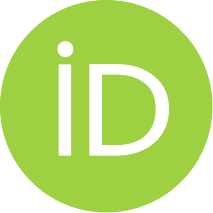}}
}

\def\authorDEIB{%
    Department of Electronics,\\
    Information, and Bioengineering\\
	Politecnico di Milano\\
	Milan, 20133, Italy
}
\def\authorPavia{%
    Department of Industrial,\\
    and Information Engineering\\
	University of Pavia\\
	Pavia, 27100, Italy
}

\author{%
        \authorORCID{https://orcid.org/0000-0002-3365-580X} Andrea Moglia \dag \\
        \authorDEIB \\
        \texttt{andrea.moglia@polimi.it} \\
	\And
        \authorORCID{https://orcid.org/0009-0002-3434-0559} Matteo Leccardi \dag \\
        \authorDEIB \\
        \texttt{matteo.leccardi@polimi.it} \\
    \And
        \authorORCID{https://orcid.org/0000-0001-9394-4913} Matteo Cavicchioli \\
        \authorDEIB \\
        \texttt{matteo.cavicchioli@polimi.it} \\
    \And
        Alice Maccarini \\
        \authorPavia \\
        \hspace{3mm} \texttt{alice.maccarini@unipv.it} \hspace{3mm} \\
    \And
        \authorORCID{https://orcid.org/0000-0001-6557-2120} Marco Marcon \\
        \authorDEIB \\
        \texttt{marco.marcon@polimi.it} \\
    \And
        \authorORCID{https://orcid.org/0000-0002-6276-6314} Luca Mainardi \\
        \authorDEIB \\
        \texttt{luca.mainardi@polimi.it} \\
    \And
    \authorORCID{https://orcid.org/0000-0003-3995-8673} Pietro Cerveri \\
	\authorPavia \\
	\texttt{pietro.cerveri@unipv.it} \\
    \authorDEIB \\
    \texttt{pietro.cerveri@polimi.it} \\
}

\title{Generalist Models in Medical Image Segmentation:\\A Survey and Performance Comparison\\with Task-Specific Approaches}

\renewcommand{\shorttitle}{
Generalist Models in Medical Image Segmentation: A Survey and Performance Comparison\\with Task-Specific Approaches\vspace{0.5mm}
}

\hypersetup{
pdftitle={Generalist Models in Medical Image Segmentation: A Survey and Performance Comparison with Task-Specific Approaches},
pdfsubject={cs.AI, cs.CV},
pdfauthor={Andrea Moglia, Matteo Leccardi, Matteo Cavicchioli, Alice Maccarini, Marco Marcon, Luca Mainardi, Pietro Cerveri},
pdfkeywords={Medical Image Segmentation, Foundation Models, Segment Anything Models, U-Net},
}

\begin{document}

\maketitle

\begin{center}
\vspace{2cm}
\parbox{15cm}{%
    \rule[2mm]{15cm}{0.1mm}\\
    \dag: Equally contributing authors.
}   
\end{center}

\newcommand{\warning}[1]{\textcolor{Orange}{#1}}

\definecolor{colorCodeUnavailable}{HTML}{ffcccc}
\definecolor{colorTablePapersSpecializedHeaderClear}{HTML}{b1ccb6}
\definecolor{colorTablePapersSpecializedHeaderDark}{HTML}{a4bfa9}
\definecolor{colorTablePapersSpecializedRowsPattern}{HTML}{e6ffe6}
\definecolor{colorTablePapersFoundationalHeaderClear}{HTML}{ffd9b3}
\definecolor{colorTablePapersFoundationalHeaderDark}{HTML}{ffcc99}
\definecolor{colorTablePapersFoundationalRowsPattern}{HTML}{ffe6cc}
\definecolor{colorTableResultsOverviewPrimary}{HTML}{c7f3ff}
\definecolor{colorTableResultsOverviewBest}{HTML}{e8bdff}


\newcolumntype{C}[1]{>{\centering\let\newline\\\arraybackslash\hspace{0pt}}m{#1}}

\setlength{\tabcolsep}{1.8pt} 
\renewcommand{\arraystretch}{2.2} 

\newcommand{\tablenoteswidth}{15cm}
\newcommand{\tablefootervspace}{5pt}


\newcommand{\makeTableHeaderContinued}[1]{
    \hiderowcolors
    \multicolumn{#1}{l}{
        \vspace{\tablefootervspace}~~~~ $\rightarrow$ continued
    } \\
    \showrowcolors
}


\newcommand{\makeTableOtherFooters}[1]{
    \hiderowcolors
    \multicolumn{#1}{r}{\vspace{\tablefootervspace} continues $\rightarrow$~~~~} \\
    \showrowcolors
}

\newcommand{\makeTableLastFooter}[2]{ 
    \hiderowcolors
    \hline \hline
    \multicolumn{#1}{l}{~~~~\parbox{\tablenoteswidth}{
        \vspace{\tablefootervspace}
        #2
    }} \\
    \showrowcolors
}

\newcommand{\outOfTableComments}[1]{
    \textcolor{black!70}{\vspace{-35pt}
        #1
    }
}

\clearpage

~\\
~\\
\begin{abstract}

Following the successful paradigm shift of large language models, leveraging pre-training on a massive corpus of data and fine-tuning on different downstream tasks, generalist models have made their foray into computer vision. The introduction of Segment Anything Model (SAM) set a milestone on segmentation of natural images, inspiring the design of a multitude of architectures for medical image segmentation. In this survey we offer a comprehensive and in-depth investigation on generalist models for medical image segmentation. We start with an introduction on the fundamentals concepts underpinning their development. Then, we provide a taxonomy on the different declinations of SAM in terms of zero-shot, few-shot, fine-tuning, adapters, on the recent SAM 2, on other innovative models trained on images alone, and others trained on both text and images. We thoroughly analyze their performances at the level of both primary research and best-in-literature, followed by a rigorous comparison with the state-of-the-art task-specific models. We emphasize the need to address challenges in terms of compliance with regulatory frameworks, privacy and security laws, budget, and trustworthy artificial intelligence (AI). Finally, we share our perspective on future directions concerning synthetic data, early fusion, lessons learnt from generalist models in natural language processing, agentic AI and physical AI, and clinical translation.

\end{abstract}

~\\
~\\


\section{Introduction}
\label{sec:introduction}

Biomedical image segmentation has undergone remarkable transformations over the past decade, evolving from simple convolutional neural network (CNN) approaches to sophisticated deep learning architectures. The field initially witnessed a significant breakthrough with the introduction of U-Net by \citet{ronneberger2015u} which revolutionized medical image segmentation and set a milestone in the field with its unique encoder-decoder architecture and skip connections. U-Net provided a robust framework for semantic segmentation, particularly in biomedical imaging, enabling precise delineation of anatomical structures with remarkable accuracy, and was later on adopted in many and diverse computer-vision tasks.
The subsequent emergence of transformer-based architectures marked another pivotal moment in image analysis and segmentation. Initially developed for natural language processing, transformers rapidly transitioned into computer vision, challenging traditional CNN paradigms. The seminal ``Attention is All You Need'' paper from Google by \citet{vaswani2017attention} and ``An Image is worth 16x16 Words'' by \citet{dosovitskiy2020image} laid the groundwork for architectural innovations that would subsequently transform medical imaging. Shortly after the Swin Transformer introduced by \citet{9710580Swin,9879380SwinV2} proposed a significant advancement by creating a hierarchical vision transformer (ViT) that could efficiently process images with improved computational complexity.
Parallel to transformer development, CNN architectures continued evolving. ConvNeXt from Meta by \citet{10205236ConvNeXtv2} re-imagined CNNs by incorporating transformer-like design principles, such as an inverted bottleneck in each block, separable depth-wise convolutions and wide convolutional kernels. Its successor, ConvNeXt V2 by \citet{10205236ConvNeXtv2} further refined these approaches by introducing architectural modifications and advanced unsupervised pre-training strategies such as masked-image reconstruction, thus peeking into (but not entering) the realm of foundation models.
The progression of these architectures and pre-training frameworks on increasingly larger datasets culminated in the development of foundation models.

Foundation models were defined as models trained on large-scale data, generally using self-supervised learning, that can be adapted, e.g., by fine-tuning, to a wide range of downstream tasks \citep{bommasani2021opportunities}. Others used the term universal models for those approaches that can be characterized by transferability and ability to handle multiple tasks, without the need of fine-tuning 33
30
\citep{10510478}.
As both categories target the concept of generalist AI, the expression of generalist models can be used when referring to them.

Generalist models in computer vision emerged following the successful paradigm shift of large language models (LLMs) like Bidirectional Encoder Representations from Transformers (BERT) - a transformer encoder - and Generative Pre-trained Transformers (GPT) - a transformer decoder - which demonstrated that self-supervised pretraining on vast datasets could lead to highly transferable representations and that scaling of models and resources was key for learning meaningful features for generalization \citep{devlin2019bert, radford2018improving}.

\begin{figure*}[!ht]
  \centering
    \includegraphics[width=1.0\linewidth]{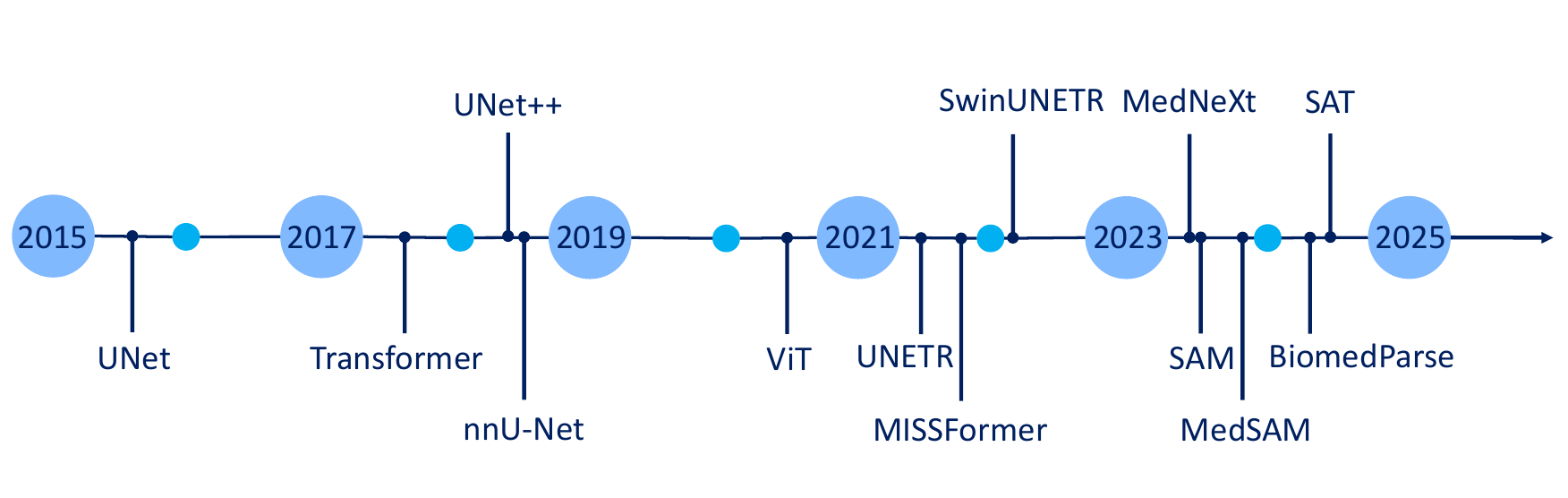}
    \caption{Timeline of key developments of generalist and task-specific models for medical image segmentation. The time is referred to the publication date of the primary work, e.g., in arXiv.}
    \label{fig:cronology}
\end{figure*}

This paradigm shift first materialized in computer vision through two major self-supervised pre-training approaches: Contrastive Language–Image Pre-training (CLIP) from OpenAI by \citet{clip2021openai} which showed how training on image-text pairs could create robust visual representations, and self-supervised distillation with no labels (DINO) from Meta by \citet{dino_cvf} which introduced self-supervised learning for pre-training ViT. 
The field then evolved toward more generalizable architectures like SEEM (Segment Everything Everywhere All at Once) by \citet{seem_arxiv,seem_neurips}, Segment Anything Model (SAM) by \citet{kirillov2023segany}, and SAM 2 by \citet{ravi2024sam2}.
This progression from language to vision generalist models has now reached medical imaging community which is transitioning from  supervised, task-specific models with limited generalization capabilities towards the new pre-train-and-adapt paradigms \citep{moor2023foundation}. Some notable examples include \textbf{adaptations of SAM}, e.g., MedSAM \citep{MedSAM_arxiv, MedSAM_nature} and Medical SAM 2 \citep{zhu2024medicalsam2segment}, and \textbf{native generalist models} such as Microsoft's BiomedParse \citep{zhao2024biomedparsebiomedicalfoundationmodel}.
The timeline of the most significant development in the field is displayed in Fig. \ref{fig:cronology}.

Contrary to prevailing discourse that often polarizes discussions around transformers versus CNN architectures, our research posits that the fundamental dichotomy in contemporary medical image segmentation rather lies between generalist and task-specific models. Generalist models, pre-trained on millions of multi-modal medical images, exhibited remarkable adaptability and consistent performance across diverse anatomical regions. They transcend the limitations of specialized, task-specific deep learning models that traditionally focus on narrow datasets, limited anatomical contexts, single task and just one imaging modality.
The emergence of generalist models in medical imaging represents more than a technological advancement; it signifies a philosophical shift in the AI-based approach. By leveraging extensive pre-training strategies and incorporating multi-modal learning techniques, these models challenge the conventional wisdom of over-specialization. They demonstrate the potential to generalize learning across complex medical imaging tasks, reducing the need for extensive, task-specific annotated datasets.





\section{Comparison with other surveys and our contributions}
\label{sec:comparison-surveys-contributions}


We extensively searched reviews published until March 31, 2025 on generalist models in medical image segmentation using PubMed, Web of Science, Scopus, IEEE Xplore, and arXiv.
We found 13 surveys, seven of which specifically focused on SAM and SAM 2 \citep{zhang2023comprehensive, zhang2024segment, zhang2024unleashing, ali2024review, jiaxing2025sam2, lee2024foundation, sun2024medical}. The review by \citet{gan2025review} provided an overview on different segmentation approaches from deep learning to generalist models like SAM based ones. \citet{li2024artificial} and \citet{liang2025vision} provided comprehensive reviews on medical imaging, including a few SAM based generalist models. \citet{he2024foundation} and  \citet{khan2025comprehensive} reviewed generalist models for the vast field of healthcare, with a limited focus on medical segmentation.  The review by \citet{bian2025artificial} partially addressed a comparison between task specific and generalist models for medical imaging segmentation. However, it did not provide details on the architectures from a technical point of view. Additionally, the performances comparison did not specify the dataset used and metrics values \citep{bian2025artificial}.\\
There is no published survey answering the following questions: 

\begin{enumerate}
    \item What are the performance gaps between generalist and state-of-the-art (SOTA) task-specific approaches for the medical image segmentation on the same dataset?
    \item Which is the best performing approach for a specific organ?
    \item What is the performance progress over time of both approaches?
    \item What are the challenges to overcome?
    \item What are directions for future research?
\end{enumerate}

By answering to these questions, in this survey we sought to contribute substantially to the ongoing discourse regarding the future of medical image segmentation and the transformative potential of generalist models, critically examining whether they truly represent a paradigm shift. Main contributions of this survey summarize as:

\begin{enumerate}
    \item we propose a unified and extensible taxonomy that integrates model architecture, fusion strategies, prompt modalities, and adaptation methods (zero-shot, few-shot, fine-tuning, PEFT), serving as a generalist reference framework for benchmarking generalist medical segmentation models;
    \item we perform a critical architectural dissection of the most advanced generalist models, identifying architectural invariants and bottlenecks that limit transferability and scalability in 3D medical imaging tasks;
    \item we construct a performance trajectory analysis by aggregating and aligning quantitative metrics across datasets, timepoints, and update versions, exposing the performance stagnation or acceleration patterns of generalist versus task-specific models;
    \item we establish a task-wise and organ-wise performance leaderboard, benchmarking generalist and specialist models under standardized protocols, and propose a robust statistical evaluation framework to quantify generalization gaps across anatomical domains and modalities;
    \item We analyze map regulatory, ethical, and practical deployment constraints, identifying unresolved challenges in adapting generalist segmentation models to real-world clinical settings, with a focus on explainability, interactivity, and human-in-the-loop dynamics;
    \item we consolidate and release the proposed taxonomy into a GitHub repository to foster reproducibility and model reproducibility auditing.
\end{enumerate}

In this survey we reviewed the published literature until April 2025 indexed by Google Scholar and arXiv using the search term \textit{generalist models medical image segmentation}. We also checked all references of the previously published surveys, above-mentioned, to retrieve further results. We included all the SOTA task-specific models used for comparison in the publications on generalist models. In order to expand the range of task-specific approaches, we recursively checked each respective publication to add other task-specific models. 
Our survey is focused on generalist models capable to process 3D radiological volumes, e.g., computed tomography (CT) and magnetic resonance imaging (MRI). Therefore, generalist approaches on 2D imaging modality, e.g., retinal images, were excluded.

\begin{figure*}[!ht]
    \begin{center}
    \includegraphics[width=\linewidth]{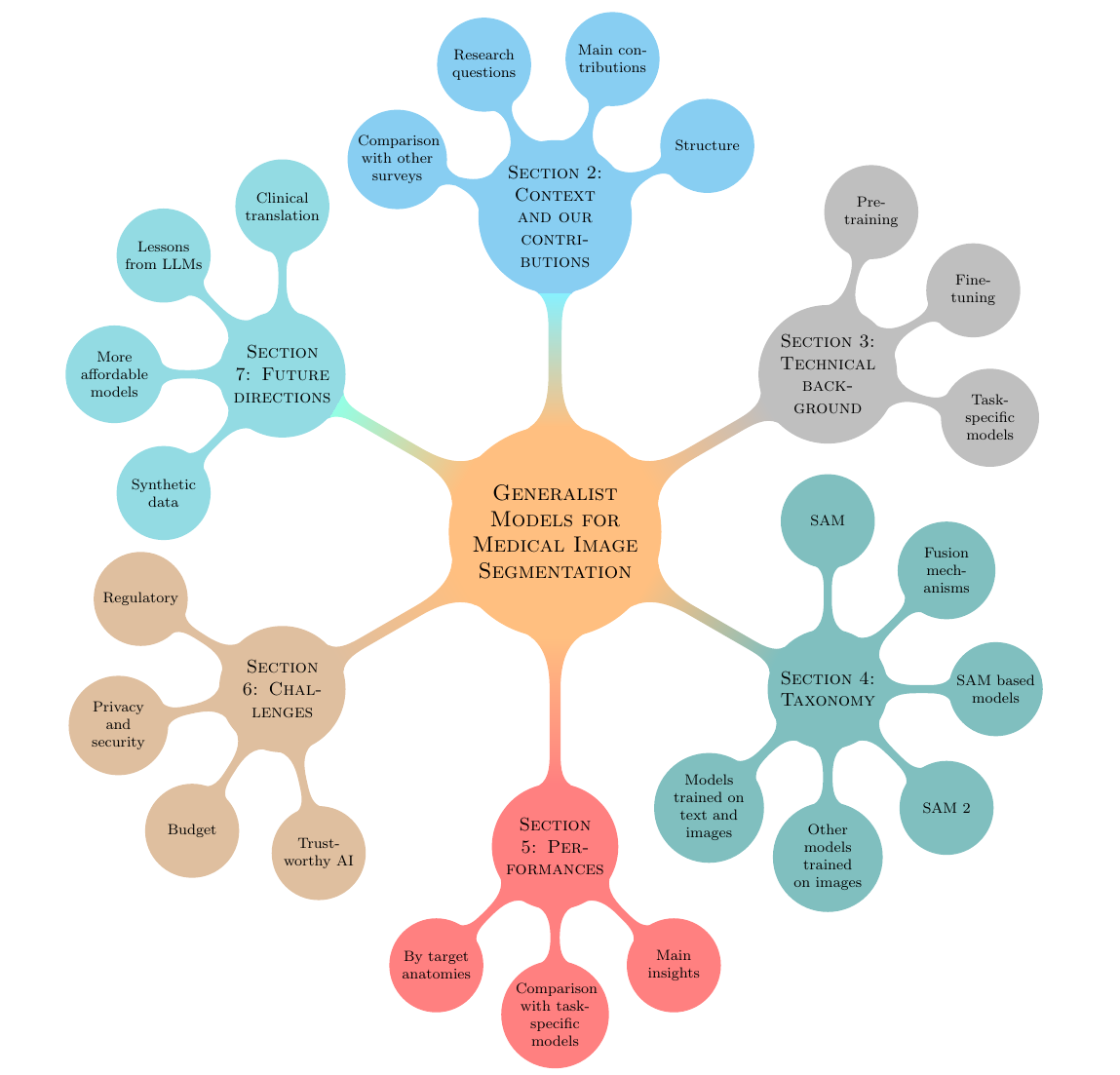}
    \caption{Outline of the survey.}
    \label{fig:outline}
    \end{center}
\end{figure*}

The outline of the survey is depicted in Fig. \ref{fig:outline}. In Section \ref{sec:background} we provided a technical background on generalist models and list the SOTA task-specific approaches for medical imaging segmentation. In Section \ref{sec:taxonomy} we classified the different generalist models. We grouped them in tables reporting details on the publication of the first and most recent version of each model, the size of the model in terms of number of parameters and floating point operations per second, and hardware resources during training. In Section \ref{sec:results} we compared the specialized and generalist models on datasets for different anatomical structures. We also reported the highest performances of each model in tabular form. In Sections \ref{sec:challenges} and \ref{sec:future-directions} we report current challenges and explore future directions. Finally, Section \ref{sec:conclusions} ends the survey.

\section{Technical background}
\label{sec:background}

\subsection{Pre-training}
\label{subsec:pre-training}

\textbf{Masked prediction:} masked-language model is an unsupervised pre-training objective introduced with BERT and consisting in predicting masked text tokens in a sentence \citep{devlin2019bert}. This approach was later adapted to computer vision, leading to masked image modeling objective to recover the original visual tokens after masking some image patches \citep{bao2021beit}. Masked image modeling has been used in transformer-based models, e.g., Bidirectional Encoder representation from Image Transformers (BEiT), ViT, and masked autoencoders (MAE) \citep{bao2021beit, dosovitskiy2020image, he2022masked}.

Since MAE mask all information of some tokens it can be regarded as a global masking approach. In contrast, \textbf{local masking} was introduced as a pre-training strategy consisting of masking only some channels of the tokens to help a network to reconstruct sharp details and learn better local context \citep{valanarasu2023disruptiveautoencodersleveraginglowlevel}.

In \textbf{image-text matching}, the purpose is to establish the correspondence between images and text, in particular find whether  image–text pairs match (positive pairs) or not (negative pairs) \citep{lu2025integrating}. Image and text representations are fused into higher level multimodal representations, followed by a softmax layer for classification. Since image-text matching only focuses on the global matching between images and text, without considering matching between image patches and text, it is not a suitable approach for downstream tasks like segmentation \citep{lu2025integrating}.

\textbf{Contrastive pre-training} exploits contrastive objective to connect text with images. It is best exemplified by CLIP, a method to optimize the vector representations of a vision encoder and a text encoder in an embedding space, ensuring that image–text pairs are aligned in a shared latent space \citep{clip2021openai}. The vision encoder can be a ResNet or a ViT or a ConvNeXt, while the text encoder is a transformer like BERT \citep{zhao2025clip}. The loss functions fosters matching image-text pairs, i.e, those with similar vector representations resulting in higher cosine similarity, while penalizes unmatching image-text-pairs, i.e. those with dissimilar vector representations, thus resulting in lower cosine similarity \citep{clip2021openai}. CLIP may seem similar to image-text matching. However, CLIP is focused on alignment in the representation space, while image-text matching fuse image and text representations into a shared visual-semantic embedding space. CLIP was applied successfully to 2D medical image segmentation from CT and MRI volumes, by taking 3D data as 2D slices, and 3D \citep{zhao2025clip}.

\textbf{Distillation}: Knowledge distillation transfers the knowledge from a model called teacher to a smaller one called student to reduce the computational costs and inference time, while keeping accuracy \citep{li2025vision}. In contrast to knowledge distillation where the teacher model is known a priori, in DINO the teacher is built from past iterations of the student model \citep{dino_cvf}. The student and teacher models have the same architecture. Stop-gradient is applied to the teacher, allowing gradients to flow only through the student network during training. The parameters of the teacher are updated by using exponential moving average applied to those of the student. This ensures that the teacher network is more stable than the student one, which in turn learns better representations by trying to match a the slowly evolving teacher \citep{dino_cvf}. DINOv2 \citep{dinov2_arxiv, oquab2024dinov} proposed an automatic pipeline to build a curated dataset of 142 million of images for self-supervised learning, based on DINO. A large ViT, pre-trained on this dataset, was then distilled into smaller ViT models outperforming the same small models trained from scratch \citep{oquab2024dinov}.

\begin{figure*}[!ht]
  \centering
    \includegraphics[width=0.5\linewidth]{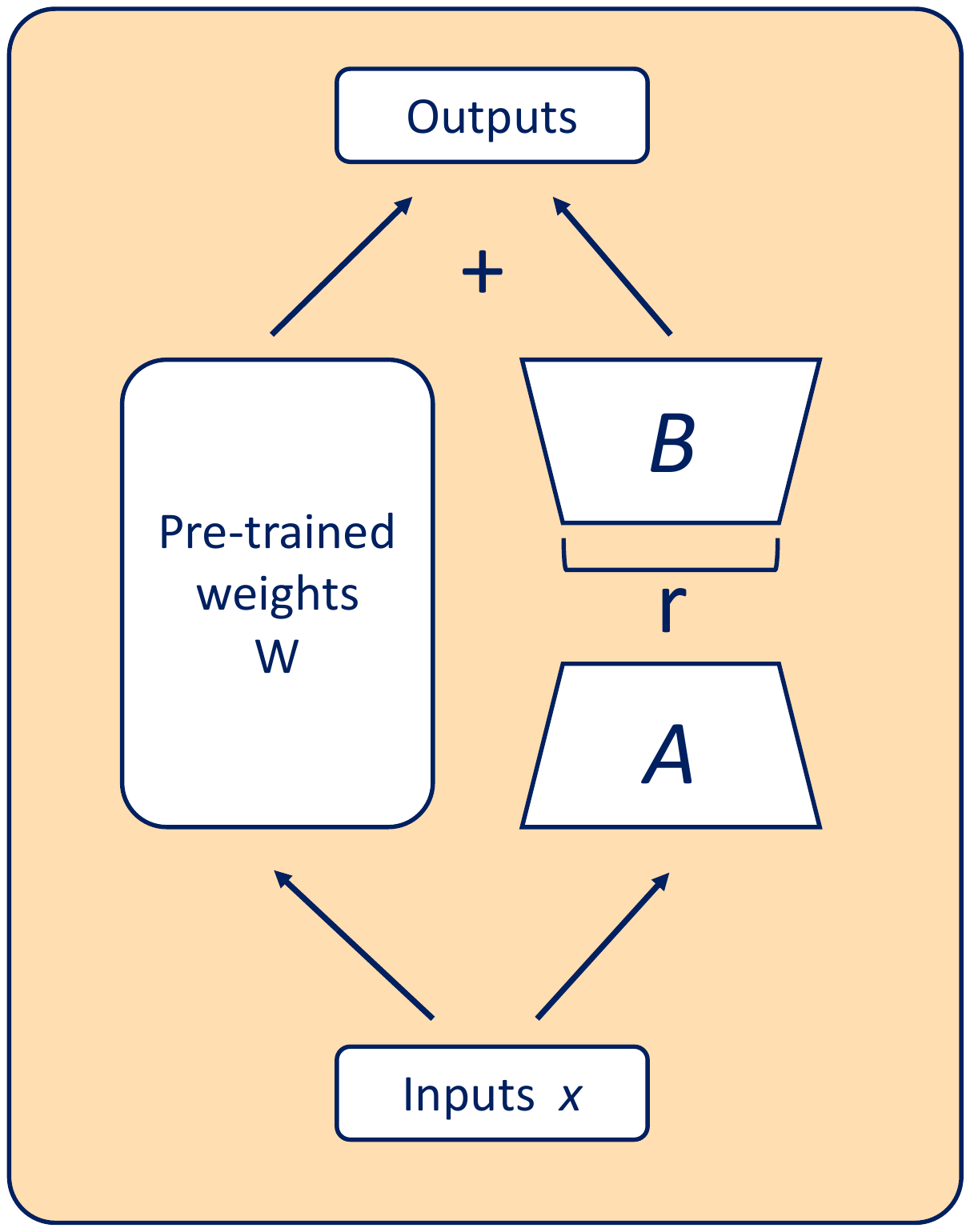}
    \caption{Architecture of LoRA. Image adapted from \citep{hu2022lora}.}
    \label{fig:lora}
\end{figure*}

\subsection{Fine-tuning}
\label{subsec:fine-training}
Transfer learning involves using a pre-trained model on a dataset and applying it to a specific task, e.g. for replacing the original head of the network with one for the particular task. Fine-tuning is a particular transfer learning approach where some layers of the pre-trained model are unfrozen to update the corresponding weights using an annotated dataset for the target task.

\textbf{Full fine-tuning} consists in updating all the weights. Since the size of the publicly available datasets in medical imaging is small, the typical approach is using a pre-trained model on a large dataset of natural images. However, with the emergence of generalist models and the consequent increase of model parameters, fine-tuning them on small datasets of medical images may lead to overfitting \citep{dutt2023parameter}. Therefore, effective and parameter-efficient methods of transfer learning become imperative.
Different fine-tuning strategies have been proposed:

\textbf{Parameter efficient fine-tuning (PEFT)} involves the update of a small number of parameters. PEFT can be performed by different strategies:
\begin{itemize}  
    \item \textbf{Low rank adaptation (LoRA)} reduces the number of parameters by applying a low-rank decomposition to the model weight updates \citep{hu2022lora}.
    Its workflow is depicted in Fig. \ref{fig:lora}. In this way, the small weights matrices \texttt{A} and \texttt{B}, downsized by the hyperparameter \texttt{r}, need to be adjusted and saved insated of the larger matrix \texttt{W}. This approach reduces computational and memory overhead while maintaining model performance. By decoupling the LoRA weight matrices from the pre-trained model, the latter can remains unchanged, thus enabling model customization without the need to store multiple full copies of the pre-trained model.
    \item \textbf{Adapters} add light modules to each transformer layer \citep{houlsby2019parameter}. An adapter consists of a linear down-projection, a nonlinear activation function, and a linear up-projection, together with a residual connection \citep{houlsby2019parameter}.
    \item \textbf{Prompt tuning} in computer vision follows the paradigm of NLP. Initial works in NLP treated prompts as prepended language instruction to the input so that a pre-trained model can be used for a specific task \citep{jia2022visual}. More recently, prompts were treated as task-specific continuous vectors to be optimized via gradient descent during fine-tuning \citep{jia2022visual}.  Visual prompt tuning was proposed in computer vision, by prepending small amount (1\%) of learnable task-specific parameters into the input of layers of ViT, while freezing the pre-trained transformer backbone \citep{jia2022visual, dosovitskiy2020image}. These prompts can be relearned for a new task while leaving the (frozen) backbone network task agnostic \citep{fischer2024prompt}. This allows to train a general purpose architecture once, and adapting to specific tasks with a minimal amount of parameters \citep{fischer2024prompt}. Prompt tuning performed very closely to full fine-tuning on segmentation of CT volumes of abdominal organs \citep{fischer2024prompt}.
    \item \textbf{Selective tuning} consists in selectively finetuning specific parameters of pre-trained models, e.g., by adjusting mean, variance, scale, and bias parameters in layer normalization or batch normalization layers \citep{lu2025integrating}. Although some works showed that this method can effectively adapt medical pre-trained models to new distributions without extensive parameter adjustments, the effectiveness of selective fine-tuning depends on the extent of the domain shift \citep{lu2025integrating}. 
\end{itemize}

\subsection{SOTA Task-specific Models}
\label{subsec:task-specific-approaches}

Task specific models for medical imaging segmentation were developed for a specific task, usually trained and tested on few datasets, achieving high performances. In this section we present the SOTA of 3D models which were used as benchmark in the reviewed publications on the generalist models. Their characteristics, and the links to the publications and code repositories are reported in Table \ref{table:models-specialized} of the appendix.

\subsubsection{UNet and other models based on local fusion}
\label{subsubsec:unet-variants-fusion}

The following architectures leveraged \textbf{local fusion} since they mostly employed convolutions to fuse semantic features of different scales locally, without accounting for global information.
\begin{itemize}
    \item \textbf{UNet} is a U-shape fully CNN with an encoder and a decoder. The encoder extracts features through convolutions, while the decoder restores the initial resolution of the input image through deconvolutions. UNet gradually \textbf{fuses} features by concatenating down-sampled features from the encoder with up-sampled features from the decoder through skip connections to improve the segmentation performance, especially for localization \citep{ronneberger2015u}. 
    \item \textbf{V-Net} extended UNet to process 3D volumes instead of 2D slices and added residual connections into the convolutional and deconvolutional layers \citep{milletari2016v}.
    \item \textbf{UNet++} added dense convolutional blocks within the skip connections of the original UNet to bridge the semantic gap between the feature maps of the encoder and decoder \citep{zhou2019unet++}. Skip connections enabled feature propagation along horizontal and vertical directions and more flexible feature \textbf{fusion} at the decoders. UNet++ addressed the need to find the optimal depth of the encoder and decoder depending on the task \citep{zhou2019unet++}.
    \item \textbf{nnU-Net} leveraged an ensemble of three distinct simple U-Net architectures: a 2D model for slice-wise processing, a 3D model for whole-volume processing, and a cascaded 3D approach \citep{Isensee2021}. The self-configuring framework autonomously determines the optimal preprocessing pipeline, architectural parameters, training protocols, and post-processing strategies by analyzing dataset-specific characteristics. The framework was validated extensively across 53 diverse segmentation tasks for a total of 23 datasets \citep{Isensee2021}.
    \item \textbf{TransBTSV} was designed for 3D segmentation of MRI of the brain \citep{li2022transbtsv2betterefficientvolumetric}. It was designed with a UNet architecture, with a 3D CNN encoder extracting the volumetric local spatial features and downsampling the input 3D images at the same time, resulting in compact volumetric feature maps, sent to a transformer to model global features, with the 3D CNN decoder performing progressive upsampling, and skip connections between the encoder and the decoder \citep{li2022transbtsv2betterefficientvolumetric}.
    \item \textbf{TransBTSV2} was designed as a hybrid U-shape network with a 3D CNN encoder to capture local information and leveraging a transformer encoder to model long-distance dependencies \citep{li2022transbtsv2betterefficientvolumetric}. To lower the size of the transformer, and hence computational complexity, they reduced the number of transformer blocks from four as in TransBTS to one, but increased the hidden dimension of feature vectors. Deformable bottleneck modules were inserted into the skip connections between the CNN encoder and decoder to capture features of lesions with irregular shape \citep{li2022transbtsv2betterefficientvolumetric}.
\end{itemize}

\subsubsection{Models with global fusion}
\label{subsubsec:other-models-fusion}

The following approaches provide \textbf{global fusion} through the attention mechanism or depthwise convolution layers with large kernels to widen the receptive field.
\begin{itemize}
    \item \textbf{CoTr} was proposed with a hybrid encoder consisting of a CNN and a deformable transformer, and a pure CNN decoder \citep{10.1007/978-3-030-87199-4_16}. To reduce the computational complexity, the deformable transformer integrated multi-scale deformable self-attention, focusing only on a small set of key sampling locations around a reference location, instead of all locations \citep{10.1007/978-3-030-87199-4_16}. In contrast to TransUNet, which processed only the low-resolution feature maps from the last stage, CoTr allowed the transformer to \textbf{fuse globally} the multi-scale feature maps from the CNN encoder and kept abundant high-resolution information for segmentation \citep{10.1007/978-3-030-87199-4_16}.
    \item \textbf{MedFormer} was designed with a hybrid encoder with CNN blocks and transformer blocks with bidirectional multi-head attention, which eliminated redundant tokens via low-rank projection and reduced the complexity of conventional self-attention from quadratic to a linear level \citep{gao2022data}. It also added a semantically and spatially \textbf{global multi-scale fusion} mechanism to improve segmentation with negligible computational overhead. MedFormer gradually restored the resolution through a series of up-sampling and bidirectional multi-head attention blocks in the decoder \citep{gao2022data}.
    \item \textbf{TransUNet} represented the first architecture based on transformers for medical image segmentation \citep{chen2021transunet}. More specifically, it combined them with a CNN encoder in a U-shape hybrid configuration, where the CNN first extracted the image features, which were then flattened and sent as patches to the transformer. TransUNet enabled a seamless \textbf{fusion} of global features from the transformer with high-resolution CNN features. The output of the transformer was then upsampled in the decoding path and concatenated with the output of the CNN encoder at different resolutions through skip connections for precise localization \citep{chen2021transunet}. TransUNet was recently updated with a CNN decoder and a transformer decoder in addition to the CNN encoder and transformer encoder as in the original work \citep{CHEN2024103280}. The transformer decoder used learnable queries, refined through cross-attention with CNN features, and employed a coarse-to-fine attention refinement approach \citep{CHEN2024103280}.
    \item \textbf{UNETR} was developed ad a U-shape architecture with a transformer as encoder to process 3D radiological volumes, a decoder connected by skip connections linking the output every three layers of a 12-layer transformer \citep{unetr}.
    \item \textbf{UNETR++} introduced an efficient paired attention block that combined spatial and channel attention mechanisms \citep{unetr_pp_arxiv}.
    \item \textbf{Swin-UNet} proposed the  Shifted Window (Swin) transformer block in the encoder, bottleneck, and decoder into a UNet-like architecture to reduce the computation complexity of transformers from quadratic to linear \citep{cao2022swin}. Swin built hierarchical feature maps by starting from small-sized patches and gradually merging neighboring patches in deeper layers. The linear computational complexity was ensured by computing self-attention locally within non-overlapping windows that partition an image. Moreover, the window in a layer was shifted from the previous one, resulting in the self-attention computation in the new window to cross the boundaries of the previous window, thus providing connections among them \citep{liu2021swin}.
    \item \textbf{SwinUNETR} replicated the UNETR architecture by inserting the Swin transformer into the encoder \citep{SwinUNETR_arxiv}.
    \item \textbf{3D UX-Net} was proposed as a CNN with an encoder where large kernels simulated the behaviour of Swin transformers to extract features with a global receptive field, by replacing the window multi-head self-attention with depth wise convolutions. The multiscale output of the encoder was connected to a decoder through connections forming a U-shape. 3D UX-Net introduced pointwise depth convolution to scale the extracted representations effectively with fewer parameters \citep{3duxnet_arxiv}.
    \item \textbf{nnFormer} introduced a novel transformer architecture combining attention layers with convolutional operations in alternating sequence in the descending path of the encoder in a U-shape architecture. A volume-based self-attention mechanism that processed 3D data both locally and globally to build feature pyramids and provide large receptive fields was added in the bottleneck. Finally, skip attention was integrated in the skip connections of the ascending path of the decoder, replacing summation and concatenation in traditional skip connections \citep{nnFormer_arxiv}.
    \item \textbf{MISSFormer} was designed as a hierarchical encoder-decoder model with transformer blocks in all encoding and decoding steps, and with a transformer context bridge between the encoder and decoder for information \textbf{fusion at multi-scale} \citep{huang2022missformer}. Each transformer block contains a convolution and a skip connection between the two fully connected layers to capture local information in addition to global dependencies. The output of the transformer blocks provided features of different scales, concatenated, and sent to the transformer context bridge to capture global dependencies. Finally, the output features are split into feature maps of different scales,   and to the transformer blocks of the decoder to mix global dependencies with local context \citep{huang2022missformer}.
    \item \textbf{LHU-Net} was designed as hybrid CNN-transformer network with a U-shape encoder-decoder structure with skip connections to connect them \citep{sadegheih2024lhunetlighthybridunet}. It exploited three different attention mechanisms. First,  spatial attention of ViT to capture local features in the first layers. Second, channel attention of ViT to capture global features in the deep layers. Third, large kernel attention with one deformable layer to capture a wide range of spatial representation to focus on the desired receptive field. Initial convolutional layers were used to reduce the size of the feature maps. Then, hybrid blocks of convolutions and \textbf{fusion} blocks were inserted in both the descending and ascending paths. For these blocks, the attention from the large kernel attention with one deformable layer was \textbf{fused} with the the ViT channel attention in the last block of the encoder and the first one of the decoder, and with the spatial ViT attention for all the other blocks \citep{sadegheih2024lhunetlighthybridunet}.
    \item \textbf{SCANeXt} combined the strengths of residual spatial and channel attention, followed by a ConvNeXt-inspired depth-wise convolution block \citep{scanext}.
    \item \textbf{MedNeXt} was proposed as a U-shape encoder-decoder network, built upon Meta AI's ConvNeXt architecture, effectively translating transformer-inspired design elements into a pure convolutional approach while preserving CNNs inherent inductive biases that are particularly valuable in data-scarce medical settings \citep{mednext_arxiv}. The encoder and the decoder leveraged MedNeXr blocks consisting of depthwise convolution layer with large kernels to replicate a large attention window of Swin-Transformers, an expansion layer, and compression layer \citep{mednext_arxiv}.
\end{itemize}

\subsubsection{Other models}
\label{subsubsec:other-models}

\begin{itemize}
    \item \textbf{SegResNet} was designed as an encoder-decoder network with each encoder layer consisting of ResNet-like blocks \citep{myronenko20183d, he2016deep}. A variational autoencoder branch was added to regularize the encoder during training \citep{myronenko20183d}.
    \item \textbf{NexToU} wad developed as a hybrid CNN-graph neural network, combining a pool graph module to identify key nodes in the global network to extract crucial topological information, and Swin graph module, adapted from Swin transforemr, to capture local information to recognize irregularly shaped vasculature \citep{shi2023nextouefficienttopologyawareunet}.
\end{itemize}

Table \ref{table:models-specialized} in the appendix lists all specialized models with their key features.

\section{Taxonomy}
\label{sec:taxonomy}

\begin{figure*}[!ht]
  \centering
    \includegraphics[width=1.1\linewidth]{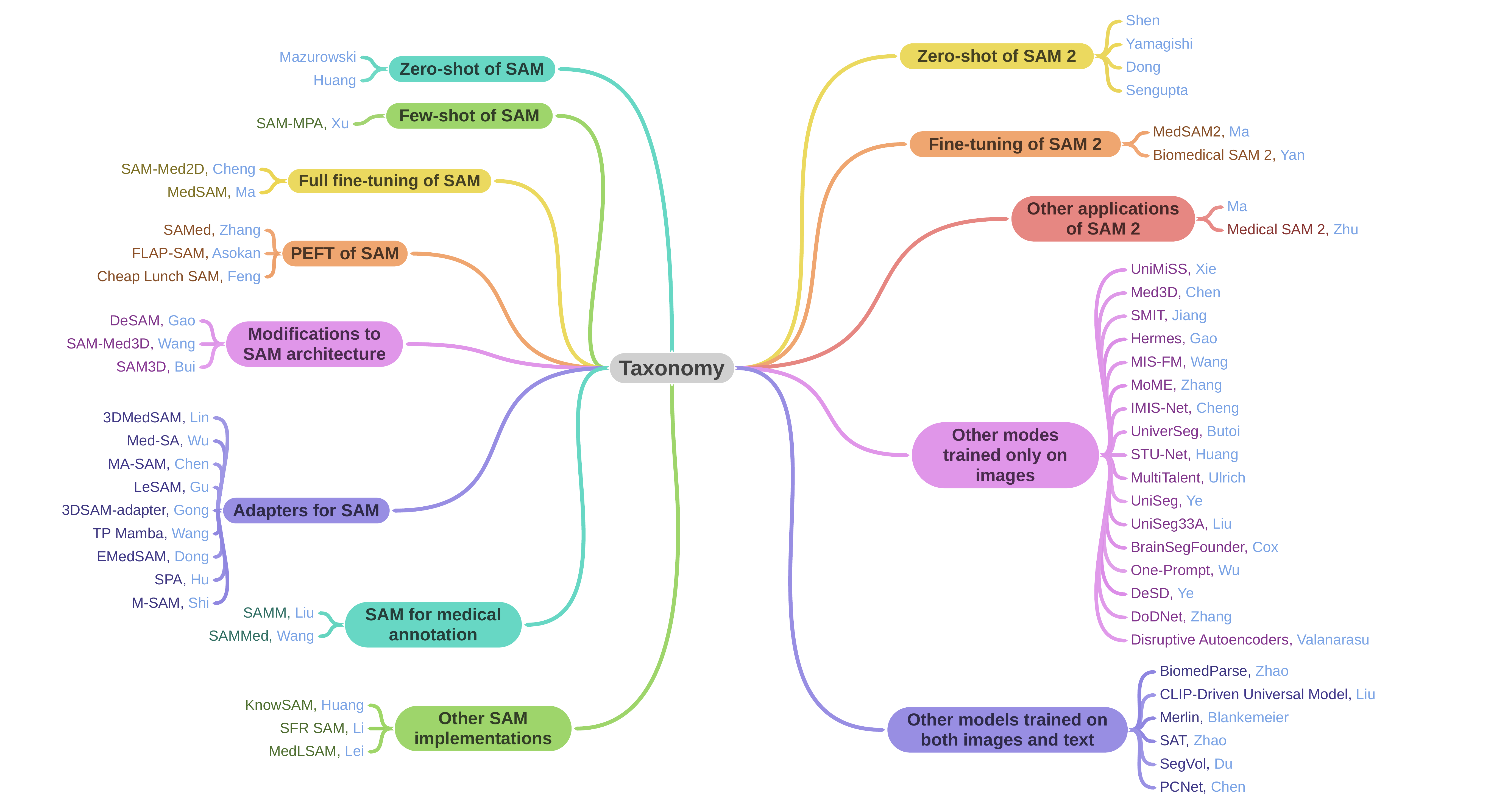}
    \caption{Proposed taxonomy for the generalist models for medical image segmentation.}
    \label{fig:taxonomy}
\end{figure*}

In this section we offer a classification of generalist models. According to their definition provided in Section \ref{sec:introduction} we included those which underwent pre-training with either self-supervised or supervised approach \citep{bommasani2021opportunities, 10510478}. We excluded models like TotalSegmentator and TotalSegmentator MRI since they did not concern either the implementation of a new generalist model or a variant of an existing one \citep{doi:10.1148/ryai.230024, doi:10.1148/radiol.241613}. Although they were evaluated on a large number of anatomical structures, 104 and 80, respectively, they represented a simple testing of nnU-Net \citep{doi:10.1148/ryai.230024, doi:10.1148/radiol.241613}.
The taxonomy is graphically illustrated in Fig. \ref{fig:taxonomy}, while their main characteristics, the links to the publications and code repositories are reported in Table \ref{table:models-foundational}. In the next sections all generalist models will be described.

\begin{figure*}[!ht]
  \centering
    \includegraphics[width=0.8\linewidth]{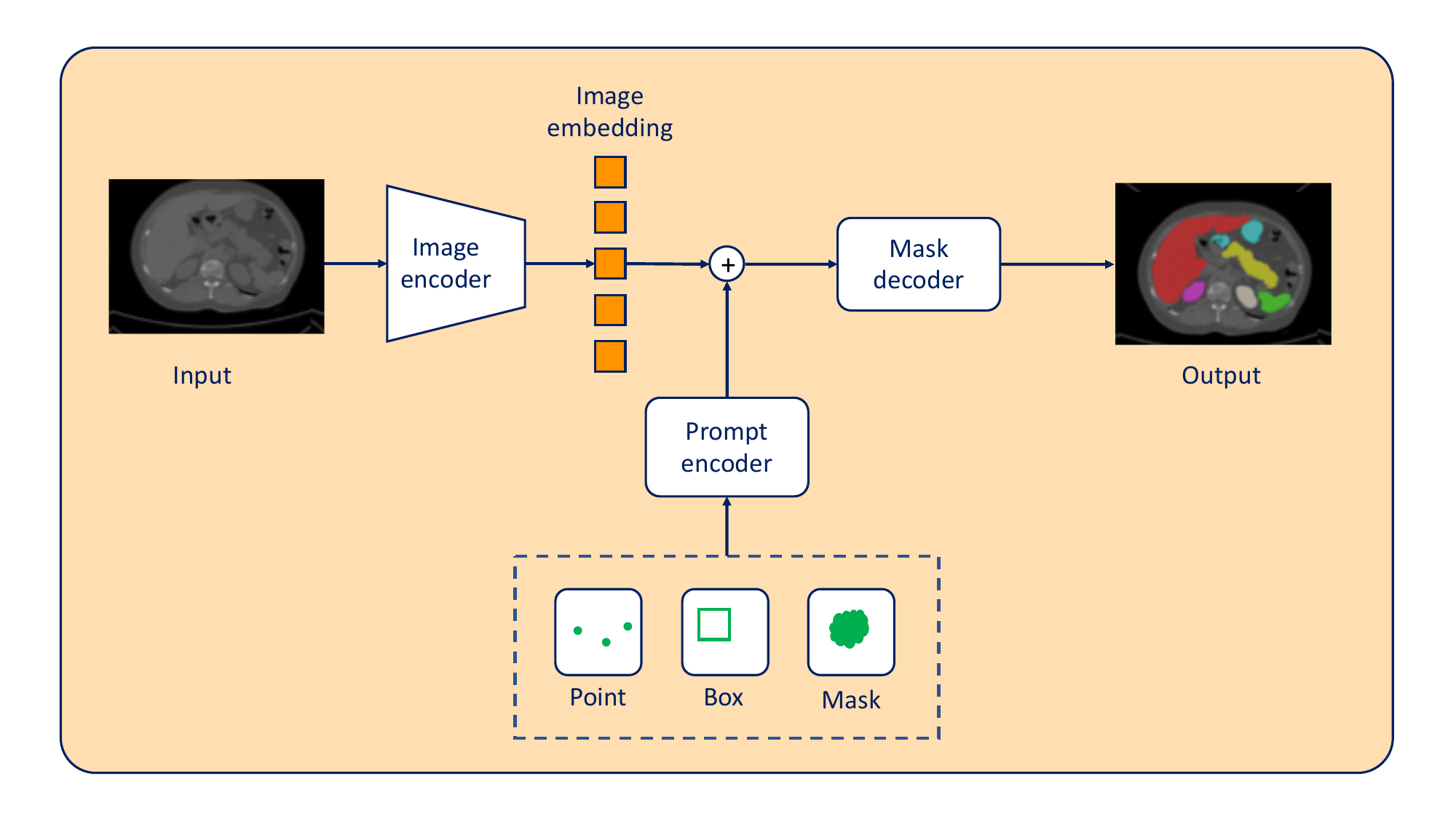}
    \caption{Architecture of SAM. Image adapted from \citep{SAM4MIS}.}
    \label{fig:sam}
\end{figure*}

\subsection{SAM}
\label{subsec:sam}

\begin{figure*}[!ht]
  \centering
    \includegraphics[width=1\linewidth]{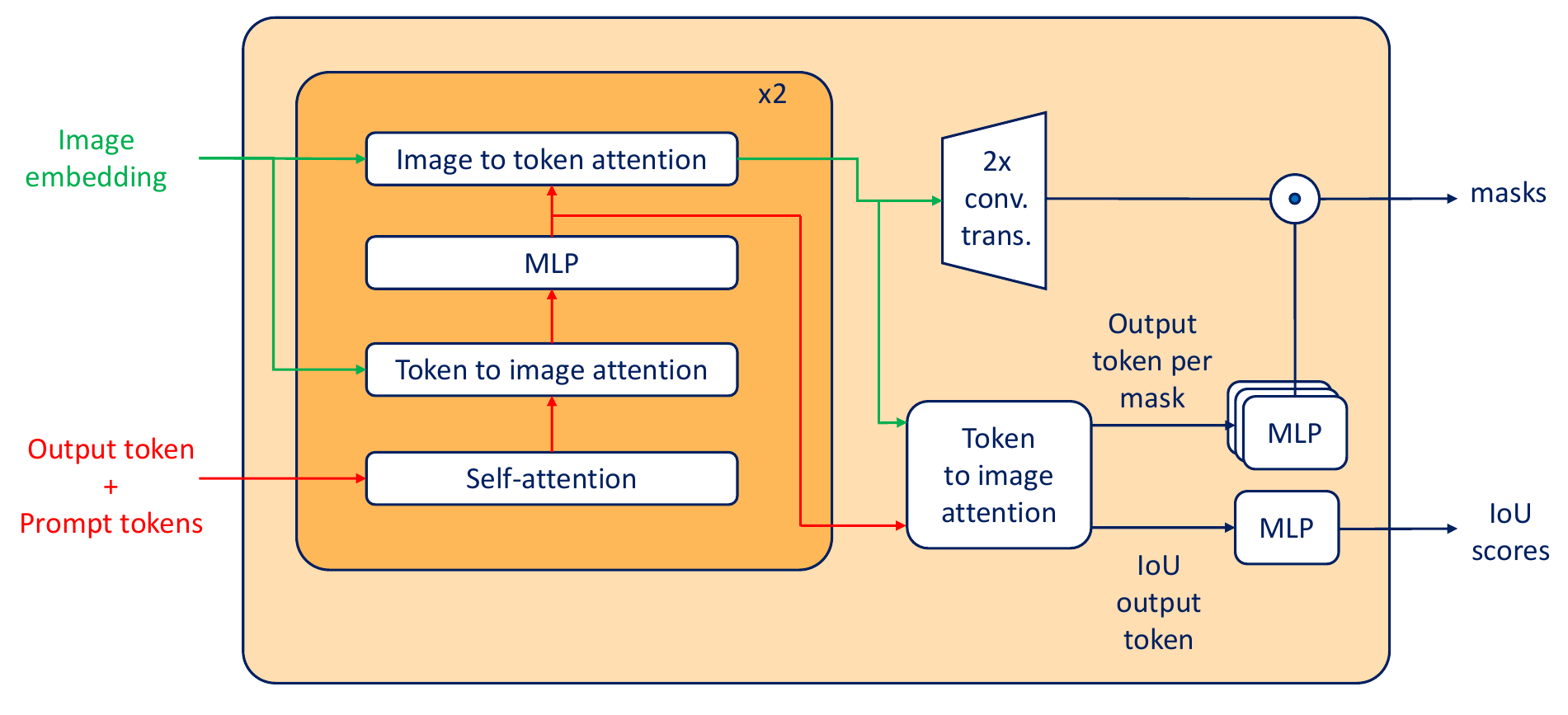}
    \caption{Architecture of the SAM mask decoder. Image adapted from \citep{kirillov2023segany}.}
    \label{fig:sam-decoder}
\end{figure*}

The SAM was the first generalist promptable model for general image segmentation, pre-trained on a large dataset \citep{kirillov2023segany}. Its architecture is illustrated in Fig. \ref{fig:sam}SAM model consists of three components: an image encoder based on an MAE pre-trained ViT; a prompt encoder to accept points, bounding boxes, masks, or text as input, and to encode them into a feature space aligned with the image features extracted by the image encoder; and a mask decoder, depicted in Fig. \ref{fig:sam-decoder}, leveraging the transformer architecture to map the image embedding and prompt embedding to an output mask. The output of the prompt encoder was enriched by an output token, analogous to the \texttt{[class]} token in ViT. Overall, the output token and the prompt tokens, were called tokens for simplicity. The decoder performed self-attention of the tokens, cross-attention from the tokens to the image embeddings, MLP, and a cross-attention from the image embedding to the tokens. Another cross-attention was performed from the tokens to the image embeddings. Finally, am MLP mapped the output of the transformer to a linear classifier to compute the segmentation masks (Fig. \ref{fig:sam-decoder}). 
SAM can operate in manual (with point, bounding boxes, or text as prompts) or automatic mode \citep{kirillov2023segany}. 
In the manual mode, the point prompt include both positive and negative points, for the foreground and background of one object, respectively. The bounding box prompt corresponds to the spatial region of the object that needs to be segmented. Finally, the text prompt indicates the text to describe the object. However, at the time of this writing it was not released  on the official GitHub repository. In the automatic mode, SAM generates segmentation masks for all the potential objects in the whole image without manual prompts. First, SAM draws a grid of uniformly spaced points on the whole image. Second, the prompt encoder will produce a point embedding and combine it with the embedding of the image encoder. Third, the mask decoder will output several potential masks for the entire image. Finally, a filtering processing removes duplicate and low-quality masks using, for instance, non-maximal suppression \citep{kirillov2023segany, sam_huang}. 
The dataset for pre-training SAM, called SA-1B, consisted on one billion on masks from 11 million of natural images \citep{kirillov2023segany}. 

\subsection{Fusion in generalist models}
\label{subsec:fusion-generalists}
The reviewed generalist models explored different mechanisms of fusion. We categorized fusion into the following levels:
\begin{itemize}
    \item \textbf{F$_1$: SAM Fusion. } For the models based on SAM fusion (cft. Section \ref{subsec:sam}).
    \item \textbf{F$_2$: Additional Fusion in SAM variants. } For those models leveraging another fusion mechanism in addition to the one of SAM.
    \item \textbf{F$_3$: SAM 2 Fusion. } For the models based on SAM 2 fusion.
    \item \textbf{F$_4$: Other Fusion. } For the models not based on SAM or SAM 2.
\end{itemize}

\subsection{Variants of SAM}
\label{subsec:sam-variants}
In the following sections we describe the various approaches on SAM in terms of zero-shot (Section \ref{subsubsec:sam-zero-shot}), few-shot (Section \ref{subsubsec:sam-few-shot}), full fine-tuning (Section \ref{subsubsec:sam-fine-tuning}), PEFT (section \ref{subsubsec:sam-peft}), design of adapters (Section \ref{subsubsec:sam-adapters}), modifications to architecture (Section \ref{subsubsec:sam-modified-architecture}), medical annotations (Section \ref{subsubsec:sam-annotations}), and other implementations (Section \ref{subsubsec:sam-others}).

\subsubsection{Zero-shot of SAM}
\label{subsubsec:sam-zero-shot}

\citet{sam_mazurowski} performed the first attempt to evaluate SAM's zero-shot performance on medical images across various 2D and 3D imaging modalities, including MRI (e.g., brain tumors), CT (e.g., liver), X-ray (e.g., chest), ultrasound (e.g., breast), and PET scans. They tested points and box prompting strategies, and compared SAM against other interactive segmentation methods. Their findings revealed that SAM's performance varies significantly across different datasets with box prompts scoring higher than points. The study highlighted that SAM performed best on well-circumscribed objects with unambiguous prompts. This means that SAM, trained on an extensive dataset of natural images, can partially transfer its abilities to the medical image domain \citep{sam_mazurowski}.
\citet{sam_huang} conducted a comprehensive evaluation of SAM on medical images by creating COSMOS, a large dataset spanning 18 modalities, 84 objects, 1050K 2D images, and 6033K masks. Similarly to Mazurowski et al., their analysis revealed SAM's variable performance - excellent on some objects but unstable or failing on others. They found that SAM with ViT-H outperformed ViT-B, and manual prompts (especially bounding boxes) yielded better results than automatic mode (result consistent with observations by Mazurowski et al.).

\textit{Fusion level: F$_1$ in both studies.}

\subsubsection{Few-shot of SAM}
\label{subsubsec:sam-few-shot}

\citet{xu2024sammpaapplyingsamfewshot} proposed SAM-MPA, by integrating mask propagation and automatic prompt generation into SAM, as a framework to adapt SAM for few-shot medical image segmentation. SAM-MPA addressed the challenges of few-shot segmentation in terms of selection of a set of labeled images as support images, propagation of mask knowledge from support images to query images, and generation of high-quality prompts. To solve these issues they clustered samples, and selected the most representative instance from each cluster to form the support image set. Then they performed unsupervised registration between support and unlabeled query images to be segmented to get a coarse mask. Finally they proposed an automatic prompt generation from the coarse mask and combining points, box, and mask as input to the prompt encoder, and the unlabeled images as input to the vision encoder. A post-refinement process was added to optimize the SAM segmentation results \citep{xu2024sammpaapplyingsamfewshot}.

\textit{Fusion level: F$_1$.}

\subsubsection{Full fine-tuning of SAM}
\label{subsubsec:sam-fine-tuning}

\citet{cheng2023sammed2d} introduced SAM-Med2D, trained on 4.6M images and 19.7M masks across various modalities, by fine-tuning the SAM encoder using an adapter, the prompt encoder, and the decoder. SAM-Med2D was tested on unseen images of nine datasets of MICCAI2023 (0.52M images and 1.31M masks) for generalizability \citet{cheng2023sammed2d}. 
MedSAM by \citet{MedSAM_arxiv,MedSAM_nature} represented a significant advancement in medical image segmentation as the first generalist model capable of universal segmentation across diverse medical imaging modalities. Trained on over 1.5 million image-mask pairs spanning 10 imaging modalities and 30+ cancer types, it adapted SAM architecture through comprehensive fine-tuning of the image encoder and mask decoder while maintaining the prompt encoder capabilities. Rather than attempting fully automatic segmentation, struggling with task variability, MedSAM used bounding box prompts to specify target regions, making it adaptable to both 2D and 3D images while maintaining precise control over the segmentation target. MedSAM was evaluated on 86 internal and 60 external validation tasks \citep{MedSAM_nature}.

\textit{Fusion level: F$_1$ in both studies.}

\subsubsection{PEFT of SAM}
\label{subsubsec:sam-peft}
SAMed, proposed by \citet{samed_zhang_liu}, was one of the first fine-tuned versions of SAM for medical imaging segmentation. In SAMed the image encoder was fine-tuned with LoRA, while the prompt encoder and mask decode were fully fine-tuned. Applying LoRA also to the SAM decoder reduced the model size, but the performances dropped. SAMed was evaluated on the Synapse dataset \citep{samed_zhang_liu}.

FLAP-SAM enabled federated learning across different centers through LORA in the attention layers of the SAM image encoder and mask decoder in addition to fine-tuning with LoRA the final layers of the decoder (upsampling and multy-layer perceptron) \citep{asokan2024federatedlearningfriendlyapproachparameterefficient}. FLAP-SAM was evaluated on three datasets (Fed-KITS2019, a six-client federated version of the KiTS19 dataset, an MRI dataset on brain from three hospitals, and Prostate MRI) \citep{asokan2024federatedlearningfriendlyapproachparameterefficient}.
\citet{feng2023cheaplunchmedicalimage} fine-tuned SAM by LoRA on the image encoder and mask decoder after synthesizing data from few exemplars of the BraTS and Synapse datasets.

\textit{Fusion level: F$_1$ in all the studies.}

\subsubsection{Adapting SAM through adapters}
\label{subsubsec:sam-adapters}

\begin{figure*}[!ht]
  \centering
    \includegraphics[width=1\linewidth]{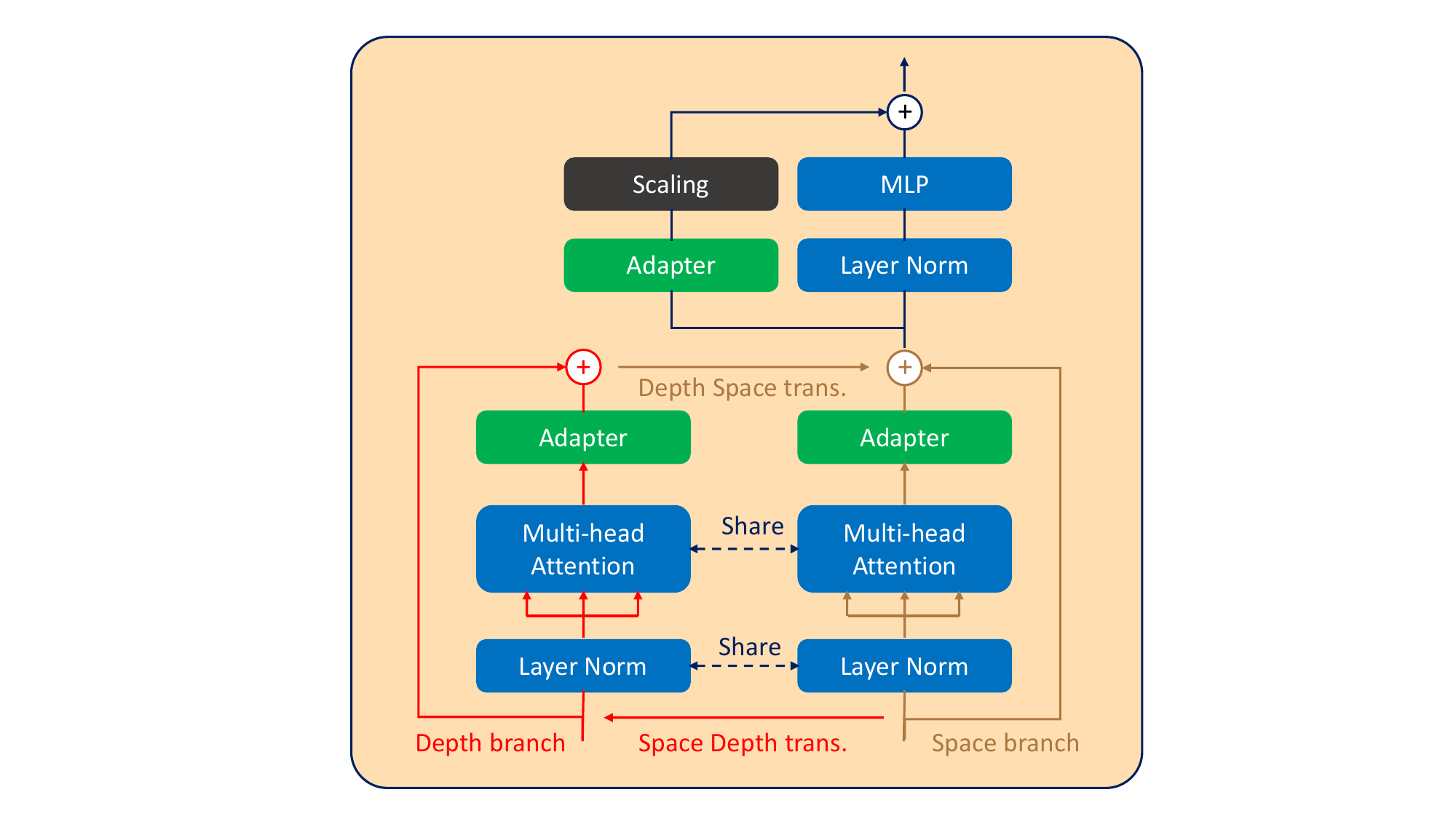}
    \caption{3D Medical image adaptation of Medical SAM Adapter. Image adapted from \citep{wu2023medicalsamadapteradapting}.}
    \label{fig:3d-medical-adapter}
\end{figure*}

Medical SAM Adapter (Med-SA) was the first attempt to inject adapters into SAM \citep{wu2023medicalsamadapteradapting}. They consisted of a down-projection, ReLU activation, and up-projection. Two adapters were inserted into each layer of the ViT SAM encoder, one after the multi-head attention, and the second in the residual path of the multi-layer perceptron after the multi-head attention.  
In the first adapter the space-depth transpose technique was introduced to adapt 2D SAM to 3D medical imaging by adding, in each block of the transformer, a parallel branch with layer normalization, multi-head attention and one adapter, as depicted in Fig. \ref{fig:3d-medical-adapter}.
Three adapters were added also to the decoder. The first integrated the prompt embedding, the second was places in the residual path of the multi-layer perceptron after the multi-head attention, and the third one after the residual connection of the image embedding-to-prompt cross-attention. The first adapter (Hyper-Prompting Adapter) consisted of a set of weight maps of the embedding from the prompt encoder \citep{wu2023medicalsamadapteradapting}.
Med-SA was evaluated on 17 tasks on different image modalities, like CT, MRI, and ultrasound \citep{wu2023medicalsamadapteradapting}.

\citet{ma-sam_cheng} proposed MA-SAM, a modality agnostic SAM adaptation framework injecting 3D adapters with 3D convolutional layers into the transformer blocks of the image encoder to capture the volumetric and  temporal information of medical images, and videos respectively. The SAM decoder was modified with a progressive up-sampling mechanism to recover the prediction resolution. The encoder was fine-tuned with factor tuning (FacT), while the decoder was fully fine-tuned. They also explored a \textbf{multi-scale fusion} in a UNet-like architecture by connecting the multi-scale feature maps of the image encoder with corresponding stages of the mask decoder using skip connections. However, during tests the progressive up-sampling approach provided better results \citep{ma-sam_cheng}. MA-SAM was evaluated on five medical image segmentation tasks, by using 11 public datasets across CT, MRI, and surgical video data \citep{ma-sam_cheng}. MA-SAM was tested for generalization on AMOS22 dataset, and MRI scans of the prostate \citep{ma-sam_cheng}.
3D Medical SAM-Adapter (3DMedSAM) introduced several adapters into SAM \citep{LIN20251}. In the first one, 3D convolutions were added after the 3D patching process. The second one was placed between each attention block, by concatenating one down-projection and a 3D convolution layer with the first adapter, followed by one up-projection layer. In the last adapter 3D convolutions replaced the 2D convolutions of the decoder. 3DMedSAM was fine-tuned on a private dataset for transthoracic echocardiography for left atrial appendage, the LiTS17, and the BTCV \citep{LIN20251}.
LeSAM was proposed for segmentation of lesions \citep{10540651}. Its architecture modified the SAM image encoder with two adapters into each transformer block to integrate task-specific knowledge, and SAM mask decoder into a UNet-like network for improved alignment with the lesion boundary \citep{10540651}. The two adapters were placed at the beginning and the end of the transformer block. The adpaters consisted of a down-projection linear layer, an GeLU activation, and an up- projection linear layer \citep{hendrycks2016gaussian}. The adapters were pre-trained by a self-supervised strategy on images from RadImageNet, a dataset with 1.35 million CT, MRI, and ultrasound scans on 11 anatomical structures and 165 pathological labels, followed by supervised training \citep{mei2022radimagenet}.
The mask decoder upsampled the mask embedding and progressively \textbf{fused} the features of the SAM vision encoder with the upsampling branch of the decoder \citep{10540651}.

Similarly, \citet{3DSAM-adapter} introduced 3DSAM-adapter, focused on tumor segmentation, by modifying the vision and prompt encoders, and decoder of SAM. The vision encoder was redesigned with embedding 3D patches, and a 3D adapter, inserted between two adjacent attention blocks, with a depth-wise 3D convolution between down-projection and up-projection layers. They proposed a visual sampler to ensure that the prompt embeddings share the same semantic features as the image embeddings, by \textbf{fusing} them using self-attention, and cross-attention. In the decoder they replaced the 2D convolutions with 3D ones, and added a multi-layer aggregation mechanism to concatenate the intermediate output of the encoder to produce a mask feature map \citep{3DSAM-adapter}.

Tri-Plane Mamba modified the ViT block of the SAM vision encoder by injecting LoRA into the self-attention, and tri-plane mamba module as adapter to capture local and global 3D features \citep{10.1007/978-3-031-72114-4_61}. This model was evaluated on the BTCV dataset \citep{10.1007/978-3-031-72114-4_61}.

EMedSAM modified the SAM encoder and decoder with adapters. Moreover, the modified SAM vision encoder underwent a distillation training strategy from the ViT-H to reduce its size \citep{Dong2024}. The vision encoder was based on TinyViT, integrating convolutions, followed by transformers. It was trained as a student model by distillation from ViT-H as a teacher model. The adapters of the encoder and decoder leveraged Medical SAM Adapter \citep{wu2023medicalsamadapteradapting}. EMedSAM was evaluated on FLARE 2022 and on a private dataset \citep{Dong2024}.

Mask-Enhanced SAM (M-SAM) proposed a coarse-to-fine segmentation approach for 3D tumor lesion segmentation \citep{shi2024maskenhancedsegmentmodeltumor}. Its architecture leveraged the SAM-Med3D. Image embeddings from the vision encoder and prompt encoder were sent to the decoder to obtain an initial coarse mask. Then, image embeddings and mask embeddings from the coarse mask were fed into the mask-enhanced adapter to update the embeddings iteratively for a mask refinement. The image embeddings and mask embeddings were \textbf{fused} into the mutual feature enhancement block, consisting of transformers, inside the mask-enhanced adapter. M-SAM was evaluated on seven datasets \citep{shi2024maskenhancedsegmentmodeltumor}.

The spatial prior adapter (SPA) was proposed as a PEFT strategy for SAM \citep{10829779}. It modified the SAM vision encoder and mask decoder. In the vision encoder a spatial prior module and a feature communication module were inserted. The former consisted of CNNs blocks to capture localized spatial information, whereas the latter \textbf{fused} the features extracted by both the ViT of the SAM vision encoder and the spatial prior module through cross-attention. The SAM decoder was modified by inserting the multiscale feature fusion module, concatenating the multi-scale features and the fused featured by cross-attention \citep{10829779}. SPA was fine-tuned on Kvasir, Promise12, and Synapse datasets with both interactive (with points or bounding boxes prompts) or end-to-end segmentation (without prompts) mode \citep{10829779}.

\textit{Fusion level: F$_1$ for all models, while F$_1$ + F$_2$ for MA-SAM, LeSAM, 3DSAM-adapter, M-SAM, and SPA}.

\subsubsection{Modifications to SAM architecture}
\label{subsubsec:sam-modified-architecture}

\citet{sam3d_bui} developed SAM3D to process 3D volumetric images instead of a sequence of 2D slices, by replacing the SAM decoder with a 3D decoder consisting of four 3D convolutional blocks with skip connections.

\citet{wang2024sammed3dgeneralpurposesegmentationmodels} introduced SAM-Med3D, a general-purpose segmentation model with a fully 3D architecture (vision and prompt encoder, and decoder) with integrated 3D positional encoding, 3D convolutions and layer normalization, enabling it to capture inter-slice context using only a single 3D prompt per volume. SAM-Med3D was trained on the expansive SA-Med3D-140K dataset—comprising over 22K 3D images and 143K masks from 70 public and 24 private datasets covering 28 modalities. It was tested on 16 public datasets. For generalization on downstream tasks it was tested on AMOS2022, TotalSegmentator, and two unseen datasets from the MICCAI 2023 Challenge \citep{wang2024sammed3dgeneralpurposesegmentationmodels}.

DeSAM was designed with a modified SAM decoder to improve the SAM performances in automatic mode \citep{gao2024desamdecoupledsegmentmodel}. It added a prompt-relevant IoU module, and a prompt-decoupled mask module to SAM. The former was designed like the SAM decoder. It consisted of cross-attention and an IoU prediction head, but discarded the mask prediction output to generate only mask embeddings from the cross-attention. The latter had a UNet-like architecture to extract multi-scale embeddings from the SAM vision encoder. The bottleneck embeddings were \textbf{fused} with the output of the prompt-relevant IoU module to generate the mask from the image and mask embeddings. DeSAM was evaluated on eight datasets (two of abdominal organs and six of prostate) \citep{gao2024desamdecoupledsegmentmodel}.

\textit{Fusion level: F$_1$ for SAM3D, SAM-Med3D, and F$_1$ + F$_2$ for DeSAM}.

\subsubsection{SAM for medical annotations}
\label{subsubsec:sam-annotations}
SAM\textsuperscript{Med} combined SAM\textsuperscript{assist} and SAM\textsuperscript{auto} modules to accelerate annotations on medical imaging \citep{wang2023mathrmsammedmedicalimageannotation}. The former leveraged prompt learning to effectively adapt SAM to the downstream medical segmentation task, while the latter enabled automatic prompt generation for the images without user interaction. For SAM\textsuperscript{assist} only the SAM prompt encoder was trained. SAM\textsuperscript{auto}, trained on a small number of images,  exploited few-shot learning for coarse segmentation useful to generate prompts more closely aligned with the target
objects. These prompts were then fed into SAM\textsuperscript{assist}. SAM\textsuperscript{Med} was evaluated on eight public datasets \citep{wang2023mathrmsammedmedicalimageannotation}.

\citet{liu2024sammsegmentmedicalmodel} presented SAMM, by integrating SAM into 3D Slicer, an open-source software tool for visualization, processing, segmentation, registration, and analysis of medical images. SAMM showed promising performance on three imaging modalities such as MRI, CT and ultra-sound on the segmentation of cerebral hemorrhages and identification of tumors within the stomach and lungs, without retraining or fine-tuning the model.

\textit{Fusion level: F$_1$ for SAM\textsuperscript{Med} and SAMM}

\subsubsection{Other SAM implementations}
\label{subsubsec:sam-others}
MedLSAM integrated SAM with the Localize Anything Model for 3D Medical Images (MedLAM), the first generalist model for 3D medical image localization trained on 14,012 CT scans from 16 different datasets \citep{lei2024medlsamlocalizesegmentmodel}.  MedLAM introduced the sub-patch localization strategy by subdividing the target organ into multiple segments, each of which with a bounding box tailored to represent more accurately the organ in each slice \citep{lei2024medlsamlocalizesegmentmodel}. MedLSAM was assessed on two datasets (StructSeg19 and WORD) \citep{lei2024medlsamlocalizesegmentmodel}.

KnowSAM was proposed to harness the generalization capabilities of SAM through distillation to improve semi-supervised medical image segmentation\\\citep{10843257}. Two subnets engaged in co-teaching to mutually correct each other within a multi-view co-training strategy. Then, a hybrid aggregation module \textbf{fused} their prediction maps with entropy and dissimilarity maps to mitigate the impact of uncertainty and inconsistency. Finally, a learnable prompt strategy generated a learnable feature prompt, fed into the SAM decoder of
SAM, along with the aggregated map from the two initial subnets. A medical SAM adapter was added to the encoder and decoder \citep{wu2023medicalsamadapteradapting}. The output of SAM decoder was used for knowledge distillation. A data augmentation strategy was applied to both labeled and unlabeled data. KnowSAM was evaluated on 11 datasets (five for colonoscopy, three for ultrasound, one for dermoscopy, the ACDC, and one for breast cancer) \citep{10843257}.

Stitching, Fine-tuning, and Re-training (SFR) was proposed as a SAM-enabled semi-supervised approach \citep{li2025stitchingfinetuningretrainingsamenabled}. A stitching module arranged each 3D radiological volume slice by slice into a 2D image, sent as input to SAM for fine-tuning with LoRA to produce high-quality pseudo labels for the unlabeled images. A retraining module with the size of V-Net was trained with with both labeled images and pseudo labels. SFR was evaluated on five datasets.
SFR+ extended SFR introducing a confidence estimation to determine how to handle each unlabeled sample, and a selective training strategy in the fine-tuning and re-training modules for more effective handling of unlabeled samples \citep{li2025stitchingfinetuningretrainingsamenabled}.

\textit{Fusion level: F$_1$ for all models, while F$_1$ + F$_2$ for KnowSAM}.

\begin{figure*}[!ht]
  \centering
    \includegraphics[width=1\linewidth]{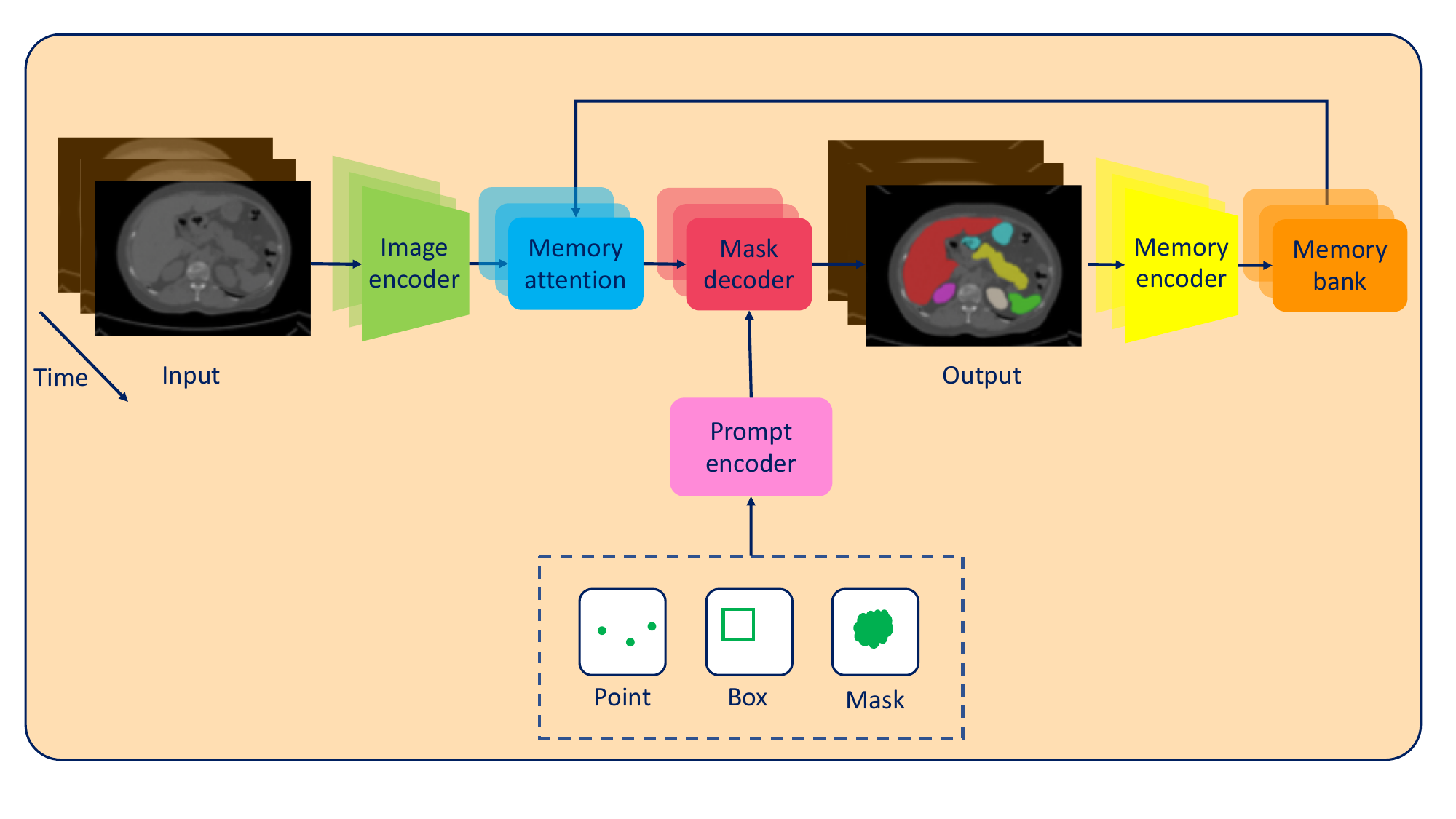}
    \caption{Architecture of SAM 2. Image adapted from \citep{ravi2024sam2}.}
    \label{fig:sam-2}
\end{figure*}

\subsection{SAM 2}
\label{subsec:sam2}

SAM 2 extended SAM to handle both images and videos, treating 3D images as a sequence of 2D frames \citep{ravi2024sam2}. Its architecture is displayed in Fig. \ref{fig:sam-2}. Compared with SAM, three new modules were added to SAM 2, namely memory attention, memory encoder, and memory bank. The memory encoder was specialized in creating memories of frames based on the prediction from the mask decoder, and store them into a memory bank for use in the following frames. The memory attention mechanism conditioned the current embeddings from the image encoder on these memories, allowing the model to track objects across frames \citep{ravi2024sam2}.
The image encoder, based on hierarchical MAE, was designed to run once for the entire interaction \citep{ryali2023hiera}.
SAM 2 accepted various types of prompts (points, boxes, or masks) on any frame and could propagate segmentations temporally while allowing interactive refinement. This allowed direct porting of SAM 2 to surgical videos and, most remarkably, to 3D medical images such as CT, MRI, PET and 3D ultrasound by treating each 2D slice of the 3D volumes as a sequence of video frames \citep{ravi2024sam2, ZHANG2022102088}.\\

\subsubsection{Zero-shot of SAM 2}
\label{subsubsec:sam2-zero-shot}

One of the first study on SAM 2 assessed its zero-shot performance on 21 datasets, covering five modalities, and three different types of surgical videos \citep{dong2024segmentmodel2application}. The results have shown similar performance between SAM and SAM 2 on single frame 2D segmentation, while there was variable performance under multi-frame 3D segmentation depending on the choices of slices to annotate, and the direction of the slice propagation \citep{dong2024segmentmodel2application}. These findings were confirmed by another study on 11 publicly available datasets \citep{sengupta2024sam2bettersam}. While SAM 2 reported improvements over SAM in some cases, particularly with MRI images, it underperformed SAM on CT and ultrasound images \citep{sengupta2024sam2bettersam}. 
Another work showed promising results of SAM 2 for segmenting larger organs with clear boundaries on the TotalSegmentator dataset \citep{yamagishi2025zeroshot3dsegmentationabdominal}, though its overall zero-shot performance still falls short of supervised methods on BraTS and MSD pancres, liver, lung, and spleen \citep{shen2025interactive3dmedicalimage}.

\textit{Fusion level: F$_3$ in both studies.}

\subsubsection{Fine-tuning of SAM 2}
\label{subsubsec:sam2-fine-tuning}
MedSAM2 was introduced as a full fine-tuning of all SAM 2 components on 455k 3D image-masks pairs from public datasets \citep{ma2025medsam2segment3dmedical}. MedSAM2 was also applied to streamline the annotation workflow with human-in-the-loop, where humans first drew a 2D bounding box at the middle slice, fed as a prompt to MedSAM2 to generate a 2D segmentation mask, later revised by humans for refinement. Then, MedSAM2 was ran again to generate a complete 3D lesion segmentation mask for all the slices. Finally, the human annotator refined the 3D segmentation. After repeating this process for dozens of new annotations, MedSAM2 was fine-tuned to improve accuracy. This pipeline was iterated multiple times to generate large-scale annotations for CT, MRI \citep{ma2025medsam2segment3dmedical}. 

Biomedical SAM-2 (BioSAM-2) fine-tuned the image encoder and mask decoder of SAM 2 on four datasets (Abdomen CT from MICCAI 2022 FLARE challenge, Abdomen MR from MICCAI 2022 AMOS Challenge, MICCAI 2017 EndoVis challenge, and NeurIPS 2022 Cell Segmentation challenge) \citep{yan2024biomedicalsam2segment}.

\textit{Fusion level: F$_3$ for both MedSAM2 and Bio-SAM-2.}

\subsubsection{Other applications of SAM 2}
\label{subsubsec:sam2-other-applications}
\citet{ma2024segmentmedicalimagesvideos-sam2} conducted a comprehensive benchmark of SAM 2 across 11 medical imaging modalities and developed a transfer learning pipeline for quick domain adaptation, also implementing their solution as a 3D Slicer plugin for practical clinical use. \citet{zhu2024medicalsam2segment} proposed Medical SAM 2, introducing a self-sorting memory bank mechanism into SAM 2 to dynamically select and retain the most informative embeddings, rather than simply using the most recent frames as in SAM 2. The self-sorting memory bank enabled one-prompt segmentation, allowing the Medical SAM 2 to handle unordered (without temporal relationships) medical images effectively. At each time frame the self-sorting memory bank was resampled to emphasize embeddings similar to the current embedding. The resampling process effectively prioritized embeddings more similar to current one, thus enhancing the relevance of the memory bank in the attention mechanism \citep{zhu2024medicalsam2segment}. Medical SAM 2 was tested on 78 datasets across various medical domains \citep{zhu2024medicalsam2segment}.

\textit{Fusion level: F$_3$ for both.}

\subsection{Other models trained only on image data}
\label{subsec:other-models-image-only-training}

The team behind nnU-Net developed a multi-dataset learning and pre-training method called MULTI daTAset LEarNing and pre-Training (MultiTalent) \citep{10.1007/978-3-031-43898-1_62}. It was proposed to address three challenges: to handle segmentation classes not present in one dataset but annotated in another one; to work with different annotation protocols for the same target structure; and to segment overlapping target structures with different level of detail. e..g, liver, liver vessel and liver tumor. Three different backbones were trained, i.e., the 3D U-Net generated by the nnU-Net, a Resenc U-Net (a UNet with with residual blocks in the encoder), and a SwinUNETR. MultiTalent was trained on 13 public abdominal CT datasets with a total of 1,477 3D images, while BTCV, AMOS, and KiTS19 datasets were used to evaluate the generalization of the MultiTalent features in a pretraining and fine-tuning setting \citep{10.1007/978-3-031-43898-1_62}. 
\\

UniSeg was designed with a vision encoder, a \textbf{fusion} and selection module, and a prompt-driven decoder \citep{ye2023unisegpromptdrivenuniversalsegmentation,10.1007/978-3-031-43898-1_49}. The UniSeg architecture was inspired by nnU-Net. The extracted features by the encoder were concatenated with a learnable prompt called universal prompt, designed to describe the correlations between the various tasks. The resulting concatenation was the input of the \textbf{fusion} module, which produced the task-specific prompts for the decoder. UniSeg was evaluated on 11 datasets \citep{ye2023unisegpromptdrivenuniversalsegmentation,10.1007/978-3-031-43898-1_49}.

\citet{huang2023stunetscalabletransferablemedical} proposed scalable and transferable UNet (STU-Net), a series of models of varying size, based on nnU-Net and pre-trained on supervised learning, using the TotalSegmentator dataset, and fine-tuned on AutoPET22, AMOS22, and FLARE22 datasets.

\citet{liu2022universalsegmentation33anatomies} developed Universal Segmentation model for 33 structures. It consisted of an encoder, a cross-patch transformer module, and a decoder. The cross-patch transformer module enabled to \textbf{fuse} more information in adjacent patches, thus enlarging the aggregated receptive field for improved segmentation performance. The model was trained on seven partially labeled datasets (BTCV, CTPelvic1K, MSD Liver, MSD Spleen, MSD Pancreas, KiTS, and CTSpine1K), totaling approximately 2'800 3D CT volumes \citep{liu2022universalsegmentation33anatomies}.

UniverSeg by \citet{butoi2023universeguniversalmedicalimage,10376558} was designed as a UNet-like network for few-shot segmentation for new tasks without retraining, by injecting a crossblock in each encoding and decoding step to transfer information from a set of example image-label pairs (the support set) to a new query image. UniverSeg was trained on MegaMedical, composed of 53 datasets for 26 medical domains with 16 imaging modalities. UniverSEg was tested on six datasets, three of which on unseen anatomy during training \citep{butoi2023universeguniversalmedicalimage}.

The IMed-361M dataset contained 6.4 million images, 87.6 million GTs, and 273.4 million interactive masks, covering 14 imaging modalities and 204 segmentation targets from 110 public datasets and several medical institutions \citep{cheng2024interactivemedicalimagesegmentation}. It was used for a fine-tuning strategy similar to SAM with an image encoder (ViT), a prompt encoder (accepting text, points, and bounding boxes), and a mask decoder based on transformer. Points and boxes were represented by the sum of positional encoding and learned embeddings as in SAM, while text was encoded using CLIP text encoder \citep{cheng2024interactivemedicalimagesegmentation}.

BrainSegFounder was designed a 3D generalist model for brain tumor and lesion segmentation, with a two-stage self-supervised pre-training strategy followed by fine-tuning \citep{COX2024103301}. Its architecture leveraged Swin-UNETR. BrainSegFounder was pre-trained on the UK Biobank dataset with 82.8K 3D MRI scans in the first stage, and on BraTS and ATLAS 2.0 datasets in the second one. During the first phase, it learned key features such as shapes and sizes of different brain structures, while in the second one disease-specific attributes, such as geometric shapes of tumors and lesions and spatial placements within the brain. Then, the pre-trained encoder was attached to a UNet decoder for fine-tuning on the BraTS and ATLAS 2.0 datasets \citep{COX2024103301}.

The Mixture of Modality Experts (MoME) was proposed as a generalist model for brain lesion segmentation on MRI \citep{10879789}. It consisted of a set of expert networks with encoder-decoder architecture producing a multi-resolution output. A hierarchical gating network was designed to \textbf{fuse} the multi-resolution output from the multiple experts, as a weighted aggregation. A curriculum strategy was applied for model training to gradually transition from specialising each individual expert to tuning the whole model to encourage expert collaboration and refinement. MoME+ extended MoME by accepting more than one single input image, each from a different modality. A trainable dispatch network was designed to address the potential mismatch between the number of input images and the one of experts, in case no image for a specific modality was fed. MoME was assessed on nine datasets (6,585 MRI scans), on eight lesions, on the five most common MRI imaging modalities (T1w, T2w, T1ce, FLAIR, and DWI).

MIS-FM was proposed as a self-supervised strategy to generate paired images and segmentation labels to pre-train 3D medical image segmentation models \citep{wang2023misfm3dmedicalimage}. MIS-FM introduced \textbf{volume fusion} where to sub-volumes cropped from two different 3D scans were merged into a new sub-volume, based on a \textbf{fusion coefficient map}. A 3D segmentaion model, based on CNNs and transformers, was pre-trained with the fused sub-volume to predict the label of each voxel. This model consisted of UNet-like architecture with convolutions for embedding, a pyramid parallel convolution and transformer module to extract local and global features in both the down-sampling and up-sampling path of UNet, and a prediction head. MIS-FM was pre-trained on 110k CTs from public and private datasets. It was tested on three datasets (MICCAI 2015 Head-Neck dataset, SegTHOR, and Synapse) as downstream tasks \citep{wang2023misfm3dmedicalimage}.

Hermes was inspired by radiology residency programs, where the radiologists expertise grows from daily exposure to a wide range of mages across body regions, diseases, and modalities \citep{10658004}. Tokens representing the task and image modality were \textbf{fused} by attention mechanism with the image features, extracted by either a CNN or a transformer. Hermes was trained on 11 datasets (2,438 3D  volumes). It was evaluated for generalization on two datasets \citep{10658004}.

One-Prompt was developed as a generalist model with one-shot learning \citep{10655138}. It consisted of an image encoder and a sequence of one-prompt former modules as decoder. The encoder was designed for three inputs (the query image to be segmented, a template image, and the prompt on the template image). It could be either a CNN or a transformer. The encoded query and template features were sent to the one-prompt former, consisting of two parallel branches of cross-attention. A final cross-attention transferred the prompted template segmentation to the query domain. Finally, a self-attention followed by feedforward neural network were employed to project the embedding to generate the segmentation mask. One-Prompt was trained and tested on 64 and 14 public datasets, respectively \citep{10655138}.

Deep Self-Distillation (DeSD) was proposed as a self-supervised approach to reformulate self-distillation by subdividing the student model into four sub-encoders, each of which was trained to match the features produced by the teacher network \citep{10.1007/978-3-031-16440-8_52}. The student and teacher encoders were based on 3D ResNet-50, with decoder blocks to restore the image resolution, and one atrous spatial pyramid pooling module between the encoder ad decoder for \textbf{multi-scale fusion} \citep{chen2017rethinking}. DeSD was pre-trained on DeepLesion dataset with 10,594 CTs, and evaluated on seven datasets \citep{10.1007/978-3-031-16440-8_52}.

Self-distilled Masked Image Transformer (SMIT) was developed as a self-distillation method with masked image modeling to perform self-supervised learning on ViT \citep{10.1007/978-3-031-16440-8_53}. Two augmented views of 3D image patches were fed to a student (with masking) and a teacher (without masking) networks. The masked image modeling tasks included masked image prediction to recover the masked image, and masked patch token distillation such that the student predicts the tokens of the teacher model. SMIT was pre-trained on 3,643 CTs, fine-tuned on BTCV, and on a dataset of MRI of abdominal organs \citep{10.1007/978-3-031-16440-8_53}.

Med3D was designed for multi-organ segmentation on 3DSeg-8, a dataset mixing eight partially labeled datasets of CT and MRI images \citep{chen2019med3dtransferlearning3d}. Its architecture consisted of a shared encoder, based on a ResNet, connected to eight decoders, based on a convolutional layer, one of which to segment a specific organ. Med3D was evaluated on new tasks, e.g., lung segmentation and pulmonary nodule segmentation \citep{chen2019med3dtransferlearning3d}.

However, a multi-head network with a shared encoder and multiple specific decoders as Med3D was not flexible since a new decoder must be attached for each new segmentation task. Therefore, other architectures have been designed.
One example was the Dynamic on-demand Network (DoDNet) to segment multiple organs and tumors on partially labeled datasets \citep{zhang2020dodnetlearningsegmentmultiorgan}. Its architecture leveraged an encoder and a decoder in a U-shape configuration, a dynamic filter generating module, and a dynamic head. The encoder and decoder consisted of 3D residual convolutional layers. The dynamic filter generating module, based on a convolutional layer, was fed with the concatenation between the extracted features by the encoder with the encoded information of the specific task, and generated a different kernel for each task, dynamically selected during inference. These dynamic kernels were sent to a dynamic head, consisting of convolutional kernels, to enable specific kernels to be assigned to each segmentation task of a specific organ and tumors. DoDNet was pre-trained on 1,155 CTs of seven partially labeled datasets and evaluated on a multi-organ dataset (BTCV) \citep{zhang2020dodnetlearningsegmentmultiorgan}.

\citet{valanarasu2023disruptiveautoencodersleveraginglowlevel} proposed Disruptive Autoencoders (DAE) as a pre-trainaed method, by integrating Swin-UNETR with a combination of downsampling, noise, and local masking to extract features from a wide range of conditions commonly found in medical imaging, e.g., low-resolution, blurring, and sharp details \citep{valanarasu2023disruptiveautoencodersleveraginglowlevel}. Downsapling and noise were initially added to 3D medical images, followed by tokenization, and local masking. The result was fed into Swin-UNETR. This model was pretrained on 10k 3D radiological volumes from CT and four modalities of MRI. A cross-modal contrastive loss function was designed to maximize features of the same modality and minimize those of different modalities. The pre-trained model was fine-tuned on two different datasets (BTCV and FeTA) \citep{valanarasu2023disruptiveautoencodersleveraginglowlevel}.

UniMiSS was designed as as a self-supervised approach for segmentation of both 2D and 3D medical images, pre-trained on a large set of both 2D images to compensate the lack of 3D images \citep{xie2022unimissuniversalmedicalselfsupervised}. Its U-shape architecture consisted of an encoder, a decoder, and skip connection between them. The encoder and the decoder were based on four and three blocks, respectively, each of which included one switchable patch embedding, converting the input images to either 2D or 3D embedding, and several transformer layers. MiSS was pre-trained with self-distillation with one teacher and one student network. UniMiSS was pre-traines on 5k CT and 109k 2D X-ray images. It was then evaluated on six datasets, two of which for segmentation from 3D volumes from CT and MRI \citep{xie2022unimissuniversalmedicalselfsupervised}. 

\textit{Fusion level: F$_4$ for UniSeg, Universal Segmentation model for 33 structures, MoME, MIS-FM, Hermes, and DeSD.}

\subsection{Models trained on both text and image data}
\label{subsec:models-image-text-training}

The generalist models discussed so far primarily leverage mask-based labels and image data alone for pre-training. These models demonstrate remarkable capabilities in learning universal image representations and achieving SOTA performance in various segmentation tasks. However, while these image-centric approaches excelled at capturing visual patterns and anatomical structures from large image datasets, they inherently operated within the visual domain. Acknowledging the rich semantic information embedded within medical texts, such as radiology reports and clinical notes, the field explored generalist models that incorporate language to further enrich their understanding and broaden their applicability. This shift recognized that medical image understanding is not solely a visual task, but deeply intertwined with textual context and expert knowledge. Or rather, that image-based and text-based tasks can benefit from cross-domain fusion. The subsequent section will delve into models that actively leverage text, often through CLIP-inspired contrastive learning frameworks, to create more semantically aware and versatile medical imaging generalist models \citep{clip2021openai}.\\

\citet{du2024segvol_arxiv, du2024segvol} developed SegVol, the first generalist model for volumetric medical image segmentation. SegVol architecture was inspired by SAM, being designed with an image encoder, a spatial encoder supporting points and bounding boxes as prompts, and a decoder. Moreover, it added a semantic encoder leveraging the text encoder of CLIP to accept textual prompts. SegVol supported multi-prompt, e.g., bounding box plus text, or point plus text prompts. The embedded features from the image, spatial, and semantic encoder were \textbf{fused} into embeddings sent as input to the mask decoder. To improve the precision of segmentation a zoom-out-zoom in mechanism was proposed. A zoom-out process, resizing a volumetric image, was initially performed for a coarse segmentation mask. During the zoom-in phase, the region of interest of the original image was cropped, using a sliding window for precise inference guided by prompts generated from the coarse segmentation mask. Finally, the region of interest of the prediction mask was be back-filled to the coarse segmentation mask to generate the final prediction \citep{du2024segvol_arxiv, du2024segvol}. SegVol was trained on 90K unlabeled CT volumes and 6K labeled CT volumes, fine-tuned on 6K labeled CTs with 150K labeled segmentation masks, and tested on 22 anatomical segmentation tasks with several large datasets, e.g., the AMOS22, the Universal Lesion Segmentation Challenge 23, and the SegTHOR \citep{du2024segvol_arxiv, du2024segvol}. 

The Prior Category Network (PCNet) was developed to exploit a prior category knowledge to enhance the segmentation \citep{10510478}. Prior category prompts were crafted to identify the specific organ and to provide information about anatomical structure and inter-category relationships. Additionally, a hierarchy category system was designed to combine organs, anatomical structures, and functional systems. A text branch within prior category prompts generated CLIP embeddings for each organ. These embeddings were combined with the image features from an image encoder through an attention module. PCNet was trained on TotalSeg dataset with different visual backbones (e.g., UNet, and VNet) and evaluated for transferability on 12 datasets \citep{10510478}.

The CLIP-Driven Universal Model by \citet{10376801} exploited the CLIP-generated text embeddings to learn semantically meaningful relationships between anatomical structures for the segmentation of partially labeled datasets. The language branch firstly generated the language embedding for each organ, taken by a multi-layer perceptron to generate a parameter for each class. The CLIP-Driven Universal Model accepted different visual backbones, both transformers like Swin UNETR and CNN. The features extracted by the visual encoder were combined with the text encoder according to the CLIP architecture, while the output of the visual decoder was combined with the parameter generated by the multi-layer perceptron to predict the segmentation map. CLIP-Driven Universal Model was trained on 14 datasets for 25 organs and six types of tumors (3,410 CT scans), and evaluated on four additional datasets (6,173 external CT scans) \citep{10376801}.

Merlin was developed as a generalist models combining CTs, electronic health records, and radiology reports for different tasks including segmentation on 20 organs \citep{blankemeier2024merlinvisionlanguagefoundation}. It was pre-trained by a supervised learning strategy consisting of CT scans encoded by an image encoder with electronic health records as labels, and contrastive learning between radiology reports and CT scans. For segmentation the Merlin architecture was adapted by matching the vision encoder (a ResNet-152) with the decoder of a UNet. Merlin was trained on 15k CTs, validated on 5k CTs, tested internally on 5k CTs and externally on 7k CTs from VerSe and Total Segmentator datasets \citep{blankemeier2024merlinvisionlanguagefoundation}.

BiomedParse was developed by Microsoft as a holisitc approach for medical imaging analysis tasks like segmentation, detection, and recognition \citep{zhao2024biomedparsebiomedicalfoundationmodel,Zhao_2024}. It enabled text-prompted segmentation, without the need for manual bounding box annotations. This was made possible by the creation of BiomedParseData, aggregating 45 publicly available biomedical segmentation datasets, encompassing 1.1 million images across nine imaging modalities and 25 anatomical sites. A key insight was the exploitation of GPT-4 to create a biomedical ontology to overcome the issue of noisy and inconsistent textual description associated with those segmentation datasets. To enhance the capability of BiomedParse to handle diverse text prompt, GPT-4 was used to synthesize synonymous text description, doubling the size of the image-mask-description to 6.8 million \citep{zhao2024biomedparsebiomedicalfoundationmodel}. BiomedParse architecture consisted of an image encoder, a text encoder, a mask decoder to generate segmentation masks from the image and text representations, and a meta-object classifier to facilitate joint training of the image encoder with object semantics. The image encoder was initialized with Focal, a SOTA fully-convolutional image encoder based on focal modulation \citep{yang2022focalmodulationnetworks,yang2022focal} (a more efficient alternative to self-attention for CNNs), while PubMedBERT was chosen for initialization of the text encoder \citep{zhao2024biomedparsebiomedicalfoundationmodel}.

\begin{figure*}[!ht]
  \centering
    \includegraphics[width=1\linewidth]{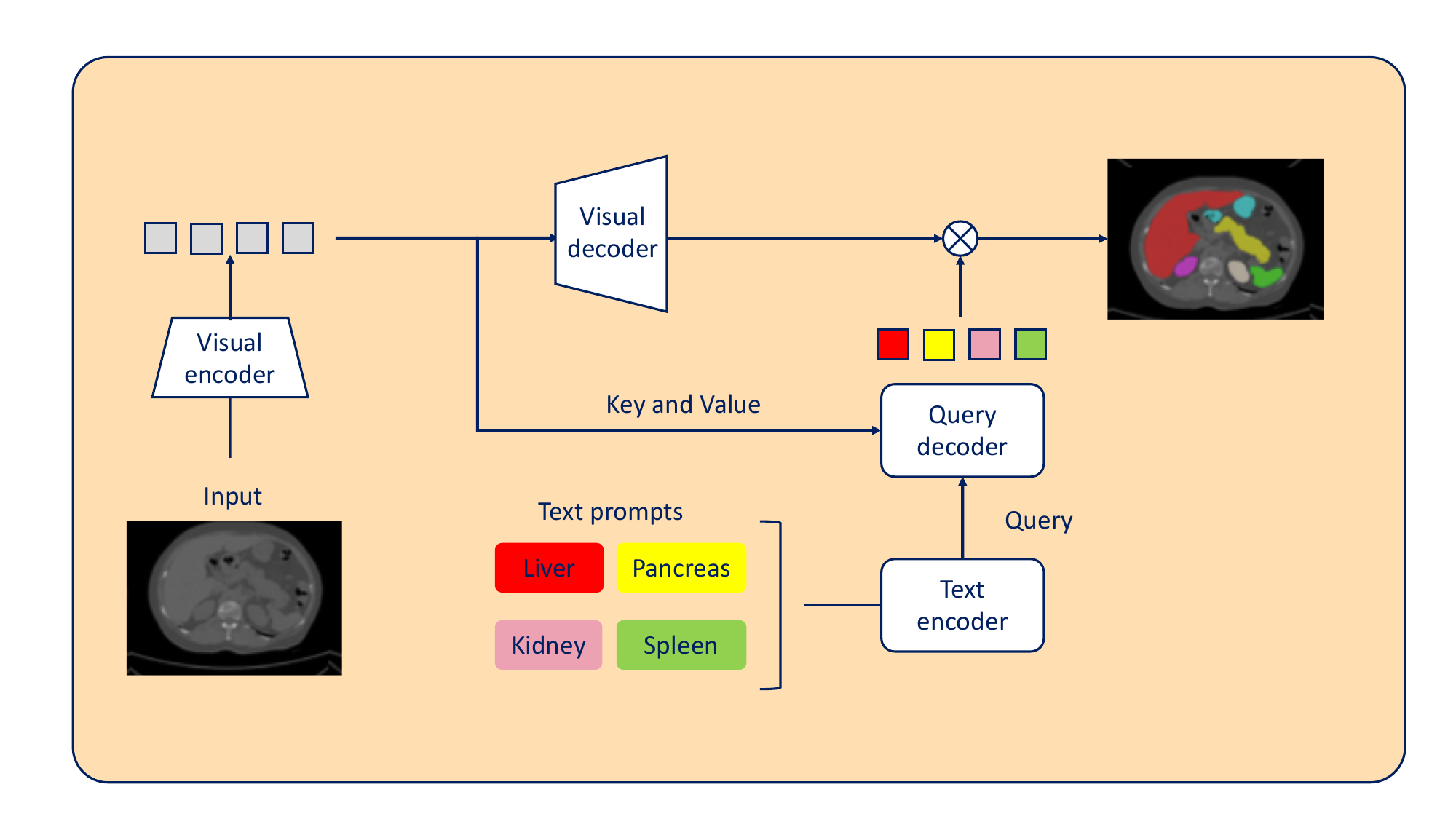}
    \caption{Architecture of SAT. Image adapted from \citep{zhao2025modelrulealluniversal}.}
    \label{fig:sat}
\end{figure*}

\citet{zhao2025modelrulealluniversal} presented Segment Anything with Text (SAT), a segmentation model for 3D medical images with text prompt. its architecture is shown in Fig. \ref{fig:sat}. SAT was pretrained on a multimodal knowledge tree on anatomy concepts and definitions, by linking the visual regions from the image dataset to the corresponding concepts represented in textual form in the text dataset. The image dataset included 22k 3D scans from 72 publicly available datasets, encompassing 497 segmentation classes across eight body regions. The text dataset was based on the the anatomical concepts and definitions collected by acquiring textual knowledge from the Unified Medical Language System a comprehensive medical knowledge graph with concept definitions their relations, and complemented by using search engines. GPT-4 was also used to extract the relations between the anatomical structures \citep{zhao2025modelrulealluniversal}. SAT architecture was designed with a 3D UNet as visual encoder and decoder linked by skip connections, a BERT text encoder pre-trained on PubMed abstracts, and a query decoder to address the visual variations among patients. The visual and text encoders were initially pre-trained by contrastive learning using the multimodal anatomical knowledge tree. The query decoder was a transformer-based query coupled with the multi-scale features from the UNet encoder as keys-values. The output of the query decoder and UNet decoder were multiplied to yield the segmentation mask \citep{zhao2025modelrulealluniversal}.

\clearpage

\newcommand{\selectTableModelsFont}{\fontfamily{cmss}\fontsize{6}{7.5}\selectfont}


\newcommand{\TMncols}{7}

\newcommand{\TMcwModel}{4.5cm}
\newcommand{\TMcwAffiliations}{1.6cm}
\newcommand{\TMcwDate}{0.8cm}
\newcommand{\TMcwPublication}{1.9cm}
\newcommand{\TMcwCit}{2.2cm}
\newcommand{\TMcwCitInner}{1.6cm}
\newcommand{\TMcwDatePublicationCit}{4.7cm}
\newcommand{\TMcwCode}{1.0cm}
\newcommand{\TMcwArchitecture}{2.9cm}
\newcommand{\TMcwParams}{1.5cm}
\newcommand{\TMcwResources}{2.3cm}


\newcommand{\TableModelsHeading}[1]{
    \rowcolor{#1}
        \parbox[c]{\TMcwModel}{
                \textbf{Model \\
                \textit{Paper Title}}
            } & 
        \parbox[c]{\TMcwAffiliations}{
                \textbf{Reseach Group\\
                Nationality}
            } &
        \parbox[c]{\TMcwDatePublicationCit}{
            \centering%
            \vspace{5pt}%
            \textbf{First Publication\\
            \textit{Last Publication}}\\
            \begin{tabular}{p{\TMcwDate} p{\TMcwPublication} p{\TMcwCit}}
                \hspace{-2pt}\textbf{Date} & \hspace{-2pt}\textbf{Publication} & \hspace{-2pt}\textbf{Reference}
            \end{tabular}
            } &
        \textbf{Code} &
        \parbox[c]{\TMcwArchitecture}{
                \textbf{Architecture \\
                \textit{(Visual Backbone)}}
            } &
        \parbox[c]{\TMcwParams}{
                \textbf{N. Params (M)\\
                \textit{GFLOPS}}
            } &
        \parbox[c]{\TMcwResources}{
                \textbf{Computing \\
                Resources}
            } \\
}

\newcommand{\makeTableModelFirstHeading}[1]{ 
    \hline
    \TableModelsHeading{#1}
    \hline
}

\newcommand{\makeTableModelOtherHeadings}[1]{ 
    \makeTableHeaderContinued{\TMncols} 
    \makeTableModelFirstHeading{#1}
}


\def\codeUnavailableSize{3.0mm}
\def\codeUnavailable{
\begin{tikzpicture}
 \begin{scope}[line width=2pt, cap=round, color=colorCodeUnavailable]
  \draw (0mm,0mm) -- (\codeUnavailableSize,\codeUnavailableSize);
  \draw (0mm,\codeUnavailableSize) -- (\codeUnavailableSize,0mm);
 \end{scope}
\end{tikzpicture}
}
\def\putIconGithub{\includegraphics[height=3.8mm]{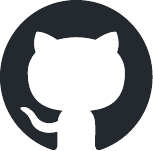}}


\newcommand{\makeTableModels}[6]{

    {
    \selectTableModelsFont
    \rowcolors{1}{white}{#2}
    \begin{center}
    \begin{longtable}[c]{
        p{\TMcwModel} 
        p{\TMcwAffiliations} 
        p{\TMcwDatePublicationCit} 
        C{\TMcwCode} 
        p{\TMcwArchitecture} 
        p{\TMcwParams} 
        p{\TMcwResources}
    }
    
        \hiderowcolors
        \caption{
            #3
        }
        \label{#4} \\
        \showrowcolors
    
        \makeTableModelFirstHeading{#1}
        \endfirsthead
        
        \makeTableModelOtherHeadings{#1}
        \endhead
        
        \makeTableOtherFooters{\TMncols}
        \endfoot
        
        \makeTableLastFooter{\TMncols}{}
        \endlastfoot
    
        #5
    
    \end{longtable}
    \outOfTableComments{
        \hspace{-1.5cm}\parbox{15cm}{#6}
    }
    \end{center}
    }
}

\newcommand{\TMcellvspacer}{\vspace{4pt}}

\newcommand{\TMformatModel}[2]{
    \parbox[c]{\TMcwModel}{
        \TMcellvspacer
        \textbf{#1} \vspace{2pt}\\
        \textit{#2}
        \TMcellvspacer
    }
}
\newcommand{\TMformatAffil}[1]{
    \parbox[c]{\TMcwAffiliations}{
        \TMcellvspacer
        #1
        \TMcellvspacer
    }
}
\newcommand{\TMformatDatePubliCit}[6]{
    {\def\arraystretch{1.2}%
        \begin{tabular}{%
            @{}p{\TMcwDate}%
            >{\raggedright\arraybackslash}p{\TMcwPublication}%
            >{\raggedright\arraybackslash}p{\TMcwCitInner}@{}%
        }
             #1 & #2 & #3 \\
            \textit{#4} & \textit{#5} & \textit{#6}\\
        \end{tabular}
    }%
}
\newcommand{\TMformatArchitecture}[1]{
    \parbox[c]{\TMcwArchitecture}{
        \TMcellvspacer
        #1
        \TMcellvspacer
    }
}
\newcommand{\TMformatParams}[1]{
    \parbox[c]{\TMcwParams}{
        \TMcellvspacer
        #1
        \TMcellvspacer
    }
}
\newcommand{\TMformatRes}[1]{
    \parbox[c]{\TMcwResources}{
        \TMcellvspacer
        #1
        \TMcellvspacer
    }
}

\newcommand{\TMflag}[1]{%
    \includegraphics[height=5pt]{#1}~%
}
\newcommand{\TMflagAustralia}{\TMflag{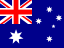}}
\newcommand{\TMflagCanada}{\TMflag{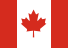}}
\newcommand{\TMflagChina}{\TMflag{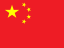}}
\newcommand{\TMflagFrance}{\TMflag{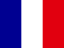}}
\newcommand{\TMflagGermany}{\TMflag{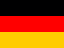}}
\newcommand{\TMflagJapan}{\TMflag{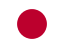}}
\newcommand{\TMflagSingapore}{\TMflag{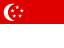}}
\newcommand{\TMflagSwitzerland}{\TMflag{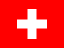}}
\newcommand{\TMflagUAE}{\TMflag{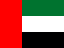}}
\newcommand{\TMflagUK}{\TMflag{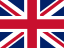}}
\newcommand{\TMflagUSA}{\TMflag{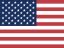}}

\newgeometry{left=1cm, right=1cm} 
    \makeTableModels{
    colorTablePapersFoundationalHeaderDark%
}{
    colorTablePapersFoundationalRowsPattern%
}{
    Reviewed generalist models for 3D medical image segmentation.
}{table:models-foundational}{
    \TMformatModel{MedSAM2}{MedSAM2: Segment Anything in 3D Medical Images and Videos} & \TMformatAffil{\TMflagCanada Canada\vspace{2pt}\\\TMflagUSA U.S.A.} & \TMformatDatePubliCit{\href{https://doi.org/10.48550/arXiv.2504.03600}{2025-04}}{\href{https://doi.org/10.48550/arXiv.2504.03600}{arXiv}}{\citet{ma2025medsam2segment3dmedical}}{-}{-}{-} & \href{https://github.com/bowang-lab/MedSAM2}{\putIconGithub} & \TMformatArchitecture{Transformer with Convolutions (SAM 2)\\\textit{(SAM 2)}} & \TMformatParams{38.9\\-} & \TMformatRes{12 Nvidia H100 80GB}\\
    \TMformatModel{SPA}{SPA: Leveraging the SAM with Spatial Priors Adapter for Enhanced Medical Image Segmentation} & \TMformatAffil{\TMflagJapan Japan} & \TMformatDatePubliCit{\href{https://doi.org/10.1109/JBHI.2025.3526174}{2025-01}}{\href{https://doi.org/10.1109/JBHI.2025.3526174}{IEEE Journal of Biomedical Health Informatics}}{\citet{10829779}}{-}{-}{-} & \codeUnavailable & \TMformatArchitecture{Transformer with Convolutions (SAM)\\\textit{(SAM)}} & \TMformatParams{6.88\\\textit{8.81}} & \TMformatRes{1 Nvidia GeForce RTX 3090 24GB}\\
    \TMformatModel{KnowSAM}{Learnable Prompting SAM-induced Knowledge Distillation for Semi-supervised Medical Image Segmentation} & \TMformatAffil{\TMflagChina China} & \TMformatDatePubliCit{\href{https://doi.org/10.48550/arXiv.2412.13742}{2024-12}}{\href{https://doi.org/10.48550/arXiv.2412.13742}{arXiv}}{\citet{huang2024learnablepromptingsaminducedknowledge}}{\href{https://doi.org/10.1109/TMI.2025.3530097}{2025-01}}{\href{https://doi.org/10.1109/TMI.2025.3530097}{IEEE Transactions on Medical Imaging}}{\citet{10843257}} & \href{https://github.com/taozh2017/KnowSAM}{\putIconGithub} & \TMformatArchitecture{Transformer with Convolutions (SAM)\\\textit{(SAM)}} & \TMformatParams{-\\-} & \TMformatRes{1 Nvidia GeForce RTX 3090 24GB}\\
    \TMformatModel{3DMedSAM}{Volumetric medical image segmentation via fully 3D adaptation of Segment Anything Model} & \TMformatAffil{\TMflagChina China\vspace{2pt}\\\TMflagUK U.K.} & \TMformatDatePubliCit{\href{https://doi.org/10.1016/j.bbe.2024.11.001}{2024-12}}{\href{https://doi.org/10.1016/j.bbe.2024.11.001}{Biocybernetics and Biomedical Engineering}}{\citet{LIN20251}}{-}{-}{-} & \codeUnavailable & \TMformatArchitecture{Transformer with Convolutions (SAM)\\\textit{(SAM)}} & \TMformatParams{-\\-} & \TMformatRes{1 Nvidia GeForce RTX 3090 Ti 24GB}\\
    \TMformatModel{IMIS-Net}{Interactive Medical Image Segmentation: A Benchmark Dataset and Baseline} & \TMformatAffil{\TMflagChina China} & \TMformatDatePubliCit{\href{https://doi.org/10.48550/arXiv.2411.12814}{2024-11}}{\href{https://doi.org/10.48550/arXiv.2411.12814}{arXiv}}{\citet{cheng2024interactivemedicalimagesegmentation}}{-}{-}{-} & \href{https://github.com/uni-medical/IMIS-Bench}{\putIconGithub} & \TMformatArchitecture{Transformer\\\textit{(2D ViT-Base)}} & \TMformatParams{29.68\\-} & \TMformatRes{72 Nvidia GeForce RTX 4090 24GB}\\
    \TMformatModel{SAM-MPA}{SAM-MPA: Applying SAM to Few-shot Medical Image Segmentation using Mask Propagation and Auto-prompting} & \TMformatAffil{\TMflagChina China} & \TMformatDatePubliCit{\href{https://openreview.net/forum?id=IjZI80PUdr}{2024-10}}{\href{https://openreview.net/forum?id=IjZI80PUdr}{NeurIPS}}{\citet{xu2024sammpa}}{\href{https://doi.org/10.48550/arXiv.2411.17363}{2024-11}}{\href{https://doi.org/10.48550/arXiv.2411.17363}{arXiv}}{\citet{xu2024sammpaapplyingsamfewshot}} & \codeUnavailable & \TMformatArchitecture{Transformer with Convolutions (SAM)\\\textit{(SAM)}} & \TMformatParams{-\\-} & \TMformatRes{2 Nvidia V100 32GB}\\
    \TMformatModel{TP-Mamba}{Tri-Plane Mamba: Efficiently Adapting Segment Anything Model for3D Medical Images} & \TMformatAffil{\TMflagChina China} & \TMformatDatePubliCit{\href{https://doi.org/10.48550/arXiv.2409.08492}{2024-09}}{\href{https://doi.org/10.48550/arXiv.2409.08492}{arXiv}}{\citet{wang2024triplanemambaefficientlyadapting}}{\href{https://doi.org/10.1007/978-3-031-72114-4_61}{2024-10}}{\href{https://doi.org/10.1007/978-3-031-72114-4_61}{MICCAI}}{\citet{10.1007/978-3-031-72114-4_61}} & \href{https://github.com/xmed-lab/TP-Mamba}{\putIconGithub} & \TMformatArchitecture{Transformer with Convolutions (SAM)\\\textit{(SAM)}} & \TMformatParams{-\\-} & \TMformatRes{-}\\
    \TMformatModel{SAM 2}{SAM 2: Segment Anything in Images and Videos} & \TMformatAffil{\TMflagUSA U.S.A.} & \TMformatDatePubliCit{\href{https://doi.org/10.48550/arXiv.2408.00714}{2024-08}}{\href{https://doi.org/10.48550/arXiv.2408.00714}{arXiv}}{\citet{ravi2024sam2segmentimages}}{\href{https://openreview.net/forum?id=Ha6RTeWMd0}{2025-01}}{\href{https://openreview.net/forum?id=Ha6RTeWMd0}{ICLR}}{\citet{ravi2025sam}} & \href{https://github.com/facebookresearch/sam2}{\putIconGithub} & \TMformatArchitecture{Transformer with Convolutions\\\textit{(SAM)}} & \TMformatParams{636.0\\-} & \TMformatRes{256 Nvidia A100 80GB}\\
    \TMformatModel{Medical SAM 2 (MedSAM-2)}{Medical SAM 2: Segment medical images as video via Segment Anything Model 2} & \TMformatAffil{\TMflagUK U.K.} & \TMformatDatePubliCit{\href{https://doi.org/10.48550/arXiv.2408.00874}{2024-08}}{\href{https://doi.org/10.48550/arXiv.2408.00874}{arXiv}}{\citet{zhu2024medicalsam2segment}}{-}{-}{-} & \href{https://github.com/SuperMedIntel/Medical-SAM2}{\putIconGithub} & \TMformatArchitecture{Transformer with Convolutions (SAM 2)\\\textit{(SAM 2)}} & \TMformatParams{-\\-} & \TMformatRes{64 Nvidia A100 80GB}\\
    \TMformatModel{EMedSAM}{An efficient segment anything model for the segmentation of medical images} & \TMformatAffil{\TMflagChina China} & \TMformatDatePubliCit{\href{https://www.nature.com/articles/s41598-024-70288-8}{2024-08}}{\href{https://www.nature.com/articles/s41598-024-70288-8}{Scientific Reports}}{\citet{Dong2024}}{-}{-}{-} & \codeUnavailable & \TMformatArchitecture{Transformer with Convolutions (SAM)\\\textit{(SAM)}} & \TMformatParams{21.0\\-} & \TMformatRes{2 Nvidia A100 80GB}\\
    \TMformatModel{Biomedical SAM-2 (BioSAM-2)}{Biomedical SAM 2: Segment Anything in Biomedical Images and Videos} & \TMformatAffil{\TMflagUSA U.S.A.} & \TMformatDatePubliCit{\href{https://doi.org/10.48550/arXiv.2408.03286}{2024-08}}{\href{https://doi.org/10.48550/arXiv.2408.03286}{arXiv}}{\citet{yan2024biomedicalsam2segment}}{\href{https://openreview.net/forum?id=xaPv4b8z2D}{2024-10}}{\href{https://openreview.net/forum?id=xaPv4b8z2D}{NeurIPS}}{\citet{yan2024biomedical}} & \href{https://github.com/ZhilingYan/Biomedical-SAM-2}{\putIconGithub} & \TMformatArchitecture{Transformer with Convolutions (SAM 2)\\\textit{(SAM 2)}} & \TMformatParams{-\\-} & \TMformatRes{-}\\
    \TMformatModel{FLAP-SAM}{A Federated Learning-Friendly Approach for Parameter-Efficient Fine-Tuning of SAM in 3D Segmentation} & \TMformatAffil{\TMflagUAE U.A.E.} & \TMformatDatePubliCit{\href{https://doi.org/10.48550/arXiv.2407.21739}{2024-07}}{\href{https://doi.org/10.48550/arXiv.2407.21739}{arXiv}}{\citet{asokan2024federatedlearningfriendlyapproachparameterefficient}}{\href{https://doi.org/10.1007/978-3-031-77610-6_21}{2025-01}}{\href{https://doi.org/10.1007/978-3-031-77610-6_21}{MICCAI}}{\citet{10.1007/978-3-031-77610-6_21}} & \href{https://github.com/BioMedIA-MBZUAI/FLAP-SAM}{\putIconGithub} & \TMformatArchitecture{Transformer with Convolutions (SAM)\\\textit{(SAM)}} & \TMformatParams{91.767\\-} & \TMformatRes{1 Nvidia A100 40GB}\\
    \TMformatModel{Merlin}{Merlin: A Vision Language Foundation Model for 3D Computed Tomography} & \TMformatAffil{\TMflagUSA U.S.A.} & \TMformatDatePubliCit{\href{https://doi.org/10.48550/arXiv.2406.06512}{2024-06}}{\href{https://doi.org/10.48550/arXiv.2406.06512}{arXiv}}{\citet{blankemeier2024merlinvisionlanguagefoundation}}{-}{-}{-} & \href{https://github.com/StanfordMIMI/Merlin}{\putIconGithub} & \TMformatArchitecture{ConvNet\\\textit{(3D ResNet-152)}} & \TMformatParams{-\\-} & \TMformatRes{1 Nvidia RTX A6000 48GB}\\
    \TMformatModel{LeSAM}{LeSAM: Adapt Segment Anything Model for Medical Lesion Segmentation} & \TMformatAffil{\TMflagChina China} & \TMformatDatePubliCit{\href{https://doi.org/10.1109/JBHI.2024.3406871}{2024-06}}{\href{https://doi.org/10.1109/JBHI.2024.3406871}{IEEE Journal of Biomedical and Health Informatics}}{\citet{10540651}}{-}{-}{-} & \codeUnavailable & \TMformatArchitecture{Transformer with Convolutions (SAM)\\\textit{(SAM)}} & \TMformatParams{-\\-} & \TMformatRes{1 Nvidia GeForce RTX 4090 24GB}\\
    \TMformatModel{BrainSegFounder}{BrainSegFounder: Towards 3D foundation models for neuroimage segmentation} & \TMformatAffil{\TMflagUSA U.S.A.} & \TMformatDatePubliCit{\href{https://doi.org/10.48550/arXiv.2406.10395}{2024-06}}{\href{https://doi.org/10.48550/arXiv.2406.10395}{arXiv}}{\citet{cox2024brainsegfounder3dfoundationmodels}}{\href{https://doi.org/10.1016/j.media.2024.103301}{2024-08}}{\href{https://doi.org/10.1016/j.media.2024.103301}{Medical Image Analysis}}{\citet{COX2024103301}} & \href{https://github.com/lab-smile/BrainSegFounder}{\putIconGithub} & \TMformatArchitecture{Transformer\\\textit{(SwinUNETR)}} & \TMformatParams{69.0\\-} & \TMformatRes{64 Nvidia A100 80GB}\\
    \TMformatModel{MoME}{A Foundation Model for Lesion Segmentation on Brain MRI with Mixture of Modality Experts} & \TMformatAffil{\TMflagChina China\vspace{2pt}\\\TMflagUK U.K.} & \TMformatDatePubliCit{\href{https://doi.org/10.48550/arXiv.2405.10246}{2024-05}}{\href{https://doi.org/10.48550/arXiv.2405.10246}{arXiv}}{\citet{zhang2024foundationmodelbrainlesion}}{\href{https://doi.org/10.1109/TMI.2025.3540809}{2025-02}}{\href{https://doi.org/10.1109/TMI.2025.3540809}{IEEE Transactions on Medical Imaging}}{\citet{10879789}} & \href{https://github.com/ZhangxinruBIT/MoME}{\putIconGithub} & \TMformatArchitecture{ConvNet\\\textit{(nnU-Net framework)}} & \TMformatParams{-\\-} & \TMformatRes{1 Nvidia A100 80GB}\\
    \TMformatModel{BiomedParse}{A foundation model for joint segmentation, detection and recognition of biomedical objects across nine modalities} & \TMformatAffil{\TMflagUSA U.S.A.} & \TMformatDatePubliCit{\href{https://doi.org/10.48550/arXiv.2405.12971}{2024-05}}{\href{https://doi.org/10.48550/arXiv.2405.12971}{arXiv}}{\citet{Zhao2025}}{\href{https://doi.org/10.1038/s41592-024-02499-w}{2024-11}}{\href{https://doi.org/10.1038/s41592-024-02499-w}{Nature Methods}}{\citet{Zhao2025}} & \href{https://microsoft.github.io/BiomedParse/}{\putIconGithub} & \TMformatArchitecture{ConvNet with Focal Modulation\\\textit{(2D FocalNet (custom size))}} & \TMformatParams{-\\-} & \TMformatRes{-}\\
    \TMformatModel{PCNet}{PCNet: Prior Category Network for CT Universal Segmentation Model} & \TMformatAffil{\TMflagChina China} & \TMformatDatePubliCit{\href{https://doi.org/10.1109/TMI.2024.3395349}{2024-04}}{\href{https://doi.org/10.1109/TMI.2024.3395349}{IEEE Transactions on Medical Imaging}}{\citet{10510478}}{-}{-}{-} & \href{https://github.com/PKU-MIPET/PCNet}{\putIconGithub} & \TMformatArchitecture{ConvNet with Attention\\\textit{(STUNet)}} & \TMformatParams{441.54\\-} & \TMformatRes{1 Nvidia A800 80GB}\\
    \TMformatModel{SFR SAM}{Stitching, Fine-tuning, Re-training: A SAM-enabled Framework for Semi-supervised 3D Medical Image Segmentation} & \TMformatAffil{\TMflagChina China} & \TMformatDatePubliCit{\href{https://doi.org/10.48550/arXiv.2403.11229}{2024-03}}{\href{https://doi.org/10.48550/arXiv.2403.11229}{arXiv}}{\citet{li2025stitchingfinetuningretrainingsamenabled}}{\href{https://doi.org/10.1109/TMI.2025.3532084}{2025-01}}{\href{https://doi.org/10.1109/TMI.2025.3532084}{IEEE Transactions on Medical Imaging}}{\citet{10847777}} & \href{https://github.com/ShumengLI/SFR}{\putIconGithub} & \TMformatArchitecture{Transformer with Convolutions (SAM)\\\textit{(SAM)}} & \TMformatParams{18.0\\-} & \TMformatRes{1 Nvidia GeForce RTX 4090 Ti 24GB}\\
    \TMformatModel{MEA M-SAM}{Mask-Enhanced Segment Anything Model forTumor Lesion Semantic Segmentation} & \TMformatAffil{\TMflagChina China} & \TMformatDatePubliCit{\href{https://doi.org/10.48550/arXiv.2403.05912}{2024-03}}{\href{https://doi.org/10.48550/arXiv.2403.05912}{arXiv}}{\citet{shi2024maskenhancedsegmentmodeltumor}}{\href{https://doi.org/10.1007/978-3-031-72111-3_38}{2024-10}}{\href{https://doi.org/10.1007/978-3-031-72111-3_38}{MICCAI}}{\citet{10.1007/978-3-031-72111-3_38}} & \codeUnavailable & \TMformatArchitecture{Transformer with Convolutions (SAM)\\\textit{(SAM)}} & \TMformatParams{118.0\\-} & \TMformatRes{1 Nvidia V100 32GB}\\
    \TMformatModel{SAT}{One Model to Rule them All: Towards Universal Segmentation for Medical Images with Text Prompts} & \TMformatAffil{\TMflagChina China} & \TMformatDatePubliCit{\href{https://arxiv.org/abs/2312.17183v1}{2023-12}}{\href{https://arxiv.org/abs/2312.17183v1}{arXiv}}{\citet{zhao2025modelrulealluniversal}}{\href{https://doi.org/10.48550/arXiv.2312.17183}{2025-02}}{\href{https://doi.org/10.48550/arXiv.2312.17183}{arXiv}}{\citet{zhao2025modelrulealluniversal}} & \href{https://github.com/zhaoziheng/SAT}{\putIconGithub} & \TMformatArchitecture{ConvNet\\\textit{(3D U-Net)}} & \TMformatParams{447.0\\-} & \TMformatRes{16 Nvidia A100 80GB}\\
    \TMformatModel{Med-SA}{Medical SAM Adapter: Adapting Segment Anything Model for Medical Image Segmentation} & \TMformatAffil{\TMflagSingapore Singapore\vspace{2pt}\\\TMflagUK U.K.} & \TMformatDatePubliCit{\href{https://doi.org/10.48550/arXiv.2304.12620}{2023-12}}{\href{https://doi.org/10.48550/arXiv.2304.12620}{arXiv}}{\citet{wu2023medicalsamadapteradapting}}{-}{-}{-} & \href{https://github.com/SuperMedIntel/Medical-SAM-Adapter}{\putIconGithub} & \TMformatArchitecture{Transformer with Convolutions (SAM)\\\textit{(SAM)}} & \TMformatParams{363.0\\-} & \TMformatRes{4 Nvidia A100 80GB}\\
    \TMformatModel{SegVol}{SegVol: Universal and Interactive Volumetric Medical Image Segmentation} & \TMformatAffil{\TMflagChina China} & \TMformatDatePubliCit{\href{https://doi.org/10.48550/arXiv.2311.13385}{2023-11}}{\href{https://doi.org/10.48550/arXiv.2311.13385}{arXiv}}{\citet{du2025segvoluniversalinteractivevolumetric}}{\href{https://openreview.net/forum?id=105ZuvpdyW}{2024-09}}{\href{https://openreview.net/forum?id=105ZuvpdyW}{NeurIPS}}{\citet{du2024segvol}} & \href{https://github.com/BAAI-DCAI/SegVol}{\putIconGithub} & \TMformatArchitecture{Transformer\\\textit{(3D ViT-Base)}} & \TMformatParams{181.0\\-} & \TMformatRes{8 Nvidia A100 40GB}\\
    \TMformatModel{SAM-Med3D}{SAM-Med3D: Towards General-purpose Segmentation Models for Volumetric Medical Images} & \TMformatAffil{\TMflagChina China} & \TMformatDatePubliCit{\href{https://doi.org/10.48550/arXiv.2310.15161}{2023-10}}{\href{https://doi.org/10.48550/arXiv.2310.15161}{arXiv}}{\citet{wang2024sammed3dgeneralpurposesegmentationmodels}}{-}{-}{-} & \href{https://github.com/uni-medical/SAM-Med3D}{\putIconGithub} & \TMformatArchitecture{Transformer with Convolutions (SAM)\\\textit{(SAM)}} & \TMformatParams{101.0\\-} & \TMformatRes{2 Nvidia A100 80GB}\\
    \TMformatModel{SAM3D}{SAM3D: Segment Anything Model in Volumetric Medical Images} & \TMformatAffil{\TMflagUSA U.S.A.} & \TMformatDatePubliCit{\href{https://doi.org/10.48550/arXiv.2309.03493}{2023-09}}{\href{https://doi.org/10.48550/arXiv.2309.03493}{arXiv}}{\citet{bui2024sam3dsegmentmodelvolumetric}}{\href{https://doi.org/10.1109/ISBI56570.2024.10635844}{2024-08}}{\href{https://doi.org/10.1109/ISBI56570.2024.10635844}{IEEE ISBI}}{\citet{10635844}} & \href{https://github.com/UARK-AICV/SAM3D}{\putIconGithub} & \TMformatArchitecture{Transformer with Convolutions (SAM)\\\textit{(SAM)}} & \TMformatParams{91.88\\-} & \TMformatRes{1 Nvidia GeForce RTX 2080 Ti 11GB}\\
    \TMformatModel{MA-SAM}{MA-SAM: Modality-agnostic SAM adaptation for 3D medical image segmentation} & \TMformatAffil{\TMflagChina China\vspace{2pt}\\\TMflagUSA U.S.A.} & \TMformatDatePubliCit{\href{https://doi.org/10.48550/arXiv.2309.08842}{2023-09}}{\href{https://doi.org/10.48550/arXiv.2309.08842}{arXiv}}{\citet{chen2023masammodalityagnosticsamadaptation}}{\href{https://doi.org/10.1016/j.media.2024.103310}{2024-08}}{\href{https://doi.org/10.1016/j.media.2024.103310}{Medical Image Analysis}}{\citet{CHEN2024103310}} & \href{https://github.com/cchen-cc/MA-SAM}{\putIconGithub} & \TMformatArchitecture{Transformer with Convolutions (SAM)\\\textit{(SAM)}} & \TMformatParams{638.3\\\textit{783.4}} & \TMformatRes{8 Nvidia A100 80GB}\\
    \TMformatModel{SAM-Med2D}{SAM-Med2D} & \TMformatAffil{\TMflagChina China} & \TMformatDatePubliCit{\href{https://doi.org/10.48550/arXiv.2308.16184}{2023-08}}{\href{https://doi.org/10.48550/arXiv.2308.16184}{arXiv}}{\citet{cheng2023sammed2d}}{-}{-}{-} & \href{https://github.com/OpenGVLab/SAM-Med2D}{\putIconGithub} & \TMformatArchitecture{Transformer with Convolutions (SAM)\\\textit{(SAM)}} & \TMformatParams{271.0\\-} & \TMformatRes{8 Nvidia A100 80GB}\\
    \TMformatModel{Cheap Lunch SAM}{Cheap Lunch for Medical Image Segmentation by Fine-tuning SAM on Few Exemplars} & \TMformatAffil{\TMflagChina China} & \TMformatDatePubliCit{\href{https://doi.org/10.48550/arXiv.2308.14133}{2023-08}}{\href{https://doi.org/10.48550/arXiv.2308.14133}{arXiv}}{\citet{feng2023cheaplunchmedicalimage}}{\href{https://doi.org/10.1007/978-3-031-76160-7_2}{2024-12}}{\href{https://doi.org/10.1007/978-3-031-76160-7_2}{BrainLes}}{\citet{10.1007/978-3-031-76160-7_2}} & \codeUnavailable & \TMformatArchitecture{Transformer with Convolutions (SAM)\\\textit{(SAM)}} & \TMformatParams{-\\-} & \TMformatRes{1 Nvidia GeForce RTX 3090 24GB}\\
    \TMformatModel{SAMMed}{SAMMed: A medical image annotation framework based on large vision model} & \TMformatAffil{\TMflagChina China\vspace{2pt}\\\TMflagUSA U.S.A.} & \TMformatDatePubliCit{\href{https://doi.org/10.48550/arXiv.2307.05617}{2023-07}}{\href{https://doi.org/10.48550/arXiv.2307.05617}{arXiv}}{\citet{wang2023mathrmsammedmedicalimageannotation}}{-}{-}{-} & \codeUnavailable & \TMformatArchitecture{Transformer with Convolutions (SAM)\\\textit{(SAM)}} & \TMformatParams{-\\-} & \TMformatRes{1 Nvidia V100 32GB}\\
    \TMformatModel{Disruptive Autoencoders}{Disruptive Autoencoders: Leveraging Low-level features for 3D Medical Image Pre-training} & \TMformatAffil{\TMflagUSA U.S.A.} & \TMformatDatePubliCit{\href{https://doi.org/10.48550/arXiv.2307.16896}{2023-07}}{\href{https://doi.org/10.48550/arXiv.2307.16896}{arXiv}}{\citet{valanarasu2023disruptiveautoencodersleveraginglowlevel}}{\href{https://proceedings.mlr.press/v250/valanarasu24a.html}{2024-07}}{\href{https://proceedings.mlr.press/v250/valanarasu24a.html}{PMLR}}{\citet{pmlr-v250-valanarasu24a}} & \href{https://github.com/Project-MONAI/research-contributions/tree/main/DAE}{\putIconGithub} & \TMformatArchitecture{Transformer\\\textit{(SwinUNETR)}} & \TMformatParams{-\\-} & \TMformatRes{8 Nvidia V100 32GB in DXG-1 Server}\\
    \TMformatModel{MIS-FM}{MIS-FM: 3D Medical Image Segmentation using Foundation Models Pretrained on a Large-Scale Unannotated Dataset} & \TMformatAffil{\TMflagChina China} & \TMformatDatePubliCit{\href{https://doi.org/10.48550/arXiv.2306.16925}{2023-06}}{\href{https://doi.org/10.48550/arXiv.2306.16925}{arXiv}}{\citet{wang2023misfm3dmedicalimage}}{-}{-}{-} & \href{https://github.com/openmedlab/MIS-FM}{\putIconGithub} & \TMformatArchitecture{Transformer with Convolutions\\\textit{(Custom)}} & \TMformatParams{-\\-} & \TMformatRes{2 Nvidia A100 80GB}\\
    \TMformatModel{MedLSAM}{MedLSAM: Localize and Segment Anything Model for 3D CT Images} & \TMformatAffil{\TMflagChina China} & \TMformatDatePubliCit{\href{https://doi.org/10.48550/arXiv.2306.14752}{2023-06}}{\href{https://doi.org/10.48550/arXiv.2306.14752}{arXiv}}{\citet{lei2024medlsamlocalizesegmentmodel}}{\href{https://doi.org/10.1016/j.media.2024.103370}{2024-10}}{\href{https://doi.org/10.1016/j.media.2024.103370}{Medical Image Analysis}}{\citet{LEI2025103370}} & \href{https://github.com/openmedlab/MedLSAM}{\putIconGithub} & \TMformatArchitecture{Transformer with Convolutions (SAM)\\\textit{(SAM)}} & \TMformatParams{-\\-} & \TMformatRes{4 Nvidia GeForce RTX 3090 Ti 24GB}\\
    \TMformatModel{HERMES}{Training Like a Medical Resident: Context-Prior Learning Toward Universal Medical Image Segmentation} & \TMformatAffil{\TMflagChina China\vspace{2pt}\\\TMflagUSA U.S.A.} & \TMformatDatePubliCit{\href{https://doi.org/10.48550/arXiv.2306.02416}{2023-06}}{\href{https://doi.org/10.48550/arXiv.2306.02416}{arXiv}}{\citet{gao2024traininglikemedicalresident}}{\href{https://doi.org/10.1109/CVPR52733.2024.01064}{2024-09}}{\href{https://doi.org/10.1109/CVPR52733.2024.01064}{IEEE/CVF CVPR}}{\citet{10658004}} & \href{https://github.com/yhygao/universal-medical-image-segmentation}{\putIconGithub} & \TMformatArchitecture{Transformer with Convolutions\\\textit{(MedFormer)}} & \TMformatParams{-\\-} & \TMformatRes{-}\\
    \TMformatModel{DeSAM}{DeSAM: Decoupled Segment Anything Model for Generalizable Medical Image Segmentation} & \TMformatAffil{\TMflagChina China} & \TMformatDatePubliCit{\href{https://doi.org/10.48550/arXiv.2306.00499}{2023-06}}{\href{https://doi.org/10.48550/arXiv.2306.00499}{arXiv}}{\citet{gao2024desamdecoupledsegmentmodel}}{\href{https://doi.org/10.1007/978-3-031-72390-2_48}{2024-10}}{\href{https://doi.org/10.1007/978-3-031-72390-2_48}{MICCAI}}{\citet{10.1007/978-3-031-72390-2_48}} & \href{https://github.com/yifangao112/DeSAM}{\putIconGithub} & \TMformatArchitecture{Transformer with Convolutions (SAM)\\\textit{(SAM)}} & \TMformatParams{-\\-} & \TMformatRes{1 Nvidia GeForce RTX 3060 12GB}\\
    \TMformatModel{3DSAM-adapter}{3DSAM-adapter: Holistic adaptation of SAM from 2D to 3D for promptable tumor segmentation} & \TMformatAffil{\TMflagChina China} & \TMformatDatePubliCit{\href{https://doi.org/10.48550/arXiv.2306.13465}{2023-06}}{\href{https://doi.org/10.48550/arXiv.2306.13465}{arXiv}}{\citet{GONG2024103324}}{\href{https://doi.org/10.1016/j.media.2024.103324}{2024-08}}{\href{https://doi.org/10.1016/j.media.2024.103324}{Medical Image Analysis}}{\citet{GONG2024103324}} & \href{https://github.com/med-air/3DSAM-adapter}{\putIconGithub} & \TMformatArchitecture{Transformer with Convolutions (SAM)\\\textit{(SAM)}} & \TMformatParams{123.8\\\textit{4551.4}} & \TMformatRes{1 Nvidia A40 48GB}\\
    \TMformatModel{One-Prompt}{One-Prompt to Segment All Medical Images} & \TMformatAffil{\TMflagSingapore Singapore\vspace{2pt}\\\TMflagUAE U.A.E.\vspace{2pt}\\\TMflagUK U.K.\vspace{2pt}\\\TMflagUSA U.S.A.} & \TMformatDatePubliCit{\href{https://doi.org/10.48550/arXiv.2305.10300}{2023-05}}{\href{https://doi.org/10.48550/arXiv.2305.10300}{arXiv}}{\citet{wu2024onepromptsegmentmedicalimages}}{\href{https://doi.org/10.1109/CVPR52733.2024.01074}{2024-09}}{\href{https://doi.org/10.1109/CVPR52733.2024.01074}{IEEE/CVF CVPR}}{\citet{10655138}} & \href{https://github.com/SuperMedIntel/one-prompt}{\putIconGithub} & \TMformatArchitecture{ConvNet\\\textit{(2D U-Net)}} & \TMformatParams{192.0\\-} & \TMformatRes{64 Nvidia A100 80GB}\\
    \TMformatModel{UniverSeg}{UniverSeg: Universal Medical Image Segmentation} & \TMformatAffil{\TMflagUSA U.S.A.} & \TMformatDatePubliCit{\href{https://doi.org/10.48550/arXiv.2304.06131}{2023-04}}{\href{https://doi.org/10.48550/arXiv.2304.06131}{arXiv}}{\citet{butoi2023universeguniversalmedicalimage}}{\href{https://doi.org/10.1109/ICCV51070.2023.01960}{2024-01}}{\href{https://doi.org/10.1109/ICCV51070.2023.01960}{IEEE/CVF ICCV}}{\citet{10376558}} & \href{https://github.com/JJGO/UniverSeg}{\putIconGithub} & \TMformatArchitecture{ConvNet\\\textit{(2D U-Net)}} & \TMformatParams{1.18\\-} & \TMformatRes{1 Nvidia V100 32GB}\\
    \TMformatModel{UniSeg}{UniSeg: A Prompt-Driven Universal Segmentation Model as Well as A Strong Representation Learner} & \TMformatAffil{\TMflagChina China} & \TMformatDatePubliCit{\href{https://doi.org/10.48550/arXiv.2304.03493}{2023-04}}{\href{https://doi.org/10.48550/arXiv.2304.03493}{arXiv}}{\citet{ye2023unisegpromptdrivenuniversalsegmentation}}{\href{https://doi.org/10.1007/978-3-031-43898-1_49}{2023-10}}{\href{https://doi.org/10.1007/978-3-031-43898-1_49}{MICCAI}}{\citet{10.1007/978-3-031-43898-1_49}} & \href{https://github.com/yeerwen/UniSeg}{\putIconGithub} & \TMformatArchitecture{ConvNet\\\textit{(nnU-Net framework)}} & \TMformatParams{-\\-} & \TMformatRes{-}\\
    \TMformatModel{STU-Net}{STU-Net: Scalable and Transferable Medical Image Segmentation Models Empowered by Large-Scale Supervised Pre-training} & \TMformatAffil{\TMflagChina China} & \TMformatDatePubliCit{\href{https://doi.org/10.48550/arXiv.2304.06716}{2023-04}}{\href{https://doi.org/10.48550/arXiv.2304.06716}{arXiv}}{\citet{huang2023stunetscalabletransferablemedical}}{-}{-}{-} & \href{https://github.com/openmedlab/STU-Net}{\putIconGithub} & \TMformatArchitecture{ConvNet\\\textit{(nnU-Net framework)}} & \TMformatParams{1457.33\\\textit{12600.0}} & \TMformatRes{1 Nvidia A100 80GB}\\
    \TMformatModel{SAMed}{Customized Segment Anything Model for Medical Image Segmentation} & \TMformatAffil{\TMflagChina China} & \TMformatDatePubliCit{\href{https://doi.org/10.48550/arXiv.2304.13785}{2023-04}}{\href{https://doi.org/10.48550/arXiv.2304.13785}{arXiv}}{\citet{zhang2023customizedsegmentmodelmedical}}{-}{-}{-} & \href{https://github.com/hitachinsk/SAMed}{\putIconGithub} & \TMformatArchitecture{Transformer with Convolutions (SAM)\\\textit{(SAM)}} & \TMformatParams{18.81\\-} & \TMformatRes{-}\\
    \TMformatModel{SAM}{Segment Anything} & \TMformatAffil{\TMflagUSA U.S.A.} & \TMformatDatePubliCit{\href{https://doi.org/10.48550/arXiv.2304.02643}{2023-04}}{\href{https://doi.org/10.48550/arXiv.2304.02643}{arXiv}}{\citet{kirillov2023segment}}{\href{https://doi.org/10.1109/ICCV51070.2023.00371}{2023-10}}{\href{https://doi.org/10.1109/ICCV51070.2023.00371}{IEEE/CVF ICCV}}{\citet{10378323}} & \href{https://github.com/facebookresearch/segment-anything}{\putIconGithub} & \TMformatArchitecture{Transformer with Convolutions\\\textit{(2D ViT-Huge)}} & \TMformatParams{636.0\\\textit{373.0}} & \TMformatRes{256 Nvidia A100 80GB}\\
    \TMformatModel{MedSAM}{Segment anything in medical images} & \TMformatAffil{\TMflagCanada Canada} & \TMformatDatePubliCit{\href{https://doi.org/10.48550/arXiv.2304.12306}{2023-04}}{\href{https://doi.org/10.48550/arXiv.2304.12306}{arXiv}}{\citet{Ma2024}}{\href{https://doi.org/10.1038/s41467-024-44824-z}{2024-01}}{\href{https://doi.org/10.1038/s41467-024-44824-z}{Nature Communications}}{\citet{Ma2024}} & \href{https://github.com/bowang-lab/MedSAM?tab=readme-ov-file}{\putIconGithub} & \TMformatArchitecture{Transformer with Convolutions (SAM)\\\textit{(SAM)}} & \TMformatParams{93.7\\\textit{82.0}} & \TMformatRes{20 Nvidia A100 80GB}\\
    \TMformatModel{MultiTalent }{MultiTalent: A Multi-dataset Approach to Medical Image Segmentation} & \TMformatAffil{\TMflagGermany Germany} & \TMformatDatePubliCit{\href{https://doi.org/10.48550/arXiv.2303.14444}{2023-03}}{\href{https://doi.org/10.48550/arXiv.2303.14444}{arXiv}}{\citet{10.1007/978-3-031-43898-1_62}}{\href{https://doi.org/10.1007/978-3-031-43898-1_62}{2023-10}}{\href{https://doi.org/10.1007/978-3-031-43898-1_62}{MICCAI}}{\citet{10.1007/978-3-031-43898-1_62}} & \href{https://github.com/MIC-DKFZ/MultiTalent}{\putIconGithub} & \TMformatArchitecture{ConvNet\\\textit{(3D U-Net, 3D U-Net with Residuals, SwinUNETR)}} & \TMformatParams{69.34\\\textit{1200.0}} & \TMformatRes{-}\\
    \TMformatModel{CLIP-Driven Universal Model}{Universal and Extensible Language-Vision Models for Organ Segmentation and Tumor Detection from Abdominal Computed Tomography} & \TMformatAffil{\TMflagUSA U.S.A.} & \TMformatDatePubliCit{\href{https://doi.org/10.48550/arXiv.2301.00785}{2023-01}}{\href{https://doi.org/10.48550/arXiv.2301.00785}{arXiv}}{\citet{Liu_2023}}{\href{https://doi.org/10.1016/j.media.2024.103226}{2024-01}}{\href{https://doi.org/10.1016/j.media.2024.103226}{Medical Image Analysis}}{\citet{LIU2024103226}} & \href{https://github.com/ljwztc/CLIP-Driven-Universal-Model}{\putIconGithub} & \TMformatArchitecture{Transformer\\\textit{(SwinUNETR)}} & \TMformatParams{62.25\\\textit{350.0}} & \TMformatRes{8 Nvidia RTX A5000 24GB}\\
    \TMformatModel{DeSD}{DeSD: Self-Supervised Learning withDeep Self-Distillation for3D Medical Image Segmentation} & \TMformatAffil{\TMflagChina China} & \TMformatDatePubliCit{\href{https://doi.org/10.1007/978-3-031-16440-8_52}{2022-09}}{\href{https://doi.org/10.1007/978-3-031-16440-8_52}{MICCAI}}{\citet{10.1007/978-3-031-16440-8_52}}{-}{-}{-} & \href{https://github.com/yeerwen/DeSD}{\putIconGithub} & \TMformatArchitecture{ConvNet\\\textit{(3D ResNet-50)}} & \TMformatParams{-\\-} & \TMformatRes{-}\\
    \TMformatModel{SMIT}{Self-supervised 3D Anatomy Segmentation Using Self-distilled Masked Image Transformer (SMIT)} & \TMformatAffil{\TMflagUSA U.S.A.} & \TMformatDatePubliCit{\href{https://doi.org/10.48550/arXiv.2205.10342}{2022-05}}{\href{https://doi.org/10.48550/arXiv.2205.10342}{arXiv}}{\citet{10.1007/978-3-031-16440-8_53}}{\href{https://doi.org/10.1007/978-3-031-16440-8_53}{2022-09}}{\href{https://doi.org/10.1007/978-3-031-16440-8_53}{MICCAI}}{\citet{10.1007/978-3-031-16440-8_53}} & \href{https://github.com/harveerar/SMIT.git}{\putIconGithub} & \TMformatArchitecture{Transformer with Convolutions\\\textit{(3D Swin-Small)}} & \TMformatParams{28.19\\-} & \TMformatRes{4 Nvidia V100 32GB}\\
    \TMformatModel{UniSeg33A}{Universal Segmentation of 33 Anatomies} & \TMformatAffil{\TMflagChina China} & \TMformatDatePubliCit{\href{https://doi.org/10.48550/arXiv.2203.02098}{2022-03}}{\href{https://doi.org/10.48550/arXiv.2203.02098}{arXiv}}{\citet{liu2022universalsegmentation33anatomies}}{-}{-}{-} & \codeUnavailable & \TMformatArchitecture{ConvNet with Attention\\\textit{(3D U-Net, Transformer Blocks)}} & \TMformatParams{-\\-} & \TMformatRes{-}\\
    \TMformatModel{UniMiSS}{UniMiSS: Universal Medical Self-supervised Learning viaBreaking Dimensionality Barrier} & \TMformatAffil{\TMflagAustralia Australia\vspace{2pt}\\\TMflagChina China} & \TMformatDatePubliCit{\href{https://doi.org/10.48550/arXiv.2112.09356}{2021-12}}{\href{https://doi.org/10.48550/arXiv.2112.09356}{arXiv}}{\citet{xie2022unimissuniversalmedicalselfsupervised}}{\href{https://doi.org/10.1007/978-3-031-19803-8_33}{2022-10}}{\href{https://doi.org/10.1007/978-3-031-19803-8_33}{ECCV}}{\citet{10.1007/978-3-031-19803-8_33}} & \href{https://github.com/YtongXie/UniMiSS-code}{\putIconGithub} & \TMformatArchitecture{Transformer\\\textit{(2D+3D PVT-Small)}} & \TMformatParams{-\\-} & \TMformatRes{8 Nvidia V100 32GB}\\
    \TMformatModel{DoDNet}{DoDNet: Learning To Segment Multi-Organ and Tumors From Multiple Partially Labeled Datasets} & \TMformatAffil{\TMflagAustralia Australia\vspace{2pt}\\\TMflagChina China} & \TMformatDatePubliCit{\href{https://doi.org/10.48550/arXiv.2011.10217}{2020-11}}{\href{https://doi.org/10.48550/arXiv.2011.10217}{arXiv}}{\citet{zhang2020dodnetlearningsegmentmultiorgan}}{\href{https://openaccess.thecvf.com/content/CVPR2021/html/Zhang_DoDNet_Learning_To_Segment_Multi-Organ_and_Tumors_From_Multiple_Partially_CVPR_2021_paper.html}{2021-06}}{\href{https://openaccess.thecvf.com/content/CVPR2021/html/Zhang_DoDNet_Learning_To_Segment_Multi-Organ_and_Tumors_From_Multiple_Partially_CVPR_2021_paper.html}{IEEE/CVF CVPR}}{\citet{Zhang_2021_CVPR}} & \href{https://github.com/aim-uofa/partially-labelled}{\putIconGithub} & \TMformatArchitecture{ConvNet\\\textit{(3D U-Net with Residuals)}} & \TMformatParams{17.3\\-} & \TMformatRes{-}\\
    \TMformatModel{Med3D}{Med3D: Transfer Learning for 3D Medical Image Analysis} & \TMformatAffil{\TMflagChina China} & \TMformatDatePubliCit{\href{https://doi.org/10.48550/arXiv.1904.00625}{2019-04}}{\href{https://doi.org/10.48550/arXiv.1904.00625}{arXiv}}{\citet{chen2019med3dtransferlearning3d}}{-}{-}{-} & \href{https://github.com/Tencent/MedicalNet}{\putIconGithub} & \TMformatArchitecture{ConvNet\\\textit{(3D ResNet-152)}} & \TMformatParams{-\\-} & \TMformatRes{-}
}{~}
\restoregeometry

\section{Contribution analysis}
\label{sec:results}

\subsection{Statistics}
\label{sec:results-statistics}

This survey reviewed 55 publications on generalist models for medical imaging segmentation, of which 48 proposed the design of new models or adaptations of existing ones (e.g., by fine-tuning). The remaining ones concerned zero-shot of SAM 2 (four), zero-shot of SAM (two), and the integration of SAM into 3D Slicer (one). In compliance with our taxonomy (Fig. \ref{fig:taxonomy}), the studies can be further grouped as follows: 25 on SAM (nine on adapters, three on modification to SAM architecture, three on PEFT, two on fully-fine tuning, two on zero-shot with SAM, two on medical annotations, one on few-shot, and three on other implementations); eight on SAM 2 (four on zero-shot, two on fine-tuning, and two on other applications); 17 on innovative models trained only on medical images; and six on new model trained with both medical images and text.
Additionally, 24 SOTA task-specific models were considered for comparison. Overall, this work analyzed 79 works.

Considering the 48 publications on new generalist models and the 24 on task-specific ones, the authors were affiliated with 60 distinct institutions across seven countries. In terms of geographical distribution, Greater China (a union of mainland China, Hong Kong, Taiwan and Macao) led the ranking with 41 works, followed by the United States with 24. Other significant contributions came from Germany (eight) and the U.K. (six). The remaining countries include the U.A.E. (three), Australia (two), Canada (two), Singapore (two) and Japan (one).
Notably, Greater China leads with 31 generalist models, U.S.A. follows with 14, Germany only counts one, the U.K. counts five, U.A.E., Australia, Canada and Singapore have two each, and Japan counts one.

Since there exists more than one publication for most models, we defined the first one as \textit{"primary"}, and the one reporting the best score as \textit{"best-in-literature"}.
ArXiv was by far the most frequent venue for the primary work, recurring in 63 models, of which 41 generalist. The time elapsed from the primary to the latest publication on a specific model was equal to 16 and 11 months, for generalist and task-specific models, respectively.

The models were evaluated on a wide range of datasets. The full list with the characteristics and a description of each dataset is reported in Table~\ref{table:datasets-all} in the appendix.
On average the models were tested on 4.3 datasets in the primary work (median of three datasets), and on 9.2 datasets for the best-in-literature. By considering primary works, the generalist models were tested on more datasets than task-specific ones (4.6 vs. 3.8 on average). However, the trend was inverted considering the best-in-literature results, with generalist models tested on average on 7.5 datasets, and task-specific models on average on 12.4 datasets. 
This could be due to several factors: open-source implementations of well-established task-specific models for segmentation might be easier to use and more resource-friendly than generalist models. Also, it has to be considered that specialized models are often used as benchmark by generalist models, as well as by other specialized models.
Considering primary works, the top-five models ranked by number of tested datasets were  nnU-Net (19), SwinUNETR (eight), MedFormer (seven), LHU-Net (six) and SwinUNETR-V2 (five) for specialized models, while SAT (32), PCNet (18), STU-Net (18), BiomedParse (14) and the CLIP-Driven Universal Model (13) for the generalist models. On the other hand, considering best-in-literature results, the trend is inverted: the top-five specialized models are the nnU-Net (48), SwinUNETR (41), nnFormer (29), UNETR (25) and 3D UX-Net (17), while the top-five generalist models are SAT (32), MedSAM (30), SAM (22), STU-Net (20) and PCNet (18).
Ranking models by their increase of tested dataset was helpful in assessing which models were preferred by research groups as baselines. For specialized models the top-five models ranked by increase in number of tested dataset from primary work to best-in-literature are the SwinUNETR (33), nnU-Net (29), nnFormer (26), UNETR (21), and the original U-Net (17), while for generalist models the ranking is MedSAM (30), SAM (22), SAM-Med3D (12), SAM-Med2D (11) and Med-SA (eight). 
It should be noted that MedSAM was actually tested on many dataset, in one of the most important testing efforts ever. However, results were reported both in the papers and in the supplementary material as Dice scores per single organ, merging many different datasets all together, while the literature standard seems to be to show results as Dice scores aggregated per dataset, plus optionally Dice scores for each class in the tested dataset. However, the key take-away remains that MedSAM was the preferred generalist model for benchmarking.

Code availability was generally high, with 61 out of 72 models (40 out of 48 generalist) releasing publicly the source code (mainly on GitHub). 

Concerning the architecture of the 48 generalist models, 22 were based on SAM, three on SAM 2, 14 on CNNs, six on pure transformers, and three on hybrid networks mixing transformers and CNNs. In contrast, the works on task-specific models showed a more balanced mix with eight CNNs, five pure Transformers, 10 hybrid models (transformer with CNNs),and one graph neural network.

In terms of model complexity, 48 out of the 72 works reported the number of parameters, while only 25 the GFLOPS (billions of floating points operations per second). Overall, the values range from a minimum of 1.18~M (millions) GFLOPS to a maximum of 1457.33~M GFLOPS, with a median of 44~M GFLOPS. When stratified, task-specific models (22 out of 24 available) showed a narrower range (range: 9~M - 97.6~M, median 38~M GFLOPS), whereas generalist models (26 out of 48) exhibited a broader spread, ranging from 1.18~M to 1457.3~M, with a median of 91.8~M GFLOPS.
It is worth noting that the value of GFLOPS depends on the hardware resources, internal compiler optimizations, input image size, and the author choice of reporting. In fact, some authors reported the GFLOPS for a single input patch, while others considered the GFLOPS to segment the whole 3D image. This heterogeneity made comparison based on GFLOPS misleading and dangerous.

Information on the hardware used to train the models highlighted disparities. Fifty-seven of the 72 works provided details on the hardware resources used for training, with the total required GPU memory ranging from 11~GB up to 5120~GB with median 56~GB. As expected generalist models were more resources avid than task-specific ones, with a median of 96~GB and 40~GB, respectively. We then categorized the memory consumption in four different tiers: 16 models (two generalists) were trained with less than 16~GB, 23 (13 generalists) required a memory ranging from 16~GB to~64~GB, four (three generalists) a memory from 64~GB to ~80~GB, and 23 (19 generalists) needed more than 80~GB.
Medical SAM 2 (MedSAM-2), BrainSegFounder, and OnePrompt all required 5120~GB for training. IMIS-Net, MedSAM, and SAT exceeded 1000~GB, with MedSAM2 following closely at 960~GB, while the rest stayed below 640~GB of. 
For comparison, SAM and SAM~2 were trained using a staggering 20,480~GB of  video RAM on high performance GPUs.

\subsection{Performances by target anatomies} 
\label{subsec:target-anatomies}

The segmentation performances of the primary work in terms of Dice score are reported, in Table~\ref{table:results-original-specialized} and Table \ref{table:results-original-foundational} of the appendix or the task-specific and generalist models, respectively. The best-in-literature scores are reported in Table \ref{table:results-best-foundational}, and in Table \ref{table:results-best-specialized} in the appendix for the task-specific models. In particular, the tables with the best-in-literature results highlight the performance gain in percentage from the primary work on a specific dataset (Table \ref{table:results-best-foundational}, and Table \ref{table:results-best-specialized} in the appendix).
The number of datasets may vary from the primary to the best-in-literature works since some models, e.g., nnU-Net, and SwinUNETR, were re-implemented from different research groups over time.

It is worth noting that the best performance was obtained by different strategies, e.g., retraining, different pre-training, or fine-tuning depending on the model as well as by direct re-implementation by part of a different research group.

By comparing the generalist and task-specific models on the different target anatomies at the level of both primary and best-in-literature works, and ranking the five best models of each type, the \textbf{winners} were:

\begin{itemize}
    \item \textbf{Brain:} Generalist models on most datasets, while task-specific on FeTA2021, MSD Ippocampus, and OASIS-3 datasets (median Dice score from primary research). Task specific on eight out of 10 datasets (BraTS, FeTA2021, ISLES, MSD Ippocampus, Multiple Sceloris Lesions, OASIS-1, OASIS-3, WMH) for median Dice score from best-in-literature (Table \ref{table:results-overview-brain} of the appendix).
    \item \textbf{Head and neck:} Task-specific (on primary research). Tie on best-in-literature with generalist models obtaining a higher median Dice score on Head and Neck Dataset, while task-specific on ToothFairy dataset (Table \ref{table:results-overview-head-and-neck} of the appendix).
    \item \textbf{Lungs:} Generalist models on both primary research and best-in-literature (Table \ref{table:results-overview-lungs} of the appendix).
    \item \textbf{Heart and thoracic vessels:} Generalist models on all datasets with the exception of ACDC, and Left Atrial Segmentation (median DSC on both primary research, and best-in-literature) (Table \ref{table:results-overview-heart-and-thoracic-vasculature} of the appendix).
    \item \textbf{Thoracic structures:} Task-specific on both primary work, and best-in-literature (Table \ref{table:results-overview-thoracic-structures-} of the appendix).
    \item \textbf{Bones:} Specialists models on primary work, while generalists on best-in-literature on two out of three datasets (TotalSegmentator Ribs Vertebrae, and TotalSegmentator Ribs Vertebrae) (Table \ref{table:results-overview-bones} of the appendix).
    \item \textbf{Muscles:} Generalist models on both primary work, and best-in-literature (Table \ref{table:results-overview-muscles} of the appendix).
    \item \textbf{Liver:} Generalist models on primary work; generalist models on best-in-literature on all datsaets except ATLAS2023 (Table \ref{table:results-overview-liver} of the appendix).
    \item \textbf{Pancreas:} Tie: generalist models on MSD dataset (primary work, and best-in-literature); task-specific models on NIH dataset (primary work, and best in literature) (Table \ref{table:results-overview-pancreas} of the appendix).
    \item \textbf{Colon:} Generalist models on both primary work, and best-in-literature (Table \ref{table:results-overview-colon} of the appendix).
    \item \textbf{Kidney:} Tie: generalist models on KiPA22 dataset (primary work, and best in literature); task-specific models on KiTS dataset (primary work, and best-in-literature) (Table \ref{table:results-overview-kidney} of the appendix).
    \item \textbf{Spleen:} Task-specific models on both primary work and best-in-literature (Table \ref{table:results-overview-spleen} of the appendix).
    \item \textbf{Prostate:} Generalist models on primary work; task-specific on MSD prostate, while generalist models on PROMISE12 (best-in-literature) (Table \ref{table:results-overview-prostate} of the appendix).
    \item \textbf{Abdominal organs – multi organ:} Tie with generalist models on eight datasets (AbdomenCT-1k, BTCV, BTCV Cervix, CHAOS MultiOrgan, MOTS, TotalSegmentor (All), TotalSegmentor (Organs), and Touchstone 1.0) on primary works, while task specific on seven datasets (AMOS2022, AbdomenCT-1k, BTCV, CHAOS MultiOrgan, FLARE MICCAI, TouchStone 1.0, and WORD) on best-in-literature (Table \ref{table:results-overview-abdominal-multi-organ} of the appendix).
    \item \textbf{Whole-body lesions:} Generalist models on both primary work, and best-in-literature (Table \ref{table:results-overview-whole-body-lesions} of the appendix).
\end{itemize}

\newcommand{\selectResultsOverallFontTable}{\fontfamily{cmss}\fontsize{6}{7}\selectfont}


\newcommand{\ndatasets}{18}
\newcommand{\ncolumns}{20}

\newcommand{\tabledatasetcolumnswidth}{25pt}


\newcommand{\spheading}[2][11.5em]{
    \rotatebox{90}{\parbox{#1}{\raggedright \vspace{-3pt}#2}}
}

\newcommand{\overallResultsTableHeading}{
    ~Model & First Publ. & \spheading{BTCV} & \spheading{BraTS} & \spheading{KiTS} & \spheading{LiTS / MSD Liver} & \spheading{MSD Pancreas Tumour} & \spheading{MSD Lung Tumors} & \spheading{Synapse} & \spheading{AMOS} & \spheading{ACDC} & \spheading{PROMISE12} & \spheading{MSD Colon Cancer} & \spheading{FLARE} & \spheading{TotalSegmentator} & \spheading{MSD Spleen} & \spheading{SegTHOR} & \spheading{MSD Hepatic Vessels} & \spheading{MSD Prostate} & \spheading{TotalSegmentator Organs} \\
}

\newcommand{\makeTableFirstHeading}[1]{ 
    \hline \rowcolor{#1}
    \overallResultsTableHeading 
    \hline
}

\newcommand{\makeTableOtherHeadings}[1]{ 
    \makeTableHeaderContinued{\ncolumns} 
    \makeTableFirstHeading{#1}
}


\newcommand{\bilFILL}{\vspace{1pt}}
\newcommand{\bilparbox}[1]{
    \parbox{25pt}{\centering #1}
}
\newcommand{\bilMultiline}[1]{
    \bilparbox{
        \bilFILL #1 \bilFILL
    }
}

\newcommand{\sTROst}[1]{
    \textbf{#1}%
}
\newcommand{\sTROnd}[1]{
    \underline{#1}%
}
\newcommand{\sTROrd}[1]{
    \textit{#1}%
}
\newcommand{\cTROst}[1]{
    #1
}
\newcommand{\cTROnd}[1]{
    #1
}
\newcommand{\cTROrd}[1]{
    #1
}

\newcommand{\textcolorTROcomment}[1]{
    \textcolor{black!40}{#1}
}

\newcommand{\stackComment}[3]{
    {
        \def\arraystretch{0.75}%
        \begin{tabular}{#1}
             #2 \\
            \scalefont{0.1}\textcolorTROcomment{(#3)}
        \end{tabular}
    }
}
\newcommand{\stackDiceComment}[2]{
    \stackComment{C{\tabledatasetcolumnswidth}}{#1}{#2}
}
\newcommand{\stackModelComment}[2]{
    \stackComment{l}{#1}{#2}
}
\newcommand{\stackDiceCitIncreaseComment}[4]{
    {%
        \def\arraystretch{0.75}%
        \begin{tabular}{C{\tabledatasetcolumnswidth}}
             #1 \\
            \scalefont{0.1}\citet{#2}
            \IfStrEq{#3}{}{%
                }{%
                \\%
                \scalefont{0.1}\textcolor{black!70!green}{(+#3\%)}
              }%
            \IfStrEq{#4}{}{%
                }{%
                \\%
                \scalefont{0.1}\textcolorTROcomment{(#4)}
              }%
        \end{tabular}
    }
}

\newcommand{\selectResultsOverallFontCaption}{\rmfamily\fontsize{10}{11}\selectfont}

\newcommand{\captionCommonPhrase}{
    Best result considering models in this table are formatted as \sTROst{first}, \sTROnd{second-best} and \sTROrd{third-best}.
}


\newcommand{\makeTableResultsOverall}[6]{

    {
    \selectResultsOverallFontTable
    \rowcolors{1}{white!1}{#2}
    \begin{center}
    \begin{longtable}[c]{%
        p{1.25cm} p{0.75cm} *{\ndatasets}{C{\tabledatasetcolumnswidth}}
    }
        \hiderowcolors
        \caption{
            #3
            \captionCommonPhrase
        }
        \label{#4} \\
        \showrowcolors
    
        \makeTableFirstHeading{#1}
        \endfirsthead
        
        \makeTableOtherHeadings{#1}
        \endhead
        
        \makeTableOtherFooters{\ncolumns}
        \endfoot
        
        \makeTableLastFooter{\ncolumns}{}
        \endlastfoot
    
        #5
    
    \end{longtable}
    \outOfTableComments{#6}
    \end{center}
    }
}
\newgeometry{left=0.6cm, right=0.6cm} 
    \makeTableResultsOverall{%
    colorTablePapersFoundationalHeaderClear%
}{
    colorTablePapersFoundationalRowsPattern%
}{%
    Highest Dice score achieved by generalist models expressed as percentage [\%]. Table cells with reference represent either a model tested on a dataset, not used in the primary publication, or an improvement over the primary work. Table cells with percentage increment in green refer to the improvement of Dice score w.r.t. to the primary publication.
}{table:results-best-foundational}{
    MedSAM2 & 2025-04 &  &  &  &  &  &  &  &  &  &  &  &  &  &  &  &  &  &  \\
    SPA & 2025-01 &  &  &  &  &  &  & \cTROst{\sTROst{92.88}} &  &  & \cTROst{\sTROst{94.29}} &  &  &  &  &  &  &  &  \\
    3DMedSAM & 2024-12 & 88.60 &  &  & 60.45 &  &  &  &  &  &  &  &  &  &  &  &  &  &  \\
    KnowSAM & 2024-12 &  &  &  &  &  &  &  &  & \sTROnd{91.13} &  &  &  &  &  &  &  &  &  \\
    IMIS-Net & 2024-11 &  &  &  &  &  &  &  &  &  &  &  &  & 79.06 &  & \sTROnd{89.27} &  &  &  \\
    SAM-MPA & 2024-10 &  &  &  &  &  &  &  &  &  &  &  &  &  &  &  &  &  &  \\
    TP-Mamba & 2024-09 & 84.80 &  &  &  &  &  &  &  &  &  &  &  &  &  &  &  &  &  \\
    EMedSAM & 2024-08 &  & 89.30 &  &  &  &  &  &  &  &  &  & \stackDiceComment{0.88}{a} &  &  &  &  &  &  \\
    Biomedical SAM-2 (BioSAM-2) & 2024-08 &  &  &  &  &  &  &  & 74.39 &  &  &  & 76.32 &  &  &  &  &  &  \\
    SAM 2 & 2024-08 & \stackDiceCitIncreaseComment{86.00}{zhu2024medicalsam2segment}{}{} & \stackDiceCitIncreaseComment{75.52}{shen2025interactive3dmedicalimage}{}{b} & \stackDiceCitIncreaseComment{64.60}{zhu2024medicalsam2segment}{}{} & \stackDiceCitIncreaseComment{81.32}{shen2025interactive3dmedicalimage}{}{b} & \stackDiceCitIncreaseComment{44.73}{shen2025interactive3dmedicalimage}{}{b} & \stackDiceCitIncreaseComment{71.61}{shen2025interactive3dmedicalimage}{}{b} &  & \stackDiceCitIncreaseComment{54.92}{yan2024biomedical}{}{} &  &  &  & \stackDiceCitIncreaseComment{47.44}{yan2024biomedical}{}{} & \stackDiceCitIncreaseComment{77.62}{cheng2024interactivemedicalimagesegmentation}{}{} & \stackDiceCitIncreaseComment{79.59}{shen2025interactive3dmedicalimage}{}{b} & \stackDiceCitIncreaseComment{85.86}{cheng2024interactivemedicalimagesegmentation}{}{} &  &  &  \\
    Medical SAM 2 (MedSAM-2) & 2024-08 & \sTROrd{89.00} & \stackDiceCitIncreaseComment{28.91}{10847777}{}{a} & 78.20 &  &  &  &  &  &  &  &  &  &  &  &  &  &  &  \\
    FLAP-SAM & 2024-07 &  &  & 60.46 &  &  &  &  &  &  & 88.67 &  &  &  &  &  &  &  &  \\
    LeSAM & 2024-06 &  & 84.95 & \cTROnd{\sTROnd{91.86}} & 70.62 & \cTROrd{\sTROrd{79.42}} & \sTROrd{79.57} &  &  &  &  & \cTROst{\sTROst{77.18}} &  &  &  &  & \cTROst{\sTROst{79.59}} &  &  \\
    Merlin & 2024-06 &  &  &  &  &  &  &  &  &  &  &  &  & 86.00 &  &  &  &  &  \\
    BrainSegFounder & 2024-06 &  & \sTROnd{91.15} &  &  &  &  &  &  &  &  &  &  &  &  &  &  &  &  \\
    MoME & 2024-05 &  & 88.86 &  &  &  &  &  &  &  &  &  &  &  &  &  &  &  &  \\
    BiomedParse & 2024-05 &  & 79.95 & 80.22 & 83.39 & 50.62 & 66.09 &  & 86.33 & \sTROst{92.26} & 89.97 & 66.51 &  &  & \sTROnd{96.86} &  & 66.03 & 72.85 &  \\
    PCNet & 2024-04 & 83.85 &  & 86.19 & \cTROst{\sTROst{96.63}} & \cTROnd{\sTROnd{79.70}} &  &  &  &  &  &  & \sTROnd{90.62} & \cTROnd{\sTROst{91.64}} & 95.77 & 87.66 &  &  & \cTROnd{\sTROst{91.09}} \\
    MEA M-SAM & 2024-03 &  & \cTROst{\sTROst{92.08}} & \cTROst{\sTROst{93.50}} & 89.95 & \cTROst{\sTROst{80.49}} & \cTROst{\sTROst{81.62}} &  &  &  &  &  &  &  &  &  &  &  &  \\
    SFR SAM & 2024-03 & 77.07 & 86.09 &  &  &  &  &  &  &  &  &  &  &  &  &  &  &  &  \\
    Med-SA & 2023-12 & 88.30 & \stackDiceCitIncreaseComment{\sTROrd{90.50}}{10.1007/978-3-031-72111-3_38}{1.40}{} & \stackDiceCitIncreaseComment{91.05}{10540651}{}{} & \stackDiceCitIncreaseComment{83.67}{10.1007/978-3-031-72111-3_38}{}{} & \stackDiceCitIncreaseComment{78.68}{10540651}{}{} & \stackDiceCitIncreaseComment{78.72}{10540651}{}{} & \stackDiceCitIncreaseComment{\cTROnd{\sTROnd{92.42}}}{10829779}{}{} &  &  & \stackDiceCitIncreaseComment{\cTROnd{\sTROnd{93.66}}}{10829779}{}{} & \stackDiceCitIncreaseComment{\cTROrd{\sTROrd{75.36}}}{10540651}{}{} &  &  &  &  &  &  &  \\
    SAT & 2023-12 & 81.60 & 55.68 & 71.53 & 78.86 & 59.23 & 61.28 &  & 84.82 & 89.64 & 87.28 & 38.45 & \sTROst{91.78} & \sTROrd{86.71} & 94.97 & \sTROrd{88.98} & 63.43 & 77.98 & \cTROrd{\sTROnd{90.42}} \\
    SegVol & 2023-11 & \stackDiceCitIncreaseComment{73.81}{wang2024sammed3dgeneralpurposesegmentationmodels}{}{} &  &  &  &  &  &  & 85.93 &  &  &  &  &  &  & 81.55 &  &  &  \\
    SAM-Med3D & 2023-10 & \stackDiceCitIncreaseComment{84.70}{zhu2024medicalsam2segment}{5.53}{} & \stackDiceCitIncreaseComment{86.45}{10.1007/978-3-031-72111-3_38}{}{} & \stackDiceCitIncreaseComment{86.65}{10.1007/978-3-031-72111-3_38}{14.59}{} & \stackDiceCitIncreaseComment{88.71}{10.1007/978-3-031-72111-3_38}{}{} & \stackDiceCitIncreaseComment{75.76}{10.1007/978-3-031-72111-3_38}{}{} & \stackDiceCitIncreaseComment{78.32}{10.1007/978-3-031-72111-3_38}{}{} &  & 75.41 &  &  &  &  & 84.68 &  & \stackDiceCitIncreaseComment{77.27}{du2024segvol}{}{} &  &  &  \\
    SAM3D & 2023-09 &  & 72.90 & \stackDiceCitIncreaseComment{80.36}{10.1007/978-3-031-72111-3_38}{}{} & \stackDiceCitIncreaseComment{82.27}{10.1007/978-3-031-72111-3_38}{}{} & \stackDiceCitIncreaseComment{71.26}{10.1007/978-3-031-72111-3_38}{}{} & 71.42 & 79.56 &  & \sTROrd{90.41} &  &  &  &  &  &  &  &  &  \\
    MA-SAM & 2023-09 & 87.20 &  & \stackDiceCitIncreaseComment{60.23}{10.1007/978-3-031-77610-6_21}{}{} &  & 40.20 &  &  &  &  & 92.60 & 47.70 &  &  &  &  &  &  &  \\
    Cheap Lunch SAM & 2023-08 &  & 85.28 &  &  &  &  & 85.95 &  &  &  &  &  &  &  &  &  &  &  \\
    SAM-Med2D & 2023-08 & \stackDiceCitIncreaseComment{50.05}{wang2024sammed3dgeneralpurposesegmentationmodels}{}{} & \stackDiceCitIncreaseComment{84.21}{10540651}{}{c} & \stackDiceCitIncreaseComment{\sTROrd{91.46}}{10540651}{11.59}{} & \stackDiceCitIncreaseComment{69.89}{10540651}{}{} & \stackDiceCitIncreaseComment{79.02}{10540651}{}{} & \stackDiceCitIncreaseComment{79.25}{10540651}{}{} &  & \stackDiceCitIncreaseComment{66.68}{du2024segvol}{}{} &  &  & \stackDiceCitIncreaseComment{\cTROnd{\sTROnd{76.45}}}{10540651}{}{} & 85.10 & \stackDiceCitIncreaseComment{75.92}{cheng2024interactivemedicalimagesegmentation}{}{} &  & \stackDiceCitIncreaseComment{86.43}{cheng2024interactivemedicalimagesegmentation}{}{} &  &  &  \\
    Disruptive Autoencoders & 2023-07 & \cTROst{\sTROst{92.10}} &  &  &  &  &  &  &  &  &  &  &  &  &  &  &  &  &  \\
    SAMMed & 2023-07 & 70.30 &  & 84.00 & 92.00 &  &  &  &  &  &  &  &  &  &  &  &  &  &  \\
    DeSAM & 2023-06 &  &  &  &  &  &  &  &  &  &  &  &  &  &  &  &  &  &  \\
    MedLSAM & 2023-06 &  &  &  &  &  &  &  &  &  &  &  &  &  &  &  &  &  &  \\
    HERMES & 2023-06 & 86.29 & \stackDiceCitIncreaseComment{88.03}{10879789}{}{} & 85.98 & 68.32 & 72.07 &  &  & \sTROrd{88.59} &  &  &  &  &  &  &  &  &  &  \\
    MIS-FM & 2023-06 &  &  &  &  &  &  & 89.11 &  &  &  &  &  &  &  & \cTROrd{\sTROst{89.56}} &  &  &  \\
    3DSAM-adapter & 2023-06 & \stackDiceCitIncreaseComment{70.80}{CHEN2024103310}{}{} &  & 81.50 & 61.25 & 66.87 &  &  &  &  & \stackDiceCitIncreaseComment{81.20}{CHEN2024103310}{}{} & 60.93 &  &  &  &  &  &  &  \\
    One-Prompt & 2023-05 &  &  & 67.30 &  &  &  &  &  &  &  &  &  &  &  &  &  &  &  \\
    SAM & 2023-04 & \stackDiceCitIncreaseComment{54.80}{zhu2024medicalsam2segment}{}{} & \stackDiceCitIncreaseComment{77.56}{10540651}{}{d} & \stackDiceCitIncreaseComment{84.73}{10540651}{}{} & \stackDiceCitIncreaseComment{62.00}{Zhao_2024}{}{} & \stackDiceCitIncreaseComment{62.54}{Zhao_2024}{}{} & \stackDiceCitIncreaseComment{75.12}{10540651}{}{} & \stackDiceCitIncreaseComment{90.15}{10829779}{}{} & \stackDiceCitIncreaseComment{66.62}{Zhao_2024}{}{} & \stackDiceCitIncreaseComment{68.86}{Zhao_2024}{}{} & \stackDiceCitIncreaseComment{89.91}{10829779}{}{} & \stackDiceCitIncreaseComment{63.21}{10540651}{}{} & \stackDiceCitIncreaseComment{58.30}{Dong2024}{}{a} & \stackDiceCitIncreaseComment{75.45}{cheng2024interactivemedicalimagesegmentation}{}{} & \stackDiceCitIncreaseComment{84.54}{Zhao_2024}{}{} & \stackDiceCitIncreaseComment{84.46}{cheng2024interactivemedicalimagesegmentation}{}{e} & \stackDiceCitIncreaseComment{30.97}{Zhao_2024}{}{} & \stackDiceCitIncreaseComment{62.68}{Zhao_2024}{}{} &  \\
    STU-Net & 2023-04 & 83.83 &  & 85.44 & \cTROnd{\sTROnd{95.88}} & 78.95 &  &  & \cTROnd{\sTROst{90.49}} &  &  &  & \sTROrd{89.87} & \cTROrd{\sTROnd{90.06}} & 95.52 & 85.91 &  &  & \sTROrd{89.82} \\
    UniSeg & 2023-04 & 84.60 & 83.30 & 88.20 & 79.10 & 70.90 & 70.90 &  &  &  &  & 55.00 &  &  & \sTROrd{96.40} &  & \cTROrd{\sTROrd{71.20}} & \cTROst{\sTROst{89.70}} &  \\
    SAMed & 2023-04 & \stackDiceCitIncreaseComment{84.40}{LIN20251}{}{} & \stackDiceCitIncreaseComment{85.52}{10829779}{}{} &  &  &  &  & \stackDiceCitIncreaseComment{\cTROrd{\sTROrd{92.33}}}{10829779}{8.03}{} &  &  & \stackDiceCitIncreaseComment{\cTROrd{\sTROrd{93.47}}}{10829779}{}{} &  &  &  &  &  &  &  &  \\
    UniverSeg & 2023-04 & \stackDiceCitIncreaseComment{84.20}{zhu2024medicalsam2segment}{}{} &  & \stackDiceCitIncreaseComment{63.80}{zhu2024medicalsam2segment}{}{} &  &  &  &  &  & 70.90 &  &  &  &  &  &  &  &  &  \\
    MedSAM & 2023-04 & \stackDiceCitIncreaseComment{80.30}{zhu2024medicalsam2segment}{}{} & \stackDiceCitIncreaseComment{75.63}{10540651}{}{f} & \stackDiceCitIncreaseComment{79.54}{Zhao_2024}{}{} & \stackDiceCitIncreaseComment{68.52}{Zhao_2024}{}{} & \stackDiceCitIncreaseComment{77.12}{Zhao_2024}{}{} & \stackDiceCitIncreaseComment{70.46}{Zhao_2024}{}{} & \stackDiceCitIncreaseComment{90.74}{10829779}{}{} & \stackDiceCitIncreaseComment{80.19}{zhao2025modelrulealluniversal}{}{g} & \stackDiceCitIncreaseComment{82.82}{Zhao_2024}{}{} & \stackDiceCitIncreaseComment{92.46}{10829779}{}{} & \stackDiceCitIncreaseComment{72.76}{Zhao_2024}{}{} & \stackDiceCitIncreaseComment{79.50}{Dong2024}{}{a} & \stackDiceCitIncreaseComment{80.11}{zhao2025modelrulealluniversal}{}{g} & \stackDiceCitIncreaseComment{95.02}{Zhao_2024}{}{} & \stackDiceCitIncreaseComment{74.90}{zhao2025modelrulealluniversal}{}{g} & \stackDiceCitIncreaseComment{37.04}{Zhao_2024}{}{} & \stackDiceCitIncreaseComment{74.09}{zhao2025modelrulealluniversal}{}{g} & \stackDiceCitIncreaseComment{82.71}{zhao2025modelrulealluniversal}{}{g} \\
    MultiTalent & 2023-03 & \sTROnd{89.07} & \stackDiceCitIncreaseComment{86.67}{10879789}{}{} & 90.45 &  &  &  &  & \sTROnd{89.81} &  &  &  &  &  &  &  &  &  &  \\
    CLIP-Driven Universal Model & 2023-01 & 86.13 & \stackDiceCitIncreaseComment{82.60}{10.1007/978-3-031-43898-1_49}{}{} & \stackDiceCitIncreaseComment{80.70}{10.1007/978-3-031-43898-1_49}{}{} & 87.39 & 72.59 & \sTROnd{80.01} &  &  &  &  & 63.14 &  &  & \cTROst{\sTROst{97.27}} &  & \cTROnd{\sTROnd{71.51}} & \stackDiceCitIncreaseComment{\sTROrd{87.60}}{10.1007/978-3-031-43898-1_49}{}{} & 88.95 \\
    DeSD & 2022-09 & \stackDiceCitIncreaseComment{83.62}{10658004}{}{} &  & 89.20 & 81.90 & 70.60 & 72.70 &  & \stackDiceCitIncreaseComment{84.46}{10658004}{}{h} &  &  & 51.90 &  &  & 96.00 &  & 68.20 &  &  \\
    SMIT & 2022-05 & 87.80 &  &  &  &  &  &  &  &  &  &  &  &  &  &  &  &  &  \\
    UniSeg33A & 2022-03 &  &  &  &  &  &  &  &  &  &  &  &  &  &  &  &  &  &  \\
    UniMiSS & 2021-12 & 88.11 &  & \stackDiceCitIncreaseComment{61.21}{10658004}{}{} & \stackDiceCitIncreaseComment{63.94}{10658004}{}{} &  &  &  & \stackDiceCitIncreaseComment{84.66}{10658004}{}{i} &  &  &  &  &  &  &  &  &  &  \\
    DoDNet & 2020-11 & 86.44 & \stackDiceCitIncreaseComment{83.20}{10.1007/978-3-031-43898-1_49}{}{} & \stackDiceCitIncreaseComment{87.20}{10.1007/978-3-031-43898-1_49}{0.15}{} & 81.17 & 71.54 & 71.25 &  &  &  &  & \stackDiceCitIncreaseComment{54.60}{10.1007/978-3-031-43898-1_49}{3.05}{} &  &  & \stackDiceCitIncreaseComment{\sTROrd{96.50}}{10.1007/978-3-031-43898-1_49}{2.59}{} &  & \stackDiceCitIncreaseComment{70.40}{10.1007/978-3-031-43898-1_49}{2.50}{} & \stackDiceCitIncreaseComment{\cTROrd{\sTROnd{89.10}}}{10.1007/978-3-031-43898-1_49}{}{} &  \\
    Med3D & 2019-04 &  &  &  & \sTROrd{94.60} &  &  &  &  &  &  &  &  &  &  &  &  &  &  \\
}{
    ~\\
    ~\\
    \begin{itemize} \setlength\itemsep{-3pt}
        \item[a] Using the 2 click prompt configuration.
        \item[b] From Table I and II of \citet{shen2025interactive3dmedicalimage} using the 5 clicks prompt configuration.
        \item[c] Average Dice score between BraTS WT, ET, TC (91.58\%, 74.84\%, 86.22\%).
        \item[d] Average Dice score between BraTS WT, ET, TC (91.58\%, 74.84\%, 86.22\%).
        \item[e] With bbox \citep{du2024segvol}.
        \item[f] Average Dice score between BraTS WT, ET, TC (80.85\%, 65.69\%, 80.35\%).
        \item[g] Used MedSAM Tight Oracle Box prompt \citep{zhao2025modelrulealluniversal}.
        \item[h] Average Dice score between AMOS CT and MRI (86.36\% and 82.56\%).
        \item[i] Average Dice score between AMOS CT and MRI (85.82\% and 83.51\%).
    \end{itemize}
}
    \clearpage
\restoregeometry


\begin{figure*}[!ht]
  \centering
    \includegraphics[width=1\linewidth]{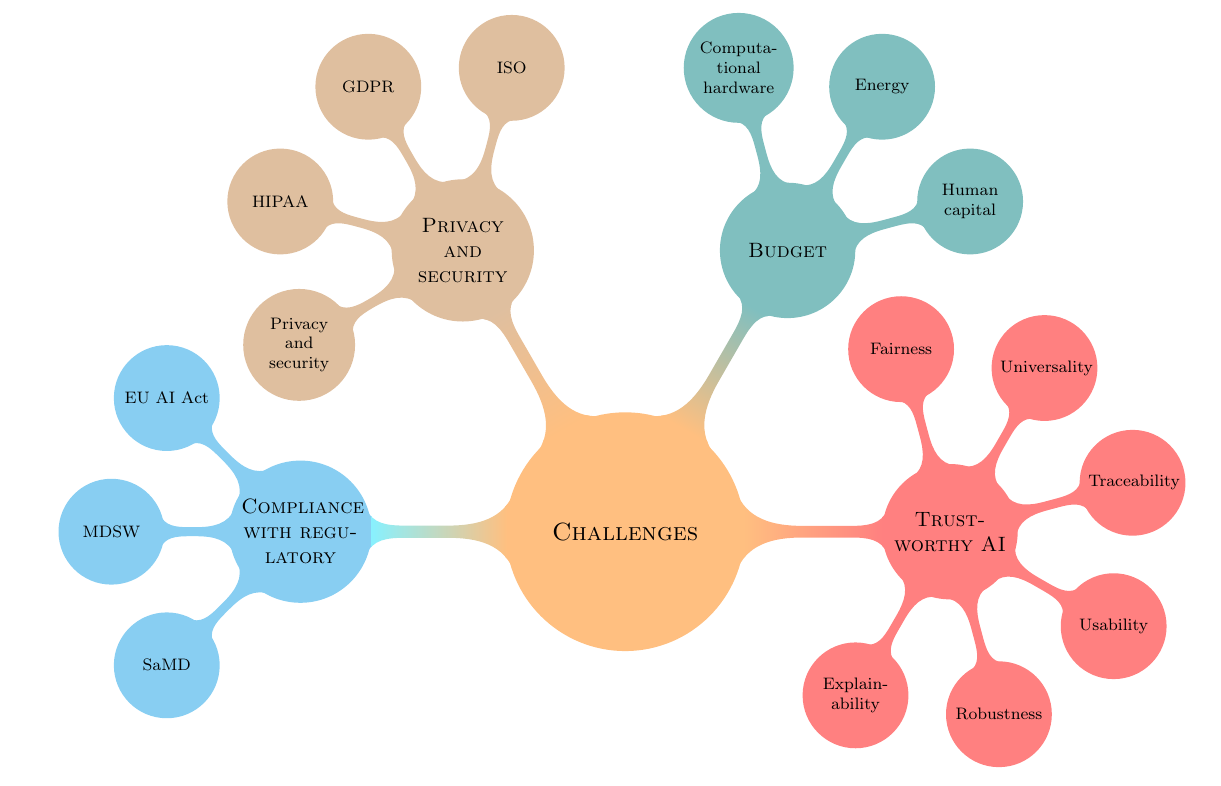}
    \caption{Challenges on current generation of generalist models for medical image segmentation.}
    \label{fig:challenges}
\end{figure*}

\section{Challenges}
\label{sec:challenges}

\subsection{Unavoidable compliance with regulatory}
The development of a medical device throughout its entire life-cycle must undergo several processes at different stages from the pre-market approval to the post-market surveillance mandated by the regulatory frameworks from different territories across the globe with their nuances and peculiarities. The regulatory frameworks have been updated in the last decade to include those medical devices where software has become a main component, especially for those based on AI. 

\textbf{Software as medical device:} In 2018 in the United States the Food and Drug Administration (FDA) recognized the prominent role of software on a vast amount of medical devices. Thus, the FDA identified certain software as medical device (SaMD) \citep{tang2025regulatory}, in line with the definition provided by the International Medical Devices Regulators Forum, defining SaMD as software intended for one or more medical purposes without being integrated into hardware medical devices \citep{tang2025regulatory, imdrf}. A regulatory framework was proposed in 2019 to extend the definition of SaMD to both the software leveraging AI for medical objectives, and to those medical devices integrating AI algorithms \citep{tang2025regulatory}. 

\textbf{Medical device software:} In Europe, medical device software (MDSW) is software that is intended to be used, alone or in combination, for a purpose as specified in the definition of a medical device in the medical devices regulation or in vitro diagnostic medical devices regulation \citep{mdsw}. 

\textbf{EU AI Act:} In 2024, the European Union published the AI Act (Regulation AI 2024/1689), the first legal framework in the world, specifically designed to address risks associated with AI, with AI medical devices defined as those with high risk, and for which the European Union Act mandates stringent regulatory measures, such as risk management protocols, data quality control, and requirements for explainability \citep{aiact}. Generalist models have not been explicitly defined nor mentioned in the EU AI Act. However, generative AI is amply regulated under the expression of general-purpose AI (GPAI) \citep{gpai}. The providers of GPAI models will have to comply with several obligations, effective from August 2, 2025: preparation of technical documentation with a general description of the model (architecture and number of parameters, modality, e.g., text and/or image, and format of inputs and outputs), and a specific description (training process, data for training, testing and validation, computational resources for model training, and estimated energy consumption). Additionally for the GPAI models with systemic risk, i.e., those  with a significant impact on the EU market e.g., with possible negative effects on public health, safety, or public security, providers of GPAI models must evaluate the model with SOTA protocols and tools, assess and mitigate possible systemic risks, report serious incidents and possible corrective measures, and ensure an adequate level of cybersecurity protection \citep{aiact, gpai}. 

In the United States the scenario on regulation of AI is much different from Europe. Although there are some frameworks and guidelines to regulate AI, there is no federal law on the development or restriction on the use of AI \citep{usai}.

\subsection{Privacy and security}
Whether SaMD or MDSW is concerned, a medical device manufacturer must also comply with privacy and security for a successful deployment. When considering SaMD or MDSW processing medical images, they must safeguard sensitive information to prevent privacy breaches compromising patient identity. Given the sensitivity of the subject matter, data security and privacy should be regulated by legislative bodies through specific laws.

\textbf{Health Insurance Portability and Accountability Act (HIPAA)}: it was published in the United States in 1996 \citep{hipaa}. It is based on several laws, of which the most important ones are the privacy and security laws. The former concerned strict rules to safeguard the privacy of patients health information, the right of patients to access their medical information and control its disclosure, and a set of specific rules which must be followed by organizations involved in the management of health data. The security law required health care organizations to implement appropriate security measures to protect electronic health data to prevent unauthorized access or security breaches.

\textbf{General Data Protection Regulation (GDPR)}: as Regulation EU 2016/679, it includes regulations that apply to organizations collecting data from EU citizens irrespective of the location of the organization \citep{gdpr}. Further, GDPR applies to the data storage of residents within the EU even if they are not EU citizens. The GDPR represents the most stringent data privacy and security law in the world.

Both HIPAA and GDPR define standards for de-identification of personal information. These measures involve different stakeholders, with patients being the most important ones. The standards ensure that their personal health data are protected. At the same time, by complying with data anonymization standards, manufacturers can process and share data for research, and commercial purposes, mitigating the risk of fines, data breaches, and damage to brand reputation.

\textbf{International Standard Organization (ISO):} Additionally, certifications like those issued by the International Standard Organization (ISO) were specifically created to protect patient data, e.g., the ISO/IEC 27559:2022 and ISO 29100:2024 report guidelines on data anonymization \citep{ISO/IEC27559:2022, ISO/IEC29100:2024}.

\subsection{Budget}
Size is a key term in the context of generalist models since it refers to both the quantity of necessary data during pre-training and the number of parameters, an indicator of the model complexity, generally correlating with performances. Therefore, size is not simply a raw number, but a measurement of the necessary investments on computational resources and of the energy consumption to develop them, fueling the debate on generalist vs. specialist models.
This survey found that the latter have a smaller size than the former in terms of number of parameters. As a consequence, specialist models can be trained with a relatively small budget in sharp contrast to generalist models, for the development of which the tech giants have a clear advantage. In fact, from ChatGPT by OpenAI for natural language processing to SAM by Meta for image segmentation, the development of generalist models has so far featured industry giants since the computational and engineering costs to train models with a massive number of parameters on massive datasets are prohibitive for academia \citep{zhang2024challenges}. 
This disparity has been raising important questions about the accessibility and democratization of research on AI, and biomedical research by considering the scope of this survey. As a result, many institutions may be constrained to fine-tuning existing pre-trained models, implemented by the big players, instead of developing their own \citep{queiroz2025fair}.

\textbf{Computational hardware}: In this survey, generalist models were very resource-hungry hardware, with Medical SAM 2 requiring 5,120 GB of GPU memory during training, a value out of reach for even for large medical centers.
Unfortunately, affordability collides with the clinical need of high end hardware for fast computation at inference time to assist clinicians, e.g., during a surgical procedure. Such hardware should not be an option but part of the basic equipment also within small and decentralized hospitals, which unfortunately may not afford investments of this magnitude. Therefore, research on high performing and cheaper hardware is strongly encouraged. Otherwise, the deployment of generalist models in small and suburban hospitals will be seriously hampered.

\textbf{Energy:} Another critical item cost for generalist model is energy. Tech giants have been close-mouthed about the energy consumption on their generalist models. Some of them admitted an increase of carbon emissions due to data centers construction \citep{chen2025much}. The AI Energy Score, a recent project hosted on HuggingFace, shows the energy requirements of several generalist models, also for computer vision tasks, but unfortunately not segmentation. As the number of data-centers increases, so does the energy to power them. Some analysts projected the energy consumption of data centers with a share of 15\% in the United States by 2028 \citep{chen2025much}. As a consequence also the expenses related to cooling the hardware will rise accordingly.

\textbf{Human capital:} Last but not least, the investment on human capital should not be omitted. By recognizing the crucial role of highly professionals with the necessary expertise, some renowned clinics have established specific departments and divisions on AI and informatics to exploit the potentialities of AI and generalist models in healthcare. Additionally, legal, and social experts are needed for a multidisciplinary collaboration with technical experts towards fairness in generalist models.

\subsection{Trustworthy AI}
As the result of a consensus among 117 interdisciplinary experts from 50 countries, the FUTURE-AI framework provides the guidelines for the development and deployment of trustworthy AI tools in healthcare based on fairness, universality, traceability, usability, robustness, and explainability \citep{lekadir2025future}. These principle can be applied also to generalist models. Therefore, they are illustrated in this section.

\textbf{Fairness} is a principle for which AI models in healthcare should perform at the same level across individuals and groups of individuals. Fairness may suffer from biases due to the differences in subjects in terms of gender, age, ethnicity, or the data, e.g., instrumentation, operators, and annotators. 
The biases in generalist models can be associated with uneven distribution of demographic data in the pre-training data. Therefore, specific patient groups are often underrepresented, resulting in datasets with distorted representations of disease prevalence and increasing the risk of AI models providing incorrect triage results and suboptimal medical treatment \citep{he2025survey, yang2025demographic}. Recent research on a visual and multimodal (text-visual) generalist models for classification on X-ray images revealed racial and gender-related bias on the model leading to disparate performance across patient subgroups \citep{glocker2023risk, yang2025demographic}. To mitigate biases in datasets composition, information on the centers where the data were acquired, the equipment used, the preprocessing and annotation processes should
be collected \citep{lekadir2025future}. Data preprocessing approaches to balance demographic representation may include resampling techniques, or data augmentation through synthetic data \citep{queiroz2025fair}.
Furthermore, there is a lack of benchmarks specifically for generalist models in contrast to task-specific ones \citep{queiroz2025fair}. 

\textbf{Universality} refers to the generalizability of AI model to external centers. Some challenges concern the differences of definition of diseases, and of medical equipment like radiological scanners. Therefore standardization of clinical definitions by medical societies, data annotation, medical data format (like DICOM for images) are encouraged \citep{lekadir2025future}.

\textbf{Traceability} refers to documenting and monitoring the entire lifetime of AI systems, e.g., generalist models, from development to deployment and usage, also in the context of continual learning, allowing them to evolve and improve with new data \citep{lekadir2025future, sun2024medical}. This monitoring 

Traceability should include a risk management plan to address risks as a consequence of data breach, or human factors leading to incorrect use of the AI tool, e.g., not following the instructions. Mitigation strategies should include proper documentation on AI system use, possible risks, and instructions for use, and technical documentation about training and testing data, evaluation metrics and benchmark used \citep{lekadir2025future}.

\textbf{Usability} refers to capability of the AI tool to reach a clinical goal efficiently and safely. Therefore the developers should design graphical user interfaces for an intuitive and effective use of the AI device, easy annotation and straightforward verification of AI inputs and results. To foster the best usage of the AI tool, reduce errors, and increase AI literacy, the developers
should provide training materials (e.g., tutorials and user manuals), and/or training activities in an accessible language, considering the diversity of users (e.g., medical doctors, nurses, technicians, etc...) \citep{lekadir2025future}.
To encourage adoption within the clinical workflow, the usability of the AI tool should be evaluated in real world settings with different end users (e.g., clinical role, technology/digital familiarity). The usability tests should provide evidence on the user satisfaction,
performance, and confidence \citep{lekadir2025future}.

\textbf{Robustness} refers to the ability to maintain the performance and accuracy despite variations in the input data. A mitigation plan includes careful selection and analysis of the training datasets, and implementation of validation studies reflecting variations of real world clinical practice, e.g., with data augmentation, and domain adaptation \citep{lekadir2025future}.

\textbf{Explainability} requires that the AI system provide information about the logic behind the AI decisions. It allows clinicians to interpret the AI model and outputs, understand the capacities and limitations of the AI tool, and intervene when necessary. The explainability should be evaluated to measure the correctness of the explanations. Limitations of the AI explanations, e.g., they are not clinically coherent, should be identified \citep{lekadir2025future}.

\section{Future directions}
\label{sec:future-directions}

\begin{figure*}[!ht]
  \centering
    \includegraphics[width=1\linewidth]{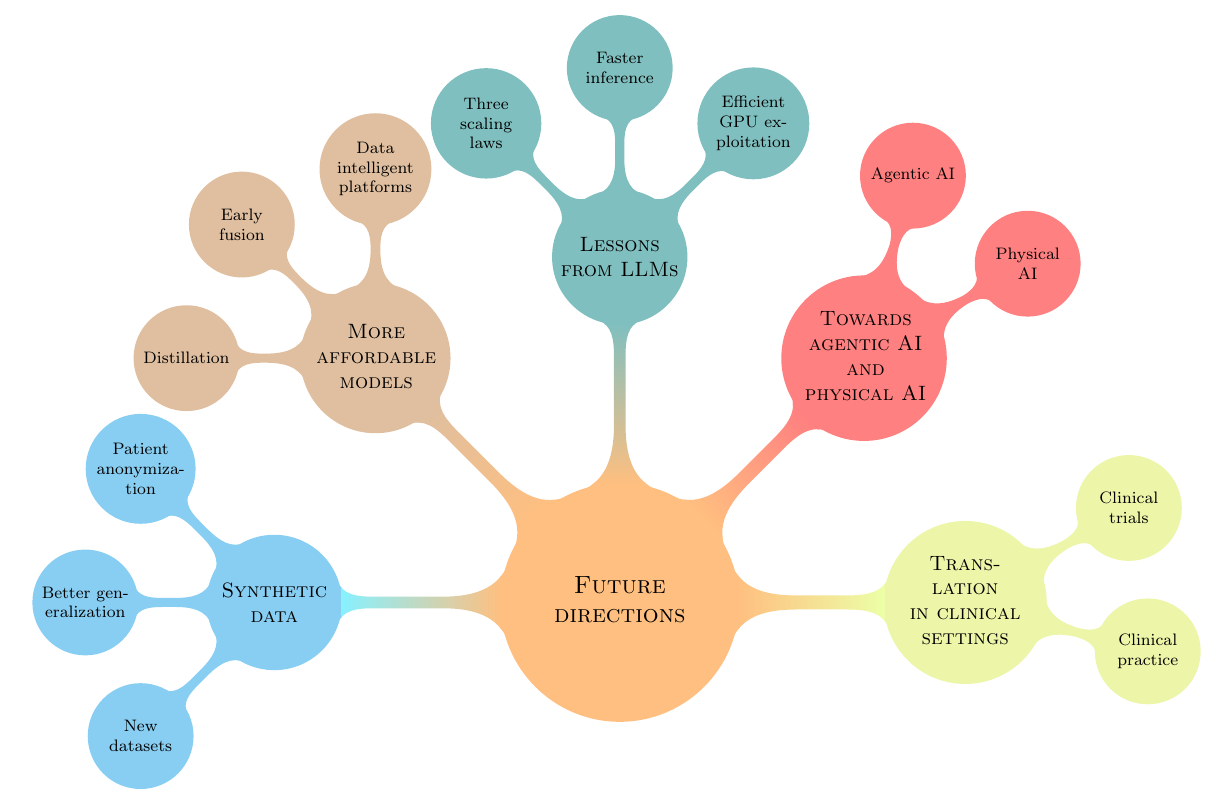}
    \caption{Future research directions on generalist models for medical image segmentation.}
    \label{fig:future-directions}
\end{figure*}

\subsection{Synthetic data}
Synthetic data is artificially generated data by capturing the statistical properties of the real data to create new data with similar properties \citep{pezoulas2024synthetic}. This approach has been attracting lot of interest since it can address several challenges in AI, especially in medical imaging field, where datasets are traditionally order of magnitude smaller than those of natural images. 
First, synthetic data can be leveraged to create \textbf{new datasets} or extend the existing ones in applications where there are no publicly available ones or where the size is limited which is frequent in medical image segmentation. They may also mitigate bias caused by uneven distribution of demographic data. Second, synthetic data may streamline the development of more diverse data \textbf{to improve generalization} of AI models on external centers. Third, Generative AI methods can be used to \textbf{streamline the anonymization} of medical images containing patient sensitive information \citep{li2025generative}. Thus, synthetic data do not violate patient privacy and security, thus respecting the provisions of HIPAA and GDPR laws \citep{pezoulas2024synthetic}. 
The implications for protection of patient privacy are remarkable. In case of successful implementation, synthetic data are not associated with any sensitive data of subjects, while preserving the statistical distribution and patterns within the dataset. This may contribute to the creation of robust datasets, and to foster a free sharing of data among institutions \citep{pezoulas2024synthetic}.
In the present survey only UniverSeg and BiomedParse generalist models generated synthetic data, although for different purposes \citep{butoi2023universeguniversalmedicalimage, zhao2024biomedparsebiomedicalfoundationmodel}. The former generated synthetic data to further increase the training task diversity, while the latter used GPT-4 to create a unifying biomedical object ontology avoiding noisy and inconsistent text description accompanying the images, and to synthesize synonymous text description from the semantic labels \citep{butoi2023universeguniversalmedicalimage, zhao2024biomedparsebiomedicalfoundationmodel}.

\subsection{More affordable models}
In order to equip small decentralized hospitals with generalist models, research should be pushed towards the development of smaller models capable to reach similar performances of larger models.\\ 
\textbf{Distillation:} The recently introduced RAD-DINO, a biomedical image encoder pre-trained only on biomedical imaging data, leveraged DINO-v2, a self-supervised method for distillation  \citep{perez2025exploring, oquab2023dinov2}. RAD-DINO, coupled with a UNet-like decoder, achieved similar performances of nn-UNet on chest x-ray images of the lung \citep{perez2025exploring}.\\
\textbf{Early fusion:} An alternative may be represented by fine-tuning on medical images early vision-language fusion-based SAM (EVF-SAM), a small SAM generalist model based on text prompt requiring only four GPUs with 24 GB of memory for training \citep{zhang2024evf}, a hardware resource within reach of several data centers.  Early fusion was used also by LLaMA 4 by Meta, the recently introduced multimodal generalist model \citep{llama4}.\\
\textbf{Data intelligent platforms:} An alternative to the optimization of AI models is represented by data intelligent platforms enabling to take full advantage of GPUs at a cost in line with the budget of small hospitals.

\subsection{Lessons from large language models}
\textbf{Three scaling laws:} Progress of performances in large language models has followed three scaling laws, each of which has led to an inflection in the degree of intelligence of AI-based systems. They are known as pretraining, post-training, and test-time scaling laws. It is interesting to point out that in the absence of a new scaling law, the performances of AI systems would improve, but a significantly lower rate. The pretraining scaling law relates hardware resources, model size, and training data \citep{scalinglaws}. It has shown that by increasing the model size and training data the performances improve, and that by increasing the power of computational hardware the model size and training data should increase in equal proportions \citep{hoffmann2022training}. According to the post-training law, the performances of a pretrained model can improve using techniques like fine-tuning, reinforcement learning, and distillation. Recent research has shown the reasoning capability of DeepSeek-R1-Zero using only reinforcement learning \citep{guo2025deepseek}.

\textbf{Faster inference:} While the first and second scaling law pertain to the training phase, the third one concerns the inference stage. It is called test-time scaling law, inference-time scaling or AI reasoning or long-thinking \citep{scalinglaws, muennighoff2025s1}. Introduced for the first time into o1 model by OpenAI, this law improves AI systems performance by allocating additional computational resources to evaluate multiple possible outcomes before selecting the best one \citep{inference-scaling}. Increasing the computation at inference time is behind the prowess of DeepSeek-R1 \citep{guo2022combination}.

\textbf{Efficient GPU exploitation:} In this regard, the successful story of DeepSeek R1 has demonstrated that a much smaller model than OpenAI-o1 could achieve performances in line with SOTA using less powerful GPU \citep{guo2025deepseek}. This was possible by a low-level programming of the GPU, optimizing inter-GPU communication, reducing latency, and using advanced algorithms to improve scalability in large clusters. These techniques allowed faster computations with lower memory requirements while maintaining acceptable accuracy levels. To optimize memory usage the model used flash attention mechanisms to reduce VRAM requirements during inference.

\subsection{Towards agentic AI, and physical AI}
What happened in the last few years in computer vision suggested that the field has followed the paradigm of natural language processing, from the design of transformers to capture relations among long range data \citep{vaswani2017attention, dosovitskiy2020image}, to the advent of generalist models, initially pretrained and demonstrating zero-shot generalizability \citep{radford2018improving, kirillov2023segany}.
In particular, the evolution of AI in medical image segmentation has been proceeding through different stages. The transition from one phase to the next one has occurred by following the above-mentioned three scaling laws which have been gradually introduced into computer vision. The first and second scaling law led from the perception stage, characterized by task-specific models, to the generative AI phase represented by the generalist models for medical image segmentation described in this survey.

\textbf{Agentic AI} represents the subsequent stage of AI, beyond generative AI, where systems can decompose complex tasks into subtasks, store and retrieve information over a long time, take action, and interact with external tools \citep{he2025survey}. Agentic AI are likely to benefit from the test-time scaling law, providing reasoned, helpful, and more accurate responses to complex questions, e.g.,  foreshadowing a future in medical image segmentation where autonomous AI agents could analyze radiological volumes of patients to predict disease progress and potential complications.

\textbf{Physical AI} embodies the fourth stage, allowing AI systems to understand the physics laws while interacting with the surrounding environment. Physical AI will benefit from the test-time scaling law, but in our opinion there is the need of a new scaling law to describe a new inflection of performances for systems which in the future will be enriched by new intelligent capabilities.
When applied to medical image segmentation, physical AI might improve tremendously the realism of digital twins. In fact, it has strong potential to realize the long-standing dream of many surgeons, i.e., to simulate realistically, according to the physics laws, a procedure on the specific anatomy of the patient, reconstructed by efficient and accurate AI segmentation models, before doing it on the real patient.

\subsection{Translation in clinical settings}
For the successful use of AI applications in healthcare, including those based on generalist models, the entire pipeline from research settings to clinical practice needs to be adequately revised. In addition to providing a solution to the previous challenges there are some specific aspects to be considered for the clinical adoption, from the perspective of research to conduct clinical trials and that of clinicians for their daily practice. 

\textbf{Clinical trials:} First, for informed consent, it may be difficult for patients to understand what they are giving consent to, e.g., how the generalist models work and the potential associated risks \citep{alsaad2024multimodal}. This poses questions on how to reframe the informed consent to include AI and generalist models. Training programs should be designed for healthcare professionals so that they can inform clearly the patients about the tools based on generalist models, their limitations, and ethical considerations \citep{alsaad2024multimodal}. Second, institutional review board and ethic committees may struggle to evaluate trials with generalist models due to the unfamiliarity with these models. For this reason academic medical institutions and teaching hospitals should integrate the existing competencies of the ethic committee with experts in data science and generative AI \citep{alsaad2024multimodal}.
The clinical trials should be performed at multi-center level to assess performance and interoperability across clinical workflows \citep{lekadir2025future}.

\textbf{Clinical practice:} Medical imaging is more demanding than natural images for AI, and to a larger extent generalist models. First, interoperability between the output formats of generalist models and medical informatics standards in use in hospitals, e.g., PACS and RIS, is required \citep{ali2024review}. Second, a false negative prediction even on few pixels may be crucial in the segmentation of challenging lesions, like early-stage tumors of the pancreas, whose timely diagnosis is of paramount importance for clinician and can make the difference on the life of patients. Third, it would be pivotal that the generalist models for medical imaging segmentation can reach the requested accuracy on multiple imaging modalities to gain a deep knowledge for the treatment of a specific disease. For instance, in pancreas cancer CT provides useful information on the organ resectability, and planning surgical interventions while MRI provides excellent soft tissue contrast, highlighting vascular and ductal details \citep{moglia2025deep}. Recent generalist models included in this survey, e.g., BiomedParse, have shown remarkable performances on different imaging modalities, envisioning their potential use in the clinical practice \citep{zhao2024biomedparsebiomedicalfoundationmodel}.
Eventually, the use of generalist models in medical imaging should translate to earlier diagnosis, better patient outcomes, increased productivity of clinicians, and healthcare organisations (e.g., reduced costs, optimised workflows) \citep{lekadir2025future}.

\section{Conclusions}
\label{sec:conclusions}

This review provided a comprehensive analysis of generalists models for medical image segmentation. Although the development of these models require massive datasets for pre-training, and huge investments for the purchase of the necessary hardware resources, the field has been ignited after the release of SAM, as witnessed by the development of numerous innovative generalist models in addition to the SAM implementations at the level of architecture or fine-tuning strategies. This triggered us to organize the literature by providing an in-depth taxonomy. Moreover, through a rigorous comparison with state-of-the-art task-specific models we highlighted which type of model is currently more accurate depending on the type of organ to be segmented. 
Finally, our review contributes to the field by emphasizing the challenges and future research directions for the adoption of generalist models in clinical practice.

\section*{Declaration of competing interest}
The authors declare that they do not have any identifiable conflicting financial interests or personal relationships that might have influenced the findings presented in this work.

\section*{Acknowledgments}
\label{sec:Acknowledgments}
This work was supported by Italian Ministry of Research and University - P.E. PE00000013-FUTURE ARTIFICIAL INTELLIGENCE RESEARCH (FAIR).

\section*{Declaration of Generative AI and AI-assisted technologies in the writing process}
During the preparation of this work, the authors did not use any tool or software based on generative AI and AI-assisted technologies.

\clearpage
\bibliographystyle{elsarticle-harv}

\clearpage

\clearpage
\appendix
\newcommand{\appendixtextbox}[1]{
    \hspace{2.4cm} 
    \parbox{15.15cm}{
        #1
    }%
}

\newgeometry{left=1cm, right=1cm} 

\hspace{-0.4cm}\appendixtextbox{
    \section{Technical background}
    \label{app:sec:technical-background}

    Table~\ref{table:models-specialized} reports the Specialized models considered in this work. It contains links to publications and online resources, and has the same layout of Table~\ref{table:models-foundational} which lists generalist models instead.
}

\makeTableModels{
    colorTablePapersSpecializedHeaderDark%
}{
    colorTablePapersSpecializedRowsPattern%
}{
    SOTA task-specific models used as benchmark for comparison with foundation models.
}{table:models-specialized}{
    \TMformatModel{LHU-Net}{LHU-Net: A Light Hybrid U-Net for Cost-Efficient, High-Performance Volumetric Medical Image Segmentation} & \TMformatAffil{\TMflagGermany Germany} & \TMformatDatePubliCit{\href{https://doi.org/10.48550/arXiv.2404.05102}{2024-04}}{\href{https://doi.org/10.48550/arXiv.2404.05102}{arXiv}}{\citet{sadegheih2024lhunetlighthybridunet}}{-}{-}{-} & \href{https://github.com/xmindflow/LHUNet}{\putIconGithub} & \TMformatArchitecture{Transformer with Convolutions\\\textit{(Custom)}} & \TMformatParams{10.5\\\textit{81.96}} & \TMformatRes{1 Nvidia A100 80GB}\\
    \TMformatModel{SCANeXt}{SCANeXt: Enhancing 3D medical image segmentation with dual attention network and depth-wise convolution} & \TMformatAffil{\TMflagChina China} & \TMformatDatePubliCit{\href{https://doi.org/10.1016/j.heliyon.2024.e26775}{2024-03}}{\href{https://doi.org/10.1016/j.heliyon.2024.e26775}{Heliyon}}{\citet{LIU2024e26775}}{-}{-}{-} & \codeUnavailable & \TMformatArchitecture{Transformer with Convolutions\\\textit{(Custom)}} & \TMformatParams{44.0\\\textit{50.53}} & \TMformatRes{1 Nvidia RTX 6000 48GB (24GB used)}\\
    \TMformatModel{SwinUNETR-V2}{SwinUNETR-V2: Stronger Swin Transformers withStagewise Convolutions for3D Medical Image Segmentation} & \TMformatAffil{\TMflagUSA U.S.A.} & \TMformatDatePubliCit{\href{https://doi.org/10.1007/978-3-031-43901-8_40}{2023-10}}{\href{https://doi.org/10.1007/978-3-031-43901-8_40}{MICCAI}}{\citet{10.1007/978-3-031-43901-8_40}}{-}{-}{-} & \href{https://github.com/Project-MONAI/research-contributions/tree/main/SwinUNETR}{\putIconGithub} & \TMformatArchitecture{Transformer with Convolutions\\\textit{(SwinUNETR with Interleaved Stage-Wise Residual Convolutions)}} & \TMformatParams{72.8\\\textit{320.0}} & \TMformatRes{-}\\
    \TMformatModel{NexToU}{NexToU: Efficient Topology-Aware U-Net for Medical Image Segmentation} & \TMformatAffil{\TMflagChina China} & \TMformatDatePubliCit{\href{https://doi.org/10.48550/arXiv.2305.15911}{2023-05}}{\href{https://doi.org/10.48550/arXiv.2305.15911}{arXiv}}{\citet{shi2023nextouefficienttopologyawareunet}}{-}{-}{-} & \href{https://github.com/PengchengShi1220/NexToU}{\putIconGithub} & \TMformatArchitecture{Graph\\\textit{(Custom)}} & \TMformatParams{23.06\\-} & \TMformatRes{1 Nvidia V100 32GB}\\
    \TMformatModel{MedNeXt}{MedNeXt: Transformer-driven Scaling of ConvNets for Medical Image Segmentation} & \TMformatAffil{\TMflagGermany Germany} & \TMformatDatePubliCit{\href{https://doi.org/10.48550/arXiv.2303.09975}{2023-03}}{\href{https://doi.org/10.48550/arXiv.2303.09975}{arXiv}}{\citet{roy2024mednexttransformerdrivenscalingconvnets}}{\href{https://doi.org/10.1007/978-3-031-43901-8_39}{2023-10}}{\href{https://doi.org/10.1007/978-3-031-43901-8_39}{MICCAI}}{\citet{10.1007/978-3-031-43901-8_39}} & \href{https://github.com/MIC-DKFZ/MedNeXt}{\putIconGithub} & \TMformatArchitecture{ConvNet\\\textit{(3D U-Net with ConvNeXt Blocks)}} & \TMformatParams{63.0\\\textit{564.0}} & \TMformatRes{-}\\
    \TMformatModel{UNETR++}{UNETR++: Delving into Efficient and Accurate 3D Medical Image Segmentation} & \TMformatAffil{\TMflagUAE U.A.E.\vspace{2pt}\\\TMflagUSA U.S.A.} & \TMformatDatePubliCit{\href{https://doi.org/10.48550/arXiv.2212.04497}{2022-12}}{\href{https://doi.org/10.48550/arXiv.2212.04497}{arXiv}}{\citet{shaker2024unetrdelvingefficientaccurate}}{\href{https://doi.org/10.1109/TMI.2024.3398728}{2024-05}}{\href{https://doi.org/10.1109/TMI.2024.3398728}{IEEE Transactions on Medical Imaging}}{\citet{10526382}} & \href{https://github.com/Amshaker/unetr_plus_plus}{\putIconGithub} & \TMformatArchitecture{Transformer\\\textit{(3D ViT (custom size))}} & \TMformatParams{42.96\\\textit{70.1}} & \TMformatRes{1 Nvidia A100 40GB}\\
    \TMformatModel{3D UX-Net}{3D UX-Net: A Large Kernel Volumetric ConvNet Modernizing Hierarchical Transformer for Medical Image Segmentation} & \TMformatAffil{\TMflagUSA U.S.A.} & \TMformatDatePubliCit{\href{https://doi.org/10.48550/arXiv.2209.15076}{2022-08}}{\href{https://doi.org/10.48550/arXiv.2209.15076}{arXiv}}{\citet{lee20233duxnetlargekernel}}{\href{https://openreview.net/forum?id=wsZsjOSytRA}{2023-05}}{\href{https://openreview.net/forum?id=wsZsjOSytRA}{ICLR}}{\citet{lee2023d}} & \href{https://github.com/MASILab/3DUX-Net}{\putIconGithub} & \TMformatArchitecture{ConvNet\\\textit{(3D U-Net with ConvNeXt Blocks)}} & \TMformatParams{53.0\\\textit{639.4}} & \TMformatRes{1 Nvidia RTX A6000 48GB}\\
    \TMformatModel{MedFormer}{A Data-scalable Transformer for Medical Image Segmentation: Architecture, Model Efficiency, and Benchmark} & \TMformatAffil{\TMflagChina China\vspace{2pt}\\\TMflagUSA U.S.A.} & \TMformatDatePubliCit{\href{https://doi.org/10.48550/arXiv.2203.00131}{2022-02}}{\href{https://doi.org/10.48550/arXiv.2203.00131}{arXiv}}{\citet{gao2023datascalabletransformermedicalimage}}{-}{-}{-} & \href{https://github.com/yhygao/CBIM-Medical-Image-Segmentation}{\putIconGithub} & \TMformatArchitecture{Transformer with Convolutions\\\textit{(Custom)}} & \TMformatParams{38.0\\\textit{460.2}} & \TMformatRes{1 Nvidia A100 40GB}\\
    \TMformatModel{TransBTSV2}{TransBTSV2: Towards Better and More Efficient Volumetric Segmentation of Medical Images} & \TMformatAffil{\TMflagChina China} & \TMformatDatePubliCit{\href{https://doi.org/10.48550/arXiv.2201.12785}{2022-01}}{\href{https://doi.org/10.48550/arXiv.2201.12785}{arXiv}}{\citet{li2022transbtsv2betterefficientvolumetric}}{-}{-}{-} & \href{https://github.com/Rubics-Xuan/TransBTS}{\putIconGithub} & \TMformatArchitecture{Transformer with Convolutions\\\textit{(3D U-Net, 3D ViT (custom size) with Deformable Attention)}} & \TMformatParams{15.3\\\textit{240.66}} & \TMformatRes{1 Nvidia TITAN RTX 24GB}\\
    \TMformatModel{SwinUNETR}{Swin UNETR: Swin Transformers for Semantic Segmentation of Brain Tumors in MRI Images} & \TMformatAffil{\TMflagUSA U.S.A.} & \TMformatDatePubliCit{\href{https://doi.org/10.48550/arXiv.2201.01266}{2022-01}}{\href{https://doi.org/10.48550/arXiv.2201.01266}{arXiv}}{\citet{hatamizadeh2022swinunetrswintransformers}}{\href{https://doi.org/10.1007/978-3-031-08999-2_22}{2022-07}}{\href{https://doi.org/10.1007/978-3-031-08999-2_22}{BrainLes}}{\citet{10.1007/978-3-031-08999-2_22}} & \href{https://github.com/Project-MONAI/research-contributions/tree/main/SwinUNETR}{\putIconGithub} & \TMformatArchitecture{Transformer\\\textit{(3D Swin-Base)}} & \TMformatParams{62.5\\\textit{295.0}} & \TMformatRes{8 Nvidia V100 32GB in DXG-1 Server}\\
    \TMformatModel{nnFormer}{nnFormer: Volumetric Medical Image Segmentation via a 3D Transformer} & \TMformatAffil{\TMflagChina China} & \TMformatDatePubliCit{\href{https://arxiv.org/abs/2109.03201v1}{2021-09}}{\href{https://arxiv.org/abs/2109.03201v1}{arXiv}}{\citet{zhou2022nnformerinterleavedtransformervolumetric}}{\href{https://ieeexplore.ieee.org/document/10183842}{2023-07}}{\href{https://ieeexplore.ieee.org/document/10183842}{IEEE Transactions on Image Processing}}{\citet{10183842}} & \href{https://github.com/282857341/nnFormer}{\putIconGithub} & \TMformatArchitecture{Transformer with Convolutions\\\textit{(3D Swin-Base)}} & \TMformatParams{37.6\\\textit{119.3}} & \TMformatRes{1 Nvidia GeForce RTX 2080 Ti 11GB}\\
    \TMformatModel{MISSFormer}{MISSFormer: An Effective Medical Image Segmentation Transformer} & \TMformatAffil{\TMflagChina China} & \TMformatDatePubliCit{\href{https://doi.org/10.48550/arXiv.2109.07162}{2021-09}}{\href{https://doi.org/10.48550/arXiv.2109.07162}{arXiv}}{\citet{huang2021missformereffectivemedicalimage}}{\href{https://doi.org/10.1109/TMI.2022.3230943}{2022-12}}{\href{https://doi.org/10.1109/TMI.2022.3230943}{IEEE Transactions on Medical Imaging}}{\citet{9994763}} & \href{https://github.com/zhifangdeng/missformer}{\putIconGithub} & \TMformatArchitecture{Transformer with Convolutions\\\textit{(2D CvT-like)}} & \TMformatParams{35.45\\\textit{36.96}} & \TMformatRes{1 Nvidia GeForce RTX 3090 24GB}\\
    \TMformatModel{Swin-Unet}{Swin-Unet: Unet-like Pure Transformer for Medical Image Segmentation} & \TMformatAffil{\TMflagChina China\vspace{2pt}\\\TMflagGermany Germany} & \TMformatDatePubliCit{\href{https://doi.org/10.48550/arXiv.2105.05537}{2021-05}}{\href{https://doi.org/10.48550/arXiv.2105.05537}{arXiv}}{\citet{cao2021swinunetunetlikepuretransformer}}{\href{https://doi.org/10.1007/978-3-031-25066-8_9}{2023-02}}{\href{https://doi.org/10.1007/978-3-031-25066-8_9}{ECCV}}{\citet{10.1007/978-3-031-25066-8_9}} & \href{https://github.com/HuCaoFighting/Swin-Unet}{\putIconGithub} & \TMformatArchitecture{Transformer\\\textit{(2D Swin-Tiny)}} & \TMformatParams{nan\\\textit{nan}} & \TMformatRes{1 Nvidia V100 32GB}\\
    \TMformatModel{UNETR}{UNETR: Transformers for 3D Medical Image Segmentation} & \TMformatAffil{\TMflagUSA U.S.A.} & \TMformatDatePubliCit{\href{https://doi.org/10.48550/arXiv.2103.10504}{2021-03}}{\href{https://doi.org/10.48550/arXiv.2103.10504}{arXiv}}{\citet{hatamizadeh2021unetrtransformers3dmedical}}{\href{https://doi.org/10.1109/WACV51458.2022.00181}{2022-02}}{\href{https://doi.org/10.1109/WACV51458.2022.00181}{IEEE/CVF WACV}}{\citet{9706678}} & \href{https://github.com/Project-MONAI/research-contributions/tree/main/UNETR}{\putIconGithub} & \TMformatArchitecture{Transformer\\\textit{(3D ViT-Base)}} & \TMformatParams{92.58\\\textit{41.19}} & \TMformatRes{8 Nvidia V100 32GB in DXG-1 Server}\\
    \TMformatModel{TransBTS}{TransBTS: Multimodal Brain Tumor Segmentation Using Transformer} & \TMformatAffil{\TMflagChina China} & \TMformatDatePubliCit{\href{https://doi.org/10.48550/arXiv.2103.04430}{2021-03}}{\href{https://doi.org/10.48550/arXiv.2103.04430}{arXiv}}{\citet{wang2021transbtsmultimodalbraintumor}}{\href{https://link.springer.com/chapter/10.1007/978-3-030-87193-2_11}{2021-09}}{\href{https://link.springer.com/chapter/10.1007/978-3-030-87193-2_11}{MICCAI}}{\citet{10.1007/978-3-030-87193-2_11}} & \href{https://github.com/Rubics-Xuan/TransBTS}{\putIconGithub} & \TMformatArchitecture{Transformer with Convolutions\\\textit{(3D U-Net, 3D ViT (custom size))}} & \TMformatParams{32.99\\\textit{333.09}} & \TMformatRes{8 Nvidia TITAN RTX 24GB}\\
    \TMformatModel{CoTr}{CoTr: Efficiently Bridging CNN and Transformer for 3D Medical Image Segmentation} & \TMformatAffil{\TMflagChina China} & \TMformatDatePubliCit{\href{https://doi.org/10.48550/arXiv.2103.03024}{2021-03}}{\href{https://doi.org/10.48550/arXiv.2103.03024}{arXiv}}{\citet{xie2021cotrefficientlybridgingcnn}}{\href{https://doi.org/10.1007/978-3-030-87199-4_16}{2021-09}}{\href{https://doi.org/10.1007/978-3-030-87199-4_16}{MICCAI}}{\citet{10.1007/978-3-030-87199-4_16}} & \href{https://github.com/YtongXie/CoTr}{\putIconGithub} & \TMformatArchitecture{Transformer with Convolutions\\\textit{(3D U-Net with Residuals, 3D ViT (custom size) with Deformable Attention)}} & \TMformatParams{41.9\\\textit{399.21}} & \TMformatRes{1 Nvidia GeForce RTX 2080 Ti 11GB}\\
    \TMformatModel{TransUNet}{TransUNet: Transformers Make Strong Encoders for Medical Image Segmentation} & \TMformatAffil{\TMflagUSA U.S.A.} & \TMformatDatePubliCit{\href{https://doi.org/10.48550/arXiv.2102.04306}{2021-02}}{\href{https://doi.org/10.48550/arXiv.2102.04306}{arXiv}}{\citet{chen2021transunettransformersmakestrong}}{\href{https://doi.org/10.1016/j.media.2024.103280}{2024-07}}{\href{https://doi.org/10.1016/j.media.2024.103280}{Medical Image Analysis}}{\citet{CHEN2024103280}} & \href{https://github.com/Beckschen/TransUNet}{\putIconGithub} & \TMformatArchitecture{Transformer with Convolutions\\\textit{(2D U-Net with Residuals, 2D ViT-Base)}} & \TMformatParams{41.4\\\textit{362.3}} & \TMformatRes{1 Nvidia Quadro RTX 8000 48GB}\\
    \TMformatModel{SETR}{Rethinking Semantic Segmentation from a Sequence-to-Sequence Perspective with Transformers} & \TMformatAffil{\TMflagChina China\vspace{2pt}\\\TMflagUK U.K.\vspace{2pt}\\\TMflagUSA U.S.A.} & \TMformatDatePubliCit{\href{https://doi.org/10.48550/arXiv.2012.15840}{2020-12}}{\href{https://doi.org/10.48550/arXiv.2012.15840}{arXiv}}{\citet{zheng2021rethinkingsemanticsegmentationsequencetosequence}}{\href{https://doi.org/10.1109/CVPR46437.2021.00681}{2021-11}}{\href{https://doi.org/10.1109/CVPR46437.2021.00681}{IEEE/CVF CVPR}}{\citet{9578646}} & \href{https://github.com/fudan-zvg/SETR}{\putIconGithub} & \TMformatArchitecture{Transformer\\\textit{(2D ViT-Large)}} & \TMformatParams{97.64\\-} & \TMformatRes{-}\\
    \TMformatModel{SegResNet}{3D MRI brain tumor segmentation using autoencoder regularization} & \TMformatAffil{\TMflagUSA U.S.A.} & \TMformatDatePubliCit{\href{https://doi.org/10.48550/arXiv.1810.11654}{2018-10}}{\href{https://doi.org/10.48550/arXiv.1810.11654}{arXiv}}{\citet{myronenko20183dmribraintumor}}{-}{-}{-} & \codeUnavailable & \TMformatArchitecture{ConvNet\\\textit{(3D U-Net with ResNet blocks)}} & \TMformatParams{-\\-} & \TMformatRes{8 Nvidia V100 32GB in DXG-1 Server}\\
    \TMformatModel{nnU-Net}{nnU-Net: a self-configuring method for deep learning-based biomedical image segmentation} & \TMformatAffil{\TMflagGermany Germany} & \TMformatDatePubliCit{\href{https://doi.org/10.48550/arXiv.1809.10486}{2018-10}}{\href{https://doi.org/10.48550/arXiv.1809.10486}{arXiv}}{\citet{isensee2018nnunetselfadaptingframeworkunetbased}}{\href{https://doi.org/10.1038/s41592-020-01008-z}{2020-12}}{\href{https://doi.org/10.1038/s41592-020-01008-z}{Nature Methods}}{\citet{Isensee2021}} & \href{https://github.com/MIC-DKFZ/nnUNet}{\putIconGithub} & \TMformatArchitecture{ConvNet\\\textit{(2D U-Net, 3D U-Net)}} & \TMformatParams{31.2\\\textit{539.7}} & \TMformatRes{-}\\
    \TMformatModel{UNet++}{UNet++: Redesigning Skip Connections to Exploit Multiscale Features in Image Segmentation} & \TMformatAffil{\TMflagUSA U.S.A.} & \TMformatDatePubliCit{\href{https://doi.org/10.48550/arXiv.1807.10165}{2018-07}}{\href{https://doi.org/10.48550/arXiv.1807.10165}{arXiv}}{\citet{zhou2018unetnestedunetarchitecture}}{\href{https://doi.org/10.1109/TMI.2019.2959609}{2019-12}}{\href{https://doi.org/10.1109/TMI.2019.2959609}{IEEE Transactions on Medical Imaging}}{\citet{8932614}} & \href{https://github.com/MrGiovanni/UNetPlusPlus}{\putIconGithub} & \TMformatArchitecture{ConvNet\\\textit{(Custom)}} & \TMformatParams{9.0\\-} & \TMformatRes{3 Nvidia GeForce GTX TITAN X 12GB}\\
    \TMformatModel{V-Net}{V-Net: Fully Convolutional Neural Networks for Volumetric Medical Image Segmentation} & \TMformatAffil{\TMflagGermany Germany} & \TMformatDatePubliCit{\href{https://doi.org/10.48550/arXiv.1606.04797}{2016-06}}{\href{https://doi.org/10.48550/arXiv.1606.04797}{arXiv}}{\citet{milletari2016vnetfullyconvolutionalneural}}{\href{https://doi.org/10.1109/3DV.2016.79}{2016-10}}{\href{https://doi.org/10.1109/3DV.2016.79}{IEEE 3DV}}{\citet{7785132}} & \href{https://github.com/faustomilletari/VNet}{\putIconGithub} & \TMformatArchitecture{ConvNet\\\textit{(Custom)}} & \TMformatParams{-\\-} & \TMformatRes{1 Nvidia Geforce GTX 1080 8GB}\\
    \TMformatModel{3D U-Net}{3D U-Net: Learning Dense Volumetric Segmentation from Sparse Annotation} & \TMformatAffil{\TMflagGermany Germany} & \TMformatDatePubliCit{\href{https://doi.org/10.48550/arXiv.1606.06650}{2016-06}}{\href{https://doi.org/10.48550/arXiv.1606.06650}{arXiv}}{\citet{cicek20163dunetlearningdense}}{\href{https://doi.org/10.1007/978-3-319-46723-8_49}{2016-10}}{\href{https://doi.org/10.1007/978-3-319-46723-8_49}{MICCAI}}{\citet{10.1007/978-3-319-46723-8_49}} & \codeUnavailable & \TMformatArchitecture{ConvNet\\\textit{(Custom)}} & \TMformatParams{19.07\\-} & \TMformatRes{1 Nvidia GeForce GTX TITAN X 12GB}\\
    \TMformatModel{U-Net}{U-Net: Convolutional Networks for Biomedical Image Segmentation} & \TMformatAffil{\TMflagGermany Germany} & \TMformatDatePubliCit{\href{https://doi.org/10.48550/arXiv.1505.04597}{2015-05}}{\href{https://doi.org/10.48550/arXiv.1505.04597}{arXiv}}{\citet{ronneberger2015unetconvolutionalnetworksbiomedical}}{\href{https://doi.org/10.1007/978-3-319-24574-4_28}{2015-11}}{\href{https://doi.org/10.1007/978-3-319-24574-4_28}{MICCAI}}{\citet{10.1007/978-3-319-24574-4_28}} & \codeUnavailable & \TMformatArchitecture{ConvNet\\\textit{(Custom)}} & \TMformatParams{31.0\\-} & \TMformatRes{1 Nvidia GeForce GTX TITAN}
}{~}

\newgeometry{left=0.6cm, right=0.6cm} 
\appendixtextbox{
    \section{Contribution analysis}
    \label{app:sec:results}

    Tables \ref{table:results-original-specialized}, \ref{table:results-original-foundational}, \ref{table:results-best-specialized}, report results from the considered works. Best-in-literature results for generalist models can be found in Table~\ref{table:results-best-foundational}.
}

\makeTableResultsOverall{%
    colorTablePapersSpecializedHeaderClear%
}{
    colorTablePapersSpecializedRowsPattern%
}{%
    Dice score achieved by task-specific models in their first publication expressed as percentage [\%].
}{table:results-original-specialized}{
    LHU-Net & 2024-04 &  & 86.05 &  &  &  &  & \sTROrd{87.49} &  & 92.66 &  &  &  &  &  &  &  &  &  \\
    SCANeXt & 2024-03 &  & 86.60 &  &  &  &  & \cTROnd{\sTROst{89.67}} &  & \cTROst{\sTROst{95.18}} &  &  &  &  &  &  &  &  &  \\
    SwinUNETR-V2 & 2023-10 &  &  &  &  & \sTROnd{64.03} & 62.03 &  &  &  &  &  & \cTROst{\sTROst{94.70}} &  &  &  &  & \sTROnd{74.05} &  \\
    NexToU & 2023-05 & \sTROnd{87.84} &  &  &  &  &  &  &  &  &  &  &  &  &  &  &  &  &  \\
    MedNeXt & 2023-03 & \sTROst{88.76} & \sTROrd{88.01} & \sTROnd{91.02} &  &  &  &  & \cTROst{\sTROst{91.77}} &  &  &  &  &  &  &  &  &  &  \\
    UNETR++ & 2022-12 & 83.28 & 82.75 &  &  &  & \cTROnd{\sTROst{80.68}} & 87.22 &  & \cTROrd{\sTROrd{92.83}} &  &  &  &  &  &  &  &  &  \\
    3D UX-Net & 2022-08 &  &  &  &  &  &  &  & \cTROrd{\sTROnd{90.00}} &  &  &  & \cTROnd{\sTROnd{93.40}} &  &  &  &  &  &  \\
    MedFormer & 2022-02 & 85.00 &  & 85.00 & 69.00 &  & \sTROnd{74.00} &  & \sTROrd{88.00} & 92.50 &  &  &  &  &  &  &  &  &  \\
    TransBTSV2 & 2022-01 &  & \stackDiceComment{85.04}{a} & \sTROrd{90.53} & \sTROst{89.85} &  &  &  &  &  &  &  &  &  &  &  &  &  &  \\
    \stackModelComment{SwinUNETR}{b} & 2022-01 & 83.48 & \sTROnd{88.96} &  &  & \sTROrd{55.49} & 56.72 &  &  &  &  &  & \cTROrd{\sTROrd{92.90}} &  &  &  &  & \sTROrd{73.32} &  \\
    nnFormer & 2021-09 &  & 86.40 &  &  &  &  & 86.57 &  & 92.06 &  &  &  &  &  &  &  &  &  \\
    MISSFormer & 2021-09 &  &  &  &  &  &  & 81.96 &  & 91.19 &  &  &  &  &  &  &  &  &  \\
    Swin-Unet & 2021-05 &  &  &  &  &  &  & 79.13 &  & 90.00 &  &  &  &  &  &  &  &  &  \\
    \stackModelComment{TransBTS}{c} & 2021-03 &  & 83.57 & 89.10 & \sTROnd{88.95} &  &  &  &  &  &  &  &  &  &  &  &  &  &  \\
    CoTr & 2021-03 & 85.00 &  &  &  &  &  &  &  &  &  &  &  &  &  &  &  &  &  \\
    UNETR & 2021-03 & \stackDiceComment{87.35}{d} & 71.10 &  &  &  &  &  &  &  &  &  &  &  & \sTROnd{96.40} &  &  &  &  \\
    TransUNet & 2021-02 &  & \cTROnd{\sTROst{91.74}} &  &  &  &  & \sTROnd{88.39} &  &  &  &  &  &  &  &  & \sTROnd{67.67} &  &  \\
    \stackModelComment{SETR}{e} & 2020-12 &  &  &  &  &  &  &  &  &  &  &  &  &  &  &  &  &  &  \\
    SegResNet & 2018-10 &  & 82.19 &  &  &  &  &  &  &  &  &  &  &  &  &  &  &  &  \\
    nnU-Net & 2018-10 & \sTROrd{87.62} & 61.00 & \cTROrd{\sTROst{91.63}} & \sTROrd{86.50} & \sTROst{67.50} & \sTROnd{74.00} &  &  & \cTROnd{\sTROnd{92.95}} & \cTROrd{\sTROst{91.94}} & \sTROst{58.00} &  &  & \cTROnd{\sTROst{97.00}} & \cTROst{\sTROst{93.00}} & \sTROst{69.00} & \cTROnd{\sTROst{83.50}} &  \\
    UNet++ & 2018-07 &  &  &  & 82.60 &  &  &  &  &  &  &  &  &  &  &  &  &  &  \\
    V-Net & 2016-06 &  &  &  &  &  &  &  &  &  & \sTROnd{86.90} &  &  &  &  &  &  &  &  \\
    U-Net & 2015-05 &  &  &  &  &  &  &  &  &  &  &  &  &  &  &  &  &  &  \\
}{
    \begin{itemize} \setlength\itemsep{0pt}
        \item[a] Average Dice score between BraTS2019 and BraTS2020 datasets.
        \item[b] Results from SwinUNETR-V2 \citep{SwinUNETRv2_miccai} that provided updated Dice scores from the same research group.
        \item[c] Results from TransBTSV2 \citep{li2022transbtsv2betterefficientvolumetric} that provided updated Dice scores from the same research group.
        \item[d] Average Dice score between BTCV \"free\" and \"standard\" competition datasets.
        \item[e] Original model applied to natural images. Included because it was used as benchmark by some other models.
    \end{itemize}
}
\clearpage

\makeTableResultsOverall{%
    colorTablePapersFoundationalHeaderClear%
}{
    colorTablePapersFoundationalRowsPattern%
}{%
    Dice score achieved by generalist models in their first publication expressed as percentage [\%].
}{table:results-original-foundational}{
    \stackModelComment{MedSAM2}{a} & 2025-04 &  &  &  &  &  &  &  &  &  &  &  &  &  &  &  &  &  &  \\
    SPA & 2025-01 &  &  &  &  &  &  & \cTROst{\sTROst{92.88}} &  &  & \cTROst{\sTROst{94.29}} &  &  &  &  &  &  &  &  \\
    3DMedSAM & 2024-12 & 88.60 &  &  & 60.45 &  &  &  &  &  &  &  &  &  &  &  &  &  &  \\
    KnowSAM & 2024-12 &  &  &  &  &  &  &  &  & \sTROnd{91.13} &  &  &  &  &  &  &  &  &  \\
    IMIS-Net & 2024-11 &  &  &  &  &  &  &  &  &  &  &  &  & \stackDiceComment{79.06}{b} &  & \cTROrd{\sTROnd{89.27}} &  &  &  \\
    \stackModelComment{SAM-MPA}{c} & 2024-10 &  &  &  &  &  &  &  &  &  &  &  &  &  &  &  &  &  &  \\
    TP-Mamba & 2024-09 & 84.80 &  &  &  &  &  &  &  &  &  &  &  &  &  &  &  &  &  \\
    EMedSAM & 2024-08 &  & \sTROrd{89.30} &  &  &  &  &  &  &  &  &  & 0.88 &  &  &  &  &  &  \\
    SAM 2 & 2024-08 &  &  &  &  &  &  &  &  &  &  &  &  &  &  &  &  &  &  \\
    Medical SAM 2 (MedSAM-2) & 2024-08 & \cTROrd{\sTROrd{89.00}} &  & 78.20 &  &  &  &  &  &  &  &  &  &  &  &  &  &  &  \\
    Biomedical SAM-2 (BioSAM-2) & 2024-08 &  &  &  &  &  &  &  & \stackDiceComment{74.39}{d} &  &  &  & \stackDiceComment{76.32}{e} &  &  &  &  &  &  \\
    FLAP-SAM & 2024-07 &  &  & 60.46 &  &  &  &  &  &  & 88.67 &  &  &  &  &  &  &  &  \\
    LeSAM & 2024-06 &  & \stackDiceComment{84.95}{f} & \cTROnd{\sTROnd{91.86}} & 70.62 & \cTROrd{\sTROrd{79.42}} & \sTROrd{79.57} &  &  &  &  & \cTROst{\sTROst{77.18}} &  &  &  &  & \cTROst{\sTROst{79.59}} &  &  \\
    Merlin & 2024-06 &  &  &  &  &  &  &  &  &  &  &  &  & 86.00 &  &  &  &  &  \\
    BrainSegFounder & 2024-06 &  & \cTROrd{\sTROnd{91.15}} &  &  &  &  &  &  &  &  &  &  &  &  &  &  &  &  \\
    \stackModelComment{MoME}{g} & 2024-05 &  & 88.86 &  &  &  &  &  &  &  &  &  &  &  &  &  &  &  &  \\
    \stackModelComment{BiomedParse}{h} & 2024-05 &  & 79.95 & 80.22 & 83.39 & 50.62 & 66.09 &  & 86.33 & \sTROst{92.26} & \sTROrd{89.97} & \cTROnd{\sTROnd{66.51}} &  &  & \cTROrd{\sTROnd{96.86}} &  & 66.03 & \sTROrd{72.85} &  \\
    \stackModelComment{PCNet}{i} & 2024-04 & 83.85 &  & 86.19 & \cTROst{\sTROst{96.63}} & \cTROnd{\sTROnd{79.70}} &  &  &  &  &  &  & \sTROnd{90.62} & \cTROst{\sTROst{91.64}} & 95.77 & 87.66 &  &  & \cTROst{\sTROst{91.09}} \\
    MEA M-SAM & 2024-03 &  & \cTROst{\sTROst{92.08}} & \cTROst{\sTROst{93.50}} & 89.95 & \cTROst{\sTROst{80.49}} & \cTROst{\sTROst{81.62}} &  &  &  &  &  &  &  &  &  &  &  &  \\
    SFR SAM & 2024-03 & 77.07 & 86.09 &  &  &  &  &  &  &  &  &  &  &  &  &  &  &  &  \\
    \stackModelComment{Med-SA}{j} & 2023-12 & 88.30 & 89.10 &  &  &  &  &  &  &  &  &  &  &  &  &  &  &  &  \\
    \stackModelComment{SAT}{k} & 2023-12 & 81.60 & \stackDiceComment{55.68}{l} & 71.53 & 78.86 & 59.23 & 61.28 &  & \stackDiceComment{84.82}{m} & 89.64 & 87.28 & 38.45 & \sTROst{91.78} & \stackDiceComment{\cTROrd{\sTROrd{86.71}}}{n} & 94.97 & \sTROrd{88.98} & 63.43 & \cTROrd{\sTROnd{77.98}} & \cTROnd{\sTROnd{90.42}} \\
    SegVol & 2023-11 &  &  &  &  &  &  &  & 85.93 &  &  &  &  &  &  & 81.55 &  &  &  \\
    \stackModelComment{SAM-Med3D}{o} & 2023-10 & 79.17 &  & 72.06 &  &  &  &  & \stackDiceComment{75.41}{p} &  &  &  &  & 84.68 &  &  &  &  &  \\
    SAM3D & 2023-09 &  & 72.90 &  &  &  & 71.42 & 79.56 &  & \sTROrd{90.41} &  &  &  &  &  &  &  &  &  \\
    MA-SAM & 2023-09 & 87.20 &  &  &  & \stackDiceComment{40.20}{q} &  &  &  &  & \stackDiceComment{\cTROnd{\sTROnd{92.60}}}{r} & \stackDiceComment{47.70}{s} &  &  &  &  &  &  &  \\
    Cheap Lunch SAM & 2023-08 &  & 85.28 &  &  &  &  & \sTROrd{85.95} &  &  &  &  &  &  &  &  &  &  &  \\
    \stackModelComment{SAM-Med2D}{t} & 2023-08 &  &  & 79.87 &  &  &  &  &  &  &  &  & 85.10 &  &  &  &  &  &  \\
    Disruptive Autoencoders & 2023-07 & \cTROst{\sTROst{92.10}} &  &  &  &  &  &  &  &  &  &  &  &  &  &  &  &  &  \\
    SAMMed & 2023-07 & 70.30 &  & 84.00 & 92.00 &  &  &  &  &  &  &  &  &  &  &  &  &  &  \\
    \stackModelComment{DeSAM}{u} & 2023-06 &  &  &  &  &  &  &  &  &  &  &  &  &  &  &  &  &  &  \\
    \stackModelComment{MedLSAM}{v} & 2023-06 &  &  &  &  &  &  &  &  &  &  &  &  &  &  &  &  &  &  \\
    \stackModelComment{HERMES}{w} & 2023-06 & 86.29 &  & 85.98 & 68.32 & \stackDiceComment{72.07}{x} &  &  & \stackDiceComment{\sTROrd{88.59}}{y} &  &  &  &  &  &  &  &  &  &  \\
    MIS-FM & 2023-06 &  &  &  &  &  &  & \cTROrd{\sTROnd{89.11}} &  &  &  &  &  &  &  & \cTROnd{\sTROst{89.56}} &  &  &  \\
    \stackModelComment{3DSAM-adapter}{z} & 2023-06 &  &  & \stackDiceComment{81.50}{a2} & \stackDiceComment{61.25}{a2} & \stackDiceComment{66.87}{b2} &  &  &  &  &  & 60.93 &  &  &  &  &  &  &  \\
    One-Prompt & 2023-05 &  &  & 67.30 &  &  &  &  &  &  &  &  &  &  &  &  &  &  &  \\
    SAM & 2023-04 &  &  &  &  &  &  &  &  &  &  &  &  &  &  &  &  &  &  \\
    \stackModelComment{MedSAM}{c2} & 2023-04 &  &  &  &  &  &  &  &  &  &  &  &  &  &  &  &  &  &  \\
    \stackModelComment{UniverSeg}{d2} & 2023-04 &  &  &  &  &  &  &  &  & 70.90 &  &  &  &  &  &  &  &  &  \\
    SAMed & 2023-04 &  &  &  &  &  &  & \stackDiceComment{84.30}{e2} &  &  &  &  &  &  &  &  &  &  &  \\
    UniSeg & 2023-04 & 84.60 & 83.30 & 88.20 & 79.10 & 70.90 & 70.90 &  &  &  &  & 55.00 &  &  & \sTROnd{96.40} &  & \cTROrd{\sTROrd{71.20}} & \cTROst{\sTROst{89.70}} &  \\
    STU-Net & 2023-04 & 83.83 &  & 85.44 & \cTROnd{\sTROnd{95.88}} & 78.95 &  &  & \cTROnd{\sTROst{90.49}} &  &  &  & \sTROrd{89.87} & \cTROnd{\sTROnd{90.06}} & 95.52 & 85.91 &  &  & \cTROrd{\sTROrd{89.82}} \\
    MultiTalent & 2023-03 & \cTROnd{\sTROnd{89.07}} &  & \stackDiceComment{\sTROrd{90.45}}{f2} &  &  &  &  & \sTROnd{89.81} &  &  &  &  &  &  &  &  &  &  \\
    CLIP-Driven Universal Model & 2023-01 & 86.13 &  &  & 87.39 & 72.59 & \cTROrd{\sTROnd{80.01}} &  &  &  &  & \cTROrd{\sTROrd{63.14}} &  &  & \cTROst{\sTROst{97.27}} &  & \cTROnd{\sTROnd{71.51}} &  & 88.95 \\
    DeSD & 2022-09 &  &  & 89.20 & 81.90 & 70.60 & 72.70 &  &  &  &  & 51.90 &  &  & 96.00 &  & 68.20 &  &  \\
    SMIT & 2022-05 & 87.80 &  &  &  &  &  &  &  &  &  &  &  &  &  &  &  &  &  \\
    \stackModelComment{UniSeg33A}{g2} & 2022-03 &  &  &  &  &  &  &  &  &  &  &  &  &  &  &  &  &  &  \\
    UniMiSS & 2021-12 & 88.11 &  &  &  &  &  &  &  &  &  &  &  &  &  &  &  &  &  \\
    \stackModelComment{DoDNet}{h2} & 2020-11 & 86.44 &  & 87.05 & 81.17 & 71.54 & 71.25 &  &  &  &  & 51.55 &  &  & 93.91 &  & 67.90 &  &  \\
    Med3D & 2019-04 &  &  &  & \cTROrd{\sTROrd{94.60}} &  &  &  &  &  &  &  &  &  &  &  &  &  &  \\
}{
    \begin{itemize} \setlength\itemsep{0pt}
        \item[a] All results in the published paper \citep{ma2025medsam2segment3dmedical} are grouped per organ or lesion mixing different dataset sources. Please refer to the original publication for more details.
        \item[b] Only results on Totalsegmntator MRI were provided.
        \item[c] Test results were provided on BreastUS and Chest XRay public datasets, that are 2D only.
        \item[d] MRI dataset only.
        \item[e] CT dataset only.
        \item[f] Average Dice score between WT, ET, TC (91.83\%, 75.98\%, 87.05\% respectively).
        \item[g] Results from Table IV of \citep{10879789}. Dice scores were averaged across imaging modalities per dataset where more imaging modalities were provided.
        \item[h] Results obtained from the original raw segmentation metrics available on \href{https://github.com/microsoft/BiomedParse/tree/main/figures/results/dataset_results}{BiomedParse's GitHub}.
        \item[i] Results from Table II of \citep{10510478}, considered \textit{STUNet-L w/ PC} that has the highest score on \textit{TotalSegmentator}. Also in Table III that model has best mean score on all datasets.
        \item[j] Reported Dice scores in 1 point prompt setting.
        \item[k] Reporting results for SAT-Ft (fine-tuned).
        \item[l] Average Dice score between BraTS2023 GLI, MEN, MET, PED, SSA.
        \item[m] Average Dice score between CHAOS CT (97.24\%) and CHAOS MRI (87.99\%).
        \item[n] TS v2 in Table 6 \citep{zhao2025modelrulealluniversal}.
        \item[o] Results reported in the 1-point prompt framework.
        \item[p] Average Dice score between AMOS2 CT (79.94\%) and AMOS2 MRI (75.41\%).
        \item[q] Automatic, no prompts, tumor Dice score only.
        \item[r] Dice score obtain on composite dataset (PRIMOSE12 + others, please refer to original manuscript). Automatic, no prompts (Best Dice score with prompts 80.3\%).
        \item[s] Automatic, no prompts (Best Dice score with prompts 81.1\%).
        \item[t] Results reported from Table 4 only \citep{cheng2023sammed2d}.
        \item[u] Results mixed-up between datasets.
        \item[v] Results reported per-organ, not per-dataset.
        \item[w] HERMES-M (MedFormer visual backbone) is considered.
        \item[x] Average Dice score between pancreas (82.73\%) and tumor (61.41\%) with convolutional backbone.
        \item[y] Average Dice score between AMOS CT and MRI (89.98\% and 87.20\%).
        \item[z] Results reported for 3 points per volume for all datasets. Table 2 not considered \citep{3DSAM-adapter}.
        \item[a2] Tumor Dice score only.
        \item[b2] Mean between only tumor segmentation (Table 1) and whole organ (pancreas+tumor as one class) segmentation Table 6 \citep{3DSAM-adapter}.
        \item[c2] All results in the Supplementary Material of the published paper \citep{MedSAM_nature} are grouped per organ or lesion mixing different dataset sources. Please refer to the original publication for more details.
        \item[d2] These results are obtained on held-out datasets and previously unseen tasks. Due to the structure of the network, during inference, the model is provided with an unseen image for segmentation along with a set of eight example image-mask pairs of the same type and task (e.g., aorta segmentation in CT scans). The network is designed to perform on-the-fly learning from these examples and apply the learned information to the new image. The reliance on few-shot learning for segmentation likely contributes to the relatively low performance scores observed.
        \item[e2] Results from the \href{https://github.com/hitachinsk/SAMed?tab=readme-ov-file}{SAMed GitHub} where SAMed\_h with vit\_h backbone was announced. Results reported in prints was 81.88\%.
        \item[f2] Average Dice score between organ (96.89\%) and tumor (84.01\%).
        \item[g2] Reported results are unclear and were not able to understand which datasets were used.
        \item[h2] When applicable, results are the average between organ and tumor scores.
    \end{itemize}
}
\clearpage

\makeTableResultsOverall{%
    colorTablePapersSpecializedHeaderClear%
}{
    colorTablePapersSpecializedRowsPattern%
}{%
    Highest Dice score achieved by task-specific models expressed as percentage [\%]. Table cells with reference represent either a model tested on a dataset, not used in the primary publication, or an improvement over the primary work. Table cells with percentage increment in green refer to the improvement of Dice score w.r.t. to the primary publication.
}{table:results-best-specialized}{
    LHU-Net & 2024-04 &  & 86.05 &  &  &  &  & 87.49 &  & 92.66 &  &  &  &  &  &  &  &  &  \\
    SCANeXt & 2024-03 &  & 86.60 &  &  &  &  & \sTROst{89.67} &  & \cTROst{\sTROst{95.18}} &  &  &  &  &  &  &  &  &  \\
    SwinUNETR-V2 & 2023-10 &  & \stackDiceCitIncreaseComment{84.31}{sadegheih2024lhunetlighthybridunet}{}{} &  &  & 64.03 & 62.03 & \stackDiceCitIncreaseComment{83.23}{CHEN2024103280}{}{} &  &  &  &  & \cTROst{\sTROst{94.70}} &  &  &  &  & 74.05 &  \\
    NexToU & 2023-05 & 87.84 &  &  &  &  &  &  &  &  &  &  &  &  &  &  &  &  &  \\
    MedNeXt & 2023-03 & 88.76 & 88.01 & \sTROnd{91.02} &  &  & \stackDiceCitIncreaseComment{\cTROrd{\sTROnd{80.14}}}{mednext_miccai}{}{} & \stackDiceCitIncreaseComment{85.97}{mednext_miccai}{}{} & \cTROst{\sTROst{91.77}} & \stackDiceCitIncreaseComment{91.43}{mednext_miccai}{}{} &  &  &  &  &  &  &  &  &  \\
    UNETR++ & 2022-12 & \stackDiceCitIncreaseComment{87.22}{shi2023nextouefficienttopologyawareunet}{3.94}{} & \stackDiceCitIncreaseComment{85.85}{sadegheih2024lhunetlighthybridunet}{3.10}{} &  &  &  & \cTROnd{\sTROst{80.68}} & 87.22 &  & \cTROrd{\sTROrd{92.83}} &  &  &  &  &  & \stackDiceCitIncreaseComment{\sTROrd{87.33}}{wang2023misfm3dmedicalimage}{}{} &  &  &  \\
    3D UX-Net & 2022-08 & \stackDiceCitIncreaseComment{82.40}{LIN20251}{}{} & \stackDiceCitIncreaseComment{90.63}{mednext_miccai}{}{} & \stackDiceCitIncreaseComment{88.39}{mednext_miccai}{}{} & \stackDiceCitIncreaseComment{75.40}{10.1007/978-3-031-43898-1_49}{}{} & \stackDiceCitIncreaseComment{60.09}{SwinUNETRv2_miccai}{}{} & \stackDiceCitIncreaseComment{59.99}{SwinUNETRv2_miccai}{}{} & \stackDiceCitIncreaseComment{86.72}{scanext}{}{} & \cTROrd{\sTROnd{90.00}} & \stackDiceCitIncreaseComment{84.07}{scanext}{}{} & \stackDiceCitIncreaseComment{85.10}{CHEN2024103310}{}{} & \stackDiceCitIncreaseComment{\sTROrd{39.80}}{10.1007/978-3-031-43898-1_49}{}{} & \cTROnd{\sTROnd{93.40}} &  & \stackDiceCitIncreaseComment{95.70}{10.1007/978-3-031-43898-1_49}{}{} &  & \stackDiceCitIncreaseComment{67.30}{10.1007/978-3-031-43898-1_49}{}{} & \stackDiceCitIncreaseComment{\sTROnd{88.80}}{10.1007/978-3-031-43898-1_49}{}{} &  \\
    MedFormer & 2022-02 & 85.00 &  & 85.00 & 69.00 &  & 74.00 &  & 88.00 & 92.50 &  &  &  &  &  &  &  &  &  \\
    SwinUNETR & 2022-01 & \stackDiceCitIncreaseComment{\cTROnd{\sTROst{91.80}}}{pmlr-v250-valanarasu24a}{8.32}{} & \stackDiceCitIncreaseComment{90.48}{mednext_miccai}{1.52}{} & \stackDiceCitIncreaseComment{88.33}{10.1007/978-3-031-43898-1_62}{}{a} & \stackDiceCitIncreaseComment{85.52}{10376801}{}{} & \stackDiceCitIncreaseComment{74.24}{10510478}{18.75}{} & \stackDiceCitIncreaseComment{76.60}{10376801}{19.88}{} & \stackDiceCitIncreaseComment{83.51}{nnFormer_ieee}{}{} & \stackDiceCitIncreaseComment{88.63}{10.1007/978-3-031-43898-1_62}{}{b} & \stackDiceCitIncreaseComment{86.49}{zhao2025modelrulealluniversal}{}{} & \stackDiceCitIncreaseComment{87.46}{zhao2025modelrulealluniversal}{}{} & \stackDiceCitIncreaseComment{\sTROst{59.45}}{10376801}{}{} & \stackDiceCitIncreaseComment{92.90}{SwinUNETRv2_miccai}{}{} & \stackDiceCitIncreaseComment{\sTROnd{88.85}}{zhao2025modelrulealluniversal}{}{} & \stackDiceCitIncreaseComment{\cTROrd{\sTROnd{96.99}}}{10376801}{}{} & \stackDiceCitIncreaseComment{\cTROnd{\sTROnd{89.92}}}{zhao2025modelrulealluniversal}{}{} & \stackDiceCitIncreaseComment{\sTROnd{68.95}}{10376801}{}{} & \stackDiceCitIncreaseComment{\sTROrd{88.30}}{10.1007/978-3-031-43898-1_49}{14.98}{} & \stackDiceCitIncreaseComment{\sTROnd{90.41}}{zhao2025modelrulealluniversal}{}{} \\
    TransBTSV2 & 2022-01 &  & 85.04 & \sTROrd{90.53} & \sTROrd{89.85} &  &  &  &  &  &  &  &  &  &  &  &  &  &  \\
    nnFormer & 2021-09 & \stackDiceCitIncreaseComment{87.80}{LIN20251}{}{} & \stackDiceCitIncreaseComment{90.42}{mednext_miccai}{4.02}{} & \stackDiceCitIncreaseComment{89.09}{mednext_miccai}{}{} & \stackDiceCitIncreaseComment{89.83}{10.1007/978-3-031-72111-3_38}{}{} & \stackDiceCitIncreaseComment{\sTROst{78.65}}{10.1007/978-3-031-72111-3_38}{}{} & \stackDiceCitIncreaseComment{\sTROrd{77.95}}{unetr_pp_ieee}{}{} & 86.57 & \stackDiceCitIncreaseComment{84.20}{mednext_miccai}{}{} & 92.06 & \stackDiceCitIncreaseComment{\sTROnd{90.40}}{CHEN2024103310}{}{} & \stackDiceCitIncreaseComment{18.80}{10.1007/978-3-031-43898-1_49}{}{} & \stackDiceCitIncreaseComment{90.60}{3duxnet_iclr}{}{} & \stackDiceCitIncreaseComment{75.37}{huang2023stunetscalabletransferablemedical}{}{} & \stackDiceCitIncreaseComment{92.20}{10.1007/978-3-031-43898-1_49}{}{} & \stackDiceCitIncreaseComment{86.35}{wang2023misfm3dmedicalimage}{}{} & \stackDiceCitIncreaseComment{66.29}{CHEN2024103280}{}{} & \stackDiceCitIncreaseComment{87.00}{10.1007/978-3-031-43898-1_49}{}{} & \stackDiceCitIncreaseComment{79.26}{huang2023stunetscalabletransferablemedical}{}{} \\
    MISSFormer & 2021-09 &  &  &  &  &  &  & 81.96 &  & 91.19 &  &  &  &  &  &  &  &  &  \\
    Swin-Unet & 2021-05 & \stackDiceCitIncreaseComment{79.13}{shi2023nextouefficienttopologyawareunet}{}{} & \stackDiceCitIncreaseComment{82.08}{li2022transbtsv2betterefficientvolumetric}{}{c} &  & \stackDiceCitIncreaseComment{79.60}{li2022transbtsv2betterefficientvolumetric}{}{} &  &  & 79.13 &  & \stackDiceCitIncreaseComment{\sTROrd{90.41}}{huang2022missformer}{0.41}{} & \stackDiceCitIncreaseComment{87.59}{10829779}{}{} &  &  &  &  &  &  &  &  \\
    TransBTS & 2021-03 & \stackDiceCitIncreaseComment{82.35}{mednext_miccai}{}{} & \stackDiceCitIncreaseComment{\sTROrd{90.66}}{mednext_miccai}{7.09}{} & \stackDiceCitIncreaseComment{89.10}{li2022transbtsv2betterefficientvolumetric}{}{} & 88.95 &  & \stackDiceCitIncreaseComment{70.38}{unetr_pp_ieee}{}{} & \stackDiceCitIncreaseComment{83.28}{unetr_pp_ieee}{}{} & \stackDiceCitIncreaseComment{86.52}{mednext_miccai}{}{} &  &  &  & \stackDiceCitIncreaseComment{90.20}{3duxnet_iclr}{}{} &  &  &  &  &  &  \\
    CoTr & 2021-03 & 85.00 & \stackDiceCitIncreaseComment{82.90}{10.1007/978-3-031-43898-1_49}{}{} & \stackDiceCitIncreaseComment{85.10}{10.1007/978-3-031-43898-1_49}{}{} & \stackDiceCitIncreaseComment{74.70}{10.1007/978-3-031-43898-1_49}{}{} & \stackDiceCitIncreaseComment{65.80}{10.1007/978-3-031-43898-1_49}{}{} & \stackDiceCitIncreaseComment{75.74}{unetr_pp_ieee}{}{} & \stackDiceCitIncreaseComment{85.72}{CHEN2024103280}{}{} &  & \stackDiceCitIncreaseComment{91.04}{unetr_pp_ieee}{}{} &  & \stackDiceCitIncreaseComment{33.80}{10.1007/978-3-031-43898-1_49}{}{} &  &  & \stackDiceCitIncreaseComment{95.20}{10.1007/978-3-031-43898-1_49}{}{} &  & \stackDiceCitIncreaseComment{67.20}{10.1007/978-3-031-43898-1_49}{}{} & \stackDiceCitIncreaseComment{88.00}{10.1007/978-3-031-43898-1_49}{}{} &  \\
    UNETR & 2021-03 & \stackDiceCitIncreaseComment{\cTROrd{\sTROnd{89.10}}}{pmlr-v250-valanarasu24a}{1.75}{} & \stackDiceCitIncreaseComment{89.65}{mednext_miccai}{18.55}{} & \stackDiceCitIncreaseComment{84.10}{mednext_miccai}{}{} & \stackDiceCitIncreaseComment{81.48}{10.1007/978-3-031-72111-3_38}{}{} & \stackDiceCitIncreaseComment{73.65}{10.1007/978-3-031-72111-3_38}{}{} & \stackDiceCitIncreaseComment{73.29}{unetr_pp_ieee}{}{} & \stackDiceCitIncreaseComment{79.56}{nnunet_nature}{}{} & \stackDiceCitIncreaseComment{81.98}{mednext_miccai}{}{} & \stackDiceCitIncreaseComment{88.61}{nnFormer_ieee}{}{} &  &  & \stackDiceCitIncreaseComment{88.60}{3duxnet_iclr}{}{} & \stackDiceCitIncreaseComment{75.05}{huang2023stunetscalabletransferablemedical}{}{} & \sTROrd{96.40} &  & \stackDiceCitIncreaseComment{53.80}{10.1007/978-3-031-43898-1_49}{}{} & \stackDiceCitIncreaseComment{85.30}{10.1007/978-3-031-43898-1_49}{}{} & \stackDiceCitIncreaseComment{77.11}{huang2023stunetscalabletransferablemedical}{}{} \\
    TransUNet & 2021-02 & \stackDiceCitIncreaseComment{83.80}{10.1007/978-3-031-16440-8_53}{}{} & \cTROnd{\sTROst{91.74}} & \stackDiceCitIncreaseComment{80.82}{mednext_miccai}{}{} & \stackDiceCitIncreaseComment{86.50}{li2022transbtsv2betterefficientvolumetric}{}{} & \stackDiceCitIncreaseComment{\sTROrd{76.30}}{10.1007/978-3-031-72111-3_38}{}{} & \stackDiceCitIncreaseComment{75.21}{10.1007/978-3-031-72111-3_38}{}{} & \sTROnd{88.39} & \stackDiceCitIncreaseComment{85.05}{mednext_miccai}{}{} & \stackDiceCitIncreaseComment{90.44}{huang2022missformer}{}{} & \stackDiceCitIncreaseComment{\sTROrd{89.16}}{10829779}{}{} &  & \stackDiceCitIncreaseComment{82.00}{Dong2024}{}{} &  &  & \stackDiceCitIncreaseComment{85.46}{wang2023misfm3dmedicalimage}{}{} & \sTROrd{67.67} &  &  \\
    SETR & 2020-12 & \stackDiceCitIncreaseComment{78.40}{10.1007/978-3-030-87199-4_16}{}{} & \stackDiceCitIncreaseComment{63.90}{nnFormer_ieee}{}{} &  &  &  &  &  &  &  &  &  &  &  &  &  &  &  &  \\
    SegResNet & 2018-10 & \stackDiceCitIncreaseComment{84.36}{10658004}{}{} & \stackDiceCitIncreaseComment{82.73}{xing2023diffunetdiffusionembeddednetwork}{0.54}{} & \stackDiceCitIncreaseComment{81.89}{10658004}{}{} & \stackDiceCitIncreaseComment{70.85}{xing2023diffunetdiffusionembeddednetwork}{}{} &  &  &  & \stackDiceCitIncreaseComment{87.20}{10658004}{}{d} &  &  &  &  & \stackDiceCitIncreaseComment{82.05}{10510478}{}{} &  &  &  &  & \stackDiceCitIncreaseComment{83.41}{10510478}{}{} \\
    nnU-Net & 2018-10 & \stackDiceCitIncreaseComment{\sTROrd{88.80}}{pmlr-v250-valanarasu24a}{1.18}{} & \stackDiceCitIncreaseComment{\cTROrd{\sTROnd{91.23}}}{mednext_miccai}{30.23}{} & \cTROrd{\sTROst{91.63}} & \stackDiceCitIncreaseComment{\cTROrd{\sTROst{95.29}}}{huang2023stunetscalabletransferablemedical}{8.79}{} & \stackDiceCitIncreaseComment{\sTROnd{76.52}}{huang2023stunetscalabletransferablemedical}{9.02}{} & \stackDiceCitIncreaseComment{74.31}{unetr_pp_ieee}{0.31}{} & \stackDiceCitIncreaseComment{\sTROrd{87.94}}{wang2023misfm3dmedicalimage}{}{} & \stackDiceCitIncreaseComment{89.46}{huang2023stunetscalabletransferablemedical}{}{} & \cTROnd{\sTROnd{92.95}} & \sTROst{91.94} & \sTROnd{58.00} & \stackDiceCitIncreaseComment{\cTROrd{\sTROrd{93.36}}}{zhao2025modelrulealluniversal}{}{} & \stackDiceCitIncreaseComment{\cTROst{\sTROst{92.39}}}{zhao2025modelrulealluniversal}{}{e} & \cTROnd{\sTROst{97.00}} & \cTROst{\sTROst{93.00}} & \sTROst{69.00} & \stackDiceCitIncreaseComment{\cTROnd{\sTROst{89.40}}}{10.1007/978-3-031-43898-1_49}{5.90}{} & \stackDiceCitIncreaseComment{\cTROst{\sTROst{93.22}}}{zhao2025modelrulealluniversal}{}{} \\
    UNet++ & 2018-07 & \stackDiceCitIncreaseComment{81.60}{CHEN2024103280}{}{} &  &  & 82.60 &  &  &  &  &  & \stackDiceCitIncreaseComment{88.08}{10829779}{}{} &  &  &  &  &  &  &  &  \\
    V-Net & 2016-06 & \stackDiceCitIncreaseComment{80.00}{10.1007/978-3-031-72114-4_61}{}{} & \stackDiceCitIncreaseComment{78.69}{10.1007/978-3-030-87193-2_11}{}{} & \stackDiceCitIncreaseComment{87.21}{li2022transbtsv2betterefficientvolumetric}{}{} & \stackDiceCitIncreaseComment{\sTROnd{93.90}}{chen2019med3dtransferlearning3d}{}{} &  &  & \stackDiceCitIncreaseComment{68.81}{cao2022swin}{}{} &  &  & 86.90 &  &  & \stackDiceCitIncreaseComment{\sTROrd{84.20}}{10510478}{}{} &  &  &  &  & \stackDiceCitIncreaseComment{\sTROrd{85.67}}{10510478}{}{} \\
    U-Net & 2015-05 & \stackDiceCitIncreaseComment{85.19}{shi2023nextouefficienttopologyawareunet}{}{} & \stackDiceCitIncreaseComment{85.93}{10847777}{}{} & \stackDiceCitIncreaseComment{89.92}{10.1007/978-3-031-43898-1_62}{}{} & \stackDiceCitIncreaseComment{79.90}{8932614}{}{} &  &  & \stackDiceCitIncreaseComment{76.85}{cao2022swin}{}{} & \stackDiceCitIncreaseComment{\sTROrd{89.60}}{10.1007/978-3-031-43898-1_62}{}{f} & \stackDiceCitIncreaseComment{87.55}{cao2022swin}{}{} & \stackDiceCitIncreaseComment{87.73}{10829779}{}{} &  & \stackDiceCitIncreaseComment{89.20}{3duxnet_iclr}{}{} & \stackDiceCitIncreaseComment{80.51}{10510478}{}{} &  &  &  &  & \stackDiceCitIncreaseComment{82.73}{10510478}{}{} \\
}{
    \begin{itemize} \setlength\itemsep{0pt}
        \item[a] Supervised approach \citep{10.1007/978-3-031-43898-1_62}, mean between organ and tumor Dice scores.
        \item[b] Supervised approach \citep{10.1007/978-3-031-43898-1_62}.
        \item[c] Average Dice score between BraTS2019 and BraTS2020 datasets.
        \item[d] Average Dice score between AMOS CT and MRI (88.97\% and 85.43\%).
        \item[e] TS v2 in Table 6 \citep{zhao2025modelrulealluniversal}.
        \item[f] Results from UNet with the MultiTalent approach \citep{10.1007/978-3-031-43898-1_62}.
    \end{itemize}
}
\clearpage

\newgeometry{left=1.8cm, right=1.8cm} 
\hspace{-1.2cm}\appendixtextbox{
    \subsection{Datasets}
    \label{app:subsec:datasets}

    Table~\ref{table:datasets-all} reports information, links and resources about 3D medical image  datasets used to benchmark foundation and specialized models. 
}

\newcommand{\selectTableDatasetsFont}{\fontfamily{cmss}\fontsize{6}{7.5}\selectfont}


\newcommand{\TDncols}{7}

\newcommand{\TDcwName}{3.0cm}
\newcommand{\TDcwRelDataset}{2.0cm}
\newcommand{\TDcwModality}{1.7cm}
\newcommand{\TDcwAnatomyRegion}{2.0cm}
\newcommand{\TDcwObjects}{4.5cm}
\newcommand{\TDcwNumImages}{1.5cm}
\newcommand{\TDcwLinks}{2.5cm}

\newcommand{\TDcwDescription}{17.45cm}


\newcommand{\TableDatasetsHeading}[1]{
    \rowcolor{#1}
        \parbox[c]{\TDcwName}{
                \vspace{4pt}%
                \textbf{Dataset \\
                \textit{Full Name} \\
                (References)}
                \vspace{4pt}%
            } & 
        \parbox[c]{\TDcwRelDataset}{
                \textbf{Related Datasets}
            } &
        \textbf{Modality}
              &
        \parbox[c]{\TDcwAnatomyRegion}{
            \raggedright%
            \textbf{Main Anatomical Structure\\
            (Region)}
            } &
        \parbox[c]{\TDcwObjects}{
                \textbf{N. Objects\\
                Objects}
            } &
        \parbox[c]{\TDcwNumImages}{
                \textbf{N. Images\\
                (with labels)}
            } &
        \parbox[c]{\TDcwLinks}{
                \textbf{Links}
            } \\
}

\newcommand{\makeTableDatasetsFirstHeading}[1]{ 
    \hline
    \TableDatasetsHeading{#1}
    \hline
}

\newcommand{\makeTableDatasetsOtherHeadings}[1]{ 
    \makeTableHeaderContinued{\TDncols} 
    \makeTableDatasetsFirstHeading{#1}
}



\newcommand{\makeTableDatasets}[6]{

    {
    \selectTableDatasetsFont
    \rowcolors{2}{CornflowerBlue!1}{#2}
    \begin{center}
    \begin{longtable}[c]{
        p{\TDcwName} 
        p{\TDcwRelDataset} 
        p{\TDcwModality} 
        p{\TDcwAnatomyRegion} 
        p{\TDcwObjects} 
        p{\TDcwNumImages} 
        p{\TDcwLinks} 
    }
    
        \hiderowcolors
        \caption{
            #3
        }
        \label{#4} \\
        \showrowcolors
    
        \makeTableDatasetsFirstHeading{#1}
        \endfirsthead
        
        \makeTableDatasetsOtherHeadings{#1}
        \endhead
        
        \makeTableOtherFooters{\TDncols}
        \endfoot
        
        \makeTableLastFooter{\TDncols}{}
        \endlastfoot
    
        #5
    
    \end{longtable}
    \outOfTableComments{
        \hspace{-1.5cm}\parbox{15cm}{#6}
    }
    \end{center}
    }
}

\newcommand{\TDcellvspacer}{\vspace{4pt}}

\newcommand{\TDcellName}[3]{%
    \parbox[t]{\TDcwName}{%
        \raggedright%
        #1 \vspace{2pt}\\
        \textit{#2} \vspace{2pt}\\
        #3 
        \TDcellvspacer
    }%
}

\newcommand{\TDcellRelDataset}[1]{%
    \parbox[t]{\TDcwRelDataset}{%
        \raggedright%
        #1
        \TDcellvspacer
    }%
}

\newcommand{\TDcellModality}[1]{%
    \parbox[t]{\TDcwModality}{%
        \raggedright%
        #1
        \TDcellvspacer
    }%
}

\newcommand{\TDcellAnatomyRegion}[2]{%
    \parbox[t]{\TDcwAnatomyRegion}{%
        \raggedright%
        #1\\
        (#2)
        \TDcellvspacer
    }%
}

\newcommand{\TDcellObjects}[2]{%
    \parbox[t]{\TDcwObjects}{%
        \raggedright%
        \textit{#1} \vspace{2pt}\\
        {#2}
        \TDcellvspacer
    }%
}

\newcommand{\TDcellNumImages}[2]{%
    \parbox[t]{\TDcwNumImages}{%
        \raggedright%
        #1 \vspace{2pt}\\
        (#2)
        \TDcellvspacer
    }%
}

\newcommand{\TDcellLinks}[1]{%
    \parbox[t]{\TDcwLinks}{%
        \raggedright%
        #1
        \TDcellvspacer
    }%
}

\newcommand{\TDrowDescription}[1]{%
    \multicolumn{\TDncols}{c}{%
        \parbox[t]{17cm}{%
            \textcolor{black!75}{#1}\\
            ~\\
            ~
        }%
    }%
}

%

\newcommand{\TDformatRow}[1]{
    \multicolumn{\TDncols}{p{\TDcwDescription}}{%
        \begin{tabular}{
            p{\TDcwName} 
            p{\TDcwRelDataset} 
            p{\TDcwModality} 
            p{\TDcwAnatomyRegion} 
            p{\TDcwObjects} 
            p{\TDcwNumImages} 
            p{\TDcwLinks} 
        }
            #1
        \end{tabular}
    }
}

\makeTableDatasets{
    CornflowerBlue!20%
}{
    CornflowerBlue!10%
}{
    Full list of the datasets used in the reviewed studies.
}{table:datasets-all}{
    \TDformatRow{
        \TDcellName{3D-IRCADb}{Liver segmentation 3D-IRCADb-01}{\citep{soler20103d}} & 
        \TDcellRelDataset{-} & 
        \TDcellModality{3D CT} & 
        \TDcellAnatomyRegion{Liver, Tumors}{Abdomen} & 
        \TDcellObjects{35}{Aorta, Artery, Biliary System, Bladder, Bone, Colon, Duodenum, Gallbladder, Heart, Hyperplasie, Inferior Vena Cava, Kidney (Left), Kidney (Right), Kidneys, Liver, Liver Cyst, Liver Tumor, Lung (Left), Lung (Right), Lungs, Lymph Nodes, Metal, Metastasectomy, Pancreas, Portal Vein and Splenic Vein, Skin, Spleen, Stomach, Stones, Surrenal Gland, Surrenal Gland (Left), Surrenal Gland (Left) Tumor, Surrenal Gland (Right) Tumor, Tumor, Venous System} & 
        \TDcellNumImages{22}{22} & 
        \TDcellLinks{\href{https://www.ircad.fr/research/data-sets/liver-segmentation-3d-ircadb-01/}{Official Website} \\\vspace{2pt} \href{https://cloud.ircad.fr/index.php/s/JN3z7EynBiwYyjy/download}{Download} \\\vspace{2pt} \href{https://www-sop.inria.fr/geometrica/events/wam/abstract-ircad.pdf}{Publication}} \\
        \TDrowDescription{The 3D-ircadb -01 database consists of 3D CT scans from 10 female and 10 male patients with a liver tumor incidence rate of 75\%. Not all classes are reported in all images or in equal proportion in the dataset, with the majority of classes pesent in just a few images.} \\
    }\\
    \TDformatRow{
        \TDcellName{AbdomenAtlas}{AbdomenAtlas}{\citep{LI2024103285}} & 
        \TDcellRelDataset{-} & 
        \TDcellModality{3D CT / CT (CE)} & 
        \TDcellAnatomyRegion{Abdominal Organs, Bones}{Abdomen, Pelvis, Thorax} & 
        \TDcellObjects{25}{Adrenal Gland (Left), Adrenal Gland (Right), Aorta, Bladder, Celiac Trunk, Colon, Duodenum, Esophagus, Femur (Left), Femur (Right), Gallbladder, Hepatic Vessels, Inferior Vena Cava, Kidney (Left), Kidney (Right), Liver, Lung (Left), Lung (Right), Pancreas, Portal and Spleenic Veins, Prostate, Rectum, Small Intestine, Spleen, Stomach} & 
        \TDcellNumImages{3410}{3410} & 
        \TDcellLinks{\href{https://www.zongweiz.com/dataset}{Official Website} \\\vspace{2pt} \href{https://github.com/MrGiovanni/AbdomenAtlas}{GitHub} \\\vspace{2pt} \href{https://doi.org/10.1016/j.media.2024.103285}{Publication} \\\vspace{2pt} \href{https://github.com/MrGiovanni/SuPreM}{Secondary Website}} \\
        \TDrowDescription{The AbdomenAtlas dataset (also AbdomenAtlas-8K) is a CT abdominal organs dataset created from other publicly available dataset. Currently, only AbdomenAtlas 1.0 Mini and 1.1 Mini are downloadable. Version 1.0 includes a subset of nine of the 1.1 classes: spleen, liver, left kidney, right kidney, stomach, gallbladder, pancreas, aorta, and inferior vena cava. The bigger annotation set, named AbdomenAtlas 1.1, uncludes 25 classes. The authors on the official website committed to release 3410 volume-mask pairs publicly out of the total 8448. Other atlases, namely AbdomenAtlas2.0 and AbdomenAtlas 3.0, will be released. The AbdomenAtlas project is an ongoing effort to build a comprehensive organ and tumors segmentation dataset, and its specifications may change frequently. Please refer to the official websites. AbdomenAtlas (Mini version) is currently the suggested training set for the Touchstone Benchmark.} \\
    }\\
    \TDformatRow{
        \TDcellName{AbdomenCT-1K}{AbdomenCT-1K}{\citep{9497733}} & 
        \TDcellRelDataset{KiTS19, LiTS, MSD Pancreas, MSD Spleen, NIH Pancreas-CT} & 
        \TDcellModality{3D CT / CT (CE)} & 
        \TDcellAnatomyRegion{Abdominal Organs}{Abdomen} & 
        \TDcellObjects{4}{Kidneys, Liver, Pancreas, Spleen} & 
        \TDcellNumImages{1112}{1000} & 
        \TDcellLinks{\href{https://github.com/JunMa11/AbdomenCT-1K}{Official Website} \\\vspace{2pt} \href{https://ieeexplore.ieee.org/document/9497733}{Publication}} \\
        \TDrowDescription{The AbdomenCT-1K is a large-scale abdominal CT dataset comprising 1112 CT scans for segmentation of abdominal organs. Data primarily come from six datasets, five of which are public datasets: LiTS (201 cases), KiTS19 (300 cases), MSD Spleen (61 cases), MSD Pancreas (420 cases), and NIH Pancreas (80 cases). There is also a new dataset from Nanjing University consisting of 50 CT scans. Every CT scan has comprehensive annotations for the four organs.} \\
    }\\
    \TDformatRow{
        \TDcellName{ACDC}{Automatic Cardiac Diagnosis Challenge}{\citep{8360453}} & 
        \TDcellRelDataset{-} & 
        \TDcellModality{3D MRI (Cine)} & 
        \TDcellAnatomyRegion{Heart}{Thorax} & 
        \TDcellObjects{3}{Heart Ventricle (Left), Heart Ventricle (Right), Myocardium} & 
        \TDcellNumImages{150}{150} & 
        \TDcellLinks{\href{https://www.creatis.insa-lyon.fr/Challenge/acdc/}{Official Website} \\\vspace{2pt} \href{https://humanheart-project.creatis.insa-lyon.fr/database/\#collection/637218c173e9f0047faa00fb}{Download} \\\vspace{2pt} \href{https://ieeexplore.ieee.org/document/8360453}{Publication}} \\
        \TDrowDescription{The ACDC (Automatic Cardiac Diagnosis Challenge) was a competition at MICCAI 2017. Cases are divided into 5 subcategories: NOR (normal), MINF (myocardial infarction with systolic heart failure), DCM (dilated cardiomyopathy), HCM (hypertrophic cardiomyopathy), and ARV (abnormal right ventricle), with 30 cases each. Each case comprises a 4D image of one cardiac cycle, with annotations for the diastolic (ED) and systolic (ES) frames, for a total of 300 annotated volumes. The data is divided by the officials into a training set of 100 cases and a test set of 50 cases, with each subclass having 20 cases in the training set and 10 cases in the test set.} \\
    }\\
    \TDformatRow{
        \TDcellName{AMOS}{Multi-Modality Abdominal Multi-Organ Segmentation Challenge 2022}{\citep{ji2022amos,ji2022amoslargescaleabdominalmultiorgan}} & 
        \TDcellRelDataset{AMOS 2022 CT, AMOS 2022 MRI} & 
        \TDcellModality{3D CT,\\3D MRI} & 
        \TDcellAnatomyRegion{Abdominal Organs}{Abdomen, Pelvis} & 
        \TDcellObjects{16}{Adrenal Gland (Left), Adrenal Gland (Right), Aorta, Bladder, Duodenum, Esophagus, Gallbladder, Inferior Vena Cava, Kidney (Left), Kidney (Right), Liver, Pancreas, Prostate, Spleen, Stomach, Uterus} & 
        \TDcellNumImages{600}{600} & 
        \TDcellLinks{\href{https://amos22.grand-challenge.org/}{Official Challenge Website} \\\vspace{2pt} \href{https://era-ai-biomed.github.io/amos/index.html}{Official Website} \\\vspace{2pt} \href{https://arxiv.org/abs/2206.08023}{Preprint} \\\vspace{2pt} \href{https://proceedings.neurips.cc/paper\_files/paper/2022/hash/ee604e1bedbd069d9fc9328b7b9584be-Abstract-Datasets\_and\_Benchmarks.html}{Publication}} \\
        \TDrowDescription{AMOS provides 500 CT and 100 MR scans from multi-centers, multi-vendors, multi-modalities, multi-phases, and multi-disease patients, each with voxel-level annotations for 15 abdominal organs. Official data split: 240 for training, 120 for validation, 240 for test.} \\
    }\\
    \TDformatRow{
        \TDcellName{ASOCA}{Automated Segmentation of Coronary Arteries}{\citep{gharleghi2022automated,gharleghi2023annotated}} & 
        \TDcellRelDataset{-} & 
        \TDcellModality{3D CT (CE)} & 
        \TDcellAnatomyRegion{Heart}{Thorax} & 
        \TDcellObjects{1}{Heart Coronary Arteries} & 
        \TDcellNumImages{60}{40} & 
        \TDcellLinks{\href{https://asoca.grand-challenge.org/}{Official Challenge Website} \\\vspace{2pt} \href{https://doi.org/10.1016/j.compmedimag.2022.102049}{Publication}} \\
        \TDrowDescription{The ASOCA dataset contains Computed Tomography Coronary Angiography (CCTA) images for automated segmentation of coronary arteries. It includes voxel-wise manual annotations of the coronary artery tree. The dataset is composed of 30 CCTA volumes for each of the Normal and Diseased categories. For each category, 20 volumes have label masks, while 10 are without labels.} \\
    }\\

    \TDformatRow{
        \TDcellName{ATLAS 2023}{A Tumour and Liver Automatic Segmentation}{\citep{data8050079}} & 
        \TDcellRelDataset{-} & 
        \TDcellModality{3D MRI (T1-CE)} & 
        \TDcellAnatomyRegion{Liver, Tumors}{Abdomen} & 
        \TDcellObjects{2}{Hepatocellular Carcinoma, Liver} & 
        \TDcellNumImages{90}{60} & 
        \TDcellLinks{\href{https://atlas-challenge.u-bourgogne.fr/}{Official Website} \\\vspace{2pt} \href{https://doi.org/10.3390/data8050079}{Publication}} \\
        \TDrowDescription{ATLAS is the MICCAI 2023 Challenge for segmentation of the liver and tumor(s), somewhat similar to the LiTS dataset. The difference is that the ATLAS dataset provides a dataset in the Contrast-Enhanced MRI modality, rather than the CT modality of LiTS. This modal difference is due to the collection of the ATLAS dataset being related to the treatment of hepatocellular carcinoma with transarterial radioembolisation (TARE), and TARE treatment requires the preoperative capture of CE-MRI images for radiometric estimation.} \\
    }\\

    \TDformatRow{
        \TDcellName{ATLAS v2.0}{ATLAS R2.0 - Anatomical Tracings of Lesions After Stroke}{\citep{liew2022large}} & 
        \TDcellRelDataset{-} & 
        \TDcellModality{3D MRI (T1)} & 
        \TDcellAnatomyRegion{Brain}{Head} & 
        \TDcellObjects{1}{Brain Ischemic Stroke Lesion} & 
        \TDcellNumImages{1271}{655} & 
        \TDcellLinks{\href{https://atlas.grand-challenge.org/}{Official Challenge Website} \\\vspace{2pt} \href{https://fcon\_1000.projects.nitrc.org/indi/retro/atlas.html}{Official Website} \\\vspace{2pt} \href{https://doi.org/10.1038/s41597-022-01401-7}{Publication}} \\
        \TDrowDescription{ATLAS v2.0 is a dataset for segmenting brain stroke lesion areas from MRI T1 weighted (T1W) single modality images, and it is related to (but does not coincide with) the MICCAI ISLES 2022 challenge (the two datasets are disjoint).} \\
    }\\

    \TDformatRow{
        \TDcellName{AutoPET}{Automated Lesion Segmentation in Whole-Body PET/CT}{\citep{gatidis2022whole}} & 
        \TDcellRelDataset{AutoPET I, AutoPET II, AutoPET III} & 
        \TDcellModality{3D CT,\\3D PET (FDG),\\3D PET (PSMA)} & 
        \TDcellAnatomyRegion{Tumors}{Whole Body} & 
        \TDcellObjects{1}{Tumor} & 
        \TDcellNumImages{1816}{1616} & 
        \TDcellLinks{\href{https://autopet.grand-challenge.org/}{Official Challenge Website (1)} \\\vspace{2pt} \href{https://autopet-ii.grand-challenge.org/}{Official Challenge Website (2)} \\\vspace{2pt} \href{https://autopet-iii.grand-challenge.org/task/}{Official Challenge Website (3)} \\\vspace{2pt} \href{https://www.nature.com/articles/s41597-022-01718-3}{Publication}} \\
        \TDrowDescription{The AutoPET dataset provides whole-body PET/CT volumes with manual tumor lesion annotations and comprises FDG-PET/CT images (1,014 cases from 900 patients) collected primarily from the University Hospital Tbingen and LMU in Munich, and PSMA-PET/CT images (597 cases from 378 patients) from the same institutions. The complete dataset (also referred to as AutoPET III) is an extension of the original AutoPET, which was expanded mutiple times from its first release (Autopet I, II and III in 2022, 2023 and 2024 MICCAI challenges respectively)} \\
    }\\

    \TDformatRow{
        \TDcellName{BraTS}{Brain Tumor Segmentation}{\citep{6975210,bakas_2024_10978907,s25061838}} & 
        \TDcellRelDataset{BraTS 2012, BraTS 2013, BraTS 2014, BraTS 2015, BraTS 2016, BraTS 2017, BraTS 2018, BraTS 2019, BraTS 2020, BraTS 2021, BraTS 2022, BraTS 2023, BraTS 2024} & 
        \TDcellModality{3D MRI (T1),\\3D MRI (T1-CE),\\3D MRI (T2),\\3D MRI (T2-FLAIR)} & 
        \TDcellAnatomyRegion{Brain, Tumors}{Head} & 
        \TDcellObjects{10}{Brain Enhancing Tumor, Brain Gross Tumor Volume, Brain Metastasis, Brain Non-enhancing Tumor Core, Brain Peritumoral Edema, Brain Resection Cavity, Brain Surrounding Non-enhancing FLAIR Superintensity, Brain Tumor Cystic Component, Glioma, Meningioma} & 
        \TDcellNumImages{7189}{6457} & 
        \TDcellLinks{\href{https://www.mdpi.com/1424-8220/25/6/1838}{BraTS Datasets Comprehensive Review} \\\vspace{2pt} \href{https://www.synapse.org/Synapse:syn53708249/wiki/626323}{BraTS 2024 Website} \\\vspace{2pt} \href{https://www.synapse.org/Synapse:syn64153130/wiki/630130}{BraTS 2025 Website}} \\
        \TDrowDescription{The BraTS challenge has evolved from a glioma segmentation task in 2012 to a comprehensive neuro-oncological AI platform, marked by a continuous expansion in dataset size (dozens to thousands of cases) and diversity (including meningioma, metastases, pediatric tumors). Beyond segmentation, BraTS diversified tasks to address clinical needs like survival prediction, radiogenomics, and post-treatment assessment. Notably, the BraTS datasets from 2016 and 2017 were adopted as the dataset for the "Brain Tumors" task within the Medical Segmentation Decathlon (MSD). Specifically, the MSD Brain Tumors dataset incliude 484 training and 266 test multimodal MRI images from patients diagniosed with glioblastoma or low-grade glioma, and the segmentation task includes three targets: enhancing tumor, peritumoral edema and necrotic core. Given the extraordinary complexity of the evolution of this dataset, here are reported the statistics of BraTS 2024, condensed  the six segmentation tasks and considering only segemntation tasks. For  a full overview of the 13 years of evolution of the BraTS dataset, please refer to \citet{s25061838}.} \\
    }\\

    \TDformatRow{
        \TDcellName{BTCV}{Multi-Atlas Labeling Beyond The Cranial Vault - Abdomen}{\citep{landman2015miccai}} & 
        \TDcellRelDataset{Synapse} & 
        \TDcellModality{3D CT (CE)} & 
        \TDcellAnatomyRegion{Abdominal Organs}{Abdomen} & 
        \TDcellObjects{13}{Adrenal Gland (Left), Adrenal Gland (Right), Aorta, Esophagus, Gallbladder, Inferior Vena Cava, Kidney (Left), Kidney (Right), Liver, Pancreas, Portal and Spleenic Veins, Spleen, Stomach} & 
        \TDcellNumImages{50}{30} & 
        \TDcellLinks{\href{https://www.synapse.org/Synapse:syn3193805/wiki/}{Official Website}} \\
        \TDrowDescription{The BTCV dataset, originating from the MICCAI 2015 Multi-Atlas Labeling Beyond The Cranial Vault workshop, is a key benchmark for abdominal organ segmentation, specifically referring to its Abdomen version. Provided by Vanderbilt University Medical Center, it comprises 50 abdominal CT scans from patients with metastatic liver cancer or postoperative abdominal wall hernia, captured during the portal venous contrast phase with varying resolutions and slice thicknesses. The BTCV is linked to the Synapse dataset, which is a label-subset of BTCV Abdomen.} \\
    }\\

    \TDformatRow{
        \TDcellName{BTCV Cervix}{Multi-Atlas Labeling Beyond The Cranial Vault - Cervix}{\citep{landman2015miccai}} & 
        \TDcellRelDataset{-} & 
        \TDcellModality{3D CT (CE)} & 
        \TDcellAnatomyRegion{Abdominal Organs}{Pelvis} & 
        \TDcellObjects{4}{Bladder, Brain Enhancing Tumor, Small Intestine, Uterus} & 
        \TDcellNumImages{50}{30} & 
        \TDcellLinks{\href{https://www.synapse.org/\#!Synapse:syn3193805/wiki/217790}{Official Website}} \\
        \TDrowDescription{The BTCV Cervix dataset is a CT segmentation dataset for cervical cancer patients, primarily used for radiation therapy planning. The name BTCV comes from the Workshop "Multi-Atlas Labeling Beyond The Cranial Vault" held at MICCAI 2015, and is also synonim of the BTCV dataset.} \\
    }\\

    \TDformatRow{
        \TDcellName{CANDI}{The Child and Adolescent NeuroDevelopment Initiative}{\citep{Kennedy2012}} & 
        \TDcellRelDataset{-} & 
        \TDcellModality{3D MRI (T1)} & 
        \TDcellAnatomyRegion{Brain}{Head} & 
        \TDcellObjects{39}{Brain 3rd Ventricle, Brain 4th Ventricle, Brain 5th Ventricle, Brain CSF, Brain Left Inferior Lateral Ventricle, Brain Left Lateral Ventricle, Brain Left Undetermined, Brain Left Vessel, Brain Right Inferior Lateral Ventricle, Brain Right Lateral Ventricle, Brain Right Undetermined, Brain Right Vessel, Brain Stem, Cerebral Cortex (Left), Cerebral Cortex (Right), Hippocampus (Left), Hippocampus (Right), Left Accumbens Area, Left Amygdala, Left Caudate, Left Cerebellum Cortex, Left Cerebellum White Matter, Left Cerebral White Matter, Left Pallidum, Left Putamen, Left Thalamus Proper, Left Ventral Diencephalon, Optic Chiasm, Right Accumbens Area, Right Amygdala, Right Caudate, Right Cerebellum Cortex, Right Cerebellum White Matter, Right Cerebral White Matter, Right Pallidum, Right Putamen, Right Thalamus Proper, Right Ventral Diencephalon, White Matter Hypointensities} & 
        \TDcellNumImages{263}{263} & 
        \TDcellLinks{\href{https://www.nitrc.org/projects/candi\_share/}{Official Website} \\\vspace{2pt} \href{https://doi.org/10.1007/s12021-011-9133-y}{Publication}} \\
        \TDrowDescription{The Child and Adolescent NeuroDevelopment Initiative (CANDI, or also CANDShare) at UMass Medical School is making available a series of structural brain images, as well as their anatomic segmentations, demographic and behavioral data and a set of related morphometric resources. Initially, the CANDI dataset featured 103 subjects, encompassing T1-weighted MRI scans and anatomic segmentations. This group included healthy controls (29), individuals with schizophrenia spectrum disorders (20), and those with bipolar disorder with psychosis (19) and bipolar disorder without psychosis (35), all aged four to seventeen. A broader collection within the CANDI initiative expanded to 263 subjects, aged three to twenty-one, adding normative subjects (70) and children with ADHD (31) to the existing diagnostic categories of bipolar disorder (130) and childhood onset schizophrenia (32).} \\
    }\\

    \TDformatRow{
        \TDcellName{CHAOS}{Combined Healthy Abdominal Organ Segmentation}{\citep{CHAOS2021,CHAOSdata2019,kavur2019}} & 
        \TDcellRelDataset{-} & 
        \TDcellModality{3D CT / CT (CE),\\3D MRI (T1),\\3D MRI (T2)} & 
        \TDcellAnatomyRegion{Abdominal Organs}{Abdomen} & 
        \TDcellObjects{4}{Kidney (Left), Kidney (Right), Liver, Spleen} & 
        \TDcellNumImages{40}{20} & 
        \TDcellLinks{\href{https://chaos.grand-challenge.org/Combined\_Healthy\_Abdominal\_Organ\_Segmentation/}{Official Challenge Website} \\\vspace{2pt} \href{https://zenodo.org/record/3431873\#.Yl\_9itpBxaQ}{Official Website} \\\vspace{2pt} \href{https://doi.org/10.1016/j.media.2020.101950}{Publication}} \\
        \TDrowDescription{The CHAOS dataset was released during the ISBI 2019 Challenge, its unique strength lies in offering paired multimodal CT and MR data with corresponding annotations. The dataset includes 40 cases of paired CT and MR scans. For training, 20 cases are fully annotated, while the remaining 20 cases are unannotated, as per the official release. A key point to note is the discrepancy in annotations between modalities: CT scans only provide liver annotations, whereas MR scans are annotated for four different organs.} \\
    }\\

    \TDformatRow{
        \TDcellName{CTSpine1K}{CTSpine1K}{\citep{deng2024ctspine1klargescaledatasetspinal}} & 
        \TDcellRelDataset{COLONOG, COVID-19, HNSCC-3DCT-RT, MSD Liver} & 
        \TDcellModality{3D CT} & 
        \TDcellAnatomyRegion{Spine}{Abdomen, Neck, Pelvis, Thorax} & 
        \TDcellObjects{25}{Vertebra C1 (Primary Vertebra), Vertebra C2 (Secondary Vertebra), Vertebra C3 (Tertiary Vertebra), Vertebra C4 (Intervertebral), Vertebra C5 (Arch Root), Vertebra C6 (Small Joint), Vertebra C7 (Upper Joint), Vertebra L1 (First Sacral), Vertebra L2 (Second Sacral), Vertebra L3 (Third Sacral), Vertebra L4 (Fourth Sacral), Vertebra L5 (Fifth Sacral), Vertebra L6 (Sixth Sacral), Vertebra T1 (First Lumbar), Vertebra T10 (Tenth Lumbar), Vertebra T11 (Eleventh Lumbar), Vertebra T12 (Twelfth Lumbar), Vertebra T2 (Second Lumbar), Vertebra T3 (Third Lumbar), Vertebra T4 (Fourth Lumbar), Vertebra T5 (Fifth Lumbar), Vertebra T6 (Sixth Lumbar), Vertebra T7 (Seventh Lumbar), Vertebra T8 (Eight Lumbar), Vertebra T9 (Ninth Lumbar)} & 
        \TDcellNumImages{1005}{1005} & 
        \TDcellLinks{\href{https://github.com/MIRACLE-Center/CTSpine1K}{Official Website} \\\vspace{2pt} \href{https://arxiv.org/abs/2105.14711}{Preprint}} \\
        \TDrowDescription{CTSpine1K is a large-scale CT dataset comprising 1005 cases specifically designed for spinal segmentation. It aggregates data from four public datasets (COLONOG, HNSCC-3DCT-RT, MSD Liver, and COVID-19), filtering for quality. The dataset provides annotations for 25 types of vertebrae (C1-C7, T1-T12, L1-L6), though the L6 vertebra is rare. To ensure data distribution consistency, the dataset is partitioned into training (610 cases), validation (197 cases), and test (198 cases) sets, maintaining proportional representation from each source dataset.} \\
    }\\

    \TDformatRow{
        \TDcellName{DLBS}{The Dallas Lifespan Brain Study}{\citep{Park2025}} & 
        \TDcellRelDataset{DLBS Epoch 1, DLBS Epoch 2, DLBS Epoch 3} & 
        \TDcellModality{3D fMRI (ASL),\\3D fMRI (BOLD),\\3D MRI (DTI),\\3D MRI (T1 MP-RAGE),\\3D MRI (T2-FLAIR),\\3D PET (Amyloid),\\3D PET (Tau)} & 
        \TDcellAnatomyRegion{Brain}{Head} & 
        \TDcellObjects{-}{[On Demand from Dataset Curators]} & 
        \TDcellNumImages{1692}{-} & 
        \TDcellLinks{\href{https://doi.org/10.18112/openneuro.ds004856.v1.2.0}{Official Website} \\\vspace{2pt} \href{https://fcon\_1000.projects.nitrc.org/indi/retro/dlbs.html}{Secondary Website} \\\vspace{2pt} \href{https://doi.org/10.1038/s41597-025-04847-7}{Publication}} \\
        \TDrowDescription{The Dallas Lifespan Brain Study is a significant longitudinal research initiative that investigated brain and cognitive changes across the adult lifespan (ages 21-89). It gathered extensive data, including detailed neuropsychological assessments, various MRI types (structural, diffusion, functional), and crucially, PET measures of amyloid and tau in cognitively normal participants. A key innovation was its robust sampling of middle-aged individuals. This rich dataset is now openly available on OpenNeuro.org In the works considered in this publication, e.g. the HERMES model, a subset of the whole dataset was used consisting of 213 3D MRI (T1) with three classes. This subset is described by \citet{doi:10.1212/WNL.0b013e318245d295}. Here, we report the condensed global statistics of the DLBS dataset. Please refer to the publication for details. Note that in reporting the total number of images, here is reported the sum of all imaging modalities for the three epochs of the longitudinal study.} \\
    }\\

    \TDformatRow{
        \TDcellName{FeTA}{Fetal Tissue Annotation and Segmentation Challenge}{\citep{Payette2021,Payette_2025}} & 
        \TDcellRelDataset{FeTA 2021, FeTA 2022, FeTA 2024} & 
        \TDcellModality{3D MRI (T2)} & 
        \TDcellAnatomyRegion{Brain}{Head} & 
        \TDcellObjects{7}{Brain Cerebellum, Brain Deep Gray Matter, Brain External Cerebrospinal Fluid, Brain Grey Matter, Brain Stem, Brain Ventricles, Brain White Matter} & 
        \TDcellNumImages{300}{120} & 
        \TDcellLinks{\href{https://fetachallenge.github.io/}{GitHub} \\\vspace{2pt} \href{https://feta.grand-challenge.org/}{Official Challenge Website} \\\vspace{2pt} \href{https://feta.grand-challenge.org/feta-2021/}{Official Challenge Website (1)}} \\
        \TDrowDescription{The FeTA is a medical imaging competition focused on the reconstruction and segmentation of the fetal brain. The 2021 and 2022 iteratuions used T2-weighted MRI and feature 120 training cases from two institutions and 160 test cases from four institutions, all with manual segmentation labels for seven different brain tissues. The 2024 iteration included an additional 20 test cases from Kings College London using a low-field Siemens machine at 0.55T. Training data are either at 1.3T or 3T.} \\
    }\\

    \TDformatRow{
        \TDcellName{FLARE}{MICCAI FLARE Challenge}{\citep{MA2022102616,ma2023unleashingstrengthsunlabeleddata,ma2024automaticorganpancancersegmentation}} & 
        \TDcellRelDataset{AMOS, AutoPET, COVID-19, DeepLesion, FLARE 2021, FLARE 2022, FLARE 2023, FLARE 2024, FLARE 2025, KiTS19, KiTS23, LIDC, LiTS, MELA, MSD, NIH Pancreas-CT, TCIA} & 
        \TDcellModality{3D CT / CT (CE),\\3D MRI,\\3D PET} & 
        \TDcellAnatomyRegion{Abdominal Organs, Tumors}{Abdomen} & 
        \TDcellObjects{14}{Adrenal Gland (Left), Adrenal Gland (Right), Aorta, Duodenum, Esophagus, Gallbladder, Inferior Vena Cava, Kidney (Left), Kidney (Right), Liver, Pancreas, Spleen, Stomach, Tumor} & 
        \TDcellNumImages{14000}{7400} & 
        \TDcellLinks{\href{https://flare.grand-challenge.org/}{FLARE 2021 Challenge Website} \\\vspace{2pt} \href{https://doi.org/10.1016/j.media.2022.102616}{FLARE 2021 Publication} \\\vspace{2pt} \href{https://flare22.grand-challenge.org/}{FLARE 2022 Challenge Website} \\\vspace{2pt} \href{https://arxiv.org/abs/2308.05862}{FLARE 2022 Preprint} \\\vspace{2pt} \href{https://codalab.lisn.upsaclay.fr/competitions/12239}{FLARE 2023 Challenge Website} \\\vspace{2pt} \href{https://arxiv.org/abs/2408.12534}{FLARE 2023 Preprint} \\\vspace{2pt} \href{https://link.springer.com/book/10.1007/978-3-031-23911-3}{FLARE 2022 Challenge Proceedings} \\\vspace{2pt} \href{https://link.springer.com/book/10.1007/978-3-031-58776-4}{FLARE 2023 Challenge Proceedings} \\\vspace{2pt} \href{https://www.codabench.org/competitions/2319}{FLARE 2024 Challenge Website Task 1} \\\vspace{2pt} \href{https://www.codabench.org/competitions/2296/}{FLARE 2024 Challenge Website Task 3} \\\vspace{2pt} \href{https://flare-medfm.github.io/}{FLARE 2025 Challenge Website}} \\
        \TDrowDescription{The FLARE challenges (originally Fast, Low-GPU-Memory Abdominal Organ Segmentation, then the name evolved over time) progressively increased in complexity and scale since their inception in 2021. Designed to test automated organ and tumor segmentation in CT and, more recently, MRI, these challenges have continuously introduced larger and more diverse datasets. A key aspect of their evolution has been the incorporation of unlabeled training data, pushing participants to develop algorithms that are not only accurate but also efficient and capable of leveraging vast amounts of unannotated information through unsupervised pre-training.  FLARE 2021 began with 511 CT images, all fully annotated (100 with hidden labels) for 4 abdominal organs (liver, spleen, pancreas, kidneys).  FLARE 2022 expanded to 2350 CT images, introducing 2000 unlabeled cases alongside only 50 annotated ones, and increased the segmentation targets to 13 abdominal organs. FLARE 2023 further escalated, providing 4500 CT images, with 1800 unlabeled and 2200 partially annotated for 13 abdominal organs and pan-cancer tumors, representing 14 categories.  The FLARE 2024 series diversified into tasks, with Task 1 offering 10,600 whole-body CT images (5000 partially annotated, 5000 unlabeled) for single tumor segmentation, and Task 3 introducing an unsupervised cross-modality domain adaptation challenge with over 5200 unlabeled MRI images and 1250 PET images (target domains) for 13 abdominal organs, relying on 2300 labeled CT images as the source domain. In all challenges, validation sets were fully labeled and provided, while test set labels were withheld for fair evaluation. The last iteration, FLARE 2025, comprises six tasks, of which four involve segmentation in CT, MRI and PET images, most of them blended with the FLARE 2025 tasks. For example, in Task 3, PET scans were only recently added.  Here are reported the condensied statistics of all challenges. The total number of images across all iterations is 16450 (10000 CT from FLARE 2024 Task 1 + 5200 MRI from Task 3, plus 1250 PET scans from the 2025 integration), while the number of annotated images is 7400 (2300 with abdominal organs partial annotations from Task 3 + 5100 pantumor annotations only from Task 1). These numbers consider likely overlaps between different tasks dataset. If no overlap is present, then the order of magnitude should be 20k-30k. It is worth noticing the important overlap with other common datasets, since the FLARE dataset was built mainly from previously-existing, publicly-available datasets.} \\
    }\\

    \TDformatRow{
        \TDcellName{HNASC}{Head and Neck Auto Segmentation}{\citep{https://doi.org/10.1002/mp.12197}} & 
        \TDcellRelDataset{-} & 
        \TDcellModality{3D CT} & 
        \TDcellAnatomyRegion{Bones, Brain, Head Glands}{Head} & 
        \TDcellObjects{9}{Brain Stem, Mandible, Optic Chiasm, Optic Nerve (Left), Optic Nerve (Right), Parotid Gland (Left), Parotid Gland (Right), Submandibular Gland (Left), Submandibular Gland (Right)} & 
        \TDcellNumImages{48}{48} & 
        \TDcellLinks{\href{https://www.imagenglab.com/newsite/pddca/}{Official Website} \\\vspace{2pt} \href{https://www.imagenglab.com/wiki/mediawiki/index.php?title=2015\_MICCAI\_Challenge}{Official Challenge Website} \\\vspace{2pt} \href{https://doi.org/10.1002/mp.12197}{Publication}} \\
        \TDrowDescription{The MICCAI 2015 Head and Neck Auto-Segmentation Challenge dataset,  also referred to as the Public Domain Database for Computational Anatomy (PDDCA), provides a benchmark for evaluating automatic segmentation algorithms for applications in radiotherapy planning. The dataset prises patients sourced from the RTOG 0522 study. The segmentation targets include nine critical head and neck structures along with manual identification of bony landmarks.} \\
    }\\

    \TDformatRow{
        \TDcellName{ISLES}{Ischemic Stroke LEsion Segmentation}{\citep{HernandezPetzsche2022}} & 
        \TDcellRelDataset{ATLAS v2.0} & 
        \TDcellModality{3D MRI (DWI),\\3D MRI (T2-FLAIR)} & 
        \TDcellAnatomyRegion{Brain}{Head} & 
        \TDcellObjects{1}{Brain Ischemic Stroke Lesion} & 
        \TDcellNumImages{250}{250} & 
        \TDcellLinks{\href{https://www.isles-challenge.org/}{Official Website} \\\vspace{2pt} \href{https://isles22.grand-challenge.org/}{Official Challenge Website} \\\vspace{2pt} \href{https://doi.org/10.1038/s41597-022-01875-5}{Publication}} \\
        \TDrowDescription{ISLES is a dataset of multimodal MRI images to automatically segment acute to subacute ischemic stroke lesions, multiple emboli and cortical infarcts, and is associated with the ISLES 2022 MICCAI challenge. The dataset is divided into a training set of 250 cases and a test set of 150 cases which is used solely for model validation and is not disclosed (not image nor segmentation mask). The ISLES challenge has been held since 2015 hosting several editions, and has grown over time both in scale and in the lesion types included (the 2015 challenge only included ischemic stroke lesions). The ATLAS v2.0 dataset is related to the MICCAI ISLES 2022 Challenge Task 2, bus is disjoint from the ISLES dataset.} \\
    }\\

    \TDformatRow{
        \TDcellName{KiPA}{Kidney Parsing}{\citep{HE2021102055}} & 
        \TDcellRelDataset{-} & 
        \TDcellModality{3D CT (CE)} & 
        \TDcellAnatomyRegion{Kidneys, Tumors}{Abdomen} & 
        \TDcellObjects{4}{Kidney Tumor, Kidneys, Renal Artery, Renal Vein} & 
        \TDcellNumImages{100}{70} & 
        \TDcellLinks{\href{https://kipa22.grand-challenge.org/}{Official Challenge Website} \\\vspace{2pt} \href{https://doi.org/10.1016/j.media.2021.102055}{Publication}} \\
        \TDrowDescription{KiPA is the dataset associated with the MICCAI KiPA 2022 challenge aimed at segmenting 3D kidneys, kidney tumors, arteries, and veins. The dataset includes 130 cases of CT scans with complete annotations. The data is officially divided into 70 cases for the training dataset, 30 cases for the open testing dataset (hidden labels), and 30 cases for the closed testing dataset (hidden image and labels). The dataset includes abnormal kidney samples and the annotation of fine renal vascular structures.} \\
    }\\

    \TDformatRow{
        \TDcellName{KiTS}{Kidney Tumor Segmentation}{\citep{HELLER2021101821}} & 
        \TDcellRelDataset{KiTS19, KiTS21, KiTS23} & 
        \TDcellModality{3D CT / CT (CE)} & 
        \TDcellAnatomyRegion{Kidneys}{Abdomen} & 
        \TDcellObjects{3}{Kidney Cyst, Kidney Tumor, Kidneys} & 
        \TDcellNumImages{599}{489} & 
        \TDcellLinks{\href{https://kits-challenge.org/}{Official Website} \\\vspace{2pt} \href{https://doi.org/10.1016/j.media.2020.101821}{KiTS19 Results Publication} \\\vspace{2pt} \href{https://arxiv.org/abs/1904.00445}{KiTS19 Challenge Data Preprint} \\\vspace{2pt} \href{https://arxiv.org/abs/2307.01984}{KiTS21 Challenge Data Preprint}} \\
        \TDrowDescription{The KiTS dataset is a collection of CT scans used for challenges in medical image segmentation, specifically focusing on kidneys and their associated pathologies. The first iteration, KiTS19, released for MICCAI 2019, focused solely on segmenting kidneys and tumors, comprising 210 training and 90 test cases. These 90 test cases were later integrated into the training sets of subsequent challenges. KiTS21, presented at MICCAI 2021, expanded upon KiTS19 by adding the segmentation of cysts to the task. It included 300 publicly available training cases, which incorporated all the data from KiTS19, along with 100 new, non-public testing cases. The most recent iteration, KiTS23, featured at MICCAI 2023, continued to build on its predecessors by encompassing 599 cases (489 for training and 110 for testing). The training set includes all previous KiTS data. A key enhancement in KiTS23 is the inclusion of cases from the "nephrogenic contrast phase" in addition to the "late arterial" phase, and its 110 testing cases are entirely new to the challenge.} \\
    }\\

    \TDformatRow{
        \TDcellName{LASC}{Left Atrial Segmentation Challenge}{\citep{10.1007/978-3-642-54268-8_1,XIONG2021101832}} & 
        \TDcellRelDataset{LASC13, LASC18} & 
        \TDcellModality{3D CT (CE),\\3D MRI (T1-CE)} & 
        \TDcellAnatomyRegion{Heart}{Thorax} & 
        \TDcellObjects{1}{Heart Atrium (Left)} & 
        \TDcellNumImages{184}{110} & 
        \TDcellLinks{\href{https://www.kaggle.com/datasets/adarshsng/heart-mri-image-dataset-left-atrial-segmentation}{LASC13 Kaggle Challenge} \\\vspace{2pt} \href{https://arxiv.org/abs/1902.09063}{LASC13 Preprint} \\\vspace{2pt} \href{https://doi.org/10.1007/978-3-642-54268-8\_1}{LASC13 Publication} \\\vspace{2pt} \href{https://www.cardiacatlas.org/atriaseg2018-challenge/}{LASC18 Official Website} \\\vspace{2pt} \href{https://ieee-dataport.org/documents/left-atrium-2018}{LASC18 IEEE Dataport} \\\vspace{2pt} \href{https://doi.org/10.1016/j.media.2020.101832}{LASC18 Publication} \\\vspace{2pt} \href{https://www.cardiacatlas.org/}{The Cardiac Atlas Project}} \\
        \TDrowDescription{The Left Atrial Segmentation Challenge (LASC) datasets focus on the segmentation of the left atrium from medical images, essential for guiding atrial fibrillation treatments and cardiac modeling. The LASC 2013 dataset, used at MICCAI 2013 (STACOM 2013), provided 30 MRI and 30 CT scans. For each modality, 10 datasets were for training with expert segmentations, and 20 for evaluation. The task focused on segmenting the LA, including parts of the LA appendage and proximal pulmonary veins. The Left Atrium 2018 dataset, used at MICCAI 2018, also involved the segmentation of the LA cavity from 154 (100 with labels) Gadolinium-Enhanced MRI (GE-MRI), crucial for understanding atrial fibrosis despite low image contrast. Here are reported the condensed startistics of the two dataset iterations, considering reuse of the MRI scans. LASC18 is part of the Cardiac Atlas Project.} \\
    }\\

    \TDformatRow{
        \TDcellName{LIDC-IDRI}{The Lung Image Database Consortium and Image Database Resource Initiative}{\citep{https://doi.org/10.1118/1.3528204}} & 
        \TDcellRelDataset{-} & 
        \TDcellModality{3D CT (LD)} & 
        \TDcellAnatomyRegion{Lung}{Thorax} & 
        \TDcellObjects{1}{Lung Nodule} & 
        \TDcellNumImages{1308}{1308} & 
        \TDcellLinks{\href{https://www.cancerimagingarchive.net/collection/lidc-idri/}{Official Website} \\\vspace{2pt} \href{https://doi.org/10.1118/1.3528204}{Publication}} \\
        \TDrowDescription{The LIDC-IDRI dataset comprises clinical thoracic CT scans from 1,010 patients. It contains 7,371 lesions identified as "nodule" by experienced thoracic radiologists. Nodule annotations include segmentation masks and characterization data.} \\
    }\\

    \TDformatRow{
        \TDcellName{LiTS / MSD Liver}{The Liver Tumor Segmentation Benchmark}{\citep{lits_benchmark}} & 
        \TDcellRelDataset{MSD Liver} & 
        \TDcellModality{3D CT} & 
        \TDcellAnatomyRegion{Liver, Tumors}{Abdomen} & 
        \TDcellObjects{2}{Liver, Liver Tumor} & 
        \TDcellNumImages{201}{131} & 
        \TDcellLinks{\href{https://competitions.codalab.org/competitions/17094}{Official Challenge Website} \\\vspace{2pt} \href{http://medicaldecathlon.com/}{MSD Website} \\\vspace{2pt} \href{https://doi.org/10.1016/j.media.2022.102680}{Publication}} \\
        \TDrowDescription{The LiTS dataset is a multi-center CT imaging dataset compiled from 7 distinct medical institutions. The dataset features diverse primary and secondary tumors with varied sizes, appearances, and lesion-to-background contrast levels. It was the basis for related competitions held at ISBI 2017, MICCAI 2017, and MICCAI 2018, and is included integrally as the Liver Tumor task in the Medical Segmentation Decathlon (MSD).} \\
    }\\

    \TDformatRow{
        \TDcellName{LUNA16}{Lung Nodule Analysis 2016}{\citep{SETIO20171}} & 
        \TDcellRelDataset{LIDC-IDRI} & 
        \TDcellModality{3D CT (LD)} & 
        \TDcellAnatomyRegion{Lung}{Thorax} & 
        \TDcellObjects{2}{Lung Nodule, Lungs} & 
        \TDcellNumImages{888}{888} & 
        \TDcellLinks{\href{https://luna16.grand-challenge.org/Home/}{Official Challenge Website} \\\vspace{2pt} \href{https://doi.org/10.1016/j.media.2017.06.015}{Publication}} \\
        \TDrowDescription{The LUNA16 dataset is a refined subset of the LIDC-IDRI database. While LIDC-IDRI comprises more than 1000 low-dose lung CT images with expert radiologist annotations including nodule outlines, LUNA16 meticulously filters this by excluding scans with slice thickness greater than 2.5mm (or 3mm) and nodules smaller than 3mm. For the challenge, LUNA16's primary tasks involve nodule detection, providing 1186 annotated nodule locations and diameters (no segmentations), and false positive reduction, which entails classifying 551,065 candidate locations as true or false positives. The reference standard for these tasks specifically uses nodules  3mm confirmed by at least three out of four radiologists from the original LIDC-IDRI annotations. Although LUNA16 also includes whole lung segmentation masks, these were provided as an auxiliary resource and were not part of the official challenge tasks. Even if technically segmentations of nodules are not included, they can be inferred from the LIDC-IDRI dataset and from the provided ROIs of the nodules, so it will be considered as if included.} \\
    }\\

    \TDformatRow{
        \TDcellName{M\&Ms}{Multi-Centre, Multi-Vendor \& Multi-Disease Cardiac Image Segmentation Challenge}{\citep{9458279}} & 
        \TDcellRelDataset{-} & 
        \TDcellModality{3D MRI (T1-CE)} & 
        \TDcellAnatomyRegion{Heart}{Thorax} & 
        \TDcellObjects{3}{Heart Ventricle (Left), Heart Ventricle (Right), Myocardium} & 
        \TDcellNumImages{375}{150} & 
        \TDcellLinks{\href{https://www.ub.edu/mnms/}{Official Website} \\\vspace{2pt} \href{https://doi.org/10.1109/TMI.2021.3090082}{Publication}} \\
        \TDrowDescription{The M\&Ms Challenge (part of MICCAI 2020) dataset is a collection of 375 images from diverse clinical centers across Spain, Germany, and Canada. It encompasses both healthy individuals and patients with various cardiac pathologies, acquired using MRI scanners from Siemens, General Electric, Philips, and Canon. Expert clinicians have meticulously segmented the left ventricle, right ventricle, and left ventricular myocardium in the images following the same standard as in the ACDC dataset. In the original challenge, training images were 175, of which 25 provided without annotations. The remaining 200 images were used for testing.} \\
    }\\

    \TDformatRow{
        \TDcellName{MM-WHS}{Multi-Modality Whole Heart Segmentation}{-} & 
        \TDcellRelDataset{-} & 
        \TDcellModality{3D CT / CT (CE),\\3D MRI (T1-CE)} & 
        \TDcellAnatomyRegion{Heart}{Thorax} & 
        \TDcellObjects{7}{Aorta, Heart Atrium (Left), Heart Atrium (Right), Heart Ventricle (Left), Heart Ventricle (Right), Myocardium (Left Ventricle), Pulmonary Artery} & 
        \TDcellNumImages{120}{40} & 
        \TDcellLinks{\href{https://zmiclab.github.io/zxh/0/mmwhs/}{Official Website}} \\
        \TDrowDescription{The MM-WHS dataset, introduced at MICCAI 2017, is is aimed at entire heart and its key substructures segmentation from various clinical imaging conditions. It comprises a total of 120 cardiac images, evenly split between 60 CT/CTA and 60 MRI scans. The dataset is divided into a training set (20 CT and 20 MRI scans) and a test set (40 CT and 40 MRI scans). The training set includes manual annotations for seven major cardiac substructures: the left and right ventricular cavities, left and right atrial cavities, left ventricular myocardium, ascending aorta, and pulmonary artery.} \\
    }\\

    \TDformatRow{
        \TDcellName{MOTS}{Multi-Organ and Tumor Segmentation}{\citep{Zhang_2021_CVPR}} & 
        \TDcellRelDataset{KiTS, LiTS / MSD Liver, MSD Colon, MSD Hepatic Vessels, MSD Lung, MSD Pancreas, MSD Spleen} & 
        \TDcellModality{3D CT / CT (CE)} & 
        \TDcellAnatomyRegion{Abdominal Organs, Tumors}{Abdomen} & 
        \TDcellObjects{11}{Colon Tumor, Hepatic Vessels, Kidney Cyst, Kidney Tumor, Kidneys, Liver, Liver Tumor, Lung Nodule, Pancreas, Pancreas Tumor, Spleen} & 
        \TDcellNumImages{1155}{920} & 
        \TDcellLinks{\href{https://github.com/jianpengz/DoDNet}{Official Website}} \\
        \TDrowDescription{The MOTS dataset was created by \citet{Zhang_2021_CVPR} for training and pre-training the DoDNet segmentation model. The dataset is an ensemble of seven publicly-available datasets, specifically from the KiTS dataset and the MSD collection of dataset involving only abdominal organs. Some images are specifically identified as test images. Dataset under direct request.} \\
    }\\

    \TDformatRow{
        \TDcellName{MSD Cardiac}{Medical Segmentation Decathlon - Cardiac}{\citep{MSD_arxiv_dataset,MSD_nature}} & 
        \TDcellRelDataset{-} & 
        \TDcellModality{3D MRI (T1-CE)} & 
        \TDcellAnatomyRegion{Heart}{Thorax} & 
        \TDcellObjects{1}{Heart Atrium (Left)} & 
        \TDcellNumImages{30}{20} & 
        \TDcellLinks{\href{http://medicaldecathlon.com/}{MSD Website} \\\vspace{2pt} \href{https://doi.org/10.1038/s41467-022-30695-9}{Publication} \\\vspace{2pt} \href{https://arxiv.org/abs/1902.09063}{Preprint}} \\
        \TDrowDescription{The MSD Cardiac (MSD Task02) dataset, also known as MSD Heart, is a sub-task of the Medical Segmentation Decathlon, focusing on left atrium segmentation from single-modality MRI images.} \\
    }\\

    \TDformatRow{
        \TDcellName{MSD Colon Cancer}{Medical Segmentation Decathlon - Colon Cancer}{\citep{MSD_arxiv_dataset,MSD_nature}} & 
        \TDcellRelDataset{-} & 
        \TDcellModality{3D CT} & 
        \TDcellAnatomyRegion{Colon, Tumors}{Abdomen} & 
        \TDcellObjects{1}{Colon Tumor} & 
        \TDcellNumImages{190}{126} & 
        \TDcellLinks{\href{http://medicaldecathlon.com/}{MSD Website} \\\vspace{2pt} \href{https://doi.org/10.1038/s41467-022-30695-9}{Publication} \\\vspace{2pt} \href{https://arxiv.org/abs/1902.09063}{Preprint}} \\
        \TDrowDescription{The MSD Colon Cancer (MSD Task10) dataset is a sub-task of the Medical Segmentation Decathlon, focusing on colon tumor segmentation from CT images. It comprises venous phase CT scans from 190 patients undergoing surgery for primary colon cancer.} \\
    }\\

    \TDformatRow{
        \TDcellName{MSD Hepatic Vessels}{Medical Segmentation Decathlon - Hepatic Vessels}{\citep{MSD_arxiv_dataset,MSD_nature}} & 
        \TDcellRelDataset{-} & 
        \TDcellModality{3D CT (CE)} & 
        \TDcellAnatomyRegion{Liver, Tumors}{Abdomen} & 
        \TDcellObjects{2}{Hepatic Vessels, Liver Tumor} & 
        \TDcellNumImages{443}{303} & 
        \TDcellLinks{\href{http://medicaldecathlon.com/}{MSD Website} \\\vspace{2pt} \href{https://doi.org/10.1038/s41467-022-30695-9}{Publication} \\\vspace{2pt} \href{https://arxiv.org/abs/1902.09063}{Preprint}} \\
        \TDrowDescription{The MSD Hepatic Vessel (MSD Task08) dataset is a sub-task of the Medical Segmentation Decathlon, with the objective of segmenting hepatic vessels and tumors from liver CT scans. It is worth noting that some subsequent research \citep{10.1007/978-3-030-87193-2_1} has raised concerns about the image annotation quality, suggesting that approximately 65.5\% of vessel pixels may be unmarked and 8.5\% of pixels mislabeled as vessels.} \\
    }\\

    \TDformatRow{
        \TDcellName{MSD Hippocampus}{Medical Segmentation Decathlon - Hippocampus}{\citep{MSD_arxiv_dataset,MSD_nature}} & 
        \TDcellRelDataset{-} & 
        \TDcellModality{3D MRI (T1 MP-RAGE)} & 
        \TDcellAnatomyRegion{Brain}{Head} & 
        \TDcellObjects{2}{Hippocampus (Anterior), Hippocampus (Posterior)} & 
        \TDcellNumImages{390}{260} & 
        \TDcellLinks{\href{http://medicaldecathlon.com/}{MSD Website} \\\vspace{2pt} \href{https://doi.org/10.1038/s41467-022-30695-9}{Publication} \\\vspace{2pt} \href{https://arxiv.org/abs/1902.09063}{Preprint}} \\
        \TDrowDescription{The MSD Hippocampus (MSD Task04) dataset is a sub-task of the Medical Segmentation Decathlon, focusing on the segmentation of the hippocampal region from single-modality MRI. This dataset contains segmentations of the two distinct anterior and posterior parts of the hippocampus. The dataset officially comprises 394 images, with 263 intended for training and 131 for testing. However, the downloadable training set contains 260 cases, and the test set contains 130 cases. Test results can be submitted to the official MSD website for evaluation.} \\
    }\\

    \TDformatRow{
        \TDcellName{MSD Lung Tumors}{Medical Segmentation Decathlon - Lung Tumours}{\citep{MSD_arxiv_dataset,MSD_nature}} & 
        \TDcellRelDataset{-} & 
        \TDcellModality{3D CT} & 
        \TDcellAnatomyRegion{Lung, Tumors}{Thorax} & 
        \TDcellObjects{1}{Lung Nodule} & 
        \TDcellNumImages{95}{63} & 
        \TDcellLinks{\href{http://medicaldecathlon.com/}{MSD Website} \\\vspace{2pt} \href{https://doi.org/10.1038/s41467-022-30695-9}{Publication} \\\vspace{2pt} \href{https://arxiv.org/abs/1902.09063}{Preprint}} \\
        \TDrowDescription{The MSD Lung Tumours (MSD Task06) dataset is a sub-task of the Medical Segmentation Decathlon, focusing on lung tumor segmentation from thin-section CT images. It includes CT scans of 96 patients with non-small cell lung cancer (NSCLC), officially divided into 64 cases for training and 32 for testing. However, 63 cases can be downloaded for the training set.} \\
    }\\

    \TDformatRow{
        \TDcellName{MSD Pancreas Tumour}{Medical Segmentation Decathlon - Pancreas Tumour}{\citep{MSD_arxiv_dataset,MSD_nature}} & 
        \TDcellRelDataset{-} & 
        \TDcellModality{3D CT (CE)} & 
        \TDcellAnatomyRegion{Pancreas}{Abdomen} & 
        \TDcellObjects{2}{Pancreas, Pancreas Tumor} & 
        \TDcellNumImages{420}{281} & 
        \TDcellLinks{\href{http://medicaldecathlon.com/}{MSD Website} \\\vspace{2pt} \href{https://doi.org/10.1038/s41467-022-30695-9}{Publication} \\\vspace{2pt} \href{https://arxiv.org/abs/1902.09063}{Preprint}} \\
        \TDrowDescription{The MSD Pancreas Tumour (MSD Task07) dataset is a sub-task of the Medical Segmentation Decathlon, focusing on segmenting both the pancreas and its tumors from CT images. It's considered one of the two most challenging tasks in MSD, alongside the Colon Cancer task. The dataset specifically includes three types of pancreatic tumors: intraductal papillary mucinous neoplasms, pancreatic neuroendocrine tumors, and pancreatic ductal adenocarcinomas.} \\
    }\\

    \TDformatRow{
        \TDcellName{MSD Prostate}{Medical Segmentation Decathlon - Colon Cancer}{\citep{MSD_arxiv_dataset,MSD_nature}} & 
        \TDcellRelDataset{-} & 
        \TDcellModality{3D MRI (T2)} & 
        \TDcellAnatomyRegion{Prostate}{Pelvis} & 
        \TDcellObjects{2}{Prostate (Peripheral Zone), Prostate (Transition Zone)} & 
        \TDcellNumImages{48}{32} & 
        \TDcellLinks{\href{http://medicaldecathlon.com/}{MSD Website} \\\vspace{2pt} \href{https://doi.org/10.1038/s41467-022-30695-9}{Publication} \\\vspace{2pt} \href{https://arxiv.org/abs/1902.09063}{Preprint}} \\
        \TDrowDescription{The MSD Prostate (MSD Task05) dataset is a sub-task of the Medical Segmentation Decathlon, focusing on segmenting two distinct prostate regions: the central gland and the peripheral zone. This dataset utilizes multi-parametric MR images (T2-weighted and ADC).} \\
    }\\

    \TDformatRow{
        \TDcellName{MSD Spleen}{Medical Segmentation Decathlon - Spleen}{\citep{MSD_arxiv_dataset,MSD_nature}} & 
        \TDcellRelDataset{-} & 
        \TDcellModality{3D CT (CE)} & 
        \TDcellAnatomyRegion{Spleen}{Abdomen} & 
        \TDcellObjects{1}{Spleen} & 
        \TDcellNumImages{61}{41} & 
        \TDcellLinks{\href{http://medicaldecathlon.com/}{MSD Website} \\\vspace{2pt} \href{https://doi.org/10.1038/s41467-022-30695-9}{Publication} \\\vspace{2pt} \href{https://arxiv.org/abs/1902.09063}{Preprint}} \\
        \TDrowDescription{The MSD Spleen (MSD Task09) dataset is a sub-task of the Medical Segmentation Decathlon, focusing on spleen segmentation from CT images. The dataset consists of portal venous phase CT scans from patients undergoing chemotherapy for liver metastases.} \\
    }\\

    \TDformatRow{
        \TDcellName{MSSEG}{Multiple Sclerosis Lesion Segmentation}{\citep{Styner_Lee_Chin_Chin_Commowick_Tran_Markovic-Plese_Jewells_Warfield2008}} & 
        \TDcellRelDataset{-} & 
        \TDcellModality{3D MRI (DTI),\\3D MRI (T1),\\3D MRI (T2),\\3D MRI (T2-FLAIR)} & 
        \TDcellAnatomyRegion{Brain}{Head} & 
        \TDcellObjects{1}{Brain Hemorrage} & 
        \TDcellNumImages{51}{20} & 
        \TDcellLinks{\href{https://www.nitrc.org/projects/msseg/}{Official Website} \\\vspace{2pt} \href{https://doi.org/10.54294/lmkqvm}{Publication}} \\
        \TDrowDescription{The MSSEG (also MSseg08) dataset, created for a MICCAI 2008 challenge, is an MRI-based dataset focused on fully automated 3D segmentation of Multiple Sclerosis (MS) lesions. The data was provided by Boston Children's Hospital and the University of North Carolina (UNC) using a Siemens 3T Allegra MRI scanner.} \\
    }\\

    \TDformatRow{
        \TDcellName{OASIS-1}{OASIS-1: Cross-sectional MRI Data in Young, Middle Aged, Nondemented and Demented Older Adults}{\citep{10.1162/jocn.2007.19.9.1498}} & 
        \TDcellRelDataset{-} & 
        \TDcellModality{3D MRI (T1 MP-RAGE)} & 
        \TDcellAnatomyRegion{Brain}{Head} & 
        \TDcellObjects{-}{[Too Many To List]} & 
        \TDcellNumImages{416}{416} & 
        \TDcellLinks{\href{https://sites.wustl.edu/oasisbrains/}{OASIS Project} \\\vspace{2pt} \href{https://sites.wustl.edu/oasisbrains/home/oasis-1/}{Official Website} \\\vspace{2pt} \href{https://doi.org/10.1162/jocn.2007.19.9.1498}{Publication}} \\
        \TDrowDescription{The Open Access Series of Imaging Studies (OASIS) project aims to provide freely available neuroimaging datasets to the scientific community. The series includes four main datasets: OASIS-1 (cross-sectional MRI data for aging and Alzheimer's), OASIS-2 (longitudinal MRI data for aging and Alzheimer's), OASIS-3 (extensive longitudinal multimodal data for aging and Alzheimers Disease), OASIS-3 Tau (OASIS-3 Flortaucipir F18 (AV1451) PET) and OASIS-4 (MR and clinical data for individuals with memory complaints).  The OASIS-1 dataset is a cross-sectional collection of MRI scans from 416 subjects aged 18 to 96. Each subject has 3 or 4 individual MRI scans from single sessions. Notably, 100 subjects over 60 years old have been clinically diagnosed with very mild to moderate Alzheimer's disease. A separate reliability dataset includes 20 non-demented subjects rescanned within 90 days.  The dataset consists of 35 label classes which are brain portions, sections, and sub-organs.} \\
    }\\

    \TDformatRow{
        \TDcellName{OASIS-3}{OASIS-3: Longitudinal Multimodal Neuroimaging, Clinical, and Cognitive Dataset for Normal Aging and Alzheimers Disease}{\citep{10.1162/jocn.2007.19.9.1498}} & 
        \TDcellRelDataset{-} & 
        \TDcellModality{3D CT,\\3D fMRI (ASL),\\3D fMRI (BOLD),\\3D MRI (DTI),\\3D MRI (SWI),\\3D MRI (T1 MP-RAGE),\\3D MRI (T2),\\3D MRI (T2-FLAIR),\\3D PET (Amyloid),\\3D PET (FDG),\\3D PET (Tau)} & 
        \TDcellAnatomyRegion{Brain}{Head} & 
        \TDcellObjects{-}{[On Demand from Dataset Curators], Brain, Cerebral Cortex, Cerebral Cortex white Matter, Subcortical Gray Matter} & 
        \TDcellNumImages{6922}{-} & 
        \TDcellLinks{\href{https://sites.wustl.edu/oasisbrains/}{OASIS Project} \\\vspace{2pt} \href{https://sites.wustl.edu/oasisbrains/home/oasis-3/}{Official Website} \\\vspace{2pt} \href{https://doi.org/10.1162/jocn.2007.19.9.1498}{Publication} \\\vspace{2pt} \href{https://doi.org/10.1101/2019.12.13.19014902}{Preprint}} \\
        \TDrowDescription{The Open Access Series of Imaging Studies (OASIS) project aims to provide freely available neuroimaging datasets to the scientific community. The series includes four main datasets: OASIS-1 (cross-sectional MRI data for aging and Alzheimer's), OASIS-2 (longitudinal MRI data for aging and Alzheimer's), OASIS-3 (extensive longitudinal multimodal data for aging and Alzheimers Disease), OASIS-3 Tau (OASIS-3 Flortaucipir F18 (AV1451) PET) and OASIS-4 (MR and clinical data for individuals with memory complaints).  OASIS-3 is a retrospective, longitudinal compilation of multimodal data collected over 30 years from 1378 participants (755 cognitively normal, 622 with cognitive decline, aged 42-95). It includes 2842 MRI sessions with diverse sequences such as T1w, T2w, FLAIR, ASL, SWI, time of flight, resting-state BOLD, and DTI. Many MRI sessions are accompanied by FreeSurfer segmentation masks. The dataset also features over 2157 raw PET imaging scans from PIB, AV45, and FDG tracers, with accompanying post-processed files from the Pet Unified Pipeline (PUP). Additionally, 451 Tau PET sessions (AV1451) are available as a sub-project. Also 1472 CT scans are available. Available labels numerosity and description is not very clear from website and publications.} \\
    }\\

    \TDformatRow{
        \TDcellName{Pancreas-CT}{NIH Pancreas-CT}{\citep{Roth2016}} & 
        \TDcellRelDataset{-} & 
        \TDcellModality{3D CT (CE)} & 
        \TDcellAnatomyRegion{Pancreas}{Abdomen} & 
        \TDcellObjects{1}{Pancreas} & 
        \TDcellNumImages{80}{80} & 
        \TDcellLinks{\href{https://www.cancerimagingarchive.net/collection/pancreas-ct/}{Official Website}} \\
        \TDrowDescription{The Pancreas-CT dataset comprises 80 images, specifically focusing on manual annotations of the pancreas. Provided by the National Institutes of Health Clinical Center, this dataset explicitly excludes pancreatic tumors, featuring 17 healthy kidney donors and 63 patients without major abdominal diseases or pancreatic cancer. The scans, acquired in the portal venous phase using Philips and Siemens scanners, have undergone meticulous manual segmentation of the pancreas. Originally 82 cases, the latest Version 2 has removed two redundant cases (25 and 70). This dataset is incorporated into larger public datasets like AbdomenCT-1K and AbdomenAtlas.} \\
    }\\

    \TDformatRow{
        \TDcellName{PROMISE12}{Prostate MRI Image Segmentation}{\citep{LITJENS2014359,dowling2009automatic}} & 
        \TDcellRelDataset{PROMISE09} & 
        \TDcellModality{3D MRI (T1),\\3D MRI (T2)} & 
        \TDcellAnatomyRegion{Prostate}{Pelvis} & 
        \TDcellObjects{1}{Prostate} & 
        \TDcellNumImages{50}{50} & 
        \TDcellLinks{\href{https://www.na-mic.org/wiki/2009\_prostate\_segmentation\_challenge\_MICCAI}{PROMISE09 Official Website} \\\vspace{2pt} \href{https://www.na-mic.org/wiki/Training\_Data\_Prostate\_Segmentation\_Challenge\_MICCAI09}{PROMISE09 Data} \\\vspace{2pt} \href{https://promise12.grand-challenge.org/}{PROMISE12 Challenge Website} \\\vspace{2pt} \href{https://doi.org/10.1016/j.media.2013.12.002}{PROMISE12 Publication}} \\
        \TDrowDescription{The PROMISE12 dataset was introduced as part of a MICCAI 2012 challenge. It provides 50 prostate MRI images along with their corresponding segmentation annotations, and is composed of multi-center, multi-vendor, and multi-protocol data. PROMISE12 is an extension of the PROMISE09 callenge dataset.} \\
    }\\

    \TDformatRow{
        \TDcellName{Prostate158}{Prostate158}{\citep{ADAMS2022105817}} & 
        \TDcellRelDataset{-} & 
        \TDcellModality{3D MRI (DWI),\\3D MRI (DWI-ADC),\\3D MRI (T2)} & 
        \TDcellAnatomyRegion{Prostate, Tumors}{Pelvis} & 
        \TDcellObjects{3}{Prostate (Central Gland), Prostate (Peripheral Zone), Prostate Cancer} & 
        \TDcellNumImages{158}{139} & 
        \TDcellLinks{\href{https://github.com/kbressem/prostate158}{Official Website} \\\vspace{2pt} \href{https://doi.org/10.1016/j.compbiomed.2022.105817}{Publication}} \\
        \TDrowDescription{The Prostate158 dataset is a curated collection of 158 expert-annotated biparametric 3 Tesla prostate MRI studies with segmentation masks for prostate anatomical zones and cancerous lesions. Each study includes T2-weighted and diffusion-weighted images with apparent diffusion coefficient maps. For cancerous lesions histopathologic confirmation is available.} \\
    }\\

    \TDformatRow{
        \TDcellName{SCD}{Sunnybrook Cardiac Data}{\citep{Radau_Lu_Connelly_Paul_Dick_Wright2009}} & 
        \TDcellRelDataset{-} & 
        \TDcellModality{3D MRI (Cine)} & 
        \TDcellAnatomyRegion{Heart}{Thorax} & 
        \TDcellObjects{2}{Heart Ventricle (Left), Myocardium (Left Ventricle)} & 
        \TDcellNumImages{45}{45} & 
        \TDcellLinks{\href{https://www.cardiacatlas.org/}{The Cardiac Atlas Project} \\\vspace{2pt} \href{https://www.cardiacatlas.org/sunnybrook-cardiac-data/}{Official Website} \\\vspace{2pt} \href{https://midasjournal.org/browse/publication/658}{Publication}} \\
        \TDrowDescription{The Sunnybrook Cardiac Data, also known as the 2009 Cardiac MR Left Ventricle Segmentation Challenge data, consist of 45 cine-MRI images from a mixed of patients and pathologies: healthy, hypertrophy, heart failure with infarction and heart failure without infarction. Subset of this data set was first used in the automated myocardium segmentation challenge from MICCAI 2009. The whole complete data set is now available.} \\
    }\\

    \TDformatRow{
        \TDcellName{SegTHOR}{Segmentation of Thoracic Organs at Risk}{\citep{9286453}} & 
        \TDcellRelDataset{-} & 
        \TDcellModality{3D CT} & 
        \TDcellAnatomyRegion{Thoracic Organs}{Thorax} & 
        \TDcellObjects{4}{Aorta, Esophagus, Heart, Trachea} & 
        \TDcellNumImages{60}{40} & 
        \TDcellLinks{\href{https://competitions.codalab.org/competitions/21145\#learn\_the\_details-overview}{Official Challenge Website} \\\vspace{2pt} \href{https://ieeexplore.ieee.org/document/9286453}{Publication} \\\vspace{2pt} \href{https://ceur-ws.org/Vol-2349/}{SegTHOR2019 Proceedings}} \\
        \TDrowDescription{The SegTHOR dataset, an official challenge of IEEE ISBI 2019, is a CT dataset specifically for the segmentation of four thoracic organs: heart, aorta, trachea, and esophagus. These organs surround tumors and require protection during radiotherapy, each presenting varying spatial and appearance characteristics. The dataset comprises 60 3D CT scans, with 40 cases for training and 20 for testing. It is worth noticing that the heart is not wholly segmented since segmentations only include the part at risk, which roughly corresponds to the lower half.} \\
    }\\

    \TDformatRow{
        \TDcellName{SpineWeb}{SpineWeb}{\citep{ZHENG2017327}} & 
        \TDcellRelDataset{Automatic 3D MRI IVD Localization and Segmentation, IVDM3Seg} & 
        \TDcellModality{3D MRI (T2 Dixon Protocol),\\3D MRI (T2)} & 
        \TDcellAnatomyRegion{Spine}{Abdomen, Pelvis} & 
        \TDcellObjects{13}{Intervertebral Disc (L1-L2), Intervertebral Disc (L2-L3), Intervertebral Disc (L3-L4), Intervertebral Disc (L4-L5), Intervertebral Disc (T11-T12), Intervertebral Disc (T12-L1), Vertebra L1 (First Sacral), Vertebra L2 (Second Sacral), Vertebra L3 (Third Sacral), Vertebra L4 (Fourth Sacral), Vertebra L5 (Fifth Sacral), Vertebra T11 (Eleventh Lumbar), Vertebra T12 (Twelfth Lumbar)} & 
        \TDcellNumImages{24}{16} & 
        \TDcellLinks{\href{https://doi.org/10.1016/j.media.2016.08.005}{MICCAI 2015 Publication} \\\vspace{2pt} \href{https://link.springer.com/book/10.1007/978-3-319-55050-3}{CSI 2016 Challenge Publication} \\\vspace{2pt} \href{https://csi2016.wordpress.com/}{CSI 2016 Challenge Website} \\\vspace{2pt} \href{https://zenodo.org/records/22304}{SpineWeb 2015 Data} \\\vspace{2pt} \href{https://ivdm3seg.weebly.com/}{MICCAI 2018 IVDM3Seg Official Website}} \\
        \TDrowDescription{The SpineWeb dataset stems from two MICCAI challengges aimed at Intervertebral Disc (IVD) analysis from MRI scans, crucial for understanding low back pain. The MICCAI 2015 challenge (Automatic 3D MRI IVD Localization and Segmentation) used 25 T2-weighted MRI cases. The later MICCAI 2018 (IVDM3Seg) challenge evolved to include 16 multi-modality MR cases (Dixon protocol), aiming for more robust algorithms in varied clinical settings. Each multi-modality MRI patient scans set contains four aligned volumes: in-phase, opposed-phase, fat and water images. In total there are 96 high resolution 3D MRI volume data. One mask volume is present for each patient. Here we report the combined statistics of the two datasets created by the same research group. Overall, SpineWeb was initiative from a canadian medical imaging research group, however the related websites have been shut down, and few indications remain of the original challenge.} \\
    }\\

    \TDformatRow{
        \TDcellName{Synapse}{Multi-Atlas Labeling Beyond The Cranial Vault - Abdomen (Label Subset of Eight)}{\citep{landman2015miccai}} & 
        \TDcellRelDataset{-} & 
        \TDcellModality{3D CT (CE)} & 
        \TDcellAnatomyRegion{Abdominal Organs}{Abdomen} & 
        \TDcellObjects{8}{Aorta, Gallbladder, Kidney (Left), Kidney (Right), Liver, Pancreas, Spleen, Stomach} & 
        \TDcellNumImages{50}{30} & 
        \TDcellLinks{\href{https://www.synapse.org/Synapse:syn3193805/wiki/}{Official Website}} \\
        \TDrowDescription{The Synapse platform, managed by Sage Bionetworks, serves as a hub for collaborative scientific research and data sharing. It famously hosted the MICCAI 2015 Multi-Atlas Labeling Beyond The Cranial Vault (BTCV) Abdomen Challenge. Consequently, the BTCV Abdomen dataset, a collection of abdominal CT scans for multi-organ segmentation, is often colloquially referred to as "the Synapse dataset" within the research community because it's distributed via this platform. While the full BTCV dataset originally features 13 or 14 distinct organ classes, a standardized subset of eight major abdominal organs became a widely adopted benchmark in subsequent research, leading to confusion between the platform, the full dataset, and its popular subset, with many works using the wrong names. The standardized label subset was not proposed by the challenge organizers, rather the community converged towards this organs subset. The dataset with this major organs subset is commonly referred to as the Synapse dataset, as opposed to the BTCV dataset that considers all 13 classes.} \\
    }\\

    \TDformatRow{
        \TDcellName{ToothFairy}{ToothFairy MICCAI 2023 Challenge Dataset}{\citep{Cipriano_2022_CVPR}} & 
        \TDcellRelDataset{-} & 
        \TDcellModality{3D CT (CB)} & 
        \TDcellAnatomyRegion{Mandible}{Head} & 
        \TDcellObjects{1}{Inferior Alveolar Nerve} & 
        \TDcellNumImages{443}{420} & 
        \TDcellLinks{\href{https://toothfairy.grand-challenge.org/toothfairy/}{Official Challenge Website} \\\vspace{2pt} \href{https://openaccess.thecvf.com/content/CVPR2022/html/Cipriano\_Improving\_Segmentation\_of\_the\_Inferior\_Alveolar\_Nerve\_Through\_Deep\_Label\_CVPR\_2022\_paper.html}{Publication}} \\
        \TDrowDescription{The ToothFairy dataset, introduced as part of a MICCAI 2023 challenge, is designed for voxel-level segmentation of the Inferior Alveolar Nerve (IAN) in Cone Beam Computed Tomography (CBCT) scans. It comprises 443 CBCT images, featuring both sparse annotations (443 cases total, 290 for training) of whihc some have dense annotations (153 cases total, 130 for training). For challenge evaluation, 8 cases are reserved for validation and 15 for testing, with additional undisclosed data provided during the evaluation phase.} \\
    }\\

    \TDformatRow{
        \TDcellName{TopCoW}{TopCoW (Topology-Aware Anatomical Segmentation of the Circle of Willis}{\citep{topcowchallenge}} & 
        \TDcellRelDataset{-} & 
        \TDcellModality{3D CT (CE),\\3D MRI (TOF-MRA)} & 
        \TDcellAnatomyRegion{Brain}{Head} & 
        \TDcellObjects{13}{Brain CoW Anterior Cerebral Artery (Left), Brain CoW Anterior Cerebral Artery (Right), Brain CoW Anterior Communicating Artery, Brain CoW Basilar Artery, Brain CoW Internal Carotid Artery (Left), Brain CoW Internal Carotid Artery (Right), Brain CoW Middle Cerebral Artery (Left), Brain CoW Middle Cerebral Artery (Right), Brain CoW Posterior Cerebral Artery (Left), Brain CoW Posterior Cerebral Artery (Right), Brain CoW Posterior Communicating Artery (Left), Brain CoW Posterior Communicating Artery (Right), Brain CoW Third A2 Artery} & 
        \TDcellNumImages{200}{130} & 
        \TDcellLinks{\href{https://topcow24.grand-challenge.org/topcow24/}{Official Challenge Website} \\\vspace{2pt} \href{https://arxiv.org/abs/2312.17670}{Preprint}} \\
        \TDrowDescription{The TopCoW dataset provides paired Magnetic Resonance Angiography (MRA) and Computed Tomography Angiography (CTA) scans. Initially launched as the TopCoW 2023 challenge, it focused on multi-class CoW vessel segmentation. The TopCoW 2024 edition significantly expands the dataset, increasing training data to 125 CTA/MRA pairs and doubling the online test set to 70 pairs with multi-center data. Labels for some 2023 data were updated for accuracy. The dataset includes 13 distinct vessel components of the CoW for segmentation. Originating from stroke patients at the University Hospital Zurich, scans were acquired using Siemens 1.5T or 3T MRI and various CT scanners.} \\
    }\\

    \TDformatRow{
        \TDcellName{TotalSegmentator}{TotalSegmentator}{\citep{doi:10.1148/radiol.241613,doi:10.1148/ryai.230024}} & 
        \TDcellRelDataset{-} & 
        \TDcellModality{3D CT / CT (CE),\\3D MRI} & 
        \TDcellAnatomyRegion{Whole Body}{Whole Body} & 
        \TDcellObjects{-}{[Too Many To List]} & 
        \TDcellNumImages{1526}{1437} & 
        \TDcellLinks{\href{https://github.com/wasserth/TotalSegmentator}{GitHub} \\\vspace{2pt} \href{https://totalsegmentator.com/}{Official Website} \\\vspace{2pt} \href{https://doi.org/10.1148/ryai.230024}{TotalSegmentator Publication} \\\vspace{2pt} \href{https://doi.org/10.1148/radiol.241613}{TotalSegmentator MRI Publication}} \\
        \TDrowDescription{TotalSegmentator is a series of publicly-available, whole-body CT and MRI datasets with comprehensively annotated anatomical structures. The evolution of TotalSegmentator has involved expansions in both modalities and annotation scope. The initial release in July 2022, TotalSegmentator (dubbed TotalSegmentator V1), introduced the largest publicly available CT segmentation dataset at the time. It comprised 1204 CT images, providing annotations for 104 distinct anatomical structures. These images were distributed as 1082 for training, 57 for validation, and 65 for testing. Subsequently, the TotalSegmentator MRI dataset was introduced. This dataset includes 298 MR images, offering segmentation annotations for up to 56 common anatomical structures. Of these, 251 MR images originate from routine clinical practice at the University Hospital Basel, while 47 images from the Imaging Data Commons (IDC) platform were included to enhance diversity. This MRI component accounts for various lesions, scanners, imaging sequences, and data from different medical institutions. An update to the CT dataset was released as TotalSegmentator V2 in September 2023, building upon the first version. This update increased the total number of CT images from 1204 to 1228, with the increment specifically in the test set, expanded from 65 to 89 images. The number of annotated categories also increased from 104 to 117. Here are reported the condensed statistics fro TotalSegmentator V2 and MRI. Cathegories are not reported as they are too many, please refer to the official websites and publications. Models benhmarked on TotalSegmentator usually provide Dice scores for the following categories of grouped classes: All (all labels), Cardiac, Muscles, Organs, Ribs, Vertebrae.} \\
    }\\

    \TDformatRow{
        \TDcellName{Touchstone}{Touchstone Benchmark}{\citep{NEURIPS2024_1b8726b5}} & 
        \TDcellRelDataset{AbdomenAtlas, TotalSegmentator} & 
        \TDcellModality{3D CT / CT (CE)} & 
        \TDcellAnatomyRegion{Abdominal Organs}{Abdomen} & 
        \TDcellObjects{9}{Aorta, Gallbladder, Inferior Vena Cava, Kidney (Left), Kidney (Right), Liver, Pancreas, Spleen, Stomach} & 
        \TDcellNumImages{6933}{0} & 
        \TDcellLinks{\href{https://github.com/MrGiovanni/Touchstone}{Official Website} \\\vspace{2pt} \href{https://proceedings.neurips.cc/paper\_files/paper/2024/hash/1b8726b572e0dfa72793f9f6590664fd-Abstract-Datasets\_and\_Benchmarks\_Track.html}{Publication}} \\
        \TDrowDescription{The Touchstone Benchmark is a collection of test images from 8 different hospitals used for testing thoracic, abdominal and pelvic organs segmentation algorithm  from CT images. The proposed training set is AbdomenAtlas, while the Touchstone Benchmark is a collection of volume-only test images. The Touchstone Benchmark is composed of two challenges: Touchstone 1.0 including 9 classes, the training set for which is the AbdomenAtlas 1.0 Mini, and the Touchstone 1.1 including all 25 classes, for which AbdomenAtlas 1.1 Mini should be used. The test sets are made of images from the publicly-available TotalSegmentator V2 and from a private dataset.  Currently, only Touchstone 1.0 leaderboards are available, for which 9 classes are considered.} \\
    }\\

    \TDformatRow{
        \TDcellName{ULS}{Universal Lesion Segmentation in Computed Tomography}{\citep{DEGRAUW2025103525}} & 
        \TDcellRelDataset{CCC18, DeepLesion, KiTS21, LIDC-IDRI, LiTS / MSD Liver, MSD Colon, MSD Lung, MSD Pancreas, NIH Lung Nodule} & 
        \TDcellModality{3D CT / CT (CE)} & 
        \TDcellAnatomyRegion{Tumors}{Abdomen, Thorax} & 
        \TDcellObjects{1}{Tumor} & 
        \TDcellNumImages{38824}{38824} & 
        \TDcellLinks{\href{https://uls23.grand-challenge.org/}{Official Challenge Website} \\\vspace{2pt} \href{https://doi.org/10.1016/j.media.2025.103525}{Publication}} \\
        \TDrowDescription{The ULS dataset was part of the ULS23 challenge and is a large-scale resource designed for lesion segmentation in chest and abdominal CT images. It compiles 6,514 fully annotated cases and 32,310 weakly annotated cases, with lesions centered in 256x256x128-sized Volumes of Interest (VOIs). ULS integrates several existing datasets (KiTS21, LIDC-IDRI, LiTS, MDS Task 6/7/10, NIH-LN, CCC18, DeepLesion) and introduces new data for skeletal lesions and extra pancreatic lesions, along with additional 3D annotations on some DeepLesion data.} \\
    }\\

    \TDformatRow{
        \TDcellName{VerSe}{Vertebrae Segmentation}{\citep{SEKUBOYINA2021102166}} & 
        \TDcellRelDataset{VerSe19, VerSe20} & 
        \TDcellModality{3D CT} & 
        \TDcellAnatomyRegion{Spine}{Abdomen, Neck, Pelvis, Thorax} & 
        \TDcellObjects{26}{Vertebra C1 (Primary Vertebra), Vertebra C2 (Secondary Vertebra), Vertebra C3 (Tertiary Vertebra), Vertebra C4 (Intervertebral), Vertebra C5 (Arch Root), Vertebra C6 (Small Joint), Vertebra C7 (Upper Joint), Vertebra L1 (First Sacral), Vertebra L2 (Second Sacral), Vertebra L3 (Third Sacral), Vertebra L4 (Fourth Sacral), Vertebra L5 (Fifth Sacral), Vertebra L6 (Sixth Sacral), Vertebra T1 (First Lumbar), Vertebra T10 (Tenth Lumbar), Vertebra T11 (Eleventh Lumbar), Vertebra T12 (Twelfth Lumbar), Vertebra T13 (Thirteenth Lumbar), Vertebra T2 (Second Lumbar), Vertebra T3 (Third Lumbar), Vertebra T4 (Fourth Lumbar), Vertebra T5 (Fifth Lumbar), Vertebra T6 (Sixth Lumbar), Vertebra T7 (Seventh Lumbar), Vertebra T8 (Eight Lumbar), Vertebra T9 (Ninth Lumbar)} & 
        \TDcellNumImages{374}{374} & 
        \TDcellLinks{\href{https://github.com/anjany/verse}{Official Website} \\\vspace{2pt} \href{https://doi.org/10.1016/j.media.2021.102166}{Publication}} \\
        \TDrowDescription{The VerSe dataset is a large-scale, multi-device, multi-center CT image spine segmentation dataset, formed by combining data from the MICCAI VerSe19 and VerSe20 challenges. It comprises 374 scans from 355 patients (accounting for 86 overlapping patients between the two original challenges). The dataset is divided into 141 scans for training, 120 for validation, and 113 for testing, with all scans and annotations publicly available. VerSe uniquely includes 26 vertebral annotation categories, encompassing the standard 24 vertebrae (C1-C7, T1-T12, L1-L5), plus the rarer T13 and L6 vertebrae. Partially visible vertebrae at scan edges were intentionally not annotated.} \\
    }\\

    \TDformatRow{
        \TDcellName{WMH}{White Matter Hyperintensity}{\citep{8669968}} & 
        \TDcellRelDataset{-} & 
        \TDcellModality{3D MRI (T1),\\3D MRI (T2-FLAIR)} & 
        \TDcellAnatomyRegion{Brain}{Head} & 
        \TDcellObjects{1}{White Matter Hypointensities} & 
        \TDcellNumImages{170}{60} & 
        \TDcellLinks{\href{https://doi.org/10.34894/AECRSD}{Official Challenge Website} \\\vspace{2pt} \href{https://doi.org/10.1109/TMI.2019.2905770}{Publication}} \\
        \TDrowDescription{The WMH dataset is a multimodal brain MRI dataset for the segmentation of white matter hyperintensities (WMH). WMHs are critical biomarkers of small vessel brain diseases and are key in assessing neurodegenerative conditions like dementia. The dataset includes 60 training cases sourced from various institutions and MRI scanners, each providing T1-weighted and FLAIR MRI sequences alongside expert manual annotations of WMHs. To ensure fair and valid evaluation of competing algorithms, an additional 110 hidden-label test cases from five different MRI scanners are included. While some data may contain annotations for other brain pathologies, these are specifically excluded from the WMH segmentation evaluation.} \\
    }\\

    \TDformatRow{
        \TDcellName{WORD}{Whole Abdominal Organ Dataset}{\citep{LIAO2023994,LUO2022102642}} & 
        \TDcellRelDataset{-} & 
        \TDcellModality{3D CT (CE)} & 
        \TDcellAnatomyRegion{Abdominal Organs}{Abdomen, Pelvis} & 
        \TDcellObjects{16}{Adrenal Glands, Bladder, Colon, Duodenum, Esophagus, Femur Head (Left), Femur Head (Right), Gallbladder, Intestine, Kidney (Left), Kidney (Right), Liver, Pancreas, Rectum, Spleen, Stomach} & 
        \TDcellNumImages{150}{150} & 
        \TDcellLinks{\href{https://github.com/HiLab-git/WORD}{GitHub} \\\vspace{2pt} \href{https://doi.org/10.1016/j.media.2022.102642}{Publication} \\\vspace{2pt} \href{https://doi.org/10.1016/j.ijrobp.2023.05.034}{Publication (2)}} \\
        \TDrowDescription{WORD is a large-scale CT dataset specifically designed for comprehensive abdominal organ segmentation. It features 150 CT scans that span the entire abdominal region, each meticulously annotated for 16 distinct abdominal organs. This dataset is officially split into 100 scans for training, 20 for validation, and 30 for testing, however all labels are provided. What sets WORD apart from other common abdominal organ segmentation datasets is its extensive coverage of intestinal categories, including detailed annotations for the colon, intestine, and rectum. Additionally, it uniquely includes annotations for the left and right femoral heads.} \\
    }\\
    
}{}

\hspace{-1.2cm}\appendixtextbox{
    \vspace{14pt}%
    Table~\ref{table:dataset-sources} lists online repositories or collections of publicly available datasets for 3D medical image segmentation and analysiss.
}

\newcommand{\selectTableDORFont}{\fontfamily{cmss}\fontsize{8}{9}\selectfont}


\newcommand{\TDORncols}{2}

\newcommand{\TDORcwName}{5.0cm}
\newcommand{\TDORcwLink}{3.5cm}


{
    \captionsetup{width=11cm}
    \selectTableDORFont
    \rowcolors{2}{CornflowerBlue!1}{CornflowerBlue!10}
    \begin{center}
    \begin{longtable}[c]{
        >{\raggedright}p{\TDORcwName} 
        p{\TDORcwLink} 
    }
        \hiderowcolors
        \caption{%
            Collection of public repositories or articles aggregating 3D medical image datasets. This collection can be used to scout for datasets and gathers the efforts of the whole research community in one place.
        }
        \label{table:dataset-sources} \\
        \showrowcolors
    
        \hline
        \rowcolor{CornflowerBlue!20}
        Name & Link \\
        \hline
        \endfirsthead
        
        \makeTableHeaderContinued{\TDORncols}
        \hline
        \rowcolor{CornflowerBlue!20}
        Name & Link \\
        \hline
        \endhead
        
        \makeTableOtherFooters{\TDORncols}
        \endfoot
        
        \makeTableLastFooter{\TDORncols}{}
        \endlastfoot
    
        CLIP-Driven Universal Model & \href{https://github.com/ljwztc/CLIP-Driven-Universal-Model/blob/main/documents/awesome.md}{GitHub} \\
        SAT-DS & \href{https://github.com/zhaoziheng/SAT-DS}{GitHub} \\
        TotalSegmentator & \href{https://github.com/wasserth/TotalSegmentator}{GitHub} \\
        AbdomenAtlas & \href{https://github.com/MrGiovanni/AbdomenAtlas}{GitHub} \\
        IMIS-Benchmark & \href{https://github.com/uni-medical/IMIS-Bench}{GitHUb} \\
        M3D & \href{https://github.com/BAAI-DCAI/M3D}{GitHub} \\
        BiomedParseData & \href{https://huggingface.co/datasets/microsoft/BiomedParseData}{Hugging Face} \\
        OpenMEDLab (Awesome-Medical-Dataset) & \href{https://github.com/openmedlab/Awesome-Medical-Dataset}{GitHub} \\
        Human Heart Project & \href{https://humanheart-project.creatis.insa-lyon.fr/database/}{Website}\\
        SA-Med3D-140K & \href{https://github.com/uni-medical/SAM-Med3D}{GitHub} \\
        MedSAM Dataset List & \href{https://medsam-datasetlist.github.io/}{GitHub} \\
    
    \end{longtable}
    \end{center}
}

\restoregeometry

\subsection{Performances by target anatomies}
\label{app:subsec:target-anatomies}

\def\tableResultsOverviewFont{\fontfamily{cmss}\fontsize{9}{10}}

\definecolor{colorTableResultsOverviewSymbolHColor}{HTML}{de8a23}
\def\tableResultsOverviewH{\textcolor{colorTableResultsOverviewSymbolHColor}{\bfseries H~}}
\definecolor{colorTableResultsOverviewSymbolVColor}{HTML}{80d9a8}
\def\tableResultsOverviewV{\textcolor{colorTableResultsOverviewSymbolVColor}{\bfseries V~}}

\newcommand{\selectResultsSpecificFontCaption}{\rmfamily\fontsize{10}{11}\selectfont}
\newcommand{\selectResultsSpecificFontTable}{\fontfamily{cmss}\fontsize{8}{11}\selectfont}

\newcommand{\ncolumnsResultsSpecific}{6}
\newcommand{\colspecResultsSpecific}{p{1.5cm} l l l l p{3cm}}


\newcommand{\makeTableResultsSpecificFirstHeading}[1]{
    \hline 
    \rowcolor{black!10}
    \multicolumn{\ncolumnsResultsSpecific}{c}{
        \textcolor{colorTableResultsOverviewBest!40!black}{\textbf{#1}}
    } \\
    \rowcolor{black!10}
    Benchmark & N. & Min. &{\fontsize{7.5}{7.5}\selectfont Median} & Max. & Top 5 overall \\
    \hline
}

\newcommand{\makeTableResultsSpecificOtherHeading}[1]{
    \makeTableHeaderContinued{\ncolumnsResultsSpecific}
    \makeTableResultsSpecificFirstHeading{#1}
}

\newcommand{\primaryLine}{
    \rowcolor{colorTableResultsOverviewPrimary}
    \multicolumn{\ncolumnsResultsSpecific}{l}{
        ~~~~\textbf{Primary}
    } \\
}

\newcommand{\bestInLiteratureLine}{
    \rowcolor{colorTableResultsOverviewBest}
    \multicolumn{\ncolumnsResultsSpecific}{l}{
        ~~~~\textbf{Best-in-literature}
    } \\
}



\newcommand{\makeColumnThreeLT}[3]{
    \makecell[lt]{
         #1 \\
         #2 \\
         #3
    }

}
\newcommand{\makeAHV}[3]{
    \makeColumnThreeLT{
        \textbf{A} \texttt{#1}
    }{
        \tableResultsOverviewH \texttt{#2}
    }{
        \tableResultsOverviewV \texttt{#3}
    }
}

\newcommand{\makeTRSline}[7]{
    \rowcolor{#1}
        #2 &
        #3 &
        #4 &
        #5 &
        #6 &
        \vspace{-6pt}
        \begin{itemize}[
            noitemsep,
            leftmargin=*,
            topsep=0pt, partopsep=0pt, parsep=0pt,
            labelsep=0pt,
            labelindent=20pt
        ]
            #7
        \end{itemize}
        
        \\
}
\newcommand{\makeTRSlinePrimary}[6]{
    \makeTRSline{colorTableResultsOverviewPrimary!75}{#1}{#2}{#3}{#4}{#5}{#6}
}
\newcommand{\makeTRSlineBest}[6]{
    \makeTRSline{colorTableResultsOverviewBest!75}{#1}{#2}{#3}{#4}{#5}{#6}
}

\newcommand{\makeTableResultsSpecific}[5]{
    {
    \selectResultsSpecificFontTable
    \begin{center}
    \begin{longtable}[c]{p{2.25cm} l l l l >{\raggedright\arraybackslash}p{4cm}}
        \hiderowcolors
        \caption{
            #1
        }
        \label{#2} \\
        \showrowcolors
    
        \makeTableResultsSpecificFirstHeading{#3}
        \endfirsthead
        
        \makeTableResultsSpecificOtherHeading{#3}
        \endhead
        
        \makeTableOtherFooters{\ncolumnsResultsSpecific}
        \endfoot
    
        \makeTableLastFooter{\ncolumnsResultsSpecific}{}
        \endlastfoot
    
    \primaryLine
    #4
    \bestInLiteratureLine
    #5
    \end{longtable}  
    \end{center}
    }
}

Here are reported statistics about the performance scores that the considered models obtained on different datasets, grouped by target anatomical region.

\subsubsection{Brain}
\label{subsubsec:res-brain}

\begin{figure}[h!]
    \centering
    \includegraphics[width=8cm]{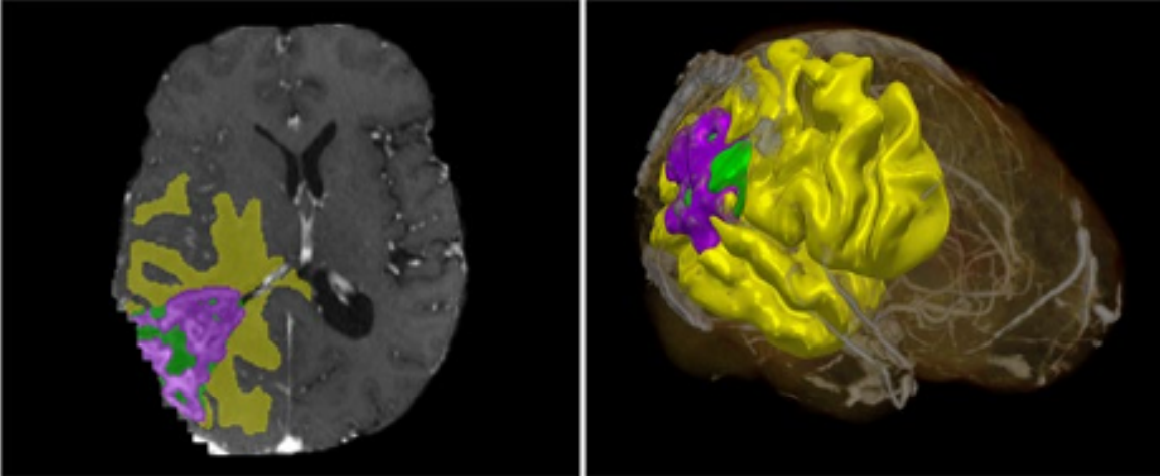}
    \caption{Brain tumor segmentation example. Courtesy of nnU-Net \citep{nnunet_nature} (on BraTS).}
    \label{fig:res-brain-tumor-example}
\end{figure}

\makeTableResultsSpecific{
    Results overview for brain datasets.
}{table:results-overview-brain}{
    Brain
}{
    \makeTRSlinePrimary{
            ATLAS v2.0
        }{
            \makeAHV{~~2}{~~2}{~~0}
        }{
            \makeColumnThreeLT{62.03}{62.03}{-}
        }{
            \makeColumnThreeLT{66.62}{66.62}{-}
        }{
            \makeColumnThreeLT{71.20}{71.20}{-}
        }{
            \item[\texttt{1.} \tableResultsOverviewH ] BrainSegFounder (71.20)  
            \item[\texttt{2.} \tableResultsOverviewH ] MoME (62.03)
        }
    \makeTRSlinePrimary{
            BraTS
        }{
            \makeAHV{~24}{~12}{~12}
        }{
            \makeColumnThreeLT{55.68}{55.68}{61.00}
        }{
            \makeColumnThreeLT{85.66}{85.69}{85.54}
        }{
            \makeColumnThreeLT{92.08}{92.08}{91.74}
        }{
            \item[\texttt{1.} \tableResultsOverviewH ] MEA M-SAM (92.08)  
            \item[\texttt{2.} \tableResultsOverviewV ] TransUNet (91.74)  
            \item[\texttt{3.} \tableResultsOverviewH ] BrainSegFounder (91.15)  
            \item[\texttt{4.} \tableResultsOverviewH ] EMedSAM (89.30)  
            \item[\texttt{5.} \tableResultsOverviewH ] Med-SA (89.10)
        }
    \makeTRSlinePrimary{
            DLBS
        }{
            \makeAHV{~~1}{~~1}{~~0}
        }{
            \makeColumnThreeLT{96.54}{96.54}{-}
        }{
            \makeColumnThreeLT{96.54}{96.54}{-}
        }{
            \makeColumnThreeLT{96.54}{96.54}{-}
        }{
            \item[\texttt{1.} \tableResultsOverviewH ] HERMES (96.54)
        }
    \makeTRSlinePrimary{
            FeTA
        }{
            \makeAHV{~~3}{~~2}{~~1}
        }{
            \makeColumnThreeLT{76.24}{76.24}{87.40}
        }{
            \makeColumnThreeLT{84.20}{80.22}{87.40}
        }{
            \makeColumnThreeLT{87.40}{84.20}{87.40}
        }{
            \item[\texttt{1.} \tableResultsOverviewV ] 3D UX-Net (87.40)  
            \item[\texttt{2.} \tableResultsOverviewH ] Disruptive Autoencoders (84.20)  
            \item[\texttt{3.} \tableResultsOverviewH ] SAT (76.24)
        }
    \makeTRSlinePrimary{
            ISLES
        }{
            \makeAHV{~~2}{~~2}{~~0}
        }{
            \makeColumnThreeLT{71.78}{71.78}{-}
        }{
            \makeColumnThreeLT{74.50}{74.50}{-}
        }{
            \makeColumnThreeLT{77.23}{77.23}{-}
        }{
            \item[\texttt{1.} \tableResultsOverviewH ] MoME (77.23)  
            \item[\texttt{2.} \tableResultsOverviewH ] IMIS-Net (71.78)
        }
    \makeTRSlinePrimary{
            MSD Hippocampus
        }{
            \makeAHV{~~3}{~~2}{~~1}
        }{
            \makeColumnThreeLT{82.40}{82.40}{89.50}
        }{
            \makeColumnThreeLT{87.62}{85.01}{89.50}
        }{
            \makeColumnThreeLT{89.50}{87.62}{89.50}
        }{
            \item[\texttt{1.} \tableResultsOverviewV ] nnU-Net (89.50)  
            \item[\texttt{2.} \tableResultsOverviewH ] SAT (87.62)  
            \item[\texttt{3.} \tableResultsOverviewH ] BiomedParse (82.40)
        }
    \makeTRSlinePrimary{
            MSSEG
        }{
            \makeAHV{~~1}{~~1}{~~0}
        }{
            \makeColumnThreeLT{56.26}{56.26}{-}
        }{
            \makeColumnThreeLT{56.26}{56.26}{-}
        }{
            \makeColumnThreeLT{56.26}{56.26}{-}
        }{
            \item[\texttt{1.} \tableResultsOverviewH ] MoME (56.26)
        }
    \makeTRSlinePrimary{
            OASIS-1
        }{
            \makeAHV{~~1}{~~1}{~~0}
        }{
            \makeColumnThreeLT{68.58}{68.58}{-}
        }{
            \makeColumnThreeLT{68.58}{68.58}{-}
        }{
            \makeColumnThreeLT{68.58}{68.58}{-}
        }{
            \item[\texttt{1.} \tableResultsOverviewH ] MoME (68.58)
        }
    \makeTRSlinePrimary{
            OASIS-3
        }{
            \makeAHV{~~1}{~~0}{~~1}
        }{
            \makeColumnThreeLT{74.19}{-}{74.19}
        }{
            \makeColumnThreeLT{74.19}{-}{74.19}
        }{
            \makeColumnThreeLT{74.19}{-}{74.19}
        }{
            \item[\texttt{1.} \tableResultsOverviewV ] NexToU (74.19)
        }
    \makeTRSlinePrimary{
            WMH
        }{
            \makeAHV{~~1}{~~1}{~~0}
        }{
            \makeColumnThreeLT{63.41}{63.41}{-}
        }{
            \makeColumnThreeLT{63.41}{63.41}{-}
        }{
            \makeColumnThreeLT{63.41}{63.41}{-}
        }{
            \item[\texttt{1.} \tableResultsOverviewH ] MoME (63.41)
        }
    
}{
    \makeTRSlineBest{
            ATLAS v2.0
        }{
            \makeAHV{~~6}{~~5}{~~1}
        }{
            \makeColumnThreeLT{33.54}{33.54}{56.45}
        }{
            \makeColumnThreeLT{56.82}{57.19}{56.45}
        }{
            \makeColumnThreeLT{71.20}{71.20}{56.45}
        }{
            \item[\texttt{1.} \tableResultsOverviewH ] BrainSegFounder (71.20)  
            \item[\texttt{2.} \tableResultsOverviewH ] MoME (62.03)  
            \item[\texttt{3.} \tableResultsOverviewH ] HERMES (57.19)  
            \item[\texttt{4.} \tableResultsOverviewV ] nnU-Net (56.45)  
            \item[\texttt{5.} \tableResultsOverviewH ] MultiTalent (55.61)
        }
    \makeTRSlineBest{
            BraTS
        }{
            \makeAHV{~42}{~23}{~19}
        }{
            \makeColumnThreeLT{28.91}{28.91}{63.90}
        }{
            \makeColumnThreeLT{85.69}{84.95}{86.05}
        }{
            \makeColumnThreeLT{92.08}{92.08}{91.74}
        }{
            \item[\texttt{1.} \tableResultsOverviewH ] MEA M-SAM (92.08)  
            \item[\texttt{2.} \tableResultsOverviewV ] TransUNet (91.74)  
            \item[\texttt{3.} \tableResultsOverviewV ] nnU-Net (91.23)  
            \item[\texttt{4.} \tableResultsOverviewH ] BrainSegFounder (91.15)  
            \item[\texttt{5.} \tableResultsOverviewV ] TransBTS (90.66)
        }
    \makeTRSlineBest{
            DLBS
        }{
            \makeAHV{~~7}{~~3}{~~4}
        }{
            \makeColumnThreeLT{86.81}{86.81}{92.01}
        }{
            \makeColumnThreeLT{94.31}{95.35}{94.27}
        }{
            \makeColumnThreeLT{96.54}{96.54}{95.13}
        }{
            \item[\texttt{1.} \tableResultsOverviewH ] HERMES (96.54)  
            \item[\texttt{2.} \tableResultsOverviewH ] UniMiSS (95.35)  
            \item[\texttt{3.} \tableResultsOverviewV ] MedFormer (95.13)  
            \item[\texttt{4.} \tableResultsOverviewV ] SegResNet (94.31)  
            \item[\texttt{5.} \tableResultsOverviewV ] nnU-Net (94.22)
        }
    \makeTRSlineBest{
            FeTA
        }{
            \makeAHV{~10}{~~3}{~~7}
        }{
            \makeColumnThreeLT{54.35}{54.35}{85.70}
        }{
            \makeColumnThreeLT{86.15}{76.24}{86.70}
        }{
            \makeColumnThreeLT{87.40}{84.20}{87.40}
        }{
            \item[\texttt{1.} \tableResultsOverviewV ] 3D UX-Net (87.40)  
            \item[\texttt{2.} \tableResultsOverviewV ] nnU-Net (87.00)  
            \item[\texttt{3.} \tableResultsOverviewV ] TransBTS (86.80)  
            \item[\texttt{4.} \tableResultsOverviewV ] SwinUNETR (86.70)  
            \item[\texttt{5.} \tableResultsOverviewV ] nnFormer (86.30)
        }
    \makeTRSlineBest{
            ISLES
        }{
            \makeAHV{~10}{~~9}{~~1}
        }{
            \makeColumnThreeLT{55.92}{55.92}{78.24}
        }{
            \makeColumnThreeLT{70.00}{68.22}{78.24}
        }{
            \makeColumnThreeLT{78.24}{77.23}{78.24}
        }{
            \item[\texttt{1.} \tableResultsOverviewV ] nnU-Net (78.24)  
            \item[\texttt{2.} \tableResultsOverviewH ] MoME (77.23)  
            \item[\texttt{3.} \tableResultsOverviewH ] MultiTalent (75.10)  
            \item[\texttt{4.} \tableResultsOverviewH ] HERMES (73.68)  
            \item[\texttt{5.} \tableResultsOverviewH ] IMIS-Net (71.78)
        }
    \makeTRSlineBest{
            MSD Hippocampus
        }{
            \makeAHV{~~6}{~~4}{~~2}
        }{
            \makeColumnThreeLT{76.81}{76.81}{89.03}
        }{
            \makeColumnThreeLT{85.01}{81.40}{89.27}
        }{
            \makeColumnThreeLT{89.50}{87.62}{89.50}
        }{
            \item[\texttt{1.} \tableResultsOverviewV ] nnU-Net (89.50)  
            \item[\texttt{2.} \tableResultsOverviewV ] SwinUNETR (89.03)  
            \item[\texttt{3.} \tableResultsOverviewH ] SAT (87.62)  
            \item[\texttt{4.} \tableResultsOverviewH ] BiomedParse (82.40)  
            \item[\texttt{5.} \tableResultsOverviewH ] MedSAM (80.39)
        }
    \makeTRSlineBest{
            MSSEG
        }{
            \makeAHV{~~5}{~~4}{~~1}
        }{
            \makeColumnThreeLT{27.44}{27.44}{58.07}
        }{
            \makeColumnThreeLT{55.60}{53.62}{58.07}
        }{
            \makeColumnThreeLT{58.07}{56.26}{58.07}
        }{
            \item[\texttt{1.} \tableResultsOverviewV ] nnU-Net (58.07)  
            \item[\texttt{2.} \tableResultsOverviewH ] MoME (56.26)  
            \item[\texttt{3.} \tableResultsOverviewH ] MultiTalent (55.60)  
            \item[\texttt{4.} \tableResultsOverviewH ] HERMES (51.64)  
            \item[\texttt{5.} \tableResultsOverviewH ] SAM-Med3D (27.44)
        }
    \makeTRSlineBest{
            OASIS-1
        }{
            \makeAHV{~~5}{~~4}{~~1}
        }{
            \makeColumnThreeLT{50.53}{50.53}{69.99}
        }{
            \makeColumnThreeLT{66.81}{65.62}{69.99}
        }{
            \makeColumnThreeLT{69.99}{68.58}{69.99}
        }{
            \item[\texttt{1.} \tableResultsOverviewV ] nnU-Net (69.99)  
            \item[\texttt{2.} \tableResultsOverviewH ] MoME (68.58)  
            \item[\texttt{3.} \tableResultsOverviewH ] MultiTalent (66.81)  
            \item[\texttt{4.} \tableResultsOverviewH ] HERMES (64.43)  
            \item[\texttt{5.} \tableResultsOverviewH ] SAM-Med3D (50.53)
        }
    \makeTRSlineBest{
            OASIS-3
        }{
            \makeAHV{~~3}{~~0}{~~3}
        }{
            \makeColumnThreeLT{72.66}{-}{72.66}
        }{
            \makeColumnThreeLT{73.59}{-}{73.59}
        }{
            \makeColumnThreeLT{74.19}{-}{74.19}
        }{
            \item[\texttt{1.} \tableResultsOverviewV ] NexToU (74.19)  
            \item[\texttt{2.} \tableResultsOverviewV ] nnU-Net (73.59)  
            \item[\texttt{3.} \tableResultsOverviewV ] nnFormer (72.66)
        }
    \makeTRSlineBest{
            WMH
        }{
            \makeAHV{~~5}{~~4}{~~1}
        }{
            \makeColumnThreeLT{44.04}{44.04}{65.09}
        }{
            \makeColumnThreeLT{61.17}{61.11}{65.09}
        }{
            \makeColumnThreeLT{65.09}{63.41}{65.09}
        }{
            \item[\texttt{1.} \tableResultsOverviewV ] nnU-Net (65.09)  
            \item[\texttt{2.} \tableResultsOverviewH ] MoME (63.41)  
            \item[\texttt{3.} \tableResultsOverviewH ] HERMES (61.17)  
            \item[\texttt{4.} \tableResultsOverviewH ] MultiTalent (61.04)  
            \item[\texttt{5.} \tableResultsOverviewH ] SAM-Med3D (44.04)
        }
    
}

\textit{Winners:} Generalist models on most datasets, while task-specific on FeTA2021, MSD Ippocampus, and OASIS-3 datasets (median Dice score from primary research). Task specific on eight out of 10 datasets (BraTS, FeTA2021, ISLES, MSD Ippocampus, Multiple Sceloris Lesions, OASIS-1, OASIS-3, WMH) for median Dice score from best in literature.

\clearpage
\subsubsection{Head and Neck}
\label{subsubsec:res-head-neck}

\begin{figure}[h!]
    \centering
    \includegraphics[width=8cm]{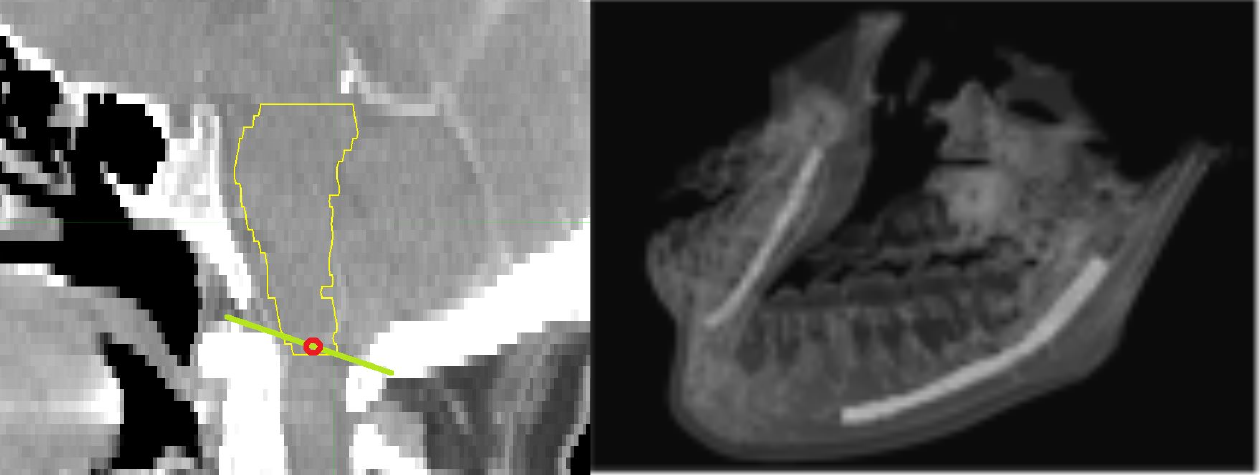}
    \caption{Brain stem (left) and Tooth Fairy (right) segmentation examples. Courtesy of Head and Neck Auto Segmentation Challenge \citep{https://doi.org/10.1002/mp.12197} and ToothFairy \citep{2024TMI}.}
    \label{fig:res-...-example}
\end{figure}

\makeTableResultsSpecific{
    Results overview for head and neck datasets.
}{table:results-overview-head-and-neck}{
    Head and Neck
}{
    \makeTRSlinePrimary{
            HNASC
        }{
            \makeAHV{~~1}{~~1}{~~0}
        }{
            \makeColumnThreeLT{82.74}{82.74}{-}
        }{
            \makeColumnThreeLT{82.74}{82.74}{-}
        }{
            \makeColumnThreeLT{82.74}{82.74}{-}
        }{
            \item[\texttt{1.} \tableResultsOverviewH ] MIS-FM (82.74)
        }
    \makeTRSlinePrimary{
            ToothFairy
        }{
            \makeAHV{~~4}{~~4}{~~0}
        }{
            \makeColumnThreeLT{61.40}{61.40}{-}
        }{
            \makeColumnThreeLT{76.73}{76.73}{-}
        }{
            \makeColumnThreeLT{80.80}{80.80}{-}
        }{
            \item[\texttt{1.} \tableResultsOverviewH ] Medical SAM 2 (MedSAM-2) (80.80)  
            \item[\texttt{2.} \tableResultsOverviewH ] SAT (78.17)  
            \item[\texttt{3.} \tableResultsOverviewH ] SAM-Med2D (75.29)  
            \item[\texttt{4.} \tableResultsOverviewH ] One-Prompt (61.40)
        }
    
}{
    \makeTRSlineBest{
            HNASC
        }{
            \makeAHV{~~5}{~~1}{~~4}
        }{
            \makeColumnThreeLT{69.10}{82.74}{69.10}
        }{
            \makeColumnThreeLT{78.66}{82.74}{74.47}
        }{
            \makeColumnThreeLT{82.74}{82.74}{80.37}
        }{
            \item[\texttt{1.} \tableResultsOverviewH ] MIS-FM (82.74)  
            \item[\texttt{2.} \tableResultsOverviewV ] UNETR++ (80.37)  
            \item[\texttt{3.} \tableResultsOverviewV ] nnU-Net (78.66)  
            \item[\texttt{4.} \tableResultsOverviewV ] nnFormer (70.27)  
            \item[\texttt{5.} \tableResultsOverviewV ] TransUNet (69.10)
        }
    \makeTRSlineBest{
            ToothFairy
        }{
            \makeAHV{~12}{~~9}{~~3}
        }{
            \makeColumnThreeLT{37.90}{47.60}{37.90}
        }{
            \makeColumnThreeLT{70.58}{65.86}{79.85}
        }{
            \makeColumnThreeLT{83.08}{80.80}{83.08}
        }{
            \item[\texttt{1.} \tableResultsOverviewV ] nnU-Net (83.08)  
            \item[\texttt{2.} \tableResultsOverviewH ] Medical SAM 2 (MedSAM-2) (80.80)  
            \item[\texttt{3.} \tableResultsOverviewV ] SwinUNETR (79.85)  
            \item[\texttt{4.} \tableResultsOverviewH ] SAT (78.17)  
            \item[\texttt{5.} \tableResultsOverviewH ] One-Prompt (76.40)
        }
    
}

\textit{Winners:} Task-specific (on primary research). Tie on best in literature with generalist models obtaining a higher median Dice score on Head and Neck Dataset, while task-specific on ToothFairy dataset.

\clearpage
\subsubsection{Lungs}
\label{res-lung}

\begin{figure}[h!]
    \centering
    \includegraphics[width=8cm]{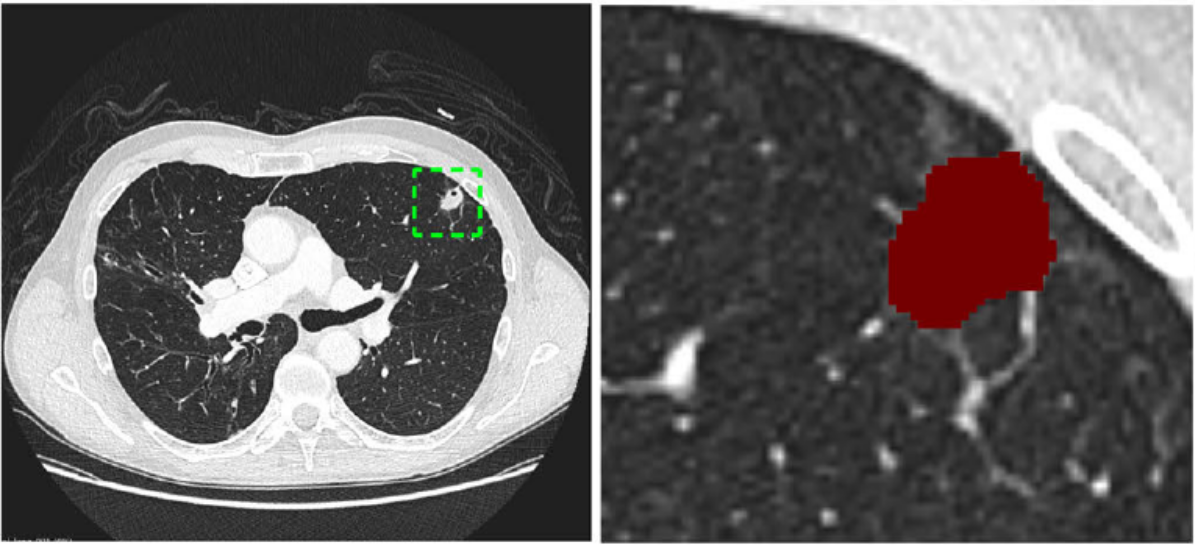}
    \caption{Lung nodule segmentation example. Courtesy of UNETR++ \cite{unetr_pp_ieee} (on MSD Lung Tumors).}
    \label{fig:res-msd-lung-example}
\end{figure}

\makeTableResultsSpecific{
    Results overview for lungs datasets.
}{table:results-overview-lungs}{
    Lungs
}{
    \makeTRSlinePrimary{
            LIDC-IDRI
        }{
            \makeAHV{~~2}{~~1}{~~1}
        }{
            \makeColumnThreeLT{77.05}{92.87}{77.05}
        }{
            \makeColumnThreeLT{84.96}{92.87}{77.05}
        }{
            \makeColumnThreeLT{92.87}{92.87}{77.05}
        }{
            \item[\texttt{1.} \tableResultsOverviewH ] Med3D (92.87)  
            \item[\texttt{2.} \tableResultsOverviewV ] UNet++ (77.05)
        }
    \makeTRSlinePrimary{
            LUNA16
        }{
            \makeAHV{~~1}{~~1}{~~0}
        }{
            \makeColumnThreeLT{97.16}{97.16}{-}
        }{
            \makeColumnThreeLT{97.16}{97.16}{-}
        }{
            \makeColumnThreeLT{97.16}{97.16}{-}
        }{
            \item[\texttt{1.} \tableResultsOverviewH ] SAT (97.16)
        }
    \makeTRSlinePrimary{
            MSD Lung Tumors
        }{
            \makeAHV{~14}{~~9}{~~5}
        }{
            \makeColumnThreeLT{56.72}{61.28}{56.72}
        }{
            \makeColumnThreeLT{72.06}{71.42}{74.00}
        }{
            \makeColumnThreeLT{81.62}{81.62}{80.68}
        }{
            \item[\texttt{1.} \tableResultsOverviewH ] MEA M-SAM (81.62)  
            \item[\texttt{2.} \tableResultsOverviewV ] UNETR++ (80.68)  
            \item[\texttt{3.} \tableResultsOverviewH ] CLIP-Driven Universal Model (80.01)  
            \item[\texttt{4.} \tableResultsOverviewH ] LeSAM (79.57)  
            \item[\texttt{5.} \tableResultsOverviewV ] MedFormer (74.00)
        }
    
}{
    \makeTRSlineBest{
            LIDC-IDRI
        }{
            \makeAHV{~~4}{~~1}{~~3}
        }{
            \makeColumnThreeLT{71.17}{92.87}{71.17}
        }{
            \makeColumnThreeLT{74.11}{92.87}{71.17}
        }{
            \makeColumnThreeLT{92.87}{92.87}{77.05}
        }{
            \item[\texttt{1.} \tableResultsOverviewH ] Med3D (92.87)  
            \item[\texttt{2.} \tableResultsOverviewV ] UNet++ (77.05)  
            \item[\texttt{3.} \tableResultsOverviewV ] V-Net (71.17)  
            \item[\texttt{4.} \tableResultsOverviewV ] U-Net (71.17)
        }
    \makeTRSlineBest{
            LUNA16
        }{
            \makeAHV{~~3}{~~1}{~~2}
        }{
            \makeColumnThreeLT{95.64}{97.16}{95.64}
        }{
            \makeColumnThreeLT{96.64}{97.16}{96.14}
        }{
            \makeColumnThreeLT{97.16}{97.16}{96.64}
        }{
            \item[\texttt{1.} \tableResultsOverviewH ] SAT (97.16)  
            \item[\texttt{2.} \tableResultsOverviewV ] nnU-Net (96.64)  
            \item[\texttt{3.} \tableResultsOverviewV ] SwinUNETR (95.64)
        }
    \makeTRSlineBest{
            MSD Lung Tumors
        }{
            \makeAHV{~27}{~15}{~12}
        }{
            \makeColumnThreeLT{59.99}{61.28}{59.99}
        }{
            \makeColumnThreeLT{74.31}{72.70}{74.76}
        }{
            \makeColumnThreeLT{81.62}{81.62}{80.68}
        }{
            \item[\texttt{1.} \tableResultsOverviewH ] MEA M-SAM (81.62)  
            \item[\texttt{2.} \tableResultsOverviewV ] UNETR++ (80.68)  
            \item[\texttt{3.} \tableResultsOverviewV ] MedNeXt (80.14)  
            \item[\texttt{4.} \tableResultsOverviewH ] CLIP-Driven Universal Model (80.01)  
            \item[\texttt{5.} \tableResultsOverviewH ] LeSAM (79.57)
        }
    
}

\textit{Winners:} Generalist models on both primary research and best in literature.

\clearpage
\subsubsection{Heart and thoracic vessels}
\label{subsubsec:res-heart}

\begin{figure}[h!]
    \centering
    \includegraphics[width=8cm]{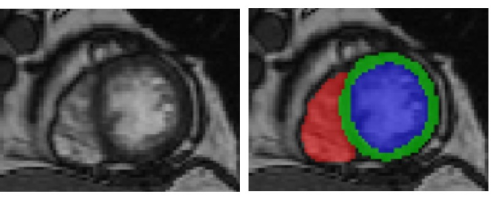}
    \caption{Heart and thoracic vasculature example. Courtesy of SCANeXt \citep{scanext} (on ACDC).}
    \label{fig:res-heart-example}
\end{figure}

\makeTableResultsSpecific{
    Results overview for heart and thoracic vasculature datasets.
}{table:results-overview-heart-and-thoracic-vasculature}{
    Heart and Thoracic Vasculature
}{
    \makeTRSlinePrimary{
            ACDC
        }{
            \makeAHV{~13}{~~5}{~~8}
        }{
            \makeColumnThreeLT{70.90}{70.90}{90.00}
        }{
            \makeColumnThreeLT{92.06}{90.41}{92.58}
        }{
            \makeColumnThreeLT{95.18}{92.26}{95.18}
        }{
            \item[\texttt{1.} \tableResultsOverviewV ] SCANeXt (95.18)  
            \item[\texttt{2.} \tableResultsOverviewV ] nnU-Net (92.95)  
            \item[\texttt{3.} \tableResultsOverviewV ] UNETR++ (92.83)  
            \item[\texttt{4.} \tableResultsOverviewV ] LHU-Net (92.66)  
            \item[\texttt{5.} \tableResultsOverviewV ] MedFormer (92.50)
        }
    \makeTRSlinePrimary{
            LASC
        }{
            \makeAHV{~~2}{~~1}{~~1}
        }{
            \makeColumnThreeLT{91.00}{91.00}{91.55}
        }{
            \makeColumnThreeLT{91.28}{91.00}{91.55}
        }{
            \makeColumnThreeLT{91.55}{91.00}{91.55}
        }{
            \item[\texttt{1.} \tableResultsOverviewV ] LHU-Net (91.55)  
            \item[\texttt{2.} \tableResultsOverviewH ] SFR SAM (91.00)
        }
    \makeTRSlinePrimary{
            M\&Ms
        }{
            \makeAHV{~~1}{~~1}{~~0}
        }{
            \makeColumnThreeLT{87.02}{87.02}{-}
        }{
            \makeColumnThreeLT{87.02}{87.02}{-}
        }{
            \makeColumnThreeLT{87.02}{87.02}{-}
        }{
            \item[\texttt{1.} \tableResultsOverviewH ] HERMES (87.02)
        }
    \makeTRSlinePrimary{
            MM-WHS
        }{
            \makeAHV{~~1}{~~1}{~~0}
        }{
            \makeColumnThreeLT{89.44}{89.44}{-}
        }{
            \makeColumnThreeLT{89.44}{89.44}{-}
        }{
            \makeColumnThreeLT{89.44}{89.44}{-}
        }{
            \item[\texttt{1.} \tableResultsOverviewH ] SAT (89.44)
        }
    \makeTRSlinePrimary{
            MSD Cardiac
        }{
            \makeAHV{~~3}{~~2}{~~1}
        }{
            \makeColumnThreeLT{89.86}{89.86}{93.00}
        }{
            \makeColumnThreeLT{93.00}{91.62}{93.00}
        }{
            \makeColumnThreeLT{93.38}{93.38}{93.00}
        }{
            \item[\texttt{1.} \tableResultsOverviewH ] SAT (93.38)  
            \item[\texttt{2.} \tableResultsOverviewV ] nnU-Net (93.00)  
            \item[\texttt{3.} \tableResultsOverviewH ] BiomedParse (89.86)
        }
    \makeTRSlinePrimary{
            TotalSegmentator Cardiac
        }{
            \makeAHV{~~4}{~~4}{~~0}
        }{
            \makeColumnThreeLT{89.57}{89.57}{-}
        }{
            \makeColumnThreeLT{91.27}{91.27}{-}
        }{
            \makeColumnThreeLT{92.52}{92.52}{-}
        }{
            \item[\texttt{1.} \tableResultsOverviewH ] SAT (92.52)  
            \item[\texttt{2.} \tableResultsOverviewH ] PCNet (91.64)  
            \item[\texttt{3.} \tableResultsOverviewH ] STU-Net (90.89)  
            \item[\texttt{4.} \tableResultsOverviewH ] CLIP-Driven Universal Model (89.57)
        }
    
}{
    \makeTRSlineBest{
            ACDC
        }{
            \makeAHV{~22}{~~7}{~15}
        }{
            \makeColumnThreeLT{68.86}{68.86}{84.07}
        }{
            \makeColumnThreeLT{90.74}{89.64}{91.19}
        }{
            \makeColumnThreeLT{95.18}{92.26}{95.18}
        }{
            \item[\texttt{1.} \tableResultsOverviewV ] SCANeXt (95.18)  
            \item[\texttt{2.} \tableResultsOverviewV ] nnU-Net (92.95)  
            \item[\texttt{3.} \tableResultsOverviewV ] UNETR++ (92.83)  
            \item[\texttt{4.} \tableResultsOverviewV ] LHU-Net (92.66)  
            \item[\texttt{5.} \tableResultsOverviewV ] MedFormer (92.50)
        }
    \makeTRSlineBest{
            LASC
        }{
            \makeAHV{~12}{~~5}{~~7}
        }{
            \makeColumnThreeLT{31.13}{31.13}{88.25}
        }{
            \makeColumnThreeLT{89.90}{57.33}{91.14}
        }{
            \makeColumnThreeLT{91.55}{91.00}{91.55}
        }{
            \item[\texttt{1.} \tableResultsOverviewV ] LHU-Net (91.55)  
            \item[\texttt{2.} \tableResultsOverviewV ] SwinUNETR-V2 (91.48)  
            \item[\texttt{3.} \tableResultsOverviewV ] SwinUNETR (91.20)  
            \item[\texttt{4.} \tableResultsOverviewV ] V-Net (91.14)  
            \item[\texttt{5.} \tableResultsOverviewV ] UNETR++ (91.05)
        }
    \makeTRSlineBest{
            M\&Ms
        }{
            \makeAHV{~~7}{~~3}{~~4}
        }{
            \makeColumnThreeLT{83.28}{85.75}{83.28}
        }{
            \makeColumnThreeLT{85.75}{86.46}{85.69}
        }{
            \makeColumnThreeLT{87.02}{87.02}{86.02}
        }{
            \item[\texttt{1.} \tableResultsOverviewH ] HERMES (87.02)  
            \item[\texttt{2.} \tableResultsOverviewH ] DeSD (86.46)  
            \item[\texttt{3.} \tableResultsOverviewV ] MedFormer (86.02)  
            \item[\texttt{4.} \tableResultsOverviewH ] UniMiSS (85.75)  
            \item[\texttt{5.} \tableResultsOverviewV ] SegResNet (85.74)
        }
    \makeTRSlineBest{
            MM-WHS
        }{
            \makeAHV{~~3}{~~1}{~~2}
        }{
            \makeColumnThreeLT{56.06}{89.44}{56.06}
        }{
            \makeColumnThreeLT{59.76}{89.44}{57.91}
        }{
            \makeColumnThreeLT{89.44}{89.44}{59.76}
        }{
            \item[\texttt{1.} \tableResultsOverviewH ] SAT (89.44)  
            \item[\texttt{2.} \tableResultsOverviewV ] nnU-Net (59.76)  
            \item[\texttt{3.} \tableResultsOverviewV ] SwinUNETR (56.06)
        }
    \makeTRSlineBest{
            MSD Cardiac
        }{
            \makeAHV{~~6}{~~4}{~~2}
        }{
            \makeColumnThreeLT{76.29}{76.29}{93.46}
        }{
            \makeColumnThreeLT{91.62}{86.73}{93.87}
        }{
            \makeColumnThreeLT{94.28}{93.38}{94.28}
        }{
            \item[\texttt{1.} \tableResultsOverviewV ] nnU-Net (94.28)  
            \item[\texttt{2.} \tableResultsOverviewV ] SwinUNETR (93.46)  
            \item[\texttt{3.} \tableResultsOverviewH ] SAT (93.38)  
            \item[\texttt{4.} \tableResultsOverviewH ] BiomedParse (89.86)  
            \item[\texttt{5.} \tableResultsOverviewH ] MedSAM (83.60)
        }
    \makeTRSlineBest{
            TotalSegmentator Cardiac
        }{
            \makeAHV{~12}{~~5}{~~7}
        }{
            \makeColumnThreeLT{75.96}{81.26}{75.96}
        }{
            \makeColumnThreeLT{86.96}{90.89}{83.77}
        }{
            \makeColumnThreeLT{93.30}{92.52}{93.30}
        }{
            \item[\texttt{1.} \tableResultsOverviewV ] nnU-Net (93.30)  
            \item[\texttt{2.} \tableResultsOverviewH ] SAT (92.52)  
            \item[\texttt{3.} \tableResultsOverviewH ] PCNet (91.64)  
            \item[\texttt{4.} \tableResultsOverviewV ] SwinUNETR (91.23)  
            \item[\texttt{5.} \tableResultsOverviewH ] STU-Net (90.89)
        }
    
}

\textit{Winners:} Generalist models on all datasets with the exception of ACDC, and Left Atrial Segmentation (median DSC on both primary research, and best in literature).

\clearpage
\subsubsection{Thoracic structures}
\label{subsubsec:res-thoracic-structures}

\begin{figure}[h!]
    \centering
    \includegraphics[width=8cm]{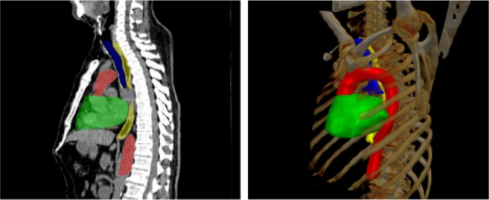}
    \caption{Thoracic structures example. Courtesy of nnU-Net \cite{nnunet_nature} (on SegTHOR).}
    \label{fig:res-thoracic-structure-examples}
\end{figure}

\makeTableResultsSpecific{
    Results overview for thoracic structures (multiorgan) datasets.
}{table:results-overview-thoracic-structures-}{
    Thoracic Structures (multiorgan)
}{
    \makeTRSlinePrimary{
            SegTHOR
        }{
            \makeAHV{~~7}{~~6}{~~1}
        }{
            \makeColumnThreeLT{81.55}{81.55}{93.00}
        }{
            \makeColumnThreeLT{88.98}{88.32}{93.00}
        }{
            \makeColumnThreeLT{93.00}{89.56}{93.00}
        }{
            \item[\texttt{1.} \tableResultsOverviewV ] nnU-Net (93.00)  
            \item[\texttt{2.} \tableResultsOverviewH ] MIS-FM (89.56)  
            \item[\texttt{3.} \tableResultsOverviewH ] IMIS-Net (89.27)  
            \item[\texttt{4.} \tableResultsOverviewH ] SAT (88.98)  
            \item[\texttt{5.} \tableResultsOverviewH ] PCNet (87.66)
        }
    
}{
    \makeTRSlineBest{
            SegTHOR
        }{
            \makeAHV{~16}{~11}{~~5}
        }{
            \makeColumnThreeLT{74.90}{74.90}{85.46}
        }{
            \makeColumnThreeLT{86.39}{85.91}{87.33}
        }{
            \makeColumnThreeLT{93.00}{89.56}{93.00}
        }{
            \item[\texttt{1.} \tableResultsOverviewV ] nnU-Net (93.00)  
            \item[\texttt{2.} \tableResultsOverviewV ] SwinUNETR (89.92)  
            \item[\texttt{3.} \tableResultsOverviewH ] MIS-FM (89.56)  
            \item[\texttt{4.} \tableResultsOverviewH ] IMIS-Net (89.27)  
            \item[\texttt{5.} \tableResultsOverviewH ] SAT (88.98)
        }
    
}

\textit{Winners:} Task-specific on both primary work, and best in literature.

\clearpage
\subsection{Bones} 
\label{subsec:res-bones}

\begin{figure}[h!]
    \centering
    \includegraphics[width=9cm]{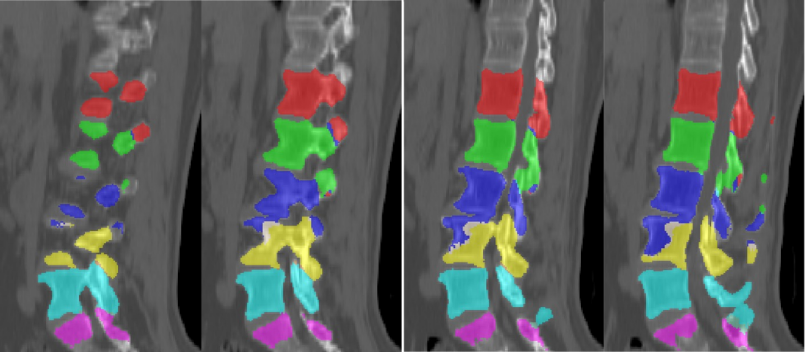}
    \caption{Vertebrae segmentation example. Courtesy of SAT \cite{zhao2025modelrulealluniversal} (on TotalSegmentator Vertebrae).}
    \label{fig:res-bones-example}
\end{figure}

\makeTableResultsSpecific{
    Results overview for bones datasets.
}{table:results-overview-bones}{
    Bones
}{
    \makeTRSlinePrimary{
            TotalSegmentator Ribs
        }{
            \makeAHV{~~3}{~~3}{~~0}
        }{
            \makeColumnThreeLT{90.29}{90.29}{-}
        }{
            \makeColumnThreeLT{91.53}{91.53}{-}
        }{
            \makeColumnThreeLT{91.66}{91.66}{-}
        }{
            \item[\texttt{1.} \tableResultsOverviewH ] PCNet (91.66)  
            \item[\texttt{2.} \tableResultsOverviewH ] SAT (91.53)  
            \item[\texttt{3.} \tableResultsOverviewH ] STU-Net (90.29)
        }
    \makeTRSlinePrimary{
            TotalSegmentator Vertebrae
        }{
            \makeAHV{~~4}{~~4}{~~0}
        }{
            \makeColumnThreeLT{86.49}{86.49}{-}
        }{
            \makeColumnThreeLT{90.43}{90.43}{-}
        }{
            \makeColumnThreeLT{91.69}{91.69}{-}
        }{
            \item[\texttt{1.} \tableResultsOverviewH ] PCNet (91.69)  
            \item[\texttt{2.} \tableResultsOverviewH ] STU-Net (90.43)  
            \item[\texttt{3.} \tableResultsOverviewH ] SAT (90.42)  
            \item[\texttt{4.} \tableResultsOverviewH ] CLIP-Driven Universal Model (86.49)
        }
    \makeTRSlinePrimary{
            VerSe
        }{
            \makeAHV{~~4}{~~4}{~~0}
        }{
            \makeColumnThreeLT{66.65}{66.65}{-}
        }{
            \makeColumnThreeLT{75.21}{75.21}{-}
        }{
            \makeColumnThreeLT{86.10}{86.10}{-}
        }{
            \item[\texttt{1.} \tableResultsOverviewH ] UniSeg (86.10)  
            \item[\texttt{2.} \tableResultsOverviewH ] SAT (81.01)  
            \item[\texttt{3.} \tableResultsOverviewH ] PCNet (69.40)  
            \item[\texttt{4.} \tableResultsOverviewH ] STU-Net (66.65)
        }
    
}{
    \makeTRSlineBest{
            TotalSegmentator Ribs
        }{
            \makeAHV{~11}{~~4}{~~7}
        }{
            \makeColumnThreeLT{74.03}{88.51}{74.03}
        }{
            \makeColumnThreeLT{88.51}{90.91}{79.45}
        }{
            \makeColumnThreeLT{92.10}{91.66}{92.10}
        }{
            \item[\texttt{1.} \tableResultsOverviewV ] nnU-Net (92.10)  
            \item[\texttt{2.} \tableResultsOverviewH ] PCNet (91.66)  
            \item[\texttt{3.} \tableResultsOverviewH ] SAT (91.53)  
            \item[\texttt{4.} \tableResultsOverviewH ] STU-Net (90.29)  
            \item[\texttt{5.} \tableResultsOverviewH ] MedSAM (88.51)
        }
    \makeTRSlineBest{
            TotalSegmentator Vertebrae
        }{
            \makeAHV{~12}{~~5}{~~7}
        }{
            \makeColumnThreeLT{73.87}{86.49}{73.87}
        }{
            \makeColumnThreeLT{88.45}{90.43}{81.29}
        }{
            \makeColumnThreeLT{95.37}{94.08}{95.37}
        }{
            \item[\texttt{1.} \tableResultsOverviewV ] nnU-Net (95.37)  
            \item[\texttt{2.} \tableResultsOverviewH ] MedSAM (94.08)  
            \item[\texttt{3.} \tableResultsOverviewV ] SwinUNETR (94.08)  
            \item[\texttt{4.} \tableResultsOverviewH ] PCNet (91.69)  
            \item[\texttt{5.} \tableResultsOverviewH ] STU-Net (90.43)
        }
    \makeTRSlineBest{
            VerSe
        }{
            \makeAHV{~12}{~~6}{~~6}
        }{
            \makeColumnThreeLT{66.65}{66.65}{84.30}
        }{
            \makeColumnThreeLT{86.05}{77.91}{87.00}
        }{
            \makeColumnThreeLT{87.20}{86.10}{87.20}
        }{
            \item[\texttt{1.} \tableResultsOverviewV ] nnU-Net (87.20)  
            \item[\texttt{2.} \tableResultsOverviewV ] 3D UX-Net (87.10)  
            \item[\texttt{3.} \tableResultsOverviewV ] CoTr (87.10)  
            \item[\texttt{4.} \tableResultsOverviewV ] SwinUNETR (86.90)  
            \item[\texttt{5.} \tableResultsOverviewH ] UniSeg (86.10)
        }
    
}

\textit{Winners:} Specialists models on primary work, while generalists on best in literature on two out of three datasets (TotalSegmentator Ribs Vertebrae, and TotalSegmentator Ribs Vertebrae).

\clearpage
\subsection{Muscles} 
\label{subsec:res-muscles}

\begin{figure}[h!]
    \centering
    \includegraphics[width=6cm]{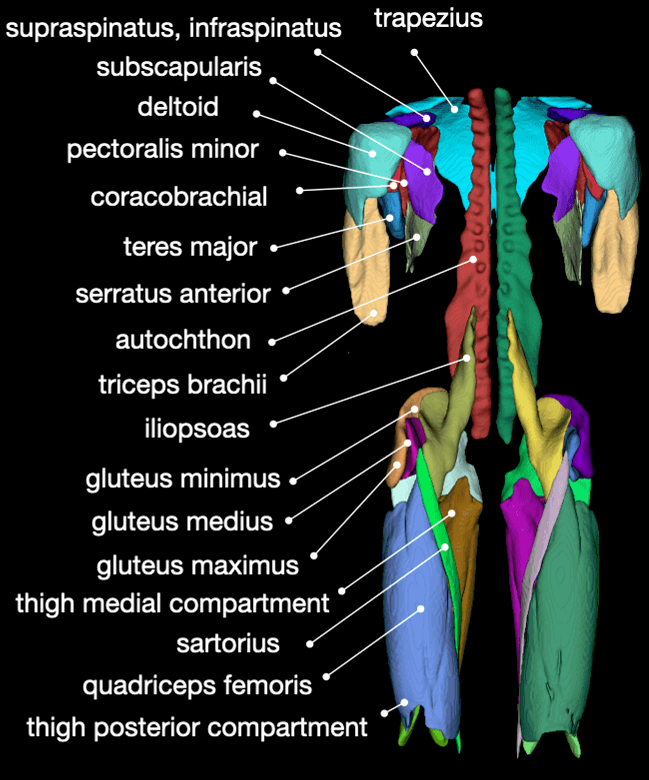}
    \caption{Muscles segmentation example. Courtesy of TotalSegmentator \citep{doi:10.1148/radiol.241613}.}
    
    \label{fig:res-muscles-example}
\end{figure}

\makeTableResultsSpecific{
    Results overview for muscles datasets.
}{table:results-overview-muscles}{
    Muscles
}{
    \makeTRSlinePrimary{
            TotalSegmentator Muscles
        }{
            \makeAHV{~~4}{~~4}{~~0}
        }{
            \makeColumnThreeLT{88.83}{88.83}{-}
        }{
            \makeColumnThreeLT{92.73}{92.73}{-}
        }{
            \makeColumnThreeLT{94.43}{94.43}{-}
        }{
            \item[\texttt{1.} \tableResultsOverviewH ] CLIP-Driven Universal Model (94.43)  
            \item[\texttt{2.} \tableResultsOverviewH ] SAT (93.33)  
            \item[\texttt{3.} \tableResultsOverviewH ] PCNet (92.13)  
            \item[\texttt{4.} \tableResultsOverviewH ] STU-Net (88.83)
        }
    
}{
    \makeTRSlineBest{
            TotalSegmentator Muscles
        }{
            \makeAHV{~12}{~~5}{~~7}
        }{
            \makeColumnThreeLT{73.29}{82.23}{73.29}
        }{
            \makeColumnThreeLT{87.39}{92.13}{84.63}
        }{
            \makeColumnThreeLT{94.43}{94.43}{91.60}
        }{
            \item[\texttt{1.} \tableResultsOverviewH ] CLIP-Driven Universal Model (94.43)  
            \item[\texttt{2.} \tableResultsOverviewH ] SAT (93.33)  
            \item[\texttt{3.} \tableResultsOverviewH ] PCNet (92.13)  
            \item[\texttt{4.} \tableResultsOverviewV ] nnU-Net (91.60)  
            \item[\texttt{5.} \tableResultsOverviewV ] SwinUNETR (90.21)
        }
    
}

\textit{Winner:} Foundation models on both primary work, and best in literature.

\clearpage
\subsubsection{Liver} 
\label{subsubsec:res-liver}

\begin{figure}[h!]
    \centering
    \includegraphics[width=9cm]{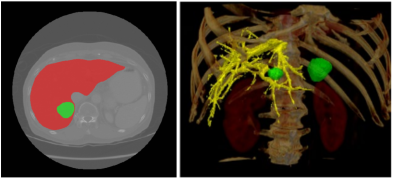}
    \caption{Liver (red), tumor (green) and hepatic vessels (yellow) segmentation example. Courtesy of TransBTS-V2 \cite{li2022transbtsv2betterefficientvolumetric} (left, on MSD Liver) and nnU-Net \cite{nnunet_nature} (right, on MSD Hepatic Vessels).}
    \label{fig:res-liver-example}
\end{figure}

\makeTableResultsSpecific{
    Results overview for liver datasets.
}{table:results-overview-liver}{
    Liver
}{
    \makeTRSlinePrimary{
            ATLAS 2023
        }{
            \makeAHV{~~4}{~~4}{~~0}
        }{
            \makeColumnThreeLT{63.80}{63.80}{-}
        }{
            \makeColumnThreeLT{71.11}{71.11}{-}
        }{
            \makeColumnThreeLT{76.26}{76.26}{-}
        }{
            \item[\texttt{1.} \tableResultsOverviewH ] SAT (76.26)  
            \item[\texttt{2.} \tableResultsOverviewH ] Medical SAM 2 (MedSAM-2) (71.80)  
            \item[\texttt{3.} \tableResultsOverviewH ] SAM-Med2D (70.42)  
            \item[\texttt{4.} \tableResultsOverviewH ] One-Prompt (63.80)
        }
    \makeTRSlinePrimary{
            MSD Hepatic Vessels
        }{
            \makeAHV{~~9}{~~7}{~~2}
        }{
            \makeColumnThreeLT{63.43}{63.43}{67.67}
        }{
            \makeColumnThreeLT{68.20}{68.20}{68.34}
        }{
            \makeColumnThreeLT{79.59}{79.59}{69.00}
        }{
            \item[\texttt{1.} \tableResultsOverviewH ] LeSAM (79.59)  
            \item[\texttt{2.} \tableResultsOverviewH ] CLIP-Driven Universal Model (71.51)  
            \item[\texttt{3.} \tableResultsOverviewH ] UniSeg (71.20)  
            \item[\texttt{4.} \tableResultsOverviewV ] nnU-Net (69.00)  
            \item[\texttt{5.} \tableResultsOverviewH ] DeSD (68.20)
        }
    \makeTRSlinePrimary{
            LiTS / MSD Liver
        }{
            \makeAHV{~20}{~15}{~~5}
        }{
            \makeColumnThreeLT{60.45}{60.45}{69.00}
        }{
            \makeColumnThreeLT{83.00}{81.90}{86.50}
        }{
            \makeColumnThreeLT{96.63}{96.63}{89.85}
        }{
            \item[\texttt{1.} \tableResultsOverviewH ] PCNet (96.63)  
            \item[\texttt{2.} \tableResultsOverviewH ] STU-Net (95.88)  
            \item[\texttt{3.} \tableResultsOverviewH ] Med3D (94.60)  
            \item[\texttt{4.} \tableResultsOverviewH ] SAMMed (92.00)  
            \item[\texttt{5.} \tableResultsOverviewH ] MEA M-SAM (89.95)
        }
    
}{
    \makeTRSlineBest{
            ATLAS 2023
        }{
            \makeAHV{~10}{~~7}{~~3}
        }{
            \makeColumnThreeLT{34.50}{53.10}{34.50}
        }{
            \makeColumnThreeLT{67.11}{63.80}{70.88}
        }{
            \makeColumnThreeLT{78.83}{76.26}{78.83}
        }{
            \item[\texttt{1.} \tableResultsOverviewV ] nnU-Net (78.83)  
            \item[\texttt{2.} \tableResultsOverviewH ] SAT (76.26)  
            \item[\texttt{3.} \tableResultsOverviewH ] Medical SAM 2 (MedSAM-2) (71.80)  
            \item[\texttt{4.} \tableResultsOverviewV ] SwinUNETR (70.88)  
            \item[\texttt{5.} \tableResultsOverviewH ] SAM-Med2D (70.42)
        }
    \makeTRSlineBest{
            MSD Hepatic Vessels
        }{
            \makeAHV{~16}{~~9}{~~7}
        }{
            \makeColumnThreeLT{30.97}{30.97}{53.80}
        }{
            \makeColumnThreeLT{67.48}{68.20}{67.30}
        }{
            \makeColumnThreeLT{79.59}{79.59}{69.00}
        }{
            \item[\texttt{1.} \tableResultsOverviewH ] LeSAM (79.59)  
            \item[\texttt{2.} \tableResultsOverviewH ] CLIP-Driven Universal Model (71.51)  
            \item[\texttt{3.} \tableResultsOverviewH ] UniSeg (71.20)  
            \item[\texttt{4.} \tableResultsOverviewH ] DoDNet (70.40)  
            \item[\texttt{5.} \tableResultsOverviewV ] nnU-Net (69.00)
        }
    \makeTRSlineBest{
            LiTS / MSD Liver
        }{
            \makeAHV{~38}{~23}{~15}
        }{
            \makeColumnThreeLT{60.45}{60.45}{69.00}
        }{
            \makeColumnThreeLT{81.69}{81.32}{82.60}
        }{
            \makeColumnThreeLT{96.63}{96.63}{95.29}
        }{
            \item[\texttt{1.} \tableResultsOverviewH ] PCNet (96.63)  
            \item[\texttt{2.} \tableResultsOverviewH ] STU-Net (95.88)  
            \item[\texttt{3.} \tableResultsOverviewV ] nnU-Net (95.29)  
            \item[\texttt{4.} \tableResultsOverviewH ] Med3D (94.60)  
            \item[\texttt{5.} \tableResultsOverviewV ] V-Net (93.90)
        }
    
}

\textit{Winner:} Foundation models on primary work; foundation models on best in literature on all datsaets except ATLAS2023.

\clearpage
\subsubsection{Pancreas} 
\label{subsubsec:res-pancreas}

\begin{figure}[h!]
    \centering
    \includegraphics[width=6cm]{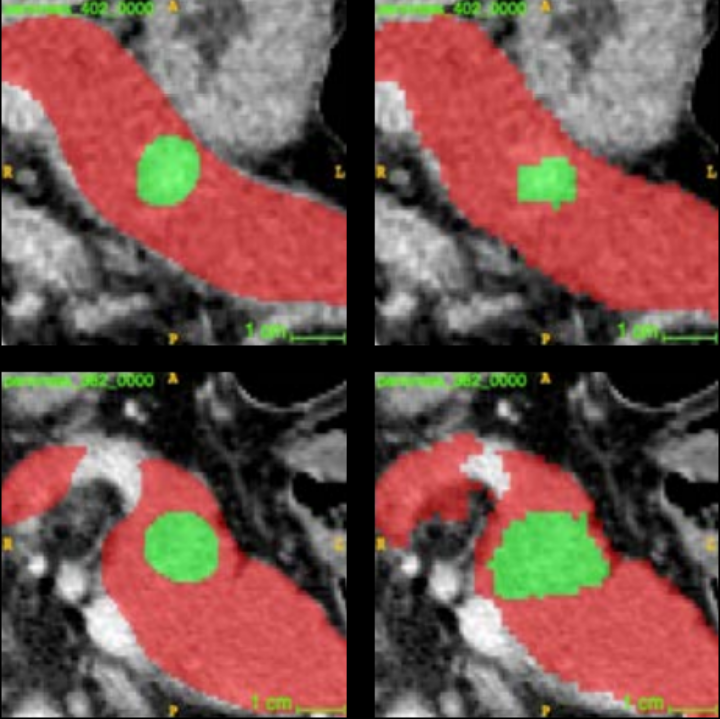}
    \caption{Pancreas (red) and tumor (green) segmentation example (ground-truth on the left, segmentation on the right). Courtesy of CLIP-Driven Universal Model \cite{10376801} (on MSD Pancreas).}
    \label{fig:res-pancreas-example}
\end{figure}

\makeTableResultsSpecific{
    Results overview for pancreas datasets.
}{table:results-overview-pancreas}{
    Pancreas
}{
    \makeTRSlinePrimary{
            MSD Pancreas Tumour
        }{
            \makeAHV{~16}{~13}{~~3}
        }{
            \makeColumnThreeLT{40.20}{40.20}{55.49}
        }{
            \makeColumnThreeLT{70.75}{71.54}{64.03}
        }{
            \makeColumnThreeLT{80.49}{80.49}{67.50}
        }{
            \item[\texttt{1.} \tableResultsOverviewH ] MEA M-SAM (80.49)  
            \item[\texttt{2.} \tableResultsOverviewH ] PCNet (79.70)  
            \item[\texttt{3.} \tableResultsOverviewH ] LeSAM (79.42)  
            \item[\texttt{4.} \tableResultsOverviewH ] STU-Net (78.95)  
            \item[\texttt{5.} \tableResultsOverviewH ] CLIP-Driven Universal Model (72.59)
        }
    \makeTRSlinePrimary{
            Pancreas-CT
        }{
            \makeAHV{~~1}{~~0}{~~1}
        }{
            \makeColumnThreeLT{81.96}{-}{81.96}
        }{
            \makeColumnThreeLT{81.96}{-}{81.96}
        }{
            \makeColumnThreeLT{81.96}{-}{81.96}
        }{
            \item[\texttt{1.} \tableResultsOverviewV ] LHU-Net (81.96)
        }
    
}{
    \makeTRSlineBest{
            MSD Pancreas Tumour
        }{
            \makeAHV{~28}{~20}{~~8}
        }{
            \makeColumnThreeLT{40.20}{40.20}{60.09}
        }{
            \makeColumnThreeLT{72.33}{71.81}{73.94}
        }{
            \makeColumnThreeLT{80.49}{80.49}{78.65}
        }{
            \item[\texttt{1.} \tableResultsOverviewH ] MEA M-SAM (80.49)  
            \item[\texttt{2.} \tableResultsOverviewH ] PCNet (79.70)  
            \item[\texttt{3.} \tableResultsOverviewH ] LeSAM (79.42)  
            \item[\texttt{4.} \tableResultsOverviewH ] SAM-Med2D (79.02)  
            \item[\texttt{5.} \tableResultsOverviewH ] STU-Net (78.95)
        }
    \makeTRSlineBest{
            Pancreas-CT
        }{
            \makeAHV{~~6}{~~0}{~~6}
        }{
            \makeColumnThreeLT{76.18}{-}{76.18}
        }{
            \makeColumnThreeLT{80.78}{-}{80.78}
        }{
            \makeColumnThreeLT{81.96}{-}{81.96}
        }{
            \item[\texttt{1.} \tableResultsOverviewV ] LHU-Net (81.96)  
            \item[\texttt{2.} \tableResultsOverviewV ] UNETR++ (81.49)  
            \item[\texttt{3.} \tableResultsOverviewV ] SwinUNETR-V2 (81.05)  
            \item[\texttt{4.} \tableResultsOverviewV ] SwinUNETR (80.52)  
            \item[\texttt{5.} \tableResultsOverviewV ] nnFormer (78.05)
        }
    
}

\textit{Winner:} Tie: foundation models on MSD dataset (primary work, and best in literature); task-specific models on NIH dataset (primary work, and best in literature).

\clearpage
\subsubsection{Colon} 
\label{subsubsec:res-colon}

\begin{figure}[h!]
    \centering
    \includegraphics[width=8cm]{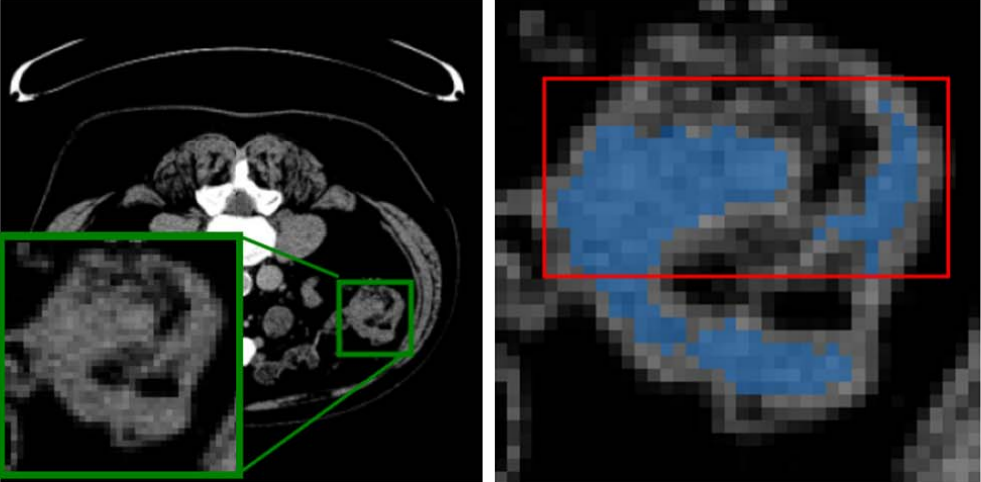}
    \caption{Colon tumor (blue) segmentation example (CT image and zoom on the left, segmentation on the right, red box prompt). Courtesy of LeSAM \citep{10540651} (MSD Colon Cancer).}
    \label{fig:res-colon-example}
\end{figure}

\makeTableResultsSpecific{
    Results overview for colon datasets.
}{table:results-overview-colon}{
    Colon
}{
    \makeTRSlinePrimary{
            MSD Colon Cancer
        }{
            \makeAHV{~10}{~~9}{~~1}
        }{
            \makeColumnThreeLT{38.45}{38.45}{58.00}
        }{
            \makeColumnThreeLT{56.50}{55.00}{58.00}
        }{
            \makeColumnThreeLT{77.18}{77.18}{58.00}
        }{
            \item[\texttt{1.} \tableResultsOverviewH ] LeSAM (77.18)  
            \item[\texttt{2.} \tableResultsOverviewH ] BiomedParse (66.51)  
            \item[\texttt{3.} \tableResultsOverviewH ] CLIP-Driven Universal Model (63.14)  
            \item[\texttt{4.} \tableResultsOverviewH ] 3DSAM-adapter (60.93)  
            \item[\texttt{5.} \tableResultsOverviewV ] nnU-Net (58.00)
        }
    
}{
    \makeTRSlineBest{
            MSD Colon Cancer
        }{
            \makeAHV{~18}{~13}{~~5}
        }{
            \makeColumnThreeLT{18.80}{38.45}{18.80}
        }{
            \makeColumnThreeLT{58.73}{63.14}{39.80}
        }{
            \makeColumnThreeLT{77.18}{77.18}{59.45}
        }{
            \item[\texttt{1.} \tableResultsOverviewH ] LeSAM (77.18)  
            \item[\texttt{2.} \tableResultsOverviewH ] SAM-Med2D (76.45)  
            \item[\texttt{3.} \tableResultsOverviewH ] Med-SA (75.36)  
            \item[\texttt{4.} \tableResultsOverviewH ] MedSAM (72.76)  
            \item[\texttt{5.} \tableResultsOverviewH ] BiomedParse (66.51)
        }
    
}

\textit{Winners:} Foundation models on both primary work, and best in literature.

\clearpage
\subsubsection{Kidneys} 
\label{subsubsec:res-kidney}

\begin{figure}[h!]
    \centering
    \includegraphics[width=8cm]{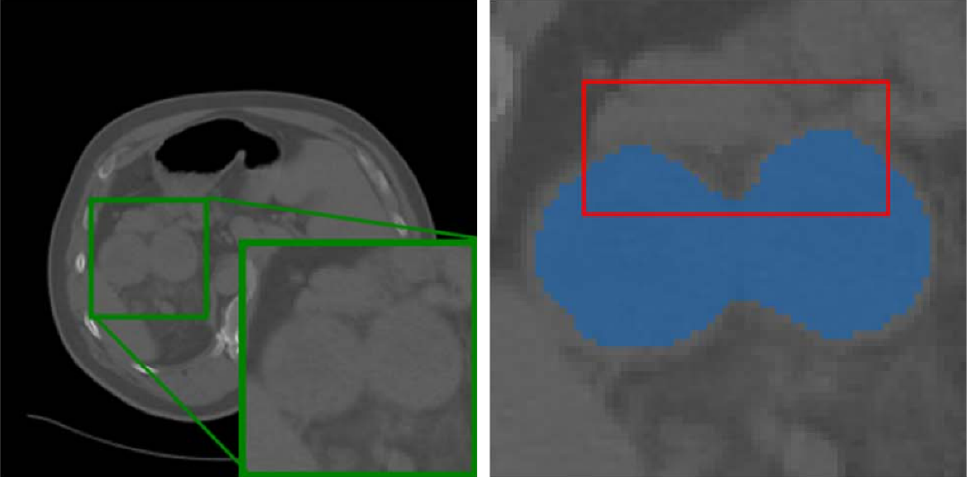}
    \caption{Kidney tumor (blue) segmentation example (CT image and zoom on the left, segmentation on the right, red box prompt). Courtesy of LeSAM \citep{10540651} (KiTS).}
    \label{fig:res-kidneys-example}
\end{figure}

\makeTableResultsSpecific{
    Results overview for kidney datasets.
}{table:results-overview-kidney}{
    Kidney
}{
    \makeTRSlinePrimary{
            KiPA
        }{
            \makeAHV{~~1}{~~1}{~~0}
        }{
            \makeColumnThreeLT{80.19}{80.19}{-}
        }{
            \makeColumnThreeLT{80.19}{80.19}{-}
        }{
            \makeColumnThreeLT{80.19}{80.19}{-}
        }{
            \item[\texttt{1.} \tableResultsOverviewH ] PCNet (80.19)
        }
    \makeTRSlinePrimary{
            KiTS
        }{
            \makeAHV{~23}{~18}{~~5}
        }{
            \makeColumnThreeLT{60.46}{60.46}{85.00}
        }{
            \makeColumnThreeLT{85.98}{84.72}{90.53}
        }{
            \makeColumnThreeLT{93.50}{93.50}{91.63}
        }{
            \item[\texttt{1.} \tableResultsOverviewH ] MEA M-SAM (93.50)  
            \item[\texttt{2.} \tableResultsOverviewH ] LeSAM (91.86)  
            \item[\texttt{3.} \tableResultsOverviewV ] nnU-Net (91.63)  
            \item[\texttt{4.} \tableResultsOverviewV ] MedNeXt (91.02)  
            \item[\texttt{5.} \tableResultsOverviewV ] TransBTSV2 (90.53)
        }
    
}{
    \makeTRSlineBest{
            KiPA
        }{
            \makeAHV{~~4}{~~2}{~~2}
        }{
            \makeColumnThreeLT{30.72}{78.44}{30.72}
        }{
            \makeColumnThreeLT{59.34}{79.31}{35.48}
        }{
            \makeColumnThreeLT{80.19}{80.19}{40.25}
        }{
            \item[\texttt{1.} \tableResultsOverviewH ] PCNet (80.19)  
            \item[\texttt{2.} \tableResultsOverviewH ] STU-Net (78.44)  
            \item[\texttt{3.} \tableResultsOverviewV ] SwinUNETR (40.25)  
            \item[\texttt{4.} \tableResultsOverviewV ] nnU-Net (30.72)
        }
    \makeTRSlineBest{
            KiTS
        }{
            \makeAHV{~41}{~27}{~14}
        }{
            \makeColumnThreeLT{60.23}{60.23}{80.82}
        }{
            \makeColumnThreeLT{85.44}{84.00}{88.36}
        }{
            \makeColumnThreeLT{93.50}{93.50}{91.63}
        }{
            \item[\texttt{1.} \tableResultsOverviewH ] MEA M-SAM (93.50)  
            \item[\texttt{2.} \tableResultsOverviewH ] LeSAM (91.86)  
            \item[\texttt{3.} \tableResultsOverviewV ] nnU-Net (91.63)  
            \item[\texttt{4.} \tableResultsOverviewH ] SAM-Med2D (91.46)  
            \item[\texttt{5.} \tableResultsOverviewH ] Med-SA (91.05)
        }
    
}

\textit{Winners:} Tie: foundation models on KiPA222 dataset (primary work, and best in literature); task-specific models on KiTS dataset (primary work, and best in literature).

\clearpage
\subsubsection{Spleen}
\label{subsubsec:res-spleen}

\begin{figure}[h!]
    \centering
    \includegraphics[width=5cm]{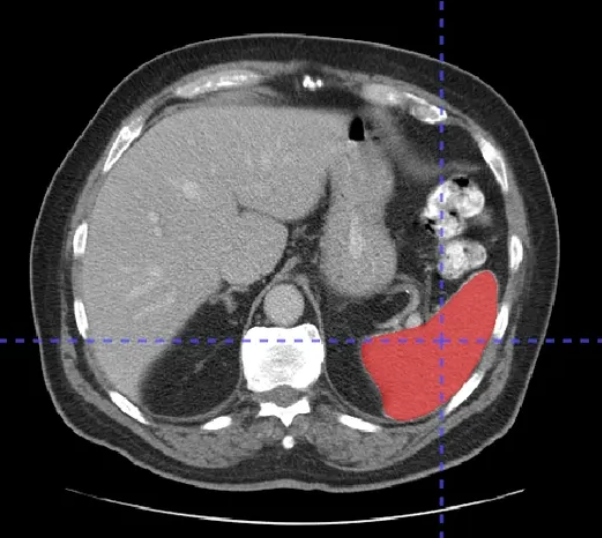}
    \caption{Spleen (red) segmentation example. Courtesy of \href{https://github.com/openmedlab/Awesome-Medical-Dataset/blob/main/resources/MSD_Spleen.md}{Awesome Medical Dataset} (MSD Spleen).}
    \label{fig:res-spleen-example}
\end{figure}

\makeTableResultsSpecific{
    Results overview for spleen datasets.
}{table:results-overview-spleen}{
    Spleen
}{
    \makeTRSlinePrimary{
            MSD Spleen
        }{
            \makeAHV{~10}{~~8}{~~2}
        }{
            \makeColumnThreeLT{93.91}{93.91}{96.40}
        }{
            \makeColumnThreeLT{96.20}{95.88}{96.70}
        }{
            \makeColumnThreeLT{97.27}{97.27}{97.00}
        }{
            \item[\texttt{1.} \tableResultsOverviewH ] CLIP-Driven Universal Model (97.27)  
            \item[\texttt{2.} \tableResultsOverviewV ] nnU-Net (97.00)  
            \item[\texttt{3.} \tableResultsOverviewH ] BiomedParse (96.86)  
            \item[\texttt{4.} \tableResultsOverviewH ] UniSeg (96.40)  
            \item[\texttt{5.} \tableResultsOverviewV ] UNETR (96.40)
        }
    
}{
    \makeTRSlineBest{
            MSD Spleen
        }{
            \makeAHV{~17}{~11}{~~6}
        }{
            \makeColumnThreeLT{79.59}{79.59}{92.20}
        }{
            \makeColumnThreeLT{95.77}{95.77}{96.05}
        }{
            \makeColumnThreeLT{97.27}{97.27}{97.00}
        }{
            \item[\texttt{1.} \tableResultsOverviewH ] CLIP-Driven Universal Model (97.27)  
            \item[\texttt{2.} \tableResultsOverviewV ] nnU-Net (97.00)  
            \item[\texttt{3.} \tableResultsOverviewV ] SwinUNETR (96.99)  
            \item[\texttt{4.} \tableResultsOverviewH ] BiomedParse (96.86)  
            \item[\texttt{5.} \tableResultsOverviewH ] DoDNet (96.50)
        }
    
}

\textit{Winners:} Task-specific models on both primary work and best in literature.

\clearpage
\subsubsection{Prostate} 
\label{subsubsec:res-prostate}

\begin{figure}[h!]
    \centering
    \includegraphics[width=8cm]{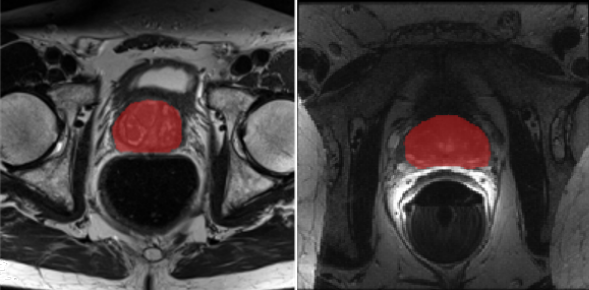}
    \caption{Prostate (red) segmentation example from two different MRI modalities. Courtesy SPA \citep{10829779} (PROMISE12).}
    \label{fig:res-prostate-example}
\end{figure}

\makeTableResultsSpecific{
    Results overview for prostate datasets.
}{table:results-overview-prostate}{
    Prostate
}{
    \makeTRSlinePrimary{
            MSD Prostate
        }{
            \makeAHV{~~6}{~~3}{~~3}
        }{
            \makeColumnThreeLT{72.85}{72.85}{73.32}
        }{
            \makeColumnThreeLT{76.02}{77.98}{74.05}
        }{
            \makeColumnThreeLT{89.70}{89.70}{83.50}
        }{
            \item[\texttt{1.} \tableResultsOverviewH ] UniSeg (89.70)  
            \item[\texttt{2.} \tableResultsOverviewV ] nnU-Net (83.50)  
            \item[\texttt{3.} \tableResultsOverviewH ] SAT (77.98)  
            \item[\texttt{4.} \tableResultsOverviewV ] SwinUNETR-V2 (74.05)  
            \item[\texttt{5.} \tableResultsOverviewV ] SwinUNETR (73.32)
        }
    \makeTRSlinePrimary{
            PROMISE12
        }{
            \makeAHV{~~7}{~~5}{~~2}
        }{
            \makeColumnThreeLT{86.90}{87.28}{86.90}
        }{
            \makeColumnThreeLT{89.97}{89.97}{89.42}
        }{
            \makeColumnThreeLT{94.29}{94.29}{91.94}
        }{
            \item[\texttt{1.} \tableResultsOverviewH ] SPA (94.29)  
            \item[\texttt{2.} \tableResultsOverviewH ] MA-SAM (92.60)  
            \item[\texttt{3.} \tableResultsOverviewV ] nnU-Net (91.94)  
            \item[\texttt{4.} \tableResultsOverviewH ] BiomedParse (89.97)  
            \item[\texttt{5.} \tableResultsOverviewH ] FLAP-SAM (88.67)
        }
    
}{
    \makeTRSlineBest{
            MSD Prostate
        }{
            \makeAHV{~14}{~~7}{~~7}
        }{
            \makeColumnThreeLT{62.68}{62.68}{74.05}
        }{
            \makeColumnThreeLT{87.30}{77.98}{88.00}
        }{
            \makeColumnThreeLT{89.70}{89.70}{89.40}
        }{
            \item[\texttt{1.} \tableResultsOverviewH ] UniSeg (89.70)  
            \item[\texttt{2.} \tableResultsOverviewV ] nnU-Net (89.40)  
            \item[\texttt{3.} \tableResultsOverviewH ] DoDNet (89.10)  
            \item[\texttt{4.} \tableResultsOverviewV ] 3D UX-Net (88.80)  
            \item[\texttt{5.} \tableResultsOverviewV ] SwinUNETR (88.30)
        }
    \makeTRSlineBest{
            PROMISE12
        }{
            \makeAHV{~19}{~10}{~~9}
        }{
            \makeColumnThreeLT{81.20}{81.20}{85.10}
        }{
            \makeColumnThreeLT{89.16}{91.22}{87.73}
        }{
            \makeColumnThreeLT{94.29}{94.29}{91.94}
        }{
            \item[\texttt{1.} \tableResultsOverviewH ] SPA (94.29)  
            \item[\texttt{2.} \tableResultsOverviewH ] Med-SA (93.66)  
            \item[\texttt{3.} \tableResultsOverviewH ] SAMed (93.47)  
            \item[\texttt{4.} \tableResultsOverviewH ] MA-SAM (92.60)  
            \item[\texttt{5.} \tableResultsOverviewH ] MedSAM (92.46)
        }
    
}

\textit{Winners:} Foundation models on primary work; task-specific on MSD prostate, while foundation models on PROMISE12 (best in the literature).

\clearpage
\subsubsection{Abdominal Organs – Multi Organ} 
\label{subsubsec:res-abdominal-multi}

\begin{figure}[h!]
    \centering
    \includegraphics[width=7.0cm]{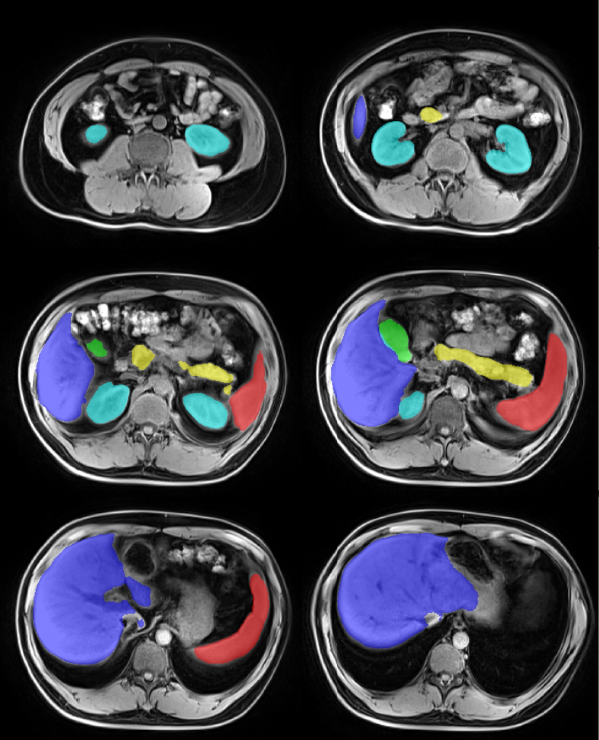}
    \caption{Abdominal multi-organ segmentation example from an MRI image. Courtesy of SAT \citep{zhao2025modelrulealluniversal}.}
    \label{fig:res-multiorgan}
\end{figure}

\makeTableResultsSpecific{
    Results overview for abdominal multi-organ datasets.
}{table:results-overview-abdominal-multi-organ}{
    Abdominal Multi-Organ
}{
    \makeTRSlinePrimary{
            AMOS
        }{
            \makeAHV{~11}{~~8}{~~3}
        }{
            \makeColumnThreeLT{74.39}{74.39}{88.00}
        }{
            \makeColumnThreeLT{88.00}{86.13}{90.00}
        }{
            \makeColumnThreeLT{91.77}{90.49}{91.77}
        }{
            \item[\texttt{1.} \tableResultsOverviewV ] MedNeXt (91.77)  
            \item[\texttt{2.} \tableResultsOverviewH ] STU-Net (90.49)  
            \item[\texttt{3.} \tableResultsOverviewV ] 3D UX-Net (90.00)  
            \item[\texttt{4.} \tableResultsOverviewH ] MultiTalent (89.81)  
            \item[\texttt{5.} \tableResultsOverviewH ] HERMES (88.59)
        }
    \makeTRSlinePrimary{
            AbdomenCT-1K
        }{
            \makeAHV{~~4}{~~4}{~~0}
        }{
            \makeColumnThreeLT{91.70}{91.70}{-}
        }{
            \makeColumnThreeLT{92.64}{92.64}{-}
        }{
            \makeColumnThreeLT{94.90}{94.90}{-}
        }{
            \item[\texttt{1.} \tableResultsOverviewH ] SAT (94.90)  
            \item[\texttt{2.} \tableResultsOverviewH ] PCNet (93.02)  
            \item[\texttt{3.} \tableResultsOverviewH ] STU-Net (92.27)  
            \item[\texttt{4.} \tableResultsOverviewH ] SFR SAM (91.70)
        }
    \makeTRSlinePrimary{
            BTCV
        }{
            \makeAHV{~27}{~19}{~~8}
        }{
            \makeColumnThreeLT{70.30}{70.30}{83.28}
        }{
            \makeColumnThreeLT{86.29}{86.29}{86.17}
        }{
            \makeColumnThreeLT{92.10}{92.10}{88.76}
        }{
            \item[\texttt{1.} \tableResultsOverviewH ] Disruptive Autoencoders (92.10)  
            \item[\texttt{2.} \tableResultsOverviewH ] MultiTalent (89.07)  
            \item[\texttt{3.} \tableResultsOverviewH ] Medical SAM 2 (MedSAM-2) (89.00)  
            \item[\texttt{4.} \tableResultsOverviewV ] MedNeXt (88.76)  
            \item[\texttt{5.} \tableResultsOverviewH ] 3DMedSAM (88.60)
        }
    \makeTRSlinePrimary{
            BTCV Cervix
        }{
            \makeAHV{~~1}{~~1}{~~0}
        }{
            \makeColumnThreeLT{90.54}{90.54}{-}
        }{
            \makeColumnThreeLT{90.54}{90.54}{-}
        }{
            \makeColumnThreeLT{90.54}{90.54}{-}
        }{
            \item[\texttt{1.} \tableResultsOverviewH ] PCNet (90.54)
        }
    \makeTRSlinePrimary{
            CHAOS
        }{
            \makeAHV{~~4}{~~3}{~~1}
        }{
            \makeColumnThreeLT{72.44}{91.36}{72.44}
        }{
            \makeColumnThreeLT{91.79}{92.22}{72.44}
        }{
            \makeColumnThreeLT{92.61}{92.61}{72.44}
        }{
            \item[\texttt{1.} \tableResultsOverviewH ] SAT (92.61)  
            \item[\texttt{2.} \tableResultsOverviewH ] HERMES (92.22)  
            \item[\texttt{3.} \tableResultsOverviewH ] UniMiSS (91.36)  
            \item[\texttt{4.} \tableResultsOverviewV ] nnU-Net (72.44)
        }
    \makeTRSlinePrimary{
            FLARE
        }{
            \makeAHV{~~9}{~~6}{~~3}
        }{
            \makeColumnThreeLT{0.88}{0.88}{92.90}
        }{
            \makeColumnThreeLT{90.62}{87.48}{93.40}
        }{
            \makeColumnThreeLT{94.70}{91.78}{94.70}
        }{
            \item[\texttt{1.} \tableResultsOverviewV ] SwinUNETR-V2 (94.70)  
            \item[\texttt{2.} \tableResultsOverviewV ] 3D UX-Net (93.40)  
            \item[\texttt{3.} \tableResultsOverviewV ] SwinUNETR (92.90)  
            \item[\texttt{4.} \tableResultsOverviewH ] SAT (91.78)  
            \item[\texttt{5.} \tableResultsOverviewH ] PCNet (90.62)
        }
    \makeTRSlinePrimary{
            MOTS
        }{
            \makeAHV{~~2}{~~2}{~~0}
        }{
            \makeColumnThreeLT{75.64}{75.64}{-}
        }{
            \makeColumnThreeLT{77.69}{77.69}{-}
        }{
            \makeColumnThreeLT{79.74}{79.74}{-}
        }{
            \item[\texttt{1.} \tableResultsOverviewH ] CLIP-Driven Universal Model (79.74)  
            \item[\texttt{2.} \tableResultsOverviewH ] DoDNet (75.64)
        }
    \makeTRSlinePrimary{
            Synapse
        }{
            \makeAHV{~12}{~~5}{~~7}
        }{
            \makeColumnThreeLT{79.13}{79.56}{79.13}
        }{
            \makeColumnThreeLT{86.89}{85.95}{87.22}
        }{
            \makeColumnThreeLT{92.88}{92.88}{89.67}
        }{
            \item[\texttt{1.} \tableResultsOverviewH ] SPA (92.88)  
            \item[\texttt{2.} \tableResultsOverviewV ] SCANeXt (89.67)  
            \item[\texttt{3.} \tableResultsOverviewH ] MIS-FM (89.11)  
            \item[\texttt{4.} \tableResultsOverviewV ] TransUNet (88.39)  
            \item[\texttt{5.} \tableResultsOverviewV ] LHU-Net (87.49)
        }
    \makeTRSlinePrimary{
            TotalSegmentator
        }{
            \makeAHV{~~6}{~~6}{~~0}
        }{
            \makeColumnThreeLT{79.06}{79.06}{-}
        }{
            \makeColumnThreeLT{86.35}{86.35}{-}
        }{
            \makeColumnThreeLT{91.64}{91.64}{-}
        }{
            \item[\texttt{1.} \tableResultsOverviewH ] PCNet (91.64)  
            \item[\texttt{2.} \tableResultsOverviewH ] STU-Net (90.06)  
            \item[\texttt{3.} \tableResultsOverviewH ] SAT (86.71)  
            \item[\texttt{4.} \tableResultsOverviewH ] Merlin (86.00)  
            \item[\texttt{5.} \tableResultsOverviewH ] SAM-Med3D (84.68)
        }
    \makeTRSlinePrimary{
            TotalSegmentator Organs
        }{
            \makeAHV{~~4}{~~4}{~~0}
        }{
            \makeColumnThreeLT{88.95}{88.95}{-}
        }{
            \makeColumnThreeLT{90.12}{90.12}{-}
        }{
            \makeColumnThreeLT{91.09}{91.09}{-}
        }{
            \item[\texttt{1.} \tableResultsOverviewH ] PCNet (91.09)  
            \item[\texttt{2.} \tableResultsOverviewH ] SAT (90.42)  
            \item[\texttt{3.} \tableResultsOverviewH ] STU-Net (89.82)  
            \item[\texttt{4.} \tableResultsOverviewH ] CLIP-Driven Universal Model (88.95)
        }
    \makeTRSlinePrimary{
            Touchstone
        }{
            \makeAHV{~10}{~~3}{~~7}
        }{
            \makeColumnThreeLT{83.30}{87.10}{83.30}
        }{
            \makeColumnThreeLT{88.40}{88.80}{88.00}
        }{
            \makeColumnThreeLT{89.20}{89.00}{89.20}
        }{
            \item[\texttt{1.} \tableResultsOverviewV ] MedNeXt (89.20)  
            \item[\texttt{2.} \tableResultsOverviewH ] STU-Net (89.00)  
            \item[\texttt{3.} \tableResultsOverviewV ] MedFormer (89.00)  
            \item[\texttt{4.} \tableResultsOverviewV ] nnU-Net (88.90)  
            \item[\texttt{5.} \tableResultsOverviewH ] UniSeg (88.80)
        }
    \makeTRSlinePrimary{
            WORD
        }{
            \makeAHV{~~5}{~~3}{~~2}
        }{
            \makeColumnThreeLT{77.42}{77.42}{86.83}
        }{
            \makeColumnThreeLT{86.83}{79.17}{87.17}
        }{
            \makeColumnThreeLT{87.92}{87.92}{87.51}
        }{
            \item[\texttt{1.} \tableResultsOverviewH ] SAT (87.92)  
            \item[\texttt{2.} \tableResultsOverviewV ] SwinUNETR-V2 (87.51)  
            \item[\texttt{3.} \tableResultsOverviewV ] SwinUNETR (86.83)  
            \item[\texttt{4.} \tableResultsOverviewH ] PCNet (79.17)  
            \item[\texttt{5.} \tableResultsOverviewH ] STU-Net (77.42)
        }
    
}{
    \makeTRSlineBest{
            AMOS
        }{
            \makeAHV{~25}{~14}{~11}
        }{
            \makeColumnThreeLT{54.92}{54.92}{81.98}
        }{
            \makeColumnThreeLT{85.93}{84.56}{88.00}
        }{
            \makeColumnThreeLT{91.77}{90.49}{91.77}
        }{
            \item[\texttt{1.} \tableResultsOverviewV ] MedNeXt (91.77)  
            \item[\texttt{2.} \tableResultsOverviewH ] STU-Net (90.49)  
            \item[\texttt{3.} \tableResultsOverviewV ] 3D UX-Net (90.00)  
            \item[\texttt{4.} \tableResultsOverviewH ] MultiTalent (89.81)  
            \item[\texttt{5.} \tableResultsOverviewV ] U-Net (89.60)
        }
    \makeTRSlineBest{
            AbdomenCT-1K
        }{
            \makeAHV{~~8}{~~5}{~~3}
        }{
            \makeColumnThreeLT{86.03}{86.03}{88.26}
        }{
            \makeColumnThreeLT{92.64}{92.27}{93.73}
        }{
            \makeColumnThreeLT{95.09}{94.90}{95.09}
        }{
            \item[\texttt{1.} \tableResultsOverviewV ] nnU-Net (95.09)  
            \item[\texttt{2.} \tableResultsOverviewH ] SAT (94.90)  
            \item[\texttt{3.} \tableResultsOverviewV ] SwinUNETR (93.73)  
            \item[\texttt{4.} \tableResultsOverviewH ] PCNet (93.02)  
            \item[\texttt{5.} \tableResultsOverviewH ] STU-Net (92.27)
        }
    \makeTRSlineBest{
            BTCV
        }{
            \makeAHV{~46}{~28}{~18}
        }{
            \makeColumnThreeLT{50.05}{50.05}{78.40}
        }{
            \makeColumnThreeLT{84.75}{84.65}{85.00}
        }{
            \makeColumnThreeLT{92.10}{92.10}{91.80}
        }{
            \item[\texttt{1.} \tableResultsOverviewH ] Disruptive Autoencoders (92.10)  
            \item[\texttt{2.} \tableResultsOverviewV ] SwinUNETR (91.80)  
            \item[\texttt{3.} \tableResultsOverviewV ] UNETR (89.10)  
            \item[\texttt{4.} \tableResultsOverviewH ] MultiTalent (89.07)  
            \item[\texttt{5.} \tableResultsOverviewH ] Medical SAM 2 (MedSAM-2) (89.00)
        }
    \makeTRSlineBest{
            BTCV Cervix
        }{
            \makeAHV{~~4}{~~2}{~~2}
        }{
            \makeColumnThreeLT{54.22}{89.79}{54.22}
        }{
            \makeColumnThreeLT{88.31}{90.17}{70.53}
        }{
            \makeColumnThreeLT{90.54}{90.54}{86.83}
        }{
            \item[\texttt{1.} \tableResultsOverviewH ] PCNet (90.54)  
            \item[\texttt{2.} \tableResultsOverviewH ] STU-Net (89.79)  
            \item[\texttt{3.} \tableResultsOverviewV ] SwinUNETR (86.83)  
            \item[\texttt{4.} \tableResultsOverviewV ] nnU-Net (54.22)
        }
    \makeTRSlineBest{
            CHAOS
        }{
            \makeAHV{~~8}{~~4}{~~4}
        }{
            \makeColumnThreeLT{88.86}{91.36}{88.86}
        }{
            \makeColumnThreeLT{91.70}{91.88}{91.63}
        }{
            \makeColumnThreeLT{92.94}{92.61}{92.94}
        }{
            \item[\texttt{1.} \tableResultsOverviewV ] nnU-Net (92.94)  
            \item[\texttt{2.} \tableResultsOverviewH ] SAT (92.61)  
            \item[\texttt{3.} \tableResultsOverviewH ] HERMES (92.22)  
            \item[\texttt{4.} \tableResultsOverviewV ] MedFormer (91.85)  
            \item[\texttt{5.} \tableResultsOverviewH ] DeSD (91.55)
        }
    \makeTRSlineBest{
            FLARE
        }{
            \makeAHV{~18}{~~9}{~~9}
        }{
            \makeColumnThreeLT{0.88}{0.88}{82.00}
        }{
            \makeColumnThreeLT{89.53}{79.50}{90.60}
        }{
            \makeColumnThreeLT{94.70}{91.78}{94.70}
        }{
            \item[\texttt{1.} \tableResultsOverviewV ] SwinUNETR-V2 (94.70)  
            \item[\texttt{2.} \tableResultsOverviewV ] 3D UX-Net (93.40)  
            \item[\texttt{3.} \tableResultsOverviewV ] nnU-Net (93.36)  
            \item[\texttt{4.} \tableResultsOverviewV ] SwinUNETR (92.90)  
            \item[\texttt{5.} \tableResultsOverviewH ] SAT (91.78)
        }
    \makeTRSlineBest{
            MOTS
        }{
            \makeAHV{~~2}{~~2}{~~0}
        }{
            \makeColumnThreeLT{75.64}{75.64}{-}
        }{
            \makeColumnThreeLT{77.69}{77.69}{-}
        }{
            \makeColumnThreeLT{79.74}{79.74}{-}
        }{
            \item[\texttt{1.} \tableResultsOverviewH ] CLIP-Driven Universal Model (79.74)  
            \item[\texttt{2.} \tableResultsOverviewH ] DoDNet (75.64)
        }
    \makeTRSlineBest{
            Synapse
        }{
            \makeAHV{~25}{~~8}{~17}
        }{
            \makeColumnThreeLT{68.81}{79.56}{68.81}
        }{
            \makeColumnThreeLT{86.57}{90.44}{85.72}
        }{
            \makeColumnThreeLT{92.88}{92.88}{89.67}
        }{
            \item[\texttt{1.} \tableResultsOverviewH ] SPA (92.88)  
            \item[\texttt{2.} \tableResultsOverviewH ] Med-SA (92.42)  
            \item[\texttt{3.} \tableResultsOverviewH ] SAMed (92.33)  
            \item[\texttt{4.} \tableResultsOverviewH ] MedSAM (90.74)  
            \item[\texttt{5.} \tableResultsOverviewH ] SAM (90.15)
        }
    \makeTRSlineBest{
            TotalSegmentator
        }{
            \makeAHV{~17}{~10}{~~7}
        }{
            \makeColumnThreeLT{75.05}{75.45}{75.05}
        }{
            \makeColumnThreeLT{82.05}{82.40}{82.05}
        }{
            \makeColumnThreeLT{92.39}{91.64}{92.39}
        }{
            \item[\texttt{1.} \tableResultsOverviewV ] nnU-Net (92.39)  
            \item[\texttt{2.} \tableResultsOverviewH ] PCNet (91.64)  
            \item[\texttt{3.} \tableResultsOverviewH ] STU-Net (90.06)  
            \item[\texttt{4.} \tableResultsOverviewV ] SwinUNETR (88.85)  
            \item[\texttt{5.} \tableResultsOverviewH ] SAT (86.71)
        }
    \makeTRSlineBest{
            TotalSegmentator Organs
        }{
            \makeAHV{~12}{~~5}{~~7}
        }{
            \makeColumnThreeLT{77.11}{82.71}{77.11}
        }{
            \makeColumnThreeLT{87.31}{89.82}{83.41}
        }{
            \makeColumnThreeLT{93.22}{91.09}{93.22}
        }{
            \item[\texttt{1.} \tableResultsOverviewV ] nnU-Net (93.22)  
            \item[\texttt{2.} \tableResultsOverviewH ] PCNet (91.09)  
            \item[\texttt{3.} \tableResultsOverviewH ] SAT (90.42)  
            \item[\texttt{4.} \tableResultsOverviewV ] SwinUNETR (90.41)  
            \item[\texttt{5.} \tableResultsOverviewH ] STU-Net (89.82)
        }
    \makeTRSlineBest{
            Touchstone
        }{
            \makeAHV{~11}{~~4}{~~7}
        }{
            \makeColumnThreeLT{73.40}{73.40}{83.30}
        }{
            \makeColumnThreeLT{88.00}{87.95}{88.00}
        }{
            \makeColumnThreeLT{89.20}{89.00}{89.20}
        }{
            \item[\texttt{1.} \tableResultsOverviewV ] MedNeXt (89.20)  
            \item[\texttt{2.} \tableResultsOverviewH ] STU-Net (89.00)  
            \item[\texttt{3.} \tableResultsOverviewV ] MedFormer (89.00)  
            \item[\texttt{4.} \tableResultsOverviewV ] nnU-Net (88.90)  
            \item[\texttt{5.} \tableResultsOverviewH ] UniSeg (88.80)
        }
    \makeTRSlineBest{
            WORD
        }{
            \makeAHV{~~9}{~~3}{~~6}
        }{
            \makeColumnThreeLT{77.42}{77.42}{79.77}
        }{
            \makeColumnThreeLT{84.66}{79.17}{85.75}
        }{
            \makeColumnThreeLT{87.92}{87.92}{87.51}
        }{
            \item[\texttt{1.} \tableResultsOverviewH ] SAT (87.92)  
            \item[\texttt{2.} \tableResultsOverviewV ] SwinUNETR-V2 (87.51)  
            \item[\texttt{3.} \tableResultsOverviewV ] nnU-Net (87.44)  
            \item[\texttt{4.} \tableResultsOverviewV ] SwinUNETR (86.83)  
            \item[\texttt{5.} \tableResultsOverviewV ] CoTr (84.66)
        }
    
}

\textit{Winners:} Tie with foundation models on eight datasets (AbdomenCT-1k, BTCV, BTCV Cervix, CHAOS MultiOrgan, MOTS, TotalSegmentor (All), TotalSegmentor (Organs), and Touchstone 1.0) on primary works, while task specific on seven datasets (AMOS2022, AbdomenCT-1k, BTCV, CHAOS MultiOrgan, FLARE MICCAI, TouchStone 1.0, and WORD) on best in the literature.

\clearpage
\subsubsection{Whole-body Lesions}
\label{subsubsec:res-whole-body-lesions}

\begin{figure}[h!]
    \centering
    \includegraphics[width=8.0cm]{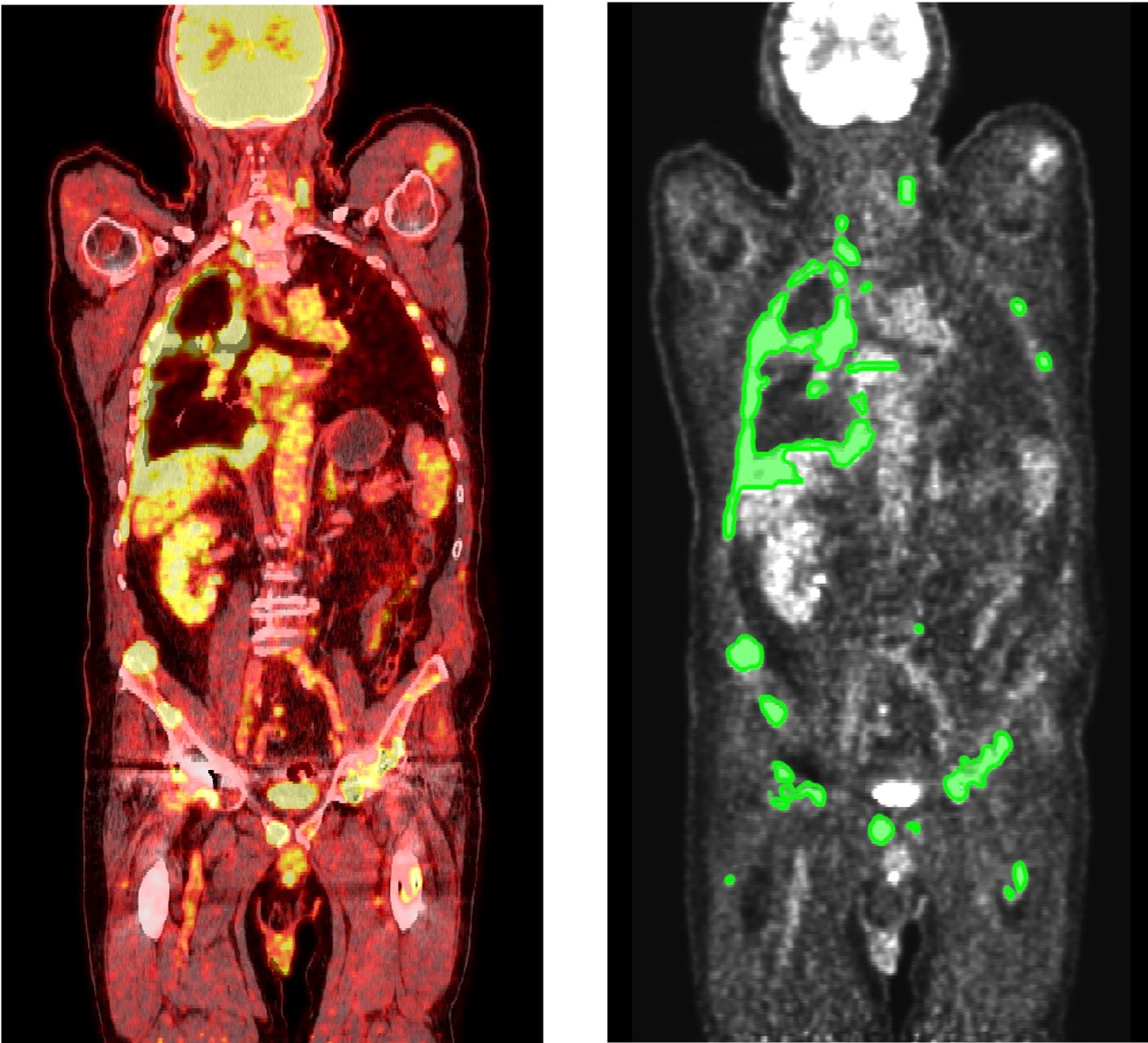}
    \caption{Whole body FDG-PET/CT image fusion (left) and over-imposed manual segmentations of FDG-avid malignant lesions (right). Courtesy of \href{https://autopet.grand-challenge.org/}{AutoPET}.}
    \label{fig:res-lesions}
\end{figure}

\makeTableResultsSpecific{
    Results overview for whole body lesions datasets.
}{table:results-overview-whole-body-lesions}{
    Whole Body Lesions
}{
    \makeTRSlinePrimary{
            AutoPET
        }{
            \makeAHV{~~1}{~~1}{~~0}
        }{
            \makeColumnThreeLT{74.04}{74.04}{-}
        }{
            \makeColumnThreeLT{74.04}{74.04}{-}
        }{
            \makeColumnThreeLT{74.04}{74.04}{-}
        }{
            \item[\texttt{1.} \tableResultsOverviewH ] HERMES (74.04)
        }
    \makeTRSlinePrimary{
            ULS
        }{
            \makeAHV{~~1}{~~1}{~~0}
        }{
            \makeColumnThreeLT{70.46}{70.46}{-}
        }{
            \makeColumnThreeLT{70.46}{70.46}{-}
        }{
            \makeColumnThreeLT{70.46}{70.46}{-}
        }{
            \item[\texttt{1.} \tableResultsOverviewH ] SegVol (70.46)
        }
    
}{
    \makeTRSlineBest{
            AutoPET
        }{
            \makeAHV{~~7}{~~3}{~~4}
        }{
            \makeColumnThreeLT{60.32}{60.32}{64.39}
        }{
            \makeColumnThreeLT{65.52}{69.02}{65.47}
        }{
            \makeColumnThreeLT{74.04}{74.04}{66.01}
        }{
            \item[\texttt{1.} \tableResultsOverviewH ] HERMES (74.04)  
            \item[\texttt{2.} \tableResultsOverviewH ] DeSD (69.02)  
            \item[\texttt{3.} \tableResultsOverviewV ] MedFormer (66.01)  
            \item[\texttt{4.} \tableResultsOverviewV ] SegResNet (65.52)  
            \item[\texttt{5.} \tableResultsOverviewV ] nnU-Net (65.43)
        }
    \makeTRSlineBest{
            ULS
        }{
            \makeAHV{~~5}{~~5}{~~0}
        }{
            \makeColumnThreeLT{35.84}{35.84}{-}
        }{
            \makeColumnThreeLT{65.74}{65.74}{-}
        }{
            \makeColumnThreeLT{70.46}{70.46}{-}
        }{
            \item[\texttt{1.} \tableResultsOverviewH ] SegVol (70.46)  
            \item[\texttt{2.} \tableResultsOverviewH ] MedSAM (70.46)  
            \item[\texttt{3.} \tableResultsOverviewH ] SAM (65.74)  
            \item[\texttt{4.} \tableResultsOverviewH ] SAM-Med3D (41.20)  
            \item[\texttt{5.} \tableResultsOverviewH ] SAM-Med2D (35.84)
        }
    
}

\textit{Winners:} Foundation models on both primary work, and best in literature.

\clearpage
\bibliographystyle{elsarticle-harv}

\begin{thebibliography}{197}
\expandafter\ifx\csname natexlab\endcsname\relax\def\natexlab#1{#1}\fi
\providecommand{\url}[1]{\texttt{#1}}
\providecommand{\href}[2]{#2}
\providecommand{\path}[1]{#1}
\providecommand{\DOIprefix}{doi:}
\providecommand{\ArXivprefix}{arXiv:}
\providecommand{\URLprefix}{URL: }
\providecommand{\Pubmedprefix}{pmid:}
\providecommand{\doi}[1]{\href{http://dx.doi.org/#1}{\path{#1}}}
\providecommand{\Pubmed}[1]{\href{pmid:#1}{\path{#1}}}
\providecommand{\bibinfo}[2]{#2}
\ifx\xfnm\relax \def\xfnm[#1]{\unskip,\space#1}\fi
\bibitem[{Akinci~DAntonoli et~al.(2025)Akinci~DAntonoli, Berger, Indrakanti, Vishwanathan, Weiss, Jung, Berkarda, Rau, Reisert, K\"{u}stner, Walter, Merkle, Boll, Breit, Nicoli, Segeroth, Cyriac, Yang and Wasserthal}]{doi:10.1148/radiol.241613}
\bibinfo{author}{Akinci~DAntonoli, T.}, \bibinfo{author}{Berger, L.K.}, \bibinfo{author}{Indrakanti, A.K.}, \bibinfo{author}{Vishwanathan, N.}, \bibinfo{author}{Weiss, J.}, \bibinfo{author}{Jung, M.}, \bibinfo{author}{Berkarda, Z.}, \bibinfo{author}{Rau, A.}, \bibinfo{author}{Reisert, M.}, \bibinfo{author}{K\"{u}stner, T.}, \bibinfo{author}{Walter, A.}, \bibinfo{author}{Merkle, E.M.}, \bibinfo{author}{Boll, D.T.}, \bibinfo{author}{Breit, H.C.}, \bibinfo{author}{Nicoli, A.P.}, \bibinfo{author}{Segeroth, M.}, \bibinfo{author}{Cyriac, J.}, \bibinfo{author}{Yang, S.}, \bibinfo{author}{Wasserthal, J.}, \bibinfo{year}{2025}.
\newblock \bibinfo{title}{Totalsegmentator mri: Robust sequence-independent segmentation of multiple anatomic structures in mri}.
\newblock \bibinfo{journal}{Radiology} \bibinfo{volume}{314}, \bibinfo{pages}{e241613}.
\newblock \URLprefix \url{https://doi.org/10.1148/radiol.241613}, \DOIprefix\doi{10.1148/radiol.241613}, \href{http://arxiv.org/abs/https://doi.org/10.1148/radiol.241613}{{\tt arXiv:https://doi.org/10.1148/radiol.241613}}. \bibinfo{note}{pMID: 39964271}.
\bibitem[{Ali et~al.(2024)Ali, Wu, Hu, Luo, Xu, Zheng, Jin, Yang and Yao}]{ali2024review}
\bibinfo{author}{Ali, M.}, \bibinfo{author}{Wu, T.}, \bibinfo{author}{Hu, H.}, \bibinfo{author}{Luo, Q.}, \bibinfo{author}{Xu, D.}, \bibinfo{author}{Zheng, W.}, \bibinfo{author}{Jin, N.}, \bibinfo{author}{Yang, C.}, \bibinfo{author}{Yao, J.}, \bibinfo{year}{2024}.
\newblock \bibinfo{title}{A review of the segment anything model (sam) for medical image analysis: Accomplishments and perspectives}.
\newblock \bibinfo{journal}{Computerized Medical Imaging and Graphics} , \bibinfo{pages}{102473}.
\bibitem[{AlSaad et~al.(2024)AlSaad, Abd-Alrazaq, Boughorbel, Ahmed, Renault, Damseh and Sheikh}]{alsaad2024multimodal}
\bibinfo{author}{AlSaad, R.}, \bibinfo{author}{Abd-Alrazaq, A.}, \bibinfo{author}{Boughorbel, S.}, \bibinfo{author}{Ahmed, A.}, \bibinfo{author}{Renault, M.A.}, \bibinfo{author}{Damseh, R.}, \bibinfo{author}{Sheikh, J.}, \bibinfo{year}{2024}.
\newblock \bibinfo{title}{Multimodal large language models in health care: applications, challenges, and future outlook}.
\newblock \bibinfo{journal}{Journal of medical Internet research} \bibinfo{volume}{26}, \bibinfo{pages}{e59505}.
\bibitem[{Asokan et~al.(2024)Asokan, Benjamin, Yaqub and Nandakumar}]{asokan2024federatedlearningfriendlyapproachparameterefficient}
\bibinfo{author}{Asokan, M.}, \bibinfo{author}{Benjamin, J.G.}, \bibinfo{author}{Yaqub, M.}, \bibinfo{author}{Nandakumar, K.}, \bibinfo{year}{2024}.
\newblock \bibinfo{title}{A federated learning-friendly approach for parameter-efficient fine-tuning of sam in 3d segmentation}.
\newblock \URLprefix \url{https://arxiv.org/abs/2407.21739}, \href{http://arxiv.org/abs/2407.21739}{{\tt arXiv:2407.21739}}.
\bibitem[{Asokan et~al.(2025)Asokan, Benjamin, Yaqub and Nandakumar}]{10.1007/978-3-031-77610-6_21}
\bibinfo{author}{Asokan, M.}, \bibinfo{author}{Benjamin, J.G.}, \bibinfo{author}{Yaqub, M.}, \bibinfo{author}{Nandakumar, K.}, \bibinfo{year}{2025}.
\newblock \bibinfo{title}{A federated learning-friendly approach for parameter-efficient fine-tuning of sam in 3d segmentation}, in: \bibinfo{editor}{Celebi, M.E.}, \bibinfo{editor}{Reyes, M.}, \bibinfo{editor}{Chen, Z.}, \bibinfo{editor}{Li, X.} (Eds.), \bibinfo{booktitle}{Medical Image Computing and Computer Assisted Intervention -- MICCAI 2024 Workshops}, \bibinfo{publisher}{Springer Nature Switzerland}, \bibinfo{address}{Cham}. pp. \bibinfo{pages}{226--235}.
\bibitem[{Bao et~al.(2021)Bao, Dong, Piao and Wei}]{bao2021beit}
\bibinfo{author}{Bao, H.}, \bibinfo{author}{Dong, L.}, \bibinfo{author}{Piao, S.}, \bibinfo{author}{Wei, F.}, \bibinfo{year}{2021}.
\newblock \bibinfo{title}{Beit: Bert pre-training of image transformers}.
\newblock \bibinfo{journal}{arXiv preprint arXiv:2106.08254} .
\bibitem[{Bian et~al.(2025)Bian, Li, Ye, Jia and Yang}]{bian2025artificial}
\bibinfo{author}{Bian, Y.}, \bibinfo{author}{Li, J.}, \bibinfo{author}{Ye, C.}, \bibinfo{author}{Jia, X.}, \bibinfo{author}{Yang, Q.}, \bibinfo{year}{2025}.
\newblock \bibinfo{title}{Artificial intelligence in medical imaging: From task-specific models to large-scale foundation models}.
\newblock \bibinfo{journal}{Chinese Medical Journal} \bibinfo{volume}{138}, \bibinfo{pages}{651--663}.
\bibitem[{Blankemeier et~al.(2024)Blankemeier, Cohen, Kumar, Veen, Gardezi, Paschali, Chen, Delbrouck, Reis, Truyts, Bluethgen, Jensen, Ostmeier, Varma, Valanarasu, Fang, Huo, Nabulsi, Ardila, Weng, Junior, Ahuja, Fries, Shah, Johnston, Boutin, Wentland, Langlotz, Hom, Gatidis and Chaudhari}]{blankemeier2024merlinvisionlanguagefoundation}
\bibinfo{author}{Blankemeier, L.}, \bibinfo{author}{Cohen, J.P.}, \bibinfo{author}{Kumar, A.}, \bibinfo{author}{Veen, D.V.}, \bibinfo{author}{Gardezi, S.J.S.}, \bibinfo{author}{Paschali, M.}, \bibinfo{author}{Chen, Z.}, \bibinfo{author}{Delbrouck, J.B.}, \bibinfo{author}{Reis, E.}, \bibinfo{author}{Truyts, C.}, \bibinfo{author}{Bluethgen, C.}, \bibinfo{author}{Jensen, M.E.K.}, \bibinfo{author}{Ostmeier, S.}, \bibinfo{author}{Varma, M.}, \bibinfo{author}{Valanarasu, J.M.J.}, \bibinfo{author}{Fang, Z.}, \bibinfo{author}{Huo, Z.}, \bibinfo{author}{Nabulsi, Z.}, \bibinfo{author}{Ardila, D.}, \bibinfo{author}{Weng, W.H.}, \bibinfo{author}{Junior, E.A.}, \bibinfo{author}{Ahuja, N.}, \bibinfo{author}{Fries, J.}, \bibinfo{author}{Shah, N.H.}, \bibinfo{author}{Johnston, A.}, \bibinfo{author}{Boutin, R.D.}, \bibinfo{author}{Wentland, A.}, \bibinfo{author}{Langlotz, C.P.}, \bibinfo{author}{Hom, J.}, \bibinfo{author}{Gatidis, S.}, \bibinfo{author}{Chaudhari, A.S.}, \bibinfo{year}{2024}.
\newblock \bibinfo{title}{Merlin: A vision language foundation model for 3d computed tomography}.
\newblock \URLprefix \url{https://arxiv.org/abs/2406.06512}, \href{http://arxiv.org/abs/2406.06512}{{\tt arXiv:2406.06512}}.
\bibitem[{Bommasani et~al.(2021)Bommasani, Hudson, Adeli, Altman, Arora, von Arx, Bernstein, Bohg, Bosselut, Brunskill et~al.}]{bommasani2021opportunities}
\bibinfo{author}{Bommasani, R.}, \bibinfo{author}{Hudson, D.A.}, \bibinfo{author}{Adeli, E.}, \bibinfo{author}{Altman, R.}, \bibinfo{author}{Arora, S.}, \bibinfo{author}{von Arx, S.}, \bibinfo{author}{Bernstein, M.S.}, \bibinfo{author}{Bohg, J.}, \bibinfo{author}{Bosselut, A.}, \bibinfo{author}{Brunskill, E.}, et~al., \bibinfo{year}{2021}.
\newblock \bibinfo{title}{On the opportunities and risks of foundation models}.
\newblock \bibinfo{journal}{arXiv preprint arXiv:2108.07258} .
\bibitem[{Bui et~al.(2024a)Bui, Hoang, Tran, Doretto, Adjeroh, Patel, Choudhary and Le}]{sam3d_bui}
\bibinfo{author}{Bui, N.T.}, \bibinfo{author}{Hoang, D.H.}, \bibinfo{author}{Tran, M.T.}, \bibinfo{author}{Doretto, G.}, \bibinfo{author}{Adjeroh, D.}, \bibinfo{author}{Patel, B.}, \bibinfo{author}{Choudhary, A.}, \bibinfo{author}{Le, N.}, \bibinfo{year}{2024}a.
\newblock \bibinfo{title}{Sam3d: Segment anything model in volumetric medical images}.
\newblock \URLprefix \url{https://arxiv.org/abs/2309.03493}, \href{http://arxiv.org/abs/2309.03493}{{\tt arXiv:2309.03493}}.
\bibitem[{Bui et~al.(2024b)Bui, Hoang, Tran, Doretto, Adjeroh, Patel, Choudhary and Le}]{bui2024sam3dsegmentmodelvolumetric}
\bibinfo{author}{Bui, N.T.}, \bibinfo{author}{Hoang, D.H.}, \bibinfo{author}{Tran, M.T.}, \bibinfo{author}{Doretto, G.}, \bibinfo{author}{Adjeroh, D.}, \bibinfo{author}{Patel, B.}, \bibinfo{author}{Choudhary, A.}, \bibinfo{author}{Le, N.}, \bibinfo{year}{2024}b.
\newblock \bibinfo{title}{Sam3d: Segment anything model in volumetric medical images}.
\newblock \URLprefix \url{https://arxiv.org/abs/2309.03493}, \href{http://arxiv.org/abs/2309.03493}{{\tt arXiv:2309.03493}}.
\bibitem[{Bui et~al.(2024c)Bui, Hoang, Tran, Doretto, Adjeroh, Patel, Choudhary and Le}]{10635844}
\bibinfo{author}{Bui, N.T.}, \bibinfo{author}{Hoang, D.H.}, \bibinfo{author}{Tran, M.T.}, \bibinfo{author}{Doretto, G.}, \bibinfo{author}{Adjeroh, D.}, \bibinfo{author}{Patel, B.}, \bibinfo{author}{Choudhary, A.}, \bibinfo{author}{Le, N.}, \bibinfo{year}{2024}c.
\newblock \bibinfo{title}{Sam3d: Segment anything model in volumetric medical images}, in: \bibinfo{booktitle}{2024 IEEE International Symposium on Biomedical Imaging (ISBI)}, pp. \bibinfo{pages}{1--4}.
\newblock \DOIprefix\doi{10.1109/ISBI56570.2024.10635844}.
\bibitem[{Butoi et~al.(2023a)Butoi, Gonzalez~Ortiz, Ma, Sabuncu, Guttag and Dalca}]{10376558}
\bibinfo{author}{Butoi, V.I.}, \bibinfo{author}{Gonzalez~Ortiz, J.J.}, \bibinfo{author}{Ma, T.}, \bibinfo{author}{Sabuncu, M.R.}, \bibinfo{author}{Guttag, J.}, \bibinfo{author}{Dalca, A.V.}, \bibinfo{year}{2023}a.
\newblock \bibinfo{title}{Universeg: Universal medical image segmentation}, in: \bibinfo{booktitle}{2023 IEEE/CVF International Conference on Computer Vision (ICCV)}, pp. \bibinfo{pages}{21381--21394}.
\newblock \DOIprefix\doi{10.1109/ICCV51070.2023.01960}.
\bibitem[{Butoi et~al.(2023b)Butoi, Ortiz, Ma, Sabuncu, Guttag and Dalca}]{butoi2023universeguniversalmedicalimage}
\bibinfo{author}{Butoi, V.I.}, \bibinfo{author}{Ortiz, J.J.G.}, \bibinfo{author}{Ma, T.}, \bibinfo{author}{Sabuncu, M.R.}, \bibinfo{author}{Guttag, J.}, \bibinfo{author}{Dalca, A.V.}, \bibinfo{year}{2023}b.
\newblock \bibinfo{title}{Universeg: Universal medical image segmentation}.
\newblock \URLprefix \url{https://arxiv.org/abs/2304.06131}, \href{http://arxiv.org/abs/2304.06131}{{\tt arXiv:2304.06131}}.
\bibitem[{Cao et~al.(2023)Cao, Wang, Chen, Jiang, Zhang, Tian and Wang}]{cao2022swin}
\bibinfo{author}{Cao, H.}, \bibinfo{author}{Wang, Y.}, \bibinfo{author}{Chen, J.}, \bibinfo{author}{Jiang, D.}, \bibinfo{author}{Zhang, X.}, \bibinfo{author}{Tian, Q.}, \bibinfo{author}{Wang, M.}, \bibinfo{year}{2023}.
\newblock \bibinfo{title}{Swin-unet: Unet-like pure transformer for medical image segmentation}, in: \bibinfo{booktitle}{Computer Vision – ECCV 2022 Workshops}, \bibinfo{publisher}{Springer Nature Switzerland}. pp. \bibinfo{pages}{205--218}.
\newblock \DOIprefix\doi{10.1007/978-3-031-25066-8_9}.
\bibitem[{Caron et~al.(2021)Caron, Touvron, Misra, Jegou, Mairal, Bojanowski and Joulin}]{dino_cvf}
\bibinfo{author}{Caron, M.}, \bibinfo{author}{Touvron, H.}, \bibinfo{author}{Misra, I.}, \bibinfo{author}{Jegou, H.}, \bibinfo{author}{Mairal, J.}, \bibinfo{author}{Bojanowski, P.}, \bibinfo{author}{Joulin, A.}, \bibinfo{year}{2021}.
\newblock \bibinfo{title}{Emerging properties in self-supervised vision transformers}, in: \bibinfo{booktitle}{2021 IEEE/CVF International Conference on Computer Vision (ICCV)}, pp. \bibinfo{pages}{9630--9640}.
\newblock \DOIprefix\doi{10.1109/ICCV48922.2021.00951}.
\bibitem[{Chen et~al.(2023)Chen, Miao, Wu, Yan, Kim, Hu, Zhong, Liu, Sun, Li, Liu, Heng and Li}]{chen2023masammodalityagnosticsamadaptation}
\bibinfo{author}{Chen, C.}, \bibinfo{author}{Miao, J.}, \bibinfo{author}{Wu, D.}, \bibinfo{author}{Yan, Z.}, \bibinfo{author}{Kim, S.}, \bibinfo{author}{Hu, J.}, \bibinfo{author}{Zhong, A.}, \bibinfo{author}{Liu, Z.}, \bibinfo{author}{Sun, L.}, \bibinfo{author}{Li, X.}, \bibinfo{author}{Liu, T.}, \bibinfo{author}{Heng, P.A.}, \bibinfo{author}{Li, Q.}, \bibinfo{year}{2023}.
\newblock \bibinfo{title}{Ma-sam: Modality-agnostic sam adaptation for 3d medical image segmentation}.
\newblock \URLprefix \url{https://arxiv.org/abs/2309.08842}, \href{http://arxiv.org/abs/2309.08842}{{\tt arXiv:2309.08842}}.
\bibitem[{Chen et~al.(2024a)Chen, Miao, Wu, Zhong, Yan, Kim, Hu, Liu, Sun, Li, Liu, Heng and Li}]{ma-sam_cheng}
\bibinfo{author}{Chen, C.}, \bibinfo{author}{Miao, J.}, \bibinfo{author}{Wu, D.}, \bibinfo{author}{Zhong, A.}, \bibinfo{author}{Yan, Z.}, \bibinfo{author}{Kim, S.}, \bibinfo{author}{Hu, J.}, \bibinfo{author}{Liu, Z.}, \bibinfo{author}{Sun, L.}, \bibinfo{author}{Li, X.}, \bibinfo{author}{Liu, T.}, \bibinfo{author}{Heng, P.A.}, \bibinfo{author}{Li, Q.}, \bibinfo{year}{2024}a.
\newblock \bibinfo{title}{Ma-sam: Modality-agnostic sam adaptation for 3d medical image segmentation}.
\newblock \bibinfo{journal}{Medical Image Analysis} \bibinfo{volume}{98}, \bibinfo{pages}{103310}.
\newblock \URLprefix \url{https://www.sciencedirect.com/science/article/pii/S1361841524002354}, \DOIprefix\doi{https://doi.org/10.1016/j.media.2024.103310}.
\bibitem[{Chen et~al.(2024b)Chen, Miao, Wu, Zhong, Yan, Kim, Hu, Liu, Sun, Li, Liu, Heng and Li}]{CHEN2024103310}
\bibinfo{author}{Chen, C.}, \bibinfo{author}{Miao, J.}, \bibinfo{author}{Wu, D.}, \bibinfo{author}{Zhong, A.}, \bibinfo{author}{Yan, Z.}, \bibinfo{author}{Kim, S.}, \bibinfo{author}{Hu, J.}, \bibinfo{author}{Liu, Z.}, \bibinfo{author}{Sun, L.}, \bibinfo{author}{Li, X.}, \bibinfo{author}{Liu, T.}, \bibinfo{author}{Heng, P.A.}, \bibinfo{author}{Li, Q.}, \bibinfo{year}{2024}b.
\newblock \bibinfo{title}{Ma-sam: Modality-agnostic sam adaptation for 3d medical image segmentation}.
\newblock \bibinfo{journal}{Medical Image Analysis} \bibinfo{volume}{98}, \bibinfo{pages}{103310}.
\newblock \URLprefix \url{https://www.sciencedirect.com/science/article/pii/S1361841524002354}, \DOIprefix\doi{https://doi.org/10.1016/j.media.2024.103310}.
\bibitem[{Chen et~al.(2021)Chen, Lu, Yu, Luo, Adeli, Wang, Lu, Yuille and Zhou}]{chen2021transunet}
\bibinfo{author}{Chen, J.}, \bibinfo{author}{Lu, Y.}, \bibinfo{author}{Yu, Q.}, \bibinfo{author}{Luo, X.}, \bibinfo{author}{Adeli, E.}, \bibinfo{author}{Wang, Y.}, \bibinfo{author}{Lu, L.}, \bibinfo{author}{Yuille, A.L.}, \bibinfo{author}{Zhou, Y.}, \bibinfo{year}{2021}.
\newblock \bibinfo{title}{Transunet: Transformers make strong encoders for medical image segmentation}.
\newblock \bibinfo{journal}{arXiv preprint arXiv:2102.04306} .
\bibitem[{Chen et~al.(2024c)Chen, Mei, Li, Lu, Yu, Wei, Luo, Xie, Adeli, Wang, Lungren, Zhang, Xing, Lu, Yuille and Zhou}]{CHEN2024103280}
\bibinfo{author}{Chen, J.}, \bibinfo{author}{Mei, J.}, \bibinfo{author}{Li, X.}, \bibinfo{author}{Lu, Y.}, \bibinfo{author}{Yu, Q.}, \bibinfo{author}{Wei, Q.}, \bibinfo{author}{Luo, X.}, \bibinfo{author}{Xie, Y.}, \bibinfo{author}{Adeli, E.}, \bibinfo{author}{Wang, Y.}, \bibinfo{author}{Lungren, M.P.}, \bibinfo{author}{Zhang, S.}, \bibinfo{author}{Xing, L.}, \bibinfo{author}{Lu, L.}, \bibinfo{author}{Yuille, A.}, \bibinfo{author}{Zhou, Y.}, \bibinfo{year}{2024}c.
\newblock \bibinfo{title}{Transunet: Rethinking the u-net architecture design for medical image segmentation through the lens of transformers}.
\newblock \bibinfo{journal}{Medical Image Analysis} \bibinfo{volume}{97}, \bibinfo{pages}{103280}.
\newblock \URLprefix \url{https://www.sciencedirect.com/science/article/pii/S1361841524002056}, \DOIprefix\doi{https://doi.org/10.1016/j.media.2024.103280}.
\bibitem[{Chen et~al.(2017)Chen, Papandreou, Schroff and Adam}]{chen2017rethinking}
\bibinfo{author}{Chen, L.C.}, \bibinfo{author}{Papandreou, G.}, \bibinfo{author}{Schroff, F.}, \bibinfo{author}{Adam, H.}, \bibinfo{year}{2017}.
\newblock \bibinfo{title}{Rethinking atrous convolution for semantic image segmentation}.
\newblock \bibinfo{journal}{arXiv preprint arXiv:1706.05587} .
\bibitem[{Chen(2025)}]{chen2025much}
\bibinfo{author}{Chen, S.}, \bibinfo{year}{2025}.
\newblock \bibinfo{title}{How much energy will ai really consume? the good, the bad and the unknown}.
\newblock \bibinfo{journal}{Nature} \bibinfo{volume}{639}, \bibinfo{pages}{22--24}.
\bibitem[{Chen et~al.(2019)Chen, Ma and Zheng}]{chen2019med3dtransferlearning3d}
\bibinfo{author}{Chen, S.}, \bibinfo{author}{Ma, K.}, \bibinfo{author}{Zheng, Y.}, \bibinfo{year}{2019}.
\newblock \bibinfo{title}{Med3d: Transfer learning for 3d medical image analysis}.
\newblock \URLprefix \url{https://arxiv.org/abs/1904.00625}, \href{http://arxiv.org/abs/1904.00625}{{\tt arXiv:1904.00625}}.
\bibitem[{Chen et~al.(2024d)Chen, Gao, Zhu, Shao, Lu, Han and Xie}]{10510478}
\bibinfo{author}{Chen, Y.}, \bibinfo{author}{Gao, Y.}, \bibinfo{author}{Zhu, L.}, \bibinfo{author}{Shao, W.}, \bibinfo{author}{Lu, Y.}, \bibinfo{author}{Han, H.}, \bibinfo{author}{Xie, Z.}, \bibinfo{year}{2024}d.
\newblock \bibinfo{title}{Pcnet: Prior category network for ct universal segmentation model}.
\newblock \bibinfo{journal}{IEEE Transactions on Medical Imaging} \bibinfo{volume}{43}, \bibinfo{pages}{3319--3330}.
\newblock \DOIprefix\doi{10.1109/TMI.2024.3395349}.
\bibitem[{Cheng et~al.(2024)Cheng, Fu, Ye, Wang, Li, Wang, Li, Yao, Chen, Li, Su, Zhu and He}]{cheng2024interactivemedicalimagesegmentation}
\bibinfo{author}{Cheng, J.}, \bibinfo{author}{Fu, B.}, \bibinfo{author}{Ye, J.}, \bibinfo{author}{Wang, G.}, \bibinfo{author}{Li, T.}, \bibinfo{author}{Wang, H.}, \bibinfo{author}{Li, R.}, \bibinfo{author}{Yao, H.}, \bibinfo{author}{Chen, J.}, \bibinfo{author}{Li, J.}, \bibinfo{author}{Su, Y.}, \bibinfo{author}{Zhu, M.}, \bibinfo{author}{He, J.}, \bibinfo{year}{2024}.
\newblock \bibinfo{title}{Interactive medical image segmentation: A benchmark dataset and baseline}.
\newblock \URLprefix \url{https://arxiv.org/abs/2411.12814}, \href{http://arxiv.org/abs/2411.12814}{{\tt arXiv:2411.12814}}.
\bibitem[{Cheng et~al.(2023)Cheng, Ye, Deng, Chen, Li, Wang, Su, Huang, Chen, Jiang, Sun, He, Zhang, Zhu and Qiao}]{cheng2023sammed2d}
\bibinfo{author}{Cheng, J.}, \bibinfo{author}{Ye, J.}, \bibinfo{author}{Deng, Z.}, \bibinfo{author}{Chen, J.}, \bibinfo{author}{Li, T.}, \bibinfo{author}{Wang, H.}, \bibinfo{author}{Su, Y.}, \bibinfo{author}{Huang, Z.}, \bibinfo{author}{Chen, J.}, \bibinfo{author}{Jiang, L.}, \bibinfo{author}{Sun, H.}, \bibinfo{author}{He, J.}, \bibinfo{author}{Zhang, S.}, \bibinfo{author}{Zhu, M.}, \bibinfo{author}{Qiao, Y.}, \bibinfo{year}{2023}.
\newblock \bibinfo{title}{Sam-med2d}.
\newblock \URLprefix \url{https://arxiv.org/abs/2308.16184}, \href{http://arxiv.org/abs/2308.16184}{{\tt arXiv:2308.16184}}.
\bibitem[{Cox et~al.(2024a)Cox, Liu, Stolte, Yang, Liu, See, Ju and Fang}]{COX2024103301}
\bibinfo{author}{Cox, J.}, \bibinfo{author}{Liu, P.}, \bibinfo{author}{Stolte, S.E.}, \bibinfo{author}{Yang, Y.}, \bibinfo{author}{Liu, K.}, \bibinfo{author}{See, K.B.}, \bibinfo{author}{Ju, H.}, \bibinfo{author}{Fang, R.}, \bibinfo{year}{2024}a.
\newblock \bibinfo{title}{Brainsegfounder: Towards 3d foundation models for neuroimage segmentation}.
\newblock \bibinfo{journal}{Medical Image Analysis} \bibinfo{volume}{97}, \bibinfo{pages}{103301}.
\newblock \URLprefix \url{https://www.sciencedirect.com/science/article/pii/S1361841524002263}, \DOIprefix\doi{https://doi.org/10.1016/j.media.2024.103301}.
\bibitem[{Cox et~al.(2024b)Cox, Liu, Stolte, Yang, Liu, See, Ju and Fang}]{cox2024brainsegfounder3dfoundationmodels}
\bibinfo{author}{Cox, J.}, \bibinfo{author}{Liu, P.}, \bibinfo{author}{Stolte, S.E.}, \bibinfo{author}{Yang, Y.}, \bibinfo{author}{Liu, K.}, \bibinfo{author}{See, K.B.}, \bibinfo{author}{Ju, H.}, \bibinfo{author}{Fang, R.}, \bibinfo{year}{2024}b.
\newblock \bibinfo{title}{Brainsegfounder: Towards 3d foundation models for neuroimage segmentation}.
\newblock \URLprefix \url{https://arxiv.org/abs/2406.10395}, \href{http://arxiv.org/abs/2406.10395}{{\tt arXiv:2406.10395}}.
\bibitem[{Devlin et~al.(2019)Devlin, Chang, Lee and Toutanova}]{devlin2019bert}
\bibinfo{author}{Devlin, J.}, \bibinfo{author}{Chang, M.W.}, \bibinfo{author}{Lee, K.}, \bibinfo{author}{Toutanova, K.}, \bibinfo{year}{2019}.
\newblock \bibinfo{title}{Bert: Pre-training of deep bidirectional transformers for language understanding}, in: \bibinfo{booktitle}{Proceedings of the 2019 conference of the North American chapter of the association for computational linguistics: human language technologies, volume 1 (long and short papers)}, pp. \bibinfo{pages}{4171--4186}.
\bibitem[{Dong et~al.(2024a)Dong, Wang, Chen, Sun, Song, Liu and Cui}]{Dong2024}
\bibinfo{author}{Dong, G.}, \bibinfo{author}{Wang, Z.}, \bibinfo{author}{Chen, Y.}, \bibinfo{author}{Sun, Y.}, \bibinfo{author}{Song, H.}, \bibinfo{author}{Liu, L.}, \bibinfo{author}{Cui, H.}, \bibinfo{year}{2024}a.
\newblock \bibinfo{title}{An efficient segment anything model for the segmentation of medical images}.
\newblock \bibinfo{journal}{Scientific Reports} \bibinfo{volume}{14}, \bibinfo{pages}{19425}.
\newblock \URLprefix \url{https://doi.org/10.1038/s41598-024-70288-8}, \DOIprefix\doi{10.1038/s41598-024-70288-8}.
\bibitem[{Dong et~al.(2024b)Dong, Gu, Chen, Yang, Chen and Mazurowski}]{dong2024segmentmodel2application}
\bibinfo{author}{Dong, H.}, \bibinfo{author}{Gu, H.}, \bibinfo{author}{Chen, Y.}, \bibinfo{author}{Yang, J.}, \bibinfo{author}{Chen, Y.}, \bibinfo{author}{Mazurowski, M.A.}, \bibinfo{year}{2024}b.
\newblock \bibinfo{title}{Segment anything model 2: an application to 2d and 3d medical images}.
\newblock \URLprefix \url{https://arxiv.org/abs/2408.00756}, \href{http://arxiv.org/abs/2408.00756}{{\tt arXiv:2408.00756}}.
\bibitem[{Dosovitskiy et~al.(2020)Dosovitskiy, Beyer, Kolesnikov, Weissenborn, Zhai, Unterthiner, Dehghani, Minderer, Heigold, Gelly, Uszkoreit and Houlsby}]{dosovitskiy2020image}
\bibinfo{author}{Dosovitskiy, A.}, \bibinfo{author}{Beyer, L.}, \bibinfo{author}{Kolesnikov, A.}, \bibinfo{author}{Weissenborn, D.}, \bibinfo{author}{Zhai, X.}, \bibinfo{author}{Unterthiner, T.}, \bibinfo{author}{Dehghani, M.}, \bibinfo{author}{Minderer, M.}, \bibinfo{author}{Heigold, G.}, \bibinfo{author}{Gelly, S.}, \bibinfo{author}{Uszkoreit, J.}, \bibinfo{author}{Houlsby, N.}, \bibinfo{year}{2020}.
\newblock \bibinfo{title}{An image is worth 16x16 words: Transformers for image recognition at scale}.
\newblock \DOIprefix\doi{10.48550/ARXIV.2010.11929}.
\bibitem[{Du et~al.(2024a)Du, Bai, Huang and Zhao}]{du2024segvol_arxiv}
\bibinfo{author}{Du, Y.}, \bibinfo{author}{Bai, F.}, \bibinfo{author}{Huang, T.}, \bibinfo{author}{Zhao, B.}, \bibinfo{year}{2024}a.
\newblock \bibinfo{title}{Segvol: Universal and interactive volumetric medical image segmentation}.
\newblock \URLprefix \url{https://arxiv.org/abs/2311.13385}, \href{http://arxiv.org/abs/2311.13385}{{\tt arXiv:2311.13385}}.
\bibitem[{Du et~al.(2024b)Du, BAI, Huang and Zhao}]{du2024segvol}
\bibinfo{author}{Du, Y.}, \bibinfo{author}{BAI, F.}, \bibinfo{author}{Huang, T.}, \bibinfo{author}{Zhao, B.}, \bibinfo{year}{2024}b.
\newblock \bibinfo{title}{Segvol: Universal and interactive volumetric medical image segmentation}, in: \bibinfo{booktitle}{The Thirty-eighth Annual Conference on Neural Information Processing Systems}.
\newblock \URLprefix \url{https://openreview.net/forum?id=105ZuvpdyW}.
\bibitem[{Du et~al.(2025)Du, Bai, Huang and Zhao}]{du2025segvoluniversalinteractivevolumetric}
\bibinfo{author}{Du, Y.}, \bibinfo{author}{Bai, F.}, \bibinfo{author}{Huang, T.}, \bibinfo{author}{Zhao, B.}, \bibinfo{year}{2025}.
\newblock \bibinfo{title}{Segvol: Universal and interactive volumetric medical image segmentation}.
\newblock \URLprefix \url{https://arxiv.org/abs/2311.13385}, \href{http://arxiv.org/abs/2311.13385}{{\tt arXiv:2311.13385}}.
\bibitem[{Dutt et~al.(2023)Dutt, Ericsson, Sanchez, Tsaftaris and Hospedales}]{dutt2023parameter}
\bibinfo{author}{Dutt, R.}, \bibinfo{author}{Ericsson, L.}, \bibinfo{author}{Sanchez, P.}, \bibinfo{author}{Tsaftaris, S.A.}, \bibinfo{author}{Hospedales, T.}, \bibinfo{year}{2023}.
\newblock \bibinfo{title}{Parameter-efficient fine-tuning for medical image analysis: The missed opportunity}.
\newblock \bibinfo{journal}{arXiv preprint arXiv:2305.08252} .
\bibitem[{{European Union}(2016)}]{gdpr}
\bibinfo{author}{{European Union}}, \bibinfo{year}{2016}.
\newblock \bibinfo{title}{General data protection regulation}.
\newblock \bibinfo{howpublished}{\url{https://eur-lex.europa.eu/eli/reg/2016/679/oj}}.
\newblock \bibinfo{note}{Accessed: 2025-04-04}.
\bibitem[{Feng et~al.(2023)Feng, Zhu and Yu}]{feng2023cheaplunchmedicalimage}
\bibinfo{author}{Feng, W.}, \bibinfo{author}{Zhu, L.}, \bibinfo{author}{Yu, L.}, \bibinfo{year}{2023}.
\newblock \bibinfo{title}{Cheap lunch for medical image segmentation by fine-tuning sam on few exemplars}.
\newblock \URLprefix \url{https://arxiv.org/abs/2308.14133}, \href{http://arxiv.org/abs/2308.14133}{{\tt arXiv:2308.14133}}.
\bibitem[{Feng et~al.(2024)Feng, Zhu and Yu}]{10.1007/978-3-031-76160-7_2}
\bibinfo{author}{Feng, W.}, \bibinfo{author}{Zhu, L.}, \bibinfo{author}{Yu, L.}, \bibinfo{year}{2024}.
\newblock \bibinfo{title}{Cheap lunch for medical image segmentation by fine-tuning sam on few exemplars}, in: \bibinfo{editor}{Baid, U.}, \bibinfo{editor}{Dorent, R.}, \bibinfo{editor}{Malec, S.}, \bibinfo{editor}{Pytlarz, M.}, \bibinfo{editor}{Su, R.}, \bibinfo{editor}{Wijethilake, N.}, \bibinfo{editor}{Bakas, S.}, \bibinfo{editor}{Crimi, A.} (Eds.), \bibinfo{booktitle}{Brainlesion: Glioma, Multiple Sclerosis, Stroke and Traumatic Brain Injuries}, \bibinfo{publisher}{Springer Nature Switzerland}, \bibinfo{address}{Cham}. pp. \bibinfo{pages}{13--22}.
\bibitem[{Fischer et~al.(2024)Fischer, Bartler and Yang}]{fischer2024prompt}
\bibinfo{author}{Fischer, M.}, \bibinfo{author}{Bartler, A.}, \bibinfo{author}{Yang, B.}, \bibinfo{year}{2024}.
\newblock \bibinfo{title}{Prompt tuning for parameter-efficient medical image segmentation}.
\newblock \bibinfo{journal}{Medical Image Analysis} \bibinfo{volume}{91}, \bibinfo{pages}{103024}.
\bibitem[{{Future of Life Institute}(2024)}]{aiact}
\bibinfo{author}{{Future of Life Institute}}, \bibinfo{year}{2024}.
\newblock \bibinfo{title}{The eu artificial intelligence act}.
\newblock \bibinfo{howpublished}{\url{https://artificialintelligenceact.eu/}}.
\newblock \bibinfo{note}{Accessed: 2025-04-04}.
\bibitem[{Gan et~al.(2025)Gan, Ramlee, Wang and Shimizu}]{gan2025review}
\bibinfo{author}{Gan, H.S.}, \bibinfo{author}{Ramlee, M.H.}, \bibinfo{author}{Wang, Z.}, \bibinfo{author}{Shimizu, A.}, \bibinfo{year}{2025}.
\newblock \bibinfo{title}{A review on medical image segmentation: Datasets, technical models, challenges and solutions}.
\newblock \bibinfo{journal}{Wiley Interdisciplinary Reviews: Data Mining and Knowledge Discovery} \bibinfo{volume}{15}, \bibinfo{pages}{e1574}.
\bibitem[{Gao(2024)}]{10658004}
\bibinfo{author}{Gao, Y.}, \bibinfo{year}{2024}.
\newblock \bibinfo{title}{Training like a medical resident: Context-prior learning toward universal medical image segmentation}, in: \bibinfo{booktitle}{2024 IEEE/CVF Conference on Computer Vision and Pattern Recognition (CVPR)}, pp. \bibinfo{pages}{11194--11204}.
\newblock \DOIprefix\doi{10.1109/CVPR52733.2024.01064}.
\bibitem[{Gao et~al.(2024a)Gao, Li, Liu, Zhou, Zhang and Metaxas}]{gao2024traininglikemedicalresident}
\bibinfo{author}{Gao, Y.}, \bibinfo{author}{Li, Z.}, \bibinfo{author}{Liu, D.}, \bibinfo{author}{Zhou, M.}, \bibinfo{author}{Zhang, S.}, \bibinfo{author}{Metaxas, D.N.}, \bibinfo{year}{2024}a.
\newblock \bibinfo{title}{Training like a medical resident: Context-prior learning toward universal medical image segmentation}.
\newblock \URLprefix \url{https://arxiv.org/abs/2306.02416}, \href{http://arxiv.org/abs/2306.02416}{{\tt arXiv:2306.02416}}.
\bibitem[{Gao et~al.(2024b)Gao, Xia, Hu, Wang and Gao}]{gao2024desamdecoupledsegmentmodel}
\bibinfo{author}{Gao, Y.}, \bibinfo{author}{Xia, W.}, \bibinfo{author}{Hu, D.}, \bibinfo{author}{Wang, W.}, \bibinfo{author}{Gao, X.}, \bibinfo{year}{2024}b.
\newblock \bibinfo{title}{Desam: Decoupled segment anything model for generalizable medical image segmentation}.
\newblock \URLprefix \url{https://arxiv.org/abs/2306.00499}, \href{http://arxiv.org/abs/2306.00499}{{\tt arXiv:2306.00499}}.
\bibitem[{Gao et~al.(2024c)Gao, Xia, Hu, Wang and Gao}]{10.1007/978-3-031-72390-2_48}
\bibinfo{author}{Gao, Y.}, \bibinfo{author}{Xia, W.}, \bibinfo{author}{Hu, D.}, \bibinfo{author}{Wang, W.}, \bibinfo{author}{Gao, X.}, \bibinfo{year}{2024}c.
\newblock \bibinfo{title}{Desam: Decoupled segment anything model for generalizable medical image segmentation}, in: \bibinfo{editor}{Linguraru, M.G.}, \bibinfo{editor}{Dou, Q.}, \bibinfo{editor}{Feragen, A.}, \bibinfo{editor}{Giannarou, S.}, \bibinfo{editor}{Glocker, B.}, \bibinfo{editor}{Lekadir, K.}, \bibinfo{editor}{Schnabel, J.A.} (Eds.), \bibinfo{booktitle}{Medical Image Computing and Computer Assisted Intervention -- MICCAI 2024}, \bibinfo{publisher}{Springer Nature Switzerland}, \bibinfo{address}{Cham}. pp. \bibinfo{pages}{509--519}.
\bibitem[{Gao et~al.(2022)Gao, Zhou, Liu, Yan, Zhang and Metaxas}]{gao2022data}
\bibinfo{author}{Gao, Y.}, \bibinfo{author}{Zhou, M.}, \bibinfo{author}{Liu, D.}, \bibinfo{author}{Yan, Z.}, \bibinfo{author}{Zhang, S.}, \bibinfo{author}{Metaxas, D.N.}, \bibinfo{year}{2022}.
\newblock \bibinfo{title}{A data-scalable transformer for medical image segmentation: architecture, model efficiency, and benchmark}.
\newblock \bibinfo{journal}{arXiv preprint arXiv:2203.00131} .
\bibitem[{Glocker et~al.(2023)Glocker, Jones, Roschewitz and Winzeck}]{glocker2023risk}
\bibinfo{author}{Glocker, B.}, \bibinfo{author}{Jones, C.}, \bibinfo{author}{Roschewitz, M.}, \bibinfo{author}{Winzeck, S.}, \bibinfo{year}{2023}.
\newblock \bibinfo{title}{Risk of bias in chest radiography deep learning foundation models}.
\newblock \bibinfo{journal}{Radiology: Artificial Intelligence} \bibinfo{volume}{5}, \bibinfo{pages}{e230060}.
\bibitem[{Gong et~al.(2024a)Gong, Zhong, Ma, Li, Wang, Zhang, Heng and Dou}]{3DSAM-adapter}
\bibinfo{author}{Gong, S.}, \bibinfo{author}{Zhong, Y.}, \bibinfo{author}{Ma, W.}, \bibinfo{author}{Li, J.}, \bibinfo{author}{Wang, Z.}, \bibinfo{author}{Zhang, J.}, \bibinfo{author}{Heng, P.A.}, \bibinfo{author}{Dou, Q.}, \bibinfo{year}{2024}a.
\newblock \bibinfo{title}{3dsam-adapter: Holistic adaptation of sam from 2d to 3d for promptable tumor segmentation}.
\newblock \bibinfo{journal}{Medical Image Analysis} \bibinfo{volume}{98}, \bibinfo{pages}{103324}.
\newblock \URLprefix \url{https://www.sciencedirect.com/science/article/pii/S1361841524002494}, \DOIprefix\doi{https://doi.org/10.1016/j.media.2024.103324}.
\bibitem[{Gong et~al.(2024b)Gong, Zhong, Ma, Li, Wang, Zhang, Heng and Dou}]{GONG2024103324}
\bibinfo{author}{Gong, S.}, \bibinfo{author}{Zhong, Y.}, \bibinfo{author}{Ma, W.}, \bibinfo{author}{Li, J.}, \bibinfo{author}{Wang, Z.}, \bibinfo{author}{Zhang, J.}, \bibinfo{author}{Heng, P.A.}, \bibinfo{author}{Dou, Q.}, \bibinfo{year}{2024}b.
\newblock \bibinfo{title}{3dsam-adapter: Holistic adaptation of sam from 2d to 3d for promptable tumor segmentation}.
\newblock \bibinfo{journal}{Medical Image Analysis} \bibinfo{volume}{98}, \bibinfo{pages}{103324}.
\newblock \URLprefix \url{https://www.sciencedirect.com/science/article/pii/S1361841524002494}, \DOIprefix\doi{https://doi.org/10.1016/j.media.2024.103324}.
\bibitem[{Gu et~al.(2024)Gu, Wu, Tang, Mai, Shu, Li and Chen}]{10540651}
\bibinfo{author}{Gu, Y.}, \bibinfo{author}{Wu, Q.}, \bibinfo{author}{Tang, H.}, \bibinfo{author}{Mai, X.}, \bibinfo{author}{Shu, H.}, \bibinfo{author}{Li, B.}, \bibinfo{author}{Chen, Y.}, \bibinfo{year}{2024}.
\newblock \bibinfo{title}{Lesam: Adapt segment anything model for medical lesion segmentation}.
\newblock \bibinfo{journal}{IEEE Journal of Biomedical and Health Informatics} \bibinfo{volume}{28}, \bibinfo{pages}{6031--6041}.
\newblock \DOIprefix\doi{10.1109/JBHI.2024.3406871}.
\bibitem[{Guo et~al.(2025)Guo, Yang, Zhang, Song, Zhang, Xu, Zhu, Ma, Wang, Bi et~al.}]{guo2025deepseek}
\bibinfo{author}{Guo, D.}, \bibinfo{author}{Yang, D.}, \bibinfo{author}{Zhang, H.}, \bibinfo{author}{Song, J.}, \bibinfo{author}{Zhang, R.}, \bibinfo{author}{Xu, R.}, \bibinfo{author}{Zhu, Q.}, \bibinfo{author}{Ma, S.}, \bibinfo{author}{Wang, P.}, \bibinfo{author}{Bi, X.}, et~al., \bibinfo{year}{2025}.
\newblock \bibinfo{title}{Deepseek-r1: Incentivizing reasoning capability in llms via reinforcement learning}.
\newblock \bibinfo{journal}{arXiv preprint arXiv:2501.12948} .
\bibitem[{Guo et~al.(2022)Guo, Geng, Geng, Wang and Dai}]{guo2022combination}
\bibinfo{author}{Guo, W.l.}, \bibinfo{author}{Geng, A.k.}, \bibinfo{author}{Geng, C.}, \bibinfo{author}{Wang, J.}, \bibinfo{author}{Dai, Y.k.}, \bibinfo{year}{2022}.
\newblock \bibinfo{title}{Combination of unet++ and resnest to classify chronic inflammation of the choledochal cystic wall in patients with pancreaticobiliary maljunction}.
\newblock \bibinfo{journal}{The British Journal of Radiology} \bibinfo{volume}{95}.
\newblock \DOIprefix\doi{10.1259/bjr.20201189}.
\bibitem[{Hatamizadeh et~al.(2022)Hatamizadeh, Nath, Tang, Yang, Roth and Xu}]{SwinUNETR_arxiv}
\bibinfo{author}{Hatamizadeh, A.}, \bibinfo{author}{Nath, V.}, \bibinfo{author}{Tang, Y.}, \bibinfo{author}{Yang, D.}, \bibinfo{author}{Roth, H.}, \bibinfo{author}{Xu, D.}, \bibinfo{year}{2022}.
\newblock \bibinfo{title}{Swin unetr: Swin transformers for semantic segmentation of brain tumors in mri images}.
\newblock \URLprefix \url{https://arxiv.org/abs/2201.01266}, \href{http://arxiv.org/abs/2201.01266}{{\tt arXiv:2201.01266}}.
\bibitem[{Hatamizadeh et~al.(2021)Hatamizadeh, Tang, Nath, Yang, Myronenko, Landman, Roth and Xu}]{unetr}
\bibinfo{author}{Hatamizadeh, A.}, \bibinfo{author}{Tang, Y.}, \bibinfo{author}{Nath, V.}, \bibinfo{author}{Yang, D.}, \bibinfo{author}{Myronenko, A.}, \bibinfo{author}{Landman, B.}, \bibinfo{author}{Roth, H.}, \bibinfo{author}{Xu, D.}, \bibinfo{year}{2021}.
\newblock \bibinfo{title}{Unetr: Transformers for 3d medical image segmentation}.
\newblock \URLprefix \url{https://arxiv.org/abs/2103.10504}, \href{http://arxiv.org/abs/2103.10504}{{\tt arXiv:2103.10504}}.
\bibitem[{He et~al.(2022)He, Chen, Xie, Li, Doll{\'a}r and Girshick}]{he2022masked}
\bibinfo{author}{He, K.}, \bibinfo{author}{Chen, X.}, \bibinfo{author}{Xie, S.}, \bibinfo{author}{Li, Y.}, \bibinfo{author}{Doll{\'a}r, P.}, \bibinfo{author}{Girshick, R.}, \bibinfo{year}{2022}.
\newblock \bibinfo{title}{Masked autoencoders are scalable vision learners}, in: \bibinfo{booktitle}{Proceedings of the IEEE/CVF conference on computer vision and pattern recognition}, pp. \bibinfo{pages}{16000--16009}.
\bibitem[{He et~al.(2025)He, Mao, Lin, Ruan, Lan, Feng and Cambria}]{he2025survey}
\bibinfo{author}{He, K.}, \bibinfo{author}{Mao, R.}, \bibinfo{author}{Lin, Q.}, \bibinfo{author}{Ruan, Y.}, \bibinfo{author}{Lan, X.}, \bibinfo{author}{Feng, M.}, \bibinfo{author}{Cambria, E.}, \bibinfo{year}{2025}.
\newblock \bibinfo{title}{A survey of large language models for healthcare: from data, technology, and applications to accountability and ethics}.
\newblock \bibinfo{journal}{Information Fusion} , \bibinfo{pages}{102963}.
\bibitem[{He et~al.(2016)He, Zhang, Ren and Sun}]{he2016deep}
\bibinfo{author}{He, K.}, \bibinfo{author}{Zhang, X.}, \bibinfo{author}{Ren, S.}, \bibinfo{author}{Sun, J.}, \bibinfo{year}{2016}.
\newblock \bibinfo{title}{Deep residual learning for image recognition}, in: \bibinfo{booktitle}{Proceedings of the IEEE conference on computer vision and pattern recognition}, pp. \bibinfo{pages}{770--778}.
\bibitem[{He et~al.(2024)He, Huang, Jiang, Nie, Wang, Wang and Chen}]{he2024foundation}
\bibinfo{author}{He, Y.}, \bibinfo{author}{Huang, F.}, \bibinfo{author}{Jiang, X.}, \bibinfo{author}{Nie, Y.}, \bibinfo{author}{Wang, M.}, \bibinfo{author}{Wang, J.}, \bibinfo{author}{Chen, H.}, \bibinfo{year}{2024}.
\newblock \bibinfo{title}{Foundation model for advancing healthcare: challenges, opportunities and future directions}.
\newblock \bibinfo{journal}{IEEE Reviews in Biomedical Engineering} .
\bibitem[{Hendrycks and Gimpel(2016)}]{hendrycks2016gaussian}
\bibinfo{author}{Hendrycks, D.}, \bibinfo{author}{Gimpel, K.}, \bibinfo{year}{2016}.
\newblock \bibinfo{title}{Gaussian error linear units (gelus)}.
\newblock \bibinfo{journal}{arXiv preprint arXiv:1606.08415} .
\bibitem[{Hoffmann et~al.(2022)Hoffmann, Borgeaud, Mensch, Buchatskaya, Cai, Rutherford, Casas, Hendricks, Welbl, Clark et~al.}]{hoffmann2022training}
\bibinfo{author}{Hoffmann, J.}, \bibinfo{author}{Borgeaud, S.}, \bibinfo{author}{Mensch, A.}, \bibinfo{author}{Buchatskaya, E.}, \bibinfo{author}{Cai, T.}, \bibinfo{author}{Rutherford, E.}, \bibinfo{author}{Casas, D.d.L.}, \bibinfo{author}{Hendricks, L.A.}, \bibinfo{author}{Welbl, J.}, \bibinfo{author}{Clark, A.}, et~al., \bibinfo{year}{2022}.
\newblock \bibinfo{title}{Training compute-optimal large language models}.
\newblock \bibinfo{journal}{arXiv preprint arXiv:2203.15556} .
\bibitem[{Houlsby et~al.(2019)Houlsby, Giurgiu, Jastrzebski, Morrone, De~Laroussilhe, Gesmundo, Attariyan and Gelly}]{houlsby2019parameter}
\bibinfo{author}{Houlsby, N.}, \bibinfo{author}{Giurgiu, A.}, \bibinfo{author}{Jastrzebski, S.}, \bibinfo{author}{Morrone, B.}, \bibinfo{author}{De~Laroussilhe, Q.}, \bibinfo{author}{Gesmundo, A.}, \bibinfo{author}{Attariyan, M.}, \bibinfo{author}{Gelly, S.}, \bibinfo{year}{2019}.
\newblock \bibinfo{title}{Parameter-efficient transfer learning for nlp}, in: \bibinfo{booktitle}{International conference on machine learning}, \bibinfo{organization}{PMLR}. pp. \bibinfo{pages}{2790--2799}.
\bibitem[{Hu et~al.(2022)Hu, Shen, Wallis, Allen-Zhu, Li, Wang, Wang, Chen et~al.}]{hu2022lora}
\bibinfo{author}{Hu, E.J.}, \bibinfo{author}{Shen, Y.}, \bibinfo{author}{Wallis, P.}, \bibinfo{author}{Allen-Zhu, Z.}, \bibinfo{author}{Li, Y.}, \bibinfo{author}{Wang, S.}, \bibinfo{author}{Wang, L.}, \bibinfo{author}{Chen, W.}, et~al., \bibinfo{year}{2022}.
\newblock \bibinfo{title}{Lora: Low-rank adaptation of large language models.}
\newblock \bibinfo{journal}{ICLR} \bibinfo{volume}{1}, \bibinfo{pages}{3}.
\bibitem[{Hu et~al.(2025)Hu, Li, Jain, Lin and Chen}]{10829779}
\bibinfo{author}{Hu, J.}, \bibinfo{author}{Li, Y.}, \bibinfo{author}{Jain, R.K.}, \bibinfo{author}{Lin, L.}, \bibinfo{author}{Chen, Y.w.}, \bibinfo{year}{2025}.
\newblock \bibinfo{title}{Spa: Leveraging the sam with spatial priors adapter for enhanced medical image segmentation}.
\newblock \bibinfo{journal}{IEEE Journal of Biomedical and Health Informatics} , \bibinfo{pages}{1--15}\DOIprefix\doi{10.1109/JBHI.2025.3526174}.
\bibitem[{Huang et~al.(2024a)Huang, Zhou, Fu, Zhang, Zhou, Gong and Liang}]{huang2024learnablepromptingsaminducedknowledge}
\bibinfo{author}{Huang, K.}, \bibinfo{author}{Zhou, T.}, \bibinfo{author}{Fu, H.}, \bibinfo{author}{Zhang, Y.}, \bibinfo{author}{Zhou, Y.}, \bibinfo{author}{Gong, C.}, \bibinfo{author}{Liang, D.}, \bibinfo{year}{2024}a.
\newblock \bibinfo{title}{Learnable prompting sam-induced knowledge distillation for semi-supervised medical image segmentation}.
\newblock \URLprefix \url{https://arxiv.org/abs/2412.13742}, \href{http://arxiv.org/abs/2412.13742}{{\tt arXiv:2412.13742}}.
\bibitem[{Huang et~al.(2025)Huang, Zhou, Fu, Zhang, Zhou, Gong and Liang}]{10843257}
\bibinfo{author}{Huang, K.}, \bibinfo{author}{Zhou, T.}, \bibinfo{author}{Fu, H.}, \bibinfo{author}{Zhang, Y.}, \bibinfo{author}{Zhou, Y.}, \bibinfo{author}{Gong, C.}, \bibinfo{author}{Liang, D.}, \bibinfo{year}{2025}.
\newblock \bibinfo{title}{Learnable prompting sam-induced knowledge distillation for semi-supervised medical image segmentation}.
\newblock \bibinfo{journal}{IEEE Transactions on Medical Imaging} \bibinfo{volume}{44}, \bibinfo{pages}{2295--2306}.
\newblock \DOIprefix\doi{10.1109/TMI.2025.3530097}.
\bibitem[{Huang et~al.(2023a)Huang, Deng, Li, Yuan and Fu}]{huang2022missformer}
\bibinfo{author}{Huang, X.}, \bibinfo{author}{Deng, Z.}, \bibinfo{author}{Li, D.}, \bibinfo{author}{Yuan, X.}, \bibinfo{author}{Fu, Y.}, \bibinfo{year}{2023}a.
\newblock \bibinfo{title}{Missformer: An effective transformer for 2d medical image segmentation}.
\newblock \bibinfo{journal}{IEEE Transactions on Medical Imaging} \bibinfo{volume}{42}, \bibinfo{pages}{1484--1494}.
\newblock \DOIprefix\doi{10.1109/tmi.2022.3230943}.
\bibitem[{Huang et~al.(2024b)Huang, Yang, Liu, Zhou, Chang, Zhou, Chen, Yu, Chen, Chen, Liu, Chi, Hu, Yue, Li, Grau, Fan, Dong and Ni}]{sam_huang}
\bibinfo{author}{Huang, Y.}, \bibinfo{author}{Yang, X.}, \bibinfo{author}{Liu, L.}, \bibinfo{author}{Zhou, H.}, \bibinfo{author}{Chang, A.}, \bibinfo{author}{Zhou, X.}, \bibinfo{author}{Chen, R.}, \bibinfo{author}{Yu, J.}, \bibinfo{author}{Chen, J.}, \bibinfo{author}{Chen, C.}, \bibinfo{author}{Liu, S.}, \bibinfo{author}{Chi, H.}, \bibinfo{author}{Hu, X.}, \bibinfo{author}{Yue, K.}, \bibinfo{author}{Li, L.}, \bibinfo{author}{Grau, V.}, \bibinfo{author}{Fan, D.P.}, \bibinfo{author}{Dong, F.}, \bibinfo{author}{Ni, D.}, \bibinfo{year}{2024}b.
\newblock \bibinfo{title}{Segment anything model for medical images?}
\newblock \bibinfo{journal}{Medical Image Analysis} \bibinfo{volume}{92}, \bibinfo{pages}{103061}.
\newblock \URLprefix \url{https://www.sciencedirect.com/science/article/pii/S1361841523003213}, \DOIprefix\doi{https://doi.org/10.1016/j.media.2023.103061}.
\bibitem[{Huang et~al.(2023b)Huang, Wang, Deng, Ye, Su, Sun, He, Gu, Gu, Zhang and Qiao}]{huang2023stunetscalabletransferablemedical}
\bibinfo{author}{Huang, Z.}, \bibinfo{author}{Wang, H.}, \bibinfo{author}{Deng, Z.}, \bibinfo{author}{Ye, J.}, \bibinfo{author}{Su, Y.}, \bibinfo{author}{Sun, H.}, \bibinfo{author}{He, J.}, \bibinfo{author}{Gu, Y.}, \bibinfo{author}{Gu, L.}, \bibinfo{author}{Zhang, S.}, \bibinfo{author}{Qiao, Y.}, \bibinfo{year}{2023}b.
\newblock \bibinfo{title}{Stu-net: Scalable and transferable medical image segmentation models empowered by large-scale supervised pre-training}.
\newblock \URLprefix \url{https://arxiv.org/abs/2304.06716}, \href{http://arxiv.org/abs/2304.06716}{{\tt arXiv:2304.06716}}.
\bibitem[{IMDRF(2025)}]{imdrf}
\bibinfo{author}{IMDRF}, \bibinfo{year}{2025}.
\newblock \bibinfo{title}{International medical devices regulators forum}.
\newblock \bibinfo{howpublished}{\url{https://www.imdrf.org/}}.
\newblock \bibinfo{note}{Accessed: 2025-04-04}.
\bibitem[{Isensee et~al.(2021)Isensee, Jaeger, Kohl, Petersen and Maier-Hein}]{Isensee2021}
\bibinfo{author}{Isensee, F.}, \bibinfo{author}{Jaeger, P.F.}, \bibinfo{author}{Kohl, S.A.A.}, \bibinfo{author}{Petersen, J.}, \bibinfo{author}{Maier-Hein, K.H.}, \bibinfo{year}{2021}.
\newblock \bibinfo{title}{nnu-net: a self-configuring method for deep learning-based biomedical image segmentation}.
\newblock \bibinfo{journal}{Nature Methods} \bibinfo{volume}{18}, \bibinfo{pages}{203--211}.
\newblock \URLprefix \url{https://doi.org/10.1038/s41592-020-01008-z}, \DOIprefix\doi{10.1038/s41592-020-01008-z}.
\bibitem[{{ISO}(2022)}]{ISO/IEC27559:2022}
\bibinfo{author}{{ISO}}, \bibinfo{year}{2022}.
\newblock \bibinfo{title}{Iso/iec 27559:2022. information security, cybersecurity and privacy protection – privacy enhancing data de-identification framework}.
\newblock \bibinfo{howpublished}{\url{https://www.iso.org/standard/71677.html}}.
\newblock \bibinfo{note}{Accessed: 2025-10-04}.
\bibitem[{{ISO}(2024)}]{ISO/IEC29100:2024}
\bibinfo{author}{{ISO}}, \bibinfo{year}{2024}.
\newblock \bibinfo{title}{Iso/iec 29100:2024. information technology — security techniques — privacy framework}.
\newblock \bibinfo{howpublished}{\url{https://www.iso.org/standard/85938.html}}.
\newblock \bibinfo{note}{Accessed: 2025-10-04}.
\bibitem[{Jia et~al.(2022)Jia, Tang, Chen, Cardie, Belongie, Hariharan and Lim}]{jia2022visual}
\bibinfo{author}{Jia, M.}, \bibinfo{author}{Tang, L.}, \bibinfo{author}{Chen, B.C.}, \bibinfo{author}{Cardie, C.}, \bibinfo{author}{Belongie, S.}, \bibinfo{author}{Hariharan, B.}, \bibinfo{author}{Lim, S.N.}, \bibinfo{year}{2022}.
\newblock \bibinfo{title}{Visual prompt tuning}, in: \bibinfo{booktitle}{European conference on computer vision}, \bibinfo{organization}{Springer}. pp. \bibinfo{pages}{709--727}.
\bibitem[{Jiang et~al.(2022)Jiang, Tyagi, Tringale, Crane and Veeraraghavan}]{10.1007/978-3-031-16440-8_53}
\bibinfo{author}{Jiang, J.}, \bibinfo{author}{Tyagi, N.}, \bibinfo{author}{Tringale, K.}, \bibinfo{author}{Crane, C.}, \bibinfo{author}{Veeraraghavan, H.}, \bibinfo{year}{2022}.
\newblock \bibinfo{title}{Self-supervised 3d anatomy segmentation using self-distilled masked image transformer (smit)}, in: \bibinfo{editor}{Wang, L.}, \bibinfo{editor}{Dou, Q.}, \bibinfo{editor}{Fletcher, P.T.}, \bibinfo{editor}{Speidel, S.}, \bibinfo{editor}{Li, S.} (Eds.), \bibinfo{booktitle}{Medical Image Computing and Computer Assisted Intervention -- MICCAI 2022}, \bibinfo{publisher}{Springer Nature Switzerland}, \bibinfo{address}{Cham}. pp. \bibinfo{pages}{556--566}.
\bibitem[{Jiaxing and Hao(2025)}]{jiaxing2025sam2}
\bibinfo{author}{Jiaxing, Z.}, \bibinfo{author}{Hao, T.}, \bibinfo{year}{2025}.
\newblock \bibinfo{title}{Sam2 for image and video segmentation: A comprehensive survey}.
\newblock \bibinfo{journal}{arXiv preprint arXiv:2503.12781} .
\bibitem[{{Kari Briski, Nvidia}(2025)}]{scalinglaws}
\bibinfo{author}{{Kari Briski, Nvidia}}, \bibinfo{year}{2025}.
\newblock \bibinfo{title}{How scaling laws drive smarter, more powerful ai}.
\newblock \bibinfo{howpublished}{\url{https://blogs.nvidia.com/blog/ai-scaling-laws/}}.
\newblock \bibinfo{note}{Accessed: 2025-04-04}.
\bibitem[{Khan et~al.(2025)Khan, Leem, See, Wong, Zhang and Fang}]{khan2025comprehensive}
\bibinfo{author}{Khan, W.}, \bibinfo{author}{Leem, S.}, \bibinfo{author}{See, K.B.}, \bibinfo{author}{Wong, J.K.}, \bibinfo{author}{Zhang, S.}, \bibinfo{author}{Fang, R.}, \bibinfo{year}{2025}.
\newblock \bibinfo{title}{A comprehensive survey of foundation models in medicine}.
\newblock \bibinfo{journal}{IEEE Reviews in Biomedical Engineering} .
\bibitem[{Kirillov et~al.(2023a)Kirillov, Mintun, Ravi, Mao, Rolland, Gustafson, Xiao, Whitehead, Berg, Lo, Doll{\'a}r and Girshick}]{kirillov2023segany}
\bibinfo{author}{Kirillov, A.}, \bibinfo{author}{Mintun, E.}, \bibinfo{author}{Ravi, N.}, \bibinfo{author}{Mao, H.}, \bibinfo{author}{Rolland, C.}, \bibinfo{author}{Gustafson, L.}, \bibinfo{author}{Xiao, T.}, \bibinfo{author}{Whitehead, S.}, \bibinfo{author}{Berg, A.C.}, \bibinfo{author}{Lo, W.Y.}, \bibinfo{author}{Doll{\'a}r, P.}, \bibinfo{author}{Girshick, R.}, \bibinfo{year}{2023}a.
\newblock \bibinfo{title}{Segment anything}.
\newblock \bibinfo{journal}{arXiv:2304.02643} .
\bibitem[{Kirillov et~al.(2023b)Kirillov, Mintun, Ravi, Mao, Rolland, Gustafson, Xiao, Whitehead, Berg, Lo, Dollr and Girshick}]{kirillov2023segment}
\bibinfo{author}{Kirillov, A.}, \bibinfo{author}{Mintun, E.}, \bibinfo{author}{Ravi, N.}, \bibinfo{author}{Mao, H.}, \bibinfo{author}{Rolland, C.}, \bibinfo{author}{Gustafson, L.}, \bibinfo{author}{Xiao, T.}, \bibinfo{author}{Whitehead, S.}, \bibinfo{author}{Berg, A.C.}, \bibinfo{author}{Lo, W.Y.}, \bibinfo{author}{Dollr, P.}, \bibinfo{author}{Girshick, R.}, \bibinfo{year}{2023}b.
\newblock \bibinfo{title}{Segment anything}.
\newblock \URLprefix \url{https://arxiv.org/abs/2304.02643}, \href{http://arxiv.org/abs/2304.02643}{{\tt arXiv:2304.02643}}.
\bibitem[{Kirillov et~al.(2023c)Kirillov, Mintun, Ravi, Mao, Rolland, Gustafson, Xiao, Whitehead, Berg, Lo, Dollr and Girshick}]{10378323}
\bibinfo{author}{Kirillov, A.}, \bibinfo{author}{Mintun, E.}, \bibinfo{author}{Ravi, N.}, \bibinfo{author}{Mao, H.}, \bibinfo{author}{Rolland, C.}, \bibinfo{author}{Gustafson, L.}, \bibinfo{author}{Xiao, T.}, \bibinfo{author}{Whitehead, S.}, \bibinfo{author}{Berg, A.C.}, \bibinfo{author}{Lo, W.Y.}, \bibinfo{author}{Dollr, P.}, \bibinfo{author}{Girshick, R.}, \bibinfo{year}{2023}c.
\newblock \bibinfo{title}{Segment anything}, in: \bibinfo{booktitle}{2023 IEEE/CVF International Conference on Computer Vision (ICCV)}, pp. \bibinfo{pages}{3992--4003}.
\newblock \DOIprefix\doi{10.1109/ICCV51070.2023.00371}.
\bibitem[{Lee et~al.(2023)Lee, Bao, Huo and Landman}]{3duxnet_arxiv}
\bibinfo{author}{Lee, H.H.}, \bibinfo{author}{Bao, S.}, \bibinfo{author}{Huo, Y.}, \bibinfo{author}{Landman, B.A.}, \bibinfo{year}{2023}.
\newblock \bibinfo{title}{3d ux-net: A large kernel volumetric convnet modernizing hierarchical transformer for medical image segmentation}.
\newblock \URLprefix \url{https://arxiv.org/abs/2209.15076}, \href{http://arxiv.org/abs/2209.15076}{{\tt arXiv:2209.15076}}.
\bibitem[{Lee et~al.(2024)Lee, Gu, Zhao, Xu, Yang, Usuyama, Wong, Wei, Landman, Huo et~al.}]{lee2024foundation}
\bibinfo{author}{Lee, H.H.}, \bibinfo{author}{Gu, Y.}, \bibinfo{author}{Zhao, T.}, \bibinfo{author}{Xu, Y.}, \bibinfo{author}{Yang, J.}, \bibinfo{author}{Usuyama, N.}, \bibinfo{author}{Wong, C.}, \bibinfo{author}{Wei, M.}, \bibinfo{author}{Landman, B.A.}, \bibinfo{author}{Huo, Y.}, et~al., \bibinfo{year}{2024}.
\newblock \bibinfo{title}{Foundation models for biomedical image segmentation: A survey}.
\newblock \bibinfo{journal}{arXiv preprint arXiv:2401.07654} .
\bibitem[{Lei et~al.(2024)Lei, Wei, Zhang, Li and Zhang}]{lei2024medlsamlocalizesegmentmodel}
\bibinfo{author}{Lei, W.}, \bibinfo{author}{Wei, X.}, \bibinfo{author}{Zhang, X.}, \bibinfo{author}{Li, K.}, \bibinfo{author}{Zhang, S.}, \bibinfo{year}{2024}.
\newblock \bibinfo{title}{Medlsam: Localize and segment anything model for 3d ct images}.
\newblock \URLprefix \url{https://arxiv.org/abs/2306.14752}, \href{http://arxiv.org/abs/2306.14752}{{\tt arXiv:2306.14752}}.
\bibitem[{Lei et~al.(2025)Lei, Xu, Li, Zhang and Zhang}]{LEI2025103370}
\bibinfo{author}{Lei, W.}, \bibinfo{author}{Xu, W.}, \bibinfo{author}{Li, K.}, \bibinfo{author}{Zhang, X.}, \bibinfo{author}{Zhang, S.}, \bibinfo{year}{2025}.
\newblock \bibinfo{title}{Medlsam: Localize and segment anything model for 3d ct images}.
\newblock \bibinfo{journal}{Medical Image Analysis} \bibinfo{volume}{99}, \bibinfo{pages}{103370}.
\newblock \URLprefix \url{https://www.sciencedirect.com/science/article/pii/S1361841524002950}, \DOIprefix\doi{https://doi.org/10.1016/j.media.2024.103370}.
\bibitem[{Lekadir et~al.(2025)Lekadir, Frangi, Porras, Glocker, Cintas, Langlotz, Weicken, Asselbergs, Prior, Collins et~al.}]{lekadir2025future}
\bibinfo{author}{Lekadir, K.}, \bibinfo{author}{Frangi, A.F.}, \bibinfo{author}{Porras, A.R.}, \bibinfo{author}{Glocker, B.}, \bibinfo{author}{Cintas, C.}, \bibinfo{author}{Langlotz, C.P.}, \bibinfo{author}{Weicken, E.}, \bibinfo{author}{Asselbergs, F.W.}, \bibinfo{author}{Prior, F.}, \bibinfo{author}{Collins, G.S.}, et~al., \bibinfo{year}{2025}.
\newblock \bibinfo{title}{Future-ai: International consensus guideline for trustworthy and deployable artificial intelligence in healthcare}.
\newblock \bibinfo{journal}{bmj} \bibinfo{volume}{388}.
\bibitem[{Li et~al.(2025a)Li, Ayache and Delingette}]{li2025generative}
\bibinfo{author}{Li, H.}, \bibinfo{author}{Ayache, N.}, \bibinfo{author}{Delingette, H.}, \bibinfo{year}{2025}a.
\newblock \bibinfo{title}{Generative medical image anonymization based on latent code projection and optimization}.
\newblock \bibinfo{journal}{arXiv preprint arXiv:2501.09114} .
\bibitem[{Li et~al.(2022)Li, Wang, Chen, Zhang, Zha, Wang and Yu}]{li2022transbtsv2betterefficientvolumetric}
\bibinfo{author}{Li, J.}, \bibinfo{author}{Wang, W.}, \bibinfo{author}{Chen, C.}, \bibinfo{author}{Zhang, T.}, \bibinfo{author}{Zha, S.}, \bibinfo{author}{Wang, J.}, \bibinfo{author}{Yu, H.}, \bibinfo{year}{2022}.
\newblock \bibinfo{title}{Transbtsv2: Towards better and more efficient volumetric segmentation of medical images}.
\newblock \URLprefix \url{https://arxiv.org/abs/2201.12785}, \href{http://arxiv.org/abs/2201.12785}{{\tt arXiv:2201.12785}}.
\bibitem[{Li et~al.(2025b)Li, Qi, Yu, Huo, Shi and Gao}]{li2025stitchingfinetuningretrainingsamenabled}
\bibinfo{author}{Li, S.}, \bibinfo{author}{Qi, L.}, \bibinfo{author}{Yu, Q.}, \bibinfo{author}{Huo, J.}, \bibinfo{author}{Shi, Y.}, \bibinfo{author}{Gao, Y.}, \bibinfo{year}{2025}b.
\newblock \bibinfo{title}{Stitching, fine-tuning, re-training: A sam-enabled framework for semi-supervised 3d medical image segmentation}.
\newblock \URLprefix \url{https://arxiv.org/abs/2403.11229}, \href{http://arxiv.org/abs/2403.11229}{{\tt arXiv:2403.11229}}.
\bibitem[{Li et~al.(2025c)Li, Qi, Yu, Huo, Shi and Gao}]{10847777}
\bibinfo{author}{Li, S.}, \bibinfo{author}{Qi, L.}, \bibinfo{author}{Yu, Q.}, \bibinfo{author}{Huo, J.}, \bibinfo{author}{Shi, Y.}, \bibinfo{author}{Gao, Y.}, \bibinfo{year}{2025}c.
\newblock \bibinfo{title}{Stitching, fine-tuning, re-training: A sam-enabled framework for semi-supervised 3d medical image segmentation}.
\newblock \bibinfo{journal}{IEEE Transactions on Medical Imaging} , \bibinfo{pages}{1--1}\DOIprefix\doi{10.1109/TMI.2025.3532084}.
\bibitem[{Li et~al.(2025d)Li, Li, Jiang, Wang, Qiao, Feng, Luo and Zhao}]{li2025vision}
\bibinfo{author}{Li, X.}, \bibinfo{author}{Li, L.}, \bibinfo{author}{Jiang, Y.}, \bibinfo{author}{Wang, H.}, \bibinfo{author}{Qiao, X.}, \bibinfo{author}{Feng, T.}, \bibinfo{author}{Luo, H.}, \bibinfo{author}{Zhao, Y.}, \bibinfo{year}{2025}d.
\newblock \bibinfo{title}{Vision-language models in medical image analysis: From simple fusion to general large models}.
\newblock \bibinfo{journal}{Information Fusion} , \bibinfo{pages}{102995}.
\bibitem[{Li et~al.(2024)Li, Zhao, Zhang, Wu, Liu, Jiang, Cao, Xu, Li, Dai et~al.}]{li2024artificial}
\bibinfo{author}{Li, X.}, \bibinfo{author}{Zhao, L.}, \bibinfo{author}{Zhang, L.}, \bibinfo{author}{Wu, Z.}, \bibinfo{author}{Liu, Z.}, \bibinfo{author}{Jiang, H.}, \bibinfo{author}{Cao, C.}, \bibinfo{author}{Xu, S.}, \bibinfo{author}{Li, Y.}, \bibinfo{author}{Dai, H.}, et~al., \bibinfo{year}{2024}.
\newblock \bibinfo{title}{Artificial general intelligence for medical imaging analysis}.
\newblock \bibinfo{journal}{IEEE Reviews in Biomedical Engineering} .
\bibitem[{Liang et~al.(2025)Liang, Pu, Huang, Li, Wang, Ma and Chang}]{liang2025vision}
\bibinfo{author}{Liang, P.}, \bibinfo{author}{Pu, B.}, \bibinfo{author}{Huang, H.}, \bibinfo{author}{Li, Y.}, \bibinfo{author}{Wang, H.}, \bibinfo{author}{Ma, W.}, \bibinfo{author}{Chang, Q.}, \bibinfo{year}{2025}.
\newblock \bibinfo{title}{Vision foundation models in medical image analysis: Advances and challenges}.
\newblock \bibinfo{journal}{arXiv preprint arXiv:2502.14584} .
\bibitem[{Lin et~al.(2025)Lin, Zou, Deng, Wong, Aviles-Rivero, Fan, Lee, Hu and Qin}]{LIN20251}
\bibinfo{author}{Lin, H.}, \bibinfo{author}{Zou, J.}, \bibinfo{author}{Deng, S.}, \bibinfo{author}{Wong, K.P.}, \bibinfo{author}{Aviles-Rivero, A.I.}, \bibinfo{author}{Fan, Y.}, \bibinfo{author}{Lee, A.P.W.}, \bibinfo{author}{Hu, X.}, \bibinfo{author}{Qin, J.}, \bibinfo{year}{2025}.
\newblock \bibinfo{title}{Volumetric medical image segmentation via fully 3d adaptation of segment anything model}.
\newblock \bibinfo{journal}{Biocybernetics and Biomedical Engineering} \bibinfo{volume}{45}, \bibinfo{pages}{1--10}.
\newblock \URLprefix \url{https://www.sciencedirect.com/science/article/pii/S0208521624000846}, \DOIprefix\doi{https://doi.org/10.1016/j.bbe.2024.11.001}.
\bibitem[{Liu et~al.(2023a)Liu, Zhang, Chen, Xiao, Lu, Landman, Yuan, Yuille, Tang and Zhou}]{10376801}
\bibinfo{author}{Liu, J.}, \bibinfo{author}{Zhang, Y.}, \bibinfo{author}{Chen, J.N.}, \bibinfo{author}{Xiao, J.}, \bibinfo{author}{Lu, Y.}, \bibinfo{author}{Landman, B.A.}, \bibinfo{author}{Yuan, Y.}, \bibinfo{author}{Yuille, A.}, \bibinfo{author}{Tang, Y.}, \bibinfo{author}{Zhou, Z.}, \bibinfo{year}{2023}a.
\newblock \bibinfo{title}{Clip-driven universal model for organ segmentation and tumor detection}, in: \bibinfo{booktitle}{2023 IEEE/CVF International Conference on Computer Vision (ICCV)}, pp. \bibinfo{pages}{21095--21107}.
\newblock \DOIprefix\doi{10.1109/ICCV51070.2023.01934}.
\bibitem[{Liu et~al.(2023b)Liu, Zhang, Chen, Xiao, Lu, Landman, Yuan, Yuille, Tang and Zhou}]{Liu_2023}
\bibinfo{author}{Liu, J.}, \bibinfo{author}{Zhang, Y.}, \bibinfo{author}{Chen, J.N.}, \bibinfo{author}{Xiao, J.}, \bibinfo{author}{Lu, Y.}, \bibinfo{author}{Landman, B.A.}, \bibinfo{author}{Yuan, Y.}, \bibinfo{author}{Yuille, A.}, \bibinfo{author}{Tang, Y.}, \bibinfo{author}{Zhou, Z.}, \bibinfo{year}{2023}b.
\newblock \bibinfo{title}{Clip-driven universal model for organ segmentation and tumor detection}, in: \bibinfo{booktitle}{2023 IEEE/CVF International Conference on Computer Vision (ICCV)}, \bibinfo{publisher}{IEEE}. p. \bibinfo{pages}{2109521107}.
\newblock \URLprefix \url{http://dx.doi.org/10.1109/ICCV51070.2023.01934}, \DOIprefix\doi{10.1109/iccv51070.2023.01934}.
\bibitem[{Liu et~al.(2024a)Liu, Zhang, Wang, Yavuz, Chen, Yuan, Li, Yang, Yuille, Tang and Zhou}]{LIU2024103226}
\bibinfo{author}{Liu, J.}, \bibinfo{author}{Zhang, Y.}, \bibinfo{author}{Wang, K.}, \bibinfo{author}{Yavuz, M.C.}, \bibinfo{author}{Chen, X.}, \bibinfo{author}{Yuan, Y.}, \bibinfo{author}{Li, H.}, \bibinfo{author}{Yang, Y.}, \bibinfo{author}{Yuille, A.}, \bibinfo{author}{Tang, Y.}, \bibinfo{author}{Zhou, Z.}, \bibinfo{year}{2024}a.
\newblock \bibinfo{title}{Universal and extensible language-vision models for organ segmentation and tumor detection from abdominal computed tomography}.
\newblock \bibinfo{journal}{Medical Image Analysis} \bibinfo{volume}{97}, \bibinfo{pages}{103226}.
\newblock \URLprefix \url{https://www.sciencedirect.com/science/article/pii/S1361841524001518}, \DOIprefix\doi{https://doi.org/10.1016/j.media.2024.103226}.
\bibitem[{Liu et~al.(2022a)Liu, Deng, Wang, Hui, Li, Li, Luo, Sun, Quan, Yang, Hao, Xiao, Zhao, Wu and Zhou}]{liu2022universalsegmentation33anatomies}
\bibinfo{author}{Liu, P.}, \bibinfo{author}{Deng, Y.}, \bibinfo{author}{Wang, C.}, \bibinfo{author}{Hui, Y.}, \bibinfo{author}{Li, Q.}, \bibinfo{author}{Li, J.}, \bibinfo{author}{Luo, S.}, \bibinfo{author}{Sun, M.}, \bibinfo{author}{Quan, Q.}, \bibinfo{author}{Yang, S.}, \bibinfo{author}{Hao, Y.}, \bibinfo{author}{Xiao, H.}, \bibinfo{author}{Zhao, C.}, \bibinfo{author}{Wu, X.}, \bibinfo{author}{Zhou, S.K.}, \bibinfo{year}{2022}a.
\newblock \bibinfo{title}{Universal segmentation of 33 anatomies}.
\newblock \URLprefix \url{https://arxiv.org/abs/2203.02098}, \href{http://arxiv.org/abs/2203.02098}{{\tt arXiv:2203.02098}}.
\bibitem[{Liu et~al.(2024b)Liu, Zhang, She, Kheradmand and Armand}]{liu2024sammsegmentmedicalmodel}
\bibinfo{author}{Liu, Y.}, \bibinfo{author}{Zhang, J.}, \bibinfo{author}{She, Z.}, \bibinfo{author}{Kheradmand, A.}, \bibinfo{author}{Armand, M.}, \bibinfo{year}{2024}b.
\newblock \bibinfo{title}{Samm (segment any medical model): A 3d slicer integration to sam}.
\newblock \URLprefix \url{https://arxiv.org/abs/2304.05622}, \href{http://arxiv.org/abs/2304.05622}{{\tt arXiv:2304.05622}}.
\bibitem[{Liu et~al.(2024c)Liu, Zhang, Yue and Guo}]{scanext}
\bibinfo{author}{Liu, Y.}, \bibinfo{author}{Zhang, Z.}, \bibinfo{author}{Yue, J.}, \bibinfo{author}{Guo, W.}, \bibinfo{year}{2024}c.
\newblock \bibinfo{title}{Scanext: Enhancing 3d medical image segmentation with dual attention network and depth-wise convolution}.
\newblock \bibinfo{journal}{Heliyon} \bibinfo{volume}{10}, \bibinfo{pages}{e26775}.
\newblock \URLprefix \url{https://www.sciencedirect.com/science/article/pii/S2405844024028068}, \DOIprefix\doi{https://doi.org/10.1016/j.heliyon.2024.e26775}.
\bibitem[{Liu et~al.(2022b)Liu, Hu, Lin, Yao, Xie, Wei, Ning, Cao, Zhang, Dong, Wei and Guo}]{9879380SwinV2}
\bibinfo{author}{Liu, Z.}, \bibinfo{author}{Hu, H.}, \bibinfo{author}{Lin, Y.}, \bibinfo{author}{Yao, Z.}, \bibinfo{author}{Xie, Z.}, \bibinfo{author}{Wei, Y.}, \bibinfo{author}{Ning, J.}, \bibinfo{author}{Cao, Y.}, \bibinfo{author}{Zhang, Z.}, \bibinfo{author}{Dong, L.}, \bibinfo{author}{Wei, F.}, \bibinfo{author}{Guo, B.}, \bibinfo{year}{2022}b.
\newblock \bibinfo{title}{Swin transformer v2: Scaling up capacity and resolution}, in: \bibinfo{booktitle}{2022 IEEE/CVF Conference on Computer Vision and Pattern Recognition (CVPR)}, pp. \bibinfo{pages}{11999--12009}.
\newblock \DOIprefix\doi{10.1109/CVPR52688.2022.01170}.
\bibitem[{Liu et~al.(2021a)Liu, Lin, Cao, Hu, Wei, Zhang, Lin and Guo}]{9710580Swin}
\bibinfo{author}{Liu, Z.}, \bibinfo{author}{Lin, Y.}, \bibinfo{author}{Cao, Y.}, \bibinfo{author}{Hu, H.}, \bibinfo{author}{Wei, Y.}, \bibinfo{author}{Zhang, Z.}, \bibinfo{author}{Lin, S.}, \bibinfo{author}{Guo, B.}, \bibinfo{year}{2021}a.
\newblock \bibinfo{title}{Swin transformer: Hierarchical vision transformer using shifted windows}, in: \bibinfo{booktitle}{2021 IEEE/CVF International Conference on Computer Vision (ICCV)}, pp. \bibinfo{pages}{9992--10002}.
\newblock \DOIprefix\doi{10.1109/ICCV48922.2021.00986}.
\bibitem[{Liu et~al.(2021b)Liu, Lin, Cao, Hu, Wei, Zhang, Lin and Guo}]{liu2021swin}
\bibinfo{author}{Liu, Z.}, \bibinfo{author}{Lin, Y.}, \bibinfo{author}{Cao, Y.}, \bibinfo{author}{Hu, H.}, \bibinfo{author}{Wei, Y.}, \bibinfo{author}{Zhang, Z.}, \bibinfo{author}{Lin, S.}, \bibinfo{author}{Guo, B.}, \bibinfo{year}{2021}b.
\newblock \bibinfo{title}{Swin transformer: Hierarchical vision transformer using shifted windows}.
\newblock \DOIprefix\doi{10.48550/ARXIV.2103.14030}.
\bibitem[{Lu and Wang(2025)}]{lu2025integrating}
\bibinfo{author}{Lu, Y.}, \bibinfo{author}{Wang, A.}, \bibinfo{year}{2025}.
\newblock \bibinfo{title}{Integrating language into medical visual recognition and reasoning: A survey}.
\newblock \bibinfo{journal}{Medical Image Analysis} , \bibinfo{pages}{103514}.
\bibitem[{Ma et~al.(2024a)Ma, He, Li, Han, You and Wang}]{MedSAM_arxiv}
\bibinfo{author}{Ma, J.}, \bibinfo{author}{He, Y.}, \bibinfo{author}{Li, F.}, \bibinfo{author}{Han, L.}, \bibinfo{author}{You, C.}, \bibinfo{author}{Wang, B.}, \bibinfo{year}{2024}a.
\newblock \bibinfo{title}{Segment anything in medical images}.
\newblock \bibinfo{journal}{Nature Communications} \bibinfo{volume}{15}.
\newblock \URLprefix \url{http://dx.doi.org/10.1038/s41467-024-44824-z}, \DOIprefix\doi{10.1038/s41467-024-44824-z}.
\bibitem[{Ma et~al.(2024b)Ma, He, Li, Han, You and Wang}]{MedSAM_nature}
\bibinfo{author}{Ma, J.}, \bibinfo{author}{He, Y.}, \bibinfo{author}{Li, F.}, \bibinfo{author}{Han, L.}, \bibinfo{author}{You, C.}, \bibinfo{author}{Wang, B.}, \bibinfo{year}{2024}b.
\newblock \bibinfo{title}{Segment anything in medical images}.
\newblock \bibinfo{journal}{Nature Communications} \bibinfo{volume}{15}, \bibinfo{pages}{654}.
\newblock \URLprefix \url{https://doi.org/10.1038/s41467-024-44824-z}, \DOIprefix\doi{10.1038/s41467-024-44824-z}.
\bibitem[{Ma et~al.(2024c)Ma, He, Li, Han, You and Wang}]{Ma2024}
\bibinfo{author}{Ma, J.}, \bibinfo{author}{He, Y.}, \bibinfo{author}{Li, F.}, \bibinfo{author}{Han, L.}, \bibinfo{author}{You, C.}, \bibinfo{author}{Wang, B.}, \bibinfo{year}{2024}c.
\newblock \bibinfo{title}{Segment anything in medical images}.
\newblock \bibinfo{journal}{Nature Communications} \bibinfo{volume}{15}, \bibinfo{pages}{654}.
\newblock \URLprefix \url{https://doi.org/10.1038/s41467-024-44824-z}, \DOIprefix\doi{10.1038/s41467-024-44824-z}.
\bibitem[{Ma et~al.(2024d)Ma, Kim, Li, Baharoon, Asakereh, Lyu and Wang}]{ma2024segmentmedicalimagesvideos-sam2}
\bibinfo{author}{Ma, J.}, \bibinfo{author}{Kim, S.}, \bibinfo{author}{Li, F.}, \bibinfo{author}{Baharoon, M.}, \bibinfo{author}{Asakereh, R.}, \bibinfo{author}{Lyu, H.}, \bibinfo{author}{Wang, B.}, \bibinfo{year}{2024}d.
\newblock \bibinfo{title}{Segment anything in medical images and videos: Benchmark and deployment}.
\newblock \URLprefix \url{https://arxiv.org/abs/2408.03322}, \href{http://arxiv.org/abs/2408.03322}{{\tt arXiv:2408.03322}}.
\bibitem[{Ma et~al.(2025)Ma, Yang, Kim, Chen, Baharoon, Fallahpour, Asakereh, Lyu and Wang}]{ma2025medsam2segment3dmedical}
\bibinfo{author}{Ma, J.}, \bibinfo{author}{Yang, Z.}, \bibinfo{author}{Kim, S.}, \bibinfo{author}{Chen, B.}, \bibinfo{author}{Baharoon, M.}, \bibinfo{author}{Fallahpour, A.}, \bibinfo{author}{Asakereh, R.}, \bibinfo{author}{Lyu, H.}, \bibinfo{author}{Wang, B.}, \bibinfo{year}{2025}.
\newblock \bibinfo{title}{Medsam2: Segment anything in 3d medical images and videos}.
\newblock \URLprefix \url{https://arxiv.org/abs/2504.03600}, \href{http://arxiv.org/abs/2504.03600}{{\tt arXiv:2504.03600}}.
\bibitem[{Mazurowski et~al.(2023)Mazurowski, Dong, Gu, Yang, Konz and Zhang}]{sam_mazurowski}
\bibinfo{author}{Mazurowski, M.A.}, \bibinfo{author}{Dong, H.}, \bibinfo{author}{Gu, H.}, \bibinfo{author}{Yang, J.}, \bibinfo{author}{Konz, N.}, \bibinfo{author}{Zhang, Y.}, \bibinfo{year}{2023}.
\newblock \bibinfo{title}{Segment anything model for medical image analysis: An experimental study}.
\newblock \bibinfo{journal}{Medical Image Analysis} \bibinfo{volume}{89}, \bibinfo{pages}{102918}.
\newblock \URLprefix \url{https://www.sciencedirect.com/science/article/pii/S1361841523001780}, \DOIprefix\doi{https://doi.org/10.1016/j.media.2023.102918}.
\bibitem[{{MDR - 2017/746 – IVDR}(2017)}]{mdsw}
\bibinfo{author}{{MDR - 2017/746 – IVDR}}, \bibinfo{year}{2017}.
\newblock \bibinfo{title}{Medical device software}.
\newblock \bibinfo{howpublished}{\url{https://health.ec.europa.eu/system/files/2020-09/md_mdcg_2019_11_guidance_qualification_classification_software_en_0.pdf}}.
\newblock \bibinfo{note}{Accessed: 2025-04-04}.
\bibitem[{Mei et~al.(2022)Mei, Liu, Robson, Marinelli, Huang, Doshi, Jacobi, Cao, Link, Yang et~al.}]{mei2022radimagenet}
\bibinfo{author}{Mei, X.}, \bibinfo{author}{Liu, Z.}, \bibinfo{author}{Robson, P.M.}, \bibinfo{author}{Marinelli, B.}, \bibinfo{author}{Huang, M.}, \bibinfo{author}{Doshi, A.}, \bibinfo{author}{Jacobi, A.}, \bibinfo{author}{Cao, C.}, \bibinfo{author}{Link, K.E.}, \bibinfo{author}{Yang, T.}, et~al., \bibinfo{year}{2022}.
\newblock \bibinfo{title}{Radimagenet: an open radiologic deep learning research dataset for effective transfer learning}.
\newblock \bibinfo{journal}{Radiology: Artificial Intelligence} \bibinfo{volume}{4}, \bibinfo{pages}{e210315}.
\bibitem[{{Meta}(2025)}]{llama4}
\bibinfo{author}{{Meta}}, \bibinfo{year}{2025}.
\newblock \bibinfo{title}{The llama 4 herd: The beginning of a new era of natively multimodal ai innovation}.
\newblock \bibinfo{howpublished}{\url{https://ai.meta.com/blog/llama-4-multimodal-intelligence/}}.
\newblock \bibinfo{note}{Accessed: 2025-11-04}.
\bibitem[{Milletari et~al.(2016)Milletari, Navab and Ahmadi}]{milletari2016v}
\bibinfo{author}{Milletari, F.}, \bibinfo{author}{Navab, N.}, \bibinfo{author}{Ahmadi, S.A.}, \bibinfo{year}{2016}.
\newblock \bibinfo{title}{V-net: Fully convolutional neural networks for volumetric medical image segmentation}, in: \bibinfo{booktitle}{2016 Fourth International Conference on 3D Vision (3DV)}, \bibinfo{publisher}{IEEE}. pp. \bibinfo{pages}{565--571}.
\newblock \DOIprefix\doi{10.1109/3dv.2016.79}.
\bibitem[{Moglia et~al.(2025)Moglia, Cavicchioli, Mainardi and Cerveri}]{moglia2025deep}
\bibinfo{author}{Moglia, A.}, \bibinfo{author}{Cavicchioli, M.}, \bibinfo{author}{Mainardi, L.}, \bibinfo{author}{Cerveri, P.}, \bibinfo{year}{2025}.
\newblock \bibinfo{title}{Deep learning for pancreas segmentation on computed tomography: a systematic review}.
\newblock \bibinfo{journal}{Artificial Intelligence Review} \bibinfo{volume}{58}, \bibinfo{pages}{220}.
\bibitem[{Moor et~al.(2023)Moor, Banerjee, Abad, Krumholz, Leskovec, Topol and Rajpurkar}]{moor2023foundation}
\bibinfo{author}{Moor, M.}, \bibinfo{author}{Banerjee, O.}, \bibinfo{author}{Abad, Z.S.H.}, \bibinfo{author}{Krumholz, H.M.}, \bibinfo{author}{Leskovec, J.}, \bibinfo{author}{Topol, E.J.}, \bibinfo{author}{Rajpurkar, P.}, \bibinfo{year}{2023}.
\newblock \bibinfo{title}{Foundation models for generalist medical artificial intelligence}.
\newblock \bibinfo{journal}{Nature} \bibinfo{volume}{616}, \bibinfo{pages}{259--265}.
\bibitem[{Muennighoff et~al.(2025)Muennighoff, Yang, Shi, Li, Fei-Fei, Hajishirzi, Zettlemoyer, Liang, Cand{\`e}s and Hashimoto}]{muennighoff2025s1}
\bibinfo{author}{Muennighoff, N.}, \bibinfo{author}{Yang, Z.}, \bibinfo{author}{Shi, W.}, \bibinfo{author}{Li, X.L.}, \bibinfo{author}{Fei-Fei, L.}, \bibinfo{author}{Hajishirzi, H.}, \bibinfo{author}{Zettlemoyer, L.}, \bibinfo{author}{Liang, P.}, \bibinfo{author}{Cand{\`e}s, E.}, \bibinfo{author}{Hashimoto, T.}, \bibinfo{year}{2025}.
\newblock \bibinfo{title}{s1: Simple test-time scaling}.
\newblock \bibinfo{journal}{arXiv preprint arXiv:2501.19393} .
\bibitem[{Myronenko(2018)}]{myronenko20183d}
\bibinfo{author}{Myronenko, A.}, \bibinfo{year}{2018}.
\newblock \bibinfo{title}{3d mri brain tumor segmentation using autoencoder regularization}, in: \bibinfo{booktitle}{International MICCAI brainlesion workshop}, \bibinfo{organization}{Springer}. pp. \bibinfo{pages}{311--320}.
\bibitem[{{Office of the Assistant Secretary for Planning and Evaluation}(1996)}]{hipaa}
\bibinfo{author}{{Office of the Assistant Secretary for Planning and Evaluation}}, \bibinfo{year}{1996}.
\newblock \bibinfo{title}{Health insurance portability and accountability act of 1996}.
\newblock \bibinfo{howpublished}{\url{https://aspe.hhs.gov/reports/health-insurance-portability-accountability-act-1996}}.
\newblock \bibinfo{note}{Accessed: 2025-04-04}.
\bibitem[{{OpenAI}(2024)}]{inference-scaling}
\bibinfo{author}{{OpenAI}}, \bibinfo{year}{2024}.
\newblock \bibinfo{title}{Learning to reason with llms}.
\newblock \bibinfo{howpublished}{\url{https://openai.com/index/learning-to-reason-with-llms/}}.
\newblock \bibinfo{note}{Accessed: 2025-08-04}.
\bibitem[{Oquab et~al.(2024a)Oquab, Darcet, Moutakanni, Vo, Szafraniec, Khalidov, Fernandez, Haziza, Massa, El-Nouby, Assran, Ballas, Galuba, Howes, Huang, Li, Misra, Rabbat, Sharma, Synnaeve, Xu, Jegou, Mairal, Labatut, Joulin and Bojanowski}]{dinov2_arxiv}
\bibinfo{author}{Oquab, M.}, \bibinfo{author}{Darcet, T.}, \bibinfo{author}{Moutakanni, T.}, \bibinfo{author}{Vo, H.}, \bibinfo{author}{Szafraniec, M.}, \bibinfo{author}{Khalidov, V.}, \bibinfo{author}{Fernandez, P.}, \bibinfo{author}{Haziza, D.}, \bibinfo{author}{Massa, F.}, \bibinfo{author}{El-Nouby, A.}, \bibinfo{author}{Assran, M.}, \bibinfo{author}{Ballas, N.}, \bibinfo{author}{Galuba, W.}, \bibinfo{author}{Howes, R.}, \bibinfo{author}{Huang, P.Y.}, \bibinfo{author}{Li, S.W.}, \bibinfo{author}{Misra, I.}, \bibinfo{author}{Rabbat, M.}, \bibinfo{author}{Sharma, V.}, \bibinfo{author}{Synnaeve, G.}, \bibinfo{author}{Xu, H.}, \bibinfo{author}{Jegou, H.}, \bibinfo{author}{Mairal, J.}, \bibinfo{author}{Labatut, P.}, \bibinfo{author}{Joulin, A.}, \bibinfo{author}{Bojanowski, P.}, \bibinfo{year}{2024}a.
\newblock \bibinfo{title}{Dinov2: Learning robust visual features without supervision}.
\newblock \URLprefix \url{https://arxiv.org/abs/2304.07193}, \href{http://arxiv.org/abs/2304.07193}{{\tt arXiv:2304.07193}}.
\bibitem[{Oquab et~al.(2023)Oquab, Darcet, Moutakanni, Vo, Szafraniec, Khalidov, Fernandez, Haziza, Massa, El-Nouby et~al.}]{oquab2023dinov2}
\bibinfo{author}{Oquab, M.}, \bibinfo{author}{Darcet, T.}, \bibinfo{author}{Moutakanni, T.}, \bibinfo{author}{Vo, H.}, \bibinfo{author}{Szafraniec, M.}, \bibinfo{author}{Khalidov, V.}, \bibinfo{author}{Fernandez, P.}, \bibinfo{author}{Haziza, D.}, \bibinfo{author}{Massa, F.}, \bibinfo{author}{El-Nouby, A.}, et~al., \bibinfo{year}{2023}.
\newblock \bibinfo{title}{Dinov2: Learning robust visual features without supervision}.
\newblock \bibinfo{journal}{arXiv preprint arXiv:2304.07193} .
\bibitem[{Oquab et~al.(2024b)Oquab, Darcet, Moutakanni, Vo, Szafraniec, Khalidov, Fernandez, HAZIZA, Massa, El-Nouby, Assran, Ballas, Galuba, Howes, Huang, Li, Misra, Rabbat, Sharma, Synnaeve, Xu, Jegou, Mairal, Labatut, Joulin and Bojanowski}]{oquab2024dinov}
\bibinfo{author}{Oquab, M.}, \bibinfo{author}{Darcet, T.}, \bibinfo{author}{Moutakanni, T.}, \bibinfo{author}{Vo, H.V.}, \bibinfo{author}{Szafraniec, M.}, \bibinfo{author}{Khalidov, V.}, \bibinfo{author}{Fernandez, P.}, \bibinfo{author}{HAZIZA, D.}, \bibinfo{author}{Massa, F.}, \bibinfo{author}{El-Nouby, A.}, \bibinfo{author}{Assran, M.}, \bibinfo{author}{Ballas, N.}, \bibinfo{author}{Galuba, W.}, \bibinfo{author}{Howes, R.}, \bibinfo{author}{Huang, P.Y.}, \bibinfo{author}{Li, S.W.}, \bibinfo{author}{Misra, I.}, \bibinfo{author}{Rabbat, M.}, \bibinfo{author}{Sharma, V.}, \bibinfo{author}{Synnaeve, G.}, \bibinfo{author}{Xu, H.}, \bibinfo{author}{Jegou, H.}, \bibinfo{author}{Mairal, J.}, \bibinfo{author}{Labatut, P.}, \bibinfo{author}{Joulin, A.}, \bibinfo{author}{Bojanowski, P.}, \bibinfo{year}{2024}b.
\newblock \bibinfo{title}{{DINO}v2: Learning robust visual features without supervision}.
\newblock \bibinfo{journal}{Transactions on Machine Learning Research} \URLprefix \url{https://openreview.net/forum?id=a68SUt6zFt}. \bibinfo{note}{featured Certification}.
\bibitem[{P{\'e}rez-Garc{\'\i}a et~al.(2025)P{\'e}rez-Garc{\'\i}a, Sharma, Bond-Taylor, Bouzid, Salvatelli, Ilse, Bannur, Castro, Schwaighofer, Lungren et~al.}]{perez2025exploring}
\bibinfo{author}{P{\'e}rez-Garc{\'\i}a, F.}, \bibinfo{author}{Sharma, H.}, \bibinfo{author}{Bond-Taylor, S.}, \bibinfo{author}{Bouzid, K.}, \bibinfo{author}{Salvatelli, V.}, \bibinfo{author}{Ilse, M.}, \bibinfo{author}{Bannur, S.}, \bibinfo{author}{Castro, D.C.}, \bibinfo{author}{Schwaighofer, A.}, \bibinfo{author}{Lungren, M.P.}, et~al., \bibinfo{year}{2025}.
\newblock \bibinfo{title}{Exploring scalable medical image encoders beyond text supervision}.
\newblock \bibinfo{journal}{Nature Machine Intelligence} , \bibinfo{pages}{1--12}.
\bibitem[{Pezoulas et~al.(2024)Pezoulas, Zaridis, Mylona, Androutsos, Apostolidis, Tachos and Fotiadis}]{pezoulas2024synthetic}
\bibinfo{author}{Pezoulas, V.C.}, \bibinfo{author}{Zaridis, D.I.}, \bibinfo{author}{Mylona, E.}, \bibinfo{author}{Androutsos, C.}, \bibinfo{author}{Apostolidis, K.}, \bibinfo{author}{Tachos, N.S.}, \bibinfo{author}{Fotiadis, D.I.}, \bibinfo{year}{2024}.
\newblock \bibinfo{title}{Synthetic data generation methods in healthcare: A review on open-source tools and methods}.
\newblock \bibinfo{journal}{Computational and structural biotechnology journal} .
\bibitem[{Queiroz et~al.(2025)Queiroz, Carlos, Anjos and Berton}]{queiroz2025fair}
\bibinfo{author}{Queiroz, D.}, \bibinfo{author}{Carlos, A.}, \bibinfo{author}{Anjos, A.}, \bibinfo{author}{Berton, L.}, \bibinfo{year}{2025}.
\newblock \bibinfo{title}{Fair foundation models for medical image analysis: Challenges and perspectives}.
\newblock \bibinfo{journal}{arXiv preprint arXiv:2502.16841} .
\bibitem[{Radford et~al.(2021)Radford, Kim, Hallacy, Ramesh, Goh, Agarwal, Sastry, Askell, Mishkin, Clark, Krueger and Sutskever}]{clip2021openai}
\bibinfo{author}{Radford, A.}, \bibinfo{author}{Kim, J.W.}, \bibinfo{author}{Hallacy, C.}, \bibinfo{author}{Ramesh, A.}, \bibinfo{author}{Goh, G.}, \bibinfo{author}{Agarwal, S.}, \bibinfo{author}{Sastry, G.}, \bibinfo{author}{Askell, A.}, \bibinfo{author}{Mishkin, P.}, \bibinfo{author}{Clark, J.}, \bibinfo{author}{Krueger, G.}, \bibinfo{author}{Sutskever, I.}, \bibinfo{year}{2021}.
\newblock \bibinfo{title}{Learning transferable visual models from natural language supervision}.
\newblock \URLprefix \url{https://arxiv.org/abs/2103.00020}, \href{http://arxiv.org/abs/2103.00020}{{\tt arXiv:2103.00020}}.
\bibitem[{Radford et~al.(2018)Radford, Narasimhan, Salimans, Sutskever et~al.}]{radford2018improving}
\bibinfo{author}{Radford, A.}, \bibinfo{author}{Narasimhan, K.}, \bibinfo{author}{Salimans, T.}, \bibinfo{author}{Sutskever, I.}, et~al., \bibinfo{year}{2018}.
\newblock \bibinfo{title}{Improving language understanding by generative pre-training} .
\bibitem[{Ravi et~al.(2024a)Ravi, Gabeur, Hu, Hu, Ryali, Ma, Khedr, R{\"a}dle, Rolland, Gustafson, Mintun, Pan, Alwala, Carion, Wu, Girshick, Doll{\'a}r and Feichtenhofer}]{ravi2024sam2}
\bibinfo{author}{Ravi, N.}, \bibinfo{author}{Gabeur, V.}, \bibinfo{author}{Hu, Y.T.}, \bibinfo{author}{Hu, R.}, \bibinfo{author}{Ryali, C.}, \bibinfo{author}{Ma, T.}, \bibinfo{author}{Khedr, H.}, \bibinfo{author}{R{\"a}dle, R.}, \bibinfo{author}{Rolland, C.}, \bibinfo{author}{Gustafson, L.}, \bibinfo{author}{Mintun, E.}, \bibinfo{author}{Pan, J.}, \bibinfo{author}{Alwala, K.V.}, \bibinfo{author}{Carion, N.}, \bibinfo{author}{Wu, C.Y.}, \bibinfo{author}{Girshick, R.}, \bibinfo{author}{Doll{\'a}r, P.}, \bibinfo{author}{Feichtenhofer, C.}, \bibinfo{year}{2024}a.
\newblock \bibinfo{title}{Sam 2: Segment anything in images and videos}.
\newblock \bibinfo{journal}{arXiv preprint arXiv:2408.00714} \URLprefix \url{https://arxiv.org/abs/2408.00714}.
\bibitem[{Ravi et~al.(2025)Ravi, Gabeur, Hu, Hu, Ryali, Ma, Khedr, R{\"a}dle, Rolland, Gustafson, Mintun, Pan, Alwala, Carion, Wu, Girshick, Dollar and Feichtenhofer}]{ravi2025sam}
\bibinfo{author}{Ravi, N.}, \bibinfo{author}{Gabeur, V.}, \bibinfo{author}{Hu, Y.T.}, \bibinfo{author}{Hu, R.}, \bibinfo{author}{Ryali, C.}, \bibinfo{author}{Ma, T.}, \bibinfo{author}{Khedr, H.}, \bibinfo{author}{R{\"a}dle, R.}, \bibinfo{author}{Rolland, C.}, \bibinfo{author}{Gustafson, L.}, \bibinfo{author}{Mintun, E.}, \bibinfo{author}{Pan, J.}, \bibinfo{author}{Alwala, K.V.}, \bibinfo{author}{Carion, N.}, \bibinfo{author}{Wu, C.Y.}, \bibinfo{author}{Girshick, R.}, \bibinfo{author}{Dollar, P.}, \bibinfo{author}{Feichtenhofer, C.}, \bibinfo{year}{2025}.
\newblock \bibinfo{title}{{SAM} 2: Segment anything in images and videos}, in: \bibinfo{booktitle}{The Thirteenth International Conference on Learning Representations}.
\newblock \URLprefix \url{https://openreview.net/forum?id=Ha6RTeWMd0}.
\bibitem[{Ravi et~al.(2024b)Ravi, Gabeur, Hu, Hu, Ryali, Ma, Khedr, Rdle, Rolland, Gustafson, Mintun, Pan, Alwala, Carion, Wu, Girshick, Dollr and Feichtenhofer}]{ravi2024sam2segmentimages}
\bibinfo{author}{Ravi, N.}, \bibinfo{author}{Gabeur, V.}, \bibinfo{author}{Hu, Y.T.}, \bibinfo{author}{Hu, R.}, \bibinfo{author}{Ryali, C.}, \bibinfo{author}{Ma, T.}, \bibinfo{author}{Khedr, H.}, \bibinfo{author}{Rdle, R.}, \bibinfo{author}{Rolland, C.}, \bibinfo{author}{Gustafson, L.}, \bibinfo{author}{Mintun, E.}, \bibinfo{author}{Pan, J.}, \bibinfo{author}{Alwala, K.V.}, \bibinfo{author}{Carion, N.}, \bibinfo{author}{Wu, C.Y.}, \bibinfo{author}{Girshick, R.}, \bibinfo{author}{Dollr, P.}, \bibinfo{author}{Feichtenhofer, C.}, \bibinfo{year}{2024}b.
\newblock \bibinfo{title}{Sam 2: Segment anything in images and videos}.
\newblock \URLprefix \url{https://arxiv.org/abs/2408.00714}, \href{http://arxiv.org/abs/2408.00714}{{\tt arXiv:2408.00714}}.
\bibitem[{Ronneberger et~al.(2015)Ronneberger, Fischer and Brox}]{ronneberger2015u}
\bibinfo{author}{Ronneberger, O.}, \bibinfo{author}{Fischer, P.}, \bibinfo{author}{Brox, T.}, \bibinfo{year}{2015}.
\newblock \bibinfo{title}{U-net: Convolutional networks for biomedical image segmentation}, in: \bibinfo{booktitle}{Medical Image Computing and Computer-Assisted Intervention – MICCAI 2015}, \bibinfo{publisher}{Springer International Publishing}. pp. \bibinfo{pages}{234--241}.
\newblock \DOIprefix\doi{10.1007/978-3-319-24574-4\_28}.
\bibitem[{Roy et~al.(2024)Roy, Koehler, Ulrich, Baumgartner, Petersen, Isensee, Jaeger and Maier-Hein}]{mednext_arxiv}
\bibinfo{author}{Roy, S.}, \bibinfo{author}{Koehler, G.}, \bibinfo{author}{Ulrich, C.}, \bibinfo{author}{Baumgartner, M.}, \bibinfo{author}{Petersen, J.}, \bibinfo{author}{Isensee, F.}, \bibinfo{author}{Jaeger, P.F.}, \bibinfo{author}{Maier-Hein, K.}, \bibinfo{year}{2024}.
\newblock \bibinfo{title}{Mednext: Transformer-driven scaling of convnets for medical image segmentation}.
\newblock \URLprefix \url{https://arxiv.org/abs/2303.09975}, \href{http://arxiv.org/abs/2303.09975}{{\tt arXiv:2303.09975}}.
\bibitem[{Ryali et~al.(2023)Ryali, Hu, Bolya, Wei, Fan, Huang, Aggarwal, Chowdhury, Poursaeed, Hoffman et~al.}]{ryali2023hiera}
\bibinfo{author}{Ryali, C.}, \bibinfo{author}{Hu, Y.T.}, \bibinfo{author}{Bolya, D.}, \bibinfo{author}{Wei, C.}, \bibinfo{author}{Fan, H.}, \bibinfo{author}{Huang, P.Y.}, \bibinfo{author}{Aggarwal, V.}, \bibinfo{author}{Chowdhury, A.}, \bibinfo{author}{Poursaeed, O.}, \bibinfo{author}{Hoffman, J.}, et~al., \bibinfo{year}{2023}.
\newblock \bibinfo{title}{Hiera: A hierarchical vision transformer without the bells-and-whistles}, in: \bibinfo{booktitle}{International conference on machine learning}, \bibinfo{organization}{PMLR}. pp. \bibinfo{pages}{29441--29454}.
\bibitem[{Sadegheih et~al.(2024)Sadegheih, Bozorgpour, Kumari, Azad and Merhof}]{sadegheih2024lhunetlighthybridunet}
\bibinfo{author}{Sadegheih, Y.}, \bibinfo{author}{Bozorgpour, A.}, \bibinfo{author}{Kumari, P.}, \bibinfo{author}{Azad, R.}, \bibinfo{author}{Merhof, D.}, \bibinfo{year}{2024}.
\newblock \bibinfo{title}{Lhu-net: A light hybrid u-net for cost-efficient, high-performance volumetric medical image segmentation}.
\newblock \URLprefix \url{https://arxiv.org/abs/2404.05102}, \href{http://arxiv.org/abs/2404.05102}{{\tt arXiv:2404.05102}}.
\bibitem[{Sengupta et~al.(2024)Sengupta, Chakrabarty and Soni}]{sengupta2024sam2bettersam}
\bibinfo{author}{Sengupta, S.}, \bibinfo{author}{Chakrabarty, S.}, \bibinfo{author}{Soni, R.}, \bibinfo{year}{2024}.
\newblock \bibinfo{title}{Is sam 2 better than sam in medical image segmentation?}
\newblock \URLprefix \url{https://arxiv.org/abs/2408.04212}, \href{http://arxiv.org/abs/2408.04212}{{\tt arXiv:2408.04212}}.
\bibitem[{Shaker et~al.(2024)Shaker, Maaz, Rasheed, Khan, Yang and Khan}]{unetr_pp_arxiv}
\bibinfo{author}{Shaker, A.}, \bibinfo{author}{Maaz, M.}, \bibinfo{author}{Rasheed, H.}, \bibinfo{author}{Khan, S.}, \bibinfo{author}{Yang, M.H.}, \bibinfo{author}{Khan, F.S.}, \bibinfo{year}{2024}.
\newblock \bibinfo{title}{Unetr++: Delving into efficient and accurate 3d medical image segmentation}.
\newblock \URLprefix \url{https://arxiv.org/abs/2212.04497}, \href{http://arxiv.org/abs/2212.04497}{{\tt arXiv:2212.04497}}.
\bibitem[{Shen et~al.(2025)Shen, Li, Shi and Wang}]{shen2025interactive3dmedicalimage}
\bibinfo{author}{Shen, C.}, \bibinfo{author}{Li, W.}, \bibinfo{author}{Shi, Y.}, \bibinfo{author}{Wang, X.}, \bibinfo{year}{2025}.
\newblock \bibinfo{title}{Interactive 3d medical image segmentation with sam 2}.
\newblock \URLprefix \url{https://arxiv.org/abs/2408.02635}, \href{http://arxiv.org/abs/2408.02635}{{\tt arXiv:2408.02635}}.
\bibitem[{Shi et~al.(2024a)Shi, Han, Huang, Liao, Li, Kong, Zhu, Wang and Liu}]{shi2024maskenhancedsegmentmodeltumor}
\bibinfo{author}{Shi, H.}, \bibinfo{author}{Han, S.}, \bibinfo{author}{Huang, S.}, \bibinfo{author}{Liao, Y.}, \bibinfo{author}{Li, G.}, \bibinfo{author}{Kong, X.}, \bibinfo{author}{Zhu, H.}, \bibinfo{author}{Wang, X.}, \bibinfo{author}{Liu, S.}, \bibinfo{year}{2024}a.
\newblock \bibinfo{title}{Mask-enhanced segment anything model for tumor lesion semantic segmentation}.
\newblock \URLprefix \url{https://arxiv.org/abs/2403.05912}, \href{http://arxiv.org/abs/2403.05912}{{\tt arXiv:2403.05912}}.
\bibitem[{Shi et~al.(2024b)Shi, Han, Huang, Liao, Li, Kong, Zhu, Wang and Liu}]{10.1007/978-3-031-72111-3_38}
\bibinfo{author}{Shi, H.}, \bibinfo{author}{Han, S.}, \bibinfo{author}{Huang, S.}, \bibinfo{author}{Liao, Y.}, \bibinfo{author}{Li, G.}, \bibinfo{author}{Kong, X.}, \bibinfo{author}{Zhu, H.}, \bibinfo{author}{Wang, X.}, \bibinfo{author}{Liu, S.}, \bibinfo{year}{2024}b.
\newblock \bibinfo{title}{Mask-enhanced segment anything model for tumor lesion semantic segmentation}, in: \bibinfo{editor}{Linguraru, M.G.}, \bibinfo{editor}{Dou, Q.}, \bibinfo{editor}{Feragen, A.}, \bibinfo{editor}{Giannarou, S.}, \bibinfo{editor}{Glocker, B.}, \bibinfo{editor}{Lekadir, K.}, \bibinfo{editor}{Schnabel, J.A.} (Eds.), \bibinfo{booktitle}{Medical Image Computing and Computer Assisted Intervention -- MICCAI 2024}, \bibinfo{publisher}{Springer Nature Switzerland}, \bibinfo{address}{Cham}. pp. \bibinfo{pages}{403--413}.
\bibitem[{Shi et~al.(2023)Shi, Guo, Yang, Ye and Ma}]{shi2023nextouefficienttopologyawareunet}
\bibinfo{author}{Shi, P.}, \bibinfo{author}{Guo, X.}, \bibinfo{author}{Yang, Y.}, \bibinfo{author}{Ye, C.}, \bibinfo{author}{Ma, T.}, \bibinfo{year}{2023}.
\newblock \bibinfo{title}{Nextou: Efficient topology-aware u-net for medical image segmentation}.
\newblock \URLprefix \url{https://arxiv.org/abs/2305.15911}, \href{http://arxiv.org/abs/2305.15911}{{\tt arXiv:2305.15911}}.
\bibitem[{Sun et~al.(2024)Sun, Xue, Sun, Sun, Luo, Wang, Wang, Guo, Liu, Zhao et~al.}]{sun2024medical}
\bibinfo{author}{Sun, K.}, \bibinfo{author}{Xue, S.}, \bibinfo{author}{Sun, F.}, \bibinfo{author}{Sun, H.}, \bibinfo{author}{Luo, Y.}, \bibinfo{author}{Wang, L.}, \bibinfo{author}{Wang, S.}, \bibinfo{author}{Guo, N.}, \bibinfo{author}{Liu, L.}, \bibinfo{author}{Zhao, T.}, et~al., \bibinfo{year}{2024}.
\newblock \bibinfo{title}{Medical multimodal foundation models in clinical diagnosis and treatment: Applications, challenges, and future directions}.
\newblock \bibinfo{journal}{arXiv preprint arXiv:2412.02621} .
\bibitem[{Tang et~al.(2025)Tang, Xi, Li and Hu}]{tang2025regulatory}
\bibinfo{author}{Tang, D.}, \bibinfo{author}{Xi, X.}, \bibinfo{author}{Li, Y.}, \bibinfo{author}{Hu, M.}, \bibinfo{year}{2025}.
\newblock \bibinfo{title}{Regulatory approaches towards ai medical devices: A comparative study of the united states, the european union and china}.
\newblock \bibinfo{journal}{Health Policy} , \bibinfo{pages}{105260}.
\bibitem[{Ulrich et~al.(2023)Ulrich, Isensee, Wald, Zenk, Baumgartner and Maier-Hein}]{10.1007/978-3-031-43898-1_62}
\bibinfo{author}{Ulrich, C.}, \bibinfo{author}{Isensee, F.}, \bibinfo{author}{Wald, T.}, \bibinfo{author}{Zenk, M.}, \bibinfo{author}{Baumgartner, M.}, \bibinfo{author}{Maier-Hein, K.H.}, \bibinfo{year}{2023}.
\newblock \bibinfo{title}{Multitalent: A multi-dataset approach to medical image segmentation}, in: \bibinfo{editor}{Greenspan, H.}, \bibinfo{editor}{Madabhushi, A.}, \bibinfo{editor}{Mousavi, P.}, \bibinfo{editor}{Salcudean, S.}, \bibinfo{editor}{Duncan, J.}, \bibinfo{editor}{Syeda-Mahmood, T.}, \bibinfo{editor}{Taylor, R.} (Eds.), \bibinfo{booktitle}{Medical Image Computing and Computer Assisted Intervention -- MICCAI 2023}, \bibinfo{publisher}{Springer Nature Switzerland}, \bibinfo{address}{Cham}. pp. \bibinfo{pages}{648--658}.
\bibitem[{Valanarasu et~al.(2023)Valanarasu, Tang, Yang, Xu, Zhao, Li, Patel, Landman, Xu, He and Nath}]{valanarasu2023disruptiveautoencodersleveraginglowlevel}
\bibinfo{author}{Valanarasu, J.M.J.}, \bibinfo{author}{Tang, Y.}, \bibinfo{author}{Yang, D.}, \bibinfo{author}{Xu, Z.}, \bibinfo{author}{Zhao, C.}, \bibinfo{author}{Li, W.}, \bibinfo{author}{Patel, V.M.}, \bibinfo{author}{Landman, B.}, \bibinfo{author}{Xu, D.}, \bibinfo{author}{He, Y.}, \bibinfo{author}{Nath, V.}, \bibinfo{year}{2023}.
\newblock \bibinfo{title}{Disruptive autoencoders: Leveraging low-level features for 3d medical image pre-training}.
\newblock \URLprefix \url{https://arxiv.org/abs/2307.16896}, \href{http://arxiv.org/abs/2307.16896}{{\tt arXiv:2307.16896}}.
\bibitem[{Valanarasu et~al.(2024)Valanarasu, Tang, Yang, Xu, Zhao, Li, Patel, Landman, Xu, He and Nath}]{pmlr-v250-valanarasu24a}
\bibinfo{author}{Valanarasu, J.M.J.}, \bibinfo{author}{Tang, Y.}, \bibinfo{author}{Yang, D.}, \bibinfo{author}{Xu, Z.}, \bibinfo{author}{Zhao, C.}, \bibinfo{author}{Li, W.}, \bibinfo{author}{Patel, V.M.}, \bibinfo{author}{Landman, B.A.}, \bibinfo{author}{Xu, D.}, \bibinfo{author}{He, Y.}, \bibinfo{author}{Nath, V.}, \bibinfo{year}{2024}.
\newblock \bibinfo{title}{Disruptive autoencoders: Leveraging low-level features for 3d medical image pre-training}, in: \bibinfo{editor}{Burgos, N.}, \bibinfo{editor}{Petitjean, C.}, \bibinfo{editor}{Vakalopoulou, M.}, \bibinfo{editor}{Christodoulidis, S.}, \bibinfo{editor}{Coupe, P.}, \bibinfo{editor}{Delingette, H.}, \bibinfo{editor}{Lartizien, C.}, \bibinfo{editor}{Mateus, D.} (Eds.), \bibinfo{booktitle}{Proceedings of The 7nd International Conference on Medical Imaging with Deep Learning}, \bibinfo{publisher}{PMLR}. pp. \bibinfo{pages}{1553--1570}.
\newblock \URLprefix \url{https://proceedings.mlr.press/v250/valanarasu24a.html}.
\bibitem[{Vaswani et~al.(2017)Vaswani, Shazeer, Parmar, Uszkoreit, Jones, Gomez, Kaiser and Polosukhin}]{vaswani2017attention}
\bibinfo{author}{Vaswani, A.}, \bibinfo{author}{Shazeer, N.}, \bibinfo{author}{Parmar, N.}, \bibinfo{author}{Uszkoreit, J.}, \bibinfo{author}{Jones, L.}, \bibinfo{author}{Gomez, A.N.}, \bibinfo{author}{Kaiser, L.}, \bibinfo{author}{Polosukhin, I.}, \bibinfo{year}{2017}.
\newblock \bibinfo{title}{Attention is all you need}.
\newblock \DOIprefix\doi{10.48550/ARXIV.1706.03762}.
\bibitem[{Wang et~al.(2023a)Wang, Li, Wang, Zhang, Wang, Liu and Yang}]{wang2023mathrmsammedmedicalimageannotation}
\bibinfo{author}{Wang, C.}, \bibinfo{author}{Li, D.}, \bibinfo{author}{Wang, S.}, \bibinfo{author}{Zhang, C.}, \bibinfo{author}{Wang, Y.}, \bibinfo{author}{Liu, Y.}, \bibinfo{author}{Yang, G.}, \bibinfo{year}{2023}a.
\newblock \bibinfo{title}{$\mathrm{SAM^{Med}}$: A medical image annotation framework based on large vision model}.
\newblock \URLprefix \url{https://arxiv.org/abs/2307.05617}, \href{http://arxiv.org/abs/2307.05617}{{\tt arXiv:2307.05617}}.
\bibitem[{Wang et~al.(2023b)Wang, Wu, Luo, Liu, Li and Zhang}]{wang2023misfm3dmedicalimage}
\bibinfo{author}{Wang, G.}, \bibinfo{author}{Wu, J.}, \bibinfo{author}{Luo, X.}, \bibinfo{author}{Liu, X.}, \bibinfo{author}{Li, K.}, \bibinfo{author}{Zhang, S.}, \bibinfo{year}{2023}b.
\newblock \bibinfo{title}{Mis-fm: 3d medical image segmentation using foundation models pretrained on a large-scale unannotated dataset}.
\newblock \URLprefix \url{https://arxiv.org/abs/2306.16925}, \href{http://arxiv.org/abs/2306.16925}{{\tt arXiv:2306.16925}}.
\bibitem[{Wang et~al.(2024a)Wang, Guo, Ye, Deng, Cheng, Li, Chen, Su, Huang, Shen, Fu, Zhang, He and Qiao}]{wang2024sammed3dgeneralpurposesegmentationmodels}
\bibinfo{author}{Wang, H.}, \bibinfo{author}{Guo, S.}, \bibinfo{author}{Ye, J.}, \bibinfo{author}{Deng, Z.}, \bibinfo{author}{Cheng, J.}, \bibinfo{author}{Li, T.}, \bibinfo{author}{Chen, J.}, \bibinfo{author}{Su, Y.}, \bibinfo{author}{Huang, Z.}, \bibinfo{author}{Shen, Y.}, \bibinfo{author}{Fu, B.}, \bibinfo{author}{Zhang, S.}, \bibinfo{author}{He, J.}, \bibinfo{author}{Qiao, Y.}, \bibinfo{year}{2024}a.
\newblock \bibinfo{title}{Sam-med3d: Towards general-purpose segmentation models for volumetric medical images}.
\newblock \URLprefix \url{https://arxiv.org/abs/2310.15161}, \href{http://arxiv.org/abs/2310.15161}{{\tt arXiv:2310.15161}}.
\bibitem[{Wang et~al.(2024b)Wang, Lin, Ding and Li}]{10.1007/978-3-031-72114-4_61}
\bibinfo{author}{Wang, H.}, \bibinfo{author}{Lin, Y.}, \bibinfo{author}{Ding, X.}, \bibinfo{author}{Li, X.}, \bibinfo{year}{2024}b.
\newblock \bibinfo{title}{Tri-plane mamba: Efficiently adapting segment anything model for 3d medical images}, in: \bibinfo{editor}{Linguraru, M.G.}, \bibinfo{editor}{Dou, Q.}, \bibinfo{editor}{Feragen, A.}, \bibinfo{editor}{Giannarou, S.}, \bibinfo{editor}{Glocker, B.}, \bibinfo{editor}{Lekadir, K.}, \bibinfo{editor}{Schnabel, J.A.} (Eds.), \bibinfo{booktitle}{Medical Image Computing and Computer Assisted Intervention -- MICCAI 2024}, \bibinfo{publisher}{Springer Nature Switzerland}, \bibinfo{address}{Cham}. pp. \bibinfo{pages}{636--646}.
\bibitem[{Wang et~al.(2024c)Wang, Lin, Ding and Li}]{wang2024triplanemambaefficientlyadapting}
\bibinfo{author}{Wang, H.}, \bibinfo{author}{Lin, Y.}, \bibinfo{author}{Ding, X.}, \bibinfo{author}{Li, X.}, \bibinfo{year}{2024}c.
\newblock \bibinfo{title}{Tri-plane mamba: Efficiently adapting segment anything model for 3d medical images}.
\newblock \URLprefix \url{https://arxiv.org/abs/2409.08492}, \href{http://arxiv.org/abs/2409.08492}{{\tt arXiv:2409.08492}}.
\bibitem[{Wasserthal et~al.(2023)Wasserthal, Breit, Meyer, Pradella, Hinck, Sauter, Heye, Boll, Cyriac, Yang, Bach and Segeroth}]{doi:10.1148/ryai.230024}
\bibinfo{author}{Wasserthal, J.}, \bibinfo{author}{Breit, H.C.}, \bibinfo{author}{Meyer, M.T.}, \bibinfo{author}{Pradella, M.}, \bibinfo{author}{Hinck, D.}, \bibinfo{author}{Sauter, A.W.}, \bibinfo{author}{Heye, T.}, \bibinfo{author}{Boll, D.T.}, \bibinfo{author}{Cyriac, J.}, \bibinfo{author}{Yang, S.}, \bibinfo{author}{Bach, M.}, \bibinfo{author}{Segeroth, M.}, \bibinfo{year}{2023}.
\newblock \bibinfo{title}{Totalsegmentator: Robust segmentation of 104 anatomic structures in ct images}.
\newblock \bibinfo{journal}{Radiology: Artificial Intelligence} \bibinfo{volume}{5}, \bibinfo{pages}{e230024}.
\newblock \URLprefix \url{https://doi.org/10.1148/ryai.230024}, \DOIprefix\doi{10.1148/ryai.230024}, \href{http://arxiv.org/abs/https://doi.org/10.1148/ryai.230024}{{\tt arXiv:https://doi.org/10.1148/ryai.230024}}.
\bibitem[{{White and Case Law Firm}(2025)}]{usai}
\bibinfo{author}{{White and Case Law Firm}}, \bibinfo{year}{2025}.
\newblock \bibinfo{title}{Laws/regulations directly regulating ai (the “ai regulations”)}.
\newblock \bibinfo{howpublished}{\url{https://www.whitecase.com/insight-our-thinking/ai-watch-global-regulatory-tracker-united-states}}.
\newblock \bibinfo{note}{Accessed: 2025-07-04}.
\bibitem[{{WilmerHale Law Firm}(2025)}]{gpai}
\bibinfo{author}{{WilmerHale Law Firm}}, \bibinfo{year}{2025}.
\newblock \bibinfo{title}{Navigating generative ai under the european union’s artificial intelligence act}.
\newblock \bibinfo{howpublished}{\href{https://www.wilmerhale.com/en/insights/blogs/wilmerhale-privacy-and-cybersecurity-law/20241002-navigating-generative-ai-under-the-european-unions-artificial-intelligence-act\#:~:text=Obligations\%20of\%20\%E2\%80\%9CProviders\%E2\%80\%9D\%20of\%20GPAI\%20Models}{https://www.wilmerhale.com/en/insights/blogs/wilmerhale-privacy-and-cybersecurity-law/20241002-navigating-generative-ai-under-the-european-unions-artificial-intelligence-act\#:~:text=Obligations\%20of\%20\%E2\%80\%9CProviders\%E2\%80\%9D\%20of \%20GPAI\%20Models}}.
\newblock \bibinfo{note}{Accessed: 2025-07-04}.
\bibitem[{Woo et~al.(2023)Woo, Debnath, Hu, Chen, Liu, Kweon and Xie}]{10205236ConvNeXtv2}
\bibinfo{author}{Woo, S.}, \bibinfo{author}{Debnath, S.}, \bibinfo{author}{Hu, R.}, \bibinfo{author}{Chen, X.}, \bibinfo{author}{Liu, Z.}, \bibinfo{author}{Kweon, I.S.}, \bibinfo{author}{Xie, S.}, \bibinfo{year}{2023}.
\newblock \bibinfo{title}{Convnext v2: Co-designing and scaling convnets with masked autoencoders}, in: \bibinfo{booktitle}{2023 IEEE/CVF Conference on Computer Vision and Pattern Recognition (CVPR)}, pp. \bibinfo{pages}{16133--16142}.
\newblock \DOIprefix\doi{10.1109/CVPR52729.2023.01548}.
\bibitem[{Wu et~al.(2023)Wu, Ji, Liu, Fu, Xu, Xu and Jin}]{wu2023medicalsamadapteradapting}
\bibinfo{author}{Wu, J.}, \bibinfo{author}{Ji, W.}, \bibinfo{author}{Liu, Y.}, \bibinfo{author}{Fu, H.}, \bibinfo{author}{Xu, M.}, \bibinfo{author}{Xu, Y.}, \bibinfo{author}{Jin, Y.}, \bibinfo{year}{2023}.
\newblock \bibinfo{title}{Medical sam adapter: Adapting segment anything model for medical image segmentation}.
\newblock \URLprefix \url{https://arxiv.org/abs/2304.12620}, \href{http://arxiv.org/abs/2304.12620}{{\tt arXiv:2304.12620}}.
\bibitem[{Wu and Xu(2024)}]{10655138}
\bibinfo{author}{Wu, J.}, \bibinfo{author}{Xu, M.}, \bibinfo{year}{2024}.
\newblock \bibinfo{title}{One-prompt to segment all medical images}, in: \bibinfo{booktitle}{2024 IEEE/CVF Conference on Computer Vision and Pattern Recognition (CVPR)}, pp. \bibinfo{pages}{11302--11312}.
\newblock \DOIprefix\doi{10.1109/CVPR52733.2024.01074}.
\bibitem[{Wu et~al.(2024)Wu, Zhu, Jin and Xu}]{wu2024onepromptsegmentmedicalimages}
\bibinfo{author}{Wu, J.}, \bibinfo{author}{Zhu, J.}, \bibinfo{author}{Jin, Y.}, \bibinfo{author}{Xu, M.}, \bibinfo{year}{2024}.
\newblock \bibinfo{title}{One-prompt to segment all medical images}.
\newblock \URLprefix \url{https://arxiv.org/abs/2305.10300}, \href{http://arxiv.org/abs/2305.10300}{{\tt arXiv:2305.10300}}.
\bibitem[{Xie et~al.(2021)Xie, Zhang, Shen and Xia}]{10.1007/978-3-030-87199-4_16}
\bibinfo{author}{Xie, Y.}, \bibinfo{author}{Zhang, J.}, \bibinfo{author}{Shen, C.}, \bibinfo{author}{Xia, Y.}, \bibinfo{year}{2021}.
\newblock \bibinfo{title}{Cotr: Efficiently bridging cnn and transformer for 3d medical image segmentation}, in: \bibinfo{editor}{de~Bruijne, M.}, \bibinfo{editor}{Cattin, P.C.}, \bibinfo{editor}{Cotin, S.}, \bibinfo{editor}{Padoy, N.}, \bibinfo{editor}{Speidel, S.}, \bibinfo{editor}{Zheng, Y.}, \bibinfo{editor}{Essert, C.} (Eds.), \bibinfo{booktitle}{Medical Image Computing and Computer Assisted Intervention -- MICCAI 2021}, \bibinfo{publisher}{Springer International Publishing}, \bibinfo{address}{Cham}. pp. \bibinfo{pages}{171--180}.
\bibitem[{Xie et~al.(2022a)Xie, Zhang, Xia and Wu}]{xie2022unimissuniversalmedicalselfsupervised}
\bibinfo{author}{Xie, Y.}, \bibinfo{author}{Zhang, J.}, \bibinfo{author}{Xia, Y.}, \bibinfo{author}{Wu, Q.}, \bibinfo{year}{2022}a.
\newblock \bibinfo{title}{Unimiss: Universal medical self-supervised learning via breaking dimensionality barrier}.
\newblock \URLprefix \url{https://arxiv.org/abs/2112.09356}, \href{http://arxiv.org/abs/2112.09356}{{\tt arXiv:2112.09356}}.
\bibitem[{Xie et~al.(2022b)Xie, Zhang, Xia and Wu}]{10.1007/978-3-031-19803-8_33}
\bibinfo{author}{Xie, Y.}, \bibinfo{author}{Zhang, J.}, \bibinfo{author}{Xia, Y.}, \bibinfo{author}{Wu, Q.}, \bibinfo{year}{2022}b.
\newblock \bibinfo{title}{Unimiss: Universal medical self-supervised learning via breaking dimensionality barrier}, in: \bibinfo{editor}{Avidan, S.}, \bibinfo{editor}{Brostow, G.}, \bibinfo{editor}{Ciss{\'e}, M.}, \bibinfo{editor}{Farinella, G.M.}, \bibinfo{editor}{Hassner, T.} (Eds.), \bibinfo{booktitle}{Computer Vision -- ECCV 2022}, \bibinfo{publisher}{Springer Nature Switzerland}, \bibinfo{address}{Cham}. pp. \bibinfo{pages}{558--575}.
\bibitem[{Xu et~al.(2024a)Xu, Li, Yue, Wang and Guo}]{xu2024sammpaapplyingsamfewshot}
\bibinfo{author}{Xu, J.}, \bibinfo{author}{Li, X.}, \bibinfo{author}{Yue, C.}, \bibinfo{author}{Wang, Y.}, \bibinfo{author}{Guo, Y.}, \bibinfo{year}{2024}a.
\newblock \bibinfo{title}{Sam-mpa: Applying sam to few-shot medical image segmentation using mask propagation and auto-prompting}.
\newblock \URLprefix \url{https://arxiv.org/abs/2411.17363}, \href{http://arxiv.org/abs/2411.17363}{{\tt arXiv:2411.17363}}.
\bibitem[{Xu et~al.(2024b)Xu, LiXiaokang, Chengyuyue, Ma, Guo and Wang}]{xu2024sammpa}
\bibinfo{author}{Xu, J.}, \bibinfo{author}{LiXiaokang}, \bibinfo{author}{Chengyuyue}, \bibinfo{author}{Ma, C.}, \bibinfo{author}{Guo, Y.}, \bibinfo{author}{Wang, Y.}, \bibinfo{year}{2024}b.
\newblock \bibinfo{title}{{SAM}-{MPA}: Applying {SAM} to few-shot medical image segmentation using mask propagation and auto-prompting}, in: \bibinfo{booktitle}{Advancements In Medical Foundation Models: Explainability, Robustness, Security, and Beyond}.
\newblock \URLprefix \url{https://openreview.net/forum?id=IjZI80PUdr}.
\bibitem[{Yamagishi et~al.(2025)Yamagishi, Hanaoka, Kikuchi, Nakao, Nakamura, Nomura, Miki, Yoshikawa and Abe}]{yamagishi2025zeroshot3dsegmentationabdominal}
\bibinfo{author}{Yamagishi, Y.}, \bibinfo{author}{Hanaoka, S.}, \bibinfo{author}{Kikuchi, T.}, \bibinfo{author}{Nakao, T.}, \bibinfo{author}{Nakamura, Y.}, \bibinfo{author}{Nomura, Y.}, \bibinfo{author}{Miki, S.}, \bibinfo{author}{Yoshikawa, T.}, \bibinfo{author}{Abe, O.}, \bibinfo{year}{2025}.
\newblock \bibinfo{title}{Zero-shot 3d segmentation of abdominal organs in ct scans using segment anything model 2: Adapting video tracking capabilities for 3d medical imaging}.
\newblock \URLprefix \url{https://arxiv.org/abs/2408.06170}, \href{http://arxiv.org/abs/2408.06170}{{\tt arXiv:2408.06170}}.
\bibitem[{Yan et~al.(2024a)Yan, Sun, Zhou, Yuan, Zhang, Li, Kim, Song, Ren, Liu, Li, Li, He and Sun}]{yan2024biomedical}
\bibinfo{author}{Yan, Z.}, \bibinfo{author}{Sun, W.}, \bibinfo{author}{Zhou, R.}, \bibinfo{author}{Yuan, Z.}, \bibinfo{author}{Zhang, K.}, \bibinfo{author}{Li, Y.}, \bibinfo{author}{Kim, S.}, \bibinfo{author}{Song, S.}, \bibinfo{author}{Ren, H.}, \bibinfo{author}{Liu, T.}, \bibinfo{author}{Li, Q.}, \bibinfo{author}{Li, X.}, \bibinfo{author}{He, L.}, \bibinfo{author}{Sun, L.}, \bibinfo{year}{2024}a.
\newblock \bibinfo{title}{Biomedical {SAM}-2: Segment anything in biomedical images and videos}, in: \bibinfo{booktitle}{Advancements In Medical Foundation Models: Explainability, Robustness, Security, and Beyond}.
\newblock \URLprefix \url{https://openreview.net/forum?id=xaPv4b8z2D}.
\bibitem[{Yan et~al.(2024b)Yan, Sun, Zhou, Yuan, Zhang, Li, Liu, Li, Li, He and Sun}]{yan2024biomedicalsam2segment}
\bibinfo{author}{Yan, Z.}, \bibinfo{author}{Sun, W.}, \bibinfo{author}{Zhou, R.}, \bibinfo{author}{Yuan, Z.}, \bibinfo{author}{Zhang, K.}, \bibinfo{author}{Li, Y.}, \bibinfo{author}{Liu, T.}, \bibinfo{author}{Li, Q.}, \bibinfo{author}{Li, X.}, \bibinfo{author}{He, L.}, \bibinfo{author}{Sun, L.}, \bibinfo{year}{2024}b.
\newblock \bibinfo{title}{Biomedical sam 2: Segment anything in biomedical images and videos}.
\newblock \URLprefix \url{https://arxiv.org/abs/2408.03286}, \href{http://arxiv.org/abs/2408.03286}{{\tt arXiv:2408.03286}}.
\bibitem[{Yang et~al.(2022a)Yang, Li, Dai and Gao}]{yang2022focal}
\bibinfo{author}{Yang, J.}, \bibinfo{author}{Li, C.}, \bibinfo{author}{Dai, X.}, \bibinfo{author}{Gao, J.}, \bibinfo{year}{2022}a.
\newblock \bibinfo{title}{Focal modulation networks}, in: \bibinfo{editor}{Oh, A.H.}, \bibinfo{editor}{Agarwal, A.}, \bibinfo{editor}{Belgrave, D.}, \bibinfo{editor}{Cho, K.} (Eds.), \bibinfo{booktitle}{Advances in Neural Information Processing Systems}.
\newblock \URLprefix \url{https://openreview.net/forum?id=ePhEbo039l}.
\bibitem[{Yang et~al.(2022b)Yang, Li, Dai, Yuan and Gao}]{yang2022focalmodulationnetworks}
\bibinfo{author}{Yang, J.}, \bibinfo{author}{Li, C.}, \bibinfo{author}{Dai, X.}, \bibinfo{author}{Yuan, L.}, \bibinfo{author}{Gao, J.}, \bibinfo{year}{2022}b.
\newblock \bibinfo{title}{Focal modulation networks}.
\newblock \URLprefix \url{https://arxiv.org/abs/2203.11926}, \href{http://arxiv.org/abs/2203.11926}{{\tt arXiv:2203.11926}}.
\bibitem[{Yang et~al.(2025)Yang, Liu, Liu, Gulhane, Mastrodicasa, Wu, Wang, Sahani and Patel}]{yang2025demographic}
\bibinfo{author}{Yang, Y.}, \bibinfo{author}{Liu, Y.}, \bibinfo{author}{Liu, X.}, \bibinfo{author}{Gulhane, A.}, \bibinfo{author}{Mastrodicasa, D.}, \bibinfo{author}{Wu, W.}, \bibinfo{author}{Wang, E.J.}, \bibinfo{author}{Sahani, D.}, \bibinfo{author}{Patel, S.}, \bibinfo{year}{2025}.
\newblock \bibinfo{title}{Demographic bias of expert-level vision-language foundation models in medical imaging}.
\newblock \bibinfo{journal}{Science Advances} \bibinfo{volume}{11}, \bibinfo{pages}{eadq0305}.
\bibitem[{Ye et~al.(2023a)Ye, Xie, Zhang, Chen and Xia}]{ye2023unisegpromptdrivenuniversalsegmentation}
\bibinfo{author}{Ye, Y.}, \bibinfo{author}{Xie, Y.}, \bibinfo{author}{Zhang, J.}, \bibinfo{author}{Chen, Z.}, \bibinfo{author}{Xia, Y.}, \bibinfo{year}{2023}a.
\newblock \bibinfo{title}{Uniseg: A prompt-driven universal segmentation model as well as a strong representation learner}.
\newblock \URLprefix \url{https://arxiv.org/abs/2304.03493}, \href{http://arxiv.org/abs/2304.03493}{{\tt arXiv:2304.03493}}.
\bibitem[{Ye et~al.(2023b)Ye, Xie, Zhang, Chen and Xia}]{10.1007/978-3-031-43898-1_49}
\bibinfo{author}{Ye, Y.}, \bibinfo{author}{Xie, Y.}, \bibinfo{author}{Zhang, J.}, \bibinfo{author}{Chen, Z.}, \bibinfo{author}{Xia, Y.}, \bibinfo{year}{2023}b.
\newblock \bibinfo{title}{Uniseg: A prompt-driven universal segmentation model as well as a strong representation learner}, in: \bibinfo{editor}{Greenspan, H.}, \bibinfo{editor}{Madabhushi, A.}, \bibinfo{editor}{Mousavi, P.}, \bibinfo{editor}{Salcudean, S.}, \bibinfo{editor}{Duncan, J.}, \bibinfo{editor}{Syeda-Mahmood, T.}, \bibinfo{editor}{Taylor, R.} (Eds.), \bibinfo{booktitle}{Medical Image Computing and Computer Assisted Intervention -- MICCAI 2023}, \bibinfo{publisher}{Springer Nature Switzerland}, \bibinfo{address}{Cham}. pp. \bibinfo{pages}{508--518}.
\bibitem[{Ye et~al.(2022)Ye, Zhang, Chen and Xia}]{10.1007/978-3-031-16440-8_52}
\bibinfo{author}{Ye, Y.}, \bibinfo{author}{Zhang, J.}, \bibinfo{author}{Chen, Z.}, \bibinfo{author}{Xia, Y.}, \bibinfo{year}{2022}.
\newblock \bibinfo{title}{Desd: Self-supervised learning with deep self-distillation for 3d medical image segmentation}, in: \bibinfo{editor}{Wang, L.}, \bibinfo{editor}{Dou, Q.}, \bibinfo{editor}{Fletcher, P.T.}, \bibinfo{editor}{Speidel, S.}, \bibinfo{editor}{Li, S.} (Eds.), \bibinfo{booktitle}{Medical Image Computing and Computer Assisted Intervention -- MICCAI 2022}, \bibinfo{publisher}{Springer Nature Switzerland}, \bibinfo{address}{Cham}. pp. \bibinfo{pages}{545--555}.
\bibitem[{Zhang et~al.(2023)Zhang, Liu, Cui, Huang, Lin, Yang and Hu}]{zhang2023comprehensive}
\bibinfo{author}{Zhang, C.}, \bibinfo{author}{Liu, L.}, \bibinfo{author}{Cui, Y.}, \bibinfo{author}{Huang, G.}, \bibinfo{author}{Lin, W.}, \bibinfo{author}{Yang, Y.}, \bibinfo{author}{Hu, Y.}, \bibinfo{year}{2023}.
\newblock \bibinfo{title}{A comprehensive survey on segment anything model for vision and beyond}.
\newblock \bibinfo{journal}{arXiv preprint arXiv:2305.08196} .
\bibitem[{Zhang et~al.(2020)Zhang, Xie, Xia and Shen}]{zhang2020dodnetlearningsegmentmultiorgan}
\bibinfo{author}{Zhang, J.}, \bibinfo{author}{Xie, Y.}, \bibinfo{author}{Xia, Y.}, \bibinfo{author}{Shen, C.}, \bibinfo{year}{2020}.
\newblock \bibinfo{title}{Dodnet: Learning to segment multi-organ and tumors from multiple partially labeled datasets}.
\newblock \URLprefix \url{https://arxiv.org/abs/2011.10217}, \href{http://arxiv.org/abs/2011.10217}{{\tt arXiv:2011.10217}}.
\bibitem[{Zhang et~al.(2021)Zhang, Xie, Xia and Shen}]{Zhang_2021_CVPR}
\bibinfo{author}{Zhang, J.}, \bibinfo{author}{Xie, Y.}, \bibinfo{author}{Xia, Y.}, \bibinfo{author}{Shen, C.}, \bibinfo{year}{2021}.
\newblock \bibinfo{title}{Dodnet: Learning to segment multi-organ and tumors from multiple partially labeled datasets}, in: \bibinfo{booktitle}{Proceedings of the IEEE/CVF Conference on Computer Vision and Pattern Recognition (CVPR)}, pp. \bibinfo{pages}{1195--1204}.
\bibitem[{Zhang and Liu(2023a)}]{samed_zhang_liu}
\bibinfo{author}{Zhang, K.}, \bibinfo{author}{Liu, D.}, \bibinfo{year}{2023}a.
\newblock \bibinfo{title}{Customized segment anything model for medical image segmentation}.
\newblock \URLprefix \url{https://arxiv.org/abs/2304.13785}, \href{http://arxiv.org/abs/2304.13785}{{\tt arXiv:2304.13785}}.
\bibitem[{Zhang and Liu(2023b)}]{zhang2023customizedsegmentmodelmedical}
\bibinfo{author}{Zhang, K.}, \bibinfo{author}{Liu, D.}, \bibinfo{year}{2023}b.
\newblock \bibinfo{title}{Customized segment anything model for medical image segmentation}.
\newblock \URLprefix \url{https://arxiv.org/abs/2304.13785}, \href{http://arxiv.org/abs/2304.13785}{{\tt arXiv:2304.13785}}.
\bibitem[{Zhang and Metaxas(2024)}]{zhang2024challenges}
\bibinfo{author}{Zhang, S.}, \bibinfo{author}{Metaxas, D.}, \bibinfo{year}{2024}.
\newblock \bibinfo{title}{On the challenges and perspectives of foundation models for medical image analysis}.
\newblock \bibinfo{journal}{Medical image analysis} \bibinfo{volume}{91}, \bibinfo{pages}{102996}.
\bibitem[{Zhang et~al.(2025)Zhang, Ou, Basaran, Visentin, Qiao, Gu, Matthews, Liu, Ye and Bai}]{10879789}
\bibinfo{author}{Zhang, X.}, \bibinfo{author}{Ou, N.}, \bibinfo{author}{Basaran, B.D.}, \bibinfo{author}{Visentin, M.}, \bibinfo{author}{Qiao, M.}, \bibinfo{author}{Gu, R.}, \bibinfo{author}{Matthews, P.M.}, \bibinfo{author}{Liu, Y.}, \bibinfo{author}{Ye, C.}, \bibinfo{author}{Bai, W.}, \bibinfo{year}{2025}.
\newblock \bibinfo{title}{A foundation model for lesion segmentation on brain mri with mixture of modality experts}.
\newblock \bibinfo{journal}{IEEE Transactions on Medical Imaging} , \bibinfo{pages}{1--1}\DOIprefix\doi{10.1109/TMI.2025.3540809}.
\bibitem[{Zhang et~al.(2024a)Zhang, Ou, Basaran, Visentin, Qiao, Gu, Ouyang, Liu, Matthew, Ye and Bai}]{zhang2024foundationmodelbrainlesion}
\bibinfo{author}{Zhang, X.}, \bibinfo{author}{Ou, N.}, \bibinfo{author}{Basaran, B.D.}, \bibinfo{author}{Visentin, M.}, \bibinfo{author}{Qiao, M.}, \bibinfo{author}{Gu, R.}, \bibinfo{author}{Ouyang, C.}, \bibinfo{author}{Liu, Y.}, \bibinfo{author}{Matthew, P.M.}, \bibinfo{author}{Ye, C.}, \bibinfo{author}{Bai, W.}, \bibinfo{year}{2024}a.
\newblock \bibinfo{title}{A foundation model for brain lesion segmentation with mixture of modality experts}.
\newblock \URLprefix \url{https://arxiv.org/abs/2405.10246}, \href{http://arxiv.org/abs/2405.10246}{{\tt arXiv:2405.10246}}.
\bibitem[{Zhang et~al.(2024b)Zhang, Cheng, Hu, Liu, Liu, Ran, Chen, Liu and Wang}]{zhang2024evf}
\bibinfo{author}{Zhang, Y.}, \bibinfo{author}{Cheng, T.}, \bibinfo{author}{Hu, R.}, \bibinfo{author}{Liu, L.}, \bibinfo{author}{Liu, H.}, \bibinfo{author}{Ran, L.}, \bibinfo{author}{Chen, X.}, \bibinfo{author}{Liu, W.}, \bibinfo{author}{Wang, X.}, \bibinfo{year}{2024}b.
\newblock \bibinfo{title}{Evf-sam: Early vision-language fusion for text-prompted segment anything model}.
\newblock \bibinfo{journal}{arXiv preprint arXiv:2406.20076} .
\bibitem[{Zhang et~al.(2022)Zhang, Liao, Ding and Zhang}]{ZHANG2022102088}
\bibinfo{author}{Zhang, Y.}, \bibinfo{author}{Liao, Q.}, \bibinfo{author}{Ding, L.}, \bibinfo{author}{Zhang, J.}, \bibinfo{year}{2022}.
\newblock \bibinfo{title}{Bridging 2d and 3d segmentation networks for computation-efficient volumetric medical image segmentation: An empirical study of 2.5d solutions}.
\newblock \bibinfo{journal}{Computerized Medical Imaging and Graphics} \bibinfo{volume}{99}, \bibinfo{pages}{102088}.
\newblock \URLprefix \url{https://www.sciencedirect.com/science/article/pii/S0895611122000611}, \DOIprefix\doi{https://doi.org/10.1016/j.compmedimag.2022.102088}.
\bibitem[{Zhang and Shen(2024)}]{zhang2024unleashing}
\bibinfo{author}{Zhang, Y.}, \bibinfo{author}{Shen, Z.}, \bibinfo{year}{2024}.
\newblock \bibinfo{title}{Unleashing the potential of sam2 for biomedical images and videos: A survey}.
\newblock \bibinfo{journal}{arXiv preprint arXiv:2408.12889} .
\bibitem[{Zhang et~al.(2024c)Zhang, Shen and Jiao}]{zhang2024segment}
\bibinfo{author}{Zhang, Y.}, \bibinfo{author}{Shen, Z.}, \bibinfo{author}{Jiao, R.}, \bibinfo{year}{2024}c.
\newblock \bibinfo{title}{Segment anything model for medical image segmentation: Current applications and future directions}.
\newblock \bibinfo{journal}{Computers in Biology and Medicine} , \bibinfo{pages}{108238}.
\bibitem[{Zhang et~al.(2024d)Zhang, Shen and Jiao}]{SAM4MIS}
\bibinfo{author}{Zhang, Y.}, \bibinfo{author}{Shen, Z.}, \bibinfo{author}{Jiao, R.}, \bibinfo{year}{2024}d.
\newblock \bibinfo{title}{Segment anything model for medical image segmentation: Current applications and future directions}.
\newblock \bibinfo{journal}{Computers in Biology and Medicine} \bibinfo{volume}{171}, \bibinfo{pages}{108238}.
\bibitem[{Zhao et~al.(2024a)Zhao, Gu, Yang, Usuyama, Lee, Kiblawi, Naumann, Gao, Crabtree, Abel, Moung-Wen, Piening, Bifulco, Wei, Poon and Wang}]{Zhao_2024}
\bibinfo{author}{Zhao, T.}, \bibinfo{author}{Gu, Y.}, \bibinfo{author}{Yang, J.}, \bibinfo{author}{Usuyama, N.}, \bibinfo{author}{Lee, H.H.}, \bibinfo{author}{Kiblawi, S.}, \bibinfo{author}{Naumann, T.}, \bibinfo{author}{Gao, J.}, \bibinfo{author}{Crabtree, A.}, \bibinfo{author}{Abel, J.}, \bibinfo{author}{Moung-Wen, C.}, \bibinfo{author}{Piening, B.}, \bibinfo{author}{Bifulco, C.}, \bibinfo{author}{Wei, M.}, \bibinfo{author}{Poon, H.}, \bibinfo{author}{Wang, S.}, \bibinfo{year}{2024}a.
\newblock \bibinfo{title}{A foundation model for joint segmentation, detection and recognition of biomedical objects across nine modalities}.
\newblock \bibinfo{journal}{Nature Methods} \bibinfo{volume}{22}, \bibinfo{pages}{166–176}.
\newblock \URLprefix \url{http://dx.doi.org/10.1038/s41592-024-02499-w}, \DOIprefix\doi{10.1038/s41592-024-02499-w}.
\bibitem[{Zhao et~al.(2025a)Zhao, Gu, Yang, Usuyama, Lee, Kiblawi, Naumann, Gao, Crabtree, Abel, Moung-Wen, Piening, Bifulco, Wei, Poon and Wang}]{Zhao2025}
\bibinfo{author}{Zhao, T.}, \bibinfo{author}{Gu, Y.}, \bibinfo{author}{Yang, J.}, \bibinfo{author}{Usuyama, N.}, \bibinfo{author}{Lee, H.H.}, \bibinfo{author}{Kiblawi, S.}, \bibinfo{author}{Naumann, T.}, \bibinfo{author}{Gao, J.}, \bibinfo{author}{Crabtree, A.}, \bibinfo{author}{Abel, J.}, \bibinfo{author}{Moung-Wen, C.}, \bibinfo{author}{Piening, B.}, \bibinfo{author}{Bifulco, C.}, \bibinfo{author}{Wei, M.}, \bibinfo{author}{Poon, H.}, \bibinfo{author}{Wang, S.}, \bibinfo{year}{2025}a.
\newblock \bibinfo{title}{A foundation model for joint segmentation, detection and recognition of biomedical objects across nine modalities}.
\newblock \bibinfo{journal}{Nature Methods} \bibinfo{volume}{22}, \bibinfo{pages}{166--176}.
\newblock \URLprefix \url{https://doi.org/10.1038/s41592-024-02499-w}, \DOIprefix\doi{10.1038/s41592-024-02499-w}.
\bibitem[{Zhao et~al.(2024b)Zhao, Gu, Yang, Usuyama, Lee, Naumann, Gao, Crabtree, Abel, Moung-Wen, Piening, Bifulco, Wei, Poon and Wang}]{zhao2024biomedparsebiomedicalfoundationmodel}
\bibinfo{author}{Zhao, T.}, \bibinfo{author}{Gu, Y.}, \bibinfo{author}{Yang, J.}, \bibinfo{author}{Usuyama, N.}, \bibinfo{author}{Lee, H.H.}, \bibinfo{author}{Naumann, T.}, \bibinfo{author}{Gao, J.}, \bibinfo{author}{Crabtree, A.}, \bibinfo{author}{Abel, J.}, \bibinfo{author}{Moung-Wen, C.}, \bibinfo{author}{Piening, B.}, \bibinfo{author}{Bifulco, C.}, \bibinfo{author}{Wei, M.}, \bibinfo{author}{Poon, H.}, \bibinfo{author}{Wang, S.}, \bibinfo{year}{2024}b.
\newblock \bibinfo{title}{Biomedparse: a biomedical foundation model for image parsing of everything everywhere all at once}.
\newblock \URLprefix \url{https://arxiv.org/abs/2405.12971}, \DOIprefix\doi{https://doi.org/10.1038/s41592-024-02499-w}, \href{http://arxiv.org/abs/2405.12971}{{\tt arXiv:2405.12971}}.
\bibitem[{Zhao et~al.(2025b)Zhao, Liu, Wu, Wang, Li, Wang, Teng, Liu, Cui, Wang et~al.}]{zhao2025clip}
\bibinfo{author}{Zhao, Z.}, \bibinfo{author}{Liu, Y.}, \bibinfo{author}{Wu, H.}, \bibinfo{author}{Wang, M.}, \bibinfo{author}{Li, Y.}, \bibinfo{author}{Wang, S.}, \bibinfo{author}{Teng, L.}, \bibinfo{author}{Liu, D.}, \bibinfo{author}{Cui, Z.}, \bibinfo{author}{Wang, Q.}, et~al., \bibinfo{year}{2025}b.
\newblock \bibinfo{title}{Clip in medical imaging: A survey}.
\newblock \bibinfo{journal}{Medical Image Analysis} , \bibinfo{pages}{103551}.
\bibitem[{Zhao et~al.(2025c)Zhao, Zhang, Wu, Zhang, Zhang, Wang and Xie}]{zhao2025modelrulealluniversal}
\bibinfo{author}{Zhao, Z.}, \bibinfo{author}{Zhang, Y.}, \bibinfo{author}{Wu, C.}, \bibinfo{author}{Zhang, X.}, \bibinfo{author}{Zhang, Y.}, \bibinfo{author}{Wang, Y.}, \bibinfo{author}{Xie, W.}, \bibinfo{year}{2025}c.
\newblock \bibinfo{title}{One model to rule them all: Towards universal segmentation for medical images with text prompts}.
\newblock \URLprefix \url{https://arxiv.org/abs/2312.17183}, \href{http://arxiv.org/abs/2312.17183}{{\tt arXiv:2312.17183}}.
\bibitem[{Zhou et~al.(2022)Zhou, Guo, Zhang, Yu, Wang and Yu}]{nnFormer_arxiv}
\bibinfo{author}{Zhou, H.Y.}, \bibinfo{author}{Guo, J.}, \bibinfo{author}{Zhang, Y.}, \bibinfo{author}{Yu, L.}, \bibinfo{author}{Wang, L.}, \bibinfo{author}{Yu, Y.}, \bibinfo{year}{2022}.
\newblock \bibinfo{title}{nnformer: Interleaved transformer for volumetric segmentation}.
\newblock \URLprefix \url{https://arxiv.org/abs/2109.03201}, \href{http://arxiv.org/abs/2109.03201}{{\tt arXiv:2109.03201}}.
\bibitem[{Zhou et~al.(2020)Zhou, Siddiquee, Tajbakhsh and Liang}]{zhou2019unet++}
\bibinfo{author}{Zhou, Z.}, \bibinfo{author}{Siddiquee, M.M.R.}, \bibinfo{author}{Tajbakhsh, N.}, \bibinfo{author}{Liang, J.}, \bibinfo{year}{2020}.
\newblock \bibinfo{title}{Unet++: Redesigning skip connections to exploit multiscale features in image segmentation}.
\newblock \bibinfo{journal}{IEEE Transactions on Medical Imaging} \bibinfo{volume}{39}, \bibinfo{pages}{1856--1867}.
\newblock \DOIprefix\doi{10.1109/tmi.2019.2959609}.
\bibitem[{Zhu et~al.(2024)Zhu, Hamdi, Qi, Jin and Wu}]{zhu2024medicalsam2segment}
\bibinfo{author}{Zhu, J.}, \bibinfo{author}{Hamdi, A.}, \bibinfo{author}{Qi, Y.}, \bibinfo{author}{Jin, Y.}, \bibinfo{author}{Wu, J.}, \bibinfo{year}{2024}.
\newblock \bibinfo{title}{Medical sam 2: Segment medical images as video via segment anything model 2}.
\newblock \URLprefix \url{https://arxiv.org/abs/2408.00874}, \href{http://arxiv.org/abs/2408.00874}{{\tt arXiv:2408.00874}}.
\bibitem[{Zou et~al.(2023a)Zou, Yang, Zhang, Li, Li, Wang, Wang, Gao and Lee}]{seem_arxiv}
\bibinfo{author}{Zou, X.}, \bibinfo{author}{Yang, J.}, \bibinfo{author}{Zhang, H.}, \bibinfo{author}{Li, F.}, \bibinfo{author}{Li, L.}, \bibinfo{author}{Wang, J.}, \bibinfo{author}{Wang, L.}, \bibinfo{author}{Gao, J.}, \bibinfo{author}{Lee, Y.J.}, \bibinfo{year}{2023}a.
\newblock \bibinfo{title}{Segment everything everywhere all at once}.
\newblock \URLprefix \url{https://arxiv.org/abs/2304.06718}, \href{http://arxiv.org/abs/2304.06718}{{\tt arXiv:2304.06718}}.
\bibitem[{Zou et~al.(2023b)Zou, Yang, Zhang, Li, Li, Wang, Wang, Gao and Lee}]{seem_neurips}
\bibinfo{author}{Zou, X.}, \bibinfo{author}{Yang, J.}, \bibinfo{author}{Zhang, H.}, \bibinfo{author}{Li, F.}, \bibinfo{author}{Li, L.}, \bibinfo{author}{Wang, J.}, \bibinfo{author}{Wang, L.}, \bibinfo{author}{Gao, J.}, \bibinfo{author}{Lee, Y.J.}, \bibinfo{year}{2023}b.
\newblock \bibinfo{title}{Segment everything everywhere all at once}, in: \bibinfo{editor}{Oh, A.}, \bibinfo{editor}{Naumann, T.}, \bibinfo{editor}{Globerson, A.}, \bibinfo{editor}{Saenko, K.}, \bibinfo{editor}{Hardt, M.}, \bibinfo{editor}{Levine, S.} (Eds.), \bibinfo{booktitle}{Advances in Neural Information Processing Systems}, \bibinfo{publisher}{Curran Associates, Inc.}. pp. \bibinfo{pages}{19769--19782}.

\end{thebibliography}

\begin{thebibliography}{136}
\expandafter\ifx\csname natexlab\endcsname\relax\def\natexlab#1{#1}\fi
\providecommand{\url}[1]{\texttt{#1}}
\providecommand{\href}[2]{#2}
\providecommand{\path}[1]{#1}
\providecommand{\DOIprefix}{doi:}
\providecommand{\ArXivprefix}{arXiv:}
\providecommand{\URLprefix}{URL: }
\providecommand{\Pubmedprefix}{pmid:}
\providecommand{\doi}[1]{\href{http://dx.doi.org/#1}{\path{#1}}}
\providecommand{\Pubmed}[1]{\href{pmid:#1}{\path{#1}}}
\providecommand{\bibinfo}[2]{#2}
\ifx\xfnm\relax \def\xfnm[#1]{\unskip,\space#1}\fi
\bibitem[{Adams et~al.(2022)Adams, Makowski, Engel, Rattunde, Busch, Asbach, Niehues, Vinayahalingam, {van Ginneken}, Litjens and Bressem}]{ADAMS2022105817}
\bibinfo{author}{Adams, L.C.}, \bibinfo{author}{Makowski, M.R.}, \bibinfo{author}{Engel, G.}, \bibinfo{author}{Rattunde, M.}, \bibinfo{author}{Busch, F.}, \bibinfo{author}{Asbach, P.}, \bibinfo{author}{Niehues, S.M.}, \bibinfo{author}{Vinayahalingam, S.}, \bibinfo{author}{{van Ginneken}, B.}, \bibinfo{author}{Litjens, G.}, \bibinfo{author}{Bressem, K.K.}, \bibinfo{year}{2022}.
\newblock \bibinfo{title}{Prostate158 - an expert-annotated 3t mri dataset and algorithm for prostate cancer detection}.
\newblock \bibinfo{journal}{Computers in Biology and Medicine} \bibinfo{volume}{148}, \bibinfo{pages}{105817}.
\newblock \URLprefix \url{https://www.sciencedirect.com/science/article/pii/S0010482522005789}, \DOIprefix\doi{https://doi.org/10.1016/j.compbiomed.2022.105817}.
\bibitem[{Akinci~DAntonoli et~al.(2025)Akinci~DAntonoli, Berger, Indrakanti, Vishwanathan, Weiss, Jung, Berkarda, Rau, Reisert, K\"{u}stner, Walter, Merkle, Boll, Breit, Nicoli, Segeroth, Cyriac, Yang and Wasserthal}]{doi:10.1148/radiol.241613}
\bibinfo{author}{Akinci~DAntonoli, T.}, \bibinfo{author}{Berger, L.K.}, \bibinfo{author}{Indrakanti, A.K.}, \bibinfo{author}{Vishwanathan, N.}, \bibinfo{author}{Weiss, J.}, \bibinfo{author}{Jung, M.}, \bibinfo{author}{Berkarda, Z.}, \bibinfo{author}{Rau, A.}, \bibinfo{author}{Reisert, M.}, \bibinfo{author}{K\"{u}stner, T.}, \bibinfo{author}{Walter, A.}, \bibinfo{author}{Merkle, E.M.}, \bibinfo{author}{Boll, D.T.}, \bibinfo{author}{Breit, H.C.}, \bibinfo{author}{Nicoli, A.P.}, \bibinfo{author}{Segeroth, M.}, \bibinfo{author}{Cyriac, J.}, \bibinfo{author}{Yang, S.}, \bibinfo{author}{Wasserthal, J.}, \bibinfo{year}{2025}.
\newblock \bibinfo{title}{Totalsegmentator mri: Robust sequence-independent segmentation of multiple anatomic structures in mri}.
\newblock \bibinfo{journal}{Radiology} \bibinfo{volume}{314}, \bibinfo{pages}{e241613}.
\newblock \URLprefix \url{https://doi.org/10.1148/radiol.241613}, \DOIprefix\doi{10.1148/radiol.241613}, \href{http://arxiv.org/abs/https://doi.org/10.1148/radiol.241613}{{\tt arXiv:https://doi.org/10.1148/radiol.241613}}. \bibinfo{note}{pMID: 39964271}.
\bibitem[{Antonelli et~al.(2022)Antonelli, Reinke, Bakas, Farahani, Kopp-Schneider, Landman, Litjens, Menze, Ronneberger, Summers, van Ginneken, Bilello, Bilic, Christ, Do, Gollub, Heckers, Huisman, Jarnagin, McHugo, Napel, Pernicka, Rhode, Tobon-Gomez, Vorontsov, Meakin, Ourselin, Wiesenfarth, Arbel{\'a}ez, Bae, Chen, Daza, Feng, He, Isensee, Ji, Jia, Kim, Maier-Hein, Merhof, Pai, Park, Perslev, Rezaiifar, Rippel, Sarasua, Shen, Son, Wachinger, Wang, Wang, Xia, Xu, Xu, Zheng, Simpson, Maier-Hein and Cardoso}]{MSD_nature}
\bibinfo{author}{Antonelli, M.}, \bibinfo{author}{Reinke, A.}, \bibinfo{author}{Bakas, S.}, \bibinfo{author}{Farahani, K.}, \bibinfo{author}{Kopp-Schneider, A.}, \bibinfo{author}{Landman, B.A.}, \bibinfo{author}{Litjens, G.}, \bibinfo{author}{Menze, B.}, \bibinfo{author}{Ronneberger, O.}, \bibinfo{author}{Summers, R.M.}, \bibinfo{author}{van Ginneken, B.}, \bibinfo{author}{Bilello, M.}, \bibinfo{author}{Bilic, P.}, \bibinfo{author}{Christ, P.F.}, \bibinfo{author}{Do, R.K.G.}, \bibinfo{author}{Gollub, M.J.}, \bibinfo{author}{Heckers, S.H.}, \bibinfo{author}{Huisman, H.}, \bibinfo{author}{Jarnagin, W.R.}, \bibinfo{author}{McHugo, M.K.}, \bibinfo{author}{Napel, S.}, \bibinfo{author}{Pernicka, J.S.G.}, \bibinfo{author}{Rhode, K.}, \bibinfo{author}{Tobon-Gomez, C.}, \bibinfo{author}{Vorontsov, E.}, \bibinfo{author}{Meakin, J.A.}, \bibinfo{author}{Ourselin, S.}, \bibinfo{author}{Wiesenfarth, M.}, \bibinfo{author}{Arbel{\'a}ez, P.}, \bibinfo{author}{Bae, B.}, \bibinfo{author}{Chen, S.}, \bibinfo{author}{Daza, L.},
  \bibinfo{author}{Feng, J.}, \bibinfo{author}{He, B.}, \bibinfo{author}{Isensee, F.}, \bibinfo{author}{Ji, Y.}, \bibinfo{author}{Jia, F.}, \bibinfo{author}{Kim, I.}, \bibinfo{author}{Maier-Hein, K.}, \bibinfo{author}{Merhof, D.}, \bibinfo{author}{Pai, A.}, \bibinfo{author}{Park, B.}, \bibinfo{author}{Perslev, M.}, \bibinfo{author}{Rezaiifar, R.}, \bibinfo{author}{Rippel, O.}, \bibinfo{author}{Sarasua, I.}, \bibinfo{author}{Shen, W.}, \bibinfo{author}{Son, J.}, \bibinfo{author}{Wachinger, C.}, \bibinfo{author}{Wang, L.}, \bibinfo{author}{Wang, Y.}, \bibinfo{author}{Xia, Y.}, \bibinfo{author}{Xu, D.}, \bibinfo{author}{Xu, Z.}, \bibinfo{author}{Zheng, Y.}, \bibinfo{author}{Simpson, A.L.}, \bibinfo{author}{Maier-Hein, L.}, \bibinfo{author}{Cardoso, M.J.}, \bibinfo{year}{2022}.
\newblock \bibinfo{title}{The medical segmentation decathlon}.
\newblock \bibinfo{journal}{Nature Communications} \bibinfo{volume}{13}, \bibinfo{pages}{4128}.
\newblock \URLprefix \url{https://doi.org/10.1038/s41467-022-30695-9}, \DOIprefix\doi{10.1038/s41467-022-30695-9}.
\bibitem[{Armato~III et~al.(2011)Armato~III, McLennan, Bidaut, McNitt-Gray, Meyer, Reeves, Zhao, Aberle, Henschke, Hoffman, Kazerooni, MacMahon, van Beek, Yankelevitz, Biancardi, Bland, Brown, Engelmann, Laderach, Max, Pais, Qing, Roberts, Smith, Starkey, Batra, Caligiuri, Farooqi, Gladish, Jude, Munden, Petkovska, Quint, Schwartz, Sundaram, Dodd, Fenimore, Gur, Petrick, Freymann, Kirby, Hughes, Vande~Casteele, Gupte, Sallam, Heath, Kuhn, Dharaiya, Burns, Fryd, Salganicoff, Anand, Shreter, Vastagh, Croft and Clarke}]{https://doi.org/10.1118/1.3528204}
\bibinfo{author}{Armato~III, S.G.}, \bibinfo{author}{McLennan, G.}, \bibinfo{author}{Bidaut, L.}, \bibinfo{author}{McNitt-Gray, M.F.}, \bibinfo{author}{Meyer, C.R.}, \bibinfo{author}{Reeves, A.P.}, \bibinfo{author}{Zhao, B.}, \bibinfo{author}{Aberle, D.R.}, \bibinfo{author}{Henschke, C.I.}, \bibinfo{author}{Hoffman, E.A.}, \bibinfo{author}{Kazerooni, E.A.}, \bibinfo{author}{MacMahon, H.}, \bibinfo{author}{van Beek, E.J.R.}, \bibinfo{author}{Yankelevitz, D.}, \bibinfo{author}{Biancardi, A.M.}, \bibinfo{author}{Bland, P.H.}, \bibinfo{author}{Brown, M.S.}, \bibinfo{author}{Engelmann, R.M.}, \bibinfo{author}{Laderach, G.E.}, \bibinfo{author}{Max, D.}, \bibinfo{author}{Pais, R.C.}, \bibinfo{author}{Qing, D.P.Y.}, \bibinfo{author}{Roberts, R.Y.}, \bibinfo{author}{Smith, A.R.}, \bibinfo{author}{Starkey, A.}, \bibinfo{author}{Batra, P.}, \bibinfo{author}{Caligiuri, P.}, \bibinfo{author}{Farooqi, A.}, \bibinfo{author}{Gladish, G.W.}, \bibinfo{author}{Jude, C.M.}, \bibinfo{author}{Munden, R.F.},
  \bibinfo{author}{Petkovska, I.}, \bibinfo{author}{Quint, L.E.}, \bibinfo{author}{Schwartz, L.H.}, \bibinfo{author}{Sundaram, B.}, \bibinfo{author}{Dodd, L.E.}, \bibinfo{author}{Fenimore, C.}, \bibinfo{author}{Gur, D.}, \bibinfo{author}{Petrick, N.}, \bibinfo{author}{Freymann, J.}, \bibinfo{author}{Kirby, J.}, \bibinfo{author}{Hughes, B.}, \bibinfo{author}{Vande~Casteele, A.}, \bibinfo{author}{Gupte, S.}, \bibinfo{author}{Sallam, M.}, \bibinfo{author}{Heath, M.D.}, \bibinfo{author}{Kuhn, M.H.}, \bibinfo{author}{Dharaiya, E.}, \bibinfo{author}{Burns, R.}, \bibinfo{author}{Fryd, D.S.}, \bibinfo{author}{Salganicoff, M.}, \bibinfo{author}{Anand, V.}, \bibinfo{author}{Shreter, U.}, \bibinfo{author}{Vastagh, S.}, \bibinfo{author}{Croft, B.Y.}, \bibinfo{author}{Clarke, L.P.}, \bibinfo{year}{2011}.
\newblock \bibinfo{title}{The lung image database consortium (lidc) and image database resource initiative (idri): A completed reference database of lung nodules on ct scans}.
\newblock \bibinfo{journal}{Medical Physics} \bibinfo{volume}{38}, \bibinfo{pages}{915--931}.
\newblock \URLprefix \url{https://aapm.onlinelibrary.wiley.com/doi/abs/10.1118/1.3528204}, \DOIprefix\doi{https://doi.org/10.1118/1.3528204}, \href{http://arxiv.org/abs/https://aapm.onlinelibrary.wiley.com/doi/pdf/10.1118/1.3528204}{{\tt arXiv:https://aapm.onlinelibrary.wiley.com/doi/pdf/10.1118/1.3528204}}.
\bibitem[{Bakas et~al.(2024)Bakas, Baid, Rudie, Calabrese, Aboian, Anazodo, Conte, Albrecht, Li, Kofler, Correia De~Verdier, Huang, LaBella, Saluja, Gagnon, Aboian, Abayazeed, Farahani, Chung, Reitman, Kirkpatrick, Wang, Villanueva-Meyer, Flanders, Aboian, Nada, Aboian, Abayazeed, Lohman, Moawad, Janas, Krantchev, Memon, Velichko, Schrickel, Link, Aneja, Maresca, Nada, Vollmuth, Prez, Pease, Godfrey, Floyd, Adewole, Dako, Toyobo, Omidiji, Gbadamosi, Ogunleye, Ojo, Iorpagher, Babatunde, Aguh, Emegoakor, Kalaiwo, Linguraru, Kazerooni, Jiang, Liu, Gandhi, Khalili, Vossough, Nabavizadeh, Ware, Menze, Johanson, Meier, Chen, Petrick, Sahiner, Chai, Wiestler, Iglesias, Anwar, Van~Leemput and Piraud}]{bakas_2024_10978907}
\bibinfo{author}{Bakas, S.}, \bibinfo{author}{Baid, U.}, \bibinfo{author}{Rudie, J.}, \bibinfo{author}{Calabrese, E.}, \bibinfo{author}{Aboian, M.}, \bibinfo{author}{Anazodo, U.}, \bibinfo{author}{Conte, G.M.}, \bibinfo{author}{Albrecht, J.}, \bibinfo{author}{Li, H.B.}, \bibinfo{author}{Kofler, F.}, \bibinfo{author}{Correia De~Verdier, M.}, \bibinfo{author}{Huang, R.}, \bibinfo{author}{LaBella, D.}, \bibinfo{author}{Saluja, R.}, \bibinfo{author}{Gagnon, L.}, \bibinfo{author}{Aboian, M.}, \bibinfo{author}{Abayazeed, A.}, \bibinfo{author}{Farahani, K.}, \bibinfo{author}{Chung, V.}, \bibinfo{author}{Reitman, Z.}, \bibinfo{author}{Kirkpatrick, J.}, \bibinfo{author}{Wang, C.}, \bibinfo{author}{Villanueva-Meyer, J.}, \bibinfo{author}{Flanders, A.}, \bibinfo{author}{Aboian, M.}, \bibinfo{author}{Nada, A.}, \bibinfo{author}{Aboian, M.}, \bibinfo{author}{Abayazeed, A.}, \bibinfo{author}{Lohman, P.}, \bibinfo{author}{Moawad, A.}, \bibinfo{author}{Janas, A.}, \bibinfo{author}{Krantchev, K.}, \bibinfo{author}{Memon, F.},
  \bibinfo{author}{Velichko, Y.}, \bibinfo{author}{Schrickel, E.}, \bibinfo{author}{Link, K.}, \bibinfo{author}{Aneja, S.}, \bibinfo{author}{Maresca, R.}, \bibinfo{author}{Nada, A.}, \bibinfo{author}{Vollmuth, P.}, \bibinfo{author}{Prez, V.M.}, \bibinfo{author}{Pease, M.W.}, \bibinfo{author}{Godfrey, D.}, \bibinfo{author}{Floyd, S.}, \bibinfo{author}{Adewole, M.}, \bibinfo{author}{Dako, F.}, \bibinfo{author}{Toyobo, O.}, \bibinfo{author}{Omidiji, O.}, \bibinfo{author}{Gbadamosi, Y.}, \bibinfo{author}{Ogunleye, A.}, \bibinfo{author}{Ojo, N.}, \bibinfo{author}{Iorpagher, K.}, \bibinfo{author}{Babatunde, G.}, \bibinfo{author}{Aguh, K.}, \bibinfo{author}{Emegoakor, A.}, \bibinfo{author}{Kalaiwo, C.}, \bibinfo{author}{Linguraru, M.G.}, \bibinfo{author}{Kazerooni, A.F.}, \bibinfo{author}{Jiang, Z.}, \bibinfo{author}{Liu, X.}, \bibinfo{author}{Gandhi, D.}, \bibinfo{author}{Khalili, N.}, \bibinfo{author}{Vossough, A.}, \bibinfo{author}{Nabavizadeh, A.}, \bibinfo{author}{Ware, J.B.}, \bibinfo{author}{Menze, B.},
  \bibinfo{author}{Johanson, E.}, \bibinfo{author}{Meier, Z.}, \bibinfo{author}{Chen, W.}, \bibinfo{author}{Petrick, N.}, \bibinfo{author}{Sahiner, B.}, \bibinfo{author}{Chai, R.}, \bibinfo{author}{Wiestler, B.}, \bibinfo{author}{Iglesias, J.E.}, \bibinfo{author}{Anwar, S.M.}, \bibinfo{author}{Van~Leemput, K.}, \bibinfo{author}{Piraud, M.}, \bibinfo{year}{2024}.
\newblock \bibinfo{title}{Brats 2024 cluster of challenges (brats + beyond- brats)}.
\newblock \URLprefix \url{https://doi.org/10.5281/zenodo.10978907}, \DOIprefix\doi{10.5281/zenodo.10978907}.
\bibitem[{Bassi et~al.(2024)Bassi, Li, Tang, Isensee, Wang, Chen, Chou, Roy, Kirchhoff, Rokuss, Huang, Ye, He, Wald, Ulrich, Baumgartner, Maier-Hein, Jaeger, Ye, Xie, Zhang, Chen, Xia, Xing, Zhu, Sadegheih, Bozorgpour, Kumari, Azad, Merhof, Shi, Ma, Du, Bai, Huang, Zhao, Wang, Li, Gu, Dong, Yang, Mazurowski, Gupta, Wu, Zhuang, Chen, Roth, Xu, Blaschko, Decherchi, Cavalli, Yuille and Zhou}]{NEURIPS2024_1b8726b5}
\bibinfo{author}{Bassi, P.R.A.S.}, \bibinfo{author}{Li, W.}, \bibinfo{author}{Tang, Y.}, \bibinfo{author}{Isensee, F.}, \bibinfo{author}{Wang, Z.}, \bibinfo{author}{Chen, J.}, \bibinfo{author}{Chou, Y.C.}, \bibinfo{author}{Roy, S.}, \bibinfo{author}{Kirchhoff, Y.}, \bibinfo{author}{Rokuss, M.}, \bibinfo{author}{Huang, Z.}, \bibinfo{author}{Ye, J.}, \bibinfo{author}{He, J.}, \bibinfo{author}{Wald, T.}, \bibinfo{author}{Ulrich, C.}, \bibinfo{author}{Baumgartner, M.}, \bibinfo{author}{Maier-Hein, K.H.}, \bibinfo{author}{Jaeger, P.}, \bibinfo{author}{Ye, Y.}, \bibinfo{author}{Xie, Y.}, \bibinfo{author}{Zhang, J.}, \bibinfo{author}{Chen, Z.}, \bibinfo{author}{Xia, Y.}, \bibinfo{author}{Xing, Z.}, \bibinfo{author}{Zhu, L.}, \bibinfo{author}{Sadegheih, Y.}, \bibinfo{author}{Bozorgpour, A.}, \bibinfo{author}{Kumari, P.}, \bibinfo{author}{Azad, R.}, \bibinfo{author}{Merhof, D.}, \bibinfo{author}{Shi, P.}, \bibinfo{author}{Ma, T.}, \bibinfo{author}{Du, Y.}, \bibinfo{author}{Bai, F.}, \bibinfo{author}{Huang, T.},
  \bibinfo{author}{Zhao, B.}, \bibinfo{author}{Wang, H.}, \bibinfo{author}{Li, X.}, \bibinfo{author}{Gu, H.}, \bibinfo{author}{Dong, H.}, \bibinfo{author}{Yang, J.}, \bibinfo{author}{Mazurowski, M.A.}, \bibinfo{author}{Gupta, S.}, \bibinfo{author}{Wu, L.}, \bibinfo{author}{Zhuang, J.}, \bibinfo{author}{Chen, H.}, \bibinfo{author}{Roth, H.}, \bibinfo{author}{Xu, D.}, \bibinfo{author}{Blaschko, M.B.}, \bibinfo{author}{Decherchi, S.}, \bibinfo{author}{Cavalli, A.}, \bibinfo{author}{Yuille, A.L.}, \bibinfo{author}{Zhou, Z.}, \bibinfo{year}{2024}.
\newblock \bibinfo{title}{Touchstone benchmark: Are we on the right way for evaluating ai algorithms for medical segmentation?}, in: \bibinfo{editor}{Globerson, A.}, \bibinfo{editor}{Mackey, L.}, \bibinfo{editor}{Belgrave, D.}, \bibinfo{editor}{Fan, A.}, \bibinfo{editor}{Paquet, U.}, \bibinfo{editor}{Tomczak, J.}, \bibinfo{editor}{Zhang, C.} (Eds.), \bibinfo{booktitle}{Advances in Neural Information Processing Systems}, \bibinfo{publisher}{Curran Associates, Inc.}. pp. \bibinfo{pages}{15184--15201}.
\newblock \URLprefix \url{https://proceedings.neurips.cc/paper_files/paper/2024/file/1b8726b572e0dfa72793f9f6590664fd-Paper-Datasets_and_Benchmarks_Track.pdf}.
\bibitem[{Bernard et~al.(2018)Bernard, Lalande, Zotti, Cervenansky, Yang, Heng, Cetin, Lekadir, Camara, Gonzalez~Ballester, Sanroma, Napel, Petersen, Tziritas, Grinias, Khened, Kollerathu, Krishnamurthi, Roh, Pennec, Sermesant, Isensee, Jger, Maier-Hein, Full, Wolf, Engelhardt, Baumgartner, Koch, Wolterink, Igum, Jang, Hong, Patravali, Jain, Humbert and Jodoin}]{8360453}
\bibinfo{author}{Bernard, O.}, \bibinfo{author}{Lalande, A.}, \bibinfo{author}{Zotti, C.}, \bibinfo{author}{Cervenansky, F.}, \bibinfo{author}{Yang, X.}, \bibinfo{author}{Heng, P.A.}, \bibinfo{author}{Cetin, I.}, \bibinfo{author}{Lekadir, K.}, \bibinfo{author}{Camara, O.}, \bibinfo{author}{Gonzalez~Ballester, M.A.}, \bibinfo{author}{Sanroma, G.}, \bibinfo{author}{Napel, S.}, \bibinfo{author}{Petersen, S.}, \bibinfo{author}{Tziritas, G.}, \bibinfo{author}{Grinias, E.}, \bibinfo{author}{Khened, M.}, \bibinfo{author}{Kollerathu, V.A.}, \bibinfo{author}{Krishnamurthi, G.}, \bibinfo{author}{Roh, M.M.}, \bibinfo{author}{Pennec, X.}, \bibinfo{author}{Sermesant, M.}, \bibinfo{author}{Isensee, F.}, \bibinfo{author}{Jger, P.}, \bibinfo{author}{Maier-Hein, K.H.}, \bibinfo{author}{Full, P.M.}, \bibinfo{author}{Wolf, I.}, \bibinfo{author}{Engelhardt, S.}, \bibinfo{author}{Baumgartner, C.F.}, \bibinfo{author}{Koch, L.M.}, \bibinfo{author}{Wolterink, J.M.}, \bibinfo{author}{Igum, I.}, \bibinfo{author}{Jang, Y.},
  \bibinfo{author}{Hong, Y.}, \bibinfo{author}{Patravali, J.}, \bibinfo{author}{Jain, S.}, \bibinfo{author}{Humbert, O.}, \bibinfo{author}{Jodoin, P.M.}, \bibinfo{year}{2018}.
\newblock \bibinfo{title}{Deep learning techniques for automatic mri cardiac multi-structures segmentation and diagnosis: Is the problem solved?}
\newblock \bibinfo{journal}{IEEE Transactions on Medical Imaging} \bibinfo{volume}{37}, \bibinfo{pages}{2514--2525}.
\newblock \DOIprefix\doi{10.1109/TMI.2018.2837502}.
\bibitem[{Bilic et~al.(2023)Bilic, Christ, Li, Vorontsov, Ben-Cohen, Kaissis, Szeskin, Jacobs, Mamani, Chartrand, Lohfer, Holch, Sommer, Hofmann, Hostettler, Lev-Cohain, Drozdzal, Amitai, Vivanti, Sosna, Ezhov, Sekuboyina, Navarro, Kofler, Paetzold, Shit, Hu, Lipkov, Rempfler, Piraud, Kirschke, Wiestler, Zhang, Hlsemeyer, Beetz, Ettlinger, Antonelli, Bae, Bellver, Bi, Chen, Chlebus, Dam, Dou, Fu, Georgescu, i~Nieto, Gruen, Han, Heng, Hesser, Moltz, Igel, Isensee, Jger, Jia, Kaluva, Khened, Kim, Kim, Kim, Kohl, Konopczynski, Kori, Krishnamurthi, Li, Li, Li, Li, Lowengrub, Ma, Maier-Hein, Maninis, Meine, Merhof, Pai, Perslev, Petersen, Pont-Tuset, Qi, Qi, Rippel, Roth, Sarasua, Schenk, Shen, Torres, Wachinger, Wang, Weninger, Wu, Xu, Yang, Yu, Yuan, Yue, Zhang, Cardoso, Bakas, Braren, Heinemann, Pal, Tang, Kadoury, Soler, {van Ginneken}, Greenspan, Joskowicz and Menze}]{lits_benchmark}
\bibinfo{author}{Bilic, P.}, \bibinfo{author}{Christ, P.}, \bibinfo{author}{Li, H.B.}, \bibinfo{author}{Vorontsov, E.}, \bibinfo{author}{Ben-Cohen, A.}, \bibinfo{author}{Kaissis, G.}, \bibinfo{author}{Szeskin, A.}, \bibinfo{author}{Jacobs, C.}, \bibinfo{author}{Mamani, G.E.H.}, \bibinfo{author}{Chartrand, G.}, \bibinfo{author}{Lohfer, F.}, \bibinfo{author}{Holch, J.W.}, \bibinfo{author}{Sommer, W.}, \bibinfo{author}{Hofmann, F.}, \bibinfo{author}{Hostettler, A.}, \bibinfo{author}{Lev-Cohain, N.}, \bibinfo{author}{Drozdzal, M.}, \bibinfo{author}{Amitai, M.M.}, \bibinfo{author}{Vivanti, R.}, \bibinfo{author}{Sosna, J.}, \bibinfo{author}{Ezhov, I.}, \bibinfo{author}{Sekuboyina, A.}, \bibinfo{author}{Navarro, F.}, \bibinfo{author}{Kofler, F.}, \bibinfo{author}{Paetzold, J.C.}, \bibinfo{author}{Shit, S.}, \bibinfo{author}{Hu, X.}, \bibinfo{author}{Lipkov, J.}, \bibinfo{author}{Rempfler, M.}, \bibinfo{author}{Piraud, M.}, \bibinfo{author}{Kirschke, J.}, \bibinfo{author}{Wiestler, B.}, \bibinfo{author}{Zhang, Z.},
  \bibinfo{author}{Hlsemeyer, C.}, \bibinfo{author}{Beetz, M.}, \bibinfo{author}{Ettlinger, F.}, \bibinfo{author}{Antonelli, M.}, \bibinfo{author}{Bae, W.}, \bibinfo{author}{Bellver, M.}, \bibinfo{author}{Bi, L.}, \bibinfo{author}{Chen, H.}, \bibinfo{author}{Chlebus, G.}, \bibinfo{author}{Dam, E.B.}, \bibinfo{author}{Dou, Q.}, \bibinfo{author}{Fu, C.W.}, \bibinfo{author}{Georgescu, B.}, \bibinfo{author}{i~Nieto, X.G.}, \bibinfo{author}{Gruen, F.}, \bibinfo{author}{Han, X.}, \bibinfo{author}{Heng, P.A.}, \bibinfo{author}{Hesser, J.}, \bibinfo{author}{Moltz, J.H.}, \bibinfo{author}{Igel, C.}, \bibinfo{author}{Isensee, F.}, \bibinfo{author}{Jger, P.}, \bibinfo{author}{Jia, F.}, \bibinfo{author}{Kaluva, K.C.}, \bibinfo{author}{Khened, M.}, \bibinfo{author}{Kim, I.}, \bibinfo{author}{Kim, J.H.}, \bibinfo{author}{Kim, S.}, \bibinfo{author}{Kohl, S.}, \bibinfo{author}{Konopczynski, T.}, \bibinfo{author}{Kori, A.}, \bibinfo{author}{Krishnamurthi, G.}, \bibinfo{author}{Li, F.}, \bibinfo{author}{Li, H.},
  \bibinfo{author}{Li, J.}, \bibinfo{author}{Li, X.}, \bibinfo{author}{Lowengrub, J.}, \bibinfo{author}{Ma, J.}, \bibinfo{author}{Maier-Hein, K.}, \bibinfo{author}{Maninis, K.K.}, \bibinfo{author}{Meine, H.}, \bibinfo{author}{Merhof, D.}, \bibinfo{author}{Pai, A.}, \bibinfo{author}{Perslev, M.}, \bibinfo{author}{Petersen, J.}, \bibinfo{author}{Pont-Tuset, J.}, \bibinfo{author}{Qi, J.}, \bibinfo{author}{Qi, X.}, \bibinfo{author}{Rippel, O.}, \bibinfo{author}{Roth, K.}, \bibinfo{author}{Sarasua, I.}, \bibinfo{author}{Schenk, A.}, \bibinfo{author}{Shen, Z.}, \bibinfo{author}{Torres, J.}, \bibinfo{author}{Wachinger, C.}, \bibinfo{author}{Wang, C.}, \bibinfo{author}{Weninger, L.}, \bibinfo{author}{Wu, J.}, \bibinfo{author}{Xu, D.}, \bibinfo{author}{Yang, X.}, \bibinfo{author}{Yu, S.C.H.}, \bibinfo{author}{Yuan, Y.}, \bibinfo{author}{Yue, M.}, \bibinfo{author}{Zhang, L.}, \bibinfo{author}{Cardoso, J.}, \bibinfo{author}{Bakas, S.}, \bibinfo{author}{Braren, R.}, \bibinfo{author}{Heinemann, V.}, \bibinfo{author}{Pal,
  C.}, \bibinfo{author}{Tang, A.}, \bibinfo{author}{Kadoury, S.}, \bibinfo{author}{Soler, L.}, \bibinfo{author}{{van Ginneken}, B.}, \bibinfo{author}{Greenspan, H.}, \bibinfo{author}{Joskowicz, L.}, \bibinfo{author}{Menze, B.}, \bibinfo{year}{2023}.
\newblock \bibinfo{title}{The liver tumor segmentation benchmark (lits)}.
\newblock \bibinfo{journal}{Medical Image Analysis} \bibinfo{volume}{84}, \bibinfo{pages}{102680}.
\newblock \URLprefix \url{https://www.sciencedirect.com/science/article/pii/S1361841522003085}, \DOIprefix\doi{https://doi.org/10.1016/j.media.2022.102680}.
\bibitem[{Bolelli et~al.(2024)Bolelli, Lumetti, Vinayahalingam, Di~Bartolomeo, Pellacani, Marchesini, van Nistelrooij, van Lierop, Xi, Liu, Xin, Yang, Wang, Wang, Xu, Cui, Wodzinski, Müller, Kirchhoff, R.~Rokuss, Maier-Hein, Han, Kim, Ahn, Szczepański, Grzeszczyk, Korzeniowski, Caselles~Ballester, Prados~Carrasco, Berge’, van Ginneken, Anesi and Grana}]{2024TMI}
\bibinfo{author}{Bolelli, F.}, \bibinfo{author}{Lumetti, L.}, \bibinfo{author}{Vinayahalingam, S.}, \bibinfo{author}{Di~Bartolomeo, M.}, \bibinfo{author}{Pellacani, A.}, \bibinfo{author}{Marchesini, K.}, \bibinfo{author}{van Nistelrooij, N.}, \bibinfo{author}{van Lierop, P.}, \bibinfo{author}{Xi, T.}, \bibinfo{author}{Liu, Y.}, \bibinfo{author}{Xin, R.}, \bibinfo{author}{Yang, T.}, \bibinfo{author}{Wang, L.}, \bibinfo{author}{Wang, H.}, \bibinfo{author}{Xu, C.}, \bibinfo{author}{Cui, Z.}, \bibinfo{author}{Wodzinski, M.}, \bibinfo{author}{Müller, H.}, \bibinfo{author}{Kirchhoff, Y.}, \bibinfo{author}{R.~Rokuss, M.}, \bibinfo{author}{Maier-Hein, K.}, \bibinfo{author}{Han, J.}, \bibinfo{author}{Kim, W.}, \bibinfo{author}{Ahn, H.G.}, \bibinfo{author}{Szczepański, T.}, \bibinfo{author}{Grzeszczyk, M.K.}, \bibinfo{author}{Korzeniowski, P.}, \bibinfo{author}{Caselles~Ballester, Vicent amd Paolo Burgos-Artizzu, X.}, \bibinfo{author}{Prados~Carrasco, F.}, \bibinfo{author}{Berge’, S.}, \bibinfo{author}{van
  Ginneken, B.}, \bibinfo{author}{Anesi, A.}, \bibinfo{author}{Grana, C.}, \bibinfo{year}{2024}.
\newblock \bibinfo{title}{Segmenting the inferior alveolar canal in cbcts volumes: the toothfairy challenge}.
\newblock \bibinfo{journal}{IEEE Transactions on Medical Imaging} , \bibinfo{pages}{1--17}\DOIprefix\doi{https://doi.org/10.1109/TMI.2024.3523096}.
\bibitem[{Bonato et~al.(2025)Bonato, Nanni and Bertoldo}]{s25061838}
\bibinfo{author}{Bonato, B.}, \bibinfo{author}{Nanni, L.}, \bibinfo{author}{Bertoldo, A.}, \bibinfo{year}{2025}.
\newblock \bibinfo{title}{Advancing precision: A comprehensive review of mri segmentation datasets from brats challenges (20122025)}.
\newblock \bibinfo{journal}{Sensors} \bibinfo{volume}{25}.
\newblock \URLprefix \url{https://www.mdpi.com/1424-8220/25/6/1838}, \DOIprefix\doi{10.3390/s25061838}.
\bibitem[{Campello et~al.(2021)Campello, Gkontra, Izquierdo, Martn-Isla, Sojoudi, Full, Maier-Hein, Zhang, He, Ma, Parreo, Albiol, Kong, Shadden, Acero, Sundaresan, Saber, Elattar, Li, Menze, Khader, Haarburger, Scannell, Veta, Carscadden, Punithakumar, Liu, Tsaftaris, Huang, Yang, Li, Zhuang, Vilads, Descalzo, Guala, Mura, Friedrich, Garg, Lebel, Henriques, Karakas, avu, Petersen, Escalera, Segu, Rodrguez-Palomares and Lekadir}]{9458279}
\bibinfo{author}{Campello, V.M.}, \bibinfo{author}{Gkontra, P.}, \bibinfo{author}{Izquierdo, C.}, \bibinfo{author}{Martn-Isla, C.}, \bibinfo{author}{Sojoudi, A.}, \bibinfo{author}{Full, P.M.}, \bibinfo{author}{Maier-Hein, K.}, \bibinfo{author}{Zhang, Y.}, \bibinfo{author}{He, Z.}, \bibinfo{author}{Ma, J.}, \bibinfo{author}{Parreo, M.}, \bibinfo{author}{Albiol, A.}, \bibinfo{author}{Kong, F.}, \bibinfo{author}{Shadden, S.C.}, \bibinfo{author}{Acero, J.C.}, \bibinfo{author}{Sundaresan, V.}, \bibinfo{author}{Saber, M.}, \bibinfo{author}{Elattar, M.}, \bibinfo{author}{Li, H.}, \bibinfo{author}{Menze, B.}, \bibinfo{author}{Khader, F.}, \bibinfo{author}{Haarburger, C.}, \bibinfo{author}{Scannell, C.M.}, \bibinfo{author}{Veta, M.}, \bibinfo{author}{Carscadden, A.}, \bibinfo{author}{Punithakumar, K.}, \bibinfo{author}{Liu, X.}, \bibinfo{author}{Tsaftaris, S.A.}, \bibinfo{author}{Huang, X.}, \bibinfo{author}{Yang, X.}, \bibinfo{author}{Li, L.}, \bibinfo{author}{Zhuang, X.}, \bibinfo{author}{Vilads, D.},
  \bibinfo{author}{Descalzo, M.L.}, \bibinfo{author}{Guala, A.}, \bibinfo{author}{Mura, L.L.}, \bibinfo{author}{Friedrich, M.G.}, \bibinfo{author}{Garg, R.}, \bibinfo{author}{Lebel, J.}, \bibinfo{author}{Henriques, F.}, \bibinfo{author}{Karakas, M.}, \bibinfo{author}{avu, E.}, \bibinfo{author}{Petersen, S.E.}, \bibinfo{author}{Escalera, S.}, \bibinfo{author}{Segu, S.}, \bibinfo{author}{Rodrguez-Palomares, J.F.}, \bibinfo{author}{Lekadir, K.}, \bibinfo{year}{2021}.
\newblock \bibinfo{title}{Multi-centre, multi-vendor and multi-disease cardiac segmentation: The m\&ms challenge}.
\newblock \bibinfo{journal}{IEEE Transactions on Medical Imaging} \bibinfo{volume}{40}, \bibinfo{pages}{3543--3554}.
\newblock \DOIprefix\doi{10.1109/TMI.2021.3090082}.
\bibitem[{Cao et~al.(2021)Cao, Wang, Chen, Jiang, Zhang, Tian and Wang}]{cao2021swinunetunetlikepuretransformer}
\bibinfo{author}{Cao, H.}, \bibinfo{author}{Wang, Y.}, \bibinfo{author}{Chen, J.}, \bibinfo{author}{Jiang, D.}, \bibinfo{author}{Zhang, X.}, \bibinfo{author}{Tian, Q.}, \bibinfo{author}{Wang, M.}, \bibinfo{year}{2021}.
\newblock \bibinfo{title}{Swin-unet: Unet-like pure transformer for medical image segmentation}.
\newblock \URLprefix \url{https://arxiv.org/abs/2105.05537}, \href{http://arxiv.org/abs/2105.05537}{{\tt arXiv:2105.05537}}.
\bibitem[{Cao et~al.(2023a)Cao, Wang, Chen, Jiang, Zhang, Tian and Wang}]{10.1007/978-3-031-25066-8_9}
\bibinfo{author}{Cao, H.}, \bibinfo{author}{Wang, Y.}, \bibinfo{author}{Chen, J.}, \bibinfo{author}{Jiang, D.}, \bibinfo{author}{Zhang, X.}, \bibinfo{author}{Tian, Q.}, \bibinfo{author}{Wang, M.}, \bibinfo{year}{2023}a.
\newblock \bibinfo{title}{Swin-unet: Unet-like pure transformer for medical image segmentation}, in: \bibinfo{editor}{Karlinsky, L.}, \bibinfo{editor}{Michaeli, T.}, \bibinfo{editor}{Nishino, K.} (Eds.), \bibinfo{booktitle}{Computer Vision -- ECCV 2022 Workshops}, \bibinfo{publisher}{Springer Nature Switzerland}, \bibinfo{address}{Cham}. pp. \bibinfo{pages}{205--218}.
\bibitem[{Cao et~al.(2023b)Cao, Wang, Chen, Jiang, Zhang, Tian and Wang}]{cao2022swin}
\bibinfo{author}{Cao, H.}, \bibinfo{author}{Wang, Y.}, \bibinfo{author}{Chen, J.}, \bibinfo{author}{Jiang, D.}, \bibinfo{author}{Zhang, X.}, \bibinfo{author}{Tian, Q.}, \bibinfo{author}{Wang, M.}, \bibinfo{year}{2023}b.
\newblock \bibinfo{title}{Swin-unet: Unet-like pure transformer for medical image segmentation}, in: \bibinfo{booktitle}{Computer Vision – ECCV 2022 Workshops}, \bibinfo{publisher}{Springer Nature Switzerland}. pp. \bibinfo{pages}{205--218}.
\newblock \DOIprefix\doi{10.1007/978-3-031-25066-8_9}.
\bibitem[{Chen et~al.(2024a)Chen, Miao, Wu, Zhong, Yan, Kim, Hu, Liu, Sun, Li, Liu, Heng and Li}]{CHEN2024103310}
\bibinfo{author}{Chen, C.}, \bibinfo{author}{Miao, J.}, \bibinfo{author}{Wu, D.}, \bibinfo{author}{Zhong, A.}, \bibinfo{author}{Yan, Z.}, \bibinfo{author}{Kim, S.}, \bibinfo{author}{Hu, J.}, \bibinfo{author}{Liu, Z.}, \bibinfo{author}{Sun, L.}, \bibinfo{author}{Li, X.}, \bibinfo{author}{Liu, T.}, \bibinfo{author}{Heng, P.A.}, \bibinfo{author}{Li, Q.}, \bibinfo{year}{2024}a.
\newblock \bibinfo{title}{Ma-sam: Modality-agnostic sam adaptation for 3d medical image segmentation}.
\newblock \bibinfo{journal}{Medical Image Analysis} \bibinfo{volume}{98}, \bibinfo{pages}{103310}.
\newblock \URLprefix \url{https://www.sciencedirect.com/science/article/pii/S1361841524002354}, \DOIprefix\doi{https://doi.org/10.1016/j.media.2024.103310}.
\bibitem[{Chen et~al.(2021)Chen, Lu, Yu, Luo, Adeli, Wang, Lu, Yuille and Zhou}]{chen2021transunettransformersmakestrong}
\bibinfo{author}{Chen, J.}, \bibinfo{author}{Lu, Y.}, \bibinfo{author}{Yu, Q.}, \bibinfo{author}{Luo, X.}, \bibinfo{author}{Adeli, E.}, \bibinfo{author}{Wang, Y.}, \bibinfo{author}{Lu, L.}, \bibinfo{author}{Yuille, A.L.}, \bibinfo{author}{Zhou, Y.}, \bibinfo{year}{2021}.
\newblock \bibinfo{title}{Transunet: Transformers make strong encoders for medical image segmentation}.
\newblock \URLprefix \url{https://arxiv.org/abs/2102.04306}, \href{http://arxiv.org/abs/2102.04306}{{\tt arXiv:2102.04306}}.
\bibitem[{Chen et~al.(2024b)Chen, Mei, Li, Lu, Yu, Wei, Luo, Xie, Adeli, Wang, Lungren, Zhang, Xing, Lu, Yuille and Zhou}]{CHEN2024103280}
\bibinfo{author}{Chen, J.}, \bibinfo{author}{Mei, J.}, \bibinfo{author}{Li, X.}, \bibinfo{author}{Lu, Y.}, \bibinfo{author}{Yu, Q.}, \bibinfo{author}{Wei, Q.}, \bibinfo{author}{Luo, X.}, \bibinfo{author}{Xie, Y.}, \bibinfo{author}{Adeli, E.}, \bibinfo{author}{Wang, Y.}, \bibinfo{author}{Lungren, M.P.}, \bibinfo{author}{Zhang, S.}, \bibinfo{author}{Xing, L.}, \bibinfo{author}{Lu, L.}, \bibinfo{author}{Yuille, A.}, \bibinfo{author}{Zhou, Y.}, \bibinfo{year}{2024}b.
\newblock \bibinfo{title}{Transunet: Rethinking the u-net architecture design for medical image segmentation through the lens of transformers}.
\newblock \bibinfo{journal}{Medical Image Analysis} \bibinfo{volume}{97}, \bibinfo{pages}{103280}.
\newblock \URLprefix \url{https://www.sciencedirect.com/science/article/pii/S1361841524002056}, \DOIprefix\doi{https://doi.org/10.1016/j.media.2024.103280}.
\bibitem[{Chen et~al.(2019)Chen, Ma and Zheng}]{chen2019med3dtransferlearning3d}
\bibinfo{author}{Chen, S.}, \bibinfo{author}{Ma, K.}, \bibinfo{author}{Zheng, Y.}, \bibinfo{year}{2019}.
\newblock \bibinfo{title}{Med3d: Transfer learning for 3d medical image analysis}.
\newblock \URLprefix \url{https://arxiv.org/abs/1904.00625}, \href{http://arxiv.org/abs/1904.00625}{{\tt arXiv:1904.00625}}.
\bibitem[{Chen et~al.(2024c)Chen, Gao, Zhu, Shao, Lu, Han and Xie}]{10510478}
\bibinfo{author}{Chen, Y.}, \bibinfo{author}{Gao, Y.}, \bibinfo{author}{Zhu, L.}, \bibinfo{author}{Shao, W.}, \bibinfo{author}{Lu, Y.}, \bibinfo{author}{Han, H.}, \bibinfo{author}{Xie, Z.}, \bibinfo{year}{2024}c.
\newblock \bibinfo{title}{Pcnet: Prior category network for ct universal segmentation model}.
\newblock \bibinfo{journal}{IEEE Transactions on Medical Imaging} \bibinfo{volume}{43}, \bibinfo{pages}{3319--3330}.
\newblock \DOIprefix\doi{10.1109/TMI.2024.3395349}.
\bibitem[{Cheng et~al.(2023)Cheng, Ye, Deng, Chen, Li, Wang, Su, Huang, Chen, Jiang, Sun, He, Zhang, Zhu and Qiao}]{cheng2023sammed2d}
\bibinfo{author}{Cheng, J.}, \bibinfo{author}{Ye, J.}, \bibinfo{author}{Deng, Z.}, \bibinfo{author}{Chen, J.}, \bibinfo{author}{Li, T.}, \bibinfo{author}{Wang, H.}, \bibinfo{author}{Su, Y.}, \bibinfo{author}{Huang, Z.}, \bibinfo{author}{Chen, J.}, \bibinfo{author}{Jiang, L.}, \bibinfo{author}{Sun, H.}, \bibinfo{author}{He, J.}, \bibinfo{author}{Zhang, S.}, \bibinfo{author}{Zhu, M.}, \bibinfo{author}{Qiao, Y.}, \bibinfo{year}{2023}.
\newblock \bibinfo{title}{Sam-med2d}.
\newblock \URLprefix \url{https://arxiv.org/abs/2308.16184}, \href{http://arxiv.org/abs/2308.16184}{{\tt arXiv:2308.16184}}.
\bibitem[{{\c{C}}i{\c{c}}ek et~al.(2016){\c{C}}i{\c{c}}ek, Abdulkadir, Lienkamp, Brox and Ronneberger}]{10.1007/978-3-319-46723-8_49}
\bibinfo{author}{{\c{C}}i{\c{c}}ek, {\"O}.}, \bibinfo{author}{Abdulkadir, A.}, \bibinfo{author}{Lienkamp, S.S.}, \bibinfo{author}{Brox, T.}, \bibinfo{author}{Ronneberger, O.}, \bibinfo{year}{2016}.
\newblock \bibinfo{title}{3d u-net: Learning dense volumetric segmentation from sparse annotation}, in: \bibinfo{editor}{Ourselin, S.}, \bibinfo{editor}{Joskowicz, L.}, \bibinfo{editor}{Sabuncu, M.R.}, \bibinfo{editor}{Unal, G.}, \bibinfo{editor}{Wells, W.} (Eds.), \bibinfo{booktitle}{Medical Image Computing and Computer-Assisted Intervention -- MICCAI 2016}, \bibinfo{publisher}{Springer International Publishing}, \bibinfo{address}{Cham}. pp. \bibinfo{pages}{424--432}.
\bibitem[{Cipriano et~al.(2022)Cipriano, Allegretti, Bolelli, Pollastri and Grana}]{Cipriano_2022_CVPR}
\bibinfo{author}{Cipriano, M.}, \bibinfo{author}{Allegretti, S.}, \bibinfo{author}{Bolelli, F.}, \bibinfo{author}{Pollastri, F.}, \bibinfo{author}{Grana, C.}, \bibinfo{year}{2022}.
\newblock \bibinfo{title}{Improving segmentation of the inferior alveolar nerve through deep label propagation}, in: \bibinfo{booktitle}{Proceedings of the IEEE/CVF Conference on Computer Vision and Pattern Recognition (CVPR)}, pp. \bibinfo{pages}{21137--21146}.
\bibitem[{{de Grauw} et~al.(2025){de Grauw}, Scholten, Smit, Rutten, Prokop, {van Ginneken} and Hering}]{DEGRAUW2025103525}
\bibinfo{author}{{de Grauw}, M.}, \bibinfo{author}{Scholten, E.}, \bibinfo{author}{Smit, E.}, \bibinfo{author}{Rutten, M.}, \bibinfo{author}{Prokop, M.}, \bibinfo{author}{{van Ginneken}, B.}, \bibinfo{author}{Hering, A.}, \bibinfo{year}{2025}.
\newblock \bibinfo{title}{The uls23 challenge: A baseline model and benchmark dataset for 3d universal lesion segmentation in computed tomography}.
\newblock \bibinfo{journal}{Medical Image Analysis} \bibinfo{volume}{102}, \bibinfo{pages}{103525}.
\newblock \URLprefix \url{https://www.sciencedirect.com/science/article/pii/S1361841525000738}, \DOIprefix\doi{https://doi.org/10.1016/j.media.2025.103525}.
\bibitem[{Deng et~al.(2024)Deng, Wang, Hui, Li, Li, Luo, Sun, Quan, Yang, Hao, Liu, Xiao, Zhao, Wu and Zhou}]{deng2024ctspine1klargescaledatasetspinal}
\bibinfo{author}{Deng, Y.}, \bibinfo{author}{Wang, C.}, \bibinfo{author}{Hui, Y.}, \bibinfo{author}{Li, Q.}, \bibinfo{author}{Li, J.}, \bibinfo{author}{Luo, S.}, \bibinfo{author}{Sun, M.}, \bibinfo{author}{Quan, Q.}, \bibinfo{author}{Yang, S.}, \bibinfo{author}{Hao, Y.}, \bibinfo{author}{Liu, P.}, \bibinfo{author}{Xiao, H.}, \bibinfo{author}{Zhao, C.}, \bibinfo{author}{Wu, X.}, \bibinfo{author}{Zhou, S.K.}, \bibinfo{year}{2024}.
\newblock \bibinfo{title}{Ctspine1k: A large-scale dataset for spinal vertebrae segmentation in computed tomography}.
\newblock \URLprefix \url{https://arxiv.org/abs/2105.14711}, \href{http://arxiv.org/abs/2105.14711}{{\tt arXiv:2105.14711}}.
\bibitem[{Dong et~al.(2024)Dong, Wang, Chen, Sun, Song, Liu and Cui}]{Dong2024}
\bibinfo{author}{Dong, G.}, \bibinfo{author}{Wang, Z.}, \bibinfo{author}{Chen, Y.}, \bibinfo{author}{Sun, Y.}, \bibinfo{author}{Song, H.}, \bibinfo{author}{Liu, L.}, \bibinfo{author}{Cui, H.}, \bibinfo{year}{2024}.
\newblock \bibinfo{title}{An efficient segment anything model for the segmentation of medical images}.
\newblock \bibinfo{journal}{Scientific Reports} \bibinfo{volume}{14}, \bibinfo{pages}{19425}.
\newblock \URLprefix \url{https://doi.org/10.1038/s41598-024-70288-8}, \DOIprefix\doi{10.1038/s41598-024-70288-8}.
\bibitem[{Dowling et~al.(2009)Dowling, Fripp, Greer, Ourselin and Salvado}]{dowling2009automatic}
\bibinfo{author}{Dowling, J.}, \bibinfo{author}{Fripp, J.}, \bibinfo{author}{Greer, P.}, \bibinfo{author}{Ourselin, S.}, \bibinfo{author}{Salvado, O.}, \bibinfo{year}{2009}.
\newblock \bibinfo{title}{Automatic atlas-based segmentation of the prostate: A miccai 2009 prostate segmentation challenge entry}.
\newblock \bibinfo{journal}{Worskshop in Med Image Comput Comput Assist Interv} \bibinfo{volume}{24}, \bibinfo{pages}{17--24}.
\bibitem[{Gao(2024)}]{10658004}
\bibinfo{author}{Gao, Y.}, \bibinfo{year}{2024}.
\newblock \bibinfo{title}{Training like a medical resident: Context-prior learning toward universal medical image segmentation}, in: \bibinfo{booktitle}{2024 IEEE/CVF Conference on Computer Vision and Pattern Recognition (CVPR)}, pp. \bibinfo{pages}{11194--11204}.
\newblock \DOIprefix\doi{10.1109/CVPR52733.2024.01064}.
\bibitem[{Gao et~al.(2023)Gao, Zhou, Liu, Yan, Zhang and Metaxas}]{gao2023datascalabletransformermedicalimage}
\bibinfo{author}{Gao, Y.}, \bibinfo{author}{Zhou, M.}, \bibinfo{author}{Liu, D.}, \bibinfo{author}{Yan, Z.}, \bibinfo{author}{Zhang, S.}, \bibinfo{author}{Metaxas, D.N.}, \bibinfo{year}{2023}.
\newblock \bibinfo{title}{A data-scalable transformer for medical image segmentation: Architecture, model efficiency, and benchmark}.
\newblock \URLprefix \url{https://arxiv.org/abs/2203.00131}, \href{http://arxiv.org/abs/2203.00131}{{\tt arXiv:2203.00131}}.
\bibitem[{Gatidis et~al.(2022)Gatidis, Hepp, Fr{\"u}h, La~Foug{\`e}re, Nikolaou, Pfannenberg, Sch{\"o}lkopf, K{\"u}stner, Cyran and Rubin}]{gatidis2022whole}
\bibinfo{author}{Gatidis, S.}, \bibinfo{author}{Hepp, T.}, \bibinfo{author}{Fr{\"u}h, M.}, \bibinfo{author}{La~Foug{\`e}re, C.}, \bibinfo{author}{Nikolaou, K.}, \bibinfo{author}{Pfannenberg, C.}, \bibinfo{author}{Sch{\"o}lkopf, B.}, \bibinfo{author}{K{\"u}stner, T.}, \bibinfo{author}{Cyran, C.}, \bibinfo{author}{Rubin, D.}, \bibinfo{year}{2022}.
\newblock \bibinfo{title}{A whole-body fdg-pet/ct dataset with manually annotated tumor lesions}.
\newblock \bibinfo{journal}{Scientific Data} \bibinfo{volume}{9}, \bibinfo{pages}{601}.
\bibitem[{Gharleghi et~al.(2022)Gharleghi, Adikari, Ellenberger, Ooi, Ellis, Chen, Gao, He, Hussain, Lee, Li, Ma, Nie, Oliveira, Qi, Skandarani, Vilaa, Wang, Yang, Sowmya and Beier}]{gharleghi2022automated}
\bibinfo{author}{Gharleghi, R.}, \bibinfo{author}{Adikari, D.}, \bibinfo{author}{Ellenberger, K.}, \bibinfo{author}{Ooi, S.Y.}, \bibinfo{author}{Ellis, C.}, \bibinfo{author}{Chen, C.M.}, \bibinfo{author}{Gao, R.}, \bibinfo{author}{He, Y.}, \bibinfo{author}{Hussain, R.}, \bibinfo{author}{Lee, C.Y.}, \bibinfo{author}{Li, J.}, \bibinfo{author}{Ma, J.}, \bibinfo{author}{Nie, Z.}, \bibinfo{author}{Oliveira, B.}, \bibinfo{author}{Qi, Y.}, \bibinfo{author}{Skandarani, Y.}, \bibinfo{author}{Vilaa, J.L.}, \bibinfo{author}{Wang, X.}, \bibinfo{author}{Yang, S.}, \bibinfo{author}{Sowmya, A.}, \bibinfo{author}{Beier, S.}, \bibinfo{year}{2022}.
\newblock \bibinfo{title}{Automated segmentation of normal and diseased coronary arteries the asoca challenge}.
\newblock \bibinfo{journal}{Computerized Medical Imaging and Graphics} \bibinfo{volume}{97}, \bibinfo{pages}{102049}.
\newblock \URLprefix \url{https://www.sciencedirect.com/science/article/pii/S0895611122000222}, \DOIprefix\doi{https://doi.org/10.1016/j.compmedimag.2022.102049}.
\bibitem[{Gharleghi et~al.(2023)Gharleghi, Adikari, Ellenberger, Webster, Ellis, Sowmya, Ooi and Beier}]{gharleghi2023annotated}
\bibinfo{author}{Gharleghi, R.}, \bibinfo{author}{Adikari, D.}, \bibinfo{author}{Ellenberger, K.}, \bibinfo{author}{Webster, M.}, \bibinfo{author}{Ellis, C.}, \bibinfo{author}{Sowmya, A.}, \bibinfo{author}{Ooi, S.}, \bibinfo{author}{Beier, S.}, \bibinfo{year}{2023}.
\newblock \bibinfo{title}{Annotated computed tomography coronary angiogram images and associated data of normal and diseased arteries}.
\newblock \bibinfo{journal}{Scientific Data} \bibinfo{volume}{10}, \bibinfo{pages}{128}.
\bibitem[{Gong et~al.(2024)Gong, Zhong, Ma, Li, Wang, Zhang, Heng and Dou}]{3DSAM-adapter}
\bibinfo{author}{Gong, S.}, \bibinfo{author}{Zhong, Y.}, \bibinfo{author}{Ma, W.}, \bibinfo{author}{Li, J.}, \bibinfo{author}{Wang, Z.}, \bibinfo{author}{Zhang, J.}, \bibinfo{author}{Heng, P.A.}, \bibinfo{author}{Dou, Q.}, \bibinfo{year}{2024}.
\newblock \bibinfo{title}{3dsam-adapter: Holistic adaptation of sam from 2d to 3d for promptable tumor segmentation}.
\newblock \bibinfo{journal}{Medical Image Analysis} \bibinfo{volume}{98}, \bibinfo{pages}{103324}.
\newblock \URLprefix \url{https://www.sciencedirect.com/science/article/pii/S1361841524002494}, \DOIprefix\doi{https://doi.org/10.1016/j.media.2024.103324}.
\bibitem[{Gu et~al.(2024)Gu, Wu, Tang, Mai, Shu, Li and Chen}]{10540651}
\bibinfo{author}{Gu, Y.}, \bibinfo{author}{Wu, Q.}, \bibinfo{author}{Tang, H.}, \bibinfo{author}{Mai, X.}, \bibinfo{author}{Shu, H.}, \bibinfo{author}{Li, B.}, \bibinfo{author}{Chen, Y.}, \bibinfo{year}{2024}.
\newblock \bibinfo{title}{Lesam: Adapt segment anything model for medical lesion segmentation}.
\newblock \bibinfo{journal}{IEEE Journal of Biomedical and Health Informatics} \bibinfo{volume}{28}, \bibinfo{pages}{6031--6041}.
\newblock \DOIprefix\doi{10.1109/JBHI.2024.3406871}.
\bibitem[{Hatamizadeh et~al.(2022a)Hatamizadeh, Nath, Tang, Yang, Roth and Xu}]{hatamizadeh2022swinunetrswintransformers}
\bibinfo{author}{Hatamizadeh, A.}, \bibinfo{author}{Nath, V.}, \bibinfo{author}{Tang, Y.}, \bibinfo{author}{Yang, D.}, \bibinfo{author}{Roth, H.}, \bibinfo{author}{Xu, D.}, \bibinfo{year}{2022}a.
\newblock \bibinfo{title}{Swin unetr: Swin transformers for semantic segmentation of brain tumors in mri images}.
\newblock \URLprefix \url{https://arxiv.org/abs/2201.01266}, \href{http://arxiv.org/abs/2201.01266}{{\tt arXiv:2201.01266}}.
\bibitem[{Hatamizadeh et~al.(2022b)Hatamizadeh, Nath, Tang, Yang, Roth and Xu}]{10.1007/978-3-031-08999-2_22}
\bibinfo{author}{Hatamizadeh, A.}, \bibinfo{author}{Nath, V.}, \bibinfo{author}{Tang, Y.}, \bibinfo{author}{Yang, D.}, \bibinfo{author}{Roth, H.R.}, \bibinfo{author}{Xu, D.}, \bibinfo{year}{2022}b.
\newblock \bibinfo{title}{Swin unetr: Swin transformers for semantic segmentation of brain tumors in mri images}, in: \bibinfo{editor}{Crimi, A.}, \bibinfo{editor}{Bakas, S.} (Eds.), \bibinfo{booktitle}{Brainlesion: Glioma, Multiple Sclerosis, Stroke and Traumatic Brain Injuries}, \bibinfo{publisher}{Springer International Publishing}, \bibinfo{address}{Cham}. pp. \bibinfo{pages}{272--284}.
\bibitem[{Hatamizadeh et~al.(2021)Hatamizadeh, Tang, Nath, Yang, Myronenko, Landman, Roth and Xu}]{hatamizadeh2021unetrtransformers3dmedical}
\bibinfo{author}{Hatamizadeh, A.}, \bibinfo{author}{Tang, Y.}, \bibinfo{author}{Nath, V.}, \bibinfo{author}{Yang, D.}, \bibinfo{author}{Myronenko, A.}, \bibinfo{author}{Landman, B.}, \bibinfo{author}{Roth, H.}, \bibinfo{author}{Xu, D.}, \bibinfo{year}{2021}.
\newblock \bibinfo{title}{Unetr: Transformers for 3d medical image segmentation}.
\newblock \URLprefix \url{https://arxiv.org/abs/2103.10504}, \href{http://arxiv.org/abs/2103.10504}{{\tt arXiv:2103.10504}}.
\bibitem[{Hatamizadeh et~al.(2022c)Hatamizadeh, Tang, Nath, Yang, Myronenko, Landman, Roth and Xu}]{9706678}
\bibinfo{author}{Hatamizadeh, A.}, \bibinfo{author}{Tang, Y.}, \bibinfo{author}{Nath, V.}, \bibinfo{author}{Yang, D.}, \bibinfo{author}{Myronenko, A.}, \bibinfo{author}{Landman, B.}, \bibinfo{author}{Roth, H.R.}, \bibinfo{author}{Xu, D.}, \bibinfo{year}{2022}c.
\newblock \bibinfo{title}{Unetr: Transformers for 3d medical image segmentation}, in: \bibinfo{booktitle}{2022 IEEE/CVF Winter Conference on Applications of Computer Vision (WACV)}, pp. \bibinfo{pages}{1748--1758}.
\newblock \DOIprefix\doi{10.1109/WACV51458.2022.00181}.
\bibitem[{He et~al.(2023a)He, Nath, Yang, Tang, Myronenko and Xu}]{10.1007/978-3-031-43901-8_40}
\bibinfo{author}{He, Y.}, \bibinfo{author}{Nath, V.}, \bibinfo{author}{Yang, D.}, \bibinfo{author}{Tang, Y.}, \bibinfo{author}{Myronenko, A.}, \bibinfo{author}{Xu, D.}, \bibinfo{year}{2023}a.
\newblock \bibinfo{title}{Swinunetr-v2: Stronger swin transformers with stagewise convolutions for 3d medical image segmentation}, in: \bibinfo{editor}{Greenspan, H.}, \bibinfo{editor}{Madabhushi, A.}, \bibinfo{editor}{Mousavi, P.}, \bibinfo{editor}{Salcudean, S.}, \bibinfo{editor}{Duncan, J.}, \bibinfo{editor}{Syeda-Mahmood, T.}, \bibinfo{editor}{Taylor, R.} (Eds.), \bibinfo{booktitle}{Medical Image Computing and Computer Assisted Intervention -- MICCAI 2023}, \bibinfo{publisher}{Springer Nature Switzerland}, \bibinfo{address}{Cham}. pp. \bibinfo{pages}{416--426}.
\bibitem[{He et~al.(2023b)He, Nath, Yang, Tang, Myronenko and Xu}]{SwinUNETRv2_miccai}
\bibinfo{author}{He, Y.}, \bibinfo{author}{Nath, V.}, \bibinfo{author}{Yang, D.}, \bibinfo{author}{Tang, Y.}, \bibinfo{author}{Myronenko, A.}, \bibinfo{author}{Xu, D.}, \bibinfo{year}{2023}b.
\newblock \bibinfo{title}{Swinunetr-v2: Stronger swin transformers with stagewise convolutions for 3d medical image segmentation}, in: \bibinfo{editor}{Greenspan, H.}, \bibinfo{editor}{Madabhushi, A.}, \bibinfo{editor}{Mousavi, P.}, \bibinfo{editor}{Salcudean, S.}, \bibinfo{editor}{Duncan, J.}, \bibinfo{editor}{Syeda-Mahmood, T.}, \bibinfo{editor}{Taylor, R.} (Eds.), \bibinfo{booktitle}{Medical Image Computing and Computer Assisted Intervention -- MICCAI 2023}, \bibinfo{publisher}{Springer Nature Switzerland}, \bibinfo{address}{Cham}. pp. \bibinfo{pages}{416--426}.
\bibitem[{He et~al.(2021)He, Yang, Yang, Ge, Kong, Zhu, Zhang, Shao, Shu, Dillenseger, Coatrieux and Li}]{HE2021102055}
\bibinfo{author}{He, Y.}, \bibinfo{author}{Yang, G.}, \bibinfo{author}{Yang, J.}, \bibinfo{author}{Ge, R.}, \bibinfo{author}{Kong, Y.}, \bibinfo{author}{Zhu, X.}, \bibinfo{author}{Zhang, S.}, \bibinfo{author}{Shao, P.}, \bibinfo{author}{Shu, H.}, \bibinfo{author}{Dillenseger, J.L.}, \bibinfo{author}{Coatrieux, J.L.}, \bibinfo{author}{Li, S.}, \bibinfo{year}{2021}.
\newblock \bibinfo{title}{Meta grayscale adaptive network for 3d integrated renal structures segmentation}.
\newblock \bibinfo{journal}{Medical Image Analysis} \bibinfo{volume}{71}, \bibinfo{pages}{102055}.
\newblock \URLprefix \url{https://www.sciencedirect.com/science/article/pii/S1361841521001018}, \DOIprefix\doi{https://doi.org/10.1016/j.media.2021.102055}.
\bibitem[{Heller et~al.(2021)Heller, Isensee, Maier-Hein, Hou, Xie, Li, Nan, Mu, Lin, Han, Yao, Gao, Zhang, Wang, Hou, Yang, Xiong, Tian, Zhong, Ma, Rickman, Dean, Stai, Tejpaul, Oestreich, Blake, Kaluzniak, Raza, Rosenberg, Moore, Walczak, Rengel, Edgerton, Vasdev, Peterson, McSweeney, Peterson, Kalapara, Sathianathen, Papanikolopoulos and Weight}]{HELLER2021101821}
\bibinfo{author}{Heller, N.}, \bibinfo{author}{Isensee, F.}, \bibinfo{author}{Maier-Hein, K.H.}, \bibinfo{author}{Hou, X.}, \bibinfo{author}{Xie, C.}, \bibinfo{author}{Li, F.}, \bibinfo{author}{Nan, Y.}, \bibinfo{author}{Mu, G.}, \bibinfo{author}{Lin, Z.}, \bibinfo{author}{Han, M.}, \bibinfo{author}{Yao, G.}, \bibinfo{author}{Gao, Y.}, \bibinfo{author}{Zhang, Y.}, \bibinfo{author}{Wang, Y.}, \bibinfo{author}{Hou, F.}, \bibinfo{author}{Yang, J.}, \bibinfo{author}{Xiong, G.}, \bibinfo{author}{Tian, J.}, \bibinfo{author}{Zhong, C.}, \bibinfo{author}{Ma, J.}, \bibinfo{author}{Rickman, J.}, \bibinfo{author}{Dean, J.}, \bibinfo{author}{Stai, B.}, \bibinfo{author}{Tejpaul, R.}, \bibinfo{author}{Oestreich, M.}, \bibinfo{author}{Blake, P.}, \bibinfo{author}{Kaluzniak, H.}, \bibinfo{author}{Raza, S.}, \bibinfo{author}{Rosenberg, J.}, \bibinfo{author}{Moore, K.}, \bibinfo{author}{Walczak, E.}, \bibinfo{author}{Rengel, Z.}, \bibinfo{author}{Edgerton, Z.}, \bibinfo{author}{Vasdev, R.}, \bibinfo{author}{Peterson, M.},
  \bibinfo{author}{McSweeney, S.}, \bibinfo{author}{Peterson, S.}, \bibinfo{author}{Kalapara, A.}, \bibinfo{author}{Sathianathen, N.}, \bibinfo{author}{Papanikolopoulos, N.}, \bibinfo{author}{Weight, C.}, \bibinfo{year}{2021}.
\newblock \bibinfo{title}{The state of the art in kidney and kidney tumor segmentation in contrast-enhanced ct imaging: Results of the kits19 challenge}.
\newblock \bibinfo{journal}{Medical Image Analysis} \bibinfo{volume}{67}, \bibinfo{pages}{101821}.
\newblock \URLprefix \url{https://www.sciencedirect.com/science/article/pii/S1361841520301857}, \DOIprefix\doi{https://doi.org/10.1016/j.media.2020.101821}.
\bibitem[{Hernandez~Petzsche et~al.(2022)Hernandez~Petzsche, de~la Rosa, Hanning, Wiest, Valenzuela, Reyes, Meyer, Liew, Kofler, Ezhov, Robben, Hutton, Friedrich, Zarth, B{\"u}rkle, Baran, Menze, Broocks, Meyer, Zimmer, Boeckh-Behrens, Berndt, Ikenberg, Wiestler and Kirschke}]{HernandezPetzsche2022}
\bibinfo{author}{Hernandez~Petzsche, M.R.}, \bibinfo{author}{de~la Rosa, E.}, \bibinfo{author}{Hanning, U.}, \bibinfo{author}{Wiest, R.}, \bibinfo{author}{Valenzuela, W.}, \bibinfo{author}{Reyes, M.}, \bibinfo{author}{Meyer, M.}, \bibinfo{author}{Liew, S.L.}, \bibinfo{author}{Kofler, F.}, \bibinfo{author}{Ezhov, I.}, \bibinfo{author}{Robben, D.}, \bibinfo{author}{Hutton, A.}, \bibinfo{author}{Friedrich, T.}, \bibinfo{author}{Zarth, T.}, \bibinfo{author}{B{\"u}rkle, J.}, \bibinfo{author}{Baran, T.A.}, \bibinfo{author}{Menze, B.}, \bibinfo{author}{Broocks, G.}, \bibinfo{author}{Meyer, L.}, \bibinfo{author}{Zimmer, C.}, \bibinfo{author}{Boeckh-Behrens, T.}, \bibinfo{author}{Berndt, M.}, \bibinfo{author}{Ikenberg, B.}, \bibinfo{author}{Wiestler, B.}, \bibinfo{author}{Kirschke, J.S.}, \bibinfo{year}{2022}.
\newblock \bibinfo{title}{Isles 2022: A multi-center magnetic resonance imaging stroke lesion segmentation dataset}.
\newblock \bibinfo{journal}{Scientific Data} \bibinfo{volume}{9}, \bibinfo{pages}{762}.
\newblock \URLprefix \url{https://doi.org/10.1038/s41597-022-01875-5}, \DOIprefix\doi{10.1038/s41597-022-01875-5}.
\bibitem[{Hu et~al.(2025)Hu, Li, Jain, Lin and Chen}]{10829779}
\bibinfo{author}{Hu, J.}, \bibinfo{author}{Li, Y.}, \bibinfo{author}{Jain, R.K.}, \bibinfo{author}{Lin, L.}, \bibinfo{author}{Chen, Y.w.}, \bibinfo{year}{2025}.
\newblock \bibinfo{title}{Spa: Leveraging the sam with spatial priors adapter for enhanced medical image segmentation}.
\newblock \bibinfo{journal}{IEEE Journal of Biomedical and Health Informatics} , \bibinfo{pages}{1--15}\DOIprefix\doi{10.1109/JBHI.2025.3526174}.
\bibitem[{Huang et~al.(2021)Huang, Deng, Li and Yuan}]{huang2021missformereffectivemedicalimage}
\bibinfo{author}{Huang, X.}, \bibinfo{author}{Deng, Z.}, \bibinfo{author}{Li, D.}, \bibinfo{author}{Yuan, X.}, \bibinfo{year}{2021}.
\newblock \bibinfo{title}{Missformer: An effective medical image segmentation transformer}.
\newblock \URLprefix \url{https://arxiv.org/abs/2109.07162}, \href{http://arxiv.org/abs/2109.07162}{{\tt arXiv:2109.07162}}.
\bibitem[{Huang et~al.(2023a)Huang, Deng, Li, Yuan and Fu}]{9994763}
\bibinfo{author}{Huang, X.}, \bibinfo{author}{Deng, Z.}, \bibinfo{author}{Li, D.}, \bibinfo{author}{Yuan, X.}, \bibinfo{author}{Fu, Y.}, \bibinfo{year}{2023}a.
\newblock \bibinfo{title}{Missformer: An effective transformer for 2d medical image segmentation}.
\newblock \bibinfo{journal}{IEEE Transactions on Medical Imaging} \bibinfo{volume}{42}, \bibinfo{pages}{1484--1494}.
\newblock \DOIprefix\doi{10.1109/TMI.2022.3230943}.
\bibitem[{Huang et~al.(2023b)Huang, Deng, Li, Yuan and Fu}]{huang2022missformer}
\bibinfo{author}{Huang, X.}, \bibinfo{author}{Deng, Z.}, \bibinfo{author}{Li, D.}, \bibinfo{author}{Yuan, X.}, \bibinfo{author}{Fu, Y.}, \bibinfo{year}{2023}b.
\newblock \bibinfo{title}{Missformer: An effective transformer for 2d medical image segmentation}.
\newblock \bibinfo{journal}{IEEE Transactions on Medical Imaging} \bibinfo{volume}{42}, \bibinfo{pages}{1484--1494}.
\newblock \DOIprefix\doi{10.1109/tmi.2022.3230943}.
\bibitem[{Huang et~al.(2023c)Huang, Wang, Deng, Ye, Su, Sun, He, Gu, Gu, Zhang and Qiao}]{huang2023stunetscalabletransferablemedical}
\bibinfo{author}{Huang, Z.}, \bibinfo{author}{Wang, H.}, \bibinfo{author}{Deng, Z.}, \bibinfo{author}{Ye, J.}, \bibinfo{author}{Su, Y.}, \bibinfo{author}{Sun, H.}, \bibinfo{author}{He, J.}, \bibinfo{author}{Gu, Y.}, \bibinfo{author}{Gu, L.}, \bibinfo{author}{Zhang, S.}, \bibinfo{author}{Qiao, Y.}, \bibinfo{year}{2023}c.
\newblock \bibinfo{title}{Stu-net: Scalable and transferable medical image segmentation models empowered by large-scale supervised pre-training}.
\newblock \URLprefix \url{https://arxiv.org/abs/2304.06716}, \href{http://arxiv.org/abs/2304.06716}{{\tt arXiv:2304.06716}}.
\bibitem[{zgn iek et~al.(2016)zgn iek, Abdulkadir, Lienkamp, Brox and Ronneberger}]{cicek20163dunetlearningdense}
\bibinfo{author}{zgn iek}, \bibinfo{author}{Abdulkadir, A.}, \bibinfo{author}{Lienkamp, S.S.}, \bibinfo{author}{Brox, T.}, \bibinfo{author}{Ronneberger, O.}, \bibinfo{year}{2016}.
\newblock \bibinfo{title}{3d u-net: Learning dense volumetric segmentation from sparse annotation}.
\newblock \URLprefix \url{https://arxiv.org/abs/1606.06650}, \href{http://arxiv.org/abs/1606.06650}{{\tt arXiv:1606.06650}}.
\bibitem[{Isensee et~al.(2021a)Isensee, Jaeger, Kohl, Petersen and Maier-Hein}]{Isensee2021}
\bibinfo{author}{Isensee, F.}, \bibinfo{author}{Jaeger, P.F.}, \bibinfo{author}{Kohl, S.A.A.}, \bibinfo{author}{Petersen, J.}, \bibinfo{author}{Maier-Hein, K.H.}, \bibinfo{year}{2021}a.
\newblock \bibinfo{title}{nnu-net: a self-configuring method for deep learning-based biomedical image segmentation}.
\newblock \bibinfo{journal}{Nature Methods} \bibinfo{volume}{18}, \bibinfo{pages}{203--211}.
\newblock \URLprefix \url{https://doi.org/10.1038/s41592-020-01008-z}, \DOIprefix\doi{10.1038/s41592-020-01008-z}.
\bibitem[{Isensee et~al.(2021b)Isensee, Jaeger, Kohl, Petersen and Maier-Hein}]{nnunet_nature}
\bibinfo{author}{Isensee, F.}, \bibinfo{author}{Jaeger, P.F.}, \bibinfo{author}{Kohl, S.A.A.}, \bibinfo{author}{Petersen, J.}, \bibinfo{author}{Maier-Hein, K.H.}, \bibinfo{year}{2021}b.
\newblock \bibinfo{title}{nnu-net: a self-configuring method for deep learning-based biomedical image segmentation}.
\newblock \bibinfo{journal}{Nature Methods} \bibinfo{volume}{18}, \bibinfo{pages}{203--211}.
\newblock \URLprefix \url{https://doi.org/10.1038/s41592-020-01008-z}, \DOIprefix\doi{10.1038/s41592-020-01008-z}.
\bibitem[{Isensee et~al.(2018)Isensee, Petersen, Klein, Zimmerer, Jaeger, Kohl, Wasserthal, Koehler, Norajitra, Wirkert and Maier-Hein}]{isensee2018nnunetselfadaptingframeworkunetbased}
\bibinfo{author}{Isensee, F.}, \bibinfo{author}{Petersen, J.}, \bibinfo{author}{Klein, A.}, \bibinfo{author}{Zimmerer, D.}, \bibinfo{author}{Jaeger, P.F.}, \bibinfo{author}{Kohl, S.}, \bibinfo{author}{Wasserthal, J.}, \bibinfo{author}{Koehler, G.}, \bibinfo{author}{Norajitra, T.}, \bibinfo{author}{Wirkert, S.}, \bibinfo{author}{Maier-Hein, K.H.}, \bibinfo{year}{2018}.
\newblock \bibinfo{title}{nnu-net: Self-adapting framework for u-net-based medical image segmentation}.
\newblock \URLprefix \url{https://arxiv.org/abs/1809.10486}, \href{http://arxiv.org/abs/1809.10486}{{\tt arXiv:1809.10486}}.
\bibitem[{Ji et~al.(2022a)Ji, Bai, Ge, Yang, Zhu, Zhang, Li, Zhanng, Ma, Wan et~al.}]{ji2022amos}
\bibinfo{author}{Ji, Y.}, \bibinfo{author}{Bai, H.}, \bibinfo{author}{Ge, C.}, \bibinfo{author}{Yang, J.}, \bibinfo{author}{Zhu, Y.}, \bibinfo{author}{Zhang, R.}, \bibinfo{author}{Li, Z.}, \bibinfo{author}{Zhanng, L.}, \bibinfo{author}{Ma, W.}, \bibinfo{author}{Wan, X.}, et~al., \bibinfo{year}{2022}a.
\newblock \bibinfo{title}{Amos: A large-scale abdominal multi-organ benchmark for versatile medical image segmentation}.
\newblock \bibinfo{journal}{Advances in neural information processing systems} \bibinfo{volume}{35}, \bibinfo{pages}{36722--36732}.
\bibitem[{Ji et~al.(2022b)Ji, Bai, Yang, Ge, Zhu, Zhang, Li, Zhang, Ma, Wan and Luo}]{ji2022amoslargescaleabdominalmultiorgan}
\bibinfo{author}{Ji, Y.}, \bibinfo{author}{Bai, H.}, \bibinfo{author}{Yang, J.}, \bibinfo{author}{Ge, C.}, \bibinfo{author}{Zhu, Y.}, \bibinfo{author}{Zhang, R.}, \bibinfo{author}{Li, Z.}, \bibinfo{author}{Zhang, L.}, \bibinfo{author}{Ma, W.}, \bibinfo{author}{Wan, X.}, \bibinfo{author}{Luo, P.}, \bibinfo{year}{2022}b.
\newblock \bibinfo{title}{Amos: A large-scale abdominal multi-organ benchmark for versatile medical image segmentation}.
\newblock \URLprefix \url{https://arxiv.org/abs/2206.08023}, \href{http://arxiv.org/abs/2206.08023}{{\tt arXiv:2206.08023}}.
\bibitem[{Jiang et~al.(2022)Jiang, Tyagi, Tringale, Crane and Veeraraghavan}]{10.1007/978-3-031-16440-8_53}
\bibinfo{author}{Jiang, J.}, \bibinfo{author}{Tyagi, N.}, \bibinfo{author}{Tringale, K.}, \bibinfo{author}{Crane, C.}, \bibinfo{author}{Veeraraghavan, H.}, \bibinfo{year}{2022}.
\newblock \bibinfo{title}{Self-supervised 3d anatomy segmentation using self-distilled masked image transformer (smit)}, in: \bibinfo{editor}{Wang, L.}, \bibinfo{editor}{Dou, Q.}, \bibinfo{editor}{Fletcher, P.T.}, \bibinfo{editor}{Speidel, S.}, \bibinfo{editor}{Li, S.} (Eds.), \bibinfo{booktitle}{Medical Image Computing and Computer Assisted Intervention -- MICCAI 2022}, \bibinfo{publisher}{Springer Nature Switzerland}, \bibinfo{address}{Cham}. pp. \bibinfo{pages}{556--566}.
\bibitem[{Kavur et~al.(2021)Kavur, Gezer, Bar, Aslan, Conze, Groza, Pham, Chatterjee, Ernst, zkan, Baydar, Lachinov, Han, Pauli, Isensee, Perkonigg, Sathish, Rajan, Sheet, Dovletov, Speck, Nrnberger, Maier-Hein, {Bozda Akar}, nal, Dicle and Selver}]{CHAOS2021}
\bibinfo{author}{Kavur, A.E.}, \bibinfo{author}{Gezer, N.S.}, \bibinfo{author}{Bar, M.}, \bibinfo{author}{Aslan, S.}, \bibinfo{author}{Conze, P.H.}, \bibinfo{author}{Groza, V.}, \bibinfo{author}{Pham, D.D.}, \bibinfo{author}{Chatterjee, S.}, \bibinfo{author}{Ernst, P.}, \bibinfo{author}{zkan, S.}, \bibinfo{author}{Baydar, B.}, \bibinfo{author}{Lachinov, D.}, \bibinfo{author}{Han, S.}, \bibinfo{author}{Pauli, J.}, \bibinfo{author}{Isensee, F.}, \bibinfo{author}{Perkonigg, M.}, \bibinfo{author}{Sathish, R.}, \bibinfo{author}{Rajan, R.}, \bibinfo{author}{Sheet, D.}, \bibinfo{author}{Dovletov, G.}, \bibinfo{author}{Speck, O.}, \bibinfo{author}{Nrnberger, A.}, \bibinfo{author}{Maier-Hein, K.H.}, \bibinfo{author}{{Bozda Akar}, G.}, \bibinfo{author}{nal, G.}, \bibinfo{author}{Dicle, O.}, \bibinfo{author}{Selver, M.A.}, \bibinfo{year}{2021}.
\newblock \bibinfo{title}{{CHAOS Challenge - combined (CT-MR) healthy abdominal organ segmentation}}.
\newblock \bibinfo{journal}{Medical Image Analysis} \bibinfo{volume}{69}, \bibinfo{pages}{101950}.
\newblock \URLprefix \url{http://www.sciencedirect.com/science/article/pii/S1361841520303145}, \DOIprefix\doi{https://doi.org/10.1016/j.media.2020.101950}.
\bibitem[{Kavur et~al.(2020)Kavur, Gezer, Bar$\imath$~s, \c{S}ahin, \"{O}zkan, Baydar, Y\"{u}ksel, K$\imath$ l$\imath$~k\c{c}$\imath$ er, Olut, Bozda\u{g}$\imath$\~Akar, \"{U}nal, Dicle and Selver}]{kavur2019}
\bibinfo{author}{Kavur, A.E.}, \bibinfo{author}{Gezer, N.S.}, \bibinfo{author}{Bar$\imath$~s, M.}, \bibinfo{author}{\c{S}ahin, Y.}, \bibinfo{author}{\"{O}zkan, S.}, \bibinfo{author}{Baydar, B.}, \bibinfo{author}{Y\"{u}ksel, U.}, \bibinfo{author}{K$\imath$ l$\imath$~k\c{c}$\imath$ er, c.}, \bibinfo{author}{Olut, c.}, \bibinfo{author}{Bozda\u{g}$\imath$\~Akar, G.}, \bibinfo{author}{\"{U}nal, G.}, \bibinfo{author}{Dicle, O.}, \bibinfo{author}{Selver, M.A.}, \bibinfo{year}{2020}.
\newblock \bibinfo{title}{Comparison of semi-automatic and deep learning based automatic methods for liver segmentation in living liver transplant donors}.
\newblock \bibinfo{journal}{Diagnostic and Interventional Radiology} \bibinfo{volume}{26}, \bibinfo{pages}{11--21}.
\newblock \URLprefix \url{https://doi.org/10.5152/dir.2019.19025}, \DOIprefix\doi{10.5152/dir.2019.19}.
\bibitem[{Kavur et~al.(2019)Kavur, Selver, Dicle, Bar and Gezer}]{CHAOSdata2019}
\bibinfo{author}{Kavur, A.E.}, \bibinfo{author}{Selver, M.A.}, \bibinfo{author}{Dicle, O.}, \bibinfo{author}{Bar, M.}, \bibinfo{author}{Gezer, N.S.}, \bibinfo{year}{2019}.
\newblock \bibinfo{title}{{CHAOS - Combined (CT-MR) Healthy Abdominal Organ Segmentation Challenge Data}}.
\newblock \URLprefix \url{https://doi.org/10.5281/zenodo.3362844}, \DOIprefix\doi{10.5281/zenodo.3362844}.
\bibitem[{Kennedy et~al.(2012)Kennedy, Haselgrove, Hodge, Rane, Makris and Frazier}]{Kennedy2012}
\bibinfo{author}{Kennedy, D.N.}, \bibinfo{author}{Haselgrove, C.}, \bibinfo{author}{Hodge, S.M.}, \bibinfo{author}{Rane, P.S.}, \bibinfo{author}{Makris, N.}, \bibinfo{author}{Frazier, J.A.}, \bibinfo{year}{2012}.
\newblock \bibinfo{title}{Candishare: A resource for pediatric neuroimaging data}.
\newblock \bibinfo{journal}{Neuroinformatics} \bibinfo{volume}{10}, \bibinfo{pages}{319--322}.
\newblock \URLprefix \url{https://doi.org/10.1007/s12021-011-9133-y}, \DOIprefix\doi{10.1007/s12021-011-9133-y}.
\bibitem[{Kuijf et~al.(2019)Kuijf, Biesbroek, De~Bresser, Heinen, Andermatt, Bento, Berseth, Belyaev, Cardoso, Casamitjana, Collins, Dadar, Georgiou, Ghafoorian, Jin, Khademi, Knight, Li, Llad, Luna, Mahmood, McKinley, Mehrtash, Ourselin, Park, Park, Park, Pezold, Puybareau, Rittner, Sudre, Valverde, Vilaplana, Wiest, Xu, Xu, Zeng, Zhang, Zheng, Chen, van~der Flier, Barkhof, Viergever and Biessels}]{8669968}
\bibinfo{author}{Kuijf, H.J.}, \bibinfo{author}{Biesbroek, J.M.}, \bibinfo{author}{De~Bresser, J.}, \bibinfo{author}{Heinen, R.}, \bibinfo{author}{Andermatt, S.}, \bibinfo{author}{Bento, M.}, \bibinfo{author}{Berseth, M.}, \bibinfo{author}{Belyaev, M.}, \bibinfo{author}{Cardoso, M.J.}, \bibinfo{author}{Casamitjana, A.}, \bibinfo{author}{Collins, D.L.}, \bibinfo{author}{Dadar, M.}, \bibinfo{author}{Georgiou, A.}, \bibinfo{author}{Ghafoorian, M.}, \bibinfo{author}{Jin, D.}, \bibinfo{author}{Khademi, A.}, \bibinfo{author}{Knight, J.}, \bibinfo{author}{Li, H.}, \bibinfo{author}{Llad, X.}, \bibinfo{author}{Luna, M.}, \bibinfo{author}{Mahmood, Q.}, \bibinfo{author}{McKinley, R.}, \bibinfo{author}{Mehrtash, A.}, \bibinfo{author}{Ourselin, S.}, \bibinfo{author}{Park, B.Y.}, \bibinfo{author}{Park, H.}, \bibinfo{author}{Park, S.H.}, \bibinfo{author}{Pezold, S.}, \bibinfo{author}{Puybareau, E.}, \bibinfo{author}{Rittner, L.}, \bibinfo{author}{Sudre, C.H.}, \bibinfo{author}{Valverde, S.}, \bibinfo{author}{Vilaplana, V.},
  \bibinfo{author}{Wiest, R.}, \bibinfo{author}{Xu, Y.}, \bibinfo{author}{Xu, Z.}, \bibinfo{author}{Zeng, G.}, \bibinfo{author}{Zhang, J.}, \bibinfo{author}{Zheng, G.}, \bibinfo{author}{Chen, C.}, \bibinfo{author}{van~der Flier, W.}, \bibinfo{author}{Barkhof, F.}, \bibinfo{author}{Viergever, M.A.}, \bibinfo{author}{Biessels, G.J.}, \bibinfo{year}{2019}.
\newblock \bibinfo{title}{Standardized assessment of automatic segmentation of white matter hyperintensities and results of the wmh segmentation challenge}.
\newblock \bibinfo{journal}{IEEE Transactions on Medical Imaging} \bibinfo{volume}{38}, \bibinfo{pages}{2556--2568}.
\newblock \DOIprefix\doi{10.1109/TMI.2019.2905770}.
\bibitem[{Lambert et~al.(2020)Lambert, Petitjean, Dubray and Kuan}]{9286453}
\bibinfo{author}{Lambert, Z.}, \bibinfo{author}{Petitjean, C.}, \bibinfo{author}{Dubray, B.}, \bibinfo{author}{Kuan, S.}, \bibinfo{year}{2020}.
\newblock \bibinfo{title}{Segthor: Segmentation of thoracic organs at risk in ct images}, in: \bibinfo{booktitle}{2020 Tenth International Conference on Image Processing Theory, Tools and Applications (IPTA)}, pp. \bibinfo{pages}{1--6}.
\newblock \DOIprefix\doi{10.1109/IPTA50016.2020.9286453}.
\bibitem[{Landman et~al.(2015)Landman, Xu, Igelsias, Styner, Langerak and Klein}]{landman2015miccai}
\bibinfo{author}{Landman, B.}, \bibinfo{author}{Xu, Z.}, \bibinfo{author}{Igelsias, J.}, \bibinfo{author}{Styner, M.}, \bibinfo{author}{Langerak, T.}, \bibinfo{author}{Klein, A.}, \bibinfo{year}{2015}.
\newblock \bibinfo{title}{Miccai multi-atlas labeling beyond the cranial vault--workshop and challenge}, in: \bibinfo{booktitle}{Proc. MICCAI Multi-Atlas Labeling Beyond Cranial VaultWorkshop Challenge}, p.~\bibinfo{pages}{12}.
\bibitem[{Lee et~al.(2023a)Lee, Bao, Huo and Landman}]{lee20233duxnetlargekernel}
\bibinfo{author}{Lee, H.H.}, \bibinfo{author}{Bao, S.}, \bibinfo{author}{Huo, Y.}, \bibinfo{author}{Landman, B.A.}, \bibinfo{year}{2023}a.
\newblock \bibinfo{title}{3d ux-net: A large kernel volumetric convnet modernizing hierarchical transformer for medical image segmentation}.
\newblock \URLprefix \url{https://arxiv.org/abs/2209.15076}, \href{http://arxiv.org/abs/2209.15076}{{\tt arXiv:2209.15076}}.
\bibitem[{Lee et~al.(2023b)Lee, Bao, Huo and Landman}]{lee2023d}
\bibinfo{author}{Lee, H.H.}, \bibinfo{author}{Bao, S.}, \bibinfo{author}{Huo, Y.}, \bibinfo{author}{Landman, B.A.}, \bibinfo{year}{2023}b.
\newblock \bibinfo{title}{3d {UX}-net: A large kernel volumetric convnet modernizing hierarchical transformer for medical image segmentation}, in: \bibinfo{booktitle}{The Eleventh International Conference on Learning Representations}.
\newblock \URLprefix \url{https://openreview.net/forum?id=wsZsjOSytRA}.
\bibitem[{Lee et~al.(2023c)Lee, Bao, Huo and Landman}]{3duxnet_iclr}
\bibinfo{author}{Lee, H.H.}, \bibinfo{author}{Bao, S.}, \bibinfo{author}{Huo, Y.}, \bibinfo{author}{Landman, B.A.}, \bibinfo{year}{2023}c.
\newblock \bibinfo{title}{3d {UX}-net: A large kernel volumetric convnet modernizing hierarchical transformer for medical image segmentation}, in: \bibinfo{booktitle}{The Eleventh International Conference on Learning Representations}.
\newblock \URLprefix \url{https://openreview.net/forum?id=wsZsjOSytRA}.
\bibitem[{Li et~al.(2022)Li, Wang, Chen, Zhang, Zha, Wang and Yu}]{li2022transbtsv2betterefficientvolumetric}
\bibinfo{author}{Li, J.}, \bibinfo{author}{Wang, W.}, \bibinfo{author}{Chen, C.}, \bibinfo{author}{Zhang, T.}, \bibinfo{author}{Zha, S.}, \bibinfo{author}{Wang, J.}, \bibinfo{author}{Yu, H.}, \bibinfo{year}{2022}.
\newblock \bibinfo{title}{Transbtsv2: Towards better and more efficient volumetric segmentation of medical images}.
\newblock \URLprefix \url{https://arxiv.org/abs/2201.12785}, \href{http://arxiv.org/abs/2201.12785}{{\tt arXiv:2201.12785}}.
\bibitem[{Li et~al.(2025)Li, Qi, Yu, Huo, Shi and Gao}]{10847777}
\bibinfo{author}{Li, S.}, \bibinfo{author}{Qi, L.}, \bibinfo{author}{Yu, Q.}, \bibinfo{author}{Huo, J.}, \bibinfo{author}{Shi, Y.}, \bibinfo{author}{Gao, Y.}, \bibinfo{year}{2025}.
\newblock \bibinfo{title}{Stitching, fine-tuning, re-training: A sam-enabled framework for semi-supervised 3d medical image segmentation}.
\newblock \bibinfo{journal}{IEEE Transactions on Medical Imaging} , \bibinfo{pages}{1--1}\DOIprefix\doi{10.1109/TMI.2025.3532084}.
\bibitem[{Li et~al.(2024)Li, Qu, Chen, Bassi, Shi, Lai, Yu, Xue, Chen, Lin, Tang, Cao, Han, Zhang, Liu, Zhang, Ma, Wang, Zhang, Yuille and Zhou}]{LI2024103285}
\bibinfo{author}{Li, W.}, \bibinfo{author}{Qu, C.}, \bibinfo{author}{Chen, X.}, \bibinfo{author}{Bassi, P.R.}, \bibinfo{author}{Shi, Y.}, \bibinfo{author}{Lai, Y.}, \bibinfo{author}{Yu, Q.}, \bibinfo{author}{Xue, H.}, \bibinfo{author}{Chen, Y.}, \bibinfo{author}{Lin, X.}, \bibinfo{author}{Tang, Y.}, \bibinfo{author}{Cao, Y.}, \bibinfo{author}{Han, H.}, \bibinfo{author}{Zhang, Z.}, \bibinfo{author}{Liu, J.}, \bibinfo{author}{Zhang, T.}, \bibinfo{author}{Ma, Y.}, \bibinfo{author}{Wang, J.}, \bibinfo{author}{Zhang, G.}, \bibinfo{author}{Yuille, A.}, \bibinfo{author}{Zhou, Z.}, \bibinfo{year}{2024}.
\newblock \bibinfo{title}{Abdomenatlas: A large-scale, detailed-annotated, \& multi-center dataset for efficient transfer learning and open algorithmic benchmarking}.
\newblock \bibinfo{journal}{Medical Image Analysis} \bibinfo{volume}{97}, \bibinfo{pages}{103285}.
\newblock \URLprefix \url{https://www.sciencedirect.com/science/article/pii/S136184152400210X}, \DOIprefix\doi{https://doi.org/10.1016/j.media.2024.103285}.
\bibitem[{Liao et~al.(2023)Liao, Luo, He, Dong, Li, Li, Zhang, Zhang, Wang and Xiao}]{LIAO2023994}
\bibinfo{author}{Liao, W.}, \bibinfo{author}{Luo, X.}, \bibinfo{author}{He, Y.}, \bibinfo{author}{Dong, Y.}, \bibinfo{author}{Li, C.}, \bibinfo{author}{Li, K.}, \bibinfo{author}{Zhang, S.}, \bibinfo{author}{Zhang, S.}, \bibinfo{author}{Wang, G.}, \bibinfo{author}{Xiao, J.}, \bibinfo{year}{2023}.
\newblock \bibinfo{title}{Comprehensive evaluation of a deep learning model for automatic organs-at-risk segmentation on heterogeneous computed tomography images for abdominal radiation therapy}.
\newblock \bibinfo{journal}{International Journal of Radiation Oncology*Biology*Physics} \bibinfo{volume}{117}, \bibinfo{pages}{994--1006}.
\newblock \URLprefix \url{https://www.sciencedirect.com/science/article/pii/S0360301623005205}, \DOIprefix\doi{https://doi.org/10.1016/j.ijrobp.2023.05.034}.
\bibitem[{Liew et~al.(2022)Liew, Lo, Donnelly, Zavaliangos-Petropulu, Jeong, Barisano, Hutton, Simon, Juliano, Suri et~al.}]{liew2022large}
\bibinfo{author}{Liew, S.L.}, \bibinfo{author}{Lo, B.P.}, \bibinfo{author}{Donnelly, M.R.}, \bibinfo{author}{Zavaliangos-Petropulu, A.}, \bibinfo{author}{Jeong, J.N.}, \bibinfo{author}{Barisano, G.}, \bibinfo{author}{Hutton, A.}, \bibinfo{author}{Simon, J.P.}, \bibinfo{author}{Juliano, J.M.}, \bibinfo{author}{Suri, A.}, et~al., \bibinfo{year}{2022}.
\newblock \bibinfo{title}{A large, curated, open-source stroke neuroimaging dataset to improve lesion segmentation algorithms}.
\newblock \bibinfo{journal}{Scientific data} \bibinfo{volume}{9}, \bibinfo{pages}{320}.
\bibitem[{Lin et~al.(2025)Lin, Zou, Deng, Wong, Aviles-Rivero, Fan, Lee, Hu and Qin}]{LIN20251}
\bibinfo{author}{Lin, H.}, \bibinfo{author}{Zou, J.}, \bibinfo{author}{Deng, S.}, \bibinfo{author}{Wong, K.P.}, \bibinfo{author}{Aviles-Rivero, A.I.}, \bibinfo{author}{Fan, Y.}, \bibinfo{author}{Lee, A.P.W.}, \bibinfo{author}{Hu, X.}, \bibinfo{author}{Qin, J.}, \bibinfo{year}{2025}.
\newblock \bibinfo{title}{Volumetric medical image segmentation via fully 3d adaptation of segment anything model}.
\newblock \bibinfo{journal}{Biocybernetics and Biomedical Engineering} \bibinfo{volume}{45}, \bibinfo{pages}{1--10}.
\newblock \URLprefix \url{https://www.sciencedirect.com/science/article/pii/S0208521624000846}, \DOIprefix\doi{https://doi.org/10.1016/j.bbe.2024.11.001}.
\bibitem[{Litjens et~al.(2014)Litjens, Toth, {van de Ven}, Hoeks, Kerkstra, {van Ginneken}, Vincent, Guillard, Birbeck, Zhang, Strand, Malmberg, Ou, Davatzikos, Kirschner, Jung, Yuan, Qiu, Gao, Edwards, Maan, {van der Heijden}, Ghose, Mitra, Dowling, Barratt, Huisman and Madabhushi}]{LITJENS2014359}
\bibinfo{author}{Litjens, G.}, \bibinfo{author}{Toth, R.}, \bibinfo{author}{{van de Ven}, W.}, \bibinfo{author}{Hoeks, C.}, \bibinfo{author}{Kerkstra, S.}, \bibinfo{author}{{van Ginneken}, B.}, \bibinfo{author}{Vincent, G.}, \bibinfo{author}{Guillard, G.}, \bibinfo{author}{Birbeck, N.}, \bibinfo{author}{Zhang, J.}, \bibinfo{author}{Strand, R.}, \bibinfo{author}{Malmberg, F.}, \bibinfo{author}{Ou, Y.}, \bibinfo{author}{Davatzikos, C.}, \bibinfo{author}{Kirschner, M.}, \bibinfo{author}{Jung, F.}, \bibinfo{author}{Yuan, J.}, \bibinfo{author}{Qiu, W.}, \bibinfo{author}{Gao, Q.}, \bibinfo{author}{Edwards, P.E.}, \bibinfo{author}{Maan, B.}, \bibinfo{author}{{van der Heijden}, F.}, \bibinfo{author}{Ghose, S.}, \bibinfo{author}{Mitra, J.}, \bibinfo{author}{Dowling, J.}, \bibinfo{author}{Barratt, D.}, \bibinfo{author}{Huisman, H.}, \bibinfo{author}{Madabhushi, A.}, \bibinfo{year}{2014}.
\newblock \bibinfo{title}{Evaluation of prostate segmentation algorithms for mri: The promise12 challenge}.
\newblock \bibinfo{journal}{Medical Image Analysis} \bibinfo{volume}{18}, \bibinfo{pages}{359--373}.
\newblock \URLprefix \url{https://www.sciencedirect.com/science/article/pii/S1361841513001734}, \DOIprefix\doi{https://doi.org/10.1016/j.media.2013.12.002}.
\bibitem[{Liu et~al.(2023)Liu, Zhang, Chen, Xiao, Lu, Landman, Yuan, Yuille, Tang and Zhou}]{10376801}
\bibinfo{author}{Liu, J.}, \bibinfo{author}{Zhang, Y.}, \bibinfo{author}{Chen, J.N.}, \bibinfo{author}{Xiao, J.}, \bibinfo{author}{Lu, Y.}, \bibinfo{author}{Landman, B.A.}, \bibinfo{author}{Yuan, Y.}, \bibinfo{author}{Yuille, A.}, \bibinfo{author}{Tang, Y.}, \bibinfo{author}{Zhou, Z.}, \bibinfo{year}{2023}.
\newblock \bibinfo{title}{Clip-driven universal model for organ segmentation and tumor detection}, in: \bibinfo{booktitle}{2023 IEEE/CVF International Conference on Computer Vision (ICCV)}, pp. \bibinfo{pages}{21095--21107}.
\newblock \DOIprefix\doi{10.1109/ICCV51070.2023.01934}.
\bibitem[{Liu et~al.(2024a)Liu, Zhang, Yue and Guo}]{LIU2024e26775}
\bibinfo{author}{Liu, Y.}, \bibinfo{author}{Zhang, Z.}, \bibinfo{author}{Yue, J.}, \bibinfo{author}{Guo, W.}, \bibinfo{year}{2024}a.
\newblock \bibinfo{title}{Scanext: Enhancing 3d medical image segmentation with dual attention network and depth-wise convolution}.
\newblock \bibinfo{journal}{Heliyon} \bibinfo{volume}{10}, \bibinfo{pages}{e26775}.
\newblock \URLprefix \url{https://www.sciencedirect.com/science/article/pii/S2405844024028068}, \DOIprefix\doi{https://doi.org/10.1016/j.heliyon.2024.e26775}.
\bibitem[{Liu et~al.(2024b)Liu, Zhang, Yue and Guo}]{scanext}
\bibinfo{author}{Liu, Y.}, \bibinfo{author}{Zhang, Z.}, \bibinfo{author}{Yue, J.}, \bibinfo{author}{Guo, W.}, \bibinfo{year}{2024}b.
\newblock \bibinfo{title}{Scanext: Enhancing 3d medical image segmentation with dual attention network and depth-wise convolution}.
\newblock \bibinfo{journal}{Heliyon} \bibinfo{volume}{10}, \bibinfo{pages}{e26775}.
\newblock \URLprefix \url{https://www.sciencedirect.com/science/article/pii/S2405844024028068}, \DOIprefix\doi{https://doi.org/10.1016/j.heliyon.2024.e26775}.
\bibitem[{Luo et~al.(2022)Luo, Liao, Xiao, Chen, Song, Zhang, Li, Metaxas, Wang and Zhang}]{LUO2022102642}
\bibinfo{author}{Luo, X.}, \bibinfo{author}{Liao, W.}, \bibinfo{author}{Xiao, J.}, \bibinfo{author}{Chen, J.}, \bibinfo{author}{Song, T.}, \bibinfo{author}{Zhang, X.}, \bibinfo{author}{Li, K.}, \bibinfo{author}{Metaxas, D.N.}, \bibinfo{author}{Wang, G.}, \bibinfo{author}{Zhang, S.}, \bibinfo{year}{2022}.
\newblock \bibinfo{title}{Word: A large scale dataset, benchmark and clinical applicable study for abdominal organ segmentation from ct image}.
\newblock \bibinfo{journal}{Medical Image Analysis} \bibinfo{volume}{82}, \bibinfo{pages}{102642}.
\newblock \URLprefix \url{https://www.sciencedirect.com/science/article/pii/S1361841522002705}, \DOIprefix\doi{https://doi.org/10.1016/j.media.2022.102642}.
\bibitem[{Ma et~al.(2024a)Ma, He, Li, Han, You and Wang}]{MedSAM_nature}
\bibinfo{author}{Ma, J.}, \bibinfo{author}{He, Y.}, \bibinfo{author}{Li, F.}, \bibinfo{author}{Han, L.}, \bibinfo{author}{You, C.}, \bibinfo{author}{Wang, B.}, \bibinfo{year}{2024}a.
\newblock \bibinfo{title}{Segment anything in medical images}.
\newblock \bibinfo{journal}{Nature Communications} \bibinfo{volume}{15}, \bibinfo{pages}{654}.
\newblock \URLprefix \url{https://doi.org/10.1038/s41467-024-44824-z}, \DOIprefix\doi{10.1038/s41467-024-44824-z}.
\bibitem[{Ma et~al.(2025)Ma, Yang, Kim, Chen, Baharoon, Fallahpour, Asakereh, Lyu and Wang}]{ma2025medsam2segment3dmedical}
\bibinfo{author}{Ma, J.}, \bibinfo{author}{Yang, Z.}, \bibinfo{author}{Kim, S.}, \bibinfo{author}{Chen, B.}, \bibinfo{author}{Baharoon, M.}, \bibinfo{author}{Fallahpour, A.}, \bibinfo{author}{Asakereh, R.}, \bibinfo{author}{Lyu, H.}, \bibinfo{author}{Wang, B.}, \bibinfo{year}{2025}.
\newblock \bibinfo{title}{Medsam2: Segment anything in 3d medical images and videos}.
\newblock \URLprefix \url{https://arxiv.org/abs/2504.03600}, \href{http://arxiv.org/abs/2504.03600}{{\tt arXiv:2504.03600}}.
\bibitem[{Ma et~al.(2022a)Ma, Zhang, Gu, An, Wang, Ge, Wang, Zhang, Wang, Xu, Gou, Thaler, Payer, tern, Henderson, McSweeney, Green, Jackson, McIntosh, Nguyen, Qayyum, Conze, Huang, Zhou, Fan, Xiong, Dong, Zhu, He and Yang}]{MA2022102616}
\bibinfo{author}{Ma, J.}, \bibinfo{author}{Zhang, Y.}, \bibinfo{author}{Gu, S.}, \bibinfo{author}{An, X.}, \bibinfo{author}{Wang, Z.}, \bibinfo{author}{Ge, C.}, \bibinfo{author}{Wang, C.}, \bibinfo{author}{Zhang, F.}, \bibinfo{author}{Wang, Y.}, \bibinfo{author}{Xu, Y.}, \bibinfo{author}{Gou, S.}, \bibinfo{author}{Thaler, F.}, \bibinfo{author}{Payer, C.}, \bibinfo{author}{tern, D.}, \bibinfo{author}{Henderson, E.G.}, \bibinfo{author}{McSweeney, D.M.}, \bibinfo{author}{Green, A.}, \bibinfo{author}{Jackson, P.}, \bibinfo{author}{McIntosh, L.}, \bibinfo{author}{Nguyen, Q.C.}, \bibinfo{author}{Qayyum, A.}, \bibinfo{author}{Conze, P.H.}, \bibinfo{author}{Huang, Z.}, \bibinfo{author}{Zhou, Z.}, \bibinfo{author}{Fan, D.P.}, \bibinfo{author}{Xiong, H.}, \bibinfo{author}{Dong, G.}, \bibinfo{author}{Zhu, Q.}, \bibinfo{author}{He, J.}, \bibinfo{author}{Yang, X.}, \bibinfo{year}{2022}a.
\newblock \bibinfo{title}{Fast and low-gpu-memory abdomen ct organ segmentation: The flare challenge}.
\newblock \bibinfo{journal}{Medical Image Analysis} \bibinfo{volume}{82}, \bibinfo{pages}{102616}.
\newblock \URLprefix \url{https://www.sciencedirect.com/science/article/pii/S1361841522002444}, \DOIprefix\doi{https://doi.org/10.1016/j.media.2022.102616}.
\bibitem[{Ma et~al.(2023)Ma, Zhang, Gu, Ge, Ma, Young, Zhu, Meng, Yang, Huang, Zhang, Liu, Pan, Huang, Wang, Sun, Xu, Jia, Choi, Alves, de~Wilde, Koehler, Wu, Wiesenfarth, Zhu, Dong, He, the FLARE Challenge~Consortium and Wang}]{ma2023unleashingstrengthsunlabeleddata}
\bibinfo{author}{Ma, J.}, \bibinfo{author}{Zhang, Y.}, \bibinfo{author}{Gu, S.}, \bibinfo{author}{Ge, C.}, \bibinfo{author}{Ma, S.}, \bibinfo{author}{Young, A.}, \bibinfo{author}{Zhu, C.}, \bibinfo{author}{Meng, K.}, \bibinfo{author}{Yang, X.}, \bibinfo{author}{Huang, Z.}, \bibinfo{author}{Zhang, F.}, \bibinfo{author}{Liu, W.}, \bibinfo{author}{Pan, Y.}, \bibinfo{author}{Huang, S.}, \bibinfo{author}{Wang, J.}, \bibinfo{author}{Sun, M.}, \bibinfo{author}{Xu, W.}, \bibinfo{author}{Jia, D.}, \bibinfo{author}{Choi, J.W.}, \bibinfo{author}{Alves, N.}, \bibinfo{author}{de~Wilde, B.}, \bibinfo{author}{Koehler, G.}, \bibinfo{author}{Wu, Y.}, \bibinfo{author}{Wiesenfarth, M.}, \bibinfo{author}{Zhu, Q.}, \bibinfo{author}{Dong, G.}, \bibinfo{author}{He, J.}, \bibinfo{author}{the FLARE Challenge~Consortium}, \bibinfo{author}{Wang, B.}, \bibinfo{year}{2023}.
\newblock \bibinfo{title}{Unleashing the strengths of unlabeled data in pan-cancer abdominal organ quantification: the flare22 challenge}.
\newblock \URLprefix \url{https://arxiv.org/abs/2308.05862}, \href{http://arxiv.org/abs/2308.05862}{{\tt arXiv:2308.05862}}.
\bibitem[{Ma et~al.(2024b)Ma, Zhang, Gu, Ge, Wang, Zhou, Huang, Lyu, He and Wang}]{ma2024automaticorganpancancersegmentation}
\bibinfo{author}{Ma, J.}, \bibinfo{author}{Zhang, Y.}, \bibinfo{author}{Gu, S.}, \bibinfo{author}{Ge, C.}, \bibinfo{author}{Wang, E.}, \bibinfo{author}{Zhou, Q.}, \bibinfo{author}{Huang, Z.}, \bibinfo{author}{Lyu, P.}, \bibinfo{author}{He, J.}, \bibinfo{author}{Wang, B.}, \bibinfo{year}{2024}b.
\newblock \bibinfo{title}{Automatic organ and pan-cancer segmentation in abdomen ct: the flare 2023 challenge}.
\newblock \URLprefix \url{https://arxiv.org/abs/2408.12534}, \href{http://arxiv.org/abs/2408.12534}{{\tt arXiv:2408.12534}}.
\bibitem[{Ma et~al.(2022b)Ma, Zhang, Gu, Zhu, Ge, Zhang, An, Wang, Wang, Liu, Cao, Zhang, Liu, Wang, Li, He and Yang}]{9497733}
\bibinfo{author}{Ma, J.}, \bibinfo{author}{Zhang, Y.}, \bibinfo{author}{Gu, S.}, \bibinfo{author}{Zhu, C.}, \bibinfo{author}{Ge, C.}, \bibinfo{author}{Zhang, Y.}, \bibinfo{author}{An, X.}, \bibinfo{author}{Wang, C.}, \bibinfo{author}{Wang, Q.}, \bibinfo{author}{Liu, X.}, \bibinfo{author}{Cao, S.}, \bibinfo{author}{Zhang, Q.}, \bibinfo{author}{Liu, S.}, \bibinfo{author}{Wang, Y.}, \bibinfo{author}{Li, Y.}, \bibinfo{author}{He, J.}, \bibinfo{author}{Yang, X.}, \bibinfo{year}{2022}b.
\newblock \bibinfo{title}{Abdomenct-1k: Is abdominal organ segmentation a solved problem?}
\newblock \bibinfo{journal}{IEEE Transactions on Pattern Analysis and Machine Intelligence} \bibinfo{volume}{44}, \bibinfo{pages}{6695--6714}.
\newblock \DOIprefix\doi{10.1109/TPAMI.2021.3100536}.
\bibitem[{Marcus et~al.(2007)Marcus, Wang, Parker, Csernansky, Morris and Buckner}]{10.1162/jocn.2007.19.9.1498}
\bibinfo{author}{Marcus, D.S.}, \bibinfo{author}{Wang, T.H.}, \bibinfo{author}{Parker, J.}, \bibinfo{author}{Csernansky, J.G.}, \bibinfo{author}{Morris, J.C.}, \bibinfo{author}{Buckner, R.L.}, \bibinfo{year}{2007}.
\newblock \bibinfo{title}{Open access series of imaging studies (oasis): Cross-sectional mri data in young, middle aged, nondemented, and demented older adults}.
\newblock \bibinfo{journal}{Journal of Cognitive Neuroscience} \bibinfo{volume}{19}, \bibinfo{pages}{1498--1507}.
\newblock \URLprefix \url{https://doi.org/10.1162/jocn.2007.19.9.1498}, \DOIprefix\doi{10.1162/jocn.2007.19.9.1498}, \href{http://arxiv.org/abs/https://direct.mit.edu/jocn/article-pdf/19/9/1498/1936514/jocn.2007.19.9.1498.pdf}{{\tt arXiv:https://direct.mit.edu/jocn/article-pdf/19/9/1498/1936514/jocn.2007.19.9.1498.pdf}}.
\bibitem[{Menze et~al.(2015)Menze, Jakab, Bauer, Kalpathy-Cramer, Farahani, Kirby, Burren, Porz, Slotboom, Wiest, Lanczi, Gerstner, Weber, Arbel, Avants, Ayache, Buendia, Collins, Cordier, Corso, Criminisi, Das, Delingette, Demiralp, Durst, Dojat, Doyle, Festa, Forbes, Geremia, Glocker, Golland, Guo, Hamamci, Iftekharuddin, Jena, John, Konukoglu, Lashkari, Mariz, Meier, Pereira, Precup, Price, Raviv, Reza, Ryan, Sarikaya, Schwartz, Shin, Shotton, Silva, Sousa, Subbanna, Szekely, Taylor, Thomas, Tustison, Unal, Vasseur, Wintermark, Ye, Zhao, Zhao, Zikic, Prastawa, Reyes and Van~Leemput}]{6975210}
\bibinfo{author}{Menze, B.H.}, \bibinfo{author}{Jakab, A.}, \bibinfo{author}{Bauer, S.}, \bibinfo{author}{Kalpathy-Cramer, J.}, \bibinfo{author}{Farahani, K.}, \bibinfo{author}{Kirby, J.}, \bibinfo{author}{Burren, Y.}, \bibinfo{author}{Porz, N.}, \bibinfo{author}{Slotboom, J.}, \bibinfo{author}{Wiest, R.}, \bibinfo{author}{Lanczi, L.}, \bibinfo{author}{Gerstner, E.}, \bibinfo{author}{Weber, M.A.}, \bibinfo{author}{Arbel, T.}, \bibinfo{author}{Avants, B.B.}, \bibinfo{author}{Ayache, N.}, \bibinfo{author}{Buendia, P.}, \bibinfo{author}{Collins, D.L.}, \bibinfo{author}{Cordier, N.}, \bibinfo{author}{Corso, J.J.}, \bibinfo{author}{Criminisi, A.}, \bibinfo{author}{Das, T.}, \bibinfo{author}{Delingette, H.}, \bibinfo{author}{Demiralp, a.}, \bibinfo{author}{Durst, C.R.}, \bibinfo{author}{Dojat, M.}, \bibinfo{author}{Doyle, S.}, \bibinfo{author}{Festa, J.}, \bibinfo{author}{Forbes, F.}, \bibinfo{author}{Geremia, E.}, \bibinfo{author}{Glocker, B.}, \bibinfo{author}{Golland, P.}, \bibinfo{author}{Guo, X.},
  \bibinfo{author}{Hamamci, A.}, \bibinfo{author}{Iftekharuddin, K.M.}, \bibinfo{author}{Jena, R.}, \bibinfo{author}{John, N.M.}, \bibinfo{author}{Konukoglu, E.}, \bibinfo{author}{Lashkari, D.}, \bibinfo{author}{Mariz, J.A.}, \bibinfo{author}{Meier, R.}, \bibinfo{author}{Pereira, S.}, \bibinfo{author}{Precup, D.}, \bibinfo{author}{Price, S.J.}, \bibinfo{author}{Raviv, T.R.}, \bibinfo{author}{Reza, S.M.S.}, \bibinfo{author}{Ryan, M.}, \bibinfo{author}{Sarikaya, D.}, \bibinfo{author}{Schwartz, L.}, \bibinfo{author}{Shin, H.C.}, \bibinfo{author}{Shotton, J.}, \bibinfo{author}{Silva, C.A.}, \bibinfo{author}{Sousa, N.}, \bibinfo{author}{Subbanna, N.K.}, \bibinfo{author}{Szekely, G.}, \bibinfo{author}{Taylor, T.J.}, \bibinfo{author}{Thomas, O.M.}, \bibinfo{author}{Tustison, N.J.}, \bibinfo{author}{Unal, G.}, \bibinfo{author}{Vasseur, F.}, \bibinfo{author}{Wintermark, M.}, \bibinfo{author}{Ye, D.H.}, \bibinfo{author}{Zhao, L.}, \bibinfo{author}{Zhao, B.}, \bibinfo{author}{Zikic, D.}, \bibinfo{author}{Prastawa, M.},
  \bibinfo{author}{Reyes, M.}, \bibinfo{author}{Van~Leemput, K.}, \bibinfo{year}{2015}.
\newblock \bibinfo{title}{The multimodal brain tumor image segmentation benchmark (brats)}.
\newblock \bibinfo{journal}{IEEE Transactions on Medical Imaging} \bibinfo{volume}{34}, \bibinfo{pages}{1993--2024}.
\newblock \DOIprefix\doi{10.1109/TMI.2014.2377694}.
\bibitem[{Milletari et~al.(2016a)Milletari, Navab and Ahmadi}]{milletari2016vnetfullyconvolutionalneural}
\bibinfo{author}{Milletari, F.}, \bibinfo{author}{Navab, N.}, \bibinfo{author}{Ahmadi, S.A.}, \bibinfo{year}{2016}a.
\newblock \bibinfo{title}{V-net: Fully convolutional neural networks for volumetric medical image segmentation}.
\newblock \URLprefix \url{https://arxiv.org/abs/1606.04797}, \href{http://arxiv.org/abs/1606.04797}{{\tt arXiv:1606.04797}}.
\bibitem[{Milletari et~al.(2016b)Milletari, Navab and Ahmadi}]{7785132}
\bibinfo{author}{Milletari, F.}, \bibinfo{author}{Navab, N.}, \bibinfo{author}{Ahmadi, S.A.}, \bibinfo{year}{2016}b.
\newblock \bibinfo{title}{V-net: Fully convolutional neural networks for volumetric medical image segmentation}, in: \bibinfo{booktitle}{2016 Fourth International Conference on 3D Vision (3DV)}, pp. \bibinfo{pages}{565--571}.
\newblock \DOIprefix\doi{10.1109/3DV.2016.79}.
\bibitem[{Myronenko(2018)}]{myronenko20183dmribraintumor}
\bibinfo{author}{Myronenko, A.}, \bibinfo{year}{2018}.
\newblock \bibinfo{title}{3d mri brain tumor segmentation using autoencoder regularization}.
\newblock \URLprefix \url{https://arxiv.org/abs/1810.11654}, \href{http://arxiv.org/abs/1810.11654}{{\tt arXiv:1810.11654}}.
\bibitem[{Park et~al.(2025)Park, Hennessee, Smith, Chan, Chen, Dakanali, Farrell, Liu, Lu, Rofsky, Sun, Tamminga, Moore, Kennedy, Rodrigue and Wig}]{Park2025}
\bibinfo{author}{Park, D.C.}, \bibinfo{author}{Hennessee, J.P.}, \bibinfo{author}{Smith, E.T.}, \bibinfo{author}{Chan, M.Y.}, \bibinfo{author}{Chen, X.}, \bibinfo{author}{Dakanali, M.}, \bibinfo{author}{Farrell, M.E.}, \bibinfo{author}{Liu, P.}, \bibinfo{author}{Lu, H.}, \bibinfo{author}{Rofsky, N.}, \bibinfo{author}{Sun, X.}, \bibinfo{author}{Tamminga, C.}, \bibinfo{author}{Moore, W.}, \bibinfo{author}{Kennedy, K.M.}, \bibinfo{author}{Rodrigue, K.}, \bibinfo{author}{Wig, G.S.}, \bibinfo{year}{2025}.
\newblock \bibinfo{title}{The dallas lifespan brain study: A comprehensive adult lifespan data set of brain and cognitive aging}.
\newblock \bibinfo{journal}{Scientific Data} \bibinfo{volume}{12}, \bibinfo{pages}{846}.
\newblock \URLprefix \url{https://doi.org/10.1038/s41597-025-04847-7}, \DOIprefix\doi{10.1038/s41597-025-04847-7}.
\bibitem[{Payette et~al.(2021)Payette, de~Dumast, Kebiri, Ezhov, Paetzold, Shit, Iqbal, Khan, Kottke, Grehten, Ji, Lanczi, Nagy, Beresova, Nguyen, Natalucci, Karayannis, Menze, Bach~Cuadra and Jakab}]{Payette2021}
\bibinfo{author}{Payette, K.}, \bibinfo{author}{de~Dumast, P.}, \bibinfo{author}{Kebiri, H.}, \bibinfo{author}{Ezhov, I.}, \bibinfo{author}{Paetzold, J.C.}, \bibinfo{author}{Shit, S.}, \bibinfo{author}{Iqbal, A.}, \bibinfo{author}{Khan, R.}, \bibinfo{author}{Kottke, R.}, \bibinfo{author}{Grehten, P.}, \bibinfo{author}{Ji, H.}, \bibinfo{author}{Lanczi, L.}, \bibinfo{author}{Nagy, M.}, \bibinfo{author}{Beresova, M.}, \bibinfo{author}{Nguyen, T.D.}, \bibinfo{author}{Natalucci, G.}, \bibinfo{author}{Karayannis, T.}, \bibinfo{author}{Menze, B.}, \bibinfo{author}{Bach~Cuadra, M.}, \bibinfo{author}{Jakab, A.}, \bibinfo{year}{2021}.
\newblock \bibinfo{title}{An automatic multi-tissue human fetal brain segmentation benchmark using the fetal tissue annotation dataset}.
\newblock \bibinfo{journal}{Scientific Data} \bibinfo{volume}{8}, \bibinfo{pages}{167}.
\newblock \URLprefix \url{https://doi.org/10.1038/s41597-021-00946-3}, \DOIprefix\doi{10.1038/s41597-021-00946-3}.
\bibitem[{Payette et~al.(2025)Payette, Steger, Licandro, Dumast, Li, Barkovich, Li, Dannecker, Chen, Ouyang, McConnell, Miron, Li, Uus, Grigorescu, Gilliland, Siddiquee, Xu, Myronenko, Wang, Huang, Ye, Aleny, Comte, Camara, Masson, Nilsson, Godard, Mazher, Qayyum, Gao, Zhou, Gao, Fu, Dong, Wang, Rieu, Yang, Lee, Potka, Grzeszczyk, Sitek, Daza, Usma, Arbelaez, Lu, Zhang, Liang, Valabregue, Joshi, Nayak, Leahy, Wilhelmi, Dndliker, Ji, Gennari, Jakovi, Klai, Adi, Markovi, Grabari, Kasprian, Dovjak, Rados, Vasung, Cuadra and Jakab}]{Payette_2025}
\bibinfo{author}{Payette, K.}, \bibinfo{author}{Steger, C.}, \bibinfo{author}{Licandro, R.}, \bibinfo{author}{Dumast, P.d.}, \bibinfo{author}{Li, H.B.}, \bibinfo{author}{Barkovich, M.}, \bibinfo{author}{Li, L.}, \bibinfo{author}{Dannecker, M.}, \bibinfo{author}{Chen, C.}, \bibinfo{author}{Ouyang, C.}, \bibinfo{author}{McConnell, N.}, \bibinfo{author}{Miron, A.}, \bibinfo{author}{Li, Y.}, \bibinfo{author}{Uus, A.}, \bibinfo{author}{Grigorescu, I.}, \bibinfo{author}{Gilliland, P.R.}, \bibinfo{author}{Siddiquee, M.M.R.}, \bibinfo{author}{Xu, D.}, \bibinfo{author}{Myronenko, A.}, \bibinfo{author}{Wang, H.}, \bibinfo{author}{Huang, Z.}, \bibinfo{author}{Ye, J.}, \bibinfo{author}{Aleny, M.}, \bibinfo{author}{Comte, V.}, \bibinfo{author}{Camara, O.}, \bibinfo{author}{Masson, J.B.}, \bibinfo{author}{Nilsson, A.}, \bibinfo{author}{Godard, C.}, \bibinfo{author}{Mazher, M.}, \bibinfo{author}{Qayyum, A.}, \bibinfo{author}{Gao, Y.}, \bibinfo{author}{Zhou, H.}, \bibinfo{author}{Gao, S.}, \bibinfo{author}{Fu, J.},
  \bibinfo{author}{Dong, G.}, \bibinfo{author}{Wang, G.}, \bibinfo{author}{Rieu, Z.}, \bibinfo{author}{Yang, H.}, \bibinfo{author}{Lee, M.}, \bibinfo{author}{Potka, S.}, \bibinfo{author}{Grzeszczyk, M.K.}, \bibinfo{author}{Sitek, A.}, \bibinfo{author}{Daza, L.V.}, \bibinfo{author}{Usma, S.}, \bibinfo{author}{Arbelaez, P.}, \bibinfo{author}{Lu, W.}, \bibinfo{author}{Zhang, W.}, \bibinfo{author}{Liang, J.}, \bibinfo{author}{Valabregue, R.}, \bibinfo{author}{Joshi, A.A.}, \bibinfo{author}{Nayak, K.N.}, \bibinfo{author}{Leahy, R.M.}, \bibinfo{author}{Wilhelmi, L.}, \bibinfo{author}{Dndliker, A.}, \bibinfo{author}{Ji, H.}, \bibinfo{author}{Gennari, A.G.}, \bibinfo{author}{Jakovi, A.}, \bibinfo{author}{Klai, M.}, \bibinfo{author}{Adi, A.}, \bibinfo{author}{Markovi, P.}, \bibinfo{author}{Grabari, G.}, \bibinfo{author}{Kasprian, G.}, \bibinfo{author}{Dovjak, G.}, \bibinfo{author}{Rados, M.}, \bibinfo{author}{Vasung, L.}, \bibinfo{author}{Cuadra, M.B.}, \bibinfo{author}{Jakab, A.}, \bibinfo{year}{2025}.
\newblock \bibinfo{title}{Multi-center fetal brain tissue annotation (feta) challenge 2022 results}.
\newblock \bibinfo{journal}{IEEE Transactions on Medical Imaging} \bibinfo{volume}{44}, \bibinfo{pages}{12571272}.
\newblock \URLprefix \url{http://dx.doi.org/10.1109/TMI.2024.3485554}, \DOIprefix\doi{10.1109/tmi.2024.3485554}.
\bibitem[{Quinton et~al.(2023)Quinton, Popoff, Presles, Leclerc, Meriaudeau, Nodari, Lopez, Pellegrinelli, Chevallier, Ginhac, Vrigneaud and Alberini}]{data8050079}
\bibinfo{author}{Quinton, F.}, \bibinfo{author}{Popoff, R.}, \bibinfo{author}{Presles, B.}, \bibinfo{author}{Leclerc, S.}, \bibinfo{author}{Meriaudeau, F.}, \bibinfo{author}{Nodari, G.}, \bibinfo{author}{Lopez, O.}, \bibinfo{author}{Pellegrinelli, J.}, \bibinfo{author}{Chevallier, O.}, \bibinfo{author}{Ginhac, D.}, \bibinfo{author}{Vrigneaud, J.M.}, \bibinfo{author}{Alberini, J.L.}, \bibinfo{year}{2023}.
\newblock \bibinfo{title}{A tumour and liver automatic segmentation (atlas) dataset on contrast-enhanced magnetic resonance imaging for hepatocellular carcinoma}.
\newblock \bibinfo{journal}{Data} \bibinfo{volume}{8}.
\newblock \URLprefix \url{https://www.mdpi.com/2306-5729/8/5/79}, \DOIprefix\doi{10.3390/data8050079}.
\bibitem[{Radau et~al.(2009)Radau, Lu, Connelly, Paul, Dick and Wright}]{Radau_Lu_Connelly_Paul_Dick_Wright2009}
\bibinfo{author}{Radau, P.}, \bibinfo{author}{Lu, Y.}, \bibinfo{author}{Connelly, K.}, \bibinfo{author}{Paul, G.}, \bibinfo{author}{Dick, A.}, \bibinfo{author}{Wright, G.}, \bibinfo{year}{2009}.
\newblock \bibinfo{title}{Evaluation framework for algorithms segmenting short axis cardiac mri.} \DOIprefix\doi{10.54294/g80ruo}.
\bibitem[{Raudaschl et~al.(2017)Raudaschl, Zaffino, Sharp, Spadea, Chen, Dawant, Albrecht, Gass, Langguth, Lthi, Jung, Knapp, Wesarg, Mannion-Haworth, Bowes, Ashman, Guillard, Brett, Vincent, Orbes-Arteaga, Crdenas-Pea, Castellanos-Dominguez, Aghdasi, Li, Berens, Moe, Hannaford, Schubert and Fritscher}]{https://doi.org/10.1002/mp.12197}
\bibinfo{author}{Raudaschl, P.F.}, \bibinfo{author}{Zaffino, P.}, \bibinfo{author}{Sharp, G.C.}, \bibinfo{author}{Spadea, M.F.}, \bibinfo{author}{Chen, A.}, \bibinfo{author}{Dawant, B.M.}, \bibinfo{author}{Albrecht, T.}, \bibinfo{author}{Gass, T.}, \bibinfo{author}{Langguth, C.}, \bibinfo{author}{Lthi, M.}, \bibinfo{author}{Jung, F.}, \bibinfo{author}{Knapp, O.}, \bibinfo{author}{Wesarg, S.}, \bibinfo{author}{Mannion-Haworth, R.}, \bibinfo{author}{Bowes, M.}, \bibinfo{author}{Ashman, A.}, \bibinfo{author}{Guillard, G.}, \bibinfo{author}{Brett, A.}, \bibinfo{author}{Vincent, G.}, \bibinfo{author}{Orbes-Arteaga, M.}, \bibinfo{author}{Crdenas-Pea, D.}, \bibinfo{author}{Castellanos-Dominguez, G.}, \bibinfo{author}{Aghdasi, N.}, \bibinfo{author}{Li, Y.}, \bibinfo{author}{Berens, A.}, \bibinfo{author}{Moe, K.}, \bibinfo{author}{Hannaford, B.}, \bibinfo{author}{Schubert, R.}, \bibinfo{author}{Fritscher, K.D.}, \bibinfo{year}{2017}.
\newblock \bibinfo{title}{Evaluation of segmentation methods on head and neck ct: Auto-segmentation challenge 2015}.
\newblock \bibinfo{journal}{Medical Physics} \bibinfo{volume}{44}, \bibinfo{pages}{2020--2036}.
\newblock \URLprefix \url{https://aapm.onlinelibrary.wiley.com/doi/abs/10.1002/mp.12197}, \DOIprefix\doi{https://doi.org/10.1002/mp.12197}, \href{http://arxiv.org/abs/https://aapm.onlinelibrary.wiley.com/doi/pdf/10.1002/mp.12197}{{\tt arXiv:https://aapm.onlinelibrary.wiley.com/doi/pdf/10.1002/mp.12197}}.
\bibitem[{Rodrigue et~al.(2012)Rodrigue, Kennedy, Devous, Rieck, Hebrank, Diaz-Arrastia, Mathews and Park}]{doi:10.1212/WNL.0b013e318245d295}
\bibinfo{author}{Rodrigue, K.}, \bibinfo{author}{Kennedy, K.}, \bibinfo{author}{Devous, M.}, \bibinfo{author}{Rieck, J.}, \bibinfo{author}{Hebrank, A.}, \bibinfo{author}{Diaz-Arrastia, R.}, \bibinfo{author}{Mathews, D.}, \bibinfo{author}{Park, D.}, \bibinfo{year}{2012}.
\newblock \bibinfo{title}{-amyloid burden in healthy aging}.
\newblock \bibinfo{journal}{Neurology} \bibinfo{volume}{78}, \bibinfo{pages}{387--395}.
\newblock \URLprefix \url{https://www.neurology.org/doi/abs/10.1212/WNL.0b013e318245d295}, \DOIprefix\doi{10.1212/WNL.0b013e318245d295}, \href{http://arxiv.org/abs/https://www.neurology.org/doi/pdf/10.1212/WNL.0b013e318245d295}{{\tt arXiv:https://www.neurology.org/doi/pdf/10.1212/WNL.0b013e318245d295}}.
\bibitem[{Ronneberger et~al.(2015a)Ronneberger, Fischer and Brox}]{ronneberger2015unetconvolutionalnetworksbiomedical}
\bibinfo{author}{Ronneberger, O.}, \bibinfo{author}{Fischer, P.}, \bibinfo{author}{Brox, T.}, \bibinfo{year}{2015}a.
\newblock \bibinfo{title}{U-net: Convolutional networks for biomedical image segmentation}.
\newblock \URLprefix \url{https://arxiv.org/abs/1505.04597}, \href{http://arxiv.org/abs/1505.04597}{{\tt arXiv:1505.04597}}.
\bibitem[{Ronneberger et~al.(2015b)Ronneberger, Fischer and Brox}]{10.1007/978-3-319-24574-4_28}
\bibinfo{author}{Ronneberger, O.}, \bibinfo{author}{Fischer, P.}, \bibinfo{author}{Brox, T.}, \bibinfo{year}{2015}b.
\newblock \bibinfo{title}{U-net: Convolutional networks for biomedical image segmentation}, in: \bibinfo{editor}{Navab, N.}, \bibinfo{editor}{Hornegger, J.}, \bibinfo{editor}{Wells, W.M.}, \bibinfo{editor}{Frangi, A.F.} (Eds.), \bibinfo{booktitle}{Medical Image Computing and Computer-Assisted Intervention -- MICCAI 2015}, \bibinfo{publisher}{Springer International Publishing}, \bibinfo{address}{Cham}. pp. \bibinfo{pages}{234--241}.
\bibitem[{Roth et~al.(2016)Roth, Farag, Turkbey, Lu, Liu and Summers}]{Roth2016}
\bibinfo{author}{Roth, H.R.}, \bibinfo{author}{Farag, A.}, \bibinfo{author}{Turkbey, E.B.}, \bibinfo{author}{Lu, L.}, \bibinfo{author}{Liu, J.}, \bibinfo{author}{Summers, R.M.}, \bibinfo{year}{2016}.
\newblock \bibinfo{title}{Data from pancreas-ct}.
\newblock \bibinfo{howpublished}{\url{https://doi.org/10.7937/K9/TCIA.2016.tNB1kqBU}}.
\newblock \bibinfo{note}{The Cancer Imaging Archive}.
\bibitem[{Roy et~al.(2024)Roy, Koehler, Ulrich, Baumgartner, Petersen, Isensee, Jaeger and Maier-Hein}]{roy2024mednexttransformerdrivenscalingconvnets}
\bibinfo{author}{Roy, S.}, \bibinfo{author}{Koehler, G.}, \bibinfo{author}{Ulrich, C.}, \bibinfo{author}{Baumgartner, M.}, \bibinfo{author}{Petersen, J.}, \bibinfo{author}{Isensee, F.}, \bibinfo{author}{Jaeger, P.F.}, \bibinfo{author}{Maier-Hein, K.}, \bibinfo{year}{2024}.
\newblock \bibinfo{title}{Mednext: Transformer-driven scaling of convnets for medical image segmentation}.
\newblock \URLprefix \url{https://arxiv.org/abs/2303.09975}, \href{http://arxiv.org/abs/2303.09975}{{\tt arXiv:2303.09975}}.
\bibitem[{Roy et~al.(2023a)Roy, Koehler, Ulrich, Baumgartner, Petersen, Isensee, J{\"a}ger and Maier-Hein}]{10.1007/978-3-031-43901-8_39}
\bibinfo{author}{Roy, S.}, \bibinfo{author}{Koehler, G.}, \bibinfo{author}{Ulrich, C.}, \bibinfo{author}{Baumgartner, M.}, \bibinfo{author}{Petersen, J.}, \bibinfo{author}{Isensee, F.}, \bibinfo{author}{J{\"a}ger, P.F.}, \bibinfo{author}{Maier-Hein, K.H.}, \bibinfo{year}{2023}a.
\newblock \bibinfo{title}{Mednext: Transformer-driven scaling of convnets for medical image segmentation}, in: \bibinfo{editor}{Greenspan, H.}, \bibinfo{editor}{Madabhushi, A.}, \bibinfo{editor}{Mousavi, P.}, \bibinfo{editor}{Salcudean, S.}, \bibinfo{editor}{Duncan, J.}, \bibinfo{editor}{Syeda-Mahmood, T.}, \bibinfo{editor}{Taylor, R.} (Eds.), \bibinfo{booktitle}{Medical Image Computing and Computer Assisted Intervention -- MICCAI 2023}, \bibinfo{publisher}{Springer Nature Switzerland}, \bibinfo{address}{Cham}. pp. \bibinfo{pages}{405--415}.
\bibitem[{Roy et~al.(2023b)Roy, Koehler, Ulrich, Baumgartner, Petersen, Isensee, J{\"a}ger and Maier-Hein}]{mednext_miccai}
\bibinfo{author}{Roy, S.}, \bibinfo{author}{Koehler, G.}, \bibinfo{author}{Ulrich, C.}, \bibinfo{author}{Baumgartner, M.}, \bibinfo{author}{Petersen, J.}, \bibinfo{author}{Isensee, F.}, \bibinfo{author}{J{\"a}ger, P.F.}, \bibinfo{author}{Maier-Hein, K.H.}, \bibinfo{year}{2023}b.
\newblock \bibinfo{title}{Mednext: Transformer-driven scaling of convnets for medical image segmentation}, in: \bibinfo{editor}{Greenspan, H.}, \bibinfo{editor}{Madabhushi, A.}, \bibinfo{editor}{Mousavi, P.}, \bibinfo{editor}{Salcudean, S.}, \bibinfo{editor}{Duncan, J.}, \bibinfo{editor}{Syeda-Mahmood, T.}, \bibinfo{editor}{Taylor, R.} (Eds.), \bibinfo{booktitle}{Medical Image Computing and Computer Assisted Intervention -- MICCAI 2023}, \bibinfo{publisher}{Springer Nature Switzerland}, \bibinfo{address}{Cham}. pp. \bibinfo{pages}{405--415}.
\bibitem[{Sadegheih et~al.(2024)Sadegheih, Bozorgpour, Kumari, Azad and Merhof}]{sadegheih2024lhunetlighthybridunet}
\bibinfo{author}{Sadegheih, Y.}, \bibinfo{author}{Bozorgpour, A.}, \bibinfo{author}{Kumari, P.}, \bibinfo{author}{Azad, R.}, \bibinfo{author}{Merhof, D.}, \bibinfo{year}{2024}.
\newblock \bibinfo{title}{Lhu-net: A light hybrid u-net for cost-efficient, high-performance volumetric medical image segmentation}.
\newblock \URLprefix \url{https://arxiv.org/abs/2404.05102}, \href{http://arxiv.org/abs/2404.05102}{{\tt arXiv:2404.05102}}.
\bibitem[{Sekuboyina et~al.(2021)Sekuboyina, Husseini, Bayat, Lffler, Liebl, Li, Tetteh, Kukaka, Payer, tern, Urschler, Chen, Cheng, Lessmann, Hu, Wang, Yang, Xu, Ambellan, Amiranashvili, Ehlke, Lamecker, Lehnert, Lirio, de~Olaguer, Ramm, Sahu, Tack, Zachow, Jiang, Ma, Angerman, Wang, Brown, Kirszenberg, lodie Puybareau, Chen, Bai, Rapazzo, Yeah, Zhang, Xu, Hou, He, Zeng, Xiangshang, Liming, Netherton, Mumme, Court, Huang, He, Wang, Ling, Hunh, Boutry, Jakubicek, Chmelik, Mulay, Sivaprakasam, Paetzold, Shit, Ezhov, Wiestler, Glocker, Valentinitsch, Rempfler, Menze and Kirschke}]{SEKUBOYINA2021102166}
\bibinfo{author}{Sekuboyina, A.}, \bibinfo{author}{Husseini, M.E.}, \bibinfo{author}{Bayat, A.}, \bibinfo{author}{Lffler, M.}, \bibinfo{author}{Liebl, H.}, \bibinfo{author}{Li, H.}, \bibinfo{author}{Tetteh, G.}, \bibinfo{author}{Kukaka, J.}, \bibinfo{author}{Payer, C.}, \bibinfo{author}{tern, D.}, \bibinfo{author}{Urschler, M.}, \bibinfo{author}{Chen, M.}, \bibinfo{author}{Cheng, D.}, \bibinfo{author}{Lessmann, N.}, \bibinfo{author}{Hu, Y.}, \bibinfo{author}{Wang, T.}, \bibinfo{author}{Yang, D.}, \bibinfo{author}{Xu, D.}, \bibinfo{author}{Ambellan, F.}, \bibinfo{author}{Amiranashvili, T.}, \bibinfo{author}{Ehlke, M.}, \bibinfo{author}{Lamecker, H.}, \bibinfo{author}{Lehnert, S.}, \bibinfo{author}{Lirio, M.}, \bibinfo{author}{de~Olaguer, N.P.}, \bibinfo{author}{Ramm, H.}, \bibinfo{author}{Sahu, M.}, \bibinfo{author}{Tack, A.}, \bibinfo{author}{Zachow, S.}, \bibinfo{author}{Jiang, T.}, \bibinfo{author}{Ma, X.}, \bibinfo{author}{Angerman, C.}, \bibinfo{author}{Wang, X.}, \bibinfo{author}{Brown, K.},
  \bibinfo{author}{Kirszenberg, A.}, \bibinfo{author}{lodie Puybareau}, \bibinfo{author}{Chen, D.}, \bibinfo{author}{Bai, Y.}, \bibinfo{author}{Rapazzo, B.H.}, \bibinfo{author}{Yeah, T.}, \bibinfo{author}{Zhang, A.}, \bibinfo{author}{Xu, S.}, \bibinfo{author}{Hou, F.}, \bibinfo{author}{He, Z.}, \bibinfo{author}{Zeng, C.}, \bibinfo{author}{Xiangshang, Z.}, \bibinfo{author}{Liming, X.}, \bibinfo{author}{Netherton, T.J.}, \bibinfo{author}{Mumme, R.P.}, \bibinfo{author}{Court, L.E.}, \bibinfo{author}{Huang, Z.}, \bibinfo{author}{He, C.}, \bibinfo{author}{Wang, L.W.}, \bibinfo{author}{Ling, S.H.}, \bibinfo{author}{Hunh, L.D.}, \bibinfo{author}{Boutry, N.}, \bibinfo{author}{Jakubicek, R.}, \bibinfo{author}{Chmelik, J.}, \bibinfo{author}{Mulay, S.}, \bibinfo{author}{Sivaprakasam, M.}, \bibinfo{author}{Paetzold, J.C.}, \bibinfo{author}{Shit, S.}, \bibinfo{author}{Ezhov, I.}, \bibinfo{author}{Wiestler, B.}, \bibinfo{author}{Glocker, B.}, \bibinfo{author}{Valentinitsch, A.}, \bibinfo{author}{Rempfler, M.},
  \bibinfo{author}{Menze, B.H.}, \bibinfo{author}{Kirschke, J.S.}, \bibinfo{year}{2021}.
\newblock \bibinfo{title}{Verse: A vertebrae labelling and segmentation benchmark for multi-detector ct images}.
\newblock \bibinfo{journal}{Medical Image Analysis} \bibinfo{volume}{73}, \bibinfo{pages}{102166}.
\newblock \URLprefix \url{https://www.sciencedirect.com/science/article/pii/S1361841521002127}, \DOIprefix\doi{https://doi.org/10.1016/j.media.2021.102166}.
\bibitem[{Setio et~al.(2017)Setio, Traverso, {de Bel}, Berens, van~den Bogaard, Cerello, Chen, Dou, Fantacci, Geurts, van~der Gugten, Heng, Jansen, {de Kaste}, Kotov, Lin, Manders, Sora-Mengana, Garca-Naranjo, Papavasileiou, Prokop, Saletta, Schaefer-Prokop, Scholten, Scholten, Snoeren, Torres, Vandemeulebroucke, Walasek, Zuidhof, van Ginneken and Jacobs}]{SETIO20171}
\bibinfo{author}{Setio, A.A.A.}, \bibinfo{author}{Traverso, A.}, \bibinfo{author}{{de Bel}, T.}, \bibinfo{author}{Berens, M.S.}, \bibinfo{author}{van~den Bogaard, C.}, \bibinfo{author}{Cerello, P.}, \bibinfo{author}{Chen, H.}, \bibinfo{author}{Dou, Q.}, \bibinfo{author}{Fantacci, M.E.}, \bibinfo{author}{Geurts, B.}, \bibinfo{author}{van~der Gugten, R.}, \bibinfo{author}{Heng, P.A.}, \bibinfo{author}{Jansen, B.}, \bibinfo{author}{{de Kaste}, M.M.}, \bibinfo{author}{Kotov, V.}, \bibinfo{author}{Lin, J.Y.H.}, \bibinfo{author}{Manders, J.T.}, \bibinfo{author}{Sora-Mengana, A.}, \bibinfo{author}{Garca-Naranjo, J.C.}, \bibinfo{author}{Papavasileiou, E.}, \bibinfo{author}{Prokop, M.}, \bibinfo{author}{Saletta, M.}, \bibinfo{author}{Schaefer-Prokop, C.M.}, \bibinfo{author}{Scholten, E.T.}, \bibinfo{author}{Scholten, L.}, \bibinfo{author}{Snoeren, M.M.}, \bibinfo{author}{Torres, E.L.}, \bibinfo{author}{Vandemeulebroucke, J.}, \bibinfo{author}{Walasek, N.}, \bibinfo{author}{Zuidhof, G.C.}, \bibinfo{author}{van
  Ginneken, B.}, \bibinfo{author}{Jacobs, C.}, \bibinfo{year}{2017}.
\newblock \bibinfo{title}{Validation, comparison, and combination of algorithms for automatic detection of pulmonary nodules in computed tomography images: The luna16 challenge}.
\newblock \bibinfo{journal}{Medical Image Analysis} \bibinfo{volume}{42}, \bibinfo{pages}{1--13}.
\newblock \URLprefix \url{https://www.sciencedirect.com/science/article/pii/S1361841517301020}, \DOIprefix\doi{https://doi.org/10.1016/j.media.2017.06.015}.
\bibitem[{Shaker et~al.(2024a)Shaker, Maaz, Rasheed, Khan, Yang and Khan}]{shaker2024unetrdelvingefficientaccurate}
\bibinfo{author}{Shaker, A.}, \bibinfo{author}{Maaz, M.}, \bibinfo{author}{Rasheed, H.}, \bibinfo{author}{Khan, S.}, \bibinfo{author}{Yang, M.H.}, \bibinfo{author}{Khan, F.S.}, \bibinfo{year}{2024}a.
\newblock \bibinfo{title}{Unetr++: Delving into efficient and accurate 3d medical image segmentation}.
\newblock \URLprefix \url{https://arxiv.org/abs/2212.04497}, \href{http://arxiv.org/abs/2212.04497}{{\tt arXiv:2212.04497}}.
\bibitem[{Shaker et~al.(2024b)Shaker, Maaz, Rasheed, Khan, Yang and Shahbaz~Khan}]{10526382}
\bibinfo{author}{Shaker, A.}, \bibinfo{author}{Maaz, M.}, \bibinfo{author}{Rasheed, H.}, \bibinfo{author}{Khan, S.}, \bibinfo{author}{Yang, M.H.}, \bibinfo{author}{Shahbaz~Khan, F.}, \bibinfo{year}{2024}b.
\newblock \bibinfo{title}{Unetr++: Delving into efficient and accurate 3d medical image segmentation}.
\newblock \bibinfo{journal}{IEEE Transactions on Medical Imaging} \bibinfo{volume}{43}, \bibinfo{pages}{3377--3390}.
\newblock \DOIprefix\doi{10.1109/TMI.2024.3398728}.
\bibitem[{Shaker et~al.(2024c)Shaker, Maaz, Rasheed, Khan, Yang and Shahbaz~Khan}]{unetr_pp_ieee}
\bibinfo{author}{Shaker, A.}, \bibinfo{author}{Maaz, M.}, \bibinfo{author}{Rasheed, H.}, \bibinfo{author}{Khan, S.}, \bibinfo{author}{Yang, M.H.}, \bibinfo{author}{Shahbaz~Khan, F.}, \bibinfo{year}{2024}c.
\newblock \bibinfo{title}{Unetr++: Delving into efficient and accurate 3d medical image segmentation}.
\newblock \bibinfo{journal}{IEEE Transactions on Medical Imaging} \bibinfo{volume}{43}, \bibinfo{pages}{3377--3390}.
\newblock \DOIprefix\doi{10.1109/TMI.2024.3398728}.
\bibitem[{Shi et~al.(2024)Shi, Han, Huang, Liao, Li, Kong, Zhu, Wang and Liu}]{10.1007/978-3-031-72111-3_38}
\bibinfo{author}{Shi, H.}, \bibinfo{author}{Han, S.}, \bibinfo{author}{Huang, S.}, \bibinfo{author}{Liao, Y.}, \bibinfo{author}{Li, G.}, \bibinfo{author}{Kong, X.}, \bibinfo{author}{Zhu, H.}, \bibinfo{author}{Wang, X.}, \bibinfo{author}{Liu, S.}, \bibinfo{year}{2024}.
\newblock \bibinfo{title}{Mask-enhanced segment anything model for tumor lesion semantic segmentation}, in: \bibinfo{editor}{Linguraru, M.G.}, \bibinfo{editor}{Dou, Q.}, \bibinfo{editor}{Feragen, A.}, \bibinfo{editor}{Giannarou, S.}, \bibinfo{editor}{Glocker, B.}, \bibinfo{editor}{Lekadir, K.}, \bibinfo{editor}{Schnabel, J.A.} (Eds.), \bibinfo{booktitle}{Medical Image Computing and Computer Assisted Intervention -- MICCAI 2024}, \bibinfo{publisher}{Springer Nature Switzerland}, \bibinfo{address}{Cham}. pp. \bibinfo{pages}{403--413}.
\bibitem[{Shi et~al.(2023)Shi, Guo, Yang, Ye and Ma}]{shi2023nextouefficienttopologyawareunet}
\bibinfo{author}{Shi, P.}, \bibinfo{author}{Guo, X.}, \bibinfo{author}{Yang, Y.}, \bibinfo{author}{Ye, C.}, \bibinfo{author}{Ma, T.}, \bibinfo{year}{2023}.
\newblock \bibinfo{title}{Nextou: Efficient topology-aware u-net for medical image segmentation}.
\newblock \URLprefix \url{https://arxiv.org/abs/2305.15911}, \href{http://arxiv.org/abs/2305.15911}{{\tt arXiv:2305.15911}}.
\bibitem[{Simpson et~al.(2019)Simpson, Antonelli, Bakas, Bilello, Farahani, van Ginneken, Kopp-Schneider, Landman, Litjens, Menze, Ronneberger, Summers, Bilic, Christ, Do, Gollub, Golia-Pernicka, Heckers, Jarnagin, McHugo, Napel, Vorontsov, Maier-Hein and Cardoso}]{MSD_arxiv_dataset}
\bibinfo{author}{Simpson, A.L.}, \bibinfo{author}{Antonelli, M.}, \bibinfo{author}{Bakas, S.}, \bibinfo{author}{Bilello, M.}, \bibinfo{author}{Farahani, K.}, \bibinfo{author}{van Ginneken, B.}, \bibinfo{author}{Kopp-Schneider, A.}, \bibinfo{author}{Landman, B.A.}, \bibinfo{author}{Litjens, G.}, \bibinfo{author}{Menze, B.}, \bibinfo{author}{Ronneberger, O.}, \bibinfo{author}{Summers, R.M.}, \bibinfo{author}{Bilic, P.}, \bibinfo{author}{Christ, P.F.}, \bibinfo{author}{Do, R.K.G.}, \bibinfo{author}{Gollub, M.}, \bibinfo{author}{Golia-Pernicka, J.}, \bibinfo{author}{Heckers, S.H.}, \bibinfo{author}{Jarnagin, W.R.}, \bibinfo{author}{McHugo, M.K.}, \bibinfo{author}{Napel, S.}, \bibinfo{author}{Vorontsov, E.}, \bibinfo{author}{Maier-Hein, L.}, \bibinfo{author}{Cardoso, M.J.}, \bibinfo{year}{2019}.
\newblock \bibinfo{title}{A large annotated medical image dataset for the development and evaluation of segmentation algorithms}.
\newblock \URLprefix \url{https://arxiv.org/abs/1902.09063}, \href{http://arxiv.org/abs/1902.09063}{{\tt arXiv:1902.09063}}.
\bibitem[{Soler et~al.(2010)Soler, Hostettler, Agnus, Charnoz, Fasquel, Moreau, Osswald, Bouhadjar and Marescaux}]{soler20103d}
\bibinfo{author}{Soler, L.}, \bibinfo{author}{Hostettler, A.}, \bibinfo{author}{Agnus, V.}, \bibinfo{author}{Charnoz, A.}, \bibinfo{author}{Fasquel, J.B.}, \bibinfo{author}{Moreau, J.}, \bibinfo{author}{Osswald, A.B.}, \bibinfo{author}{Bouhadjar, M.}, \bibinfo{author}{Marescaux, J.}, \bibinfo{year}{2010}.
\newblock \bibinfo{title}{3d image reconstruction for comparison of algorithm database}.
\newblock \bibinfo{journal}{URL: https://www. ircad. fr/research/data-sets/liver-segmentation-3d-ircadb-01} .
\bibitem[{Styner et~al.(2008)Styner, Lee, Chin, Chin, Commowick, Tran, Markovic-Plese, Jewells and Warfield}]{Styner_Lee_Chin_Chin_Commowick_Tran_Markovic-Plese_Jewells_Warfield2008}
\bibinfo{author}{Styner, M.}, \bibinfo{author}{Lee, J.}, \bibinfo{author}{Chin, B.}, \bibinfo{author}{Chin, M.}, \bibinfo{author}{Commowick, O.}, \bibinfo{author}{Tran, H.}, \bibinfo{author}{Markovic-Plese, S.}, \bibinfo{author}{Jewells, V.}, \bibinfo{author}{Warfield, S.}, \bibinfo{year}{2008}.
\newblock \bibinfo{title}{3d segmentation in the clinic: A grand challenge ii: Ms lesion segmentation} \DOIprefix\doi{10.54294/lmkqvm}.
\bibitem[{Tobon-Gomez et~al.(2014)Tobon-Gomez, Peters, Weese, Pinto, Karim, Schaeffter, Razavi and Rhode}]{10.1007/978-3-642-54268-8_1}
\bibinfo{author}{Tobon-Gomez, C.}, \bibinfo{author}{Peters, J.}, \bibinfo{author}{Weese, J.}, \bibinfo{author}{Pinto, K.}, \bibinfo{author}{Karim, R.}, \bibinfo{author}{Schaeffter, T.}, \bibinfo{author}{Razavi, R.}, \bibinfo{author}{Rhode, K.S.}, \bibinfo{year}{2014}.
\newblock \bibinfo{title}{Left atrial segmentation challenge: A unified benchmarking framework}, in: \bibinfo{editor}{Camara, O.}, \bibinfo{editor}{Mansi, T.}, \bibinfo{editor}{Pop, M.}, \bibinfo{editor}{Rhode, K.}, \bibinfo{editor}{Sermesant, M.}, \bibinfo{editor}{Young, A.} (Eds.), \bibinfo{booktitle}{Statistical Atlases and Computational Models of the Heart. Imaging and Modelling Challenges}, \bibinfo{publisher}{Springer Berlin Heidelberg}, \bibinfo{address}{Berlin, Heidelberg}. pp. \bibinfo{pages}{1--13}.
\bibitem[{Ulrich et~al.(2023)Ulrich, Isensee, Wald, Zenk, Baumgartner and Maier-Hein}]{10.1007/978-3-031-43898-1_62}
\bibinfo{author}{Ulrich, C.}, \bibinfo{author}{Isensee, F.}, \bibinfo{author}{Wald, T.}, \bibinfo{author}{Zenk, M.}, \bibinfo{author}{Baumgartner, M.}, \bibinfo{author}{Maier-Hein, K.H.}, \bibinfo{year}{2023}.
\newblock \bibinfo{title}{Multitalent: A multi-dataset approach to medical image segmentation}, in: \bibinfo{editor}{Greenspan, H.}, \bibinfo{editor}{Madabhushi, A.}, \bibinfo{editor}{Mousavi, P.}, \bibinfo{editor}{Salcudean, S.}, \bibinfo{editor}{Duncan, J.}, \bibinfo{editor}{Syeda-Mahmood, T.}, \bibinfo{editor}{Taylor, R.} (Eds.), \bibinfo{booktitle}{Medical Image Computing and Computer Assisted Intervention -- MICCAI 2023}, \bibinfo{publisher}{Springer Nature Switzerland}, \bibinfo{address}{Cham}. pp. \bibinfo{pages}{648--658}.
\bibitem[{Valanarasu et~al.(2024)Valanarasu, Tang, Yang, Xu, Zhao, Li, Patel, Landman, Xu, He and Nath}]{pmlr-v250-valanarasu24a}
\bibinfo{author}{Valanarasu, J.M.J.}, \bibinfo{author}{Tang, Y.}, \bibinfo{author}{Yang, D.}, \bibinfo{author}{Xu, Z.}, \bibinfo{author}{Zhao, C.}, \bibinfo{author}{Li, W.}, \bibinfo{author}{Patel, V.M.}, \bibinfo{author}{Landman, B.A.}, \bibinfo{author}{Xu, D.}, \bibinfo{author}{He, Y.}, \bibinfo{author}{Nath, V.}, \bibinfo{year}{2024}.
\newblock \bibinfo{title}{Disruptive autoencoders: Leveraging low-level features for 3d medical image pre-training}, in: \bibinfo{editor}{Burgos, N.}, \bibinfo{editor}{Petitjean, C.}, \bibinfo{editor}{Vakalopoulou, M.}, \bibinfo{editor}{Christodoulidis, S.}, \bibinfo{editor}{Coupe, P.}, \bibinfo{editor}{Delingette, H.}, \bibinfo{editor}{Lartizien, C.}, \bibinfo{editor}{Mateus, D.} (Eds.), \bibinfo{booktitle}{Proceedings of The 7nd International Conference on Medical Imaging with Deep Learning}, \bibinfo{publisher}{PMLR}. pp. \bibinfo{pages}{1553--1570}.
\newblock \URLprefix \url{https://proceedings.mlr.press/v250/valanarasu24a.html}.
\bibitem[{Wang et~al.(2023)Wang, Wu, Luo, Liu, Li and Zhang}]{wang2023misfm3dmedicalimage}
\bibinfo{author}{Wang, G.}, \bibinfo{author}{Wu, J.}, \bibinfo{author}{Luo, X.}, \bibinfo{author}{Liu, X.}, \bibinfo{author}{Li, K.}, \bibinfo{author}{Zhang, S.}, \bibinfo{year}{2023}.
\newblock \bibinfo{title}{Mis-fm: 3d medical image segmentation using foundation models pretrained on a large-scale unannotated dataset}.
\newblock \URLprefix \url{https://arxiv.org/abs/2306.16925}, \href{http://arxiv.org/abs/2306.16925}{{\tt arXiv:2306.16925}}.
\bibitem[{Wang et~al.(2024)Wang, Lin, Ding and Li}]{10.1007/978-3-031-72114-4_61}
\bibinfo{author}{Wang, H.}, \bibinfo{author}{Lin, Y.}, \bibinfo{author}{Ding, X.}, \bibinfo{author}{Li, X.}, \bibinfo{year}{2024}.
\newblock \bibinfo{title}{Tri-plane mamba: Efficiently adapting segment anything model for 3d medical images}, in: \bibinfo{editor}{Linguraru, M.G.}, \bibinfo{editor}{Dou, Q.}, \bibinfo{editor}{Feragen, A.}, \bibinfo{editor}{Giannarou, S.}, \bibinfo{editor}{Glocker, B.}, \bibinfo{editor}{Lekadir, K.}, \bibinfo{editor}{Schnabel, J.A.} (Eds.), \bibinfo{booktitle}{Medical Image Computing and Computer Assisted Intervention -- MICCAI 2024}, \bibinfo{publisher}{Springer Nature Switzerland}, \bibinfo{address}{Cham}. pp. \bibinfo{pages}{636--646}.
\bibitem[{Wang et~al.(2021a)Wang, Chen, Ding, Li, Yu and Zha}]{wang2021transbtsmultimodalbraintumor}
\bibinfo{author}{Wang, W.}, \bibinfo{author}{Chen, C.}, \bibinfo{author}{Ding, M.}, \bibinfo{author}{Li, J.}, \bibinfo{author}{Yu, H.}, \bibinfo{author}{Zha, S.}, \bibinfo{year}{2021}a.
\newblock \bibinfo{title}{Transbts: Multimodal brain tumor segmentation using transformer}.
\newblock \URLprefix \url{https://arxiv.org/abs/2103.04430}, \href{http://arxiv.org/abs/2103.04430}{{\tt arXiv:2103.04430}}.
\bibitem[{Wang et~al.(2021b)Wang, Chen, Ding, Yu, Zha and Li}]{10.1007/978-3-030-87193-2_11}
\bibinfo{author}{Wang, W.}, \bibinfo{author}{Chen, C.}, \bibinfo{author}{Ding, M.}, \bibinfo{author}{Yu, H.}, \bibinfo{author}{Zha, S.}, \bibinfo{author}{Li, J.}, \bibinfo{year}{2021}b.
\newblock \bibinfo{title}{Transbts: Multimodal brain tumor segmentation using transformer}, in: \bibinfo{editor}{de~Bruijne, M.}, \bibinfo{editor}{Cattin, P.C.}, \bibinfo{editor}{Cotin, S.}, \bibinfo{editor}{Padoy, N.}, \bibinfo{editor}{Speidel, S.}, \bibinfo{editor}{Zheng, Y.}, \bibinfo{editor}{Essert, C.} (Eds.), \bibinfo{booktitle}{Medical Image Computing and Computer Assisted Intervention -- MICCAI 2021}, \bibinfo{publisher}{Springer International Publishing}, \bibinfo{address}{Cham}. pp. \bibinfo{pages}{109--119}.
\bibitem[{Wasserthal et~al.(2023)Wasserthal, Breit, Meyer, Pradella, Hinck, Sauter, Heye, Boll, Cyriac, Yang, Bach and Segeroth}]{doi:10.1148/ryai.230024}
\bibinfo{author}{Wasserthal, J.}, \bibinfo{author}{Breit, H.C.}, \bibinfo{author}{Meyer, M.T.}, \bibinfo{author}{Pradella, M.}, \bibinfo{author}{Hinck, D.}, \bibinfo{author}{Sauter, A.W.}, \bibinfo{author}{Heye, T.}, \bibinfo{author}{Boll, D.T.}, \bibinfo{author}{Cyriac, J.}, \bibinfo{author}{Yang, S.}, \bibinfo{author}{Bach, M.}, \bibinfo{author}{Segeroth, M.}, \bibinfo{year}{2023}.
\newblock \bibinfo{title}{Totalsegmentator: Robust segmentation of 104 anatomic structures in ct images}.
\newblock \bibinfo{journal}{Radiology: Artificial Intelligence} \bibinfo{volume}{5}, \bibinfo{pages}{e230024}.
\newblock \URLprefix \url{https://doi.org/10.1148/ryai.230024}, \DOIprefix\doi{10.1148/ryai.230024}, \href{http://arxiv.org/abs/https://doi.org/10.1148/ryai.230024}{{\tt arXiv:https://doi.org/10.1148/ryai.230024}}.
\bibitem[{Xie et~al.(2021a)Xie, Zhang, Shen and Xia}]{xie2021cotrefficientlybridgingcnn}
\bibinfo{author}{Xie, Y.}, \bibinfo{author}{Zhang, J.}, \bibinfo{author}{Shen, C.}, \bibinfo{author}{Xia, Y.}, \bibinfo{year}{2021}a.
\newblock \bibinfo{title}{Cotr: Efficiently bridging cnn and transformer for 3d medical image segmentation}.
\newblock \URLprefix \url{https://arxiv.org/abs/2103.03024}, \href{http://arxiv.org/abs/2103.03024}{{\tt arXiv:2103.03024}}.
\bibitem[{Xie et~al.(2021b)Xie, Zhang, Shen and Xia}]{10.1007/978-3-030-87199-4_16}
\bibinfo{author}{Xie, Y.}, \bibinfo{author}{Zhang, J.}, \bibinfo{author}{Shen, C.}, \bibinfo{author}{Xia, Y.}, \bibinfo{year}{2021}b.
\newblock \bibinfo{title}{Cotr: Efficiently bridging cnn and transformer for 3d medical image segmentation}, in: \bibinfo{editor}{de~Bruijne, M.}, \bibinfo{editor}{Cattin, P.C.}, \bibinfo{editor}{Cotin, S.}, \bibinfo{editor}{Padoy, N.}, \bibinfo{editor}{Speidel, S.}, \bibinfo{editor}{Zheng, Y.}, \bibinfo{editor}{Essert, C.} (Eds.), \bibinfo{booktitle}{Medical Image Computing and Computer Assisted Intervention -- MICCAI 2021}, \bibinfo{publisher}{Springer International Publishing}, \bibinfo{address}{Cham}. pp. \bibinfo{pages}{171--180}.
\bibitem[{Xing et~al.(2023)Xing, Wan, Fu, Yang and Zhu}]{xing2023diffunetdiffusionembeddednetwork}
\bibinfo{author}{Xing, Z.}, \bibinfo{author}{Wan, L.}, \bibinfo{author}{Fu, H.}, \bibinfo{author}{Yang, G.}, \bibinfo{author}{Zhu, L.}, \bibinfo{year}{2023}.
\newblock \bibinfo{title}{Diff-unet: A diffusion embedded network for volumetric segmentation}.
\newblock \URLprefix \url{https://arxiv.org/abs/2303.10326}, \href{http://arxiv.org/abs/2303.10326}{{\tt arXiv:2303.10326}}.
\bibitem[{Xiong et~al.(2021)Xiong, Xia, Hu, Huang, Bian, Zheng, Vesal, Ravikumar, Maier, Yang, Heng, Ni, Li, Tong, Si, Puybareau, Khoudli, Graud, Chen, Bai, Rueckert, Xu, Zhuang, Luo, Jia, Sermesant, Liu, Wang, Borra, Masci, Corsi, {de Vente}, Veta, Karim, Preetha, Engelhardt, Qiao, Wang, Tao, Nuez-Garcia, Camara, Savioli, Lamata and Zhao}]{XIONG2021101832}
\bibinfo{author}{Xiong, Z.}, \bibinfo{author}{Xia, Q.}, \bibinfo{author}{Hu, Z.}, \bibinfo{author}{Huang, N.}, \bibinfo{author}{Bian, C.}, \bibinfo{author}{Zheng, Y.}, \bibinfo{author}{Vesal, S.}, \bibinfo{author}{Ravikumar, N.}, \bibinfo{author}{Maier, A.}, \bibinfo{author}{Yang, X.}, \bibinfo{author}{Heng, P.A.}, \bibinfo{author}{Ni, D.}, \bibinfo{author}{Li, C.}, \bibinfo{author}{Tong, Q.}, \bibinfo{author}{Si, W.}, \bibinfo{author}{Puybareau, E.}, \bibinfo{author}{Khoudli, Y.}, \bibinfo{author}{Graud, T.}, \bibinfo{author}{Chen, C.}, \bibinfo{author}{Bai, W.}, \bibinfo{author}{Rueckert, D.}, \bibinfo{author}{Xu, L.}, \bibinfo{author}{Zhuang, X.}, \bibinfo{author}{Luo, X.}, \bibinfo{author}{Jia, S.}, \bibinfo{author}{Sermesant, M.}, \bibinfo{author}{Liu, Y.}, \bibinfo{author}{Wang, K.}, \bibinfo{author}{Borra, D.}, \bibinfo{author}{Masci, A.}, \bibinfo{author}{Corsi, C.}, \bibinfo{author}{{de Vente}, C.}, \bibinfo{author}{Veta, M.}, \bibinfo{author}{Karim, R.}, \bibinfo{author}{Preetha, C.J.},
  \bibinfo{author}{Engelhardt, S.}, \bibinfo{author}{Qiao, M.}, \bibinfo{author}{Wang, Y.}, \bibinfo{author}{Tao, Q.}, \bibinfo{author}{Nuez-Garcia, M.}, \bibinfo{author}{Camara, O.}, \bibinfo{author}{Savioli, N.}, \bibinfo{author}{Lamata, P.}, \bibinfo{author}{Zhao, J.}, \bibinfo{year}{2021}.
\newblock \bibinfo{title}{A global benchmark of algorithms for segmenting the left atrium from late gadolinium-enhanced cardiac magnetic resonance imaging}.
\newblock \bibinfo{journal}{Medical Image Analysis} \bibinfo{volume}{67}, \bibinfo{pages}{101832}.
\newblock \URLprefix \url{https://www.sciencedirect.com/science/article/pii/S1361841520301961}, \DOIprefix\doi{https://doi.org/10.1016/j.media.2020.101832}.
\bibitem[{Xu et~al.(2021)Xu, Lu, Wang, Luo, Jayender, Ma, Zheng and Li}]{10.1007/978-3-030-87193-2_1}
\bibinfo{author}{Xu, Z.}, \bibinfo{author}{Lu, D.}, \bibinfo{author}{Wang, Y.}, \bibinfo{author}{Luo, J.}, \bibinfo{author}{Jayender, J.}, \bibinfo{author}{Ma, K.}, \bibinfo{author}{Zheng, Y.}, \bibinfo{author}{Li, X.}, \bibinfo{year}{2021}.
\newblock \bibinfo{title}{Noisy labels are treasure: Mean-teacher-assisted confident learning for hepatic vessel segmentation}, in: \bibinfo{editor}{de~Bruijne, M.}, \bibinfo{editor}{Cattin, P.C.}, \bibinfo{editor}{Cotin, S.}, \bibinfo{editor}{Padoy, N.}, \bibinfo{editor}{Speidel, S.}, \bibinfo{editor}{Zheng, Y.}, \bibinfo{editor}{Essert, C.} (Eds.), \bibinfo{booktitle}{Medical Image Computing and Computer Assisted Intervention -- MICCAI 2021}, \bibinfo{publisher}{Springer International Publishing}, \bibinfo{address}{Cham}. pp. \bibinfo{pages}{3--13}.
\bibitem[{Yang et~al.(2024)Yang, Musio, Ma, Juchler, Paetzold, Al-Maskari, Hher, Li, Hamamci, Sekuboyina, Shit, Huang, Prabhakar, de~la Rosa, Waldmannstetter, Kofler, Navarro, Menten, Ezhov, Rueckert, Vos, Ruigrok, Velthuis, Kuijf, Hmmerli, Wurster, Bijlenga, Westphal, Bisschop, Colombo, Baazaoui, Makmur, Hallinan, Wiestler, Kirschke, Wiest, Montagnon, Letourneau-Guillon, Galdran, Galati, Falcetta, Zuluaga, Lin, Zhao, Zhang, Ra, Hwang, Park, Chen, Wodzinski, Mller, Shi, Liu, Ma, Yalin, Hamadache, Salvi, Llado, Estrada, Abramova, Giancardo, Oliver, Liu, Huang, Cui, Lin, Liu, Zhu, Patel, Tutino, Orouskhani, Wang, Mossa-Basha, Zhu, Rokuss, Kirchhoff, Disch, Holzschuh, Isensee, Maier-Hein, Sato, Hirsch, Wegener and Menze}]{topcowchallenge}
\bibinfo{author}{Yang, K.}, \bibinfo{author}{Musio, F.}, \bibinfo{author}{Ma, Y.}, \bibinfo{author}{Juchler, N.}, \bibinfo{author}{Paetzold, J.C.}, \bibinfo{author}{Al-Maskari, R.}, \bibinfo{author}{Hher, L.}, \bibinfo{author}{Li, H.B.}, \bibinfo{author}{Hamamci, I.E.}, \bibinfo{author}{Sekuboyina, A.}, \bibinfo{author}{Shit, S.}, \bibinfo{author}{Huang, H.}, \bibinfo{author}{Prabhakar, C.}, \bibinfo{author}{de~la Rosa, E.}, \bibinfo{author}{Waldmannstetter, D.}, \bibinfo{author}{Kofler, F.}, \bibinfo{author}{Navarro, F.}, \bibinfo{author}{Menten, M.}, \bibinfo{author}{Ezhov, I.}, \bibinfo{author}{Rueckert, D.}, \bibinfo{author}{Vos, I.}, \bibinfo{author}{Ruigrok, Y.}, \bibinfo{author}{Velthuis, B.}, \bibinfo{author}{Kuijf, H.}, \bibinfo{author}{Hmmerli, J.}, \bibinfo{author}{Wurster, C.}, \bibinfo{author}{Bijlenga, P.}, \bibinfo{author}{Westphal, L.}, \bibinfo{author}{Bisschop, J.}, \bibinfo{author}{Colombo, E.}, \bibinfo{author}{Baazaoui, H.}, \bibinfo{author}{Makmur, A.}, \bibinfo{author}{Hallinan, J.},
  \bibinfo{author}{Wiestler, B.}, \bibinfo{author}{Kirschke, J.S.}, \bibinfo{author}{Wiest, R.}, \bibinfo{author}{Montagnon, E.}, \bibinfo{author}{Letourneau-Guillon, L.}, \bibinfo{author}{Galdran, A.}, \bibinfo{author}{Galati, F.}, \bibinfo{author}{Falcetta, D.}, \bibinfo{author}{Zuluaga, M.A.}, \bibinfo{author}{Lin, C.}, \bibinfo{author}{Zhao, H.}, \bibinfo{author}{Zhang, Z.}, \bibinfo{author}{Ra, S.}, \bibinfo{author}{Hwang, J.}, \bibinfo{author}{Park, H.}, \bibinfo{author}{Chen, J.}, \bibinfo{author}{Wodzinski, M.}, \bibinfo{author}{Mller, H.}, \bibinfo{author}{Shi, P.}, \bibinfo{author}{Liu, W.}, \bibinfo{author}{Ma, T.}, \bibinfo{author}{Yalin, C.}, \bibinfo{author}{Hamadache, R.E.}, \bibinfo{author}{Salvi, J.}, \bibinfo{author}{Llado, X.}, \bibinfo{author}{Estrada, U.M.L.T.}, \bibinfo{author}{Abramova, V.}, \bibinfo{author}{Giancardo, L.}, \bibinfo{author}{Oliver, A.}, \bibinfo{author}{Liu, J.}, \bibinfo{author}{Huang, H.}, \bibinfo{author}{Cui, Y.}, \bibinfo{author}{Lin, Z.}, \bibinfo{author}{Liu,
  Y.}, \bibinfo{author}{Zhu, S.}, \bibinfo{author}{Patel, T.R.}, \bibinfo{author}{Tutino, V.M.}, \bibinfo{author}{Orouskhani, M.}, \bibinfo{author}{Wang, H.}, \bibinfo{author}{Mossa-Basha, M.}, \bibinfo{author}{Zhu, C.}, \bibinfo{author}{Rokuss, M.R.}, \bibinfo{author}{Kirchhoff, Y.}, \bibinfo{author}{Disch, N.}, \bibinfo{author}{Holzschuh, J.}, \bibinfo{author}{Isensee, F.}, \bibinfo{author}{Maier-Hein, K.}, \bibinfo{author}{Sato, Y.}, \bibinfo{author}{Hirsch, S.}, \bibinfo{author}{Wegener, S.}, \bibinfo{author}{Menze, B.}, \bibinfo{year}{2024}.
\newblock \bibinfo{title}{Benchmarking the cow with the topcow challenge: Topology-aware anatomical segmentation of the circle of willis for cta and mra}.
\newblock \URLprefix \url{https://arxiv.org/abs/2312.17670}, \href{http://arxiv.org/abs/2312.17670}{{\tt arXiv:2312.17670}}.
\bibitem[{Ye et~al.(2023)Ye, Xie, Zhang, Chen and Xia}]{10.1007/978-3-031-43898-1_49}
\bibinfo{author}{Ye, Y.}, \bibinfo{author}{Xie, Y.}, \bibinfo{author}{Zhang, J.}, \bibinfo{author}{Chen, Z.}, \bibinfo{author}{Xia, Y.}, \bibinfo{year}{2023}.
\newblock \bibinfo{title}{Uniseg: A prompt-driven universal segmentation model as well as a strong representation learner}, in: \bibinfo{editor}{Greenspan, H.}, \bibinfo{editor}{Madabhushi, A.}, \bibinfo{editor}{Mousavi, P.}, \bibinfo{editor}{Salcudean, S.}, \bibinfo{editor}{Duncan, J.}, \bibinfo{editor}{Syeda-Mahmood, T.}, \bibinfo{editor}{Taylor, R.} (Eds.), \bibinfo{booktitle}{Medical Image Computing and Computer Assisted Intervention -- MICCAI 2023}, \bibinfo{publisher}{Springer Nature Switzerland}, \bibinfo{address}{Cham}. pp. \bibinfo{pages}{508--518}.
\bibitem[{Zhang et~al.(2021)Zhang, Xie, Xia and Shen}]{Zhang_2021_CVPR}
\bibinfo{author}{Zhang, J.}, \bibinfo{author}{Xie, Y.}, \bibinfo{author}{Xia, Y.}, \bibinfo{author}{Shen, C.}, \bibinfo{year}{2021}.
\newblock \bibinfo{title}{Dodnet: Learning to segment multi-organ and tumors from multiple partially labeled datasets}, in: \bibinfo{booktitle}{Proceedings of the IEEE/CVF Conference on Computer Vision and Pattern Recognition (CVPR)}, pp. \bibinfo{pages}{1195--1204}.
\bibitem[{Zhang et~al.(2025)Zhang, Ou, Basaran, Visentin, Qiao, Gu, Matthews, Liu, Ye and Bai}]{10879789}
\bibinfo{author}{Zhang, X.}, \bibinfo{author}{Ou, N.}, \bibinfo{author}{Basaran, B.D.}, \bibinfo{author}{Visentin, M.}, \bibinfo{author}{Qiao, M.}, \bibinfo{author}{Gu, R.}, \bibinfo{author}{Matthews, P.M.}, \bibinfo{author}{Liu, Y.}, \bibinfo{author}{Ye, C.}, \bibinfo{author}{Bai, W.}, \bibinfo{year}{2025}.
\newblock \bibinfo{title}{A foundation model for lesion segmentation on brain mri with mixture of modality experts}.
\newblock \bibinfo{journal}{IEEE Transactions on Medical Imaging} , \bibinfo{pages}{1--1}\DOIprefix\doi{10.1109/TMI.2025.3540809}.
\bibitem[{Zhao et~al.(2025)Zhao, Zhang, Wu, Zhang, Zhang, Wang and Xie}]{zhao2025modelrulealluniversal}
\bibinfo{author}{Zhao, Z.}, \bibinfo{author}{Zhang, Y.}, \bibinfo{author}{Wu, C.}, \bibinfo{author}{Zhang, X.}, \bibinfo{author}{Zhang, Y.}, \bibinfo{author}{Wang, Y.}, \bibinfo{author}{Xie, W.}, \bibinfo{year}{2025}.
\newblock \bibinfo{title}{One model to rule them all: Towards universal segmentation for medical images with text prompts}.
\newblock \URLprefix \url{https://arxiv.org/abs/2312.17183}, \href{http://arxiv.org/abs/2312.17183}{{\tt arXiv:2312.17183}}.
\bibitem[{Zheng et~al.(2017)Zheng, Chu, Belav, Ibragimov, Korez, Vrtovec, Hutt, Everson, Meakin, Andrade, Glocker, Chen, Dou, Heng, Wang, Forsberg, Neubert, Fripp, Urschler, Stern, Wimmer, Novikov, Cheng, Armbrecht, Felsenberg and Li}]{ZHENG2017327}
\bibinfo{author}{Zheng, G.}, \bibinfo{author}{Chu, C.}, \bibinfo{author}{Belav, D.L.}, \bibinfo{author}{Ibragimov, B.}, \bibinfo{author}{Korez, R.}, \bibinfo{author}{Vrtovec, T.}, \bibinfo{author}{Hutt, H.}, \bibinfo{author}{Everson, R.}, \bibinfo{author}{Meakin, J.}, \bibinfo{author}{Andrade, I.L.}, \bibinfo{author}{Glocker, B.}, \bibinfo{author}{Chen, H.}, \bibinfo{author}{Dou, Q.}, \bibinfo{author}{Heng, P.A.}, \bibinfo{author}{Wang, C.}, \bibinfo{author}{Forsberg, D.}, \bibinfo{author}{Neubert, A.}, \bibinfo{author}{Fripp, J.}, \bibinfo{author}{Urschler, M.}, \bibinfo{author}{Stern, D.}, \bibinfo{author}{Wimmer, M.}, \bibinfo{author}{Novikov, A.A.}, \bibinfo{author}{Cheng, H.}, \bibinfo{author}{Armbrecht, G.}, \bibinfo{author}{Felsenberg, D.}, \bibinfo{author}{Li, S.}, \bibinfo{year}{2017}.
\newblock \bibinfo{title}{Evaluation and comparison of 3d intervertebral disc localization and segmentation methods for 3d t2 mr data: A grand challenge}.
\newblock \bibinfo{journal}{Medical Image Analysis} \bibinfo{volume}{35}, \bibinfo{pages}{327--344}.
\newblock \URLprefix \url{https://www.sciencedirect.com/science/article/pii/S1361841516301530}, \DOIprefix\doi{https://doi.org/10.1016/j.media.2016.08.005}.
\bibitem[{Zheng et~al.(2021a)Zheng, Lu, Zhao, Zhu, Luo, Wang, Fu, Feng, Xiang, Torr and Zhang}]{9578646}
\bibinfo{author}{Zheng, S.}, \bibinfo{author}{Lu, J.}, \bibinfo{author}{Zhao, H.}, \bibinfo{author}{Zhu, X.}, \bibinfo{author}{Luo, Z.}, \bibinfo{author}{Wang, Y.}, \bibinfo{author}{Fu, Y.}, \bibinfo{author}{Feng, J.}, \bibinfo{author}{Xiang, T.}, \bibinfo{author}{Torr, P.H.}, \bibinfo{author}{Zhang, L.}, \bibinfo{year}{2021}a.
\newblock \bibinfo{title}{Rethinking semantic segmentation from a sequence-to-sequence perspective with transformers}, in: \bibinfo{booktitle}{2021 IEEE/CVF Conference on Computer Vision and Pattern Recognition (CVPR)}, pp. \bibinfo{pages}{6877--6886}.
\newblock \DOIprefix\doi{10.1109/CVPR46437.2021.00681}.
\bibitem[{Zheng et~al.(2021b)Zheng, Lu, Zhao, Zhu, Luo, Wang, Fu, Feng, Xiang, Torr and Zhang}]{zheng2021rethinkingsemanticsegmentationsequencetosequence}
\bibinfo{author}{Zheng, S.}, \bibinfo{author}{Lu, J.}, \bibinfo{author}{Zhao, H.}, \bibinfo{author}{Zhu, X.}, \bibinfo{author}{Luo, Z.}, \bibinfo{author}{Wang, Y.}, \bibinfo{author}{Fu, Y.}, \bibinfo{author}{Feng, J.}, \bibinfo{author}{Xiang, T.}, \bibinfo{author}{Torr, P.H.S.}, \bibinfo{author}{Zhang, L.}, \bibinfo{year}{2021}b.
\newblock \bibinfo{title}{Rethinking semantic segmentation from a sequence-to-sequence perspective with transformers}.
\newblock \URLprefix \url{https://arxiv.org/abs/2012.15840}, \href{http://arxiv.org/abs/2012.15840}{{\tt arXiv:2012.15840}}.
\bibitem[{Zhou et~al.(2023a)Zhou, Guo, Zhang, Han, Yu, Wang and Yu}]{10183842}
\bibinfo{author}{Zhou, H.Y.}, \bibinfo{author}{Guo, J.}, \bibinfo{author}{Zhang, Y.}, \bibinfo{author}{Han, X.}, \bibinfo{author}{Yu, L.}, \bibinfo{author}{Wang, L.}, \bibinfo{author}{Yu, Y.}, \bibinfo{year}{2023}a.
\newblock \bibinfo{title}{nnformer: Volumetric medical image segmentation via a 3d transformer}.
\newblock \bibinfo{journal}{IEEE Transactions on Image Processing} \bibinfo{volume}{32}, \bibinfo{pages}{4036--4045}.
\newblock \DOIprefix\doi{10.1109/TIP.2023.3293771}.
\bibitem[{Zhou et~al.(2023b)Zhou, Guo, Zhang, Han, Yu, Wang and Yu}]{nnFormer_ieee}
\bibinfo{author}{Zhou, H.Y.}, \bibinfo{author}{Guo, J.}, \bibinfo{author}{Zhang, Y.}, \bibinfo{author}{Han, X.}, \bibinfo{author}{Yu, L.}, \bibinfo{author}{Wang, L.}, \bibinfo{author}{Yu, Y.}, \bibinfo{year}{2023}b.
\newblock \bibinfo{title}{nnformer: Volumetric medical image segmentation via a 3d transformer}.
\newblock \bibinfo{journal}{IEEE Transactions on Image Processing} \bibinfo{volume}{32}, \bibinfo{pages}{4036--4045}.
\newblock \DOIprefix\doi{10.1109/TIP.2023.3293771}.
\bibitem[{Zhou et~al.(2022)Zhou, Guo, Zhang, Yu, Wang and Yu}]{zhou2022nnformerinterleavedtransformervolumetric}
\bibinfo{author}{Zhou, H.Y.}, \bibinfo{author}{Guo, J.}, \bibinfo{author}{Zhang, Y.}, \bibinfo{author}{Yu, L.}, \bibinfo{author}{Wang, L.}, \bibinfo{author}{Yu, Y.}, \bibinfo{year}{2022}.
\newblock \bibinfo{title}{nnformer: Interleaved transformer for volumetric segmentation}.
\newblock \URLprefix \url{https://arxiv.org/abs/2109.03201}, \href{http://arxiv.org/abs/2109.03201}{{\tt arXiv:2109.03201}}.
\bibitem[{Zhou et~al.(2018)Zhou, Siddiquee, Tajbakhsh and Liang}]{zhou2018unetnestedunetarchitecture}
\bibinfo{author}{Zhou, Z.}, \bibinfo{author}{Siddiquee, M.M.R.}, \bibinfo{author}{Tajbakhsh, N.}, \bibinfo{author}{Liang, J.}, \bibinfo{year}{2018}.
\newblock \bibinfo{title}{Unet++: A nested u-net architecture for medical image segmentation}.
\newblock \URLprefix \url{https://arxiv.org/abs/1807.10165}, \href{http://arxiv.org/abs/1807.10165}{{\tt arXiv:1807.10165}}.
\bibitem[{Zhou et~al.(2020)Zhou, Siddiquee, Tajbakhsh and Liang}]{8932614}
\bibinfo{author}{Zhou, Z.}, \bibinfo{author}{Siddiquee, M.M.R.}, \bibinfo{author}{Tajbakhsh, N.}, \bibinfo{author}{Liang, J.}, \bibinfo{year}{2020}.
\newblock \bibinfo{title}{Unet++: Redesigning skip connections to exploit multiscale features in image segmentation}.
\newblock \bibinfo{journal}{IEEE Transactions on Medical Imaging} \bibinfo{volume}{39}, \bibinfo{pages}{1856--1867}.
\newblock \DOIprefix\doi{10.1109/TMI.2019.2959609}.

\end{thebibliography}

\end{document}